\documentclass[usenatbib,usegraphicx]{mn2e}
\usepackage{fleqn}
\usepackage{float}
\usepackage{color}
\usepackage{amsmath}
\usepackage{adjustbox}
\usepackage{multirow}
\usepackage{longtable}

\usepackage{url}

\usepackage{ae,aecompl}
\usepackage{amssymb}	
\usepackage{amsbsy}
\usepackage{color}
\usepackage{txfonts}
\usepackage{times}
\usepackage{morefloats}

\usepackage{amssymb}
\newcommand{\be}{\begin{equation}}
\newcommand{\ee}{\end{equation}}

\renewcommand{\vec}[1]{ {\bf #1} }
\newcommand{\dd}{{\rm d}}

\title[The GADGET-4 simulation code]{Simulating cosmic structure
  formation with the GADGET-4 code} \author[V.~Springel
et al.]  {\parbox{17cm}{Volker~Springel$^{1}$\thanks{E-mail: vspringel@mpa-garching.mpg.de}, 
    R\"udiger~Pakmor$^1$, 
    Oliver Zier$^1$, and 
    Martin Reinecke$^1$
    }\vspace*{0.2cm}\\
  $^1$Max-Planck-Institut f\"{u}r Astrophysik,
  Karl-Schwarzschild-Stra\ss{}e 1, 85740 Garching
  bei M\"{u}nchen, Germany\\
}

\begin{document}
\pagerange{\pageref{firstpage}--\pageref{lastpage}}
\pubyear{2013}

\maketitle

\label{firstpage}

\begin{abstract}
  Numerical methods have become a powerful tool for research in
  astrophysics, but their utility depends critically on the
  availability of suitable simulation codes. This calls for continuous
  efforts in code development, which is necessitated also by the
  rapidly evolving technology underlying today's computing
  hardware. Here we discuss recent methodological progress in the
  {\small GADGET} code, which has been widely applied in cosmic
  structure formation over the past two decades. The new version
  offers improvements in force accuracy, in time-stepping, in
  adaptivity to a large dynamic range in timescales, in computational
  efficiency, and in parallel scalability through a special
  MPI/shared-memory parallelization and communication strategy, and a
  more-sophisticated domain decomposition algorithm.  A manifestly
  momentum conserving fast multipole method (FMM) can be employed as
  an alternative to the one-sided TreePM gravity solver introduced in
  earlier versions. Two different flavours of smoothed particle
  hydrodynamics, a classic entropy-conserving formulation and a
  pressure-based approach, are supported for dealing with gaseous
  flows.  The code is able to cope with very large problem sizes, thus
  allowing accurate predictions for cosmic structure formation in
  support of future precision tests of cosmology, and at the same time
  is well adapted to high dynamic range zoom-calculations with extreme
  variability of the particle number density in the simulated
  volume. The {\small GADGET-4} code is publicly released to the
  community and contains infrastructure for on-the-fly group and
  substructure finding and tracking, as well as merger tree building,
  a simple model for radiative cooling and star formation, a high
  dynamic range power spectrum estimator, and an initial conditions
  generator based on second-order Lagrangian perturbation theory.
\end{abstract}


\section{Introduction}

Numerical simulations allow detailed studies of non-linear structure
formation and connect the simple high-redshift Universe with the
complex structure we see around us today \citep{Efstathiou:1985aa,
  Navarro:1997aa, Jenkins:2001aa, Springel:2005ab}. This powerful
technique has become an important pillar in astrophysical research,
decisively shaping our understanding of the coupled dynamics of dark
matter and baryonic physics \citep[see][for recent
reviews]{Naab:2017aa, Vogelsberger:2020aa}. Consequently, substantial
work has been invested in developing efficient numerical methods and
appropriate codes to model cosmic structure formation.

Sharing such codes publicly within the community has accelerated the
widespread adoption of numerical techniques and lowered the entry
barrier for new researchers or research groups entering the field. At
the same time, it is clear that code development needs to continue to
improve the accuracy and physical fidelity of the modelling
techniques. In fact, it can well be argued that efforts in this
direction need to be stepped up further, otherwise the rapidly growing
power of modern high performance computing facilities can not be used
in full for future astrophysical research. Modern astrophysical codes
have also grown in complexity to a point where single person efforts,
which traditionally have often been the mode of creation of new codes,
become increasingly more difficult, and thus are best replaced by more
open, team-driven development models.

In this spirit, we here discuss a major update of the {\small GADGET}
code \citep{Springel:2001aa}, which has seen wide-spread use in
structure formation and galaxy formation over the past two
decades. The last extensive description of the code in the literature
has been given for {\small GADGET-2} \citep{Springel:2005aa}, even
though the newer version, {\small GADGET-3}, first written for the
Aquarius project \citep{Springel:2008aa}, has been used for a large
amount of simulation work, and has also been the basis for the
development of numerous modified codes spawned from it, such as
{\small AREPO} \citep{Springel:2010ab, Weinberger:2020aa}, {\small
  L-GADGET-3} \citep{Angulo:2012aa, Angulo:2020aa}, {\small
  StellarGADGET} \citep{Pakmor:2012aa}, {\small MG-GADGET}
\citep{Puchwein:2013aa}, {\small GIZMO} \citep{Hopkins:2015aa},
{\small KETJU} \citep{Rantala:2017ab}, {\small ME-GADGET}
\citep{Zhang:2018ab}, {\small AX-GADGET} \citep{Nori:2018aa}, {\small
  MP-GADGET} \citep{Huang:2018aa}, {\small SPHGAL} \citep{Hu:2014aa},
{\small OpenACC-GADGET3} \citep{Ragagnin:2020aa}, and others. This
flurry of development going back to {\small GADGET-3}, combined with
the lack of a comprehensive description of this particular version in
the literature, now makes it actually hard to precisely define what
one means when referring to {\small GADGET-3}.

Our main motivation for the present paper on {\small GADGET-4}, which
in part is meant to address this ambiguity, lies in the growing
scientific need for more accurate and larger simulations, which call
for calculations with larger statistical power and much higher
resolution.  These are in principle possible on modern machines,
especially due to the availability of higher degrees of parallel
execution capabilities.  However, simulations that address an ever
larger dynamic range necessitate more flexible and scalable
integration schemes that can better deal with multi-physics,
multi-scale calculations.

We thus want to improve the basic scalability of the {\small GADGET}
code, remove barriers to larger simulation sizes, and allow the code
to perform better under conditions of a high dynamic range in
timescales, thereby allowing more extreme zoom simulations to be
carried out efficiently. At the same time, we would like to improve
the accuracy where possible in the gravitational and hydrodynamical
force calculations. As a fourth, more practical goal, we want to
modernize the code architecture, getting rid of a number of obsolete
constructs and improving the readability and modularity of the code,
such that scientific users can more easily develop extensions for it.
Of course, many other development efforts in the field follow quite
similar goals. For example, cosmological N-body codes that have
recently been pushed into regimes of extremely large particle number
include {\small HACC} \citep{Habib:2016aa, Heitmann:2019ab}, {\small
  PKDGRAV} \citep{Potter:2017aa}, and {\small GREEM}
\citep{Ishiyama:2020aa}. Many other sophisticated N-body codes are
actively developed and used in the field to study cosmic structure
formation and galaxy evolution, for example {\small ART}
\citep{Kravtsov:1997aa}, {\small RAMSES} \citep{Teyssier:2002aa},
{\small GYRFALCON} \citep{Dehnen:2002aa}, {\small GOTPM}
\citep{Dubinski:2004aa}, {\small ENZO} \citep{Bryan:2014aa}, {\small
  CHANGA} \cite{Menon:2015aa}, {\small SWIFT} \citep{Schaller:2016aa},
{\small GASOLINE} \citep{Wadsley:2017ab}, and {\small ABACUS}
\citep{Garrison:2019aa}.  Some of them also feature advanced
hydrodynamical solvers and treatments of other physics besides
ordinary gravity.

A special emphasis for {\small GADGET-4}, that perhaps sets it apart
from other codes, is to push for a flexible multi-purpose code that is
not restricted to a narrow range of simulation types, but rather
prioritizes flexibility over optimisation for a special application
type.  To make progress towards these goals, we have implemented a
number of new numerical methods relative to the previous versions of
the {\small GADGET} code, including a new domain decomposition
algorithm that can better balance the computational work load, both
for gravity and hydrodynamics, while at the same time yielding good
memory-balance. The time integration can optionally employ a
hierarchical scheme that is much more elastic in time and allows an
effective decoupling of short-time scale dynamics embedded in a more
slowly evolving larger system. We have also added an optional
Cartesian Fast Multipole Method (FMM), which can be used as an
alternative to the one-sided tree approach in {\small GADGET}. In the
hydrodynamics sector, the code offers classic entropy-conserving
smoothed particle hydrodynamics (SPH) as a reference implementation,
but we have also included one of the recent proposals for formulations
of SPH that alleviate its problems at contact discontinuities. Note
that while particle-based hydrodynamics is in general less accurate
than mesh-based methods, its simplicity and robustness make it still
interesting and adequate for certain applications.  In particular, to
help researchers getting started with simulations of galaxy formation
physics, we include in {\small GADGET-4} simple cooling and star
formation prescriptions.

To allow more efficient use of shared memory nodes with a large number
of compute cores, parallelization in terms of a hybrid mode with a
mixture of MPI and direct shared-memory access (based on MPI-3) is
supported.  We also add convenient processing functionality in this
code, in the form of friends-of-friends (FOF) and {\small SUBFIND}
\citep{Springel:2001ac} group and substructure finders, and an
on-the-fly merger tree generator. A new variant of the substructure
finder, {\small SUBFIND-HBT}, takes past subhalo information into
account \citep[similar to][]{Han:2018aa}, thereby allowing a very
robust and computationally efficient tracking of substructures over
time, long after they have fallen into other halos. Importantly, these
implementations are capable of processing very large simulation sizes
with up to trillions of particles, as well as zoom calculations with
billions of particles per object. The group finding can not only be
done on ordinary particle time-slices (so-called snapshots), but also
on the particles encompassing the backwards lightcone of a fiducial
observer. The code can create initial conditions on the fly with 2nd
order Lagrangian perturbation theory (2LPT), or with the Zeldovich
approximation. There is also a built-in power spectrum estimator, as
well as an array of special features, such as the ability to simulate
arbitrarily stretched periodic boxes, or boxes with periodic
boundaries in the gravitational forces only in two dimensions.

This paper is meant to provide a comprehensive description and
evaluation of these methods, making it fairly technical in nature. But
we view this detail as a valuable reference for documenting the
numerical methodology used for simulations with {\small GADGET-4},
which has already seen its first applications in the literature
\citep{Schmidt:2018aa, Wang:2019aa, Mazzarini:2020aa}.

The paper is structured as follows. In Section~\ref{secgrav}, we begin
by discussing the gravitational force calculation, which is the
backbone of cosmic structure formation. We discuss tests of the
delivered force accuracy and its computational cost in
Section~\ref{secforceacc}, and turn to the implemented
time-integration algorithms for collisionless matter in
Section~\ref{sectime}. We then specify the discretization of
hydrodynamics in Section~\ref{sechydro}. The parallelization
techniques adopted by the code are discussed in
Section~\ref{secParallel}, and the implemented processing tools are
described in Section~\ref{secprocessing}. The accuracy implications of
different algorithmic choices and numerical parameter settings are
examined in Section~\ref{seconvergence}, and a number of exemplary
test problems for the code are presented in Section~\ref{sectests}.
The scalability of the code is assessed in
Section~\ref{secscalability}. A brief general description of the
public release of the code is given in Section~\ref{secpublic}, while
the full technical description is relegated to a manual released
alongside the source code.  Finally, Section~\ref{secdiscussion} gives
a summary and conclusions, and Appendix~\ref{secappendixA} lists, for
reference, a few lengthy expressions used in the gravitational
multipole expansions.

\section{Gravity calculation} \label{secgrav}

\begin{figure}
\resizebox{8cm}{!}{\includegraphics{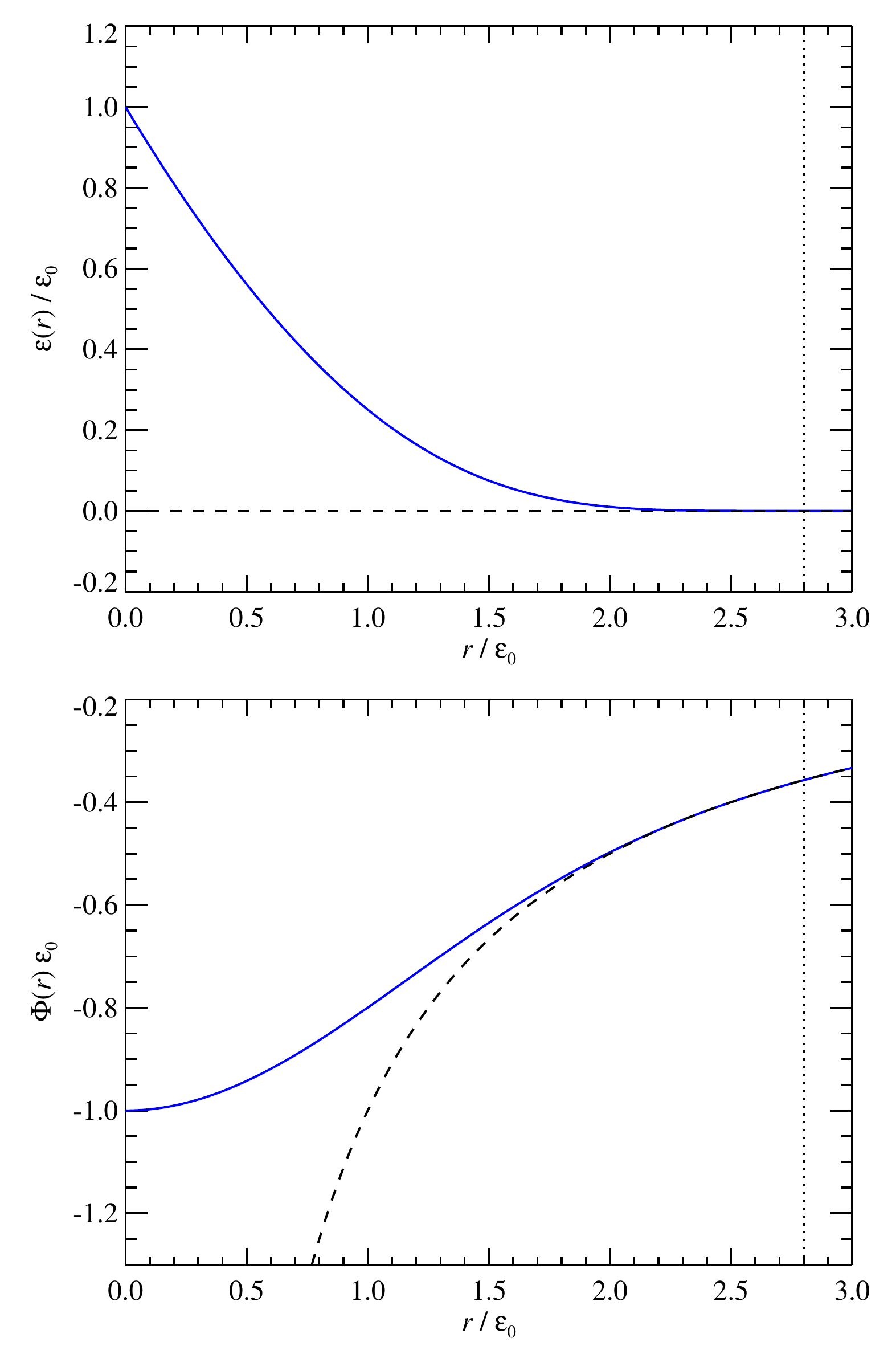}}
\caption{Shape of the softened force law in {\small GADGET-4}. The
  upper panel shows the run of the effective softening length given in
  equation~(\ref{eqnsoft}) as a function of distance. At zero lag,
  this becomes equal to the nominal softening length $\epsilon_0$.
The lower panel gives the potential of
a point mass of unit mass as a function of distance. The dashed lines
show the corresponding values for Newtonian gravity. While the  
softening vanishes completely only for distances $r\le 2.8 \epsilon$
(dotted vertical lines),  the very gradual onset of the
softening means that that this distance is not an appropriate length scale to
describe deviations from Newtonian gravity, which only become
substantial for $r \sim 1.0-1.5\,\epsilon_0$.
\label{FigSoftening}}
\end{figure}

\begin{figure}
\resizebox{8cm}{!}{\includegraphics{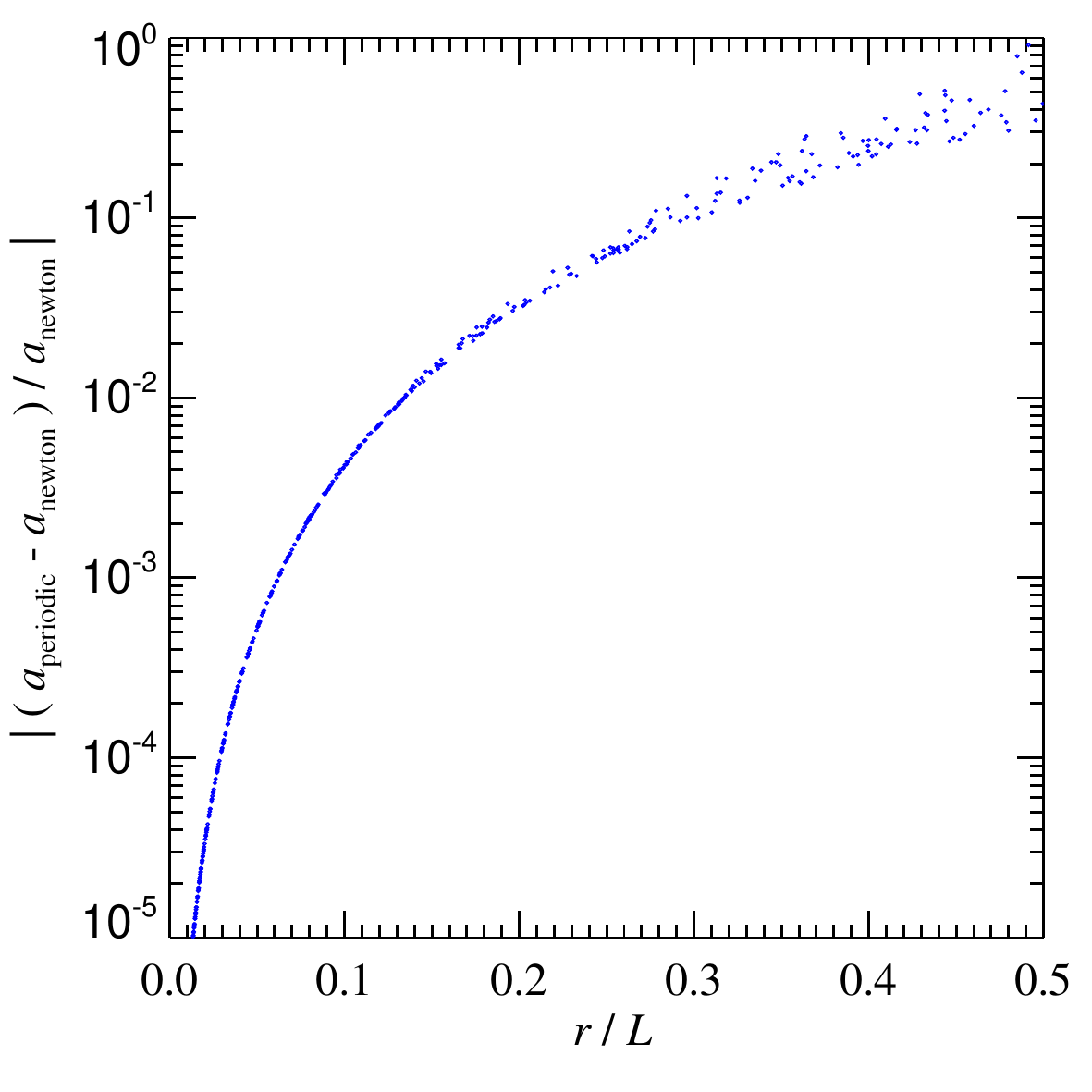}}
\caption{Force law for an
  isolated point mass in a cubic periodic box of size $L$ relative to the
  Newtonian force. The force is \textit{nowhere} equal to the
  Newtonian force, but it becomes close to it in the immediate vicinity of
  a point mass. But even there, using the Newtonian force at
  distances above $\sim 1$ percent of the box size already produces
  a relative force error in excess of $10^{-5}$. This deviation due to the
  infinite periodic grid of image masses quickly grows at larger
  distances, and
  reaches of order unity at separations approaching half the box
  size. There, the effective force law also becomes decidedly
  non-isotropic, reflected in the scatter seen at fixed $r$.
  \label{FigPeriodicForceLaw}}
\end{figure}

\subsection{Softened potential in periodic systems}

The peculiar gravitational potential $\phi(\vec{x})$ produced by $N$
particles with masses $m_j$ at coordinates $\vec{x}_j$ in a domain of
dimensions $L_x \times L_y \times L_z$ that is periodically replicated
in all three directions is given by
\begin{equation}
\phi(\vec{x}) = -\sum_{j=1}^{N}\sum_{\vec{n}=-\infty}^\infty \left\{\frac{m_j}{|\vec{x}_j - \vec{x} +
  \vec{q}_\vec{n}| + \epsilon(|\vec{x}_j - \vec{x} +
  \vec{q}_\vec{n}|)} - m_j\varphi_{\vec{n}} (\vec{x}) \right\}. \label{eqnpo0}
\end{equation}
Here $\vec{q}_\vec{n}$ denotes periodic displacement vectors given by
$\vec{q}_\vec{n} = (n_x L_x, n_yL_y, n_zL_z)$, where
$\vec{n} = (n_x, n_y, n_z)$ are integer triplets and the sum over
$\vec{n}$ extends over all these triplets. The gravitational constant
$G$ has been omitted for simplicity.  The potential contribution
$\varphi_\vec{n}(\vec{x})$ is that of a homogenous cuboid of unit mass
and extension $L_x \times L_y \times L_z$ at displacement position
$\vec{q}_\vec{n}$. This term is needed to allow for convergence of the
infinite sum over the periodic system by effectively establishing
gravitational charge neutrality. Despite adding this term, the sum
over $\vec{n}$ is still only conditionally convergent, such that the
answer can depend on the order of how one sums over the periodic
grid. For example, summing in spherical shells up to some maximum
radius produces an additional force component if the dipole moment of
the fundamental box is non-zero \citep{de-Leeuw:1980aa}, which is
however absent if one manages to truly sum the infinite grid via
Ewald summation (see below). Note that in the latter case, the added
homogeneous density contribution cannot generate any force due to
translational symmetry.

Note that one would in principle be free to add any constant of choice
to the potential given in equation~(\ref{eqnpo0}), allowing a shift of
the zero-point of the potential. Our choice corresponds to the
convention that the potential is zero for vanishing density
fluctuations. For isolated boundary conditions, we instead use the
familiar choice of putting the zero point of the potential at
infinity, making it negative everywhere. Note that for periodic
boundaries, `infinity' does not really exist. In either case, this
does not affect any of the physics, as only the gradient of the
potential enters the dynamics, and for the gravitational unbinding
computations in the {\small SUBFIND} algorithm, only potential
differences between two points are considered, such that the zero
point of the potential drops out.

The function $\epsilon(r)$ describes a gravitational softening law. We
shall assume that the softening has finite range, with
$\epsilon(r) = 0$ for $r \ge r_0$, and with $r_0$ being smaller than
half the smallest box dimension. Specifically, we adopt the same
softening law as in previous {\small GADGET} versions, in which the
potential of a point particle is replaced by the potential of a
smoothed\footnote{We here use the cubic spline kernel often used in
  SPH. Other kernel shapes, in particular ones that can be evaluated
  without a branch, are in principle possible as well, but have the
  disadvantage to complicate comparisons with the large body of
  literature results that employed the form we use here.}  mass
distribution with an outer edge at $h=2.8\, \epsilon_0$, where
$\epsilon_0$ is the softening length.  The latter gives the depth of
the potential, $-G m/\epsilon_0$, at zero lag. In the above notation,
this softening law can be expressed as
\begin{equation}
\epsilon(r; \epsilon_0) = -\frac{2.8 \epsilon_0}{W_2(r/2.8\epsilon_0)}
- r ,
\label{eqnsoft}
\end{equation}
where the function $W_2(u)$ is given in \citet[][their
eqn. 71]{Springel:2001aa}. In Figure~\ref{FigSoftening}, we show the
shape of the softening law $\epsilon(r; \epsilon_0)$. Note that while
the softening only vanishes completely for distances above
$2.8\,\epsilon_0$, the onset of the softening happens very gradually
and only starts to become significant for $r\sim 1.0\, \epsilon_0$, hence we
identify $\epsilon_0$ as the length scale characterizing the
gravitational softening length.

If we define $\vec{q}^\star_j(\vec{x})$ (or $\vec{q}_j^\star$ for
short) as the periodic displacement that minimizes the distance of
$\vec{x}_j+\vec{q}_\vec{n}$ to $\vec{x}$ (in other words,
$\vec{q}_j^\star (\vec{x})$ selects the nearest periodic image
location of particle $j$ to the position $\vec{x}$), we can write the
potential as
\begin{eqnarray}
\phi(\vec{x}) & = & -\sum_{j=1}^N \frac{m_j}{|\vec{x}_j - \vec{x} +
  \vec{q}_j^\star| + \epsilon(|\vec{x}_j - \vec{x} +
  \vec{q}_j^\star|)} \nonumber \\
& &  + \sum_{j=1}^N m_j \psi (\vec{x}_j - \vec{x} + \vec{q}_j^\star) \label{eqnpo1},
\end{eqnarray}
where we have introduced a correction potential given by
\begin{equation}
\psi(\vec{x})  = \frac{1}{|\vec{x}|}  
-
\sum_{\vec{n} =-\infty}^{\infty} \left\{ \frac{1}{|\vec{x} +
  \vec{q}_{\vec{n}}|} - \varphi_{\vec{q}_{\vec{n}}} (\vec{x}) \right\}, \label{eqnpo2}
\end{equation}
and made use of the fact that the softening of distant images
vanishes. The potential can thus be computed as the ordinary
(softened) potential from the nearest periodic image of each particle,
modified by a correction term which is the difference between the
negative point potential of this nearest image and an infinite
periodic grid of points. Note that the selection of the nearest image
for the calculation of the direct interaction ensures that it is this
interaction that is affected by softening (if any), keeping the
softening length out of the definition of $\psi(\vec{x})$.

It is important to realize that taking just the Newtonian force of the
nearest image is nowhere adequate.  Figure~\ref{FigPeriodicForceLaw}
shows the difference between the full periodic force and just the
Newtonian force of the nearest image. Even for distances as small as
$\sim 1$ percent of the box size, the correct force differs
systematically by $\simeq 10^{-5}$ from the Newtonian force, a
deviation that quickly grows for larger separations. Approximating the
force as Newtonian below a certain distance therefore leads to a
biased force. For distances of order half the box size, the difference
to the Newtonian force reaches order unity and the real periodic
force shows significant deviations from spherical symmetry.

The sum in equation~(\ref{eqnpo2}) is only conditionally
convergent. Summing it in spherical shells is possible with a
convergence factor\footnote{For example by multiplying with
  $\exp(-\lambda|x|)$, carrying out the now converging sum, and then
  considering the limit $\lambda \to 0$.}, but convergence is
extremely slow nevertheless. However, using Ewald's technique \citep{Ewald:1921aa}
the sum can be decomposed into two sums that individually converge very
rapidly and absolutely.  This decomposition can be written as
\begin{eqnarray}
\psi(\vec{x}) & = & \frac{1}{|\vec{x}|} + \frac{\pi}{\alpha^2 V}
- \sum_{\vec{n}} \frac{{\rm
erfc}(\alpha|\vec{x}+\vec{q}_\vec{n}|)}{|\vec{x}+\vec{q}_{\vec{n}}|} -
\nonumber\\
& & \frac{4\pi}{V}\sum_{\vec{k}\ne 0}  \frac{\exp[-{|\vec{k}|^2}/({4\alpha^2)}]}{ |\vec{k}|^2}
\cos\left(\vec{k}\cdot\vec{x}\right),
\label{eqnewal}
\end{eqnarray}
where $V=L_x L_y L_z$ is the volume of the domain. The first sum
still extends over all periodic images $\vec{q}_\vec{n}$, given by
$\vec{q} = (n_x L_x, n_y L_y, n_z L_z)$ with
$(n_x, n_y, n_z) \in \mathbb{Z}^3$, but is now rapidly converging in
real space thanks to the cut-off introduced by the complementary error
function. The second sum extends over
$\vec{k}\in \{ 2\pi (n_x/L_x, n_y/L_y, n_z/L_z)\; :\; (n_x, n_y,
n_z)\in \mathbb{Z}^3\}$ in frequency space and is just the discrete
periodic Fourier transform of the $4\pi/\vec{k}^2$ Green's function of
Poisson's equation, decorated with an exponential short range cut-off
factor $\exp[-r_s^2 |\vec{k}|^2]$, where $r_s = 1/(2\alpha)$ can be
interpreted as a force split scale between the short-range part given
by the first sum, and the long-range part given by the second sum.

Physically, Ewald summation can be understood as adding a negative,
spatially extended screening mass with a Gaussian profile around every
point mass. This leads to vanishing contributions of distant point
masses of the infinite grid, so that the corresponding sum now
converges rapidly. Of course, the added smooth negative mass has to be
compensated by adding the same positive contribution. But this
corresponding potential contribution can now be summed efficiently as
a rapidly converging Fourier series, because the grid of Gaussian mass
distributions is very smooth.  Note that the correction potential has
a finite value at the origin; for a cubic box of size $L$ we have
$\psi(\vec{x} \to 0) = 2.8372975/L$.

With the Ewald formula in hand, we have a method that allows the
direct summation computation of the potential for an arbitrary
periodic point distribution to machine precision. Note that this
provides the one and only correct solution to the problem -- all other
force calculation algorithms should be benchmarked against this
reference. Such alternative algorithms are of course in critical
demand, as direct summation is much too slow for large particle
numbers.

There are many different possible approaches for approximately
calculating equation~(\ref{eqnpo1}) in practice. Particle-mesh methods
create a density field on a grid and discretize Poisson's equation
directly, which is then either solved with Fourier methods, or in real
space with iterative multi-grid solvers. The primary disadvantage of
these methods is that the force softening law cannot be freely
chosen. Rather it is directly tied to the mesh geometry, with no/poor
resolution below the grid scale and anisotropic force errors at the
grid scale, i.e.~the error in solving for the force can be relatively
large. Adaptive mesh refinement and particle-particle corrections can
partially address these limitations but the effective resulting force
law is usually still not particularly clean, and it will in general be
extremely hard if not impossible to guarantee force errors below any
prescribed level for general particle distributions, in particular
highly clustered ones.

We therefore prefer so-called tree algorithms that evaluate the
short-range forces via hierarchical multipole expansions. We have
implemented in {\small GADGET-4} different flavours of such schemes,
these are (1) a classic \citet{Barnes:1986aa} style tree combined with
a new type of Ewald correction when periodic boundaries are used, (2)
a TreePM approach where long-range forces are computed with Fast
Fourier techniques and short-range forces are evaluated with the tree,
(3) a Fast Multipole Method (FMM) where the tree is accelerated by
multipole expansion both at the source and sink sides, and an Ewald
correction is used to implement periodic boundaries if present, and
(4) a FMM-PM-approach where the PM approach is combined with FMM for
the short-range forces.  Both for the classic one-sided tree and the
FMM, the expansion order can be varied in the present implementation
up to triakontadipole order in the potential, and hexadecupole order
in the force.  We now discuss these methods in turn.

\subsection{Tree algorithm}

\subsubsection{Geometry of the tree}

As standard tree algorithm, we employ a classic \citet{Barnes:1986aa}
oct-tree that is constructed through hierarchical subdivision of tree
nodes. All particles are mapped onto a finely resolved space-filling
Peano-Hilbert curve, which can be naturally cut into a hierarchy of
nested cubes which are commensurable with the geometric structure of
the oct-tree \citep[see][for a discussion of this space-filling
curve]{Springel:2005aa}. We exploit this property also in the domain
decomposition to effectively distribute branches of a fiducial global
tree to local MPI ranks, i.e.~the fact that the tree is distributed in
memory neither affects the geometry of any of the tree nodes nor the
interactions lists we process.

In {\small GADGET-2/3}, we employed a fully refined tree where a new
cubical node of half the parent's size is introduced whenever more
than one particle fall into the same octant of a node. This
eliminates the need for an arbitrary threshold for the maximum number
of particles in a node, but it has the disadvantage that two particles
may never be at the same location, otherwise the tree construction
would fail as these particles could not be split apart. We therefore
now allow for the setting of a certain threshold particle number, and
only when a tree leaf node contains more particles than this value,
the tree node is split, similar to the approach adopted by other tree
codes \citep[e.g.~{\small PKDGRAV},][]{Potter:2017aa}.  As a welcome
side effect, this can reduce the total number of nodes substantially,
thus reducing the memory requirements. It can also have a mild
positive speed impact if the threshold is in the range of a few,
because then the evaluation of some of the comparatively complicated
node-particle interactions is traded in for a larger number of simpler
particle-particle interactions, which can still end-up being slightly
faster.

\subsubsection{One-sided multipole expansion}

\begin{figure}
\begin{center}
\resizebox{6cm}{!}{\includegraphics{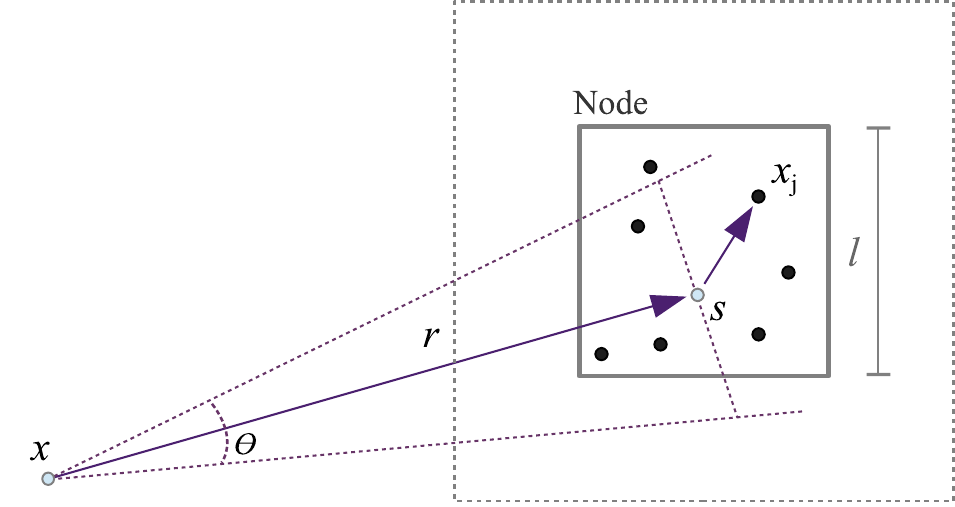}}
\end{center}
\caption{Sketch of a tree node and the geometry of our adopted opening angle
  definition and  minimal distance zone around the node. The
  centre-of-mass position $\vec{s}$ of the particles  in the cubical
  box of size $l$ has a distance $r$ to the force evaluation point
  $\vec{x}$, so that the node is seen under an angle of $\theta =
  l/r$. Comparison of this angle to a critical opening angle
  determines whether the multipole expansion of the node may be used
  or not. Additionally, we typically require that $\vec{x}$ lies outside an
  exclusion region  of side length $2\,l$ centred on the cube, in order to protect against
  pathological force errors that could otherwise arise if $\vec{s}$
  lies on the far side of the cube while $\vec{x}$ comes close to a
  particle on its near side.
  \label{FigTREESketch}}
\end{figure}

The gravitational potential generated by the points inside a  node is
\begin{equation}
\Phi(\vec{x}) = -G \sum_{j\, \in \, {\rm node}} m_j\, g(\vec{x}_j - \vec{x}),
\end{equation}
where $g(\vec{x}) = g(-\vec{x})$ is the Green's function of the
interaction, i.e.~$g(\vec{x}) = 1/|\vec{x}|$ for the Newtonian
case, or the more complicated kernel of equation~(\ref{eqnpo1}) for
the periodic case.  We now select the centre-of-mass $\vec{s}$ of the
points in the node for a Taylor expansion of the potential around this
point, yielding
\begin{equation}
\Phi(\vec{x}) = -G 
\sum_{n=0}^{p}
\frac{1}{n!}
\vec{Q}_n
\cdot
\vec{D}_{n} (\vec{s}-\vec{x})  \; + \; \mathcal{O}(\theta^{p+1})
\label{eqonesidetreepotential}
\end{equation}
up to order $p$ of the expansion, with $\theta$ being the characteristic
angular extension under which the particle group is seen.
Here we introduced the Cartesian
multipole moments
\begin{equation}
\vec{Q}_n \equiv  \sum_{j \, \in\, {\rm node}} m_j (\vec{x}_j- \vec{s})^{(n)},
\end{equation}
and derivative tensors
\begin{equation}
\vec{D}_{n} =  \vec{\nabla}^{(n)}g(\vec{x}).
\end{equation}
The notation $\vec{x}^{(n)}$ refers to the $n$-th outer product of
the vector (or tensor) with itself, while $\vec{x}\cdot\vec{y}$
denotes the inner product (i.e.~contraction) of two vectors or
tensors.  Currently, {\small GADGET-4} supports a selection of $p=1$, 2, 3, 4,
or 5 at compile time, i.e.~the multipole expansion of the potential field of a
node is carried out to dipole, quadrupole, octupole, hexadecupole, or
triakontadipole order, respectively. Since we expand around the
center-of-mass of each node, the dipole moment always vanishes, and
$p=1$ means in practice that only monopole terms need to be
considered, i.e.~$p=1$ is equivalent to $p=0$ for the potential.

Similarly, the  acceleration exerted on a test mass at location
$\vec{x}$ can be written as
\begin{equation}
\vec{a}(\vec{x}) = -\nabla\Phi(\vec{x}) = -G 
\sum_{n=0}^{p-1}
\frac{1}{n!}
\vec{Q}_n 
\cdot
\vec{D}_{n+1} (\vec{s}-\vec{x}) \; + \; \mathcal{O}(\theta^{p})
\label{eqonesidetreeacc}
\end{equation}
to order $p-1$ of the expansion.  Compared to a direct differentiation
of equation~(\ref{eqonesidetreepotential}), we here dropped the last
term in the expansion in order to end up with the same highest order
in the derivative tensor for a specified $p$, both for potential and
force, and to make the one-sided tree behave more similar to the FMM
algorithm (where the accuracy of the force is one order lower than
that of the potential, see later on). Note that a side effect of this
is that the force accuracies obtained for $p=1$ and $p=2$ are equal in
the one-sided tree.  As we shall see later on, which order $p$ is most
efficient can be problem dependent.  Using higher order is in general
more efficient if one aims very low force errors, but it also entails a
higher peak memory usage because more multipole moments need to be
stored for the nodes.

Note that the multipole tensors $\vec{Q}_n$ are totally symmetric
(i.e.~elements where two indices are interchanged are equal), so that
they feature only $(n+2)(n+1)/2$ independent elements at order $n$.
If the interaction kernel is isotropic like in the ordinary Newtonian
case, or if the periodic boundary conditions impose cubical symmetry,
the derivative tensors are totally symmetric as well. This reduces the
storage requirements substantially and also simplifies many of the
algorithmic operations with them, which we try to exploit as much as
possible. If the interaction potential is a simple power-law in
radius, like for Newtonian gravity, it turns out be sufficient to
store the multipole moments in a trace-free form, where the number of
independent elements is reduced further to $2n+1$, which can save a
substantial amount of memory for high expansion orders
\citep{Coles:2020aa}.  In this case, the Cartesian formulation is
closely analogous to an expansion into spherical harmonics, as used in
the original formulation of FMM \citep{Greengard:1987aa}. However,
since this optimisation is not viable for periodic boundary
conditions, which is the most important case for cosmology, we refrain
from considering it here. It would however still be of interest as a
possible future extension for isolated boundary conditions.

\subsubsection{Tree walk and opening criterion}

In the force calculation for a given particle, the tree is walked
top-down, starting at the root node.  For deciding whether or not a
multipole expansion of the current node is acceptable, different
criteria can be used.  We either adopt a straightforward classical
geometric opening criterion, as sketched in
Figure~\ref{FigTREESketch}, according to which the multipole expansion
of a node can be used if
\begin{equation}
\frac{l}{r} < \theta_c ,
\end{equation}
where $l$ is the side-length of the cubical tree node, $r$ is the
distance of the target coordinate to the node's centre of mass, and
$\theta_c$ is a critical angle controlling the accuracy (as well as
computational cost) of the force approximation. Alternatively, a
relative opening criterion as first proposed by
\citet{Springel:2001aa} can be used, where a rough approximation of
the expected force error is compared with the magnitude of the total
force,
\begin{equation}
\frac{M}{r^2} \left(\frac{l}{r}\right)^{p} < \alpha |\vec{a}|,
\label{eqnrelcriterion}
\end{equation}
where $|\vec{a}|$ is the total acceleration, usually estimated for a
particle from the force obtained in the previous timestep\footnote{For
  the very first force calculation, $|\vec{a}|$ is first estimated by
  carrying out a tree walk with the geometric opening criterion and a
  conservatively chosen opening angle.}. The
parameter $\alpha$ controls the force accuracy.

This second criterion aims at achieving a more or less constant force
error in every individual interaction, and the size of this error is
chosen in relation to the magnitude of the final force. This means
that nodes that are more important for the final force are evaluated
with higher relative force accuracy than less important contributions,
which gives a more economical scheme overall. Also, the final relative
force accuracy is roughly kept constant for all particles, independent
of clustering state. This is especially useful for cosmological
simulations where the peculiar forces at high redshift are small due
to near cancellation of forces in all different directions. In this
situation, the relative opening criterion automatically invests more
effort, whereas in strongly clustered regions most of the effort is
concentrated in evaluating the dominant force components. This
automatic adjustment to the clustering state is a welcome feature in
these types of calculations. Indeed, in \citet{Springel:2005aa} it was
shown that the relative criterion is typically more efficient than the
geometric one in the sense of delivering higher force accuracy than
the geometric criterion at a given computational cost, something that
we will verify here later on as well.

We typically augment both of the above criteria with an additional
geometric test that prevents the use of multipole expansions if the
distance to the geometric center of a tree node is less than $l$. This
corresponds to an exclusion zone around the tree node's geometric
centre, as sketched in Figure~\ref{FigTREESketch}. This is a simple
conservative means to protect against pathological cases that could
occur if one allows evaluations of multipoles for reference points
very close to or even within nodes \citep[see
also][]{Salmon:1994aa}. It also prevents a complete break-down of the
convergence of the Taylor expansion if the opening criterion would
otherwise allow an extremely large opening angle.

\subsubsection{Periodic boundary conditions}

If periodic boundaries are desired for the pure tree method,
additional complications arise as we have to cope with the infinite
sum of equation~(\ref{eqnpo0}). The approach followed in previous
versions of {\small GADGET} consists of exploiting equation~(\ref{eqnpo1})
directly for each node or particle interaction.  This means that
during the tree walk every node (or single particle when encountered)
is mapped to its nearest periodic image relative to the coordinate of
the force evaluation.  If the node does not have to be opened, the
first term in equation~(\ref{eqnpo1}) can be evaluated by means of a
multipole expansion with the normal Newtonian Green's function. The
second term can also be evaluated with a multipole expansion, but now
relies on the Green's function $\psi(\vec{x})$ defined in
equation~(\ref{eqnewal}), which does not have a closed form
expression. One can compute it with the rapidly converging Ewald sum,
but as this needs to be evaluated for every interaction, a substantial
computational cost is incurred.

Another solution is to tabulate the
correction potential (which is quite smooth) and its derivatives in a
three-dimensional pre-computed look-up table, from which one then
calculates $\psi$ and its derivatives as needed by tri-linear
interpolation.  This is what we had previously done in {\small
  GADGET}, following a similar approach as in \citet{Hernquist:1991aa}
Our default look-up table size has used $64^3$ entries (for one octant
only, the rest can be inferred due to the symmetries involved), and we
used separate tables for the potential and the three spatial
components of the force, which is sufficient for $p=1$.  Perhaps the
worst disadvantage of this approach is that its accuracy is
fundamentally limited by the interpolation accuracy of the look-up
table, making it hard to push the relative force errors below
$\sim 10^{-3}$. If this is nevertheless desired, one either has to make the
lookup table sufficiently fine, which soon becomes prohibitive due the
unacceptably large amounts of memory consumed by a three-dimensional
table with many entries per dimension, or come up with a different
approach that allows a more accurate but still fast way to compute the
interaction tensors.

To obtain the desired high accuracy  for
 $\vec{D}_n$, we have replaced the
trilinear
interpolation
from a look-up table
 with a scheme based on Taylor 
expansions. We still precompute a grid for the $\vec{D}_n$, but
to obtain the value at a given coordinate $\vec{x}$, we first locate the
closest point  $\vec{x}_0$ to $\vec{x}$ in our grid, and then approximate
the target value through the expansion
\begin{eqnarray}
\vec{D}_n(\vec{x}) & \simeq & \vec{D}_n(\vec{x}_0) +
\vec{D}_{n+1}(\vec{x}_0) \cdot \Delta \vec{x} + \frac{1}{2}
\vec{D}_{n+2}(\vec{x}_0) \cdot \Delta \vec{x}^{(2)} \nonumber \\
& & + \;\frac{1}{6}
\vec{D}_{n+3}(\vec{x}_0) \cdot \Delta \vec{x}^{(3)},
\end{eqnarray}
of the interaction kernel, where
$\Delta\vec{x} = \vec{x} - \vec{x}_0$.  Since we want $\vec{D}_n$ up
to order $n=5$, we actually tabulate the tensors up to order $n=8$ on
the grid. The pre-computation of these tensors is done based on
analytic derivatives of the Ewald summation formula, and hence can be
easily made exact up to machine precision.  The Taylor expansion
approach is much more accurate than tri-linear interpolation.  It
gives us the ability to recover all needed derivative tensors to a
relative accuracy of $\simeq 10^{-10}$ for arbitrary interaction
distance $\vec{x}$; only if this is achieved, the high-order versions
of the multipole methods ultimately make sense and converge properly
to the direct summation result also for difficult situations such as
cosmological initial conditions at high redshift.

\subsection{TreePM approach}

An alternative to the pure tree algorithm that avoids the need for
explicit Ewald corrections is the so-called TreePM approach
\citep{Xu:1995aa, Bagla:2002aa, Springel:2005aa}. In this method, we
adopt in equation (\ref{eqnewal}) a value of $r_s = 1/(2 \alpha)$ that
is very small against the box size, i.e.~$r_s\ll \min(L_x, L_y,
L_z)$. In this case, we can restrict the real-space sum over the
nearest periodic images exclusively to the $\vec{q} = 0$ term,
resulting in a total short-range potential of the form
\begin{equation}
\phi_{\rm sr}(\vec{x})  =  \sum_j m_j \left\{\frac{-1}{|\vec{r}_j| +
                             \epsilon(|\vec{r}_j|)} \label{eqnpo4} 
+ 
 \frac{1}{|\vec{r}_j|} + \frac{\pi}{\alpha^2 V}
- \frac{{\rm erfc}(\alpha|\vec{r}_j|)}{|\vec{r}_j|} \right\} ,
\end{equation}
where $\vec{r}_j = |\vec{x} - \vec{x}_j + \vec{q}^\star|$ denotes the
nearest periodic distance of $j$ to the reference point.  This can be
evaluated with a tree algorithm in real space, while
the frequency sum gives a long range potential of the form, 
\begin{equation}
  \phi_{\rm lr}(\vec{x})
  = \sum_{j=1}^N m_j \, \frac{4\pi}{V}\sum_{\vec{k}\ne 0}  \frac{\exp[-{|\vec{k}|^2}/({4\alpha^2)}]}{ |\vec{k}|^2}
  \cos\left[\vec{k}\cdot(\vec{x}-\vec{x}_j)\right],
\label{eqnksum}
\end{equation}
which can be computed through Fourier methods, as described
below. TreePM is thus best understood as a special version of the
Ewald summation technique, where the total potential of
eqn.~(\ref{eqnpo1}) is obtained as
$\phi(\vec{x}) = \phi_{\rm sr}(\vec{x}) +\phi_{\rm lr}(\vec{x})$.  We
note that it would in principle be possible to employ a different
cut-off function to separate short-range and long-range forces. Using
the complementary error function in real space is associated with a
rapidly and monotonically declining exponential filter in $k$-space,
which is essentially as good as it gets. Still more sharply declining
real-space cut-off functions will in general require more complicated
$k$-space filters that may need empirical calibration but could then
also be viable.

For calculating the force corresponding to $\phi_{\rm sr}$ with a tree
walk we proceed as in the ordinary tree algorithm, except that we use
different expressions for evaluating the multipole contributions that
reflect the modification of the inverse distance potential due to the
short-range cut-off.  We keep in principle the same opening criteria
as used in the ordinary tree walk\footnote{When the relative opening
  criterion is in use, the acceleration $|\vec{a}|$ in
  equation~(\ref{eqnrelcriterion}) is the total acceleration on the
  particle, i.e. the sum of the PM and tree components.}, except that
we add the further criterion that a tree node is dropped entirely from
consideration if its nearest edge is further away than a distance
$r_{\rm cut}$, where the force has become unsoftened and the cut-off
factor ${\rm erfc}(\alpha \,r)$ is so large that the ordinary
Newtonian potential is heavily suppressed.  We set
$r_{\rm cut}= \max[ R_{\rm cut} \times r_s, 2.8 \epsilon_0]$, where
the precise value adopted for the dimensionless parameter
$R_{\rm cut} \simeq 4.5 - 7.0$ affects both the computational cost and
the residual force errors in the matching region between the
short-range and long-range forces. If a maximum relative force error
of order $10^{-2}$ is sufficient, one can choose $R_{\rm cut}$ as
small as $\simeq 4.5$ for maximum computational speed, but if lower
maximum force errors are desired, larger cut-off values need to be
chosen. For $R_{\rm cut} = 7$, the suppression factor for the
potential and the force relative to the Newtonian value reach around
$10^{-5}$ at the cut-off radius.  The cut-off radius effectively
restricts the tree walk to a small spherical region around the
evaluation point, yielding a significant gain in performance compared
to a full tree walk. In addition, no Ewald corrections need to be
computed for the short-range force. These two factors can make TreePM
faster than the ordinary tree algorithm, provided the speed-up is not
offset by the cost of the additionally required PM calculation.

For efficiently evaluating the short range force kernel of
equation~(\ref{eqnpo4}) and its derivatives, we use linear
interpolation from a small one-dimensional look-up table. This avoids
the costly computation of the complementary error function for every
interaction. Again, depending on the desired accuracy goals, the
length of this table can be varied. A finer look-up table yields
smaller residual interpolation errors but reduces the cache efficiency
of the processor. If maximum relative force errors as low as $10^{-5}$
are desired, we use a table with 256 entries, for more commonly
employed force accuracies such as $10^{-3}$, already 48 entries are in
principle sufficient.

We note that if the softening length is much smaller than the split
scale $r_s$, which is almost always the case, equation (\ref{eqnpo4})
can be approximated as
\begin{equation}
\phi_{\rm sr}(\vec{x}) \simeq - \sum_j m_j \frac{{\rm
    erfc}(\alpha|\vec{r}_j|)}{|\vec{r}_j| + \epsilon(|\vec{r}_j|)},
\label{eqnpo5}
\end{equation}
which had been adopted in {\small GADGET-2} (omitting also the
constant $\pi m_j/(\alpha^2 V)$ term in the potential for simplicity,
which does not contribute to the dynamics). However, since one may not
necessarily always have $\epsilon_0 \ll r_s$, {\small GADGET-4} uses the
more accurate expression~(\ref{eqnpo4}).

The Fourier sum of equation~(\ref{eqnksum}) can be recognized as the
inverse Fourier transform of the product of the Green's function
$\exp(-r_s^2 \vec{k}^2)/\vec{k}^2$ with the Fourier transform of a
density field in which each particle is represented by a Dirac
$\delta$-function. This can be solved with a standard particle-mesh
approach \citep[e.g.][]{Hockney:1988aa} in which we first create a
binned density field, for example through clouds-in-cell (CIC)
assignment, calculate a discrete Fast Fourier Transform (FFT),
multiply with the Green's function, and then transform back to obtain
the potential. The forces can then be obtained on the same grid
through finite differencing, followed by interpolating them to the
particle positions. The smoothing effects of the assignment and
interpolation operators can be compensated in the mean by deconvolving
with their corresponding Fourier transforms,
which in the case of CIC corresponds to a division with a
\emph{sinc}-function. We use a 4-th order accurate finite differencing
formula on the potential grid to derive the forces. Using spectral
differencing instead would be possible as well, but this is
computationally more costly due to the larger number of Fourier
transforms required (and larger memory needs for retaining an
auxiliary copy of the field) while it does not provide significant
accuracy advantages.

The resulting long-range forces are accurate provided the Nyquist
frequency of the employed mesh is large enough to include all
frequencies significantly contributing to the long-range
potential. Fortunately, the exponential cut-off factor in the Green's
function introduces a well-defined scale beyond which the Fourier
spectrum of the potential declines rapidly in a smooth and isotropic
fashion, an important advantage over ordinary PM and P$^3$M
schemes. For high accuracy we need $\exp(-k^2_{\rm Nq} r_s^2) \ll 1$,
where $k_{\rm Nq} = (2\pi / L) (N_{\rm grid}/2)$ and $N_{\rm grid}$ is the number of
grid points per dimension. In practice, we define the dimensionless
parameter $A_{\rm smth} = r_s / (L/N_{\rm grid})$ and use it to parameterize
the transition scale for a given mesh size $N_{\rm grid}$, yielding
$\exp(-k^2_{\rm Nq} r_s^2) = \exp(-\pi^2 A_{\rm smth}^2)$. The value
of this factor drops to about $2\times 10^{-7}$ for our ``fastest
runtime'' but least conservative setting of $A_{\rm smth}=1.25$.
Larger values of $A_{\rm smth}$ for a given mesh size $N_{\rm grid}$ imply that
more Fourier modes are invested to sample the high-frequency decline
of the long-range force. This makes the long-range force more
accurate, but also implies that the cut-off $r_{\rm cut}$ moves to
larger scales, so that the short-range tree force has to be evaluated
over a larger region, increasing its relative computational cost. By
varying $A_{\rm smth}$ one can thus smoothly change the force accuracy
of the long-range force, and shift the balance in the computational
cost between PM and tree. Depending on the desired final force
accuracy, different sweet spots for maximum computational efficiency
exist.  We will return to this point in Section~\ref{secforceacc} when
we systematically investigate the force accuracy and relative
computational cost of the different gravity solvers available in
{\small GADGET-4}.

For the evaluation of the Fourier transforms in the TreePM approach we
employ the Fast Fourier Transform algorithm as implemented in the
{\small FFTW}\footnote{{http://www.fftw.org}} library
\citep{Frigo:2005aa}. We employ the current version 3 of {\small
  FFTW}\footnote{{\small GADGET-2} used version 2 of FFTW, which has a
  different API and is now deprecated by the developers of FFTW.} but
refrain from using its built-in MPI-parallelized functions, instead we
employ our own parallelization layer on top of one-dimensional
real-to-complex and complex-to-complex FFTs provided by the library.
This was done to avoid that dynamic memory allocation is used outside
of {\small GADGET-4}'s direct control, which can create problems if
one operates very close to the physical memory limit and does large
FFTs nevertheless. It also allows us to seamlessly switch to a
column-based data decomposition when required for scalability, which
is presently not supported by the MPI interface of {\small
  FFTW-3}. More details on this, in particular the column-based data
decomposition, are given in Section~\ref{secparfft} on
parallelization.

\subsection{Fast multipole method}  \label{subsecfmm}

The so-called fast multipole method (FMM) has been originally
introduced by \citet{Greengard:1987aa} and is arguably the fastest
known approach for hierarchical multipole expansion when high force
accuracy is required, improving upon the `one-sided' tree algorithm of
\citet{Barnes:1986aa}.  A variant of the original harmonic FMM
approach that treats all expansions in Cartesian coordinates instead
of using spherical harmonics has first been introduced by
\citet{Dehnen:2000aa, Dehnen:2002aa}.  The FMM approach can be faster
than a classic tree because the multipole expansion is not only
carried out at the source side but also on the sink side where the
target particles are located.  This allows one to calculate the
expansion between two well-separated ``source'' and ``sink'' nodes
only once, and effectively reuse it for all particles making up the
nodes, whereas in the one-sided tree it is computed many times over,
more or less for each of the constituent particles on the sink side
yet again with only slightly shifted expansion
centres. Barnes'~(\citeyear{Barnes:1990aa}) grouping algorithm for the
one-sided tree, in which a common interaction list for small particle
groups is computed, represents another approach to mitigate this
behaviour of the tree algorithm.

Furthermore, a symmetric Cartesian expansion delivers, at the same
time, the force field of the sink node onto the source node and vice
versa, yielding a manifestly momentum conserving formulation where the
vector sum of all force errors adds up to zero. This is not the case
for an ordinary tree algorithm, where in fact surprisingly large
errors in the total momentum of a system can occur.

The speed advantage expected from FMM relative to the tree code has
not been confirmed in all practical implementations
\citep[e.g.][]{Capuzzo-Dolcetta:1998aa}, but this may well originate
in the complexity of the spherical harmonics expansion approach used
in the originally proposed method. However, \citet{Dehnen:2000aa} has
shown that in the low-accuracy regime relevant for collisionless
dynamics the Cartesian expansion variant of FMM represents an
attractive formulation that can offer a significant performance
advantage compared with ordinary tree methods.  Still, the FMM method
has thus far not been applied nearly as widely in astrophysics as
classical tree algorithms, even though this has begun to change
recently.  The reasons presumably lie in FFM's somewhat higher
algorithmic complexity, difficulties in using it efficiently with
local timestepping, in harder parallelization, and in intricacies
related to the inclusion of periodic boundaries. For the present work,
we have outfitted {\small GADGET-4} with an FMM module as an
alternative to the one-sided classic tree, allowing us to explore to
what extent these issues can be overcome in our code in practice.

Our basic FMM approach closely follows the algorithm first presented
by \citet{Dehnen:2000aa}, but is augmented with parallelization on
distributed memory machines, with treatments of multiple gravitational
softening lengths and periodic boundary conditions, and with the
possibility to couple FMM with a long-range PM approach, creating a
FMM-PM method in analogy to the TreePM approach discussed above. 

\begin{figure}
\resizebox{8cm}{!}{\includegraphics{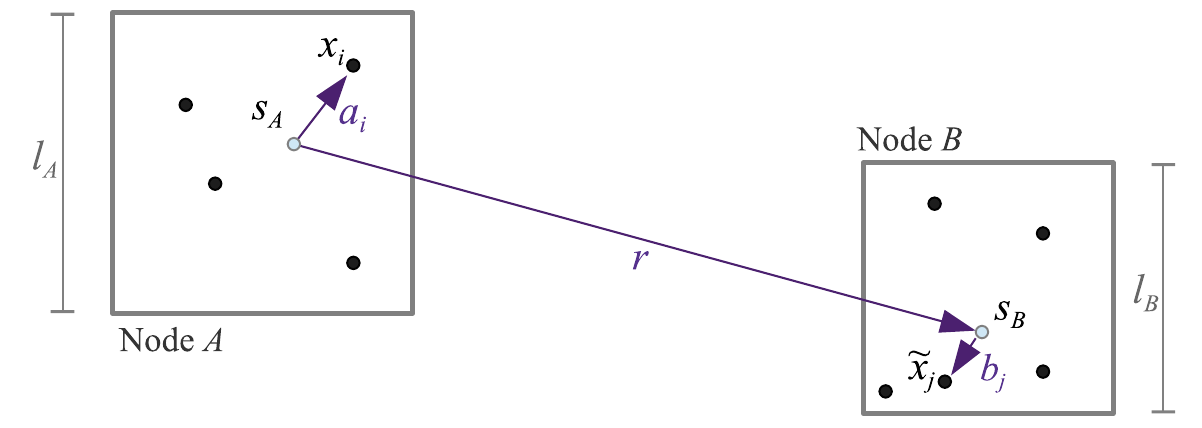}}
\caption{Sketch of the geometry and nomenclature we use to describe an
  FMM interaction between two nodes $A$ and $B$ of sizes $l_{A}$ and
  $l_{B}$. The particles in the nodes have
  positions $\vec{x}_i$ and $\vec{\tilde{x}}_j$, giving rise to
  centre-of-mass positions $\vec{s}_{A}$ and $\vec{s}_{ B}$ of the nodes, respectively, 
  with $\vec{r} = \vec{s}_{B} - \vec{s}_{A}$ being their
  distance vector, and $\vec{a}_i$ and $\vec{b}_j$ denoting the
  relative positions of the particles in the nodes to their corresponding
  center of mass locations.
  \label{FigFMMSketch}}
\end{figure}

The method starts by considering the interaction potential between the
particles of two nodes $A$ and $B$, as sketched in
Figure~\ref{FigFMMSketch}, with $\vec{x}_i$ denoting the positions of
the particles in $A$, and $\vec{\tilde x}_i$ those in $B$. We can
write the gravitational potential generated by the points of $B$ at a
particle coordinate in node $A$ as
\begin{equation}
\Phi(\vec{x}_i) = -G \sum_j m_j\, g(\vec{\tilde x}_j - \vec{x}_i) ,
\end{equation}
where $g(\vec{x}) = g(-\vec{x})$ is the symmetric Green's function of
the interaction.  In the non-periodic case without gravitational
softening, we have the simple Newtonian $g(\vec{x}) = 1/|\vec{x}|$
interaction. With periodic boundary conditions, $g(\vec{x})$ also
depends on the direction of $\vec{x}$.

We now Taylor-expand the potential by introducing expansion centres
both in $A$ and $B$. Obvious choices for them are either the geometric
centres of the nodes, or their centres of mass; we adopt the latter as
this leads to vanishing dipole moments. Let $\vec{s}_A$ and
$\vec{s}_B$ denote the centres of mass in the nodes, and
$\vec{r}= \vec{s}_B - \vec{s}_A$ be their relative distance
vector. Further, let $\vec{a}_i = \vec{x}_i - \vec{s}_A$ be the
relative position of the points in $A$ with respect to node $A$'s
center of mass, and likewise
$\vec{b}_i = \vec{\tilde x}_i - \vec{s}_B$ for node $B$. We then have
$\vec{\tilde x}_j - \vec{x}_i = \vec{r} + \vec{b}_j - \vec{a}_i$, and
obtain up to $p$-th order for the interaction potential
\begin{equation}
g(\vec{\tilde x}_j - \vec{x}_i) \simeq 
\sum_{n=0}^p \frac{1}{n!} \vec{\nabla}_{\vec{r}}^{(n)}g(\vec{r}) \cdot
(\vec{b}_j - \vec{a}_i)^{(n)} , 
\end{equation}
assuming $|\vec{a}_i| \ll |\vec{r}|$ and $|\vec{b}_j| \ll |\vec{r}|$
(well separatedness of the nodes). Again, here the notation $\vec{x}^{(n)}$
means the $n$-th outer product of the vector with itself, while
$\vec{x}\cdot\vec{y}$ denotes the inner product (i.e.~contraction) of
two vectors or tensors. Expanding the term involving the relative
positions into a binomial
series, 
\begin{equation}
(\vec{b}_j - \vec{a}_i)^{(n)} = \sum_{k = 0}^{n} \left(
\begin{array}{c}
n \\ 
k \\
\end{array}
 \right) \vec{b}_j^{(n-k)}  (-\vec{a}_i)^{(k)},
\end{equation}
and rearranging the sums yields
\begin{equation}
g(\vec{\tilde x}_j - \vec{x}_i) \simeq 
\sum_{k=0}^p \frac{(-1)^k}{k!} \vec{a}_i^{(k)} \cdot \sum_{n=0}^{p-k} \frac{1}{n!} \vec{\nabla}^{(n+k)}g(\vec{r}) \cdot
\vec{b}_j^{(n)} .
\end{equation}
By introducing the multipole moments
\begin{equation}
\vec{Q}^{B}_n \equiv  \sum_j m_j \vec{b}_j^{(n)},
\end{equation}
we hence arrive at
\begin{equation}
\Phi(\vec{x}_i) \simeq -G 
\sum_{k=0}^p \frac{(-1)^k}{k!} \vec{a}_i^{(k)} \cdot \sum_{n=0}^{p-k}
\frac{1}{n!}\vec{D}_{n+k} 
\cdot
\vec{Q}^{B}_n 
\end{equation}
for the potential,
where we have defined the $m$-th order tensors
\begin{equation}
\vec{D}_{m} =  \vec{\nabla}^{(m)}g(\vec{r}),
\end{equation}
just like for the tree algorithm.
Similarly, the potential 
$\tilde\Phi(\vec{\tilde x}_i)$ at a location of a point in $B$ due to
the particles in $A$ is given to the same order of the expansion as
\begin{eqnarray}
\tilde\Phi(\vec{\tilde x}_i) & \simeq & -G 
\sum_{k=0}^p \frac{(-1)^k}{k!} \vec{Q}_k^{A} \cdot \sum_{n=0}^{p-k}
\frac{1}{n!}\vec{D}_{n+k} 
\cdot
\vec{b}^{(n)}_j \\
& = &
-G 
\sum_{k=0}^p \frac{(-1)^k}{k!} \vec{b}_i^{(k)} \cdot \sum_{n=0}^{p-k}
\frac{(-1)^{n+k}}{n!}\vec{D}_{n+k} 
\cdot
\vec{Q}^{A}_n , \nonumber
\end{eqnarray}
and hence manifest momentum conservation,
\begin{equation}
\sum_{i\in A} m_i\nabla \Phi(\vec{x}_i) + \sum_{i\in B}
m_i\nabla \tilde \Phi(\vec{\tilde x}_i) = 0,
\end{equation}
is retained if these expressions are used to approximate the force
from $A$ on $B$, and vice versa.  Also note that the highest order
multipole moments $\vec{Q}_p^{A/B}$ of the expansion for a given order
$p$ contribute only as  constants to the potential and hence do not
affect the force. As only the latter is required for the particle
dynamics, we typically follow \citet{Dehnen:2000aa} and omit the
calculations of these multipoles in practice, simply dropping them in the
potential, unless we explicitly retain them through a code compile-time
option.

In {\small GADGET-4}, we either consider order $p=1$, $p=2$, $p=3$,
$p=4$, or $p=5$. For example, for the $p=2$ case, the explicit
expression for the potential expansion becomes
\begin{equation}
\Phi_{p=2}(\vec{x}) \simeq -G\left[ {Q}_0{D}_0 +  \frac{1}{2}  \vec{Q}_2
  \cdot \vec{D}_2 -   {Q}_0 \vec{D}_1 \cdot\vec{a}
+\frac{1}{2}  {Q}_0 \vec{D}_2 \cdot \vec{a}^{(2)} \right],
\end{equation}
but as mentioned above, the spatially constant term proportional to
$\vec{Q}_2 $ may be dropped because it does not contribute to the
dynamics, and hence one can avoid calculating the second moment tensor
at this order if one primarily cares about the forces only.  For
an expansion up to quadrupole order, $p=4$, we instead have
\begin{eqnarray}
\Phi_{p=4}(\vec{x})  & \simeq & -G \left[{Q}_0 {D}_0 + \frac{1}{2} \vec{Q}_2
                          \cdot \vec{D}_2 +\frac{1}{6} \vec{Q}_3 \cdot
                                \vec{D}_3  +\frac{1}{24} \vec{Q}_4 \cdot \vec{D}_4
  \right. \nonumber \\
& & - ({Q}_0 \vec{D}_1  +\frac{1}{2}\vec{Q}_2 \cdot \vec{D}_3  +\frac{1}{6}\vec{Q}_3 \cdot \vec{D}_4) \cdot \vec{a}
                        \nonumber \\
& & + \frac{1}{2} \left ( {Q}_0\, \vec{D}_2   + \frac{1}{2} \vec{Q}_2\, \vec{D}_4     \right)\cdot \vec{a}^{(2)}
\nonumber\\
& & \left.
 - \frac{1}{6}
    {Q}_0 \vec{D}_3 \cdot \vec{a}^{(3)}   + \frac{1}{24}
    {Q}_0 \vec{D}_4 \cdot \vec{a}^{(4)}
\right].
\end{eqnarray}
Again, we typically drop in practice the term involving the
hexadecupole moment $\vec{Q}_4$. The expansions for other orders are
given in Appendix~\ref{secappendixA}, for definiteness.

For non-periodic boundary conditions, or for the spatial part of Ewald
sums, the short-range interaction depends only on the norm $r$ of the
distance vector, $g(\vec{r}) = g(r)$. Defining
\begin{equation}
g_n(r) \equiv \left(\frac{1}{r}\frac{{\rm d}}{{\rm d}r}\right)^n g(r),
\end{equation}
the derivative tensors $\vec{D}_m$ can be
expressed in this case as
\begin{eqnarray}
{D}_0 & = & g_0  \label{eqngderiv0} \\
(\vec{D}_1)_i & = & g_1\, r_i\\
(\vec{D}_2)_{ij} & = & g_1\,\delta_{ij} + g_2
                {r_i r_j}  \\
(\vec{D}_3)_{ijk} & = & g_2 \,\left(
                        {\delta_{ij}r_k + \delta_{jk}r_i +
                        \delta_{ik}r_j} \right)  + \nonumber \\
& &  g_3 \,{r_i r_j r_k} \\
(\vec{D}_4)_{ijkl} & = &  g_2 \left( {\delta_{ij}\delta_{kl} + \delta_{jk}\delta_{il} +
                        \delta_{ik}\delta_{jl}} \right)  + \nonumber\\
                    & &   g_3 \, \left( \delta_{ij}r_kr_l + \delta_{jk}r_ir_l +
                        \delta_{ik}r_jr_l \, + \right.  \nonumber \\
      & &         \left.  \; \; \; \; \; \delta_{il}r_jr_k + \delta_{jl}r_ir_k +
                        \delta_{kl}r_ir_j   \right) + \nonumber \\
& &      g_4\, {r_i r_j r_k r_l}, \\
(\vec{D}_5)_{ijklm} & = &  g_3 \left[ 
r_m\left(\delta_{ij}\delta_{kl} + \delta_{jk}\delta_{il} +
                          \delta_{ik}\delta_{jl}\right) + \right. \nonumber\\
 & & \; \; \; \; \,  r_l\left(\delta_{ij}\delta_{km} + \delta_{jk}\delta_{im} +
                          \delta_{ik}\delta_{jm}\right) +   \nonumber\\
 & & \; \; \; \; \,  r_k\left(\delta_{ij}\delta_{lm} + \delta_{jm}\delta_{il} +
                          \delta_{im}\delta_{jl}\right) +   \nonumber\\
 & & \; \; \; \; \,  r_j\left(\delta_{im}\delta_{kl} + \delta_{km}\delta_{il} +
                          \delta_{ik}\delta_{lm}\right) +   \nonumber\\
 & & \left. \; \; \; \; \,  r_i\left(\delta_{jm}\delta_{kl} + \delta_{jk}\delta_{lm} +
                          \delta_{km}\delta_{jl}\right) 
 \right]  + \nonumber\\
                    & &   g_4 \, \left( \delta_{ij}r_kr_lr_m + \delta_{jk}r_ir_lr_m +
                        \delta_{ik}r_jr_lr_m \, + \right.  \nonumber \\
      & &           \; \; \; \; \; \delta_{il}r_jr_kr_m + \delta_{jl}r_ir_kr_m +
                        \delta_{kl}r_ir_jr_m    + \nonumber \\
      & &           \; \; \; \; \; \delta_{im}r_jr_kr_l + \delta_{jm}r_ir_kr_l +
                        \delta_{km}r_ir_jr_l    + \nonumber \\
      & &         \left.  \; \; \; \; \; \delta_{lm}r_ir_jr_k 
                          \right) + \nonumber \\
& &      g_5\, {r_i r_j r_k r_l r_m},
\label{eqngderiv}
\end{eqnarray}
where $r_i$ denotes the $i$-th component of the vector $\vec{r}$. We
spell out $\vec{D}_6$ and $\vec{D}_7$, and give explicit forms for the
derivatives $g_n$ of $g(r)$ in Appendix~\ref{secappendixA}. While all
these tensors are fully symmetric in their indices, they nevertheless
rapidly grow in size and complexity. For orders two to five, they have
6, 10, 15, and 21 independent components, respectively.

As stressed earlier, when periodic boundary conditions are desired
without the PM approach, Ewald summation is needed to realize accurate
periodic boundary conditions. The full relevant interaction potential
$g(\vec{r})$ is not an isotropic function in this case. We here
proceed similarly to the one-sided tree algorithm by splitting the
interaction into a part between the nearest periodic images of two
nodes, and an Ewald-correction of this interaction. The former can be
calculated with the spherically symmetric Newtonian interaction
potential, whereas for the latter we again employ a Taylor-series based
look up from precomputed tables, absorbing the lack of
rotational symmetry into this correction. 

Once the grouping of particles into a hierarchy of nodes is completed
in the tree construction, we precompute the multipole moments of the
tree nodes recursively from the multipole moments of their daughter
nodes, which constitutes the first phase of the FMM method (up to this
point there is no difference to the ordinary tree code). In the second
phase, we use a dual tree walk similar to \cite{Dehnen:2000aa} to
evaluate multipole expansions both at the source and sink side for
interacting pairs of nodes or particles. This is done in a symmetric
fashion where each interaction is accounted for in a manifestly
momentum-conserving way.

To this end, we define a function that accounts for the interaction
between two cells $A$ and $B$, meaning that all particles in $A$ must
receive the force from all particles in $B$ and vice versa. Further,
we invoke a criterion of well-separateness (the generalization of the
opening criterion) that, if fulfilled, says that the relevant
interactions can be accounted for by a simultaneous multipole
expansion of the mutual interaction field both at $A$ and $B$. In this
case, the corresponding field expansions are stored in $A$ and $B$
(actually, they are added to whatever field expansion the nodes may
already have). Otherwise, one of the two nodes (or both, which is what
we will end up doing) is split into all its daughter nodes and paired
up with the unsplit node, followed by calling the function again for
all the newly formed pairs. Self interactions between two identical
nodes (or particles) are always ignored; instead, the self-interaction
is treated by calling the interaction function for all possible pairs
of daughter cells in the node.

For the node opening decisions, we employ slightly modified variants
of the criteria we use for the one-sided tree algorithm. If the
geometric criterion is in use, a node-node interaction is evaluated if
\begin{equation}
  \frac{l_1 + l_2}{r} < \theta_c ,
  \label{eqnNodeDec}
\end{equation}
where $l_1$ and $l_2$ are the side-lengths of the two nodes
involved. For equal node sizes and the same value of $\theta_c$, this
means that the nodes see each other typically under a smaller
effective angle than in the one-sided tree algorithm, but as we see
later, this is also necessary for achieving roughly comparable
accuracy due to the error of the field expansion on the sink's side
that is
additionally present in FMM.
Similarly, we require 
\begin{equation}
\frac{M_{\rm max}}{r^2} \left(\frac{l_{\rm max}}{r}\right)^{p-1} < \alpha
|\vec{a}|_{\rm min},
\end{equation}
for a node-node interaction when the relative opening criterion is
selected in FMM.  Here, $M_{\rm max} = \max (M_1, M_2)$ is the maximum
of the two nodes' masses and $l_{\rm max} = \max (l_1, l_2)$ the
maximum of their side lengths. Such a simple power-law
  characterization of the force errors gives a typically close but not
  strict characterization of the errors, see \citet{Dehnen:2014aa},
  who also shows how the error estimates can be sharpened in principle
  by exploiting the lower order multipole moments.  We define
$|\vec{a}|_{\rm min} = \min(|\vec{a}|_{1}, |\vec{a}|_{2})$ as the
minimum acceleration occurring among any of the particles in the two
nodes.  The corresponding minimum $|\vec{a}|$ for every tree node is
determined during tree construction as an additional node
property. Note that the power with which the effective opening angle
$\theta \approx l_{\rm max}/r$ enters is reduced by one unit for a
given expansion order $p$ compared to the criterion used in the
one-sided tree. Again, this accounts for the additional error induced
by the sink-side expansion in FMM.  If enabled, the use of a spatial
exclusion zone around a node is the same as in the tree. In practice
this means that we do not allow for interactions of nodes that
directly touch.

The calculation of all interactions can simply be initiated by calling
the interaction routine with two copies of the root node, with the
processing of required daughter node interactions done recursively.
When distributed memory parallelization is used, we nevertheless need
to also employ a stack on which pairs of nodes that still need processing
are temporarily stored in order to hide communication times.
 
Finally, in the third phase of the FMM algorithm, the field
expansions are passed down the tree, translated to new expansion
centres if needed, and summed in nodes, until they are eventually
evaluated at particle coordinates, delivering the total potential and
force for all particles.  To this end, we recursively walk the tree in
a top-down fashion, accumulating field expansions for the nodes by
shifting the expansion centre of a node's parent node to that of the
current node (this operation is straightforward for an expansion in
Cartesian coordinates), and then adding the expansion coefficients,
until one arrives at single particles for which the field is evaluated
and added to the force/potential the particle may already have
acquired through individual cell-particle interactions in phase two
above.

As noted earlier, one advantage of FMM lies in the manifest momentum
conservation possible for this method, meaning that all the force
errors add up to zero to machine precision. This is in general not the
case for the ordinary tree algorithm.  There are however also some
disadvantages of FMM. One is that partial force calculations, where the
number of sinks is much smaller than the number of source particles
and which occur in many standard local timestep integration schemes,
are not well matched to the FMM approach, because the latter is
designed to compute the forces for {\em all} particles in a globally
efficient way. If forces only for, say, 5\% of the particles are
desired, the method will not just take 5\% of the computational
effort, but rather something still close to the full effort.  In
contrast, the one-sided tree algorithm delivers 
``linear elasticity'' in the computational cost. But we note that
this disadvantage of FMM completely disappears when a hierarchical
time-integration scheme (see Section~\ref{sechierarchicaltime}) is
used, because here all force calculations always involve equal sets of
source and sink particles.

Another slight technical complication lies in the treatment of
different gravitational softening lengths for the particles making up
nodes. Finally, the parallelization of FMM for distributed memory
machines is more involved than for an ordinary tree code,
even though recently some successful implementations have been
described \citep[e.g.][]{Potter:2017aa}. Our approach for dealing with
distributed memory parallelisation of FMM will be described in
Section~\ref{secParallel}.

\subsection{The FMM-PM approach}

We can readily generalize the TreePM approach to a FMM-PM
approach. The PM part is the same in both cases, only the evaluation
of the short range force is now carried out with the FMM method
instead of the one-sided tree. This requires a change of the
real-space Green's function $g(\vec{r})$ with respect to the Newtonian
case $g(r) = 1/r$, just as in TreePM.  Again, we use a cut-off
distance $r_{\rm cut}$ and restrict the evaluation of FMM interactions
to distances $r < r_{\rm cut}$.  Over this range, we tabulate the
interaction potential and all its higher derivatives -- which are
needed now for FMM -- in a one-dimensional look-up table from which we
interpolate with bi-linear interpolation, very similar to the TreePM
approach.

Interactions between two nodes are dropped (and hence also not further
refined) when the distance of their nearest sides exceeds the cut-off
distance $r_{\rm cut}$. We do not modify the opening criteria
themselves, however, and when the relative opening criterion
  is used, the estimated force errors are always compared against the total
  gravitational accelerations of the particles, including the PM
  contributions. As we will show later on explicitly, the multipole
approximations remain sufficiently accurate also in the transition
region where the short-range force law drops off very rapidly.  note
that recently, \citet{Wang:2021aa} has also described a combination of
FMM and PM within their code {\small PHOTONS-2}, which differs however
in a number of points from our approach.

\subsection{Accelerating short-range force calculations through a
  secondary mesh}

In so-called ``zoom-simulations'' the particle density can vary
extremely strongly within a simulation domain. In this case, a single
grid covering the full simulation box may yield only a limited
speed-up of the TreePM or FMM-PM approaches, simply because most
high-resolution particles can then be contained within one or a couple
of cells of the ``low-resolution'' PM-grid of size $N_{\rm grid}^{\rm LR}$
covering the full periodic box, and hence lie all within a single
short-range tree walk region $r_{\rm cut}^{\rm LR}$ of the
corresponding force split.  To remedy this, {\small GADGET-2}
introduced the possibility to place a secondary high-resolution mesh
 onto a certain ``high-resolution zone'' in order to further
enhance the efficiency of the TreePM approach in such simulations.

\begin{figure}
  \begin{center}
    \resizebox{6cm}{!}{\includegraphics{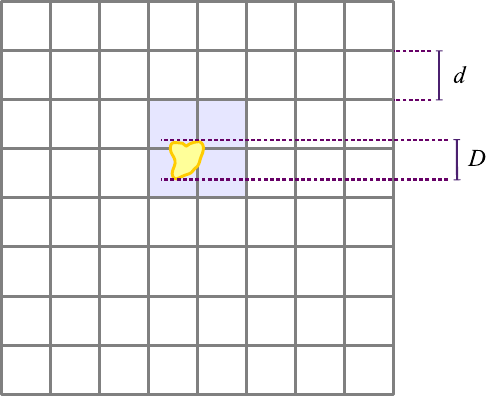}}
    \end{center}
    \caption{Placement of a
      secondary PM mesh (shaded) over
      a high-resolution particle region (yellow) such that
      an alignment with the underlying oct-tree geometry results. 
      To determine the spatial extent of a commensurable secondary PM mesh,
      we first
measure the maximum diameter $D$ of the high-resolution zone, which
in turn is determined by the spatial positions of all particles of a designated
particle type. Next, we compute the smallest power-of-two subdivision
$d$ of the full box-size that is still larger than $D$. This
identifies a fiducial regular Cartesian mesh with cell size $d$ covering the
simulation domain, lining up with one level of the global oct-tree.
All cells in this mesh that are touched by the high-resolution
region become part of the region covered by the secondary PM
mesh. The union of these cells is shrink-wrapped (imposing a cubical
footprint) to determine the final 
extension and location of the secondary PM mesh.
\label{FigSecondaryMeshPlacement}}
\end{figure}

{\small GADGET-4} supports this in a similar spirit as {\small
  GADGET-2}, but with a number of improvements and changes.  In
practice, one or several particle types are designated to identify the
spatial region where a PM calculation with a finer grid size is
desired. The code automatically determines the smallest region
enclosing these particles, and its maximum diameter $D$ in any of the
principal axes directions.  We then determine the size of the next
larger node of size $d \ge D$ in the oct-tree covering the whole
volume. The high-resolution region is now enlarged such that it lines
up with the node boundaries in the corresponding level of the
oct-tree, and has an overall cubical shape, as sketched in the example
shown in Figure~\ref{FigSecondaryMeshPlacement}.  Note that as a
result, all tree nodes of this corresponding level of the tree as well
as deeper ones (i.e. with smaller nodes) either fully overlap with the
high-resolution region, or do not overlap at all. This is a feature we
exploit in carrying out a clean force split that maintains manifest
momentum conservation also in case a secondary PM mesh is used.

The enlarged high-resolution region is now covered with a secondary
PM mesh, which is used to compute intermediate-scale forces for the
mass contained within in it, using a zero-padded FFT to realize vacuum
boundary conditions. The interaction kernel of this intermediate force
is that of the short-range force corresponding to the coarse
background mesh that covers the full box, minus the short-range force
corresponding to the high-resolution mesh.

Forces for particles  inside the high-resolution region are now
computed as follows. Their total force is the sum of the ordinary
background PM-mesh and  the intermediate mesh force, augmented with a
tree (or FMM) force that uses different short-range kernels depending on
whether the interacting partner node is contained inside the high-res
tree node, or whether it is outside. In the former case, the shorter
interaction region $r_{\rm cut}^{\rm HR}$ of the high-res PM mesh is used,
in the latter case, the more extended region $r_{\rm cut}^{\rm LR}$ of the
background PM-mesh is used. For particles outside the high-resolution
tree node, nothing changes compared to the plain TreePM or FMM-PM
algorithms, i.e.~their long-range force is simply the force of the
background PM-mesh and their short-range force is computed with the
normal, more extended cut-off $r_{\rm cut}^{\rm LR}$ for all other particles.

Note that in this tree algorithm there may never be an interaction of
a particle inside the high-res region with a tree node that {\em
  partly} covers the high-res region, or in other words such nodes
must always either fully contained or fully outside the high-res
region. We achieve this because we geometrically align our high-res
region with a level of the oct-tree, and we disallow interactions
where a node partially overlaps with the geometric bounds of the
high-res region, regardless of the opening criterion. Likewise, for
the FMM algorithm, we only allow interactions between nodes that are
either fully nested inside the high-res region, or lie fully
outside. Effectively, this hence imposes an additional opening
criterion according to which nodes that partly overlap with the
high-resolution region are always opened.  This allows interactions
between any pair of particles where both are inside the high-res
region to be treated with a sum of two PM forces and a short cut-off
$r_{\rm cut}^{\rm HR}$ for the multipole expansions, whereas all other
particle pairs are treated just with a single PM force and the
corresponding background cut-off $r_{\rm cut}^{\rm LR}$. Note that
this approach retains manifest force antisymmetry when FMM is used,
because the PM algorithm itself also delivers antisymmetric forces.

Because we have $r_{\rm cut}^{\rm HR} < r_{\rm cut}^{\rm LR}$, the
tree walks can be terminated earlier for the high-resolution particles when
the secondary mesh is used, reducing the number of interactions that
have to be evaluated for them\footnote{Technically, one could violate
  the condition $r_{\rm cut}^{\rm HR} < r_{\rm cut}^{\rm LR}$ by
  choosing a very fine mesh for the background computation, and a very
  coarse mesh for the high-resolution mesh, negating any advantage the
  secondary mesh is supposed to bring. The code rejects such a
  choice.}. This constitutes the opportunity to speed up this part
of the calculation. On the other hand, one needs to invest extra work
for a further PM calculation, which due to the required zero-padding
tends to be more expensive for a given grid resolution than the
periodic FFT that can be used for the full box. And in the tree walk,
we have to open extra nodes to avoid using nodes that overlap both
with the high-res and the low-res regions. Whether or not this is
worthwhile overall depends strongly on the particular setup
encountered in practice.  A minimum prerequisite for seeing a
beneficial impact on the run-time is that the region that still needs to be
covered by the tree walks around high-res particles,
$(4\pi/3) ( R_{\rm cut} r_s)^3 $, needs to be much smaller than the
volume $(L^{\rm HR})^3$ of the full high-resolution region, otherwise substantial
savings from being able to discard tree nodes can hardly be expected. 
Because of zero padding, the grid resolution is $d^{\rm HR} = 2 L^{\rm
  HR} / N_{\rm grid}^{\rm HR}$, so that this corresponds to the condition
\begin{equation}
N_{\rm grid}^{\rm HR} >> 2 \left(\frac{4\pi}{3}\right)^{1/3} A_{\rm smth} R_{\rm
  cut},
\end{equation}
where $N_{\rm grid}^{\rm HR}$ is the full size (including zero padding) of the
high-res PM mesh, and $r_s = A_{\rm smth}\, d^{\rm HR}$. Evidently, this
tends to be easier to fulfil if one has a large problem size (which
makes using a large $N_{\rm grid}^{\rm HR}$ worthwhile), and if one does not
aim for very high accuracy (which means that a low value for
$A_{\rm smth}$ is sufficient, as we will discuss later). We stress
that this criterion provides only a rough indication for
the regime in which a secondary mesh could in principle be beneficial.

We note that in contrast to the approach above, {\small GADGET-2} required that
the high-res particles interacted with {\em all} other particles using
the short range kernel, which meant that the high-resolution PM-mesh
needed to cover also a buffer region of size $r_{\rm cut}^{\rm LR}$
outside the designated high-res region.  As for extreme zooms one can
have $L_{\rm HR} << r_{\rm cut}^{\rm LR}$, this meant that in this
limit the effective extension of the high-resolution mesh was actually
not determined any more by the size of the high-resolution region,
limiting the dynamic range that could be bridged by the use of a
single secondary mesh. In addition, not all interacting pairs of
particles were treated symmetrically in this approach. Both
disadvantages are eliminated by the refined procedure in {\small GADGET-4}.
Another practical improvement is that the high-resolution region may
now overlap with the periodic box boundaries in the initial conditions
or during the evolution, which was previously not supported.

\subsection{Gravitational softening}

In collisionless dynamics, gravitational softening needs to be
introduced to prevent the formation of bound particle pairs, and to
protect against the occurrence of large forces and large angle deflections
(which on top require short timesteps for proper orbit integration)
when particles pass by close to each other. In practice, two
questions have to be clarified in the force computation when
individual softenings are assigned to particles and the potential is
modified at short distance scales. First, how should the effective
softening length of each interacting particle pair be obtained from
the values assigned to the two particles? Second, how should the
softening be treated in the tree algorithm, and in the calculation of
the multipole expansions?

For the first question, we always adopt the conservative choice that the
softening of any interacting pair of particles should be the larger of
the two, i.e.~we adopt 
\begin{equation}
\epsilon_{ij} = \max (\epsilon_i, \epsilon_j).
\end{equation}
This preserves force antisymmetry and respects the idea that all
interactions of a particle should be softened at least with the
softening value assigned to it.

With respect to the second question, we store in the tree algorithm
for each tree node the largest $\epsilon_i^{\rm max}$ and smallest
softening $\epsilon_i^{\rm min}$ occurring among its constituent
particles.  In case the
distance $r$ between a node and a particle (or a node and another node
in the case of FMM) is larger than the maximum
$\epsilon_{ij} = \max (\epsilon_i^{\rm max}, \epsilon_j^{\rm max})$ of
the largest softenings of the two nodes, the interaction is treated
unsoftened, i.e.~the multipole expansion is computed for the Newtonian
interaction kernel. If the opposite is the case, we apply the
softening $\epsilon_{ij}$ to the interaction of the two nodes, {\em
  provided} one can be certain that individual particles in the nodes
are not treated with an excessively large softening because of this.
In particular, one needs to protect against applying a softened
multipole expansion to nodes with a mix of particles with different
softening lengths, containing particle pairs between the nodes with
smaller symmetrized softening than the node-level
$\epsilon_{ij}$. This can only happen if both
$\epsilon_i^{\rm min} < \epsilon_{ij}$ and
$\epsilon_j^{\rm min} < \epsilon_{ij}$ are true. This is never the
case for particle-particle interactions, and also not for interactions
involving a node for which both $\epsilon^{\rm max}$ and
$\epsilon^{\rm min}$ are equal to $\epsilon_{ij}$. Hence, if the
conditions are not both true, we use the softened interaction,
otherwise we proceed in the tree walk by opening the interaction.

In order to minimize the storage needs for $\epsilon_i^{\rm min}$ and
$\epsilon_i^{\rm max}$ in the nodes as well as in the particle data
itself, we use a discrete set of allowed softening values. Each
particle then selects one of these softenings by being assigned a
``softening class'', which is simply a one-byte value that indexes a
global table with the available softening values. Correspondingly, only
a single byte is needed for $\epsilon_i^{\rm min}$ and
$\epsilon_i^{\rm max}$.

We note that in {\small GADGET-3}, softened interactions between cells
and particles were never used and such nodes were always opened,
i.e.~softening was only allowed among particle pairs.  However, this
has the severe disadvantage that in regions where the particle density
is high and the gravitational softening is comparable to the mean particle
spacing, or if it is chosen deliberately much larger than this, the
tree calculation locally degenerates to a $N^2$ behaviour and ends up
computing a large number of two-particle pair-wise forces. In
practical applications, this can happen, for example, in so-called
zoom simulations when relatively large softenings are applied to heavy
boundary particles, and those happen to contaminate the densely
sampled high resolution regions.  While our allowance for softened
tree nodes alleviates this problem, it cannot eliminate it in all
situations. For example, it is easy to see that mixing two particle
types and assigning to one a considerably larger softening than the
mean particle spacing can slow down the tree algorithm
tremendously. If possible, it is thus advantageous to refrain from
using very different softenings for particles that are spatially well mixed.

\subsection{Periodicity only in two dimensions}

A new feature in {\small GADGET-4} is the option to have periodicity
of gravity only in two spatial dimensions, while the third dimension
remains non-periodic. This is intended mostly for simulations of
stratified or shearing boxes that are frequently used to study small
patches of the interstellar medium of disk galaxies at high resolution
\citep[e.g.][]{Walch:2015aa, Simpson:2016aa}. Self-gravity with
periodic boundary conditions only in the directions parallel to the
plane of the disk is needed in such simulations.

For this setup, an Ewald decomposition of the slowly converging
potential sum can be
derived \citep{Grzybowski:2000aa}, in the form
\begin{eqnarray}
\psi(\vec{r}) & = & \frac{1}{|\vec{r}|} + \frac{2\alpha}{\sqrt{\pi}}
- \sum_{\vec{\tilde{p}}} \frac{{\rm
erfc}(\alpha|\vec{r}-\vec{\tilde{p}}|)}{|\vec{r}-\vec{\tilde{p}}|} 
\\
& & \hspace*{-1cm}-\frac{\pi}{L_x L_y}\sum_{\vec{k}\ne 0}  
\frac{\exp\left(i\,\vec{k}\cdot\vec{r}\right)}{ |\vec{k}|} 
\left[ \exp(k z)\, {\rm erfc}\left(\frac{k}{2\alpha} + \alpha z\right)
\right. \nonumber \\
& & \left. +\exp(-k z)\, {\rm erfc}\left(\frac{k}{2\alpha} - \alpha z\right)
\right] \nonumber \\
& & +\frac{2\sqrt{\pi}}{L_x L_y}
    \left(\frac{\exp(-\alpha^2z^2)}{\alpha} + \sqrt{\pi} \,z \,
{\rm erf}(\alpha z)\right) , \nonumber  
\end{eqnarray}
where for definiteness we have assumed that the $z$-direction is the
non-periodic dimension (our implementation is flexible in this
regard). Here $\vec{r}=(x, y, z)$, and the sum over $\vec{\tilde{p}}$
extends only over the periodic replicas in the $x$- and
$y$-dimensions. Similarly, the frequency integral over ${\vec{k}}$
extends only over the first two dimensions. We note that recently
\citet{Wunsch:2018aa} implemented a tree-based gravity solver with
mixed boundary conditions in the {\small FLASH} \citep{Fryxell:2000aa}
code. They derive an approximate expression for the potential of the
mixed boundary case which differes from the analytic form above, but
which appears to give sufficiently accurate results in practice.

Similarly to the ordinary periodic case, we calculate the force in the
one-sided tree algorithm by taking account of the nearest periodic
image in the $x-$ and $y$-directions with the ordinary Newtonian
(softened) kernel, and by supplementing this with a correction force
obtained from the above expression, evaluated through a Taylor
expansion around the nearest point in a precomputed look-up
table. Alternatively, we also support the TreePM and FMM-PM methods
for mixed boundary conditions. Here the PM calculation is modified
accordingly. In particular, we determine the Green's function in
Fourier space by first setting it up in real space with zero padding
in the non-periodic dimension, and then transforming it to $k$-space.

\subsection{Stretched boxes}

Related to the above, we have generalized the gravitational algorithms
of {\small GADGET-4} such that they also support periodic boundary
conditions for ``stretched'' boxes, i.e.~the employed simulation
domain does not have to be cubical in shape but may be stretched by
different factors in each of the dimensions. This feature is not only
available for periodicity in two dimensions as discussed above, but
also for ordinary gravity with periodic boundary conditions in three
dimensions. It works both for the one-sided Tree and for FMM, and also
for TreePM and FMM-PM, respectively.

However, in the latter case, certain constraints for the stretch
factors are enforced such that fully filled tree nodes remain cubical,
and that there is an integer number of PM grid cells in each spatial
dimension.  The group finders and the hydrodynamical solver are
operational for stretched boxed as well, but the placement of a
secondary high-resolution PM mesh is not supported.

\subsection{Integer coordinates}

Ordinary floating point numbers are not ideal for representing the
particle coordinates in our typical simulation setups, because they
feature variable absolute positional accuracy throughout a simulation
box (unless one restricts the numbers to specific factor of two
ranges, for example by using the $[1,2]$ interval, such that the
mantissa alone linearly encodes the position, as exploited in the Voronoi
mesh construction algorithm of the {\small AREPO} code). Because our
coordinates have bounded minimum and maximum values, at least a subset
of the bits reserved for the floating point exponent are unused.

As an alternative it is attractive to consider integers for storing
the coordinates in the box, as this delivers uniform resolution and
can make optimum use of the bits assigned for the storage. When using
32-bit variables, the use of integers gives a relative positional
accuracy within the box equal to $1/2^{32}\sim 2.33\times 10^{-10}$
instead of the machine precision of single precision floating point
numbers, $\approx 10^{-7}$. This substantially enlarges the range of
science applications where 32-bit values for the coordinates are
sufficient. When double precision positions are replaced by 64-bit
integer storage, the available relative accuracy in positional storage
goes up to $5.42\times 10^{-20}$, enough to resolve Earth's diameter
in a box one Gigaparsec on a side, while for double precision floating
point numbers it would be a thousand times worse.

A further advantage of integers is that it becomes particularly easy
to determine both, a periodic mapping into the simulation box as well
as finding the (signed) distance vector to the nearest periodic image
of a point. Both just arise automatically without any branching in the
code from the properties of the 2's-complement used to represent
signed integers, and the way integer overflows are treated on virtually
all microprocessors used in high-performance computing. For example,
imagine we use one unsigned byte to represent 256 possible positions
along one periodic dimension of a simulation box. If a particle is at
a position $x_1=200$ and another one at position $x_2=30$, then taking
the distance of particle 1 relative to particle 2 one gets
$x_1 - x_2 = 170$, but interpreting the result as a signed integer,
one gets $-86$, which is the distance of the nearest periodic image of
$x_1$ to $x_2$ (which in principle lies at coordinate
$200-256 = -56$). Similarly, if we add a displacement to $x_1$, say a
shift by $+100$ units, we do not obtain $300$ as new coordinate, but
rather $44$ due to overflow -- but this is exactly the correct
coordinate we should get due to periodic wrap-around.

Integer coordinates also simplify the oct-tree construction, as they
allow the use of fast bit-shift operations to determine in which
(daughter) node a particle coordinate falls. The absence of numerical
floating point round-off also eliminates any ambiguities in geometric
predicates related to the tree construction.  Likewise, the mapping
onto Peano-Hilbert keys can be elegantly done in a `loss-less' fashion
due to the absence of floating-point integer conversions between
coordinates and Peano-Hilbert keys.

For all these reasons, coordinates are internally stored as either
32-bit, 64-bit or even 128-bit integers in {\small GADGET-4}. Relative
distances (i.e. coordinate differences) are first computed in integer
and then converted to floating point values for subsequent
calculations (such as determination of multipole moments). We also
convert back to floating point values on input and output of particle
data from initial conditions and snapshot files, respectively.  A
further possibility would be to reduce the number of significant bits
that require storage by considering only the integer offset of a
particle to a nearby neighbour, as proposed by \citet{Yu:2018aa}. As
we anyhow order the particles internally along a space-filling
Peano-Hilbert curve, this could naturally allow a loss-less
compression scheme that can substantially reduce the memory
requirements for particle strorage. \citet{Yu:2018aa} also discuss how
this idea can be extended to coarse meshes in velocity space. So far,
such optimizations are not yet considered in {\small GADGET-4}, except
for the possibility to use reduced precision in snapshot outputs for
velocity data (half-precision format), and to apply loss-less
compression to integer position data and particle IDs.

\subsection{Non-periodic potentials}

For simulations with non-periodic (vacuum) boundary conditions, the
tree and FMM calculations simplify considerably, because mapping to
nearest periodic images and Ewald corrections are not
required. However, here the required size of the root node, which
shrink wraps to the problem, may change during the system's
evolution. We determine the root-node automatically by shrink-wrapping
the initial particle distribution, and by mapping it onto a subregion of
our integer coordinates such that accidental nearest neighbour
wrapping is avoided.  Also, the code detects if a particle moves out
of the boundaries of the original root node, in which case a new root
node extension is determined.

Also in the isolated case, the Tree or FMM calculation can be combined
with a PM acceleration for the long-range forces. In this case we
implement non-periodic boundary conditions in the PM solver through
zero-padding, i.e.~the FFT size effectively used in each dimension is
doubled relative to the periodic case. The Green's function is set-up
in real space, Fourier-transformed to obtain it in $k$-space, and then
stored for subsequent use.

\section{Force accuracy tests} \label{secforceacc}

\begin{table}
\begin{tabular}{c|cc}
\hline
\multirow{2}{*}{basic algorithm}       &  \multicolumn{2}{c}{expansion order} \\
                                       &    from dipole & to triakontadipole \\
\hline
\multirow{2}{*}{just multipole expansion} &  Tree-O1 &  Tree-O5 \\
                           &  FMM-O1 & FMM-O5 \\
\hline
\multirow{2}{*}{multipoles with mesh} & Tree-PM-O1 &  Tree-PM-O5 \\
                                & FMM-PM-O1 & FMM-PM-O5 \\
\hline
{zoom runs: multipoles, mesh,} &  Tree-HRPM-O1 &  Tree-HRPM-O5 \\
{and additional high-resh mesh}                        & FMM-HRPM-O1 & FMM-HRPM-O5 \\
\hline
\end{tabular}
\caption{Overview of different gravitational force calculation
  algorithms supported in {\small GADGET-4}, wich are selected at
  compile-time.  Tree and FMM designate that the hierarchical
  multipole calculation is done either with a classic one-sided Barnes
  \& Hut tree code, or with a fast multipole method,
  respectively. These methods can be combined with a mesh covering the
  full simulation domain (PM~schemes), where Fourier methods are used
  to accelerate the calculation of the long-range forces. For zoom
  simulations, it is possible to place a second mesh on a
  high-resolution region (HRPM~schemes) that covers a fraction of the
  simulation domain where the particle density is particularly high.
  Each of these combinations can be run with different multipole
  expansion order, designated with O1 to O5, where the digits refer to
  the order parameter $p$. We also note that each of the methods can
  be run either with periodic or non-periodic boundary conditions, and
  with two different node opening criteria (a geometric opening angle,
  or our relative opening criterion). For periodic boundaries, also
  non-cubical domains are possible, but then the placing of an
  additional high-resolution mesh is not supported. Mixed boundary
  conditions with two dimensions periodic and one non-periodic are
  also supported, but in this case again a high-resolution mesh can
  not be used.  In the older {\small GADGET-2} and {\small GADGET-3}
  codes, only the schemes Tree, Tree-PM and Tree-HRPM at order $p=2$
  were available, while {\small GADGET-1} only offered
  Tree-O3. \label{TabGravSchemes} }
\end{table}

\begin{figure}
 \resizebox{8cm}{!}{\includegraphics{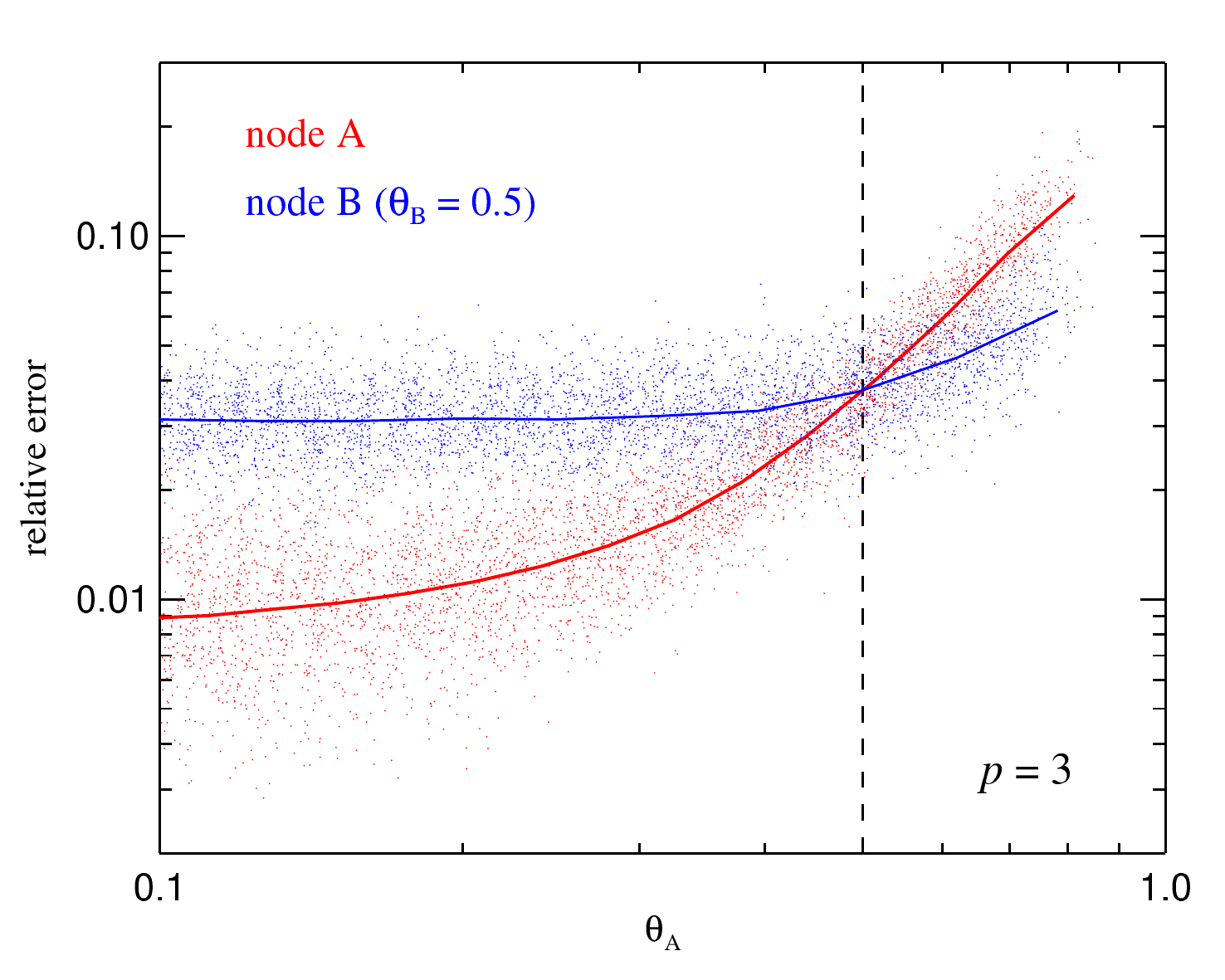}}
\caption{Relative force errors (individual points) and median force
  error (solid lines)
  for particles randomly distributed
  within two nodes if the interaction is computed with FMM (here for
  expansion order $p=3$), as a function of
  opening angle $\theta_{\rm A}$, for a fixed opening angle  $\theta_{\rm B} = 0.5$. Hence for
  $\theta_{\rm A} < 0.5$, node A is smaller than node B, whereas for
  $\theta_{\rm A} > 0.5$, node A is the larger of the two (the dashed
  vertical line indicates $\theta_{\rm A} = 0.5$). The errors
  for the particles in node A are shown in red, while those for node B
  are in blue. For asymmetric node sizes, the smaller node experiences
  the smaller relative force errors.
  \label{FigMedianForceTowNodesFixedBSize}}
\end{figure}

The extensive discussion in the previous section shows that quite a
number of different algorithms for the gravitational force calculation
are available in {\small GADGET-4}. Table~\ref{TabGravSchemes}
provides a schematic overview of these methods. It is now important
to test their accuracy systematically, and to establish how the force
accuracy depends on the multipole order and other parameters of the
chosen schemes.  Once the basic force accuracy is verified, we can
investigate the relative computational efficiency of the different
methods. We are also interested in the important question of
which algorithm delivers a given target accuracy with the smallest
computational effort.  Answering this question is non-trivial in
general as it can be problem-size and machine dependent, but given the
many choices that are possible in {\small GADGET-4}, it is important
to develop at least a basic understanding of the performance
implications of different algorithmic choices to facilitate the
adoption of close to optimum settings in practical applications.

\subsection{The FMM force accuracy between two interacting nodes}

It is instructive to begin by considering the force approximation accuracy of the
FMM approach for the interaction between two isolated cubical nodes as
sketched in Figure~\ref{FigFMMSketch}. To this end we place 10
particles of equal mass randomly in each of two cubical nodes, which
are themselves randomly placed with respect to each other, but with a
certain distance $r$ between the resulting centers of mass of each
node. We characterize the side lengths $l_{\rm A}$ and $l_{\rm B}$ of the two
nodes through their respective opening angles as seen from the other
node, i.e.~$\theta_{\rm A} = l_{\rm A}/r$ and
$\theta_{\rm B} = l_{\rm B}/r$. We then measure the force accuracy
delivered by FMM for each of the particles in A due to those in B, and
vice versa, for different opening angles and different multipole expansion
order. For every realization of this setup, we determine the mean
force error of the particles in each of the two nodes. To
accumulate statistics, we repeat the random setup and the measurement
many times over.

In Figure~\ref{FigMedianForceTowNodesFixedBSize}, we show the force
accuracy for $p=3$ as a function of opening angle $\theta_{\rm A}$ for a
\emph{fixed} setting of $\theta_{\rm B} = 0.5$, which effectively
varies the sizes of the two nodes relative to each other. The results reveal several
interesting trends (and they are qualitatively the same for different
expansion orders $p$). For $\theta_{\rm A} < \theta_{\rm B} =0.5$, the
node A is the smaller of the two, and its particles exhibit then also
a smaller relative force error than those in the larger node B. For
$\theta_{\rm A}> 0.5 $, the situation reverses, and now node B is
smaller and has the smaller relative errors. We hence conclude that
nodes should be of the same size in order to avoid an asymmetry in the
induced relative force errors.

Also note that the relative force errors for node A asymptote to a
value~$\sim 0.009$ in the limit $\theta_{\rm A} \to 0$. This is
because one then ends up seeing the effect of approximating the field
created by node B with a finite multipole order $p$. Notice that this
error is exactly the error one would get if the forces on particles in
node A would be computed with an ordinary tree algorithm and
node B is seen under an angle $\theta =0.5$. In the same limit
of $\theta_{\rm A} \to 0$, the forces on the particles in node B also
reach only a finite precision, even though node A now effectively
shrinks to a point mass. This is because the resulting point mass
field is still approximated within the volume of node B with a Taylor
expansion, creating a finite error at the location of the particles
within B. This type of error is absent in the ordinary tree
algorithm. Importantly, it typically \emph{dominates} the error budget
in FMM, because it tends to be larger than the error induced by the
source-side expansion.

\begin{figure}
  \resizebox{8cm}{!}{\includegraphics{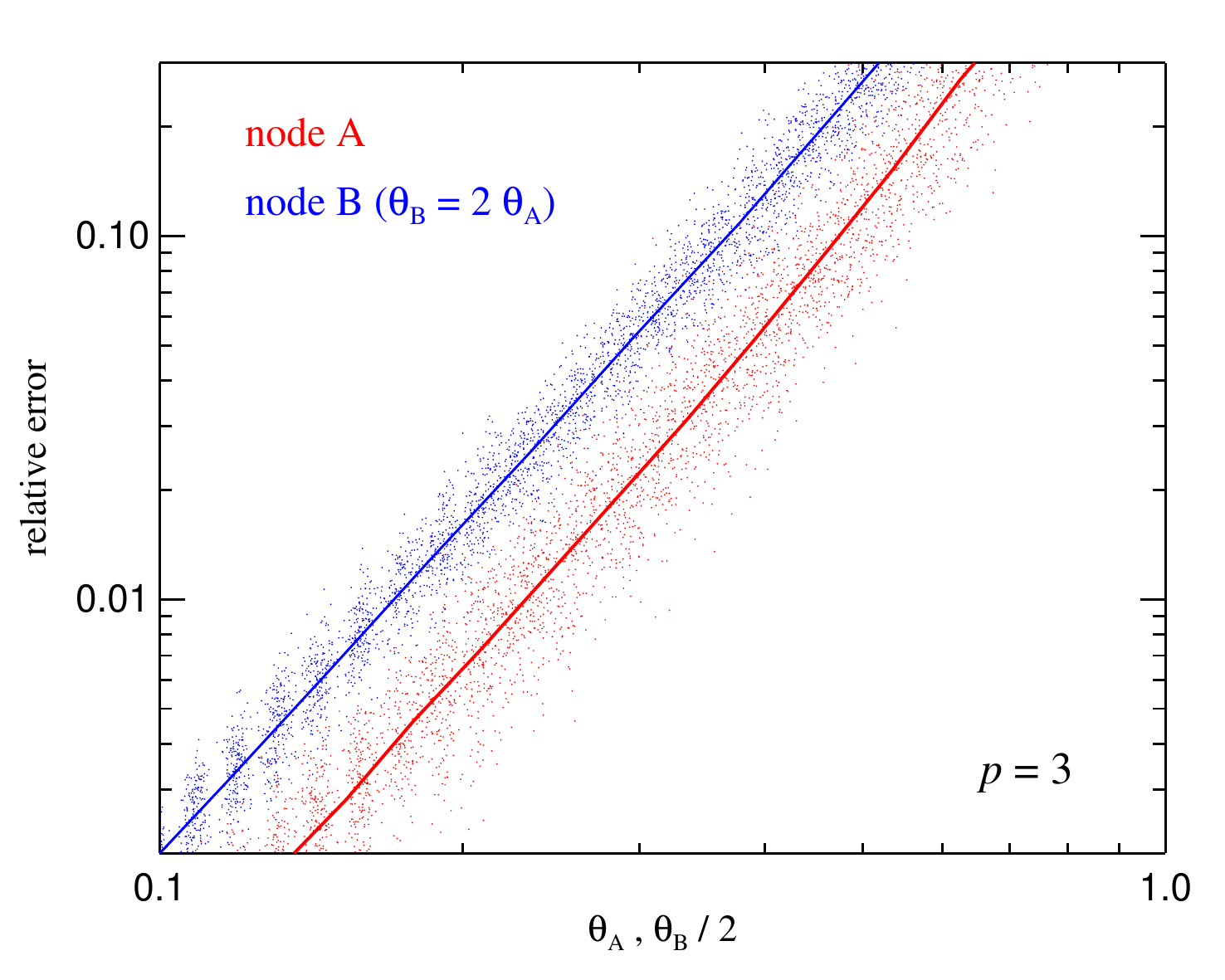}}\\%
  \resizebox{8cm}{!}{\includegraphics{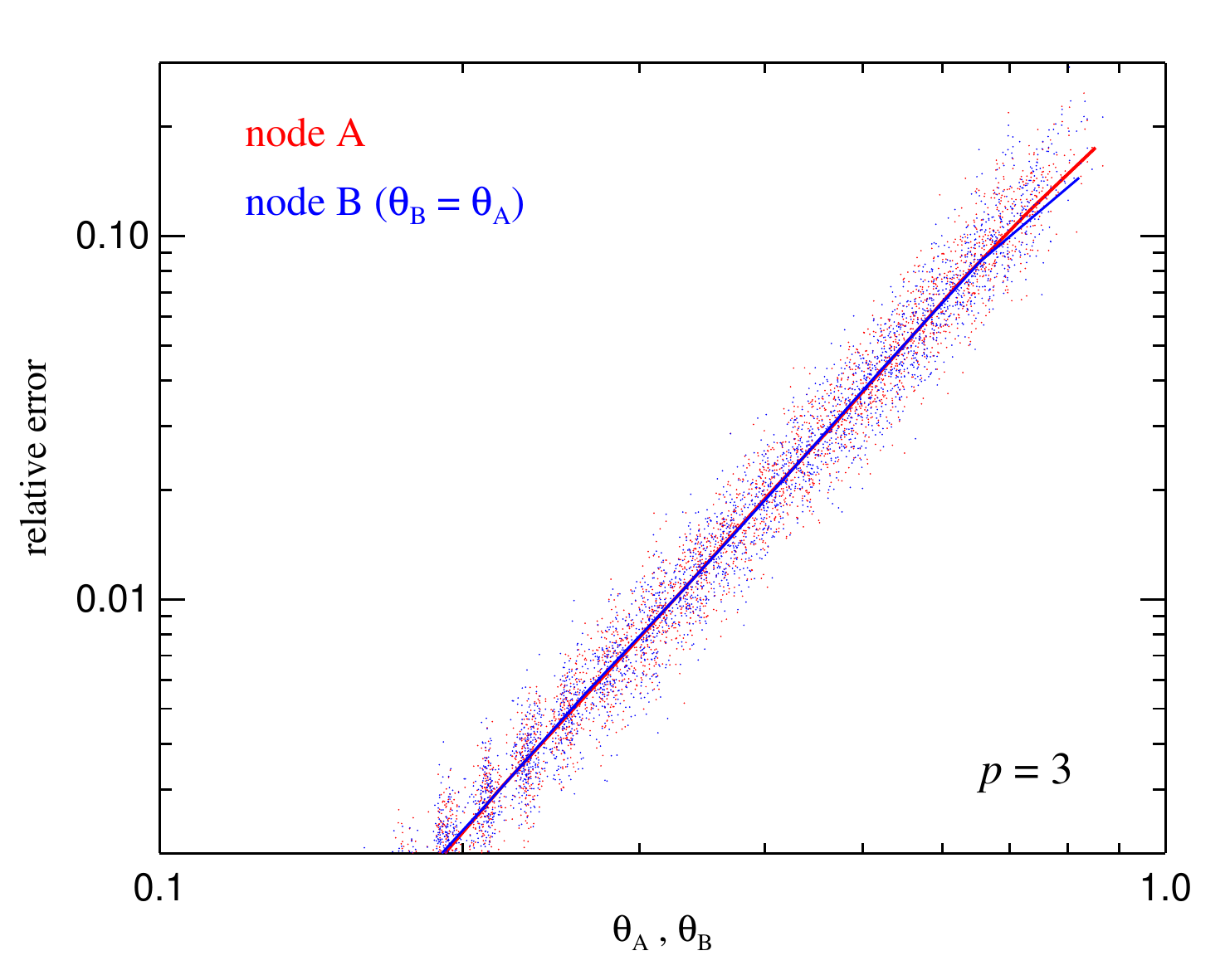}}
\caption{Relative force errors for particles randomly distributed
  within two nodes if the interaction is computed with FMM (for
  definiteness, with expansion order $p=3$),  as a
  function of opening angles. Points show individual force errors,
  solid lines medians. In the top panel, node B is always
  chosen twice as large as node A (with results for B plotted also at
  the corresponding $\theta_{\rm A}$ value), whereas in the bottom panel, an
  equal node size is enforced. Only in the latter case, the relative
  force errors show a symmetric and equal distribution at given
  opening angle -- apart from being
  smaller overall than for asymmetric node sizes.
  \label{FigMedianForceTwoNodesEqualNodeSizes}}
\end{figure}

The recommendation that equal node sizes are preferable is reinforced
by the results of Figure~\ref{FigMedianForceTwoNodesEqualNodeSizes},
where we show in the top panel measurements of the force error scaling
when one of the two nodes is twice the size of the other. Note that
this situation would quite commonly arise when applying FMM to an
oct-tree and always opening the larger of two nodes, as done in
original algorithm by \citet{Dehnen:2000aa}.  While the force
errors still scale as power-laws with $\theta^{p}$ in this case, they
are highly asymmetric in their size distributions between the two
nodes, which is undesirable. We therefore always \textit{open both nodes
simultaneously} in our FMM implementation if the opening criterion is
triggered, guaranteeing a situation of equal node sizes whenever two
nodes actually interact.  The lower panel of
Fig.~\ref{FigMedianForceTwoNodesEqualNodeSizes} displays the
corresponding error sizes in this case, which are symmetric and lowest
overall for the given opening angle.

\begin{figure}
  \resizebox{8cm}{!}{\includegraphics{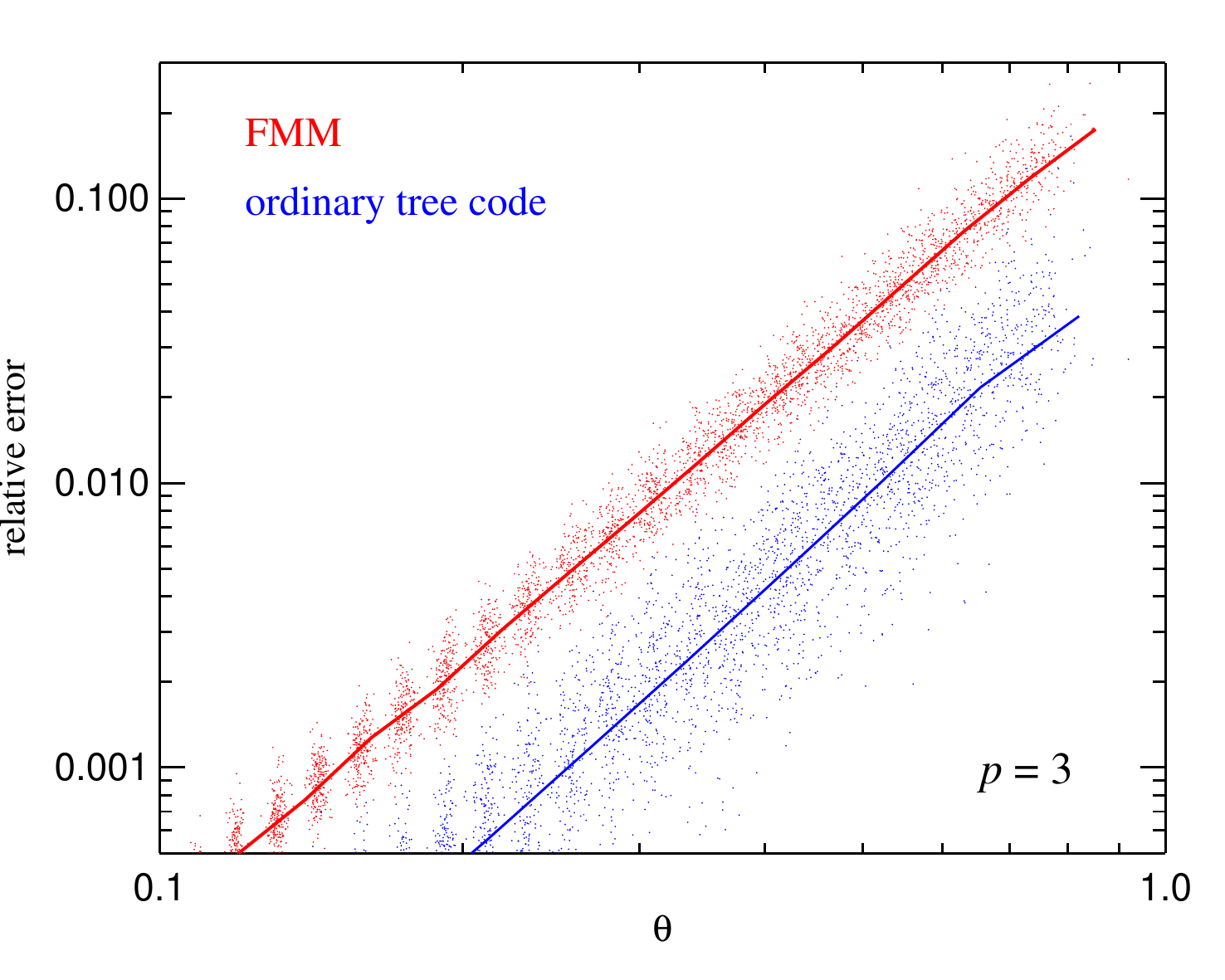}}
  \caption{Relative force errors for particles randomly distributed
    within two equal-sized nodes as a function of the opening angle if
    the interaction is computed with FMM (red) or an ordinary tree
    (blue), here for expansion order $p=3$ (the results for different
    multipole order are qualitatively similar). Points show individual
    errors, while solid lines give medians.  As expected, for equal
    expansion order and opening angle, the tree errors are lower than
    those of FMM because the latter needs to do an additional
    expansion at the sink side. However, the force errors for single
    node-node interactions scale with the same order in both
    approaches (i.e.~the slopes of the solid lines are equal), as
    expected based on the truncation error of the underlying multipole
    expansion.
    \label{FigMedianForceTowNodesTreevsFMM}}
\end{figure}

Having established that we prefer to always use
$\theta_{\rm A}= \theta_{\rm B}$ for FMM in practice, it is
interesting to compare the error sizes in this situation to those of
the ordinary tree, as a function of the same opening angle. This is
shown in Figure~\ref{FigMedianForceTowNodesTreevsFMM}, for expansion
order $p=3$ (other orders behave qualitatively very similarly). In
both cases, the errors decline as power-laws with smaller opening
angle, following the expected scalings based on the underlying Taylor
expansion. The FMM error is however offset relative to the tree by an
approximately constant factor of $\simeq 4$. This compares well to the
difference in the error level seen in
Fig.~\ref{FigMedianForceTowNodesFixedBSize} for node A between
$\theta_{\rm A}=0.5$ (i.e.~the intersection of the two lines), and for
$\theta_{\rm A} \to 0$. It corresponds to the accuracy price one pays
in FMM for doing a Taylor expansion at the sink side as well. Of
course, this loss of precision can be compensated for by using a
correspondingly smaller value for the opening angle $\theta$ in FMM
than for the tree code. In fact, based on these results, we would
expect that $\theta_{\rm FMM} \simeq \theta_{\rm Tree} /2 $ may
deliver a comparable force accuracy for FMM and tree for $p=3$.  But
such a rule of thumb can be problem dependent, hence it is important
to measure the achieved force accuracies for particle distributions
that are of interest in practical applications.  Note that our
geometric opening criterion for FMM given in eqn.~(\ref{eqnNodeDec})
will actually imply $\left< l/r\right>_{\rm FMM} \simeq
\left< l/r \right>_{\rm Tree}/2$
for a  prescribed numerical
value of the maximum allowed opening $\theta_c$ in the code's
parameterfile, so the typical opening angles for FMM will
automatically be smaller by a factor of 2 compared to the Tree, so
that roughly comparable force accuracy should result for an
unchanged setting of  $\theta_c$.

\begin{figure}
  \resizebox{8cm}{!}{\includegraphics{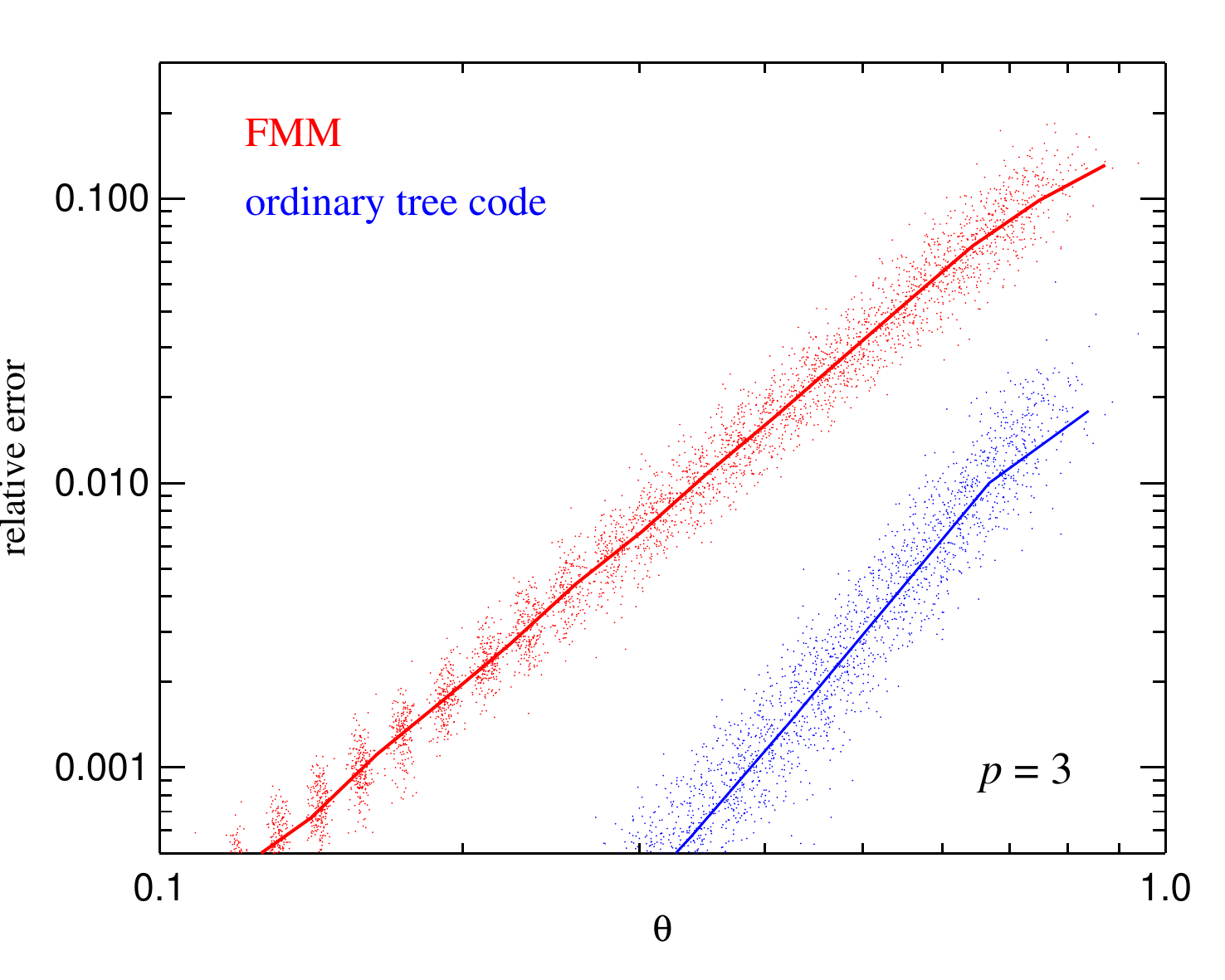}}\\%
  \caption{Relative force errors for the sum of the forces computed
    for 100 repeated calculations of the interaction of two equal
    sized nodes A and B, with randomly placed particles in them (and
    expansion order $p=3$, for definiteness).  We
    consider only the force errors for particles in node A (with
    points marking individual particles, solid lines showing medians),
    which are kept fixed for the repetitions, whereas each time we
    replace the particles in node B with a newly drawn set of
    particles. In this experiment, the force errors measured for the
    tree are reduced compared to a single node-node interaction
    (compare with
    Fig.~\ref{FigMedianForceTowNodesTreevsFMM}) due to a
    cancellation of force errors in different random
    directions. This 
    effect is absent in FMM, because here the force errors are
    dominated by the sink-side expansion within node A, which stays
    equal in this experiment. For a full particle distribution, a
    given node will interact with a number of different nodes of similar size and at
    similar distance, akin to our set of 100 repeated realizations of
    a stationary  node B, suggesting that cancellation effects of
    force errors should occur in more
  beneficial ways for the tree algorithm than for FMM.   
  \label{FigMedianForceTowNodesSpecial}}
\end{figure}

In fact, depending on the particle distribution of an N-body system,
FMM will often require an even more aggressive reduction
of the opening angle in practice to exhibit a similar final force error as
the tree algorithm. The reason for this is somewhat subtle and has to
do with the way the force errors of different node-node (for FMM) or
particle-node (for tree) interactions combine. As the total force on a
particle in an N-body system arises from the sum of the forces of many
multipole interactions, so does the total force error. If the
individual force errors are all \emph{uncorrelated}, they tend to average out
in part. In particular, if one has $N$ similar partial forces pointing
in a similar direction and making up the total force, each with
uncorrelated relative force error, then the relative error for the
total resulting force would see a $1/\sqrt{N}$ reduction compared with
the relative errors in the partial forces.

We can check whether such an effect is present with a variant of our
earlier node accuracy test. To this end, we repeat the computation of
the forces for a given geometry and placement of nodes A and B a
number of times (say 100 times), but each time we draw a new set of
particles within B. We are then interested in the total force exerted
on the particles in A (i.e. the sum of the forces of the 100 trials),
and consider the corresponding force
error. Figure~\ref{FigMedianForceTowNodesSpecial} shows the result,
comparing FMM and the Tree-based calculation. Comparing to the
corresponding result shown in
Fig.~\ref{FigMedianForceTowNodesTreevsFMM} for a single particle
realization, we see that the force errors for the tree have gone
indeed down significantly, by nearly a factor of 10 for small opening
angles, as expected by a $1/\sqrt{N}$-scaling where $N$ is the number
of forces that are added. In contrast, the FMM result for particles in
node A has \emph{not} changed, and no significant cancellation among
the added force errors is seen. The reason is that the FMM force
errors are dominated by the expansion on the sink side, and these
errors are correlated with each other and are not random in direction,
because a target particle in A has always the same location with
respect to A's center-of-mass. As we will see in the next section,
this effect is present in realistic particle distributions as well,
giving the tree algorithm an extra accuracy boost from cancellation of
uncorrelated errors that is largely absent in FMM.

\subsection{Force errors for an isolated particle distribution}

\begin{figure*}
\resizebox{8cm}{!}{\includegraphics{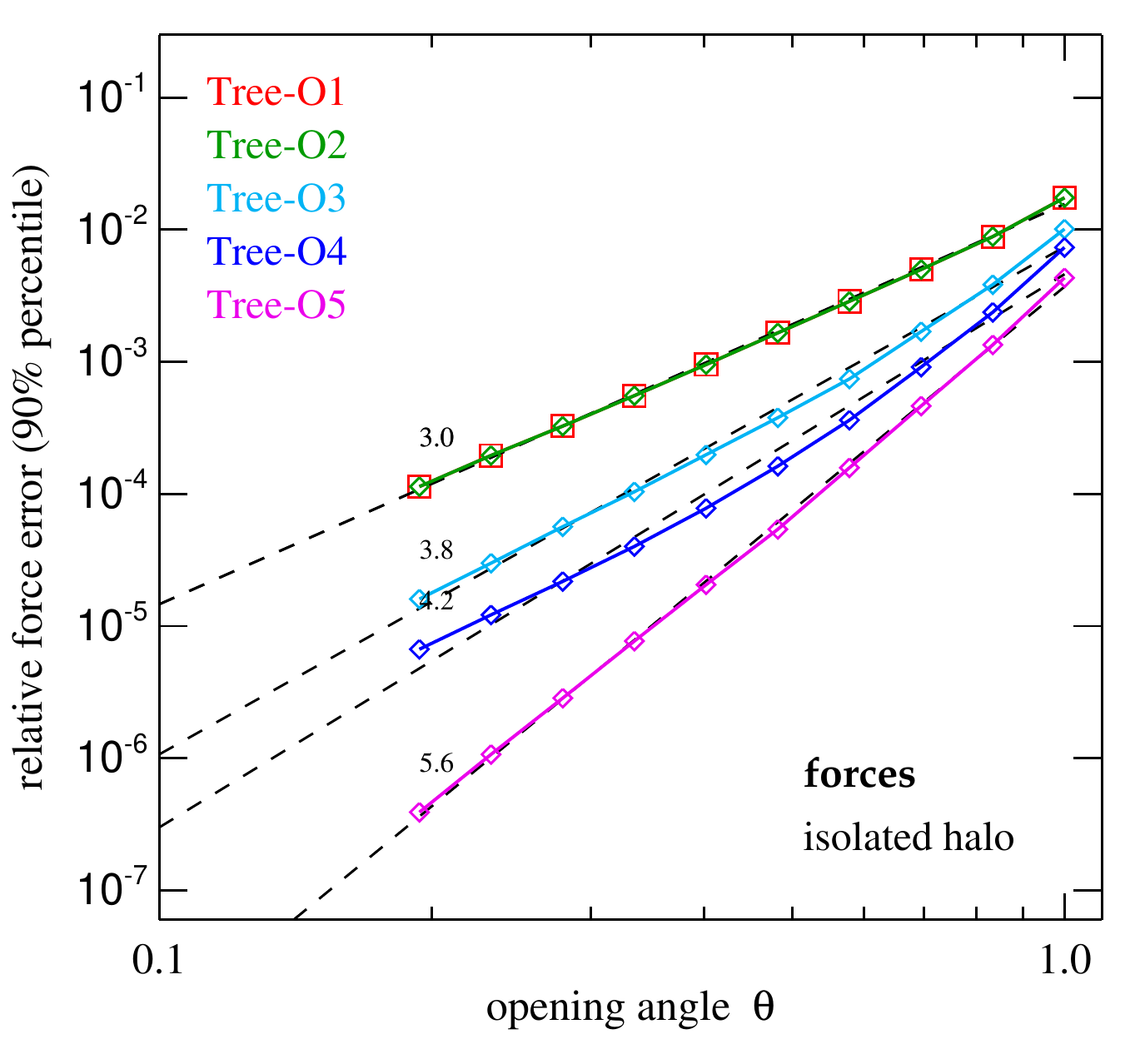}}%
\resizebox{8cm}{!}{\includegraphics{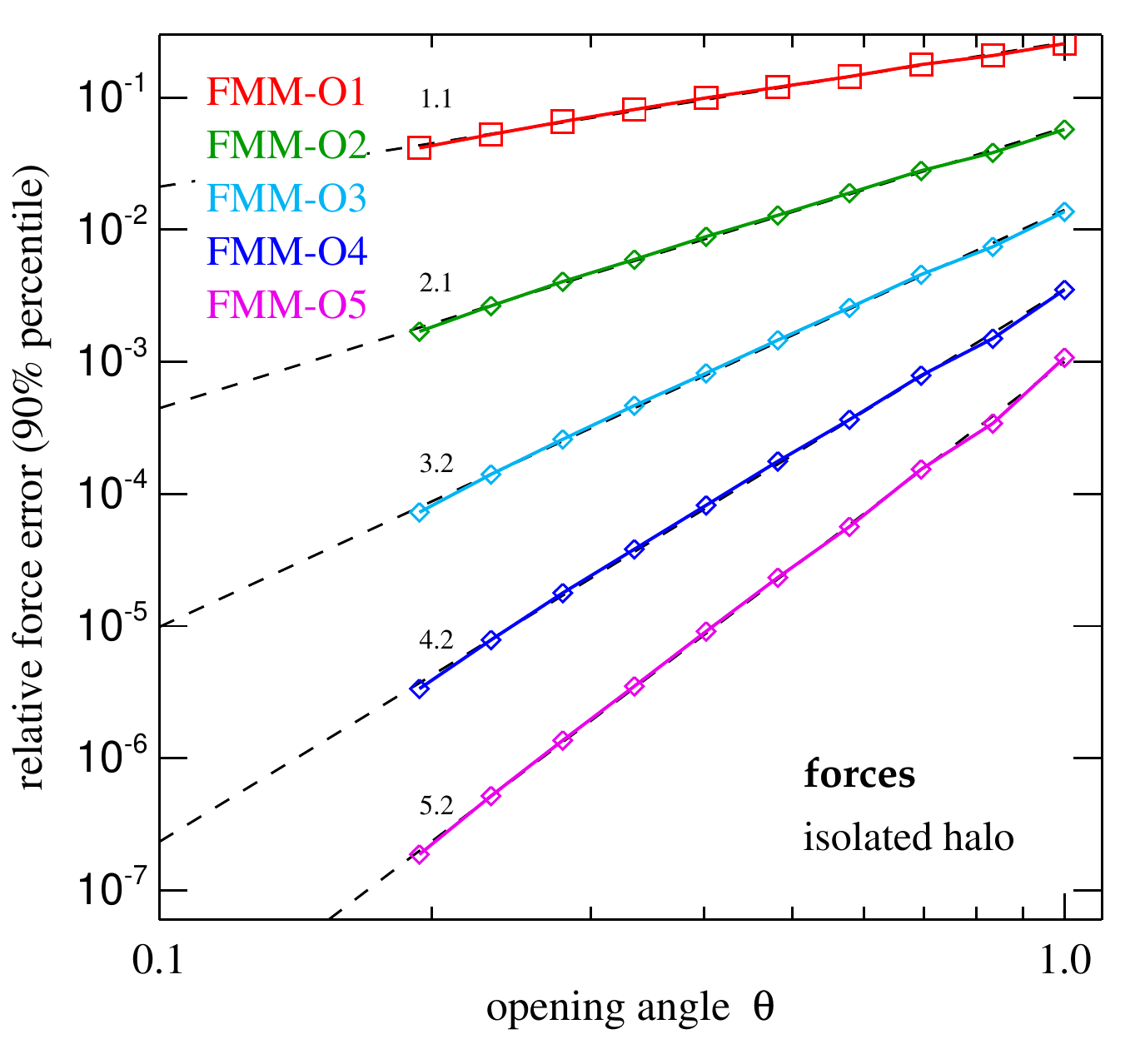}}\\%
\resizebox{8cm}{!}{\includegraphics{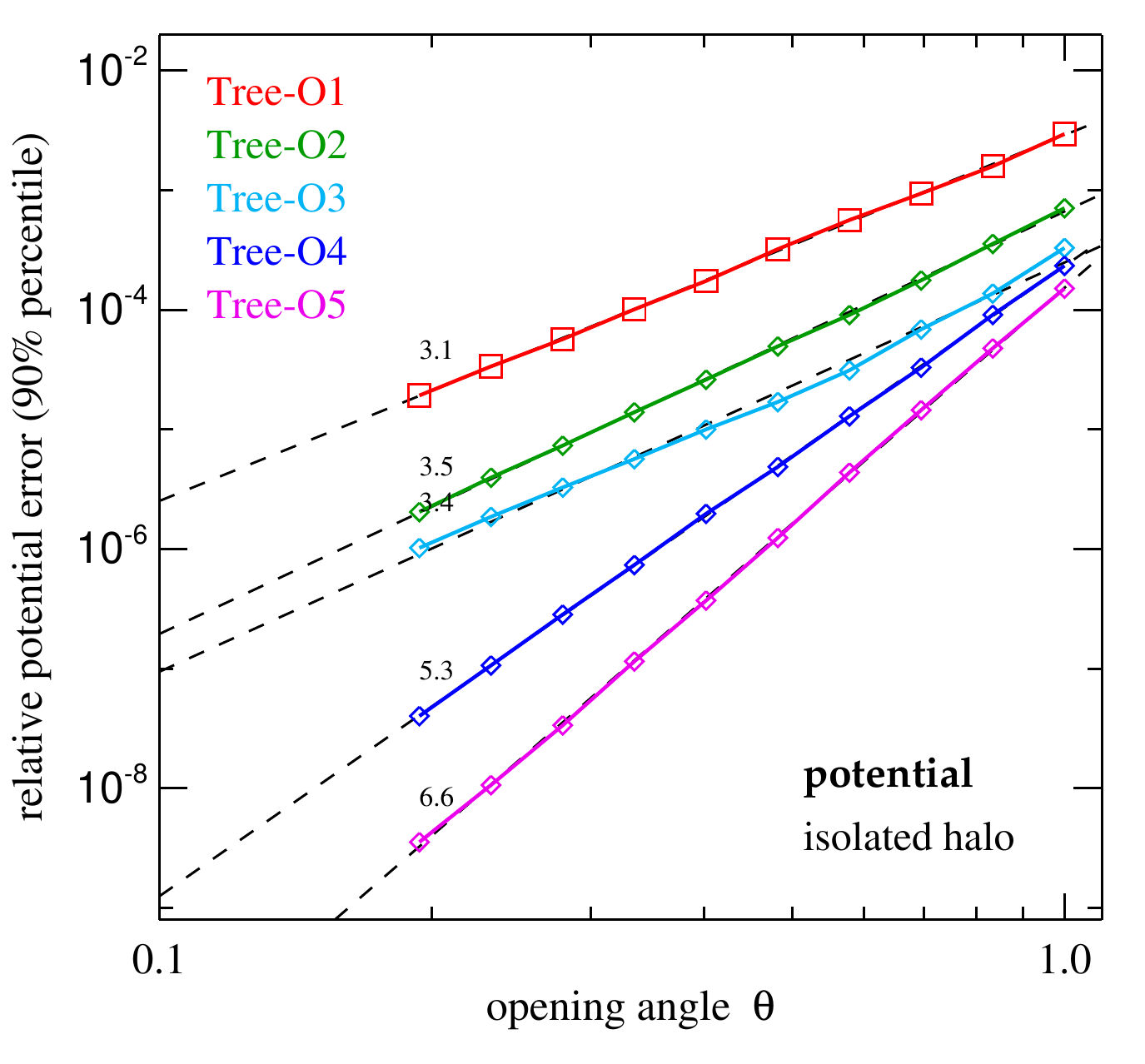}}%
\resizebox{8cm}{!}{\includegraphics{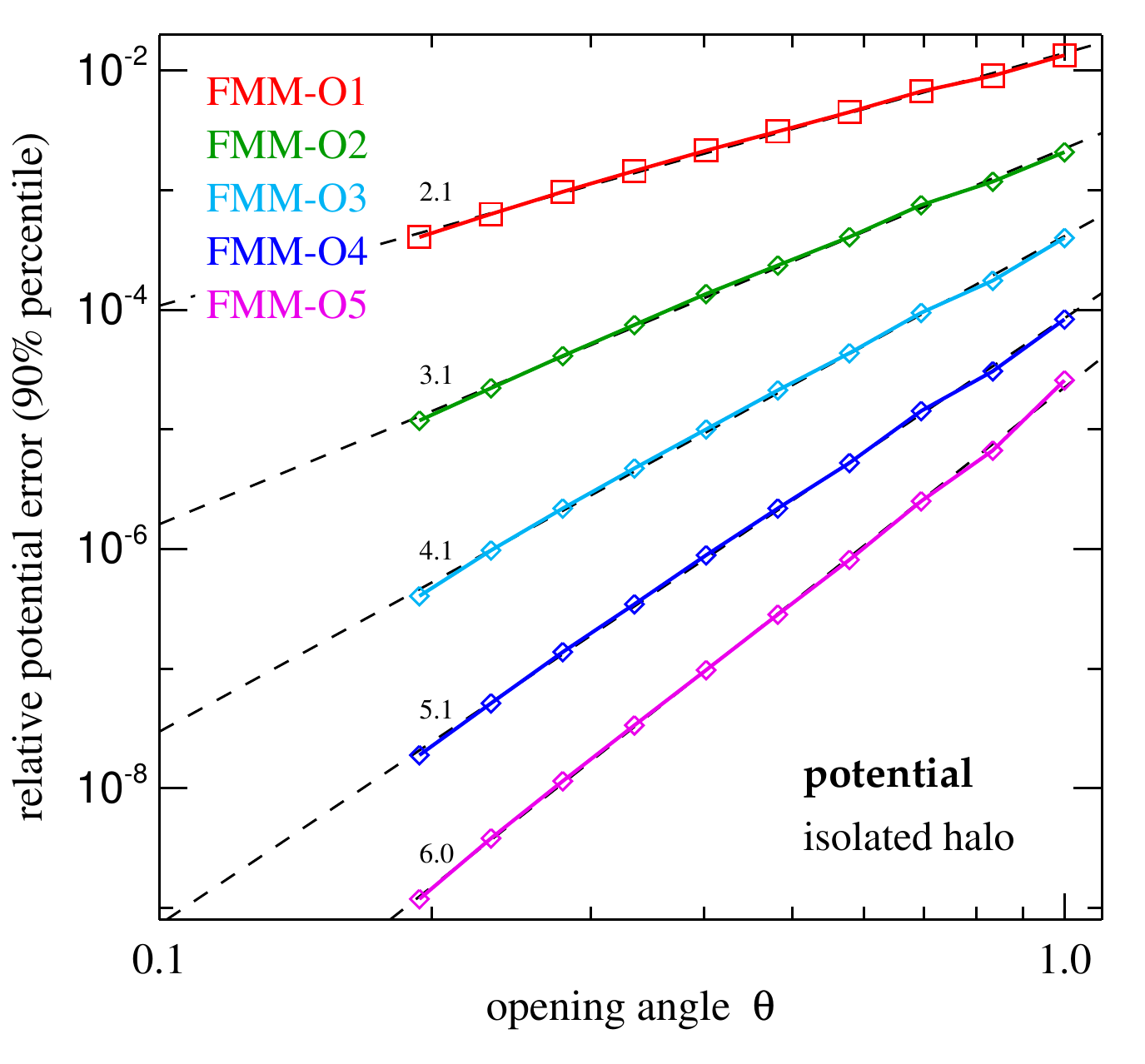}}\\%
\caption{Characteristic force (top panels) and gravitational
  potential errors (bottom panels) as a function of
  opening angle $\theta$
  for an isolated halo of 1 million particles, using  the Tree and FMM
  algorithms in {\small GADGET-4} at different multipole order, as labelled. Symbols mark our
  individual measurements for the 90 percentile errors, i.e.~90\% of
  the particles have relative errors less than the quoted values.
The scalings of the force and potential errors with opening angle are well described
by power-laws. Fits to the individual measurements are given as dashed
lines, and the values of the fitted logarithmic slopes are indicated in the different panels. 
FMM accurately follows the expected $\theta^{p+1}$ scaling of
the potential error based on the leading order term truncated in the
potential expansion, which translates to an expected  $\theta^{p}$
scaling of the force errors. Similar scalings are obtained for the
Tree algorithm, except that they are noticeably steeper than naively
expected, an effect that we attribute to progressively larger cancellations of random errors
as the interaction count increases towards smaller opening angles. For
the tree algorithm, the orders $p=1$ and $p=2$ end up being equal for
the force, due to the vanishing dipole moments and our convention for
the acceleration expansion as given in equation (\ref{eqonesidetreeacc}). 
\label{FigForceErrVsThetaIsolatedHalo}}
\end{figure*}

\begin{figure*}
\resizebox{8cm}{!}{\includegraphics{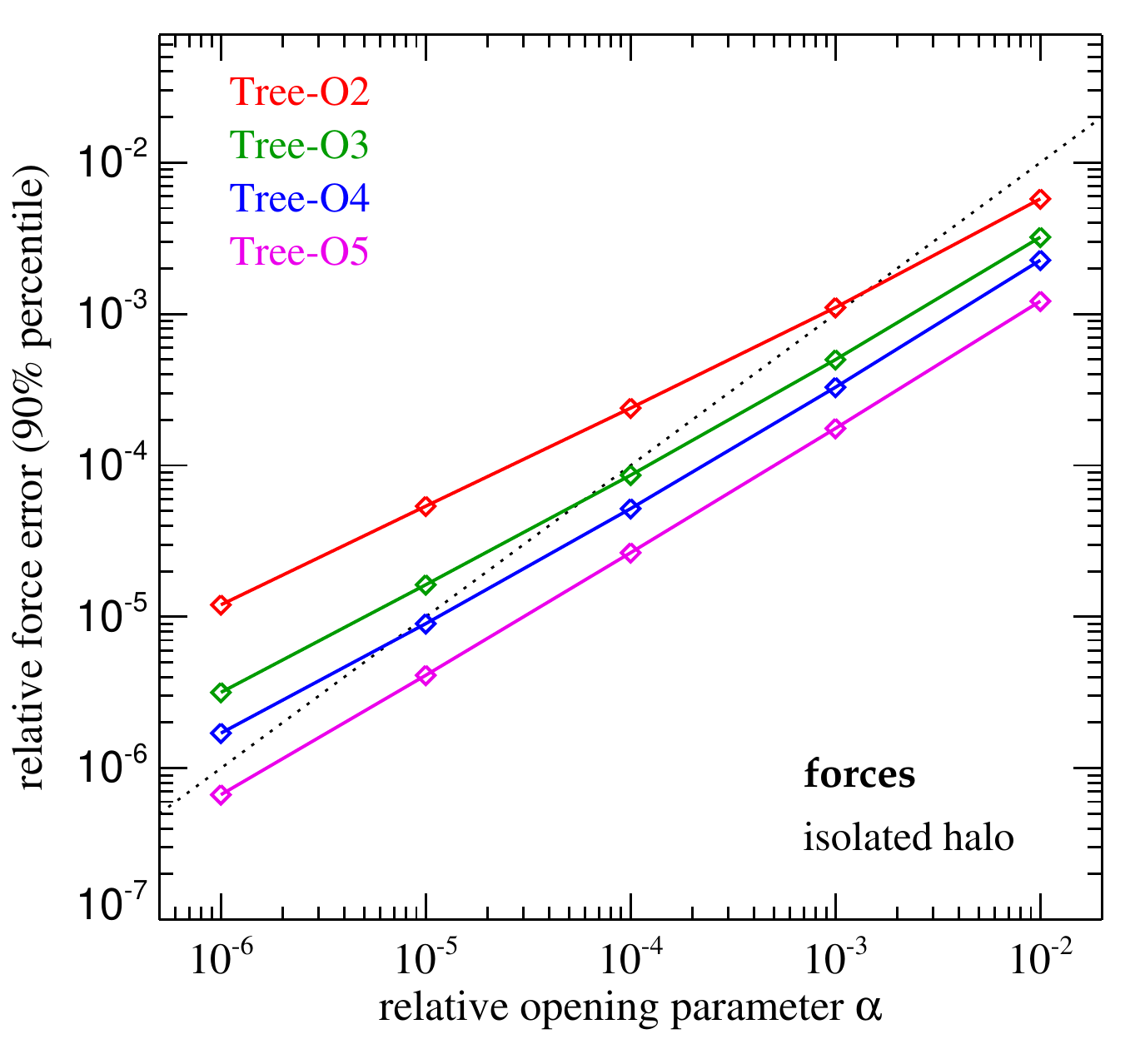}}%
\resizebox{8cm}{!}{\includegraphics{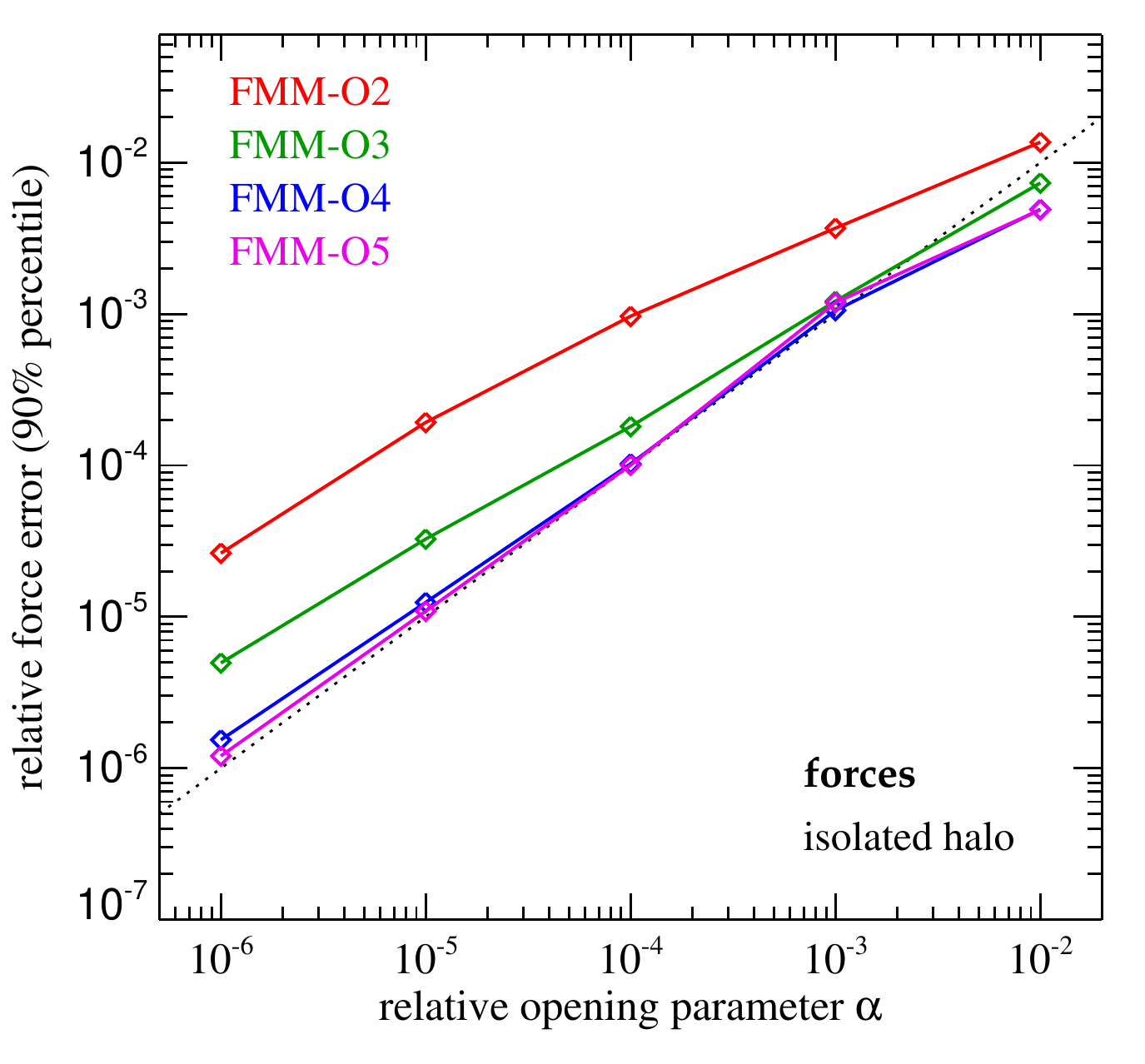}}\\%
 \caption{Characteristic force errors as a function of
  the relative opening parameter $\alpha$
  for an isolated halo of 1 million particles, using  the Tree and FMM
  algorithms at different expansion order $p$, as labelled.
Note that for fixed $\alpha$, the 
force errors are of similar size when Tree or
FMM are used, and they show only a  weak residual dependence on
multipole order $p$. This is the desired outcome, as this allows
$\alpha$ to be used as a direct control of the desired force
accuracy, approximately independent of the specific multipole
algorithm that is chosen. The dotted line in the background gives the
one-to-one relation between $\alpha$  and the 90\% force error
percentile, showing that this convenient behaviour is particularly
well realized for FMM-O4 and FMM-O5.  But note that the different
methods may vary substantially
in computational cost despite yielding similar final force errors. 
\label{FigForceErrVsAlphaIsolatedHalo}}
\end{figure*}

\begin{figure*}
\begin{center}
\resizebox{8cm}{!}{\includegraphics{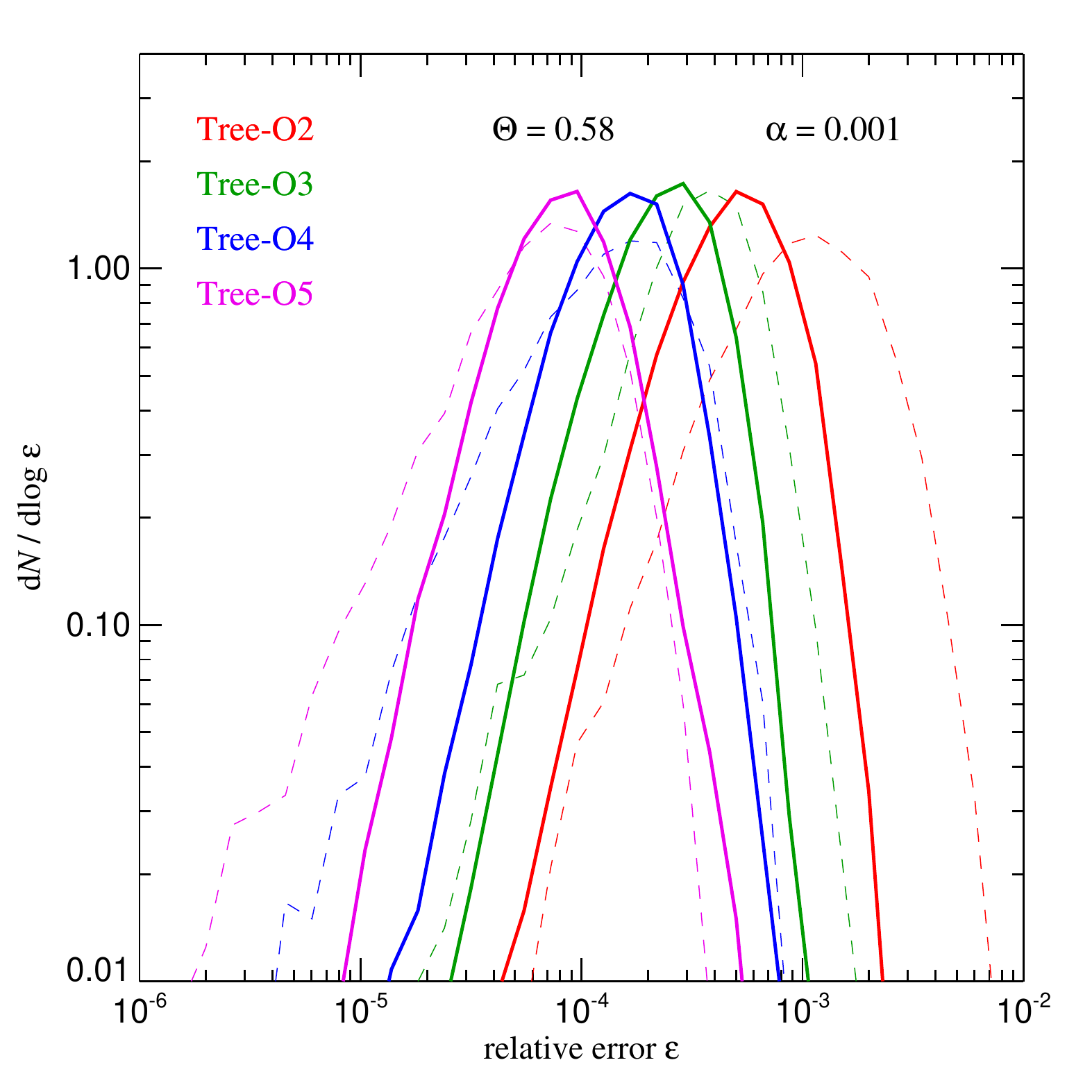}}%
\resizebox{8cm}{!}{\includegraphics{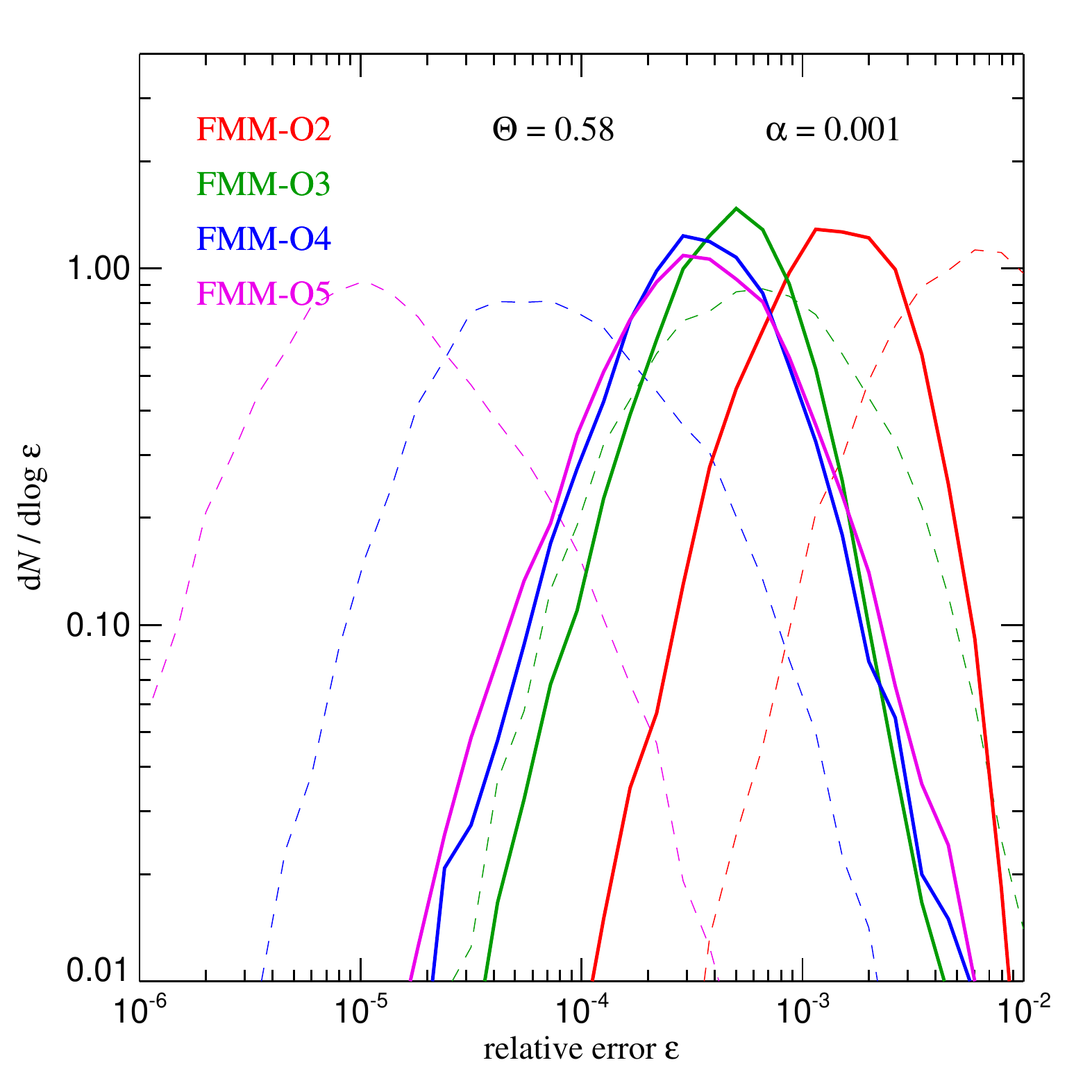}}%
\caption{Distribution of the sizes of relative force errors for an isolated dark matter
  halo (1 million particles). In the left panel, we show results for the Tree algorithm, comparing the
  force error distributions for the relative opening criterion with fixed $\alpha=0.001$ at different expansion
  order $p$ (solid lines) with the corresponding force error
  distributions obtained for a fixed geometric opening angle $\theta=0.58$ (dashed lines).  The right
  panel gives the corresponding results for the FMM algorithm. In both
  cases, we see that the geometric opening criterion produces a
  considerably broader error distribution  than the relative
  criterion, and that its median error varies much more strongly with
  expansion order. Targeting an approximately constant force error
  is thus easier and more efficient with the relative criterion.
    \label{FigForceErrorDistributionIsolatedHalo}}
\end{center}
\end{figure*}

\begin{figure*}
\resizebox{15cm}{!}{\includegraphics{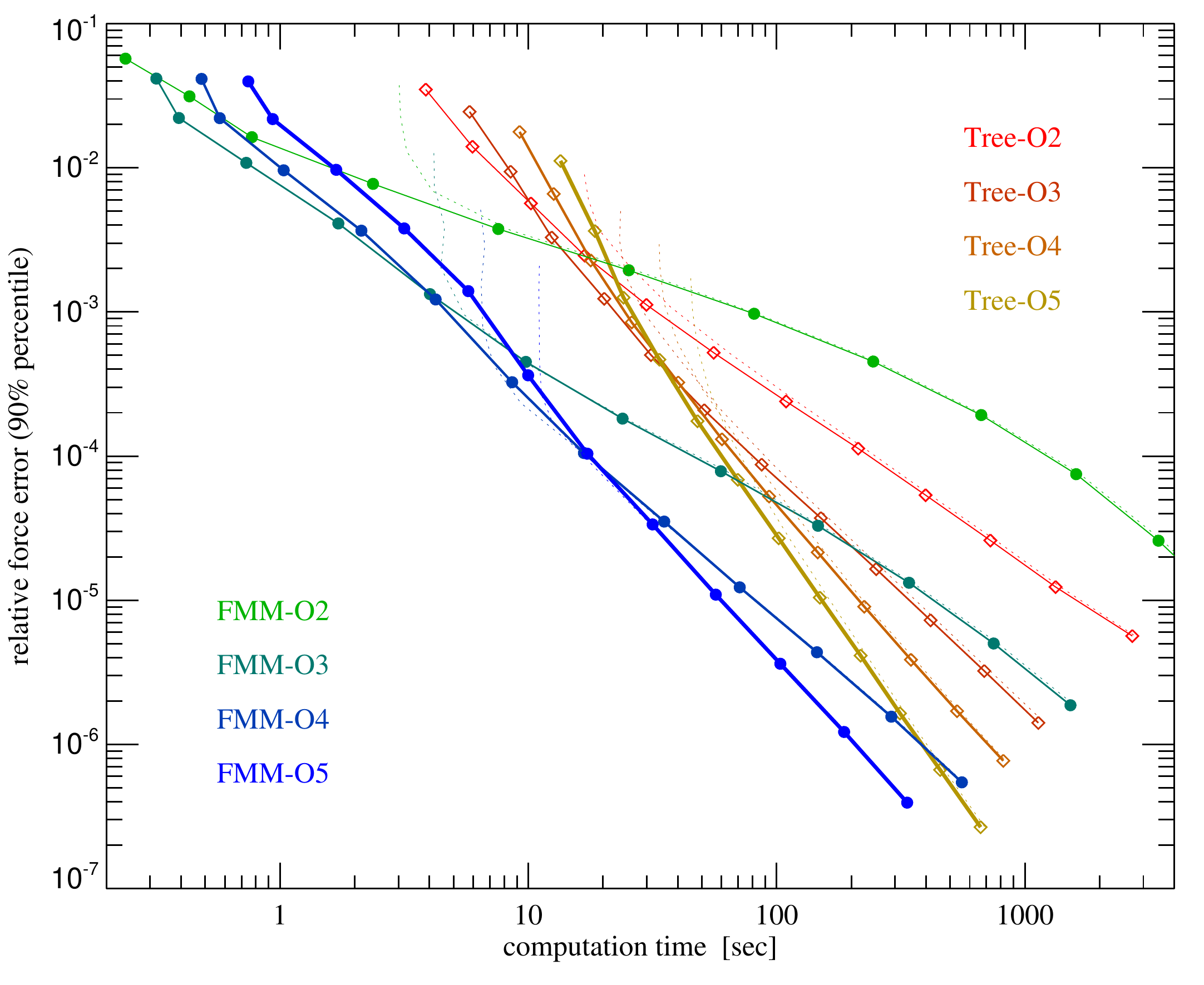}}\\%
\caption{CPU time versus achieved force accuracy, comparing the Tree
  and FMM algorithms at different order $p$ for an isolated halo of 1
  million particles, using the relative opening criterion. The
  geometric opening angle criterion always requires more computational
  time to reach a prescribed accuracy for a given order $p$ and
  algorithm, hence we omit to show results for it. In these tests, FMM
  is typically faster at the same accuracy than the Tree approach,
  except for FMM-O2, which becomes quite inefficient once a high force
  accuracy is requested.  Which order $p$ is most efficient depends on
  the absolute value of the targeted force accuracy. If very high
  force accuracy is desired, it is always favourable to use high-order
  expansions, but if a lower accuracy is sufficient, this can
  eventually be computed in less time with a lower order scheme. This
  behavior is fairly general, but the exact transition points that
  delineate the regimes where different orders $p$ are optimal depend
  on details such as the clustering of the system under study, the
  boundary conditions, and the particle number.  The measurements
  reported here are for an opening criterion where an exclusion region
  around every node (see Fig.~\ref{FigTREESketch}) is not enforced for
  the one-sided tree, and interactions between neighbouring nodes in
  FMM are allowed. If our more conservative default node opening
  criterion is adopted, the code will not venture into regimes of very
  low force accuracy and exhibit a floor in execution time, with the
  corresponding results shown as thin dotted lines.  The times
  reported here are for execution on a single core on an Intel Xeon
  Gold 6138 CPU at 2.0 GHz. We note that the relative
    contributions of the CPU time spent in different types of
    interactions is a function of the specific particle distribution,
    the order of the multipole expansion, and somewhat more weakly the
    force accuracy that is targetted.  For example, for FMM-O5 in this
    test problem, around 27.0\% of the computed interactions are
    particle-particle, 26.4\% are particle-node, while 46.6\% are
    node-node interactions. For FMM-O3, this shifts to 20.4\% for
    particle-particle, 40.4\% for particle-node, and 39.2\% for
    node-node interactions. For assessing the relative CPU cost of
    these categories, one additionally needs to keep in mind that
    particle-particle interactions are always equally cheap, while
    interactions involving nodes become more expensive at higher
    multipole order. 
\label{FigCpuVsForceErrorIsolatedHalo}}
\end{figure*}

We now consider force error measurements for the Tree and FMM
algorithms for more realistic particle distributions. To this end we
consider an isolated \citet{Hernquist:1990aa} halo consisting of 1
million particles and study the obtained force and potential accuracy
when compared to exact results computed by direct summation for a
random subset of the particles.  In
Figure~\ref{FigForceErrVsThetaIsolatedHalo}, we show the errors for
Tree and FMM algorithms as a function of opening angle $\theta$, and
for different multipole expansion order $p$. In each case, we show the
errors for the 90 percentile, i.e.~a fraction of 0.9 of the particles
has lower errors than given in the plots. Plotting median or mean
errors instead yields qualitatively very similar results. We also
include power-law fits to the measurements to highlight their scaling
with opening angle.

Reassuringly, for Tree and FMM algorithms alike, the potential and
force errors decline to good accuracy as power-laws with decreasing
opening angle.  The slopes reflect the leading order of the truncation
error of the underlying Taylor expansions, as expected. However, the
Tree-based errors decline noticeably faster in general than expected
based on the multipole approximation error of a single node-particle
interaction. Even when using only monopole moments in the force (this
corresponds to $p=1$ and $p=2$, which are equal for the tree force as
the dipole vanishes in our formulation), the error scales nearly as
$\propto \theta^3$, and when quadrupole moments are included in the
force ($p=3$), the scaling is $\propto \theta^{3.8}$. This can be
understood as a consequence of the increasing interaction count with
smaller opening angle, leading to more cancelation. For this particle
distribution, the average number of multipole interactions per
particle scales approximately as $\propto \theta^{-2.6}$ with opening
angle. If all the partial forces were aligned, similar in size, and
featured random errors of comparable size, we would expect to gain at
most a steepening of 1.3 in the slope of the total force error scaling
when the tree algorithm is used. In practice, we see a bit less, but
still about $\sim 1.0$ for $p=2$, and slightly less for $p=3$. In contrast,
little if anything of such a `cancellation-boost' is seen for FMM, as
we already anticipated based on our analysis of node-node
interactions. This suggests that some of the speed advantage of FFM is
eaten away by its less favorable behaviour with respect to error
cancellations compared to the one-sided Tree method. We will return to
the question of how CPU-time consumption as a function of achieved
accuracy compares for different algorithms later on in this section.

The Tree-O3 result for the potential, and the Tree-O4
result for the forces, shown in Figure~\ref{FigForceErrVsThetaIsolatedHalo}
look a bit peculiar. They noticeably deviate from a power-law and show a
flatter scaling than expected based on the other results. We think
this is because of the particular particle distribution examined here,
where due to the spherical symmetry contributions sourced by the
octupole moment are unusually small. Other, more irregular particle
distributions show a much more regular run of this multipole order,
confirming that this is a feature of the particular particle setup examined here.

In Figure~\ref{FigForceErrVsAlphaIsolatedHalo}, we show force error
measurements for the same particle system, but this time as a function
of the relative opening parameter $\alpha$. Again, we give
measurements both for Tree and FMM, at different expansion order. It
is reassuring that now the final force errors vary approximately in
direct proportion to $\alpha$, and show only a comparatively weak
dependence on the multipole order, or on whether the FMM or Tree
method is used. This is the desired outcome for this opening
criterion, which thus allows the use of $\alpha$ to conveniently set
the desired force accuracy level, while the expansion order can
be independently chosen according to performance considerations or memory use.

The relative opening criterion also affects the distribution of force
errors, making them narrower in general. This is shown explicitly in
Figure~\ref{FigForceErrorDistributionIsolatedHalo}, where we give
force error distributions for $\theta=0.58$ for both Tree and FMM for
different expansion order $p$. This can be compared to corresponding
error distributions obtained for $\alpha=0.001$. It is clearly seen
that the medians of the force error distributions vary much less with
expansion order when $\alpha$ is used, an effect that is particularly
prominent for FMM. Also, the error distributions become narrower when
$\alpha$ is used, particularly the low-force error tail of the
geometric opening criterion is reduced. We also anticipate that this is
reflected in a reduction of the computational cost at fixed median
force error.

That this is indeed the case is becoming clear in
Figure~\ref{FigCpuVsForceErrorIsolatedHalo}, where we now consider the
force accuracy delivered by various ways to run the gravity solver
against the CPU time invested. We show results both for Tree and FMM,
for different expansion orders. As we find that the relative opening
criterion is always more efficient than the geometric one, i.e.~for
fixed $p$ and algorithm, a given force accuracy is achieved with less
computational time if the relative opening criterion is used, we only
show results for the relative opening criterion in order to avoid
making the plot overly busy.  FMM is normally faster than the Tree
approach, but there can be exceptions. For example, FMM-O2 is quite
inefficient if a high force accuracy is targeted. In fact, for $p=2$ it
would actually be better to use Tree-O2 in this regime. Note that the
answer to the question which algorithm is the fastest depends on the
desired force accuracy. For relative errors less than $10^{-4}$ this
would be FMM-O5 (with the relative opening criterion), for errors
between $10^{-4}$ and $10^{-3}$, it is FMM-O4, and for errors between
$10^{-3}$ and $10^{-2}$ this is FMM-O3. If still larger errors are
permissible, it could even be FMM-O2. This illustrates a general
pattern that can also be observed for the results obtained with the
Tree algorithm. Higher expansion orders become computationally
worthwhile if high accuracy is demanded, but they are not necessarily
the most cost effective if low accuracy is sufficient.

\subsection{Force errors in periodic cosmological simulation boxes using Ewald corrections}

\begin{figure*}
\resizebox{8cm}{!}{\includegraphics{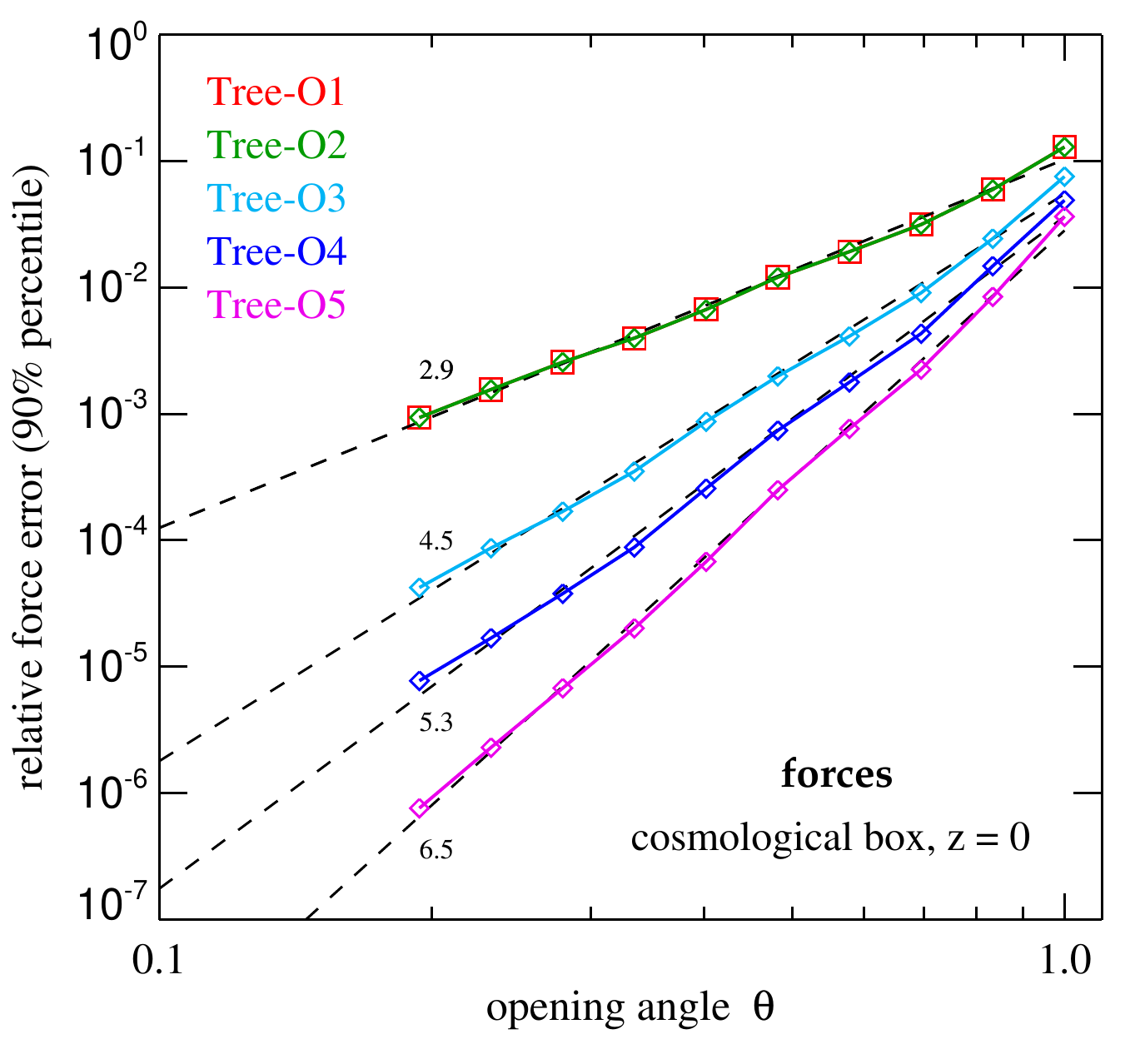}}%
\resizebox{8cm}{!}{\includegraphics{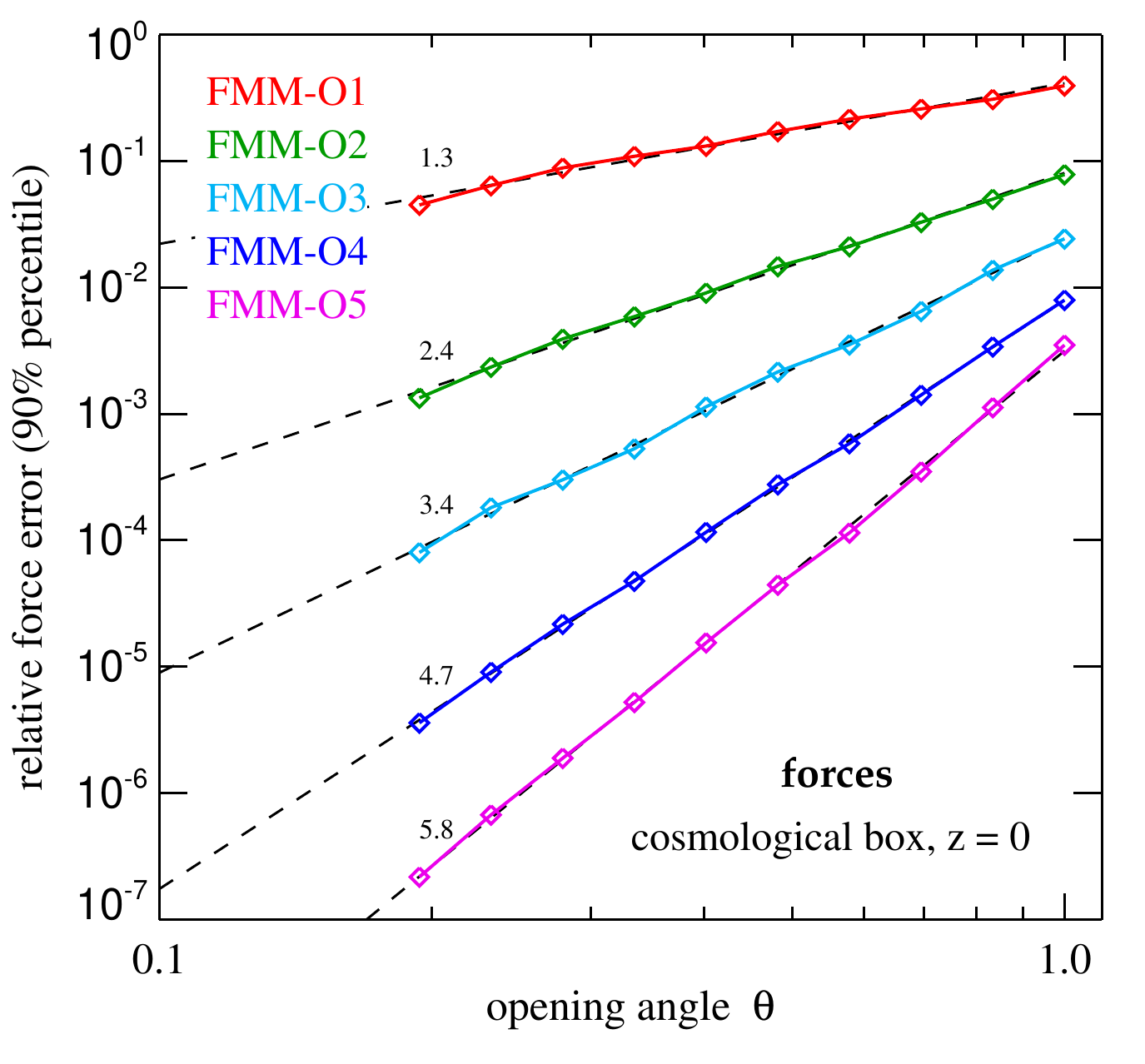}}\\%
\resizebox{8cm}{!}{\includegraphics{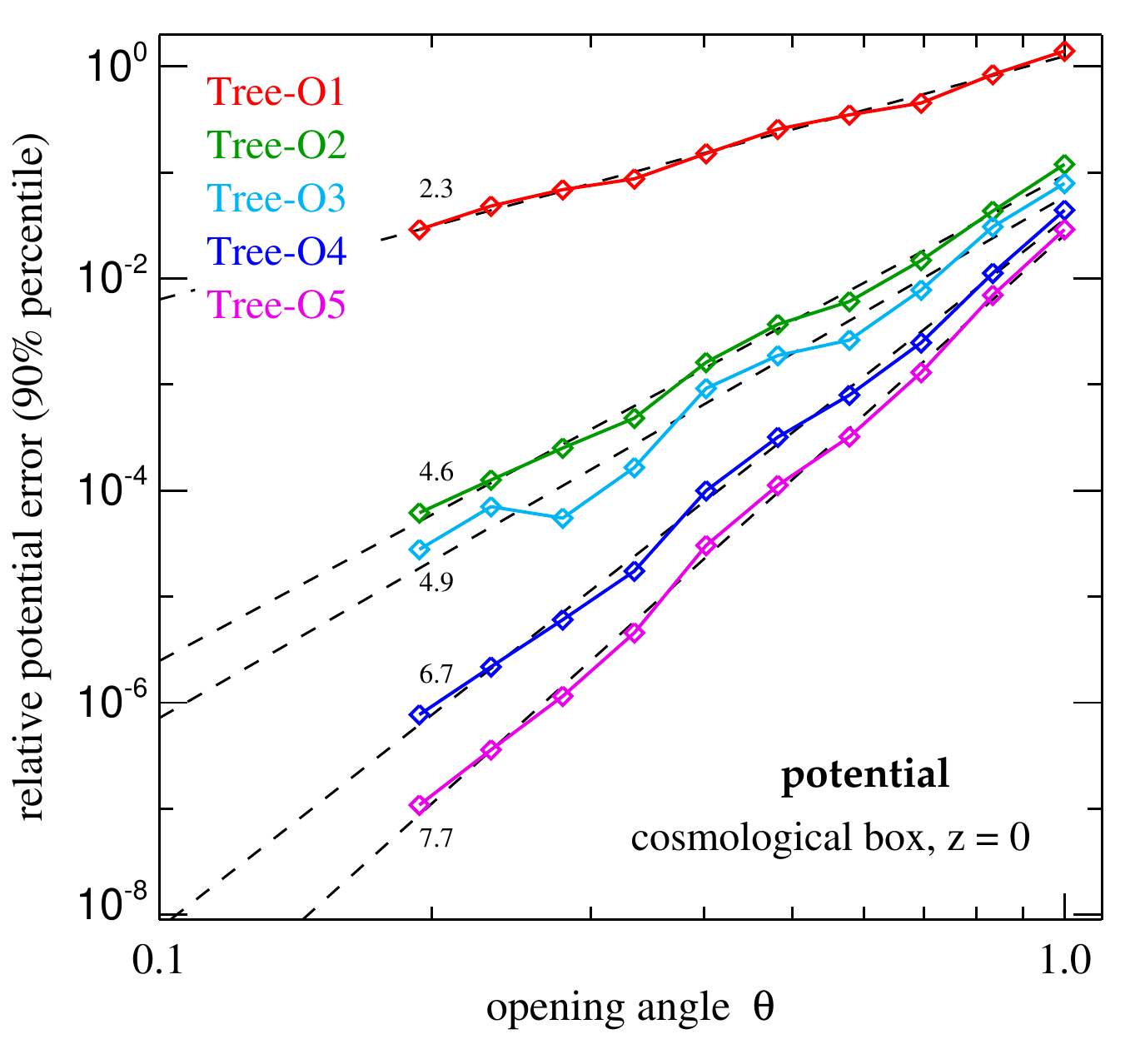}}%
\resizebox{8cm}{!}{\includegraphics{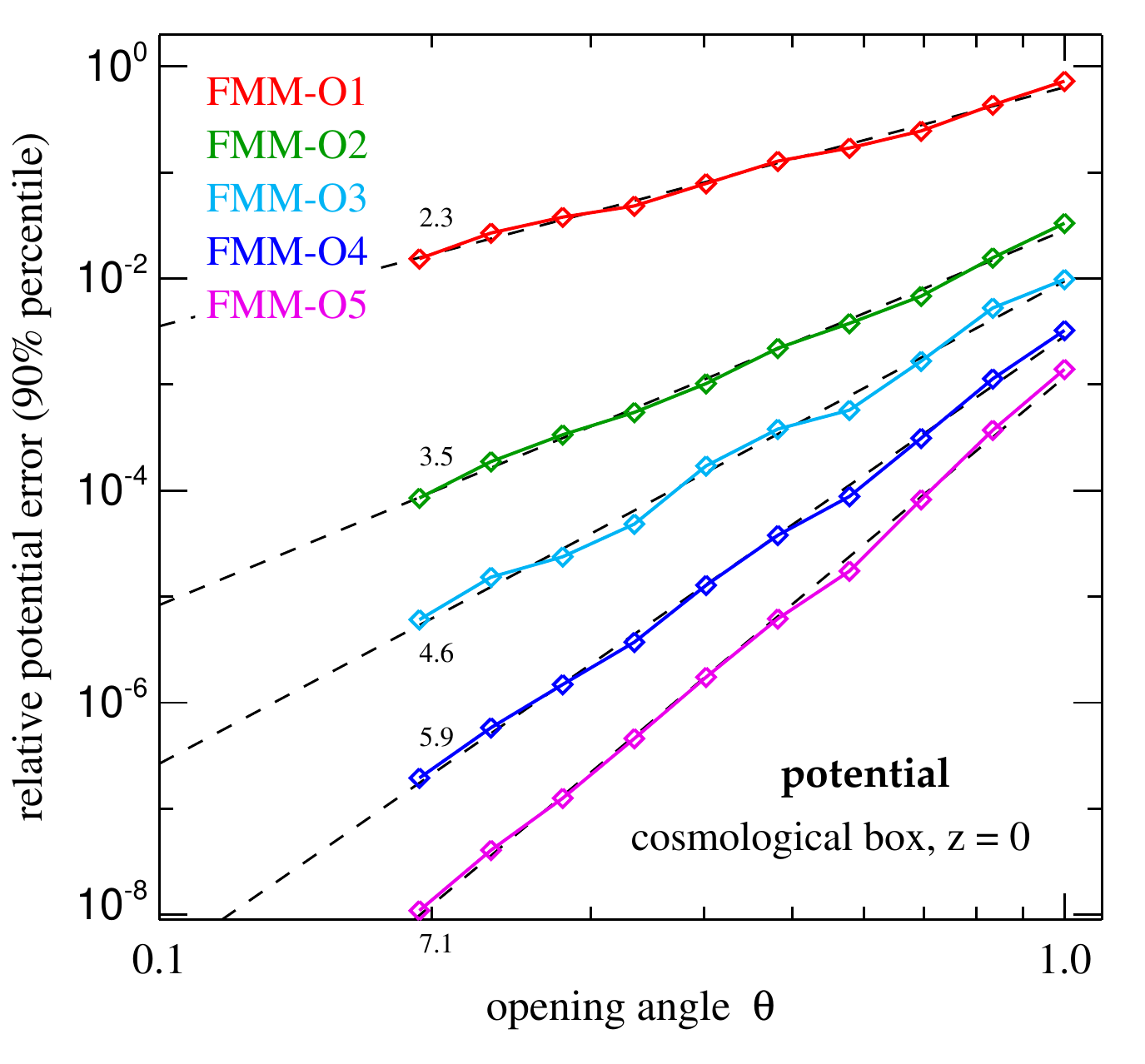}}\\%
\caption{Relative gravitational force (top panels) and potential (bottom panels) errors as a function of
  the opening angle $\theta$
  for a cosmological simulation box at redshift $z=0$ (using  $128^3$ particles
  in a periodic box
  $30\,h^{-1}{\rm Mpc}$ on a side, with $5\,h^{-1}{\rm kpc}$ softening) 
using  the Tree  (left panels) and FMM
  algorithms (right panels), for  different expansion orders as
  labelled.
The scaling of the force and potential errors with opening angle is well described
by power-laws (dashed lines); the values of the fitted slopes are inlined in the panels.
The FMM approach follows the expected $\theta^{p+1}$ and $\theta^{p}$
scalings
for potential and force errors quite accurately, whereas the Tree
method performs typically somewhat better than expected due to
progressively more relevant cancellations of random errors with
increasing interaction count. Note that for our definition of the
one-sided Tree algorithm, the $p=1$ and $p=2$ orders for
the Tree-based acceleration are equal by construction.
\label{FigMedianForceErrVsThetaBoxSimZ0}}
\end{figure*}

\begin{figure*}
\resizebox{8cm}{!}{\includegraphics{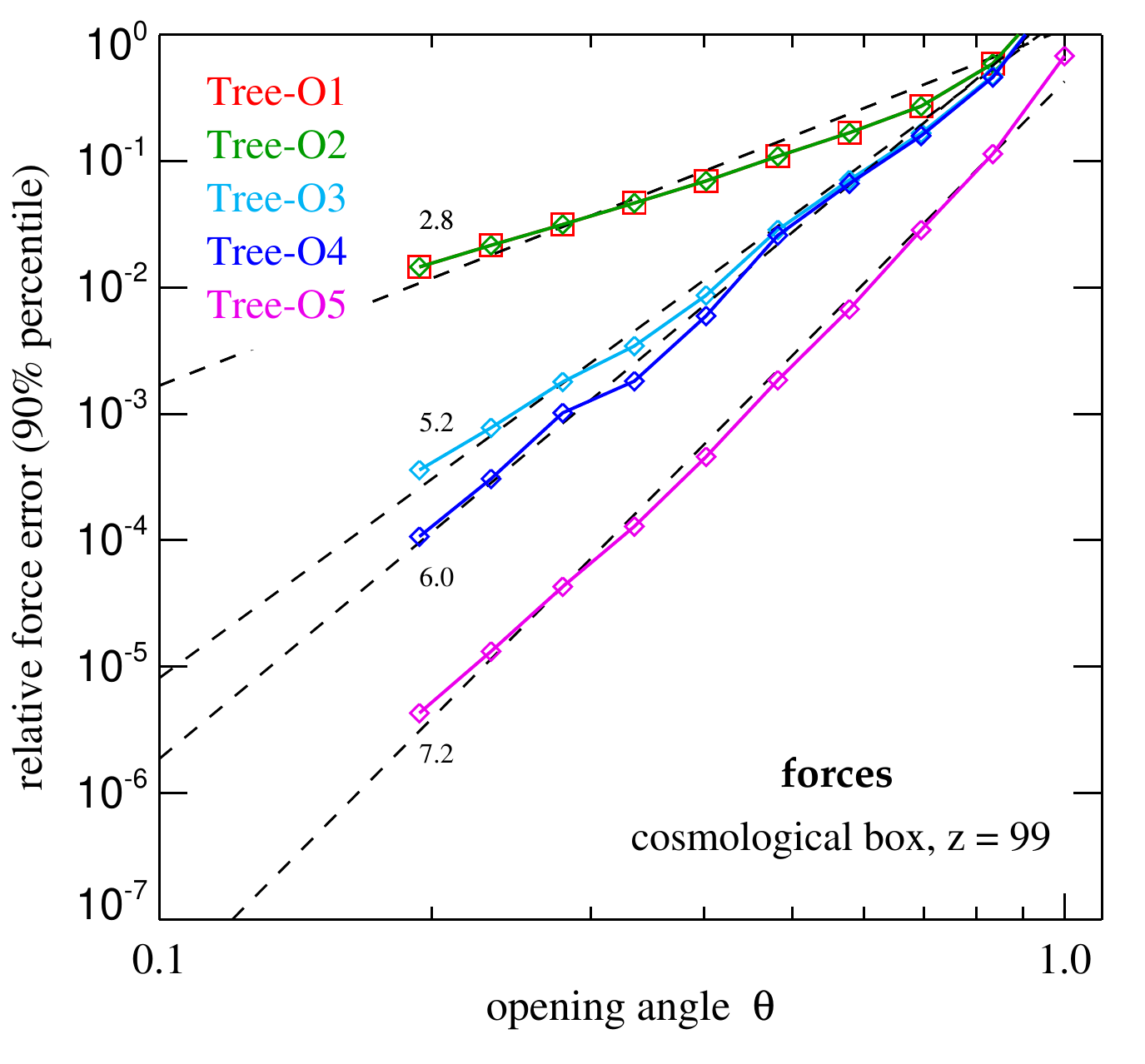}}%
\resizebox{8cm}{!}{\includegraphics{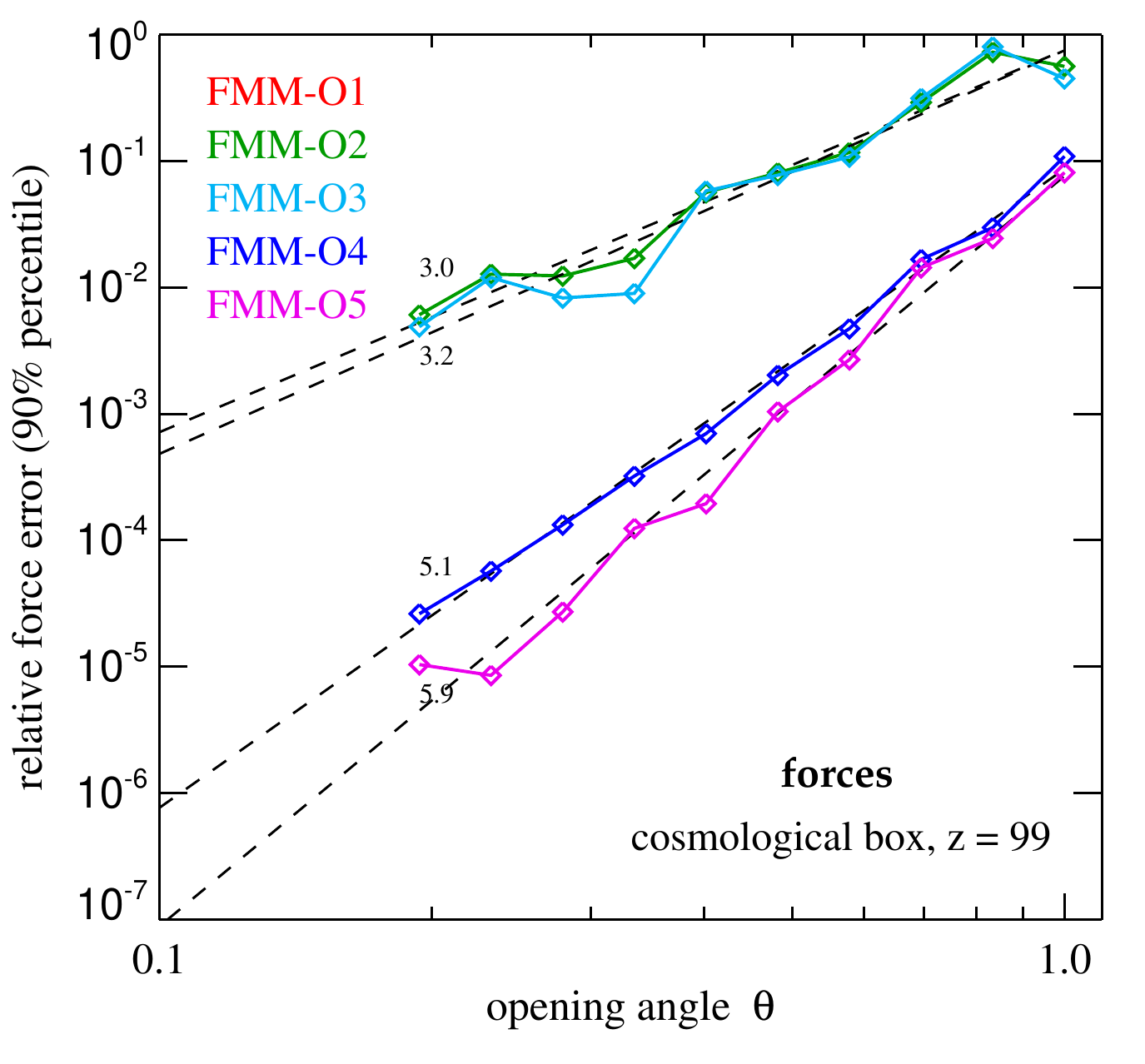}}\\%
\resizebox{8cm}{!}{\includegraphics{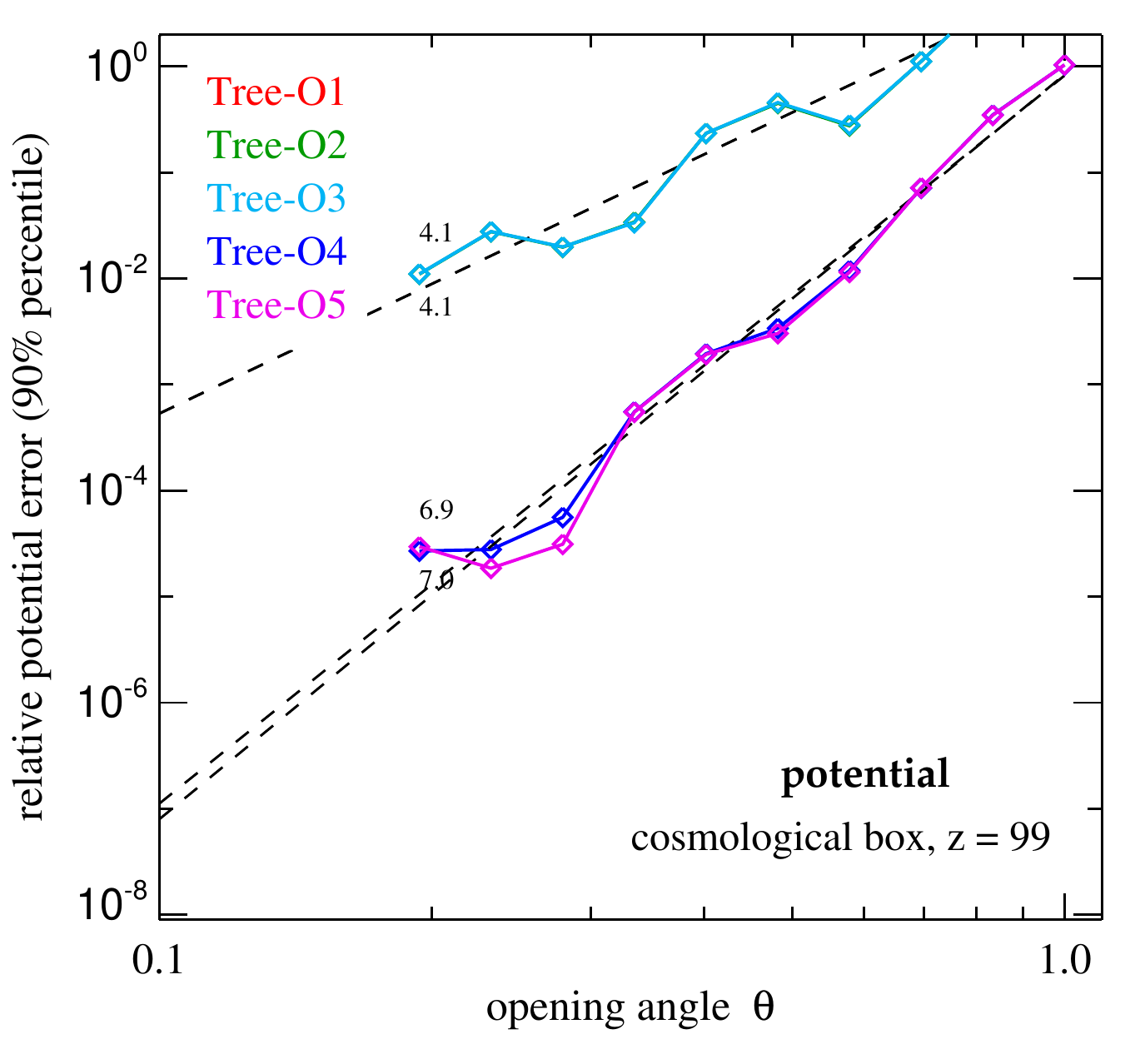}}%
\resizebox{8cm}{!}{\includegraphics{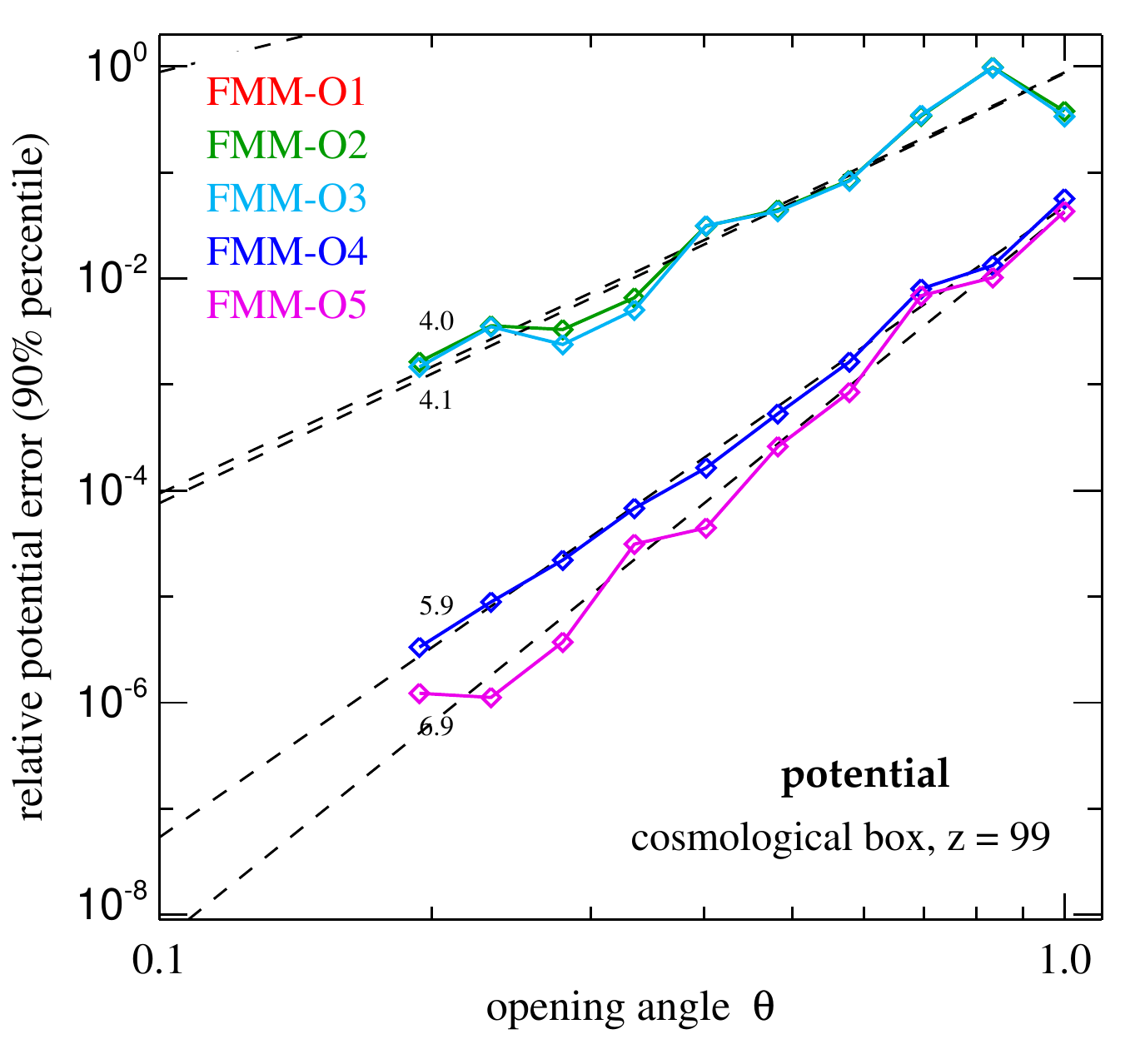}}\\%
\caption{Characteristic force and potential errors
as in Fig.~\ref{FigMedianForceErrVsThetaBoxSimZ0}, but now 
focusing on a cosmological simulation box at redshift $z=99$, where the
particle distribution is nearly uniform and only slightly perturbed by
the cosmological initial conditions.
The scaling of the force and potential errors with opening angle is
still reasonably well described
by power-laws, although the values of the fitted slopes (as indicated
in the panels) are less straightforward to understand, and the
scalings feature some residual structure. The latter is presumably
related to aliasing effects between the mildly
perturbed Cartesian grid of particles, and the nested Cartesian
oct-tree mesh used by the multi-pole expansions, causing relatively
large discrete jumps in accuracy at certain opening angles. For the
same reason, the octupole and triakontadipole moments of most tree
nodes are expected to be very small, explaining why Tree-O2/O3 and Tree-O4/O5 are
similar for the potential (in fact, Tree-O2 and Tree-O3 lie virtually
on top of each other), and Tree-O3/O4 is very similar for the force. Likewise, the
similarities of FMM-O2/O3 and FMM-O4/O5 are explained by these
symmetry effects. We note
that the accuracy of FMM-O1 for force and potential, and that of
Tree-O1 for the potential, is so bad that the relative errors exceed
unity and lie off-scale on the plots, hence these schemes are
completely inadequate
in the high redshift regime.
\label{FigMedianForceErrVsThetaBoxSimZ99}}
\end{figure*}

We now turn to the more demanding problem of \emph{accurately}
calculating N-body forces in a periodic space, which lends itself to a
clean mathematical model of ``infinite'' space, giving a direct and
accurate connection to linear perturbation theory, and thus to the
standard formalism of large-scale structure theory.  At high-redshift,
the problem is particularly acute as here only small peculiar forces
arise, which are the result of the addition of many forces from
different matter perturbations on all scales. The small net force
remains after cancellation of large forces in all directions, making
this regime demanding for methods that work only in real space.

In Figure~\ref{FigMedianForceErrVsThetaBoxSimZ0}, we show measurements
of the accuracy of force and potential values delivered by the Tree
and FMM algorithms for different expansion order, as a function of the
opening angle. The test system is here that of an evolved $z=0$ dark
matter density field created by $128^3$ particles in a box of
$100\,h^{-1}{\rm Mpc}$ across. This is thus representative of typical
cosmic structure expected for $\sigma_8 \simeq 0.85$ in today's
universe, at moderate mass resolution. We can see that both the Tree
and FMM algorithms converge well towards the exact forces, which have
here been computed with direct summation and Ewald correction for a
random subset of the particles. As for the isolated halo, the Tree
approach tends to converge with slightly steeper power-laws as the FMM
method, a benefit that arises from more favourable error cancellation
due to fewer correlations among the partial force errors. Also, we see
that the forces for Tree orders $p=1$ and $p=2$ are identical, as
expected. Only for $p=3$, the quadrupole moments of the tree nodes
start affecting the forces.

The corresponding results for high redshift, $z=99$, which are representative
for the epoch where normally initial conditions for such simulations
are created, are shown in
Figure~\ref{FigMedianForceErrVsThetaBoxSimZ99}. These results feature
some interesting effects that look peculiar at first. First of all,
the relative errors one obtains for a given opening angle are quite a
bit larger at $z=99$ than at $z=0$. This reflects the small sizes of
the peculiar potential and the peculiar forces for the only slightly
perturbed density field at high redshift, which arises from the
cancellation of large contributions to both quantities. This regime
hence requires smaller opening angles for accurate results than
needed for the highly clustered state at low redshift. Note that the
accuracy delivered by Tree-O1 and FMM-O1 at
high redshift  (i.e. without PM mesh,
as shown here) is particularly bad and essentially unusable for
anything. This also holds true for the FMM-O1 force.

Another striking result of
Fig.~\ref{FigMedianForceErrVsThetaBoxSimZ99} is the fact that the
potential for Tree-O3 is basically the same as that delivered by
Tree-O2, and that the potential for Tree-O5 is almost the same as that
for Tree-O4. This is also seen in a similar way for the
FMM-algorithm. It appears that in this regime the addition of the
octupole and triakontadipole moments is not really helping much. This
is perhaps not too surprising, given that we use cubical nodes. If
they can be approximated as nodes of uniform density, all uneven
multipole modes vanish, and as \citet{Barnes:1989aa} has shown, even
the quadrupole moment vanishes in this case exactly, making the
hexadecupole moment the first non-trivial contribution. This is why we
see a big jump in accuracy when this multipole moment is included in
the potential calculation.

Another interesting effect is that the force accuracies of Tree-O3 and
Tree-O4 are rather similar, while this is not the case for the
FMM-algorithm. There the FMM-O3 rather lines up with FMM-O2, and not
with FMM-O4. This is due to the importance of the sink-side expansion
in FMM and its associated errors, which are absent in the Tree
algorithm.

\begin{figure*}
\resizebox{9.0cm}{!}{\includegraphics{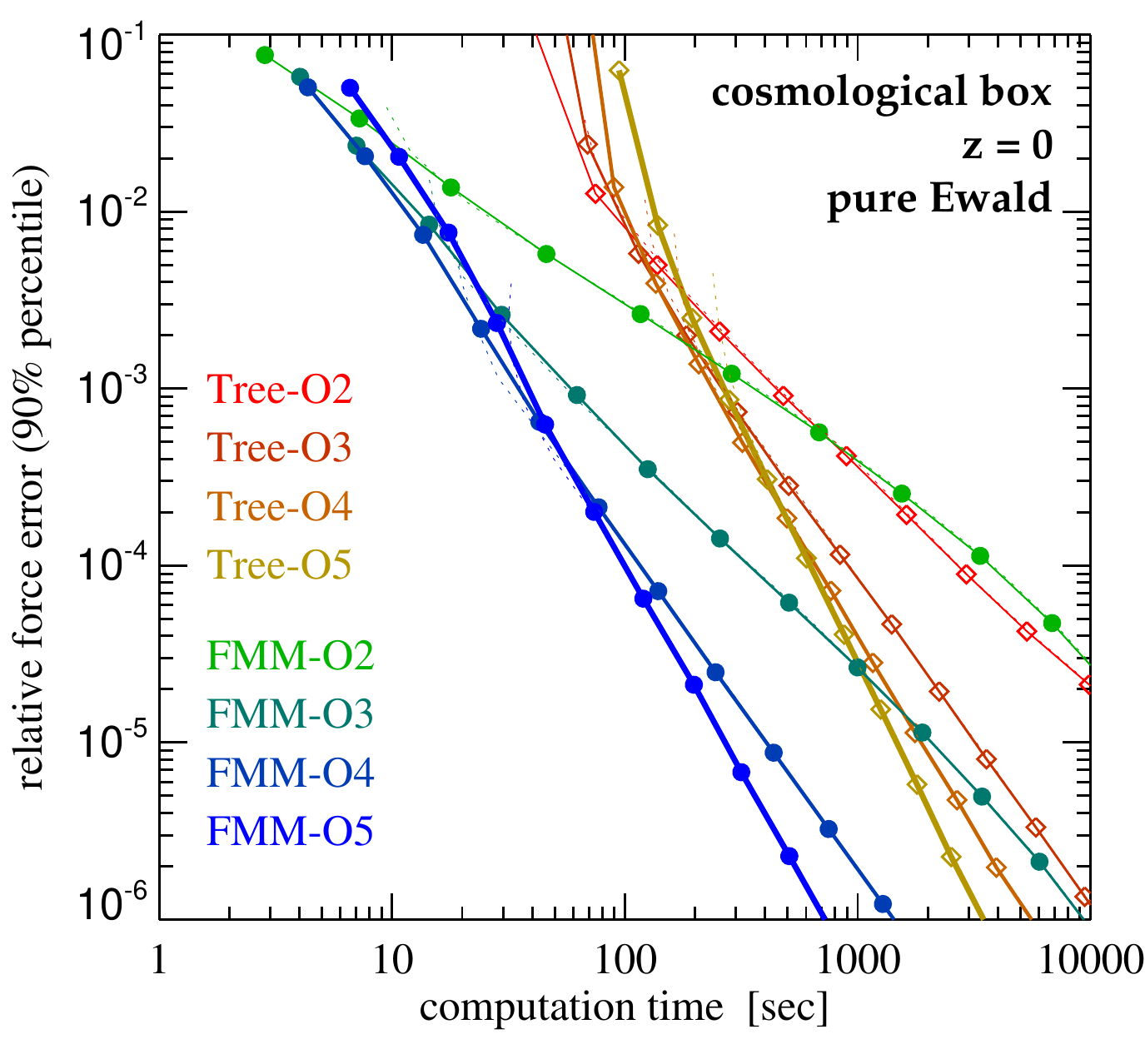}}%
\resizebox{9.0cm}{!}{\includegraphics{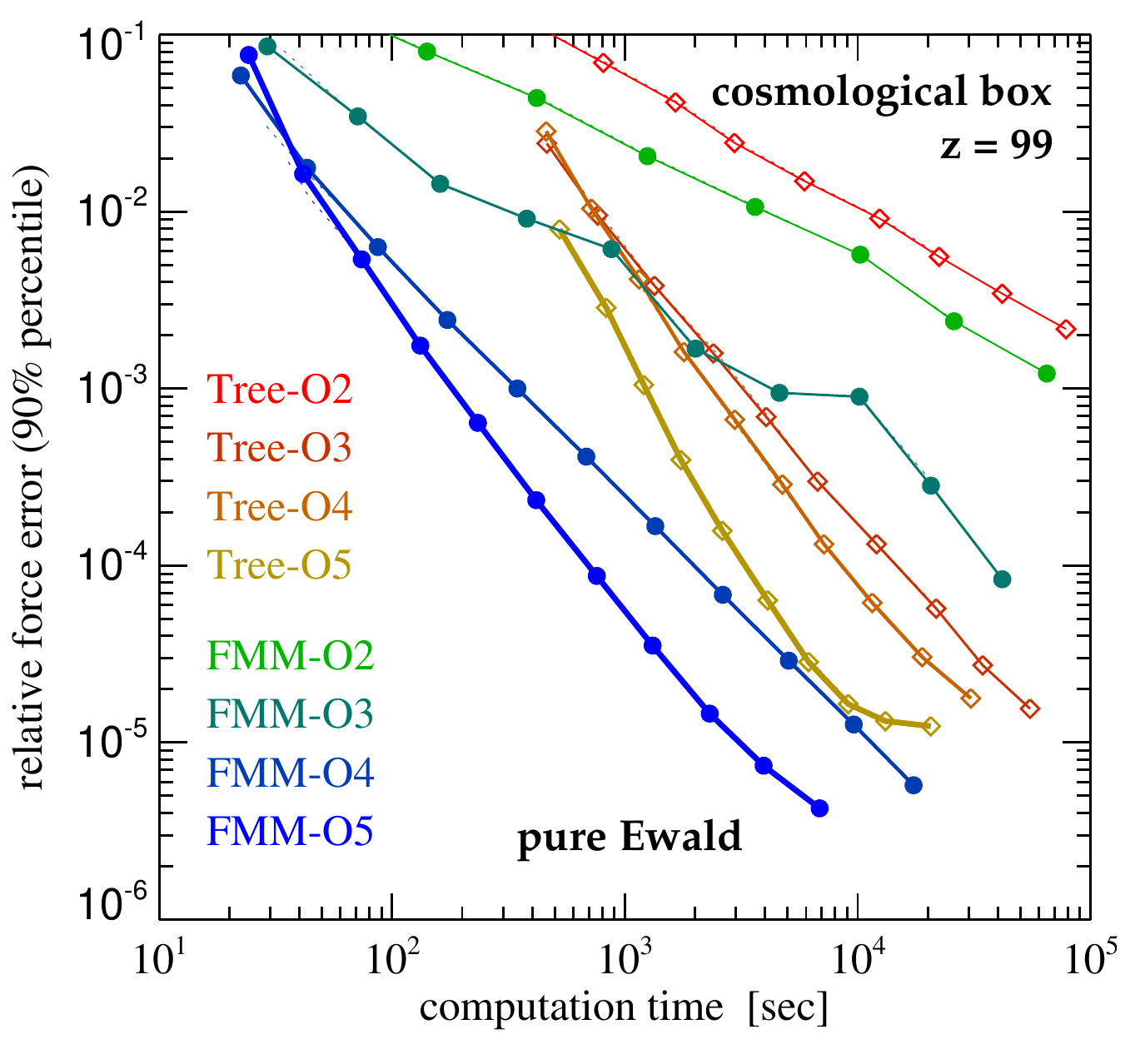}}
\caption{Force accuracy as a function of the invested computing time
  for our Tree and FMM algorithms when applied with different
  expansion order to a cosmological simulation with periodic boundary
  conditions, using Ewald summation. The particle number is $128^3$
  with box-size $L=30\,h^{-1}{\rm Mpc}$ and $5\,h^{-1}{\rm kpc}$
  softening, as in Figures~\ref{FigMedianForceErrVsThetaBoxSimZ0} and
  \ref{FigMedianForceErrVsThetaBoxSimZ99}.
  The left panel shows results for
  the evolved density field at $z=0$, while the right panel is for
  the initial conditions at $z=99$, which are more demanding to
  calculate in real space due to
  the small peculiar forces at that time (note the change in range on
  the $x$-axes in the two panels). The solid lines give results
  for different settings of our relative opening criterion; the
  geometric opening criterion is not shown for clarity as it always
  has a   worse performance (particularly at low redshift, while at high redshift it is nearly
  on par). The reported times are for execution on a single
  processor core, and for the use of Ewald correction in all
  interactions to correct for periodic boundary conditions. Thin dotted lines give equivalent results when our
  conservative near-node exclusion zone around every tree node is
  enabled, which otherwise has not been used in the results reported as
  solid lines. The higher order schemes, in particularly for FMM, are
  generally the most efficient, with few exceptions at low force
  accuracy. 
  \label{FigSpeedTreeFMMBoxSimEwald}}
\end{figure*}

\begin{figure*}
\resizebox{8cm}{!}{\includegraphics{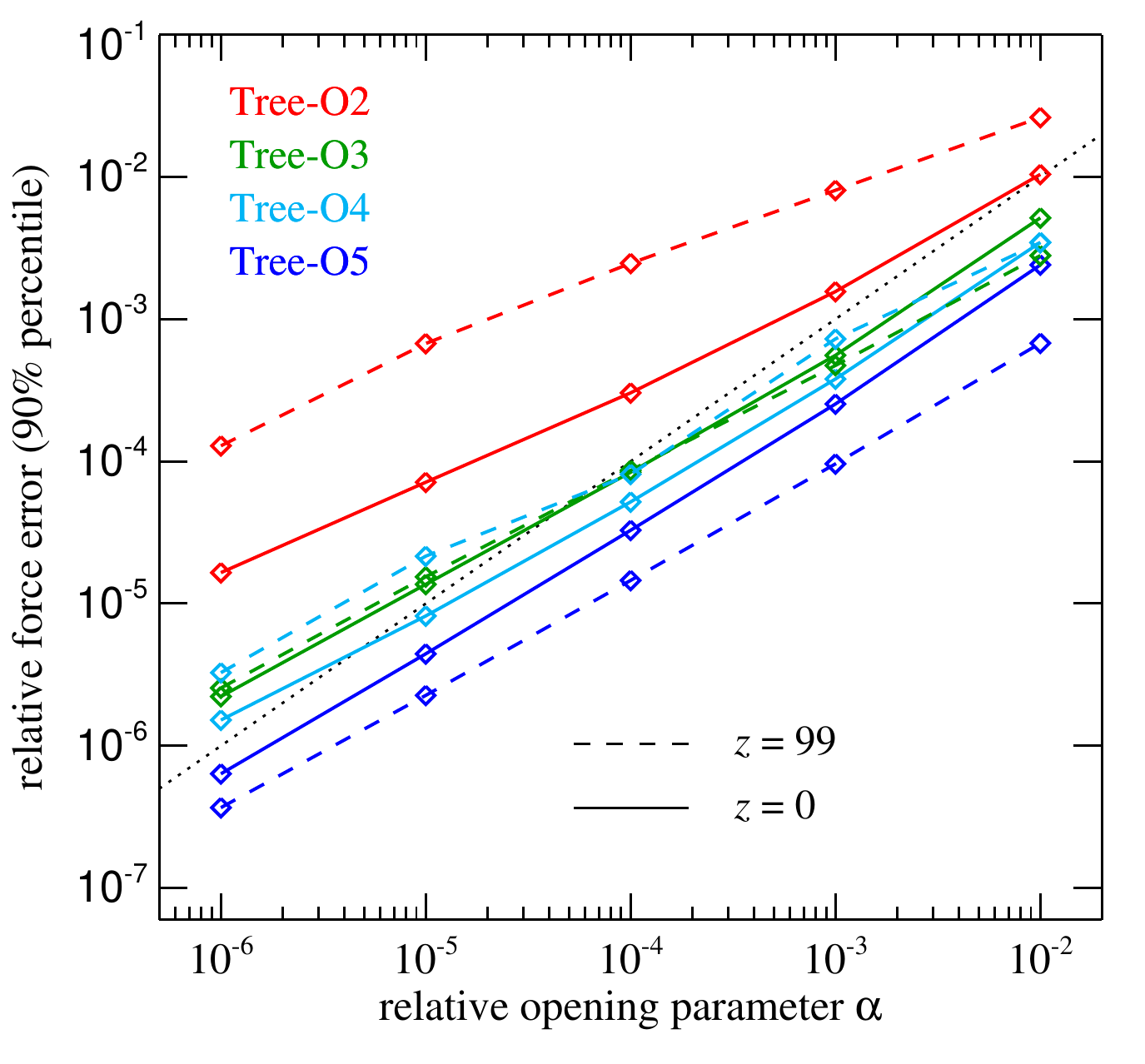}}%
\resizebox{8cm}{!}{\includegraphics{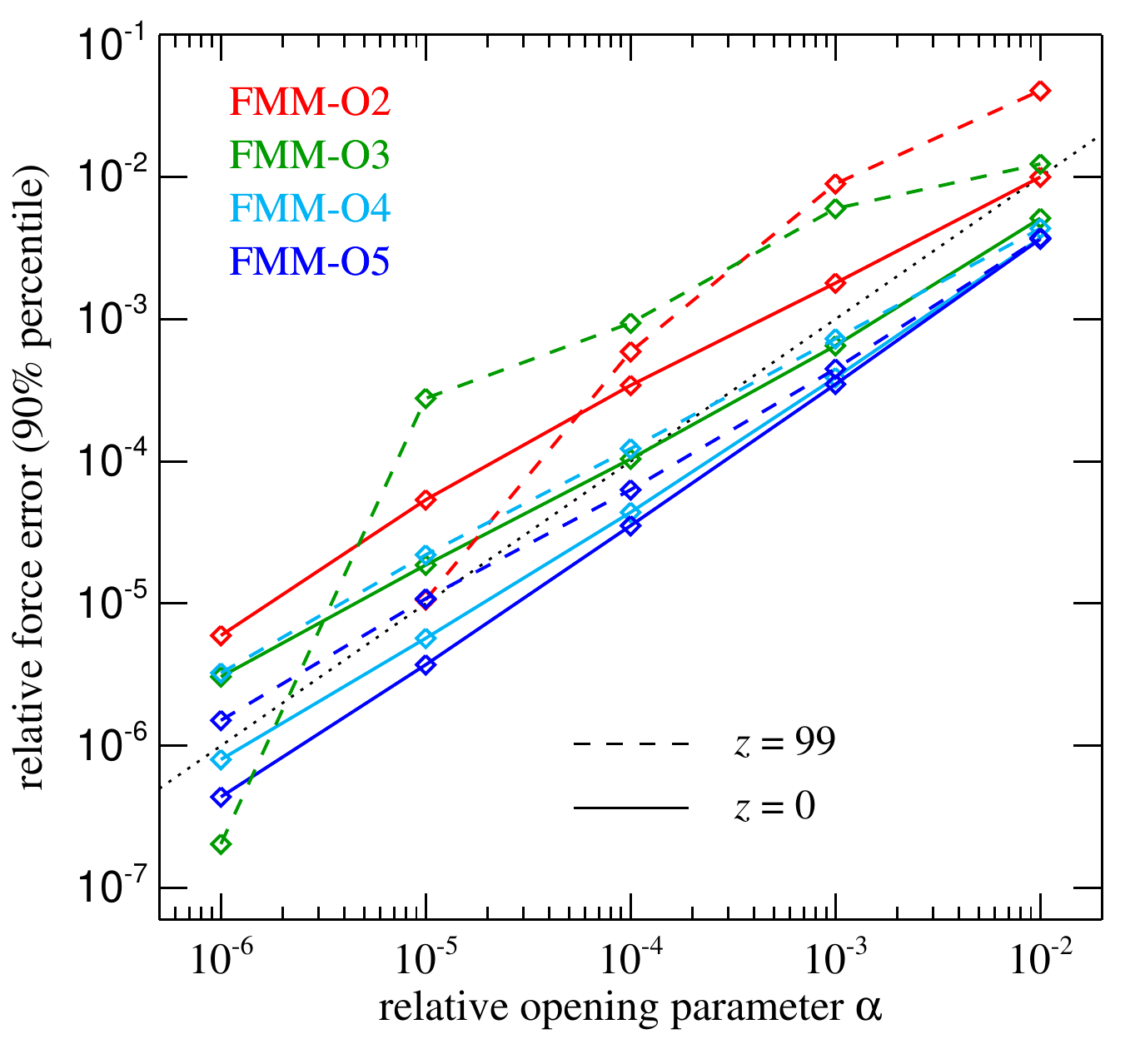}}
\caption{Force accuracy as a function of the relative opening
  parameter $\alpha$ for cosmological simulation boxes, shown both for
  the Tree (left panel) and FMM (right panel) algorithms. In each case, we
  compare results for $z=99$ (dashed) with those obtained for the
  clustered state at $z=0$ (solid), for different expansion order.
  The dotted line in the background gives a
  one-to-one relation between $\alpha$ and the 90\% force accuracy
  percentile, which may also serve as a reference for comparing Tree
  and FMM with each other. Reassuringly, the measured force accuracies
  for fixed $\alpha$ are reasonably independent of
  the   employed algorithm,  the expansion order, and the clustering
  state of the particles, especially so if one refrains from using the
  lower orders.
  This means that the parameter $\alpha$ of the relative opening
  criterion
  can used as a convenient
  and reliable control of the desired 
  relative force accuracy throughout the course of a cosmological
  simulation.
  \label{FigTreeFMMVsRelativeOpening}}
\end{figure*}

In Figure~\ref{FigSpeedTreeFMMBoxSimEwald}, we compare the relative
computational efficiencies of these schemes with each other by now
looking at the achieved force accuracy as a function of consumed CPU
time. We again focus on results for the relative opening criterion as
this is always at least as efficient as the geometric one. Note that
here all interactions are Ewald-corrected, which makes them notably
more expensive than for non-periodic boundaries. But at least for the
clustered state at $z=0$, the relative efficiencies of the different
schemes closely resemble the results shown in
Fig.~\ref{FigCpuVsForceErrorIsolatedHalo} for an isolated halo.  At
the high redshift of $z=99$, we see however that the calculations
become much more expensive. Through the use of the relative criterion
a roughly constant force accuracy for specified opening parameter
$\alpha$ is maintained independent of clustering state, something that
we demonstrate explicitly in Figure~\ref{FigTreeFMMVsRelativeOpening}.
To reach the same small relative force error at high redshift with a
pure multipole-based approach thus requires more nodes to be opened,
and hence a more costly calculation. Interestingly, in this regime
especially the low-order schemes struggle substantially, and it really
pays off to use one of the higher-order methods, particularly FMM-O5
looks attractive if relative errors of $10^{-3}$ or smaller are
desired with a pure real-space method at high redshift.

\subsection{Effective force law for Tree-PM
  and FMM-PM}

\begin{figure*}
\resizebox{16cm}{!}{\includegraphics{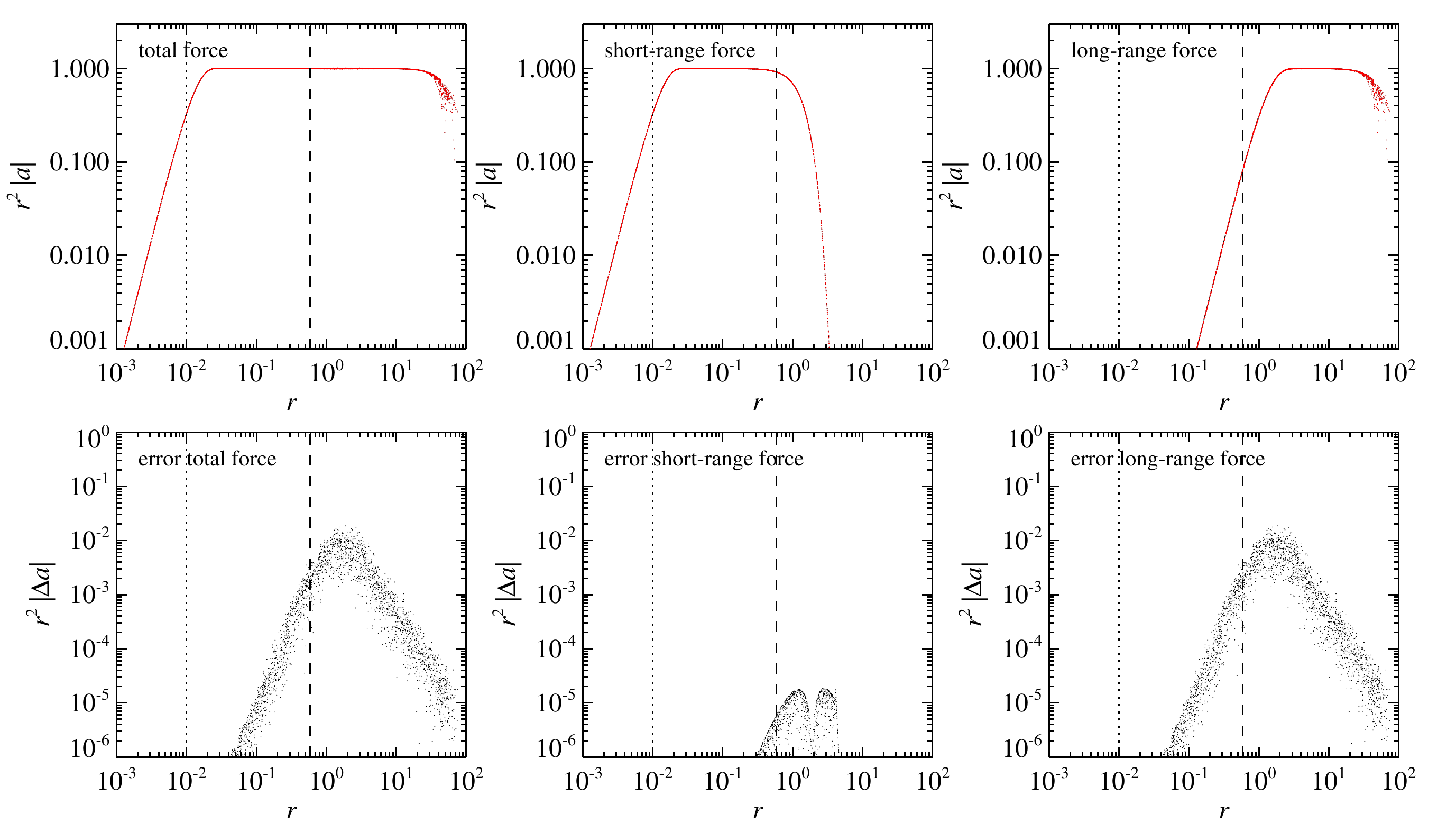}}
\resizebox{16cm}{!}{\includegraphics{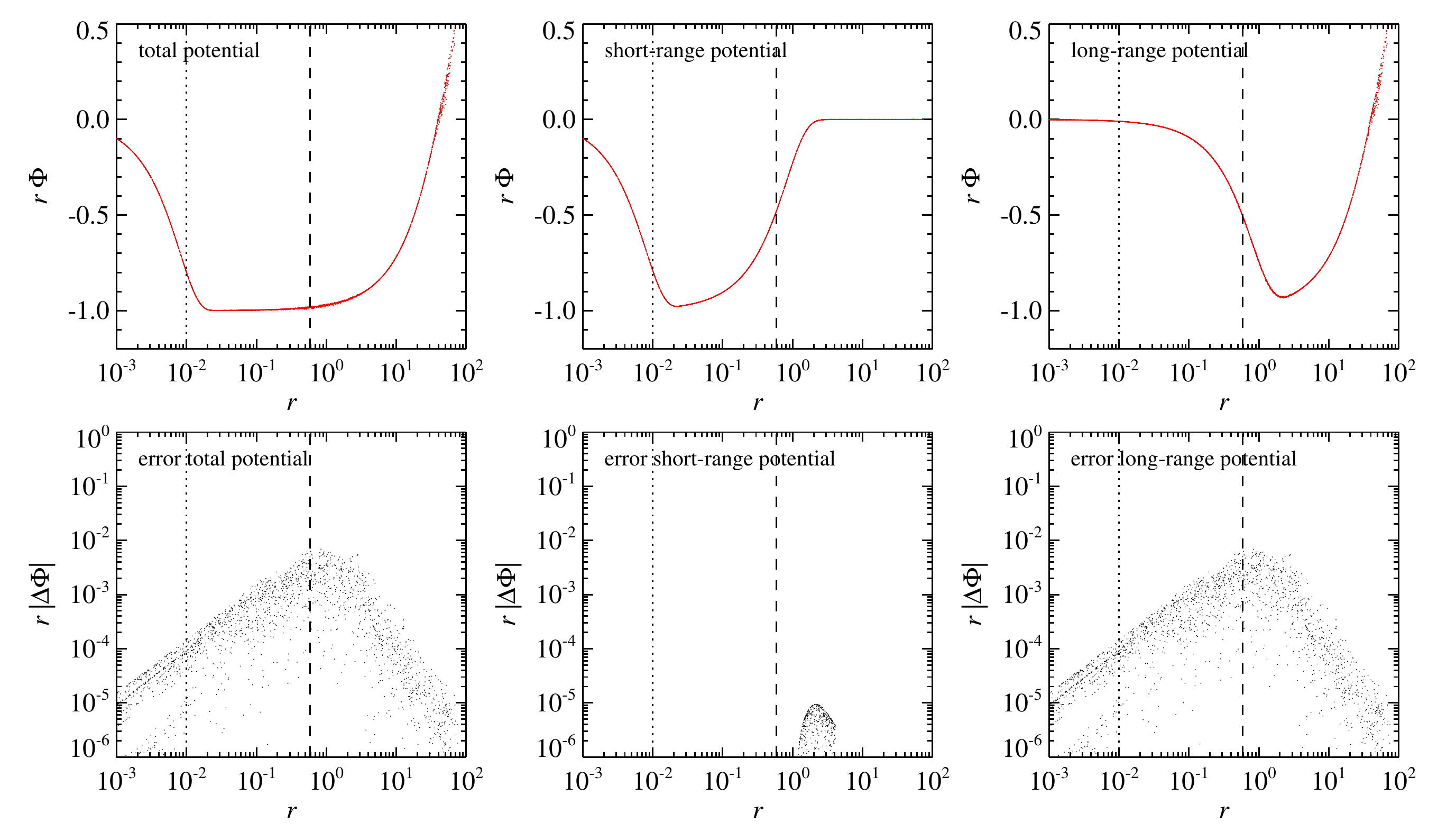}}
\caption{Force and potential laws arising in GADGET-4's TreePM and
  FMM-PM approaches to compute the total force as a sum of a
  short-range force delivered by a Tree- or FMM-approach, and a
  long-range force computed with a PM-method. The forces and
  potentials in this test are computed by placing a point particle of
  unit mass randomly into a periodic box and by choosing random
  evaluation coordinates. Note that for this setup, the results for
  Tree-PM and FMM-PM are identical by construction.  The box size is
  $100\,h^{-1}{\rm Mpc}$, with $r$ given in units of
  $h^{-1}{\rm Mpc}$.  In the top row of panels, we show the total
  force (top left) as a function of distance to the point mass, and
  decomposed into the short-range (top middle) and long-range (top
  right) components.  In the three cases, the force has been
  multiplied by $r^2$ to highlight deviations from a Newtonian force
  (which corresponds to a horizontal line in these panels). The
  increase for small $r$ reflects the gravitational softening law (of
  size $\epsilon = 10\,h^{-1}{\rm kpc}$, marked by dotted lines),
  while the decline and scatter at large $r$ are due to the periodic
  boundary conditions (see also Fig.~\ref{FigPeriodicForceLaw}). In
  the second row of panels, we show the relative force error arising
  in this approach, again as a function of distance from the point
  source and separately for the total force, and its short-range and
  long-range components. The short-range force error is higher than
  machine precision in the force matching region, due to our use of a
  look-up table to avoid costly evaluations of complementary error
  functions. However, the associated error is generally much smaller
  than the finite error of the PM approach itself, which originates
  from the use of a finite grid size (here $N_{\rm grid}= 256$) and
  grid smoothing scale (here determined by $A_{\rm smth}=1.5$,
  yielding
  $r_s = A_{\rm smth} L_{\rm box} / N_{\rm grid} = 0.59 \,h^{-1}{\rm
    Mpc}$ marked by dashed lines), from CIC binning and interpolation,
  as well as from finite differencing to get the force from the
  potential. Hence, in the limit of small tree opening angle, the
  total error is dominated by the PM-force, and peaks at the force
  matching scale. Similarly, the third and fourth rows show the
  corresponding decomposition of the total gravitational potential of
  a point mass into short-range and long-range contributions (here
  multiplied by $r$ to compress the vertical dynamic range), and the
  corresponding relative errors, respectively. Note that the total
  peculiar potential becomes positive at separations that approach
  half the box-size.
  \label{FigForceLawTreePM}}
\end{figure*}

\begin{figure}
\resizebox{8cm}{!}{\includegraphics{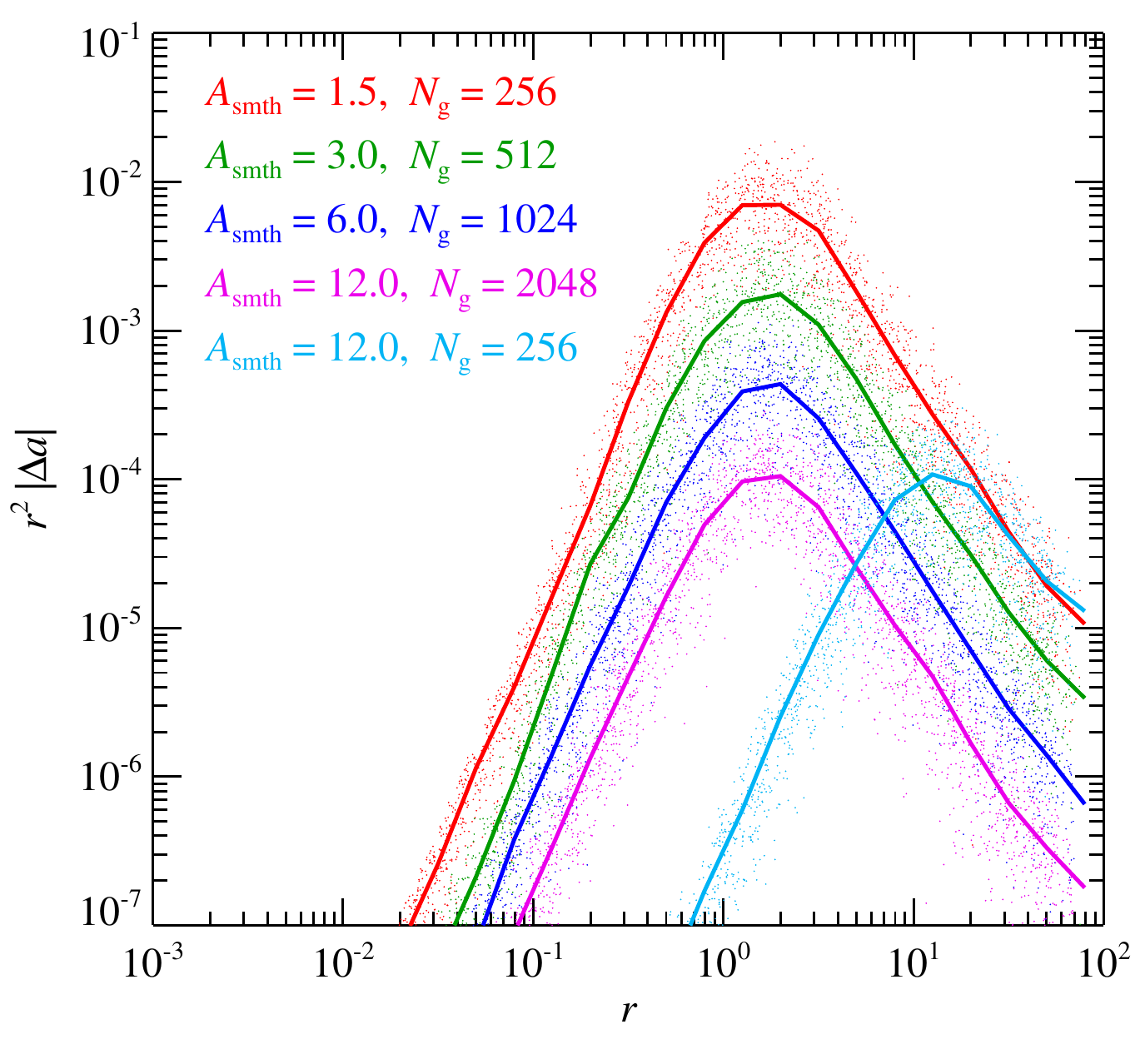}}
\caption{Relative force error for a point mass as a function of
  distance in the TreePM or FMM-PM schemes, for different choices
  of PM-mesh size $N_{\rm grid}$ and grid smoothing parameter $A_{\rm smth}$, as labelled. The
  maximum error occurs at the force matching scale, and is originating
  in the PM force. It can be arbitrarily reduced in size for fixed
  force matching scale by increasing $N_{\rm grid}$ and $A_{\rm smth}$ in
  lock-step. If only   $A_{\rm smth}$ is increased, the error declines
  as well, but then the force matching scale becomes larger, such that
  the Tree or FMM calculations of the short-range force need to cover
  a larger volume and thus become more expensive.
\label{FigAsmthTreePMVariation}}
\end{figure}

As described in Section~\ref{secgrav}, an alternative approach to
treat periodic boundaries is obtained by combining the Tree or FMM
methods with a PM-based calculation.  We begin by examining the basic
force law of the code for different scenarios in this case, in
particular to study the importance of the force matching region in
this case.  In Figure~\ref{FigForceLawTreePM} we show the effective
force law delivered by the code for a randomly placed point mass of
unit mass within a periodic box of size $L=100$ in dimensionless
units, with $G=1$. We compare the forces obtained by the short-range
tree and the long-range PM calculations with the exact forces for the
corresponding parts, determined by evaluating the relevant Ewald sums
with high precision.

The decline of the force for small $r$ shows the effect of the
gravitational softening, whereas the flat part in the top left panel
corresponds to the $\sim 1/r^2$ regime of the force. Once the distance
reaches an appreciable fraction of the box-size, the force law
deviates noticeably from this, with the force dropping to zero at a
distance approaching half the box-size. Likewise, the peculiar
potential changes sign and becomes positive in this regime.

Note that if the softening length $\epsilon_0$ exceeds about
$1.75\, r_s$, the short range force becomes {\em repulsive} as it then
needs to compensate the excess PM force which is computed disregarding
the softening length. Running the code in this mode is not very
efficient, as $r_{\rm cut}$ then needs to be made relatively large
(recall that $r_{\rm cut}$ should be larger than
$2.8\,\epsilon_0$). But if instructed to do so, {\small GADGET-4} can
still be operated in this regime.

We see from Fig.~\ref{FigForceLawTreePM} that the total error of the
force and the potential due to the Tree-PM/FMM-PM approach is
generally very small, except at around the force matching scale, where
in this example the relative errors reach a maximum of around
1\%. There are two sources for these errors. In the short-range force,
they can arise from interpolation errors in the look-up table we use
to avoid expensive evaluations of complementary error functions, or
from the use of a too small value of $r_{\rm cut}$ that causes distant
contributions in the short-range force to be neglected. In the
long-range force, the force errors arise from residual anisotropies of
the PM-force and from truncating the Fourier series at some finite
Nyquist frequency.

These errors can be reduced easily, essentially arbitrarily. In the
former case this can be done by using a finer look-up table and/or a
larger $r_{\rm cut}$, in the latter case by enlarging $A_{\rm smth}$.
Our short-range look-up table has a default length of 48 entries out
to $R_{\rm cut}$. Making the table finer reduces the achievable
short-range force errors, but this is pointless unless the residual PM
error is reduced to a similar or smaller level.  The PM-force error on
the other hand reflects the residual discreteness and finite size of
the grid used for this part of the calculation. The size of the
corresponding errors can be reduced, if desired, by resolving the PM
force with more cells (or in part by using higher-order kernels, such
as TSC instead of CIC), i.e.~keeping the transition scale $r_s$ fixed
but using a finer PM-mesh, which effectively means to increase the
grid size $N_{\rm grid}$ and the parameter $A_{\rm smth}$ in lock-step. One may
also increase $A_{\rm smth}$ alone without increasing the PM-mesh size
itself. This will also reduce the maximum error of the PM-force while
keeping its computational cost constant, but now the short-range force
becomes more expensive as then $r_s$ grows and fewer tree nodes can be
discarded. Note however that there is a limit to all this, dictated by
the requirement that we should see only one nearest neighbour within
the short-range cut-off region, otherwise the Tree-PM or FMM-PM
approach will break down. If we conservatively set
$r_{\rm cut} = 8 \,r_s$, this translates to the important condition
$A_{\rm smth} < N_{\rm grid} / 16$.

In Figure~\ref{FigAsmthTreePMVariation} we show how the maximum
long-range force error varies with the chosen $A_{\rm smth}$ and the
mesh size. Changing from $A_{\rm smth}=1.5$ to a much more conservative
value of $A_{\rm smth}=12.0$ reduces the maximum error by about two
orders of magnitude, to a level of $10^{-4}$. Provided one stays
within the constraint $A_{\rm smth} < N_{\rm grid} / 16$, this maximum error is
then independent of the grid-size itself.

\subsection{Force accuracy for the Tree-PM and FMM-PM approaches}

\begin{figure*}
\resizebox{8cm}{!}{\includegraphics{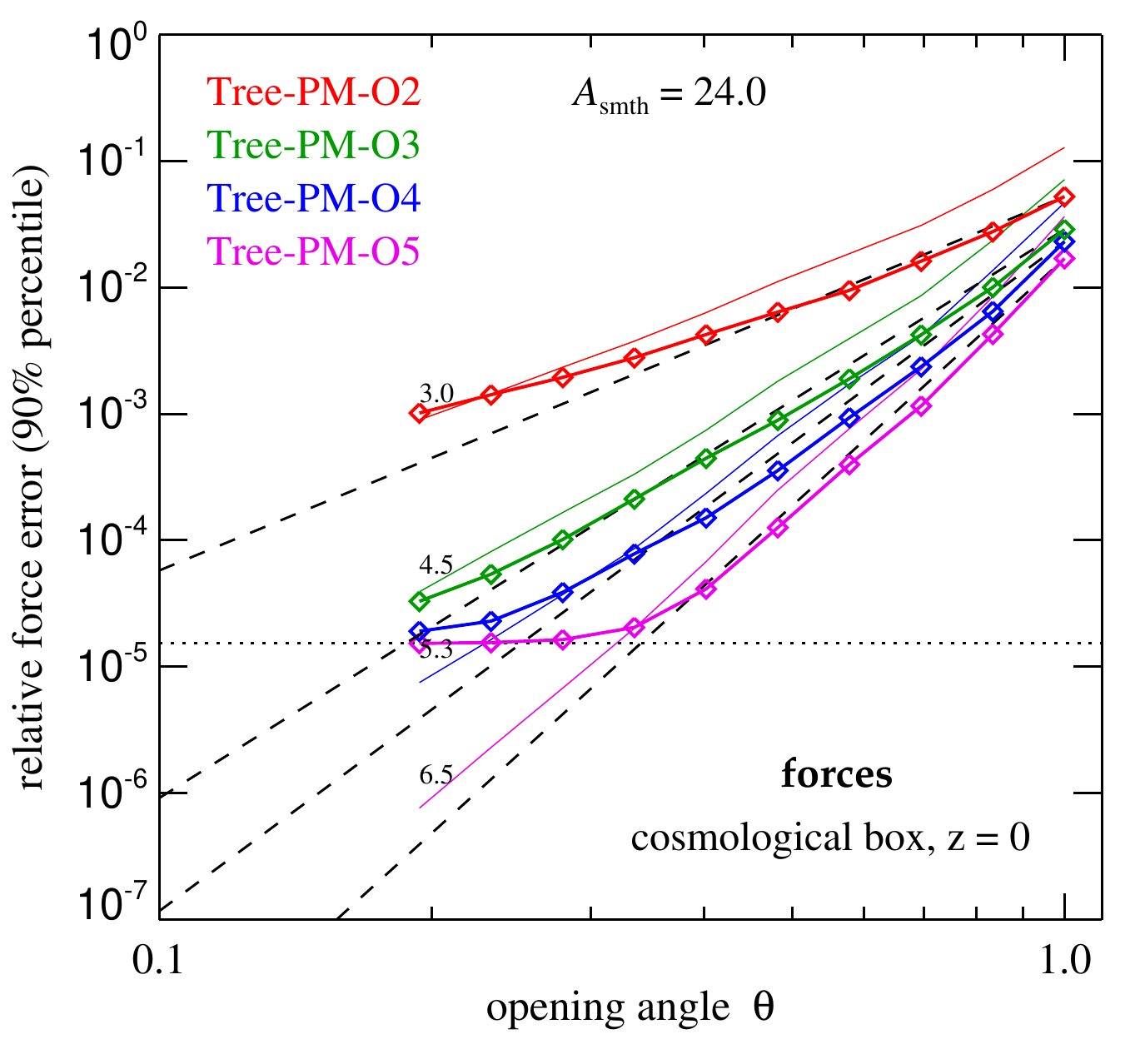}}%
\resizebox{8cm}{!}{\includegraphics{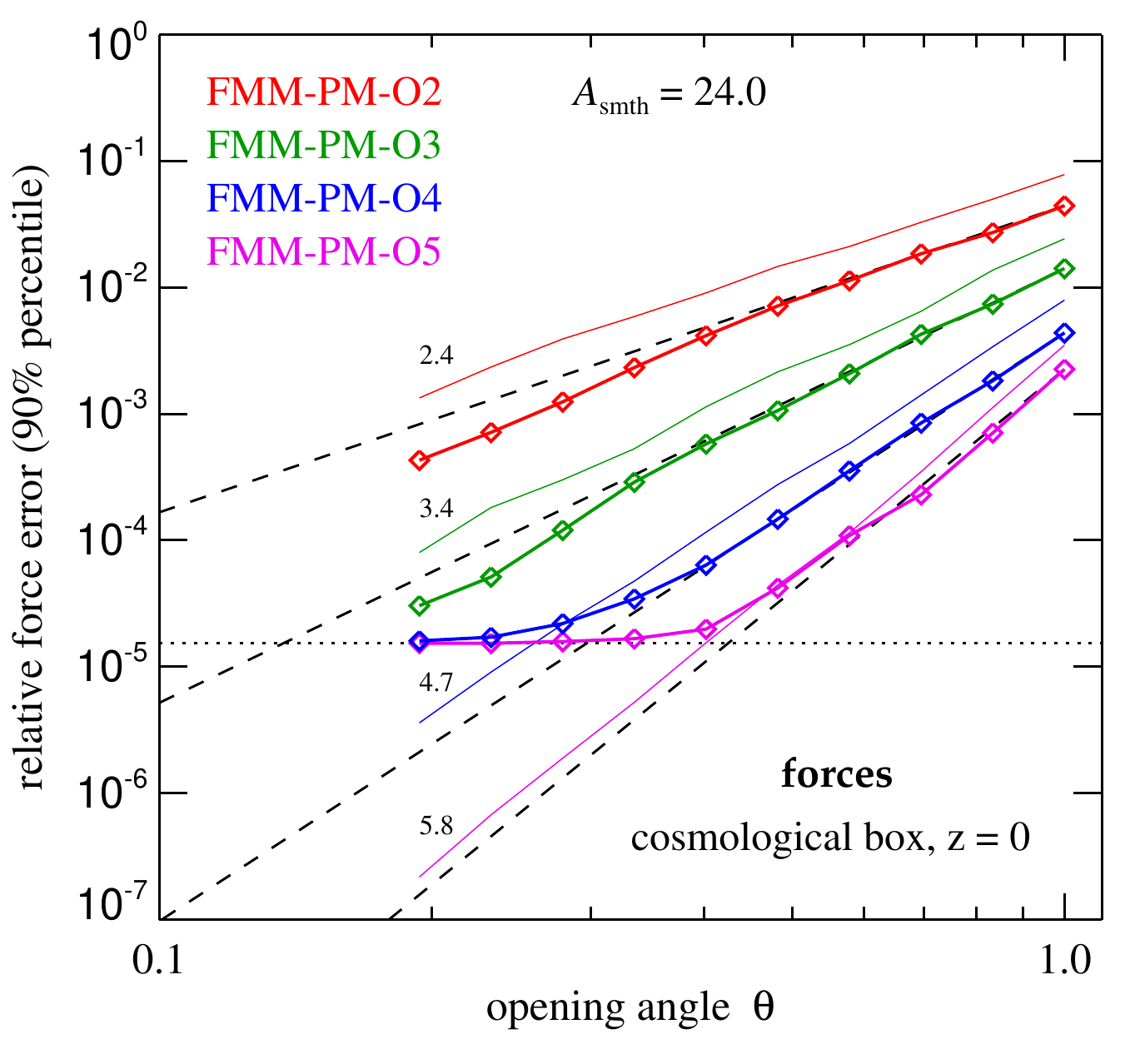}}\\
\resizebox{8cm}{!}{\includegraphics{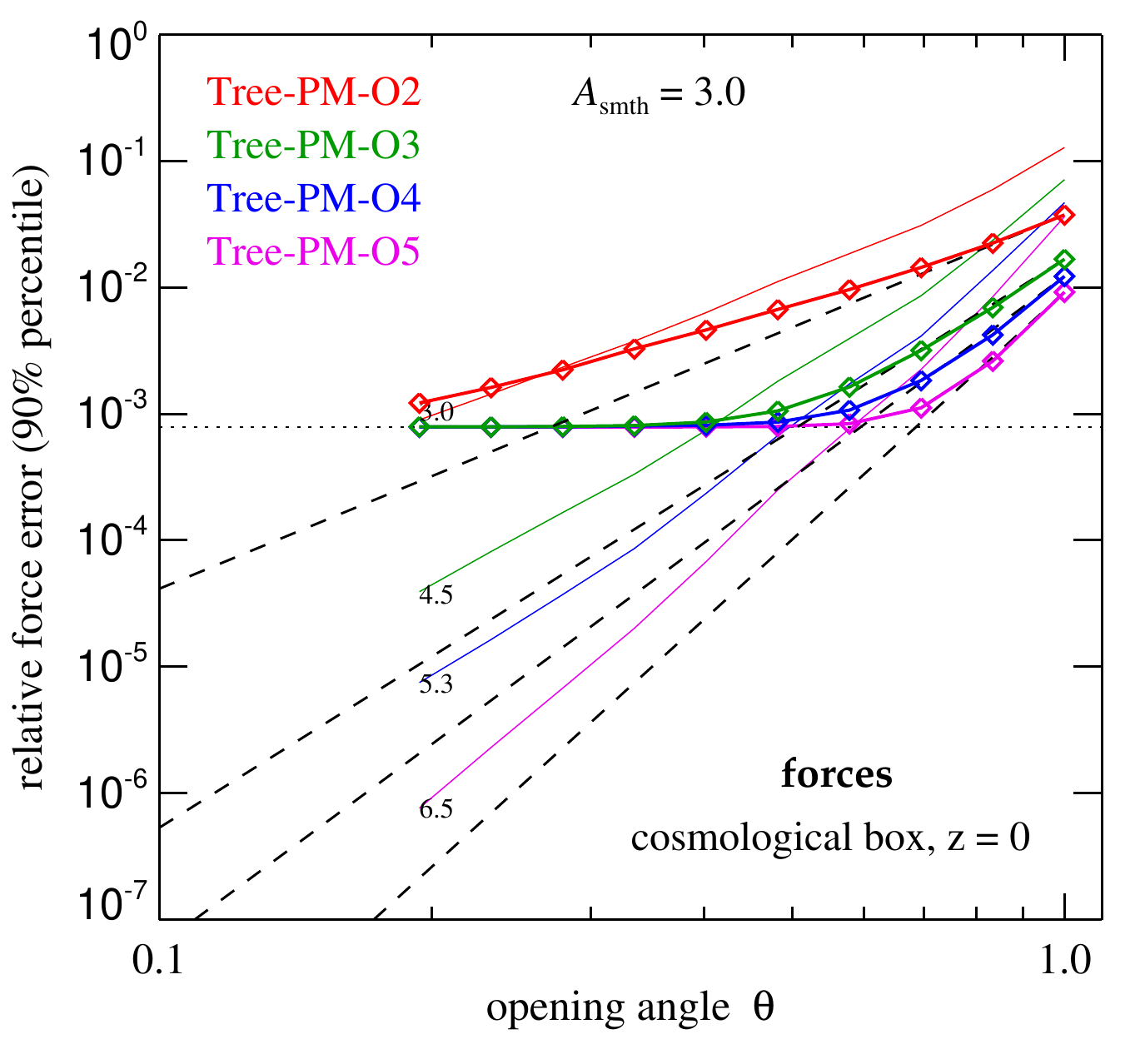}}%
\resizebox{8cm}{!}{\includegraphics{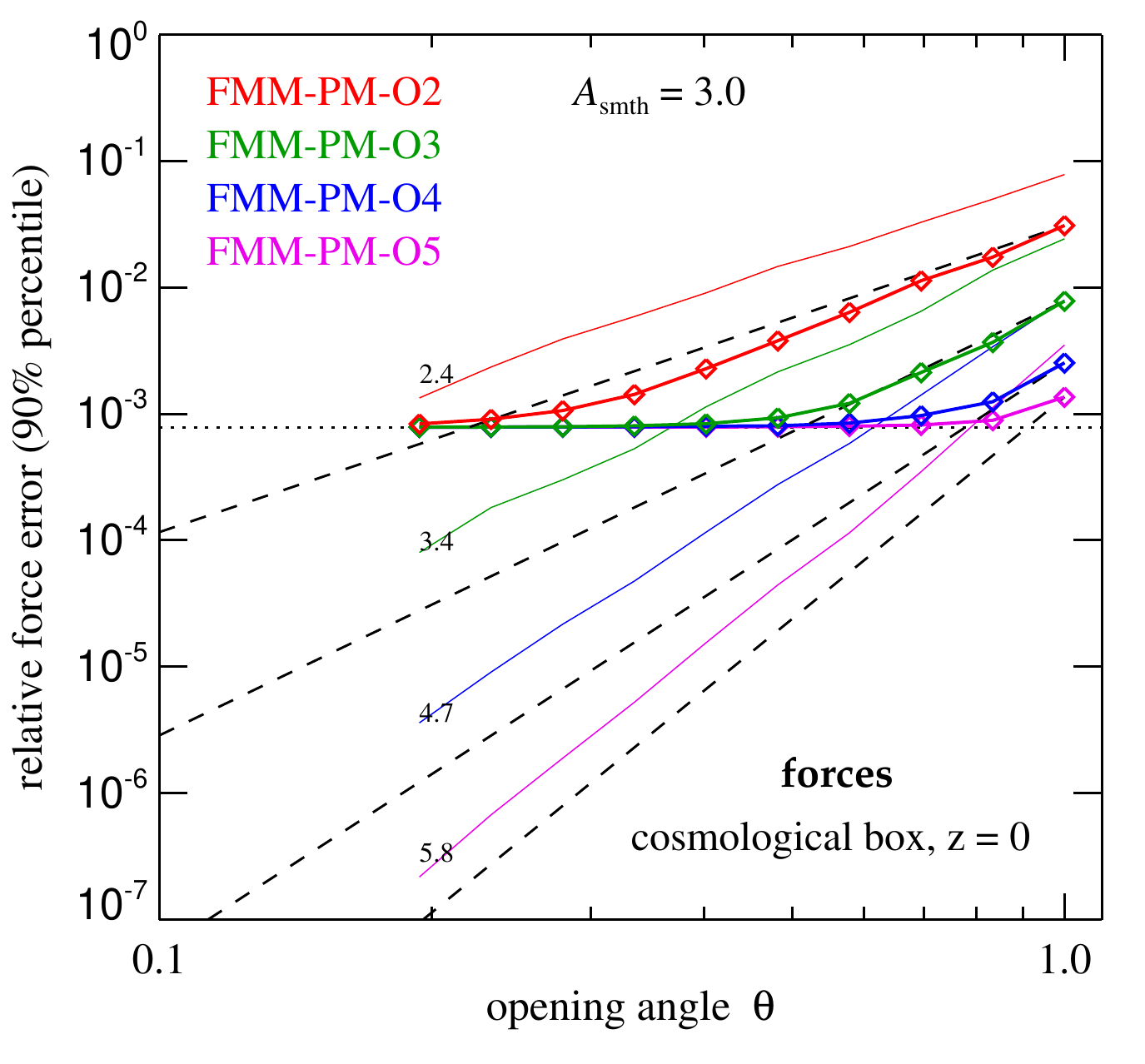}}\\
\caption{Force accuracy as a function of geometric opening angle
  $\theta$ for a cosmological simulation box at $z=0$ when the Tree-PM
  (left column) or FMM-PM (right column) algorithms are used with
  different expansion orders, and for different choices of the grid
  smoothing parameter $A_{\rm smth}$. In the top panels, a comparatively large
  force-split scale corresponding to $A_{\rm smth}=24$ with $N_{\rm grid}=512$ has been used,
  while the two bottom panels used $A_{\rm smth}=3$ with $N_{\rm grid} = 256$,
  i.e.~a four times smaller split scale. The examined
  particle distribution is the same as used in
  Figure~\ref{FigMedianForceErrVsThetaBoxSimZ0}, where equivalent
  results for pure multipole expansions with Ewald correction were
  shown. In fact, these former results are here repeated as thin solid
  lines, for easier comparison with the results obtained for the Tree-PM and
  FMM-PM algorithms, which are shown with thick solid lines. The
  dashed lines show the power-law scalings inferred from these pure
  multipole results, but normalized to the errors obtained for the
  Tree-PM or FMM-PM methods for $\theta = 1$. For large
  $\theta$, the errors are actually slightly reduced when the PM
  approach is enabled, due to the
  more accurate computation of the long-range force components this entails. In this regime, the errors
  are then dominated by the short-range force, and decline towards smaller opening
  angle as
  expected for the corresponding multipole order.
  However, the errors eventually hit an irreducible floor (marked by
  dotted horizontal lines),
  which originates in the finite accuracy of the PM calculation. Further
  reducing $\theta$ is futile in this case, but increasing
  $A_{\rm smth}$ can make the total  error still smaller, if desired.
  \label{FigTreePM-FMMPM-VsOpening}}
\end{figure*}

\begin{figure*}
\resizebox{8cm}{!}{\includegraphics{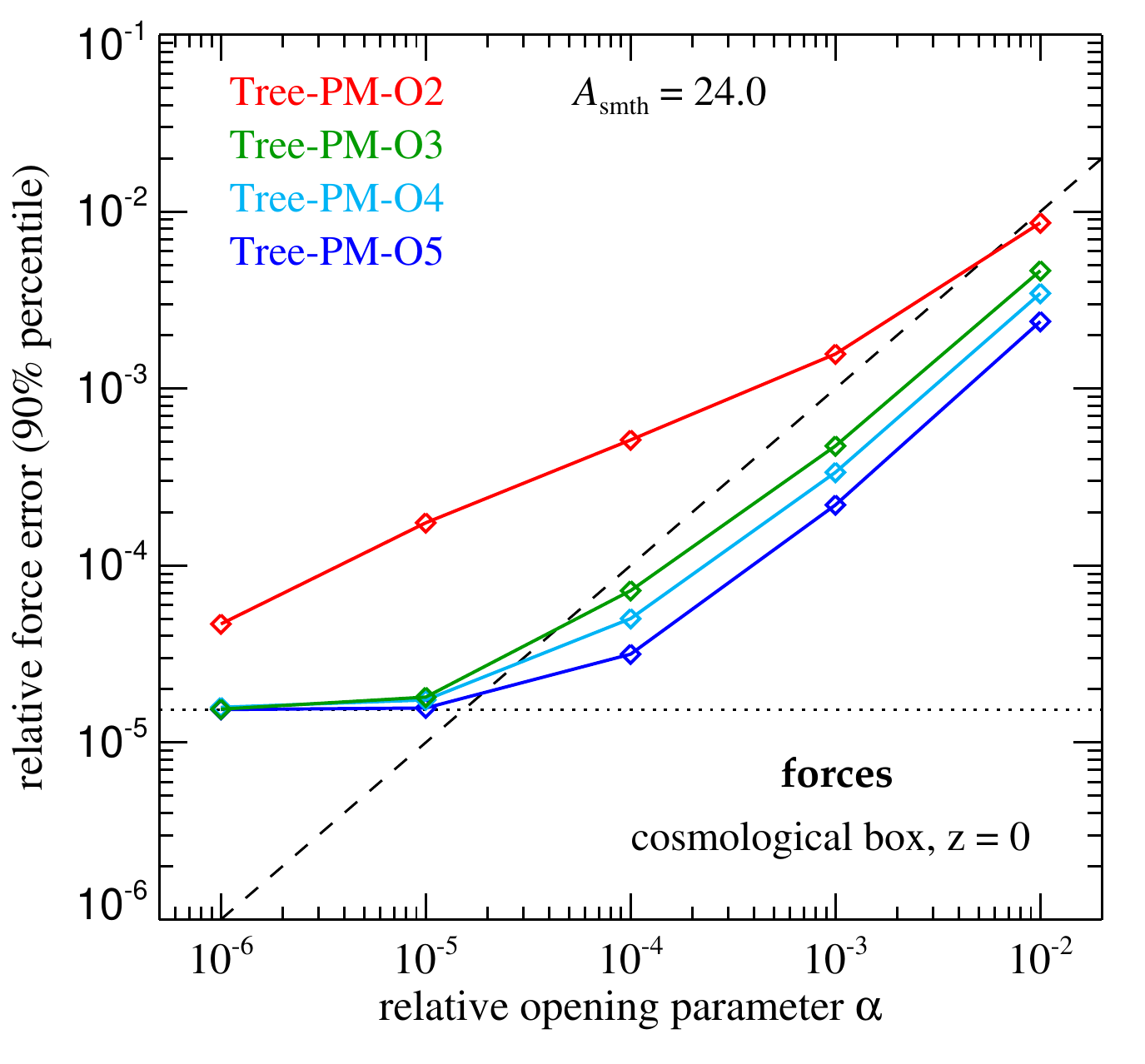}}%
\resizebox{8cm}{!}{\includegraphics{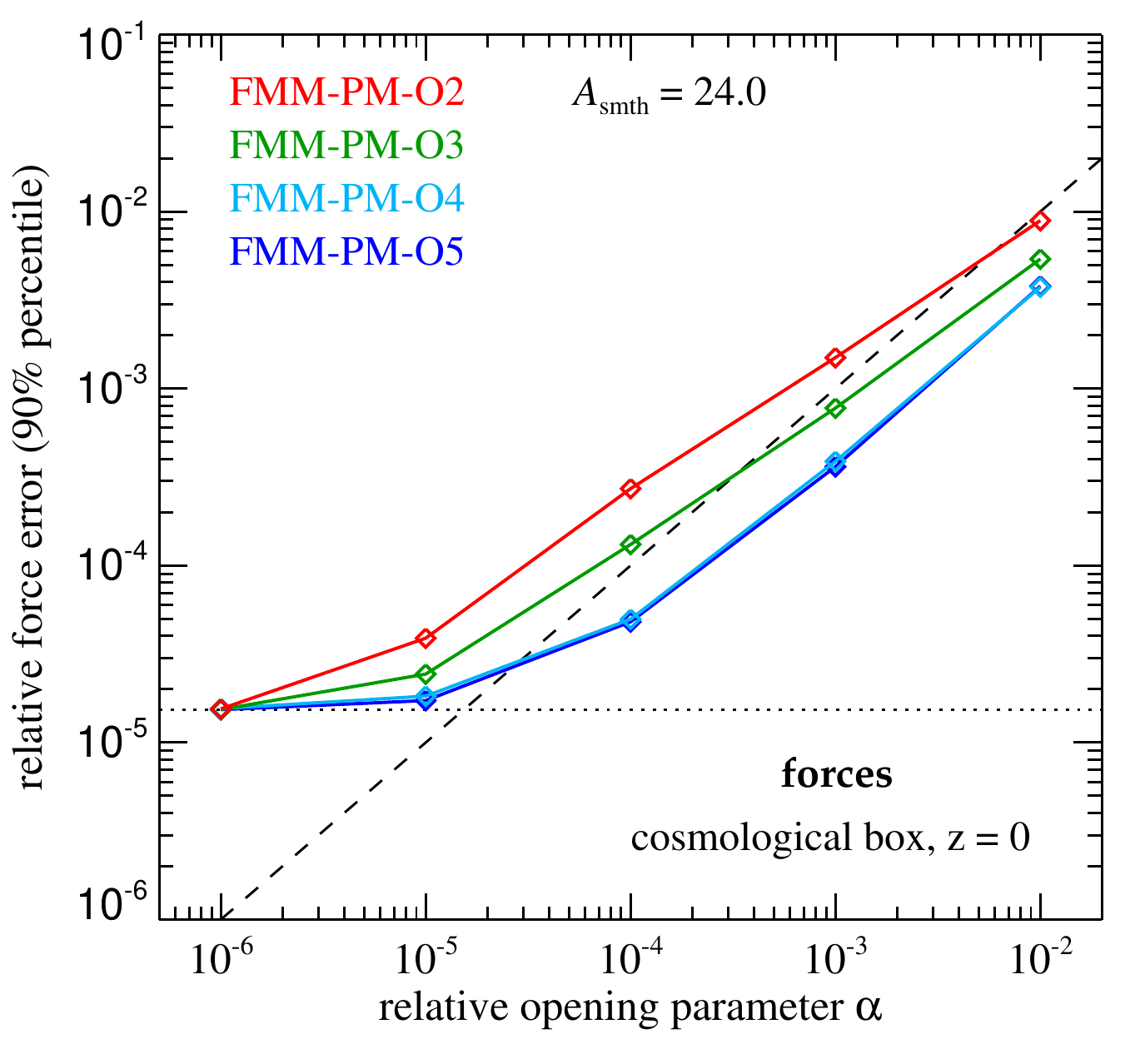}}
\caption{Force accuracy
 for a cosmological simulation box at $z=0$ when the
 Tree-PM (left panel) or FMM-PM (right panel) algorithms are used with
 the relative opening
 criterion, for different multipole orders $p$.
 The diagonal dashed line has a
 slope of unity and corresponds to a one-to-one relation between the
 opening parameter $\alpha$ and the 90-percentile force error. This
 line is followed reasonably well for large values of $\alpha$,
 to good accuracy independent of multipole order and multipole
 scheme. In this regime, the force errors are  dominated by the  tree-based
  component. Similarly to the approach with Ewald-correction, one
  can thus use  $\alpha$ to directly set the desired force accuracy level.
However,
  for
  small settings of $\alpha$, a floor in the force accuracy (dotted
  line) is reached which is set by the finite precision of the PM
  force (see also Fig.~\ref{FigTreePM-FMMPM-VsOpening}). In this case,
  a further improvement of the relative force error can not be reached by
  reducing $\alpha$; instead it requires a reduction of the force
  error from the PM computation by increasing $A_{\rm smth}$. 
  \label{FigTreePM-FMMPM-VsRelativeOpening}}
\end{figure*}

\begin{figure*}
\resizebox{8cm}{!}{\includegraphics{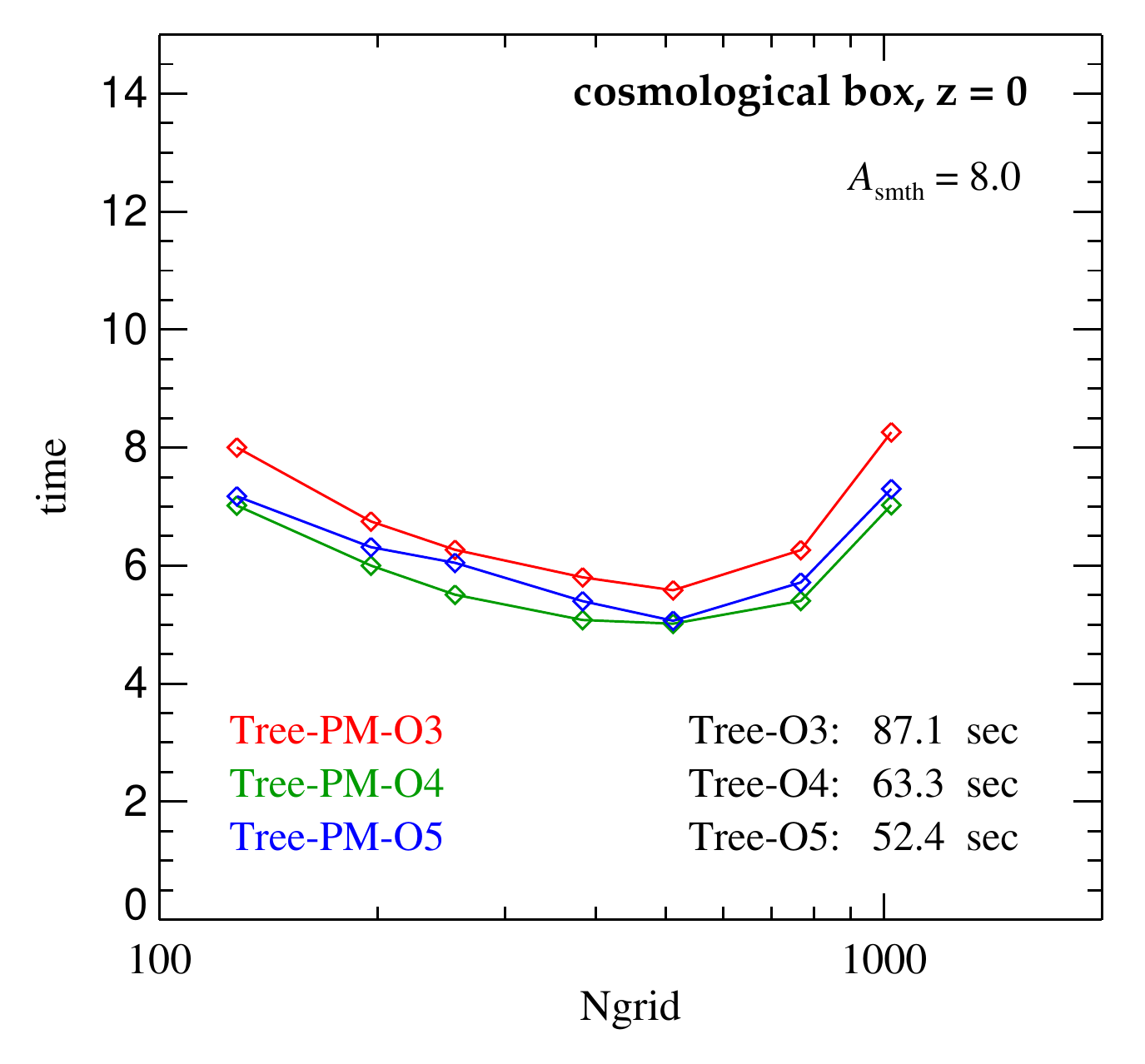}}%
\resizebox{8cm}{!}{\includegraphics{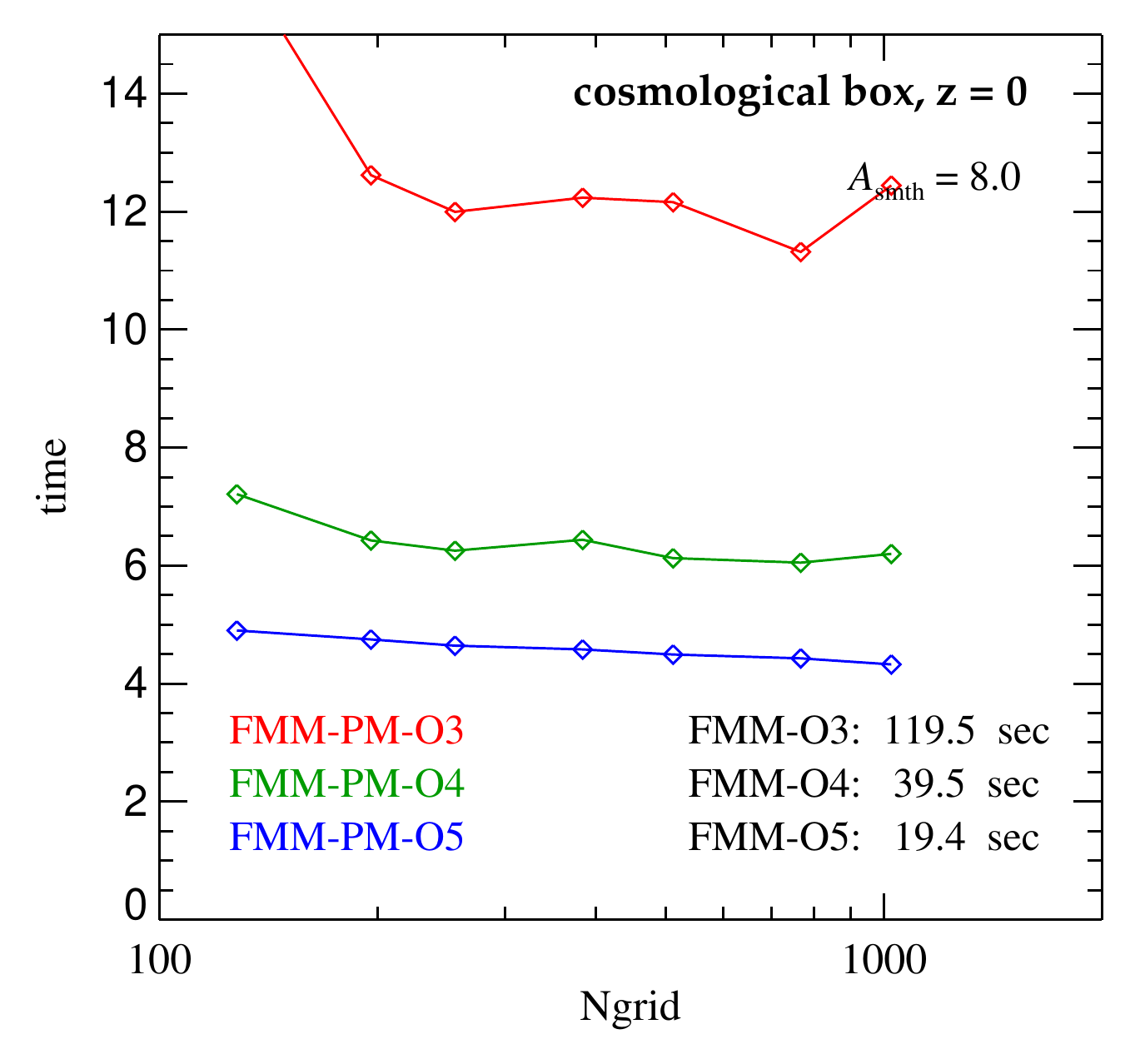}}\\
\resizebox{8cm}{!}{\includegraphics{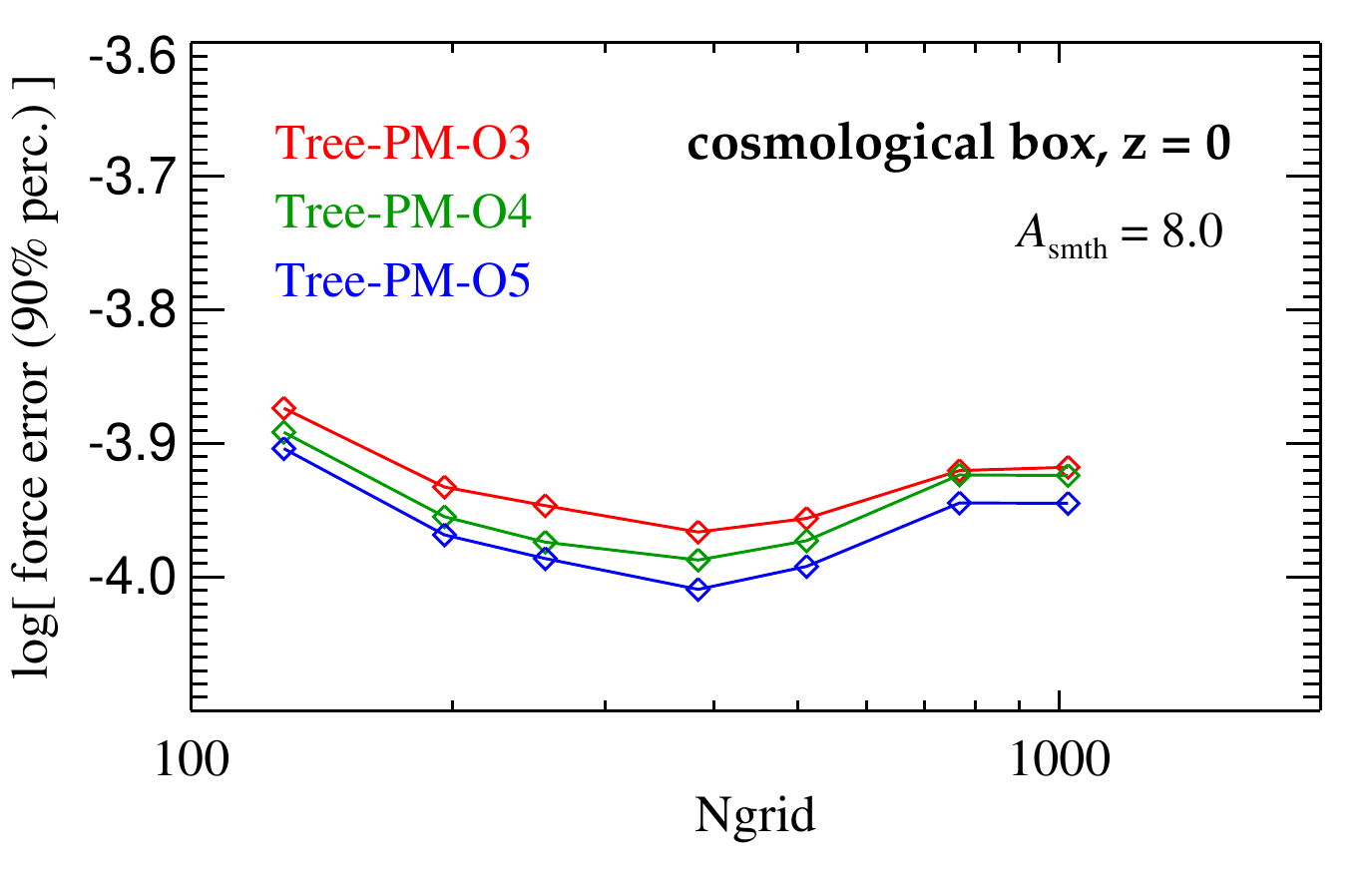}}%
\resizebox{8cm}{!}{\includegraphics{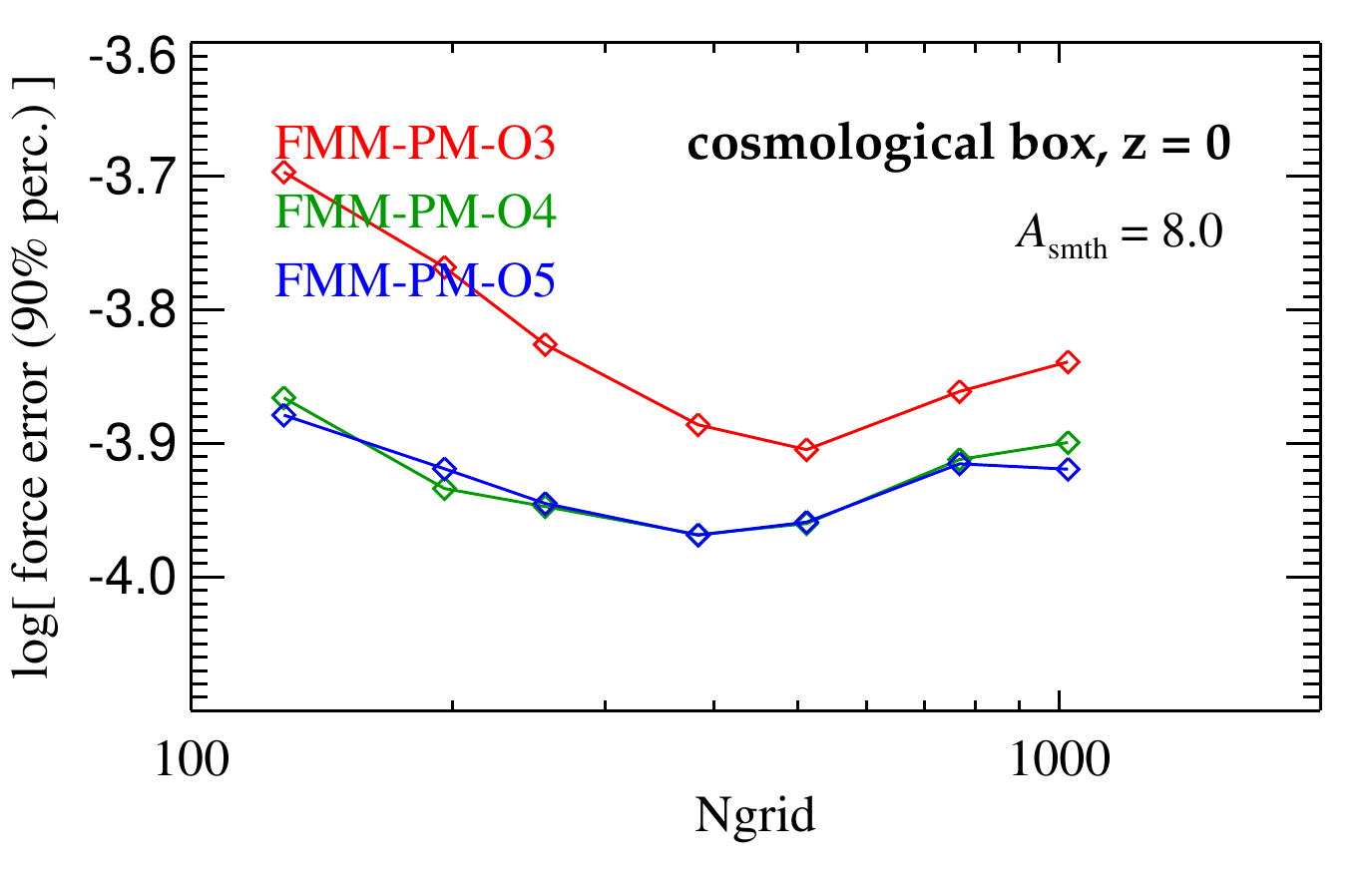}}\\
\caption{Force calculation time and force accuracy in the Tree-PM and
  FMM-PM algorithms as a function of the mesh size employed.  We here
  adopt a constant value of $\alpha$ (for definiteness chosen
  conservatively as $\alpha=0.0001$, and with a high
  $A_{\rm smth}=8.0$ to suppress PM force errors to correspondingly
  low levels), but vary the grid resolution $N_{\rm grid}$ over a wide
  range. In particular, we would like to establish whether the
  execution time depends strongly on $N_{\rm grid}$. The panels on top
  thus show the runtime for a full force evaluation of a $128^3$ run
  at $z=99$ in a $30\,h^{-1}{\rm Mpc}$ box, carried out with $N_{\rm
    cpu}=40$ cores for different $N_{\rm grid}$,
  with the TreePM algorithm on the left, and the FMM-PM on the
  right. Coloured lines give results for different multipole expansion
  orders, as labelled. Reassuringly, the force accuracy shows only
  negligible variation for different mesh choices, and also Tree-PM
  and FMM-PM are quite similar overall. The bottom panel shows the corresponding force
  accuracy measurements.  With respect to run-time, CPU-time consumption is shifted from the tree-based
  calculations to the Fourier-based computations for growing
  $N_{\rm grid}$, but their sum varies
  only comparatively weakly with $N_{\rm grid}$. There is an optimum
  for a grid resolution about $ 2-3$ times the particle grid, and this
  is somewhat more pronounced for TreePM
  compared to FMM-PM. Interestingly, FMM-PM on the other hand shows a
  stronger sensitivity of its runtime to expansion order (note that
  this is at fixed force accuracy). Clearly, only for
  high order our FMM-PM implementation begins to (moderately) outperform
  Tree-PM. Note that if the multipole methods are however run without the
  PM-acceleration for the periodic box,
  they are indeed quite a bit slower; this is demonstrated by the
  inlined numbers in the top panels, which give the corresponding
  execution times if the Tree or FMM methods are used, which have to
  rely on Ewald
  correction instead.  
  \label{FigPerformanceVsNgrid}}
\end{figure*}

We now repeat our measurements of the force accuracy in cosmological
boxes, but this time employing combinations of the PM algorithm with
the Tree or FMM methods. In Figure~\ref{FigTreePM-FMMPM-VsOpening}, we
show results for the same $z=0$ dataset considered earlier, this time
employing either Tree-PM or FMM-PM methods for different expansion order. We
also consider the PM algorithm for two different values of
$A_{\rm smth}$. In general, we see that for large opening angles, 
Tree-PM and FMM-PM show slightly better accuracy than the corresponding pure
multipole algorithms, whereas for small opening angle, the accuracy
asymptotes to a finite value that can not be improved any further by
reducing the opening angle. Clearly, in this regime the accuracy is
limited by the PM part of the algorithm, and improving it
further requires increasing $A_{\rm smth}$.

In contrast, in the opposite regime, for large opening angles, the
accuracy is limited by the short-range Tree and FMM parts,
respectively. Here the errors scale with the same power-law as in the
plain multipole algorithms, but the amplitudes are a bit smaller because
part of the force is supplied by the (in this regime) more accurate PM
computation.

This raises the question which settings are optimal as far as the
computational cost of the Tree-PM and FMM-PM algorithms is
concerned. For a fixed split scale $r_s$, the optimum will be reached
if the contribution to the force errors of PM and Tree/FMM are roughly
equal. Because if the PM was more accurate than the Tree/FMM, one
could typically save computational time by using a smaller $N_{\rm grid}$,
combined with a smaller $A_{\rm smth}$ at fixed $r_s$. This would
leave the Tree/FMM errors invariant and increase the PM error, but
since the former dominates, the final result would not change
significantly.  Conversely, if the Tree/FMM part would produce
subdominant errors, one could increase the opening angle while saving
CPU time until the errors of both force components are roughly equal.
What is not immediately clear, however, is which value of $r_s$ is
best for a given problem.

In Figure~\ref{FigTreePM-FMMPM-VsRelativeOpening} we first verify that
the relative opening criterion still roughly works in TreePM/FMM-PM as
a convenient way to set the overall relative force accuracy of the
multipole part of the force, independent of expansion order.  A simple
strategy for finding the optimum run-time regime can then be as
follows. One first decides about the desired force accuracy level, and
then sets $A_{\rm smth}$ to the smallest value such that this can be
reached. Likewise, one uses a corresponding $\alpha$ value. One can
now vary $N_{\rm grid}$ systematically, subject to the constraint
$N_{\rm grid} > 16\, A_{\rm smth}$, picking the value that minimizes the
runtime for a given expansion order $p$. Of course, the optimum thus
identified will in general depend on the particular system that is
studied, on details of how the algorithm is actually coded up, whether
or not single precision is used in parts of the calculations, how
large possible parallelization losses are, and on technical details of
the platform that is used to execute the test. But the qualitative
findings should be robust to these details, and thus can ideally allow
the identification of general recommendations that then lead to
practical settings that are not too far away from optimum ones.

Figure~\ref{FigPerformanceVsNgrid} provide some pointers along these
lines. We here examine how the performance of the Tree-PM and FMM-PM
schemes varies as a function of the chosen mesh size, at approximately
fixed force error. Interestingly, we find that the optimum is fairly
broad as a function of $N_{\rm grid}$, i.e.~provided one choses
$N_{\rm grid} \sim (2\pm 1)\, N_p$, where $N_p^3$ is the particle number, the
PM-accelerated multipole algorithms give good performance.  We also
find that going to high order tends to be more important for FMM than
for the one-sided Tree in order to reach the highest
speeds. Interestingly, FMM-PM however does not show a clear
speed-advantage over Tree-PM, unlike the case for plain FMM vs. plain
Tree. There appears to be even a slight advantage for TreePM, except
for order $p=5$, where FMM-PM-O5 comes out narrowly on top.  In the
figure, we also report the times taken for the plain multipole
algorithms when applied with Ewald correction to the same
problem. They show that the hybrid PM approaches are considerably
faster than the plain multipole algorithms with Ewald summation, hence
for cosmological simulations, Tree-PM and FMM-PM are the most
efficient solvers in {\small GADGET-4}.

\subsection{Secondary Fourier mesh for a high-resolution zoom region}

\begin{figure*}
\resizebox{18.3cm}{!}{\includegraphics{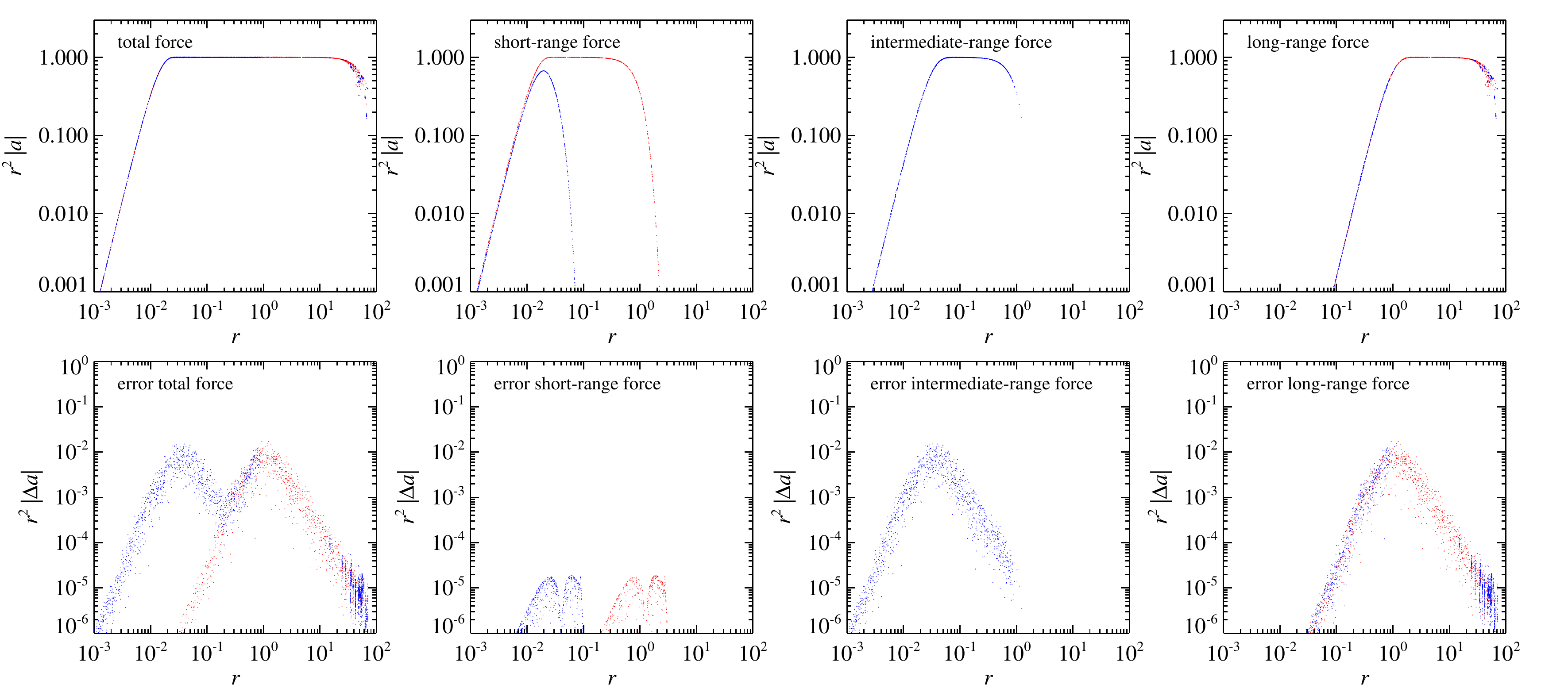}}
\resizebox{18.3cm}{!}{\includegraphics{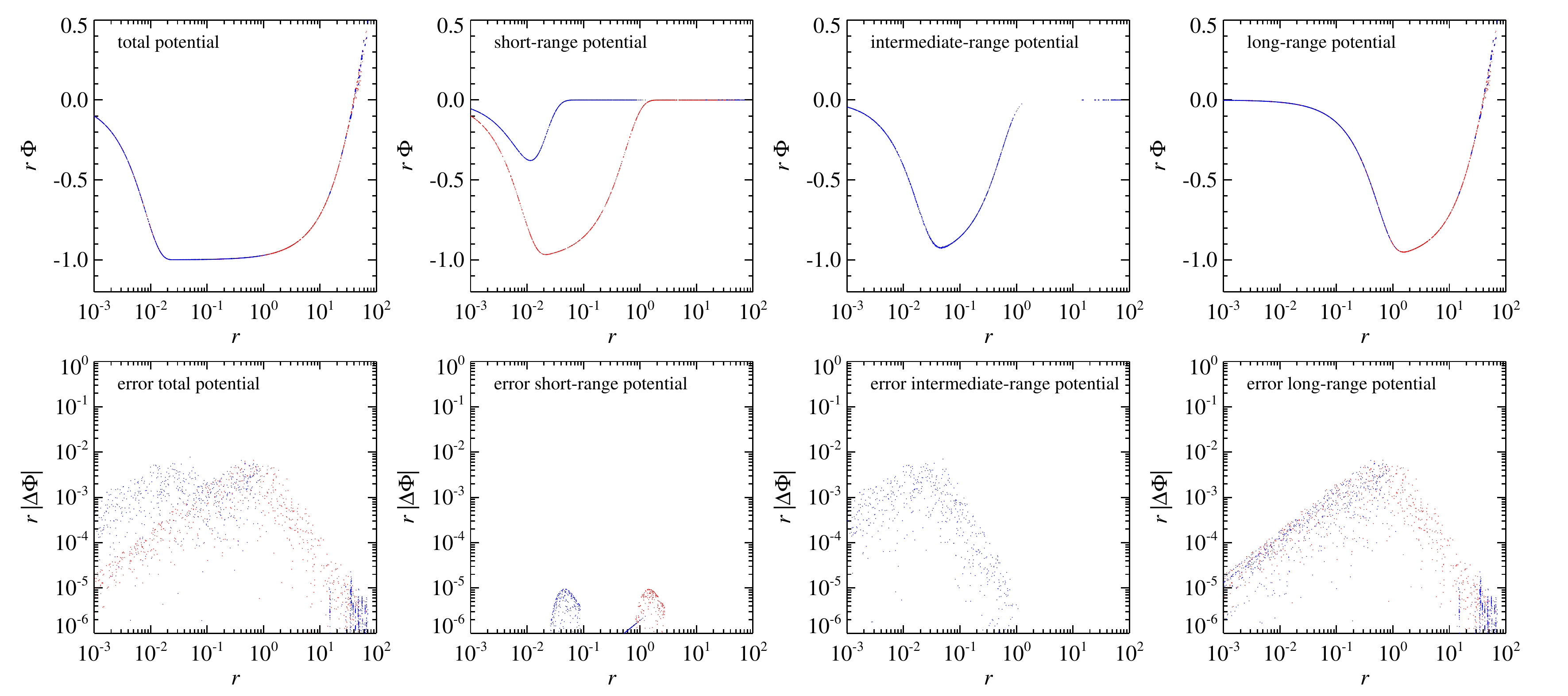}}
\caption{Force and potential laws resulting from GADGET-4's 
Tree-HRPM or FMM-HRPM approaches, where a secondary high-resolution PM
mesh is used to accelerate the force calculation in cosmological
`zoom' simulations. This plot is similar to Fig.~\ref{FigForceLawTreePM},
except that here also a secondary high-resolution PM mesh is present.
Whether or not the force between two particles is computed with the
ordinary Tree/FMM-PM approach, or with one where two Fourier meshes
are used, depends on the spatial locations of the particle pair. If
{\em both of them} are inside a cubical high-resolution region
(which includes all particles of a certain pre-defined type),
then a secondary force-split scale is introduced in the short-range
part of the usual Tree/FMM-PM method, so that the region that is
computed with the multipole approach shrinks further, and a
secondary PM mesh (with zero-padding to realize non-periodic boundary
conditions) can be used to complement the force with an
intermediate-range contribution. In the present plots, particles where this
applies are plotted in blue. If at least one particle of a pair is
outside the high-resolution region, no secondary PM mesh is used and
only the single
split scale of the ordinary Tree/FMM-PM approach applies. These are
the red particles in the plots. For the short-range force, there are
thus two different force and potential laws for blue and red
particles. The intermediate-range force is only relevant for blue
particles, while the long-range force law applies to both in the same way.
The largest errors in this approach occur in the two force
matching regions and originate in the finite accuracy of the PM
approach. The latter can however be made more accurate, if desired, by
using an appropriately larger smoothing $A_{\rm smth}$ and mesh size $N_{\rm
  g}$. Note that the number of cells per dimension for the background PM mesh covering the
full box and for the high-resolution patch can in principle be chosen
independently. 
\label{FigForceLawTreeHRPM}}
\end{figure*}

In Figure~\ref{FigForceLawTreeHRPM}, we extend the previous test to the case
of an active high-resolution placement. First we show the effective 
force law for an illustrative test case where the high-res region covers
1/10 of the periodic box. Again, we place a source particle randomly
inside the box, sometimes inside and sometimes outside the region
covered by the secondary high-resolution PM grid, and then measure the
force accuracy delivered by the code relative to the analytic expectation. This
in particular allows us to characterize the force accuracy in the
transition regions between the PM meshes and the short-range tree
force. Recall that the high-res patch is aligned with the geometry of
the oct-tree at a suitable grid level.

Similar to the ordinary TreePM and FMM-PM schemes with just a single,
periodic FFT, we obtain for $A_{\rm smth}=1.5$ a total force error
that is of order 1\% error in the force matching regions. If desired,
this force error can be reduced by using a larger value for
$A_{\rm smth}$. For a fixed PM-mesh size, this however means a larger
$r_s$ and thus more work for the Tree or FMM algorithm.

\begin{figure*}
\resizebox{8cm}{!}{\includegraphics{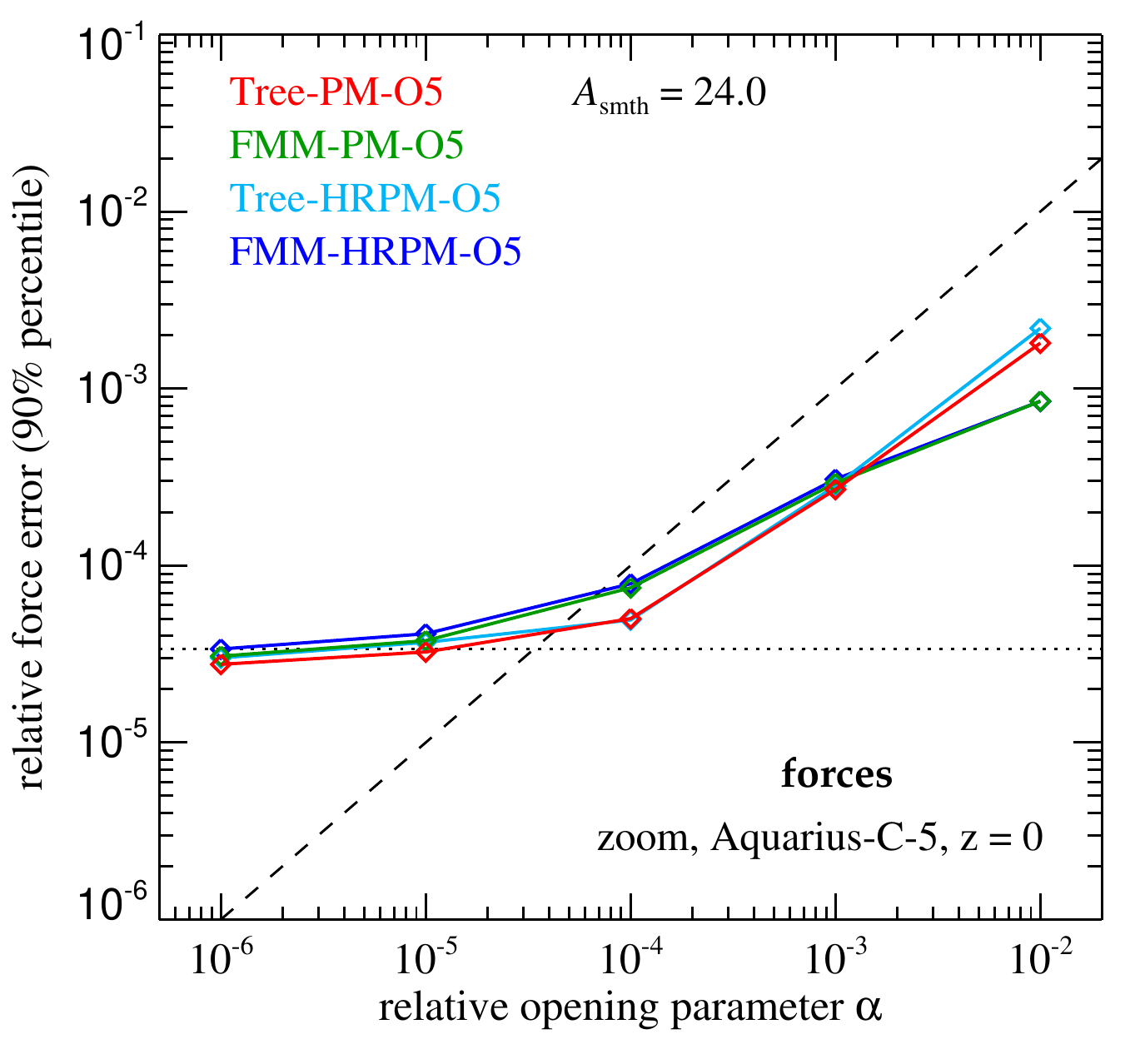}}%
\resizebox{8cm}{!}{\includegraphics{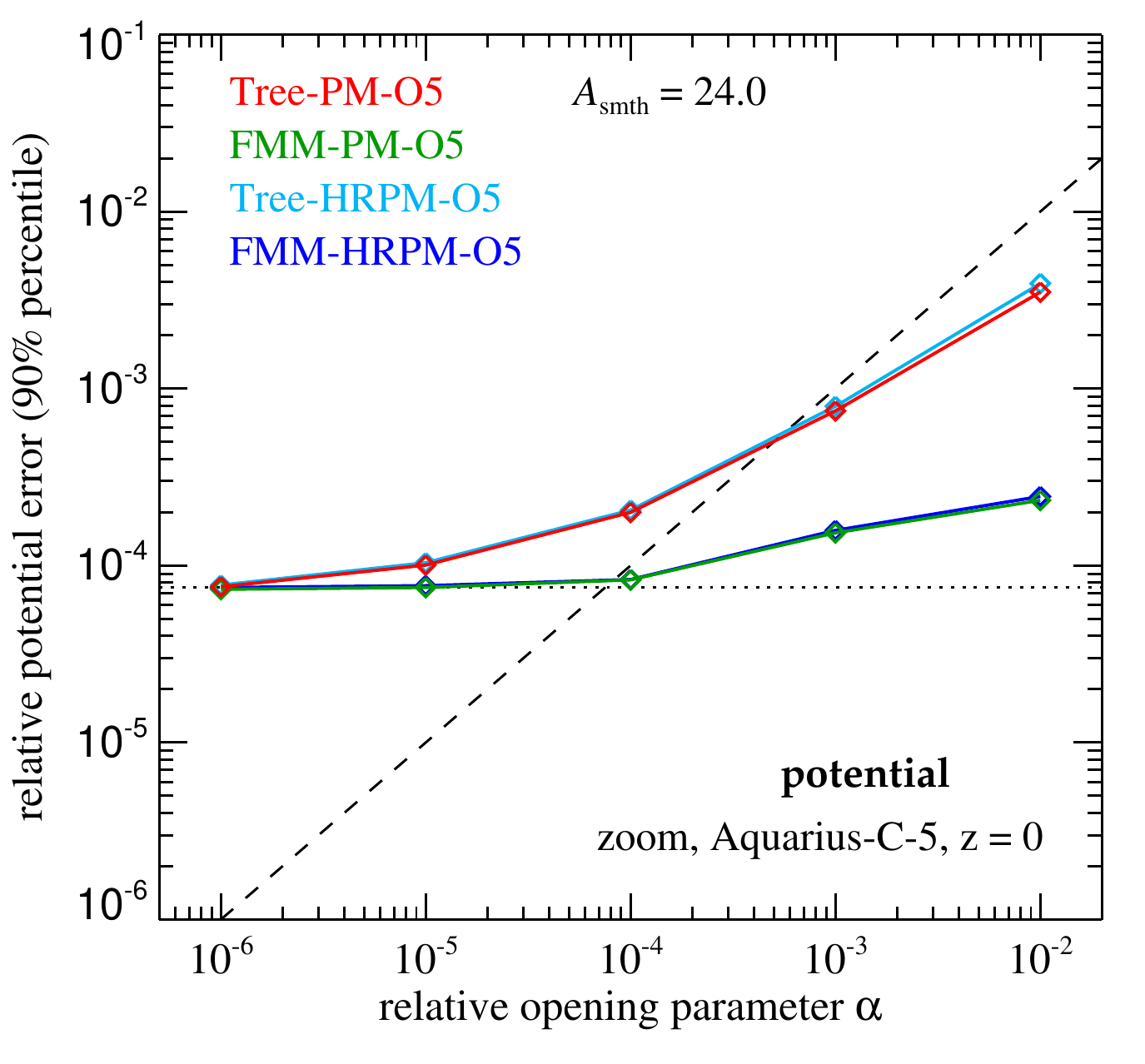}}\\%
\resizebox{8cm}{!}{\includegraphics{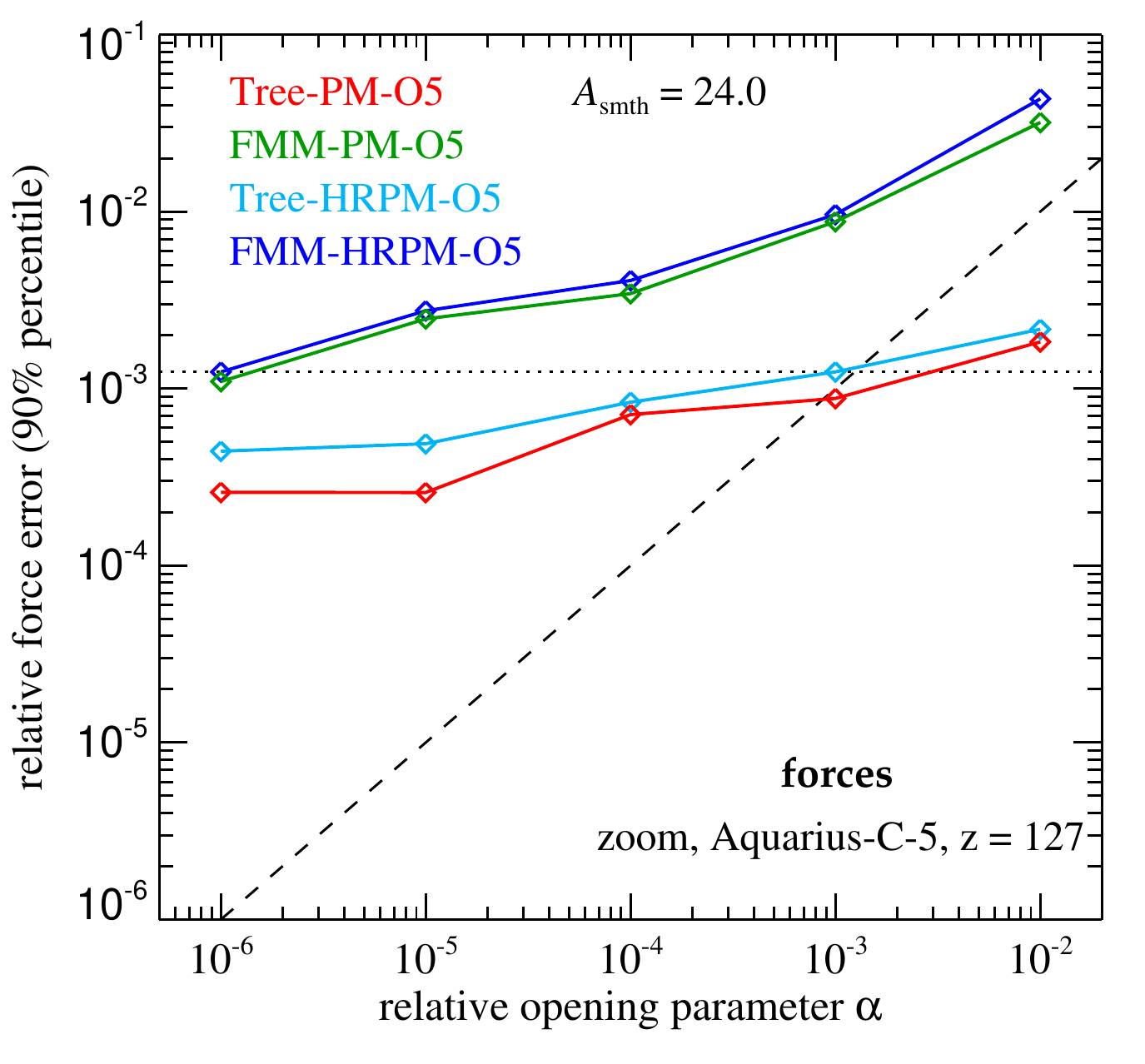}}%
\resizebox{8cm}{!}{\includegraphics{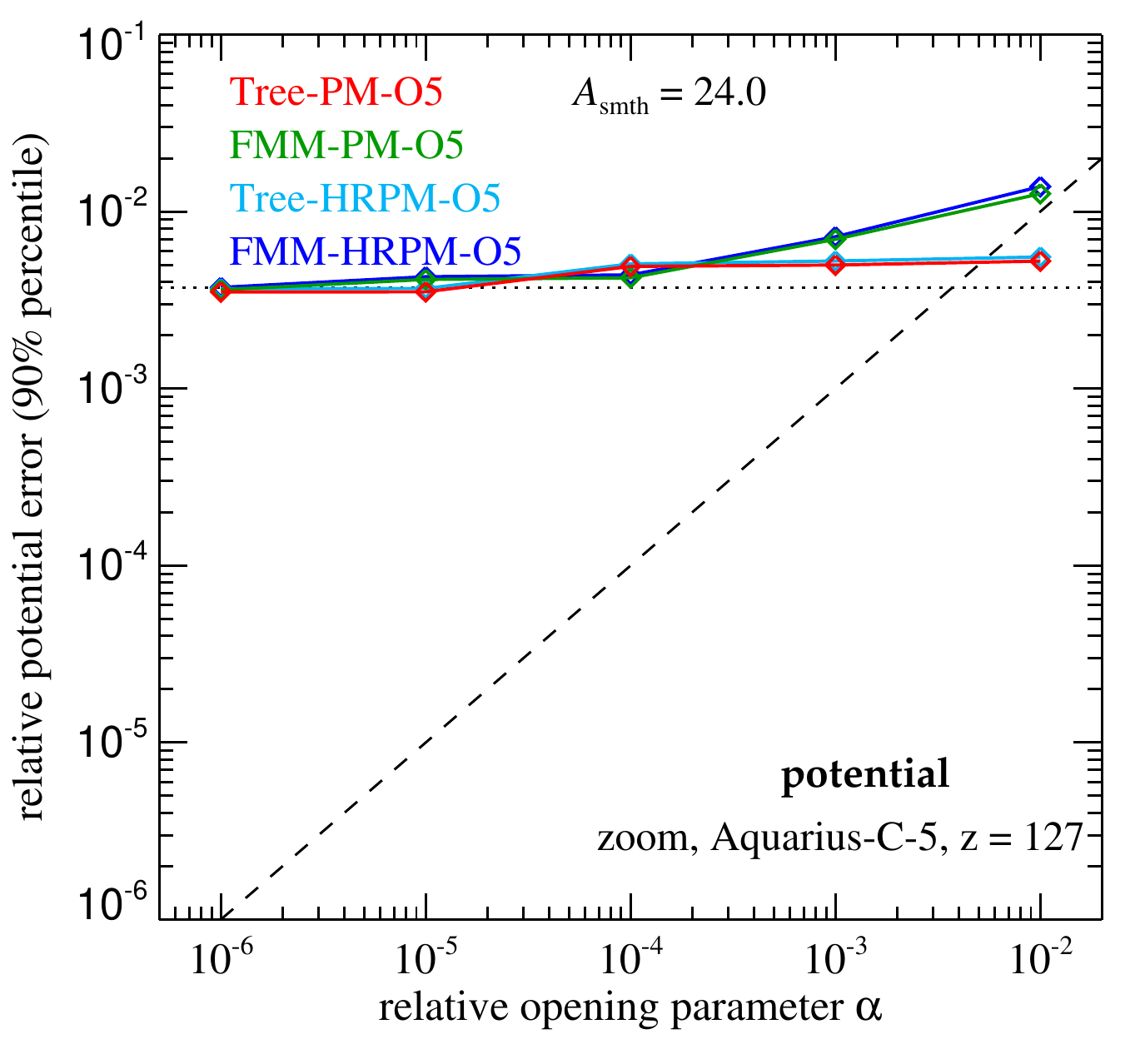}}\\
\caption{Test of the force and potential accuracy
  for a typical zoom simulation setup,
calculated with the Tree-PM and FMM-PM approaches, both with and
without the placement of a secondary high-resolution Fourier mesh.
The initial conditions and the evolved simulation correspond to the
run Aq-C-5 of the Aquarius project \citep{Springel:2008aa}.
We show results as a function of the specified relative opening
parameter $\alpha$, and restrict ourselves here to order $p=5$ for
simplicity. Other expansion orders look qualitatively very similar,
but may differ more significantly in run time. The top panels give
results for $z=0$, the bottom ones for the initial conditions at
$z=127$, with force accuracy  shown on the left and potential accuracy
on the right. A
setting of $A_{\rm smth} = 24.0$
for high accuracy of the
PM-mesh calculations is used, allowing the code to deliver small
relative errors despite the complicated force and potential matching
entailed in Tree-PM and FMM-PM, respectively. At high redshift, it is
comparatively difficult to get very low relative force and potential
errors, not only because of the smallness of the force and potential
values, but also 
due to the rapid coarsening of the sampling of the boundary region and
the associated drastic increase of the particle masses,
making this a more challenging set-up than for uniform sampling. In all cases, enabling
a secondary PM mesh hardly changes the accuracy of the final
results, as desired. Bot the Tree and FMM methods deliver very similar
accuracy, and this is effectively set by $\alpha$ in the low-accuracy
regime (approximately following the dashed line), while for low
$\alpha$ the accuracy is 
limited by the PM-approach (horizontal dotted lines). 
\label{FigFrcAccuracyHRPM}}
\end{figure*}

In Figure~\ref{FigFrcAccuracyHRPM} we show results comparing the
accuracy for different combinations of Tree-PM and Tree-HRPM, as well
as FMM-PM and FMM-HRPM. For conciseness, we here restrict ourselves to
order $p=5$, and go for relatively high accuracy facilitated by a
setting of $A_{\rm smth} = 10^{-5}$. Results for other orders or
lower values of $A_{\rm smth}$ look as expected based on our other
earlier tests.

\subsection{Force error distributions for mixed boundary conditions}

The variety of different force computation algorithms in {\small
  GADGET-4} that we have discussed thus far is further extended by
algorithms that employ stretched periodic boxes, or (stretched) boxes
in which only two dimensions are treated with periodic boundaries,
while the third has open boundaries.  It is of course important to
check the typical force accuracies delivered by the code in these
situations as well. We have done this in a dedicated suite of test
calculations, but refrain from going through all the combinations here
in the interest of brevity, because the results are qualitatively very
similar to what we have discussed thus far. We will however come back
to the problem of a stretched tall box in one of our example problems
later on.

\subsection{Correlated force errors}

\begin{figure*}
\begin{center}
\resizebox{8cm}{!}{\includegraphics{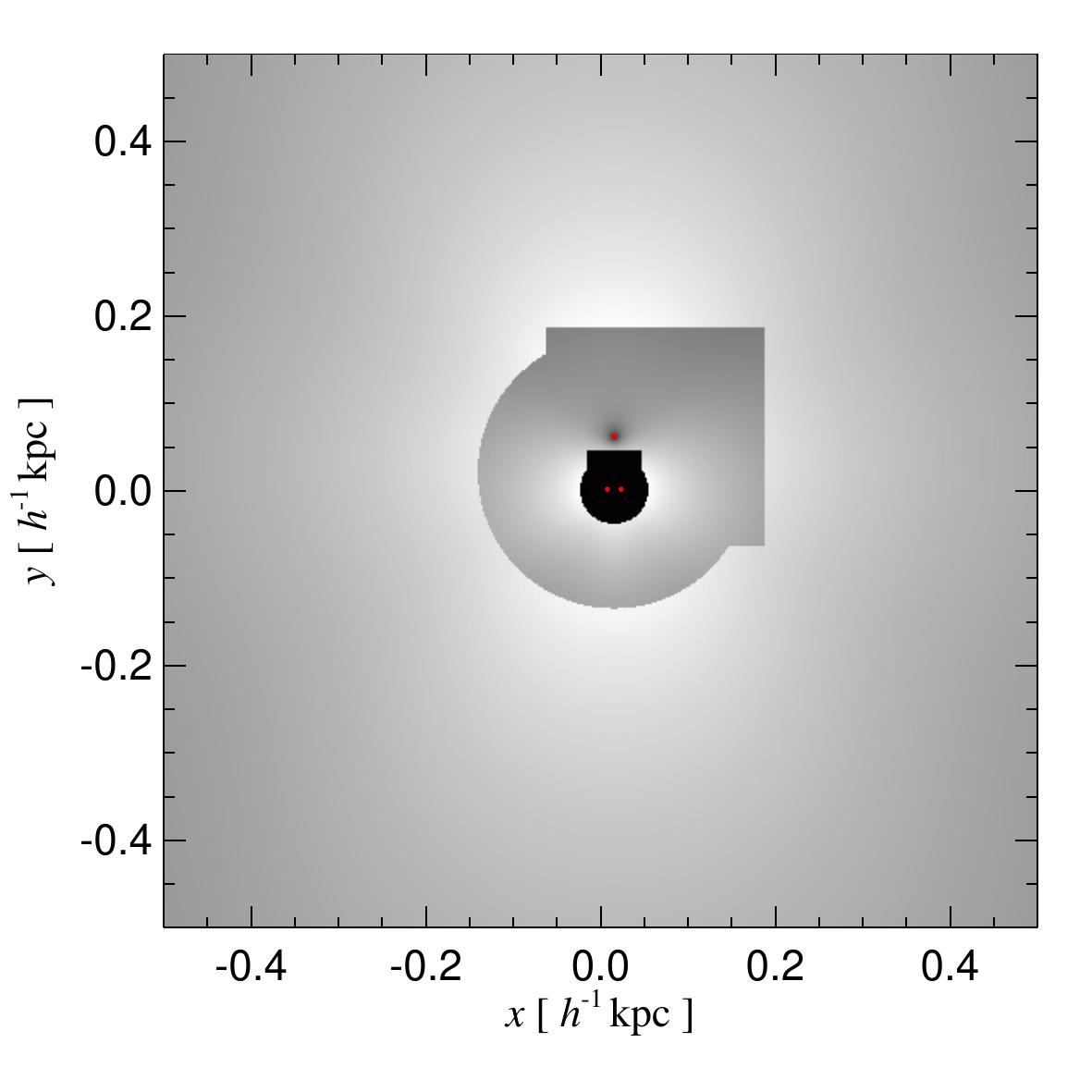}}%
\resizebox{8cm}{!}{\includegraphics{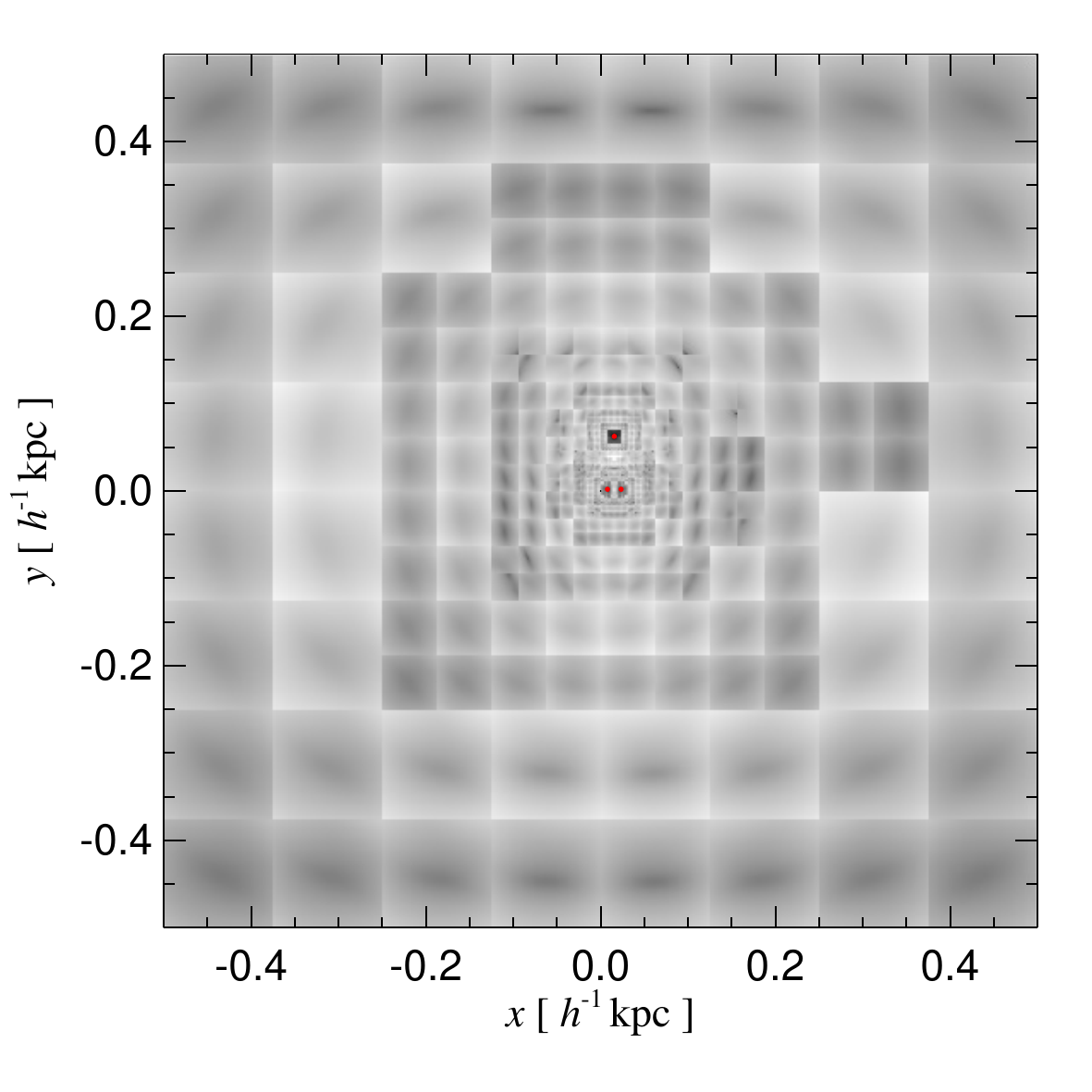}}%
\raisebox{0.83cm}{\resizebox{1.34cm}{!}{\includegraphics{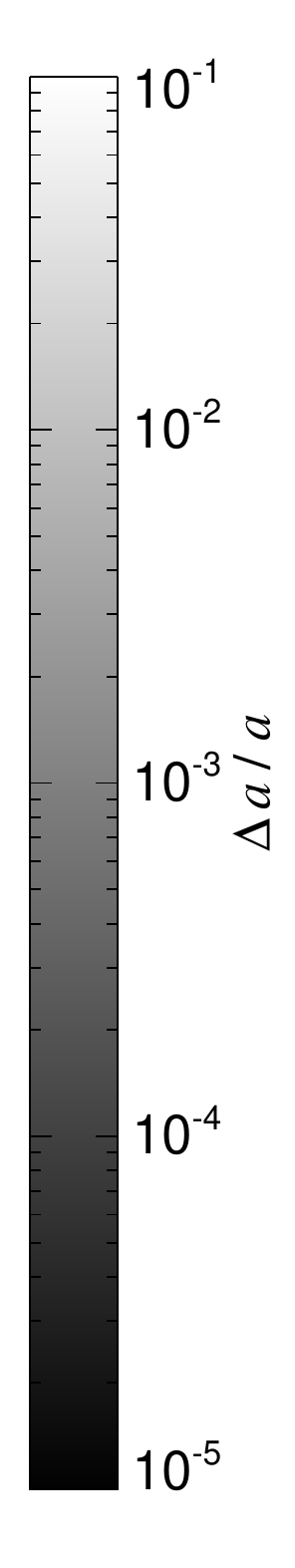}}}%
\end{center}
\caption{Maps of the relative force error in the $xy$-plane (encoded
  in grey-scale) for a distribution of three equal point masses.  The points are
  located at the coordinates marked with red dots in the maps.
The panel on the left gives the error map for a one-sided tree
algorithm, whereas the map on the right refers to the FMM approach.
  Non-periodic boundary conditions are applied; the root node of the
  underlying oct-tree is aligned with the coordinate axes
  and has a side-length of 3 length units. In the Tree-based result on
  the left, discrete jumps in the force accuracy result when an adjacent
  location experiences a flip of one of the cell-opening decisions (which are based on
  a purely geometric criterion here). In the FMM-result,
  discontinuities in the force accuracy are much more widespread, as
  they are now additionally created at the perimeter
  of local, sink-side field expansions. 
 Note that the asymmetry of both maps comes about due to the
  placement of the three points, which is asymmetric relative to the
  geometry of the nested oct-tree. 
  \label{FigCorrelatedFrcErrors}}
\end{figure*}

\begin{figure*}
\begin{center}
\resizebox{8cm}{!}{\includegraphics{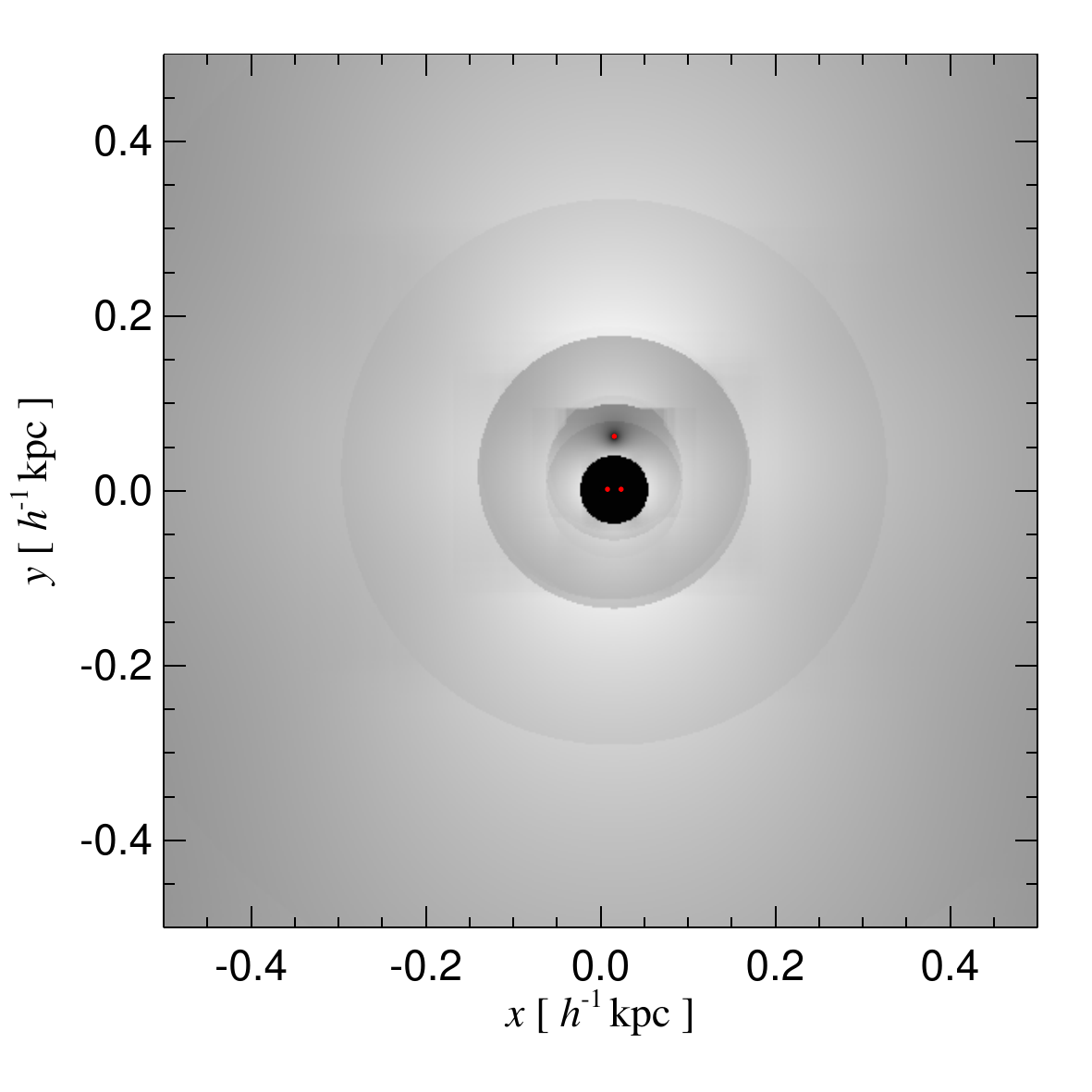}}%
\resizebox{8cm}{!}{\includegraphics{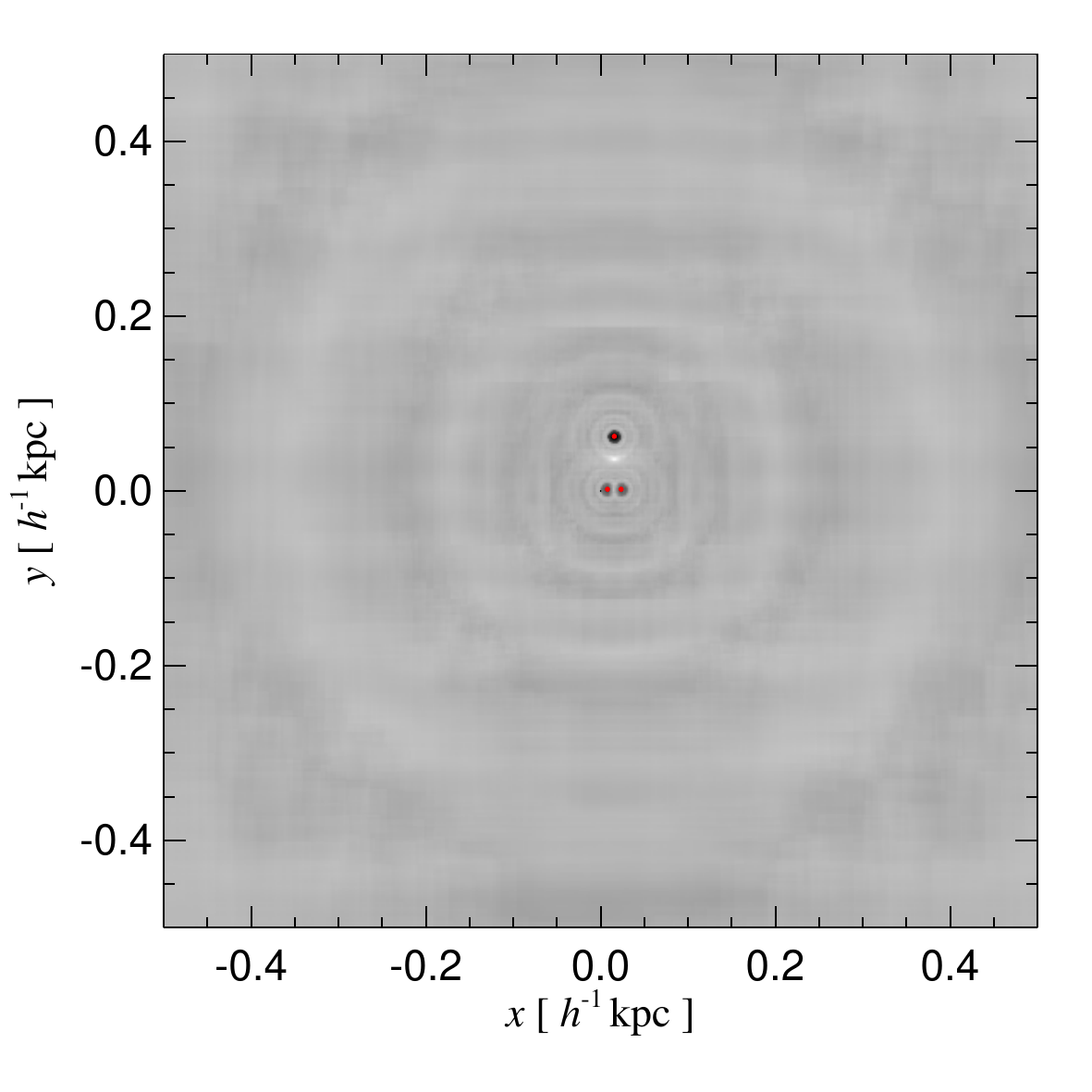}}%
\raisebox{0.83cm}{\resizebox{1.34cm}{!}{\includegraphics{plots/error_legend.pdf}}}%
\end{center}
\caption{Maps of the relative force error in the $xy$-plane for a
  distribution of three points of equal mass, averaged over 100 force
  calculations where each time the placing of the particle
  configuration relative to the geometry of the underlying oct-tree is
  randomized. Except for this translational randomizing and averaging,
  the test is equivalent to that of Fig.~\ref{FigCorrelatedFrcErrors}.
  In particular, the left panel shows the result for the Tree
  algorithm, the right panel for the FMM method.  We see that the
  effective force law after averaging becomes symmetric and largely
  eliminates discontinuities in the force accuracy (more so for FMM
  than for the Tree). This is achieved by reducing correlations of
  force errors in subsequent evaluations of the force at essentially
  no overhead.  \label{FigCorrelatedFrcErrorsWithRandomDisplacements}}
\end{figure*}

\begin{figure*}
\resizebox{8cm}{!}{\includegraphics{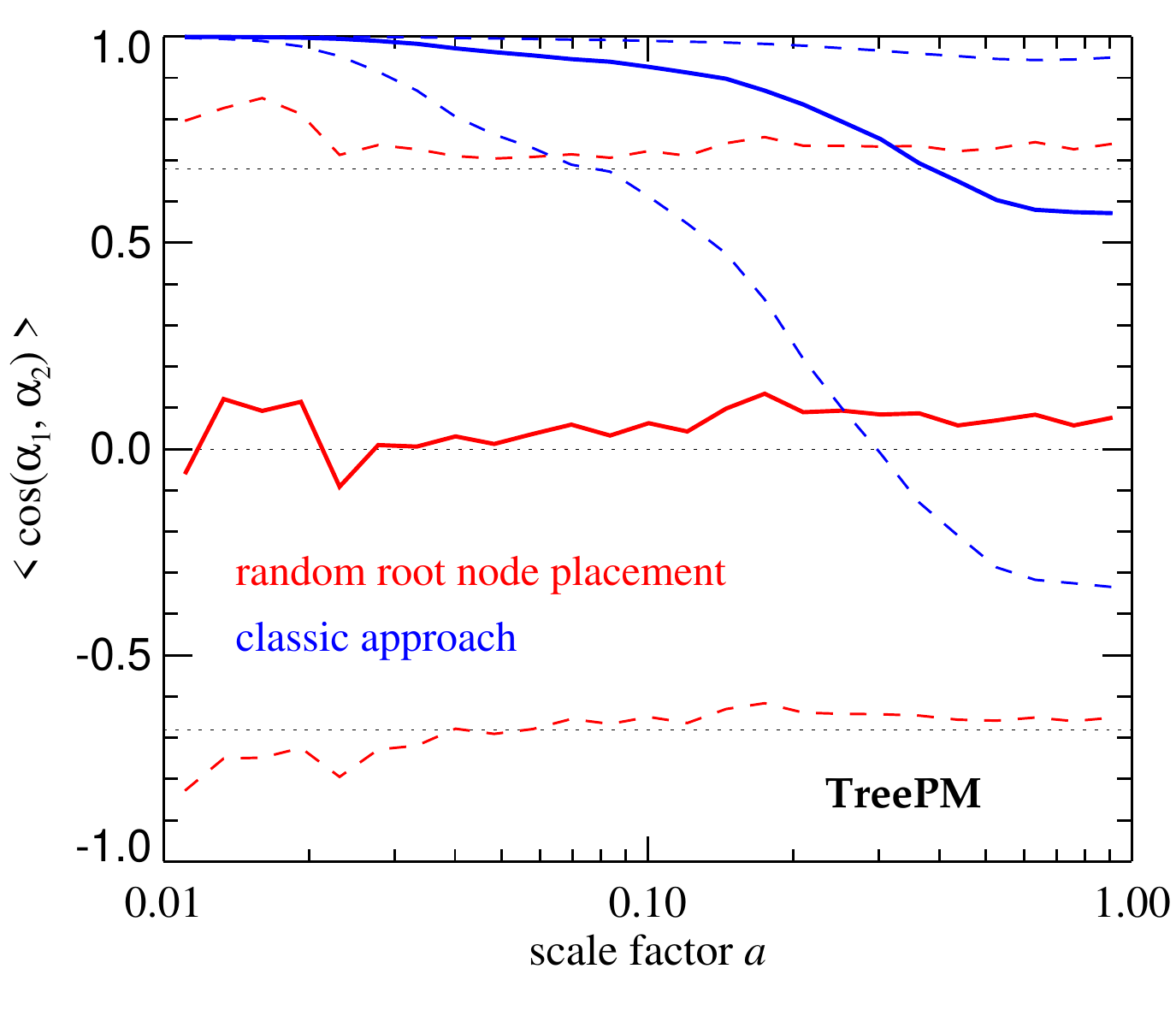}}%
\resizebox{8cm}{!}{\includegraphics{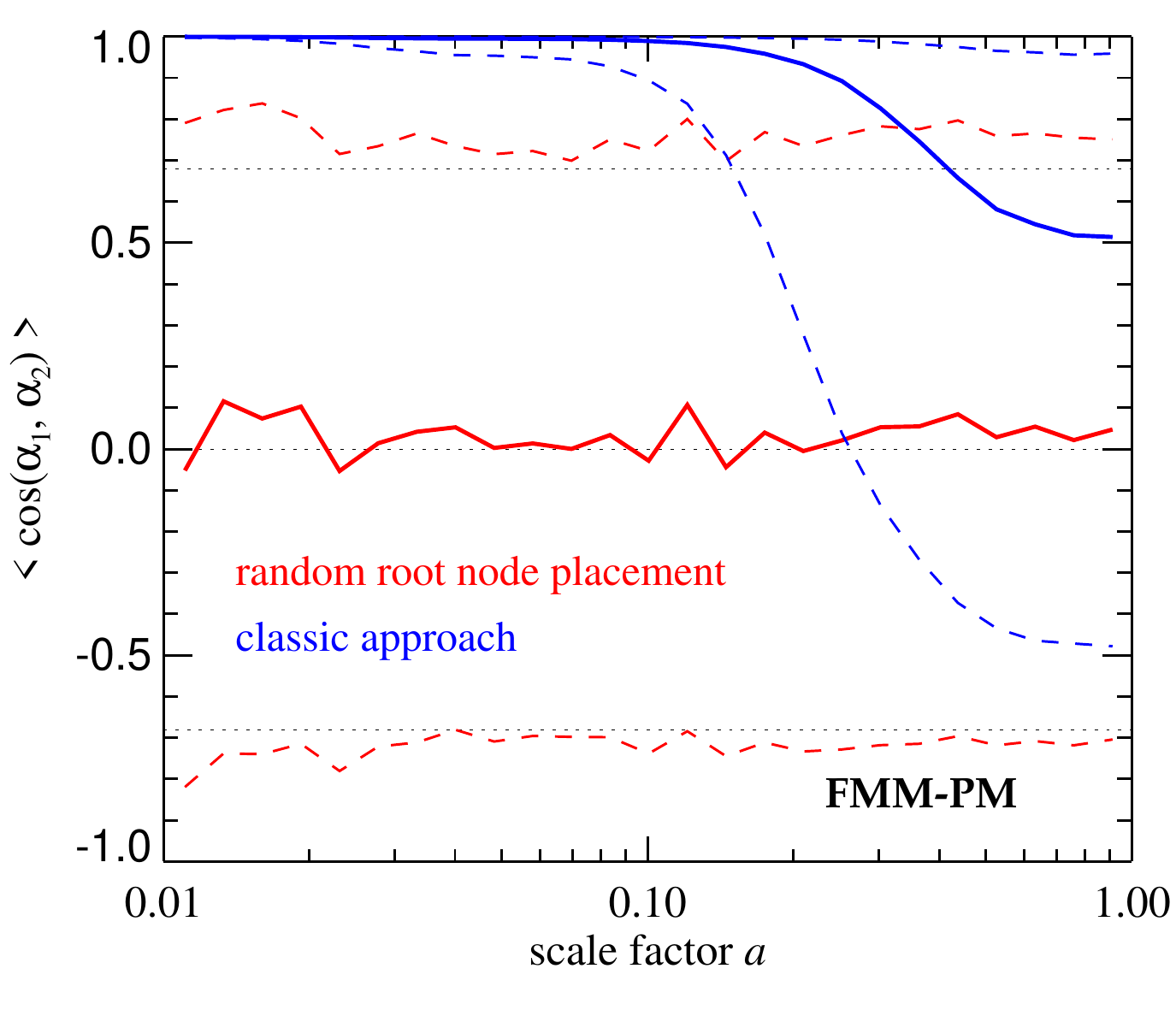}}\\
\caption{Temporal correlations in the {\em direction of force errors} in
  simulations computed with the TreePM (left panel) or FMM-PM
  algorithms (right panel), from
  high redshift to the present. The particular model examined
  here follows $256^3$ particles in a $20\,h^{-1}{\rm Mpc}$ box and is
  started at $z=99$. Force errors for a random set of particles are
  determined on the fly at times spaced $\Delta \ln a = 0.009$ apart
  (i.e.~at 512 times over the course of the simulation). We then
  determine the distribution of the cosine between the directions of
  the force errors at subsequent evaluation times, and plot the median
  (solid lines) and the 16\% and 84\% percentiles (dashed lines) as a
  function of scale factor. Results for the ordinary traditional simulation
  approach are shown in blue, while our randomization approach, where
  the whole particle set is randomly translated relative to the root
  node after each full timestep, is shown in red. Clearly, the force errors in the classical method are
  strongly correlated in time, whereas the randomization eliminates
  this successfully. Here an expansion order
  $p=3$ and a mesh size of $N_{\rm grid} = 512$ have been used; other variants of our
  gravitational solvers give qualitatively very similar
  results.
    \label{FigCorrelatedFrcErrorsinTime}}
\end{figure*}

For all algorithms discussed thus far, the residual force errors of
particles are correlated in space. In the case of the Tree algorithm,
this arises due to the geometric oct-tree pattern of the tree combined
with the fact that neighbouring particles will have similar
interaction lists, and thus experience similar errors in the specific
multipole expansions they see. For the FMM algorithm, these effects
can be expected to be substantially more severe, because the FFM
potential expansions will generally discontinuously `jump' at node
boundaries, and as we have seen earlier, the errors tend to be
dominated by the sink-side expansion.  Similarly, we expect the grid
pattern of the PM algorithm to be directly imprinted on maps of the
force error distribution, and hence to introduce spatial force error
correlations.

We show examples of this in Figure~\ref{FigCorrelatedFrcErrors}, both
for the tree and the FMM algorithms. Of course, as the force errors
are overall small, one may deem them to be inconsequential and
disregard the fact that they are spatially correlated. A random force
error without spatial and temporal correlations should indeed be
largely inconsequential in collisionless dynamics, as for example
\citet{Hernquist:1993aa} argue. But this ideal situation is not what
we encounter in these standard force calculation algorithms. In
particular, if a particle distribution is nearly stationary and
extremely cold, as it happens in a comoving cosmological simulation at
high redshift, discontinuous force errors along boundaries of large
nodes, such as they occur in FMM, can easily imprint step-like
artefacts in the phase-space distribution of particles that may become
visible at sufficiently high mass resolution, something that we in
fact did encounter when targeting simulations with extreme dynamic
range in the dark matter \citep{Wang:2019aa}, prompting us to adopt
the randomization scheme described below.

Of course, one approach to mitigate such effects is to try to reduce
the absolute size of the force errors as much as possible. This can be
done through the use of very small opening angles, at the price of a
high computational cost. Another approach is to at least eliminate the
temporal correlation of the force errors. We achieve this by
randomizing the relative location of the particle set with respect to
the computational box by adding one random displacement vector to all
particle positions each time a new domain decomposition is carried
out. Note that due to our use of integer coordinates, translating the
particles is a loss-less operation that is not affected by floating
point round-off errors, and in particular, it is fully reversible.
This randomization vector can be as large as the boxsize in each of
its dimensions. This means that typically all particles move to
another MPI rank in a domain decomposition, or in other words, there
is a complete reshuffling of the particle data. However, already for a
moderate number of MPI ranks, our domain decomposition algorithm tends
to produce new solutions every timestep that have very little
correlation with the previous domain decomposition anyway, so that
even without randomization an effectively all-to-all exchange of the
particle data in a domain decomposition could not be avoided.

The randomization may seem like a crude fix but proves actually rather
effective in practice. (Recall that randomizations are also known to
do wonders in other contexts, famously for example in
Glimm's~\citeyear{Glimm:1965aa} method for hydrodynamics.)  We
demonstrate this in
Figure~\ref{FigCorrelatedFrcErrorsWithRandomDisplacements}, where we
now show the average force error distribution after repeating it many
times. We see that we now recover a purely distance-based,
translationally invariant force law as desired.

The same phenomenon is also present in typical N-body simulations. In
Figure~\ref{FigCorrelatedFrcErrorsinTime} we show the correlations
between subsequent force evolutions that are $\Delta \ln a = 0.009$
apart, as a function of cosmological scale factor in a $256^3$ dark
matter only simulation evolved from $z=99$ to $z=0$. We measure the
expectation value of the cosine of the angle between the force error
vector occurring for a particle at time $a$ and $a + \Delta a$. We try
a TreePM and FMM-PM setup simulated with $\alpha=0.0025$, expansion
order $p=3$, and a grid size $N_{\rm grid}^3 =512^3$, and for each of these two
setups, we run a variant in which the particle set is randomly
translated with respect to the box for each timestep, and compare it
to the classic treatment where no such randomization is done.

Interestingly, the TreePM and FMM-PM schemes show strong correlations
of the force errors in time, as expected based on the above
discussion. The correlations become weaker at low redshifts once more
particles live on shorter dynamical times in halos, but even at $z=0$
remain very significant. This invokes the danger that force errors
build up over time for certain particles, reducing the accuracy of
their orbit integration. However, the randomization approach basically
eliminates the correlation of the force errors between subsequent
timesteps. In this case, the distribution function of the cosine
between subsequent force error vectors is very close to uniform at any
given time, meaning that while individual particles still experience
force errors of the same magnitude as before, the errors tend to point
into different directions for every timestep, consistent with basic
requirements of collisionless dynamics \citep{Hernquist:1993aa}.

\section{Time integration of collisionless particles}  \label{sectime}

Similar to {\small GADGET-2}, we use a kick-drift-kick formulation of
a simple leapfrog integrator to advance particle orbits in time.  This
arises from the Hamiltonian of the particle system by considering its
kinetic and potential parts alternatingly to obtain the time evolution
through operator splitting. Because the two separate evolutions
corresponding to the kick and drift operations are exact solutions of
partial Hamiltonians, some beneficial properties of symplectic
integration result, such as preservation of phase-space density, and
the prevention of the build-up of long-term secular integration errors
in the energy, at least as long as the timestep size stays fixed
\citep{Saha:1992aa, Hairer:2003aa, Hernandez:2018aa}.

The symplectic second-order accurate leapfrog integration can be
generalized to the cosmological case by introducing a conjugate
momentum equal to $\vec{p} = a^2 \vec{\dot {x}}$ for every particle as
its velocity variable, where
$\vec{\dot {x}} = {\rm d} \vec{x} / {\rm d}t$ is the comoving velocity
\citep{Quinn:1997aa}. In the cosmological case it is advantageous to
discretize time in terms of the time variable $\tau = \ln a$, because
a constant $\Delta \tau$ in this case corresponds to steps that are a
fixed fraction of the current Hubble time, which in turn closely
tracks the dynamical time at mean density. The kick operation for a
timestep $\Delta \tau$ is then given by
\begin{equation}
\vec{p}(\tau_n + \Delta \tau)  = \vec{p} (\tau_n) + \vec{a}_n(\tau_n)
\, \int_{\tau_n}^{\tau_n + \Delta \tau}  \frac{ {\rm d} \tau}{a\,H(a)} ,
\end{equation}
where $\vec{a}_n(\tau_n) $ is the comoving acceleration experienced by
the particle at time $\tau_n$ (and this acceleration does not change
during the time interval because the comoving positions are held
fixed).  Similarly, the drift operation becomes a displacement in
comoving space with constant conjugate momentum, given by
\begin{equation}
\vec{x}(\tau_n + \Delta \tau) = \vec{x}(\tau_n) + \vec{p}_n(\tau_n)
\, \int_{\tau_n}^{\tau_n + \Delta \tau} \frac{ {\rm d} \tau}{a^2H(a)} .
\end{equation}
The integrations over the Hubble rate $H(a)$ can be carried out to machine
precision with a high-accuracy numerical integrator for an arbitrary
Friedmann cosmology. Because the resulting timestep factors can be reused for many
particles when a synchronized timestep hierarchy is used, the
cost associated with these integrations is negligible.

A single timestep is then carried out as a sequence of a half-step kick,
followed by a full-step drift, and a further half-step kick, i.e.~the
time evolution operator $E$ of one step is
\begin{equation}
  E(\Delta t) = K\left(\frac{\Delta t}{2}\right) \circ D(\Delta t)
  \circ  K\left(\frac{\Delta t}{2}\right),
\end{equation}
yielding a second-order accurate time integration scheme. Note that
only one force calculation per timestep is required, as the force
calculation at the end of a timestep can be reused for the first kick
operation of the subsequent step. The kick operations of two
subsequent time-steps can in principle be combined into one kick, an
approach followed in {\small GADGET-2}.  However, in {\small GADGET-4}
we refrain from doing this merge, thereby allowing outputs to be
created in between two half-step kicks, thus retaining the formal
second-order accuracy in the output velocities. Also, this allows a
cleaner nesting of the hydrodynamical evaluation inside the gravity
steps based on operator-split principles, and it simplifies the
treatment of timestep changes in the local time integration schemes we
discuss next.

\subsection{Traditional nested time integration}

The high dynamic range in density and spatial scales that occurs in
astrophysical systems is complemented by equally large variations in
timescales. For example, the gravitational dynamical timescales in the
centres of dark matter halos are about a factor $\sim 1000$ smaller
than at the edge of halos, which are in turn nearly a factor 10
shorter than at mean density. Integrating the whole system on the
shortest dynamical timescale is highly inefficient in this case. In
fact, for hydrodynamical cosmological simulations, the corresponding
spread in timescales can become much larger still, making it imperative that
adaptive time integration schemes are used that take the local
needs of each particle into account.

A popular approach to implement local time integration is to subdivide
the simulation timespan into a power-of-two hierarchy, and then to
require that permissable timesteps are nested within this hierarchy. This means
that all timesteps are power of two multiples of the smallest
occurring timestep. Particles with the same timestep size occupy a
certain timestep bin in this scheme, and may change to shorter or
longer timesteps when they have completed their current timestep. To
maximize the amount of synchronization between different particles,
one allows an increase of the timestep size only when the new target
timebin is synchronized with the current time\footnote{Sometimes one
  may also impose the condition that the timestep may increase at most
  by a factor 2 in one go, although we do not find this essential in
  our code and thus do not impose this restriction.}, whereas reductions
of the timestep size can happen at the end of every timestep.

\begin{figure*}
  \begin{center}
    \resizebox{18cm}{!}{\includegraphics{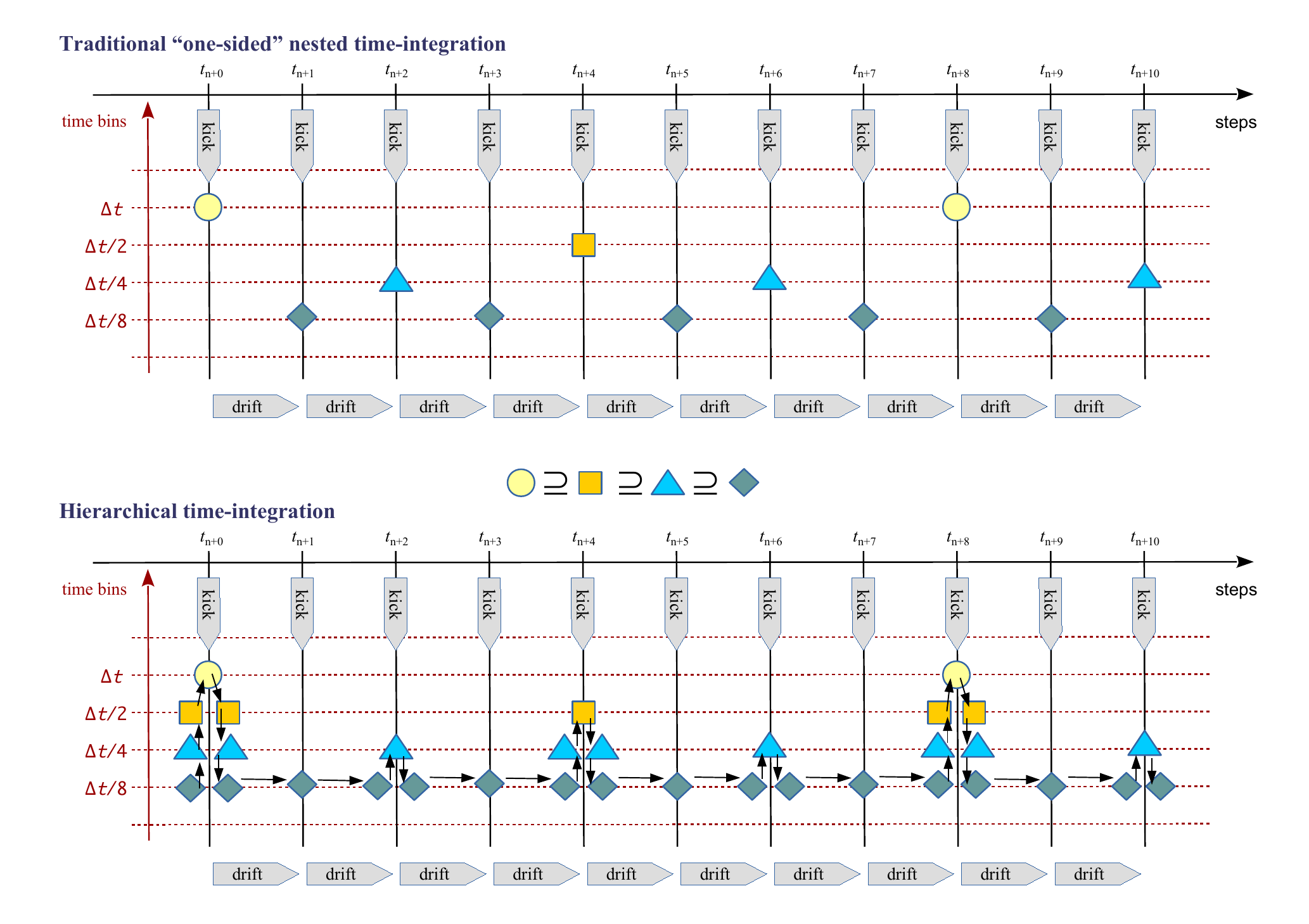}}%
    \end{center}
\caption{Sketch of the nested local time integration scheme with four occupied
  timebins. The top panel illustrates the
``one-sided'' approach where all active particles receive a force
calculation from \emph{all other particles}. At different
synchronization times $t_n$ (which are spaced apart by the smallest
timestep occurring in the system, here $\Delta t/8$), different sets of
particles are active and require a force calculation, namely exactly
those particles whose timesteps begin or end at
the current synchronization time. For example, the orange squares symbolize
particles with timestep $\Delta t /2$ or lower; this group includes
the particles with timestep  $\Delta t /4$ or lower, which in turn are
represented with a blue triangle in the sketch. In between subsequent
(partial) force calculations, the particle coordinates are pushed with
constant momenta by
cheap drift operations. The lower panel sketches
the
hierarchical time integration algorithm instead. At each synchronization
point, it may here be necessary to compute several partial forces to carry
out the correct kick operation. Each partial force only involves force
calculations where source and sink sets are equal, i.e.~here the
occurrence of a symbol in the sketch means that a force calculation of
the corresponding subset of particles with itself as source (as
opposed to 
the whole particle distribution) is required. All momentum changes
in this scheme are based on pair-wise forces that manifestly preserve
total
momentum.\label{figintschemes}}
\end{figure*}

A sketch of the resulting integration scheme for an example with four
active timebins is shown in the top panel of
Figure~\ref{figintschemes}.  We call this scheme of local time
integration ``one-sided'', because every active particle receives a
force from the full particle distribution (which therefore also needs
to be drifted to the current time). Momentum conservation is here not
manifest between all pairs of interacting particles, even in the
absence of any force errors. One may nevertheless hope that no
significant momentum error builds up in practice due to this, because
time averaging should largely eliminate the build-up of sizeable
momentum errors from one-sided interactions in the limit of small
timesteps.

\subsection{Hierarchical time integration}  \label{sechierarchicaltime}

Two issues remain unsatisfactory with the above. First, active
particles couple asymmetrically with the whole system, spoiling
manifest momentum conservation. Secondly, the computational cost does
not go down linearly with the fraction of active particles, because to
construct the tree, one still needs to synchronize (i.e.~drift)
\emph{all} particles and make them part of the tree in the first
place.  This behaviour can be improved by a systematic,
hierarchical decomposition of the Hamiltonian dynamics, which we implement
similarly as described in \citet{Pelupessy:2012aa}, and used already
successfully within the {\small AREPO} code in production for the IllustrisTNG
simulations \citep{Marinacci:2018aa, Nelson:2018aa, Naiman:2018aa,
  Pillepich:2018aa, Springel:2018aa}, as well as for the `Void-in-Voids-in-Voids'
simulations of \citet{Wang:2019aa}. Recently, another implementation of this approach
has been described by \cite{Zhu:2020aa}, who also combined this with
FMM for the first time.

First, it is perhaps useful to recall second-order operator split
methods to integrate a Hamiltonian system of the form \be H = H_1 +
H_2, \ee which can for example be done through \be E(H, \Delta t)
\simeq E\left(H_1, \frac{\Delta t}{2}\right) \circ E(H_2, {\Delta t})
\circ E\left(H_1, \frac{\Delta t}{2}\right), \ee where $E(H, \tau)$ is
the time evolution operator of the system under a Hamiltonian $H$ over
time $\tau$. This is what we already used to construct our basic
leapfrog scheme, which for fixed timesteps is symplectic. The schemes
constructed below are also symplectic because they still arise as
exact solutions of partial Hamiltonians that are alternatingly applied
to advance the system. Their symplectic nature is however lost once
particles change their timesteps, just like it happens in the ordinary
leapfrog.

Consider now a particle system P described by the Hamiltonian: \be H =
H_{\rm kin} + H_{\rm pot}. \ee Suppose we are given a timestep
$\Delta t$. Then we can subdivide the particle system into a ``slow''
and a ``fast'' part, ${\rm P = S + F}$, based on the notion that the
slow system S contains the particles for which an integration with the
given timestep size would be sufficiently accurate, while the fast
system F gets the particles where the integration timestep should be
chosen smaller. After the subdivision, we can write the Hamiltonian as
\be H = H_{\rm kin}^{\rm S} + H_{\rm pot}^{\rm S} + H_{\rm kin}^{\rm
  F} + H_{\rm pot}^{\rm F} + H_{\rm pot}^{\rm FS}, \ee or equivalently
\be H = H^{\rm S} + H^{\rm F}+ H_{\rm pot}^{\rm FS}, \ee where now
$H^{\rm S} = H_{\rm kin}^{\rm S} + H_{\rm pot}^{\rm S}$ and
$H^{\rm F} = H_{\rm kin}^{\rm F} + H_{\rm pot}^{\rm F}$ are
Hamiltonians that only involve the slow and fast particles,
respectively. The interaction term $H_{\rm pot}^{\rm FS}$ accounts for
all potential terms that involve mixed particle pairs.

The time evolution of the system could now for example be approximated
to second-order in $\Delta t$ as: 
\begin{eqnarray} E(H, \Delta t) & \simeq &
E\left(H_{\rm pot}^{\rm FS}, \frac{\Delta t}{2}\right) \circ
E\left(H^{\rm F}, \frac{\Delta t}{2}\right) \circ E(H^{\rm S}, {\Delta
  t}) \circ \nonumber \\
& & E\left(H^{\rm F}, \frac{\Delta t}{2}\right) \circ
E\left(H_{\rm pot}^{\rm FS}, \frac{\Delta t}{2}\right).
\label{eqnt1} 
\end{eqnarray}
Here the slow system is evolved on a timestep $\Delta t$, which by
construction was deemed sufficient to integrate the corresponding
particles. This step can hence be replaced with, e.g., a single
leap-frog step of size $\Delta t$. In contrast, the fast part is
evolved on half the timestep by the corresponding operators, and a
single timestep may not necessarily be sufficient for doing this
accurately. But in this case, we can reduce the timestep for this part
by subcycling (see below). For the interaction component, we shall
assume that the timestep of $\Delta t/2$ is sufficient to do it in a
single step with sufficient accuracy, on the grounds that all terms
involve one of the slow particles.

We now specialize to an N-body system and employ the kick- and drift
operators for evolution under the kinetic and potential terms,
respectively. Then the time evolution described by equation
(\ref{eqnt1}) can be written as:
\begin{eqnarray}
E(H, \Delta t) & \simeq &  
K_{\rm S}^{\rm F}\left( \frac{\Delta t}{2}\right)
K_{\rm F}^{\rm S}\left( \frac{\Delta t}{2}\right) \nonumber \\
& & E \left( H_{\rm F}, \frac{\Delta t}{2} \right) \nonumber \\
& & K_{\rm S}^{\rm S}\left( \frac{\Delta t}{2}\right)
D_{\rm S}\left( {\Delta t}\right)  
K_{\rm S}^{\rm S}\left( \frac{\Delta t}{2}\right) \nonumber \\
& & E \left( H_{\rm F}, \frac{\Delta t}{2} \right) \nonumber \\
& & K_{\rm S}^{\rm F}\left( \frac{\Delta t}{2}\right)
K_{\rm F}^{\rm S}\left( \frac{\Delta t}{2}\right) .
\label{eqnt2}
\end{eqnarray}
In this notation, the kick operator $K_{\rm A}^{\rm B}$ means that
forces due to particles in set B are applied to the particles in A.
Note that $K_{\rm S}^{\rm S}$ commutes with the time evolution
operator for the fast part of the Hamiltonian, $H^{\rm F}$, simply because
they operate on disjoint particle sets. We are hence free to move the
first occurrence of $K_{\rm S}^{\rm S}$ in front of the first $E(H_F)$ operator, and
the second  $K_{\rm S}^{\rm S}$ behind the second  $E(H_F)$. The
individual kick operators also commute, and $D_{\rm S}$ could as well
be interchanged with operators that only involve the fast
particles. We are thus free to alternatively express
equation~(\ref{eqnt2}) as
\begin{eqnarray}
E(H, \Delta t) & \simeq &  
K_{\rm P}^{\rm P}\left( \frac{\Delta t}{2}\right)
K_{\rm F}^{\rm F}\left( -\frac{\Delta t}{2}\right) \nonumber \\
& & E \left( H_{\rm F}, \frac{\Delta t}{2} \right) 
D_{\rm S}\left( {\Delta t}\right)  
E \left( H_{\rm F}, \frac{\Delta t}{2} \right) \nonumber \\
& & K_{\rm F}^{\rm F}\left( -\frac{\Delta t}{2}\right)
K_{\rm P}^{\rm P}\left( \frac{\Delta t}{2}\right),
\end{eqnarray}
where we used
$K_{\rm S}^{\rm F}( {\Delta t})
K_{\rm F}^{\rm S}( {\Delta t})
K_{\rm S}^{\rm S}( {\Delta t}) = 
K_{\rm P}^{\rm P}( {\Delta t}) K_{\rm F}^{\rm F}(- {\Delta t})$.

This is now essentially in the form we recursively apply it in
practice. We first start out by calculating the forces for the whole
system P. This in particular allows us to apply the
$K_{\rm P}^{\rm P}$ operator. Also, we can then use the size of the
forces calculated for P to distinguish between the F and S particles
by means of a timestep criterion. For the particles ending up in F,
new forces are then calculated, sourced just by those in the set F,
allowing us to partially undo the previous kick, through the
$K_{\rm F}^{\rm F}$ operator. These F particles are then treated as a
new system P', for which we now need to carry out the time evolution
over a timestep $\Delta t /2$. This can be dealt with recursively by
applying the same procedure yet again, with P' becoming the new P.  In
doing this, the first task, namely the calculation of all forces for
$P'$ has already just been done previously (as force calculation for
the set F), and thus the corresponding force calculations can be
recycled.  Also note that the forces entering the timestep criterion
for deciding between fast and slow particles in deeper levels of the
hierarchy become partial forces originating in ever smaller subsets P'
of the system.  The recursive subdivision into ever smaller particle
sets ends once the set F becomes empty. At this point, a drift for the
current set F is carried out, followed by a force calculation for this
set and a closing half-step kick. Then a drift of the slow system is
done (which may however also be delayed to a later time, just before
the next force calculation that involves any of the slow particles),
followed again by a kick-drift-kick sequence of the fast
part. Finally, one closes off the step by partially undoing the kick of
the fast system (with already known forces at this point), followed by
a force calculation for P at the end of the step, and a full kick of
the full system. The recursive application of the scheme means that
depending on where the closing step occurs one may have to back out of
multiple levels of subdivisions, since all the levels are nested into
each other in a power-of-two timestep hierarchy. The lower panel of
Figure~\ref{figintschemes} shows a sketch of this scheme for an
example involving four active timebins.

One important feature of this time integration method is that the fast
system that is split off is self-contained, i.e.~its evolution does
not rely on any residual coupling with the slow particles. In
particular, one does not have to construct the tree for all particles
when the forces on F are required; only the particles from F are
needed. Another aspect is that all the forces that are applied are for
operators of the form $K_{\rm A}^{\rm A}$ only, where source and sink
sets are equal. This implies manifest momentum conservation for the
local time integration scheme, something that is not the case in
the traditional method to implement local timesteps where
`one-sided' forces between particle pairs occur. Furthermore, when a
fast multipole moment method is applied to compute the gravitational
forces, the momentum conservation will be manifest to machine
precision despite the presence of finite force approximation errors.

However, this advantage comes at the price of \emph{additional} force
computations relative to the ordinary ``one-sided'' nested time
integration. To quantify this, let us index the highest occupied timebin
with $k=0$, and call $N_k$ the number of particles with timestep
$\Delta t = \Delta t_{\rm max} / 2^k$ or lower, where  $\Delta t_{\rm
  max}$ is the maximum occupied timestep. Note that $N_0$ is thus the
total particle number.

For the ordinary nested time integration, at any given synchronization
point of the timestep hierarchy only one force calculation is
necessary. To advance the whole system over $\Delta t_{\rm max}$,
a number of
\begin{equation}
K_{\rm onesided} = N_0 +\sum_{k=1}^{k_{\rm max}} 2^{k-1}\, N_k
  \end{equation}
  force computations is necessary. If the simulated timespan is
  $T_{\rm sim}$, the total number of force computations for the
  simulation is then
  $T_{\rm sim} / \Delta t_{\rm max} \times K_{{\rm onesided}}$. In
  contrast, for the hierarchical time integration scheme, the cost
 is characterized by
\begin{equation}
K_{\rm hierarchical} = N_0 +3 \times \sum_{k=1}^{k_{\rm max}} 2^{k-1}\, N_k
  \end{equation}
force computations. At face value this can be substantially larger
than the one-sided scheme, in the worst
case by something that approaches a factor of 3.

However, there are several reasons that mitigate this difference in
practice, and can make the hierarchical integration nevertheless
worthwhile. First of all, in many cases the hierarchical integration
can deliver improved accuracy at equal timestep size, meaning that
some of the extra cost can be absorbed into a somewhat larger
timestep. Depending on the timestep criterion and simulation set-up,
there can also be situations where the hierarchical timestep criterion
leads to a less deep timestep hierarchy (for example when large-scale
accelerations are caused by particles on long timesteps, which then do
not enter any more in relative motions on small-scales), making the
hierarchical integration more efficient by altering the occupancy
$N_k$ on the different timestep bins.

Leaving this aside, we also see that for the hierarchical integration
it can make sense to restrict the maximum allowed timestep to
something smaller than $\Delta t_{\rm max}$. This happens when we
would otherwise have $N_1 > N_0/3$, i.e.~whenever the first time bin
below the maximum timestep holds more than one third of the
particles. In this case, it is actually beneficial to rather use a
shorter maximum timestep $\Delta t_{\rm max}/2$ for all particles, as
this reduces the cost sum $K_{\rm hierarchical}$ while at the same
time it increases the time integration accuracy. The code
automatically detects this condition and modifies the timesteps of the
particles accordingly.

Importantly, whether or not the cost increase of hierarchical
integration is substantial depends sensitively on the depth of the
timestep hierarchy, and how the occupancy $N_k$ varies with the
timebin $k$. If we always have $N_{k+1} < N_k / 2$ (which is common in
cosmological simulations, in fact, quite typically the occupancy
declines more steeply than this), then the total cost is dominated by
timebins with the longest timesteps, despite the short timesteps being
executed (much) more frequently. If we additionally have
$N_1 \ll N_0/3$, then there is no significant cost difference between
hierarchical and ordinary time integration left when measured
in terms of the number of force calculations.\footnote{For
  simplicity we assume for this estimate that all force computations
  incur a similar computational expense, which is only a rough
  approximation for realistic set-ups.}

However, as noted earlier, there is an important further difference
between the two approaches. For the one-sided time integration, the
force computations require a tree construction for the full particle
set, as well as drifting these particles to the current
synchronization point. In contrast, for the hierarchical integration
scheme, the tree has to be constructed only for the particle set
involved in the current force computation. If the timebin hierarchy is
very deep, the one-sided scheme can then become dominated by the
(constant) tree construction costs, whereas this cost declines for the
hierarchical integration alongside the smaller occupancy at ever
deeper timebin levels. For very deep timestep hierarchies, it is
therefore the hierarchical integration that can outperform the
ordinary time integration, because it can avoid becoming dominated by
tree construction overhead and particle drift costs.  The hierarchical
approach has therefore better potential for simulations with an
extreme dynamic range in timescales.

\subsection{Timestepping for TreePM and FMM-PM}

A problem shared by the Tree-PM and FMM-PM approaches when local time
integration is used is that the PM part of the algorithm does not
require significantly reduced computational time if forces for only a
subset of the particles need to be calculated.  To avoid an
inefficiency of TreePM/FMM-PM for local time integration, {\small
  GADGET-2} introduced a scheme (which is optionally still supported
in {\small GADGET-4}) where the TreePM force-split also induces a
split in the associated time integration, based again on the concept of
operator splitting. When one PM mesh covers the whole simulation
domain, the gravitational part of the Hamiltonian can be split into a
long-range and a short-range part,
\begin{equation}
  H_{\rm pot} = H_{\rm long}^{\rm PM} + H_{\rm short}^{\rm {tree}}
  \end{equation}
with a single split scale,
allowing us to write the evolution of the system over one timestep
$\Delta t$ as
\begin{equation}
E(H, \Delta t)  = K_{\rm long}^{\rm PM}\left(\frac{\Delta t}{2}\right)   
\circ E(H_{\rm short}, \Delta
t) \circ K_{\rm long}^{\rm PM}\left(\frac{\Delta t}{2}\right),
\end{equation}
where $K_{\rm long}^{\rm PM}$ refers to kicks due to the long-range PM
force, and $H_{\rm short} = H_{\rm kin} + H_{\rm short}^{\rm {tree}}$
describes the short-range dynamics, which now can be treated either with the
hierarchical scheme as introduced above, or with the one-sided nested
time-stepping approach. The timestep $\Delta t_{\rm PM}$
chosen as appropriate for the long-range PM forces is then also equal to the maximum
timestep in the system. 

If a secondary high-resolution PM mesh is used following the approach
outlined earlier, the potential energy
can still be subdivided into two terms, but the
split-scale applied to each particle pair now
depends on their spatial coordinates. The associated time-scales on
 which the long range forces change can then also differ significantly
 between the high-resolution and the background regions, which may
 require considerably more conservative settings for  $\Delta t_{\rm
   PM}$.

 Another approach to avoid the PM overhead in thinly populated
 timebins is to decide on-the-fly for every force calculation whether
 it should be carried out through the Tree-PM/FMM-PM approach, or
 through a pure Tree/FMM with Ewald correction when periodic
 boundaries are used. This time-integration option is newly supported
 in {\small GADGET-4}, and we refer to it as the `time-unsplit
 Tree-PM/FMM-PM scheme', because here the PM algorithm is used to
 speed up force computations without influencing the time
 integration. Because of this property, we in principle prefer this
 approach on conceptual grounds. In practice, we define a run-time
 parameter that controls the hand-over point between force
 computations carried out with Tree-PM/FMM-PM, or with a pure Tree/FMM
 instead.  If the fraction of active particles is larger than a
 specified value, the PM approach is used to accelerate the force
 calculations, otherwise the pure multipole algorithms are applied. In
 this way, one can benefit from the higher speed of the PM-based
 approach for reasonably full timesteps, while thinly populated
 timebins can be dealt with without the overhead of doing large FFTs
 for only a small number of particles.

\subsection{Summary of time-integration and force computation schemes}

{\small GADGET-4} supports a perhaps bewildering variety of different
combinations of force computation and time integration schemes. For
definiteness, let us illustrate this with the common case of a
cosmological simulation with periodic boundary conditions integrated
in comoving coordinates. Such a simulation can in principle be carried
out either with a pure Tree with Ewald corrections, a Tree-PM scheme
with an induced split in the timesteps of long-range and short-range
forces, or with a Tree-PM scheme without such a split. Each of these
three possibilities can be either combined with a classic one-sided
local timestepping scheme, or with the new hierarchical time
integration method introduced in this paper. This yields 6 different
basic combinations. Furthermore, for each of these 6 combinations the
Tree part of the calculations can be replaced with the FMM method,
doubling this to already 12 different variations, as schematically
summarized in Table~\ref{TabIntSchemes}. Finally, one can run each of
these cases with different multipole oder $p$. If one considers each
of these choices as different force calculation algorithms, one
arrives at 60 possible ways to do the calculation. In
zoom-simulations, the possibility to use an additional secondary
Fourier mesh covering the high-resolution region adds even further
possibilities.

All of these approaches should give the same results when carried out
with conservative settings for time integration and force accuracy.
It is an important validation test to explicitly verify that this is
actually the case, something we have tried to ensure during the tests
of the code. However, it is of course expected that the different
methods may differ quite drastically in the required CPU-time for a
prescribed accuracy level.  Especially for cosmological simulations of
structure formation, the pure Tree and FMM algorithms at low
order $p$ should be slowest, both due to the high force accuracy
demands at high-$z$, which are costly to reach with a low-order
multipole expansion algorithm, and the need to Ewald-correct all the
force contributions.

The PM-accelerated versions and the use of higher order $p$ are
expected to do a lot better in terms of CPU cost, but where the
optimum lies is unfortunately problem-dependent.  Similarly, whether
or not the use of hierarchical timestepping is an advantage in terms
of performance depends critically on the depth of the timestep
hierarchy. If the latter is very deep, it can yield a performance
advantage, because the code can save on tree construction time, and
also does not need to synchronize passive particles. If it is shallow,
these advantages will be overwhelmed by the additional force
calculations and tree constructions that are necessary in the
hierarchical scheme.  In most scenarios, we expect the FMM approach to
be faster than the one-sided Tree, at least for global timesteps, with
its main downside being the somewhat higher memory need, the somewhat
less ideal parallel scalability, and the poor performance of FMM for
thinly populated timebins in local one-sided time integration schemes,
so that one is more or less forced to use hierarchical time
integration for FMM. Note that in our implementation FMM needs to
compute interactions between nodes that are stored on two different
processors on both processors, i.e.~twice, so that the associated
overhead slowly grows for fixed problem size with the number of
employed distributed-memory compute nodes.

To what extent all of these general expectations are borne out in
practice with the current implementation of {\small GADGET-4} can only
be decided through explicit tests of the performance for
scientifically relevant setups.  We report the results of a set of
tests along these lines in Section~\ref{secscalability}, providing
some guidance for the selection of the most efficient algorithm for a
particular problem. But we stress that we want to refrain from general
advice about which set-up is best, as such statements can easily be
misleading due to the complicated dependence of performance on the
problem type and the desired accuracy, and because details of the
technical computing environment such as the type of processor and the
speed of the communication backplane play a role as well.

\begin{table}
\begin{tabular}{c|rrr}
\hline
& \multicolumn{1}{c}{pure tree} &  \multicolumn{2}{c}{with mesh acceleration} \\
& \multicolumn{1}{c}{with Ewald} &  no Ewald & sometimes Ewald \\
\hline
\multirow{2}{*}{Tree-Based} & {Tree} & {TreePM} & {TreePM-NoS} \\
& {H-Tree} & {H-TreePM} & {H-TreePM-NoS} \\
\hline
\multirow{2}{*}{FMM-Based} & {FMM} & {FMM-PM} & {FMM-PM-NoS} \\
& {H-FMM} & {H-FMM-PM} & {H-FMM-PM-NoS} \\
\hline
\end{tabular}
\caption{Different combinations of gravitational force calculation and
  time integration schemes supported in {\small GADGET-4}. Variations
  with a prefix `{H-}' refer to use of the hierarchical time
  integration scheme. {Tree} and {FMM} designate that the hierarchical
  multipole calculation is done either with a classic one-sided Barnes
  \& Hut tree code, or with a fast multipole method. We note that each
  of these combinations can be run with different multipole order,
  $p=1$ to $p=5$. For the {PM} schemes, Fourier methods are used to
  accelerate the calculation of long-range forces. In the plain
  {TreePM} and {FMM-PM} approaches this is also inducing a split of
  the time integration into long- and short range dynamics. In the
  {NoS} variants of these schemes, no such split in the time
  integration is done, instead the forces are either computed with a
  pure multipole method, or with a PM acceleration, depending on the
  active particle fraction. Note that in zoom simulations, the PM variants of
  the algorithms can optionally be combined with the placement of
  additional Fourier mesh on the high-resolution region.\label{TabIntSchemes} }
\end{table}

\section{Hydrodynamical discretization} \label{sechydro}

A versatile and conceptually simple approach for simulating
hydrodynamics is the smoothed particle hydrodynamics (SPH) technique
\citep{Gingold:1977aa, Monaghan:1992aa}. It has a long history in
astrophysics, where some of its main advantages, such as automatic
spatial adaptivity, robustness, ability to deal with nearly empty
space, and ease of coupling to collisionless dynamics used for dark
matter and stars are particularly attractive points \citep[see][for
reviews]{Springel:2010aa, Price:2012aa, Rosswog:2015aa}.

However, over the past decade, accuracy problems of SPH in certain
regimes have been pointed out and seen intense scrutiny
\citep{Agertz:2007aa, Bauer:2012aa}. In certain regimes, standard
formulations of SPH have been found to be very noisy or suppress
fluid instabilities, causing concerns about the general applicability
of this technique. In turn, this triggered intense efforts to improve
or resolve these problems. Proposed solutions range from relatively
small refinements such as better artificial viscosity prescriptions or
the use of different smoothing kernels, over different formulations of
the equations of motion, to more radical changes of the underlying
discretization principle \citep[e.g.][]{Ritchie:2001aa, Price:2008aa,
  Hess:2010aa, Gaburov:2011aa, Murante:2011aa, McNally:2012aa,
  Valdarnini:2012aa, Read:2012aa, Hopkins:2013aa, Hopkins:2015aa,
  Beck:2016aa, Frontiere:2017aa}.  Also new types of quasi-Lagrangian
discretizations that are mesh rather than particle-based, such as the
moving-mesh code {\small AREPO} \citep{Springel:2010ab}, have been
proposed.

However, SPH remains useful as a low-order technique because it often
gives surprisingly robust and qualitatively correct results, even when
the particle number per object is small \citep{Schaller:2015aa,
  Huang:2019ab}, although there are also regimes where important
systematic effects due to its numerical inaccuracies show up in galaxy
formation simulations \citep[e.g.][]{Sijacki:2012aa, Nelson:2013aa}.
We have thus improved the original implementation of SPH in {\small
  GADGET-2}. The implementation included in {\small GADGET-4} offers
enhanced performance and features a few of the proposals made for a
more `modern' flavour of SPH. Which of the many improvements suggested
in recent years for SPH will ultimately prevail is however still an
unsettled issue, as illustrated by diversity of approaches in the
recent SPH literature. Contributing to this debate is beyond the scope
of this work, hence we restrict ourselves to contribute a modernized
code base that allows SPH simulations to be pushed to large sizes and
which should be a useful platform to easily make further improvements
to the basic formulation.

In {\small GADGET-4}, we presently support both `vanilla SPH', as
originally defined in terms of the conservative entropy formulation
\citep{Springel:2002ab} of standard SPH \citep{Monaghan:1983aa}, as
well as the `pressure-based' variant P-SPH \citep{Hopkins:2013aa},
which behaves better at contact discontinuities. Besides these two
basic formulations of the equations of motion, also the role of the
kernel has been emphasized in the recent SPH literature. In
particular, the \citet{Wendland:1995aa} kernels suggested by
\citet{Dehnen:2012aa} provide protection against the so-called
clumping instability, and thus allow a larger neighbours number to
mitigate the inherent noise in SPH. We offer both the traditional
cubic spline kernel, and the Wendland kernels at order 2, 4, and 6. We
note however that the Wendland kernels require substantially larger
number of neighbours for a given resolution, and it is problem
dependent whether their use is advantageous \citep[see
also][]{Zhu:2015ab}.

Finally, a third area where substantial improvements have been made
concerns the use of sophisticated parameterizations of artificial
viscosity. Here we include only one of the simpler recent
suggestions. We do not include an implementation of explicit mixing
terms between SPH particles \citep{Price:2008aa}, such as artificial
heat conduction, which has been shown to be beneficial for certain
problems. This can however be easily added to the code base if
desired.
 
\subsection{Basic SPH formulation}

\subsubsection{Standard (`vanilla') SPH}

As discussed by \citet{Springel:2002ab}, the inviscid part of the
hydrodynamics of an ideal gas discretized with SPH can be generated by
an interaction Hamiltonian of the form
\begin{equation}
H_{\rm therm} = \sum_i m_i A_i \frac{\rho_i^{\gamma-1}}{\gamma-1},
\end{equation}
where $A_i$ denotes the entropic function of particle $i$, and
$\rho_i$ is its density. The pressure is given by the ideal gas law
$P_i = A_i \rho_i^\gamma$ with adiabatic index $\gamma$. 
The densities are estimated with adaptive kernel estimation, 
\begin{equation}
  \rho_i = \sum_j m_j W(\vec{r}_{ij}, h_i) ,
  \label{eqDensityEstimate}
\end{equation}
where $W(\vec{r}, h)$ is a spherically symmetric smoothing kernel,
and the smoothing lengths $h_i$ are determined through the implicit
constraint $\rho_i h_i^3 /m_i= const.$

The Euler-Lagrange equations of motion then create antisymmetric
pair-wise pressure forces
\begin{equation}
\frac{\dd \vec{v}_i}{\dd t} = -
\sum_{j=1}^N m_j \left[ f_i \frac{P_i}{\rho_i^2} \nabla_i W_{ij}(h_i)
+ f_j \frac{P_j}{\rho_j^2} \nabla_i W_{ij}(h_j) \right], \label{eqEOMDensity}
\end{equation}
 where the $f_i$ are defined by
\begin{equation}
 f_i = \left[ 1 +
\frac{h_i}{3\rho_i}\frac{\partial \rho_i}{\partial h_i} \right]^{-1} ,
\label{eqA5}
\end{equation}
and the abbreviation $W_{ij}(h)= W(|\vec{r}_{i}-\vec{r}_{j}|, h)$ has
been used.

\subsubsection{Pressure based formulation}

The above formulation conserves both energy and entropy, but shows a
spurious surface tension at phase boundaries. This tends to suppress
or slow the growth of fluid instabilities at contact
discontinuities. The `pressure-based' variational formulation of
\citet{Hopkins:2013aa} can significantly improve on this, at the price
of a somewhat higher noise in other parts of the flow. 
Instead of estimating the density, the pressure is estimated by
\begin{equation}
P_i = \left[ \sum_j m_j A_j^{1/\gamma} W(\vec{r}_{ij}, h_i) \right]^\gamma.
\end{equation}
The equation of motion is given by
\begin{equation}
\frac{\dd \vec{v}_i}{\dd t} = -
\sum_{j=1}^N m_j \left(A_i A_j \right)^{1/\gamma}\left[ f_{ij} \frac{P_i}{P_i^{2/\gamma}} \nabla_i W_{ij}(h_i)
+ f_{ji} \frac{P_j}{P_j^{2/\gamma}} \nabla_j W_{ij}(h_j) \right], \label{eqEOMPressure}
\end{equation}
with the correction term
\begin{equation}
f_{ij} =  1- \left(\frac{h_i}{3A_j^{1/\gamma} \rho_i} \frac{\partial P_i^{1/\gamma}}{\partial h_i} \right)\left[1+ \frac{h_i}{3\rho_i} \frac{\partial \rho_i}{\partial h_i} \right]^{-1}
\end{equation}
for variable smoothing lengths.
As in our density formulation of SPH we determine the smoothing length by the condition $h_i^3\rho_i /m_i = const.$, which is equivalent to the condition of a fixed amount of particles in the kernel if the mass of all particles is the same.

Although for the equations of motion no estimate of the density is
needed, other modules of {\small GADGET-4}, e.g.~the artificial
viscosity or radiative cooling, might still depend on it.  We
additionally calculate the density according to
equation~(\ref{eqDensityEstimate}) for these purposes, rather than
estimating it through
$\rho_i = \left( P_i/ A_i\right)^{1/\gamma}$ from the smoothed
pressure.

\subsection{Kernel functions}

Much of the SPH-based literature of the last decades has been using
the $M_4$ cubic spline kernel from the B-spline family. When using it,
we set in $\nu$ dimensions
$W(r, h)= h^{-\nu} w_{\rm M4}\left(\frac{r}{h}\right)$, with
\begin{equation}
w_{\rm M4}(q) =\frac{8}{\pi} \left\{
\begin{array}{ll}
1-6 q^2 + 6 q^3, &
0\le  q \le\frac{1}{2} ,\\
2\left(1-q\right)^3, & \frac{1}{2}< q \le 1 ,\\
0 , & q>1 ,
\end{array}
\right.
\end{equation}
in three-dimensional normalization. Note that in the convention we use
here, the kernel drops to zero at a distance of $r=h$, i.e.~the
smoothing length $h$ is equal to the finite support of the kernel
(parts of the SPH literature instead define $h$ such that the kernel
drops to zero at $2h$).

Occasionally, higher order kernels like quartic or quintic kernels
from the same family have been used as well. Much of the recent
discussion about kernel choices has however been started by
\citet{Dehnen:2012aa} who pointed out that the three-dimensional
\citet{Wendland:1995aa} kernels protect against the clumping
instability, thus allowing the larger neighbour numbers required for
reducing the kernel-interpolation errors. This comes however at the
price of a significantly larger base neighbour number needed to reduce
the inherent bias of the Wendland kernels to acceptable levels, making
these kernels fairly expensive.  We include the $C_2$ Wendland kernel,
given by
\begin{equation}
w_{\rm C2}(q) =\frac{21}{\pi} \left\{
\begin{array}{ll}
(1-q)^4 (4q + 1), &
0\le  q \le1 ,\\
0 , & q>1 ,
\end{array}
\right.
\end{equation}
the  $C_4$ kernel defined by
\begin{equation}
w_{\rm C4}(q) =\frac{495}{32\pi} \left\{
\begin{array}{ll}
(1-q)^6 \left(\frac{35}{3} q^2+ 6q + 1\right), &
0\le  q \le1 ,\\
0 , & q>1 ,
\end{array}
\right.
\end{equation}
and the $C_6$ kernel given by
\begin{equation}
w_{\rm C6}(q) =\frac{1365}{64\pi} \left\{
\begin{array}{ll}
(1-q)^8 (32q^3+25q^2+8q + 1), &
0\le  q \le1 ,\\
0 , & q>1 ,
\end{array}
\right.
\end{equation} 
as alternatives to our standard kernel choice. For these kernels, from
$92$ (for $C_2$) up to $356$ (for $C_6$) neighbouring particles are
recommended, respectively.  We note that one small technical advantage
of the Wendland kernels is that they are not defined in a piecewise
fashion. This makes it slightly easier to apply vector instructions
for an on-the-fly evaluation of the kernel because if-conditions that
branch on the value of $r$ are not needed.  For completeness we also
implemented the one-dimensional and two-dimensional forms of all
kernel functions defined above following \citet{Dehnen:2012aa}.

\subsection{Artificial viscosity}

To provide irreversible dissipation of kinetic energy into heat at
shock fronts, an artificial viscosity needs to be added to the ideal
gas discretization introduced above. A difficulty is to find a
parameterization that is both sensitive to the presence of even weak
shocks, but at the same time does not apply viscous forces outside of
shocks, which would make the scheme unnecessarily diffusive
\citep{Cullen:2010aa, Read:2012aa, Hosono:2016aa}.

We add the viscous force to the equation of motion as 
\begin{equation}
 \left. \frac{\dd \vec{v}_i}{\dd t}\right|_{\rm visc} 
=
-\sum_{j=1}^N m_j \Pi_{ij} \nabla_i\overline{W}_{ij} \, ,
\label{eqnvisc}
\end{equation}
where
\begin{equation}
 \overline{W}_{ij}= \frac{1}{2}\left[ W_{ij}(h_i) +
W_{ij} (h_j)\right] 
\end{equation} denotes
a symmetrized kernel between the two particles involved.
In order to conserve total energy, we need to
compensate the work done against the viscous force through generation
of heat, which we do in terms of an entropy increase with a rate:
\begin{equation}
\left. \frac{\dd A_i}{\dd t}\right|_{\rm visc} =
\frac{1}{2}\frac{\gamma-1}{\rho_i^{\gamma-1}}\sum_{j=1}^N m_j \Pi_{ij}
\vec{v}_{ij}\cdot\nabla_i \overline{W}_{ij} \,,
\label{eqnentropy}
\end{equation}
where $\vec{v}_{ij}= \vec{v}_i - \vec{v}_j$. 

There is a lot of 
freedom in the detailed parameterization of the viscosity $\Pi_{ij}$.
Most commonly, some variant of
the form introduced 
by \citet{Monaghan:1983aa} is employed,
which is 
\begin{equation}
\label{eqvisc}
\Pi_{ij}=\left\{
\begin{tabular}{cl}
${\left[-\alpha c_{ij} \mu_{ij} +\beta \mu_{ij}^2\right]}/{\rho_{ij}}$ &
\mbox{if
$\vec{v}_{ij}\cdot\vec{r}_{ij}<0$} \\
0 & \mbox{otherwise}, \\
\end{tabular}
\right.  \end{equation}
 with 
\begin{equation}
\mu_{ij}=\frac{h_{ij}\,\vec{v}_{ij}\cdot\vec{r}_{ij} }
{\left|\vec{r}_{ij}\right|^2 + \epsilon h_{ij}^2},\label{egnMu} 
\end{equation}
and which can be viewed as a combination of a
bulk and a von Neumann-Richtmyer viscosity.
Here $h_{ij}$ and $\rho_{ij}$ denote arithmetic means of the
corresponding quantities for the two particles $i$ and $j$, with
$c_{ij}$ giving the mean sound speed, whereas
$\vec{r}_{ij}\equiv \vec{r}_i - \vec{r}_j$ is the particle distance
vector.  The strength of the viscosity is regulated by the parameters
$\alpha$ and $\beta$, with typical values in the range
$\alpha\simeq 0.5-1.0$ and the frequent choice of $\beta=2\,\alpha$.
The parameter $\epsilon\simeq 0.01$ is introduced to protect against
singularities if two particles happen to get very close.  Because the
viscosity factor $\Pi_{ij}$ is symmetric in $i$ and $j$, the viscous
force between any pair of interacting particles is antisymmetric and
along the line joining the particles, hence linear momentum and
angular momentum remain preserved.  The viscosity only acts for
particles that rapidly approach each other, and the entropy production
associated with it is always positive definite.

As discussed for {\small GADGET-2} \citep{Springel:2005aa}, we have a
slight preference for a related variant of the viscosity
parameterization, as proposed by \citet{Monaghan:1997aa} based on an
analogy to the Riemann problem. This is given by
\begin{equation}
 \Pi_{ij} =
-\frac{\alpha}{2} \frac{ v_{ij}^{\rm sig} w_{ij}}
{\rho_{ij}} , \label{eqnViscNew} 
\end{equation} 
where $v_{ij}^{\rm sig} = \left[ c_{i} + c_{j} - 3 w_{ij} \right]$ is
an estimate of the signal velocity between two particles $i$ and $j$,
and
$w_{ij}={\vec{v}_{ij}\cdot\vec{r}_{ij}} / {\left|\vec{r}_{ij}\right|}$
is the relative velocity projected onto the separation vector. This
parameterization of the viscosity is used by {\small GADGET-4} as
default.

The viscosity vanishes for solid-body rotation, but not for pure
shear flows. To cure this problem in shear flows, \citet{Balsara:1995aa}
suggested adding a heuristic correction factor to the viscosity that reduces its
strength when the shear is strong. We include this by multiplying
$\Pi_{ij}$ with a prefactor $(f_i^{\rm AV}+f_j^{\rm AV})/2$, where the
factors
\begin{equation}
f_i^{\rm AV} = \frac{ |\nabla\cdot \vec{v}|_i}{ |\nabla\cdot \vec{v}|_i + 
|\nabla\times \vec{v}|_i}
\end{equation}
are characterizing the rate of compression in relation to the local
shear. The velocity divergence and curl are computed alongside the
density estimates.

Another popular modification are time-dependent viscosity schemes that
aim to ramp up the viscosity only in regions of shocks while in other
regions the numerical viscosity should be reduced as much as possible
\citep{Morris:1997aa, Dolag:2005aa, Cullen:2010aa}.
We here use a formulation proposed by \citet{Hu:2014aa} to which we
refer for full details.  For every particle we define a target
viscosity parameter
\begin{equation}
\alpha_{{\rm tar},i}(t) = \alpha_{\rm max} \frac{h_i^2 S_i}{h_i^2 S_i
  +c_i^2} ,
\end{equation}
with a shock indicator $S_i = \max(0,- \dot{\nabla}\cdot \vec{v}_i)$. 
The actual viscosity parameter $\alpha_i(t)$ applied to a particle is evolved as
\begin{equation}
\alpha_i(t+{\rm d}t) = \left\{
\begin{array}{ll}
\xi_i\, \alpha_{{\rm tar},i}&
\alpha_i \leq \alpha_{{\rm tar},i} ,\\
\xi_i \left[\alpha_{{\rm tar},i} + (\alpha_i - \alpha_{{\rm tar},i})
  \exp(-{\rm d}t/\tau_i)\right] , & \alpha_i> \alpha_{{\rm tar},i} ,
\end{array}
\right.
\end{equation}
where ${\rm d}t$  gives the time step, and $\tau_i = 10 h_i / v_{\rm
  dec}$ is a decay time with the decay velocity
defined as
\begin{equation}
v_{{\rm dec}} =\max_{\vec{r}_{ij} \leq h_i} \left[c_i +c_j - \min(0,\vec{v}_{ij} \cdot \vec{r}_{ij}/r_{ij})\right].
\end{equation}
$\xi_i$ is a limiter defined by:
\begin{equation}
\xi_i = \frac{ |\nabla\cdot \vec{v}_i|^2}{ |\nabla\cdot \vec{v}_i|^2 + 
|\nabla\times \vec{v}_i|^2 + 0.0001(c_i/h_i)^2}.
\end{equation} 
To improve the accuracy of the time-dependent viscosity we have also
implemented higher order velocity gradient estimates using matrix
inversions, as discussed in the appendix of \citet{Hu:2014aa}.

\subsection{Cooling and star formation}

One of the areas where SPH has been particular popular is the simulation
of galaxy and star formation, as here the automatic adaptivity and the
ease with which source functions and subgrid models can be coupled to
SPH come in particularly handy. Current physics modelling in SPH has
reached a high degree of complexity and diversity, as can be inferred
from recent work, for example on simulating the interstellar medium
\citep{Hu:2016aa} or on galaxy formation in large cosmological volumes
such as the EAGLE project \citep{Schaye:2015aa}.  Clearly, there is no
single ``correct'' implementation for treating star formation and
feedback, rather the development of models faithful to the physics
remains a matter of very active research. We hence refrain from
incorporating a complex implementation into the public version of
{\small GADGET-4} at this point.

However, to help users get started with simulations of galaxy
formation, we have included a very basic radiative cooling module, and
a treatment of star formation based on the coarse-grained subgrid
model for the ISM described in \citet{Springel:2003aa}.  The cooling
routines account only for atomic cooling processes of hydrogen and
helium in collisional ionization equilibrium, as described in
\citet{Katz:1995aa}. Star formation is treated by creating new
collisionless particles based on a stochastic model from the star
formation rates estimated for the dense gas.  We note that slightly
refined versions of this treatment are still in use in today's
cosmological simulations of galaxy formation.

\subsection{Evaluation of SPH sums}

The computations required for SPH are primarily composed of two steps,
each requiring loops over neighbouring particles to compute the
corresponding kernel-weighted sums. The initial step computes the
densities, subject to the constraint of maintaining a certain number
of neighbours, i.e.~it also sets the SPH smoothing lengths for all
particles. We carry this out in the usual iterative way by
Newton-Raphson root finding \citep{Springel:2002ab}.  Along with the
density and smoothing length, we also compute a few auxiliary
quantities in the corresponding loop over neighbours, such as the
velocity divergence. The neighbouring particles are found with a range
searching technique, as described in more detail below.

Once the densities have been found, the hydrodynamical forces due
to pressure forces and viscous accelerations are calculated in the
second step, again looping over neighbours. This also produces a
corresponding rate of dissipation, i.e.~an increase of the particle
entropies. To this we may also add external source functions, such as
radiative cooling or forces from an external gravitational field.

During the hydrodynamic force loop, we also compute the quantity
\begin{equation}
  v_{i}^{\rm sig,max} = \max_{j} (c^{\rm snd}_i + c^{\rm snd}_j)
  \label{ensignv}
\end{equation}
where the maximum is taken over all interacting neighbours, and
$c_i^{\rm snd}$ is the soundspeed of particle $i$. This signal
velocity is useful for one of our timestepping criteria.

\subsection{Timestep criteria}

We define a local particle-based Courant-Friedrichs-Lewy (CFL) hydrodynamic timestep through 
\begin{equation}
\Delta t_i^{\rm cfl} =  C_{\rm CFL}\, 2 h_i / v_{i}^{\rm sig,max},
\end{equation}
where $v_{i}^{\rm sig,max}$ is defined as in equation~(\ref{ensignv}).
We further generalize the concept of signal velocity to include
the relative particle motions and to consider all possible particles,
including remote ones. This is meant to anticipate incoming
hydrodynamic waves, and to reduce the
timestep early enough just before arrival. To this end we define
\begin{equation}
\Delta t_i^{\rm signal} =  C_{\rm CFL} \min_j \frac{2 h_i +
  |\vec{r}_{ij}|}{c^{\rm snd}_i + c^{\rm snd}_j -
  \vec{r}_{ij}\cdot\vec{v}_{ij} / |\vec{r}_{ij}|} ,
\end{equation}
where the minimum is now taken over {\em all} other SPH particles.
This quantity can be efficiently calculated through a special tree
walk, as described in \citet{Springel:2010ab}.

Further, we define a
timestep criterion that restricts the allowed rate of change of the
smoothing length (or equivalently density) per timestep, through
\begin{equation}
\Delta t_i^{\rm dens} =  C_{\rm CFL} \frac{h_i}{|{\rm d}h_i / {\rm
    d}t|},
\end{equation}
where the rate of change of $h_i$ is estimated based on the local
velocity dispersion. Finally, we define a kinematic timestep based on
the total acceleration of the SPH particle (including pressure and gravity
forces, if present), as follows:
\begin{equation}
\Delta t_i^{\rm kin} =  (2\,\eta\, h_i / |\vec{a}_i|)^{1/2} ,
\end{equation}
where $\eta$ is the same parameter as used in the gravitational timestep
criterion.
The maximum timestep $\Delta t_i^{\rm sph}$
adopted for the particle is then determined as the
minimum of all these criteria, 
\begin{equation}
\Delta t_i^{\rm sph} = \min(\Delta t_i^{\rm cfl}, \Delta
t_i^{\rm signal}, \Delta t_i^{\rm dens}, \Delta t_i^{\rm kin}, \Delta t^{\rm
  global}, \Delta t_i^{\rm grav}).
\end{equation}
This also accounts for a prescribed global maximum timestep
$\Delta t^{\rm global}$, if present, and for a local gravitational
timestep if self-gravity is simulated. Hydrodynamical timesteps are
hence always at least as small as gravitational timesteps, and in case we
allow the time integration of the two to be decoupled, hydrodynamics
can subcycle a gravitational step, but not vice versa. As for gravity,
the actual timestep used for hydrodynamics is adopted as the
largest power-of-two subdivision of the total simulated timespan that
is still smaller or equal to $\Delta t_i^{\rm sph}$.

\subsection{Time integration of SPH particles}

In principle, it would be attractive to treat the time integration of
hydrodynamic interactions similar to gravity by splitting off the
locally ``fast'' particles from the ``slow'' ones, and forming a
subsystem with them for which only mutual forces among them are
computed. This could then be recursively applied to create deep timestep
hierarchies. There is however an important caveat. The computation of
the hydrodynamical forces requires an estimate of the total densities,
which in turn rely on a full set of neighbouring particles within the
smoothing length. Among these neighbours, there can easily be
particles on longer timesteps, {\em outside} the ``fast'' set
currently considered. It is therefore not straightforwardly possible
to fully decouple a subset of SPH particles from the rest of the
system and evolve it independently for a number of shorter
timesteps.

For the time being, we therefore stick with classic local SPH
timestepping, where a subset of SPH particles is active at the
current system time, and ``one-sided'' forces between particle pairs
may occur when not all interacting particles are synchronized at the
current time.

\subsubsection{On demand drifts and neighbour searches}

All particles participating in an SPH force calculation will have to
be at least predicted to the current time if they are not active
themselves.  We address this by splitting up the drift operator of
passive particles that are on longer timesteps into several drifts, as
needed. Due to the linear nature of the drift operator, multiple
applications for sub-intervals are equivalent to a single
application for the whole time interval.

To allow high adaptivity in time, it is however important to
only drift those particles to the current time that are actually
needed.  To facilitate this, each particle carries a local time
variable that informs about how far a particle has been drifted in
time. If a particle is touched while it lies in the past relative to
the current time, the SPH particle is drifted forward by the
difference in time, at which point the (passive) particle becomes
synchronized in time. Besides the drift of the particle coordinate, we
also advance a first order prediction of density and smoothing length,
as well as entropy and current velocity, in a similar fashion.

In order to arrive at a scheme that can still run efficiently when
only a tiny fraction of particles is active, we need to avoid being
forced to reconstruct the full neighbour tree every timestep. As the
gravity tree may contain only a small subset of particles when
hierarchical gravity is used, and can anyway change its geometry
rapidly due to a potentially high velocity dispersion of collisionless
particles, we use a separate search tree just for the SPH
particles. This allows to retrieve current coordinates of all
particles (for the density estimation), or to retrieve only those
neighbours below a certain timestep size. If we had to construct the
neighbour tree in every timestep, we would risk to become dominated by
tree construction overheads for thinly populated timesteps. We
therefore use a dynamic neighbour tree in which each tree node stores
the maximum and minimum coordinates of its particles in each
dimension, as well as the maximum and minimum velocities of each of
its gas particles, at the last time this node information has been
updated. This allows one to obtain a conservative bound for the
extension of the node under particle drifts at future times, which can
be used in a suitable node range search criterion. Furthermore, we
store the minimum timebin occurring for particles in the node,
allowing nodes that are irrelevant for the search because they do not
contain active particles to be quickly discarded.

\subsubsection{Detailed time stepping scheme}

Suppose we have the particle positions, velocities and entropies
$\left\{\vec{r}_i, \vec{v}_i, A_i\right\}$
in hand at time $t^{(n)}$.
First, the density is determined,
\begin{equation}
\rho_i^{(n)} = \hat\rho[\vec{r}_j^{(n)}] ,
\end{equation}
where $\hat\rho[\ldots]$ is the SPH kernel estimate,
followed by updating the pressure
\begin{equation}
P_i = A_i \rho_i^\gamma .
\end{equation} 
Next, we determine the SPH forces and dissipation rates.  If there is
no artificial viscosity, the SPH equations of motion are velocity
independent and the $A_i$ stay constant.  Then the SPH accelerations
are effectively only a function of the positions and independent of
the velocities, so that the equations can be integrated with a KDK
leapfrog just like applied to gravity.

But what if we have
viscous forces and dissipation? Here we apply the following
integration scheme
\begin{eqnarray}
\vec{a}_i^{(n)} & = & \hat a[\vec{r}_j^{(n)}, \vec{v}_j^{(n)}, A_j^{(n)}] \\
\dot{A}_i^{(n)} & = & \hat{\dot{A}}[\vec{r}_j^{(n)}, \vec{v}_j^{(n)}, A_j^{(n)}] \\
\vec{v}_i^{(n+1/2)} & = & \vec{v}_i^{(n)} + \vec{a}_i^{(n)} \frac{\Delta t}{2} \\
{A}_i^{(n+1/2)} & = & {A}_i^{(n)} + \dot{A}_i^{(n)} \frac{\Delta t}{2} \\
\vec{r}_i^{(n+1)} & = & \vec{r}_i^{(n)} + \vec{v}_i^{(n+1/2)} \Delta t  \\
\tilde{\vec{v}}_i^{(n+1)} & = & \vec{v}_i^{(n)} + \vec{a}_i^{(n)} \Delta t  \\
\tilde{A}_i^{(n+1)} & = & {A}_i^{(n)} + \dot{A}_i^{(n)} \Delta t  \\
\tilde{\vec{a}}_i^{(n+1)} & = & \hat a[\vec{r}_j^{(n+1)}, \tilde{\vec{v}}_j^{(n+1)}, \tilde{A}_j^{(n+1)}] \\
\tilde{\dot{A}}_i^{(n+1)} & = & \hat{\dot{A}}[\vec{r}_j^{(n+1)}, \tilde{\vec{v}}_j^{(n+1)}, \tilde{A}_j^{(n+1)}] \\
\vec{v}_i^{(n+1)} & = & \vec{v}_i^{(n+1/2)} + \tilde{\vec{a}}_i^{(n+1)} \frac{\Delta t}{2} \\
{A}_i^{(n+1)} & = & {A}_i^{(n+1/2)} + \tilde{\dot{A}}_i^{(n+1)} \frac{\Delta t}{2} \\
\end{eqnarray}
which yields second-order accurate estimates for
$\{ \vec{r}_i^{(n+1)}, \vec{v}_i^{(n+1)}, A_i^{(n+1)}\}$. Note,
however, that this requires in principle two hydrodynamical force
calculations per step, as for the next timestep the accelerations and
rate of entropy production need to be recomputed as
\begin{eqnarray}
\vec{a}_i^{(n+1)} & = & \hat a[\vec{r}_j^{(n+1)}, \vec{v}_j^{(n+1)},
                        A_j^{(n+1)}],
  \label{recomp1}\\
\dot{A}_i^{(n+1)} & = & \hat{\dot{A}}[\vec{r}_j^{(n+1)},
                        \vec{v}_j^{(n+1)}, A_j^{(n+1)}] . \label{recomp2}
\end{eqnarray}
This can be avoided if one sets
\begin{eqnarray}
\vec{a}_i^{(n+1)} & = & \tilde{\vec{a}}_i^{(n+1)} \label{eqsphapprox1} ,\\
\dot{A}_i^{(n+1)} & = & \tilde{\dot{A}}_i^{(n+1)} \label{eqsphapprox2},
\end{eqnarray}
which is an approximation made by default in {\small GADGET-2/3}.
This formally destroys the second-order accuracy of the subsequent
time-steps, but the inaccuracy arises exclusively from the dissipative
effects of the viscosity, i.e.~in regions of the flow which are
inherently irreversible and hence not prone to the build up of
long-term secular integration errors in time. For vanishing viscosity
this approximation becomes the ordinary KDK leapfrog appropriate for
the reversible part of hydrodynamics. It is therefore not surprising
that refraining from recomputing the accelerations after updating the
velocity and entropy at the end of a timestep does usually a very good job
in practice. In {\small GADGET-4}, the recomputation (\ref{recomp1})
and (\ref{recomp2}) can optionally be enabled, otherwise the
approximations (\ref{eqsphapprox1}) and (\ref{eqsphapprox2}) are
employed as default.

\subsubsection{Treatment of source terms}

We in general couple source terms in the hydrodynamics, such as the
gravitational field or a cooling function, through operator splitting
to the SPH dynamics. In particular, when gravity is included, the
hydrodynamical KDK step is bracketed by two gravitational half-step
kicks, or in other words, the drift operator involved in the
gravitational dynamics is replaced with a full SPH step, which
involves additional kicks due to the pressure forces.

We have further generalized this following \citet{Saitoh:2010aa} by
also allowing substepping of the gravitational timesteps with multiple
hydrodynamic steps, or in other words, the SPH timesteps may be
smaller than the gravity timestep, but can at most be as large as
the gravity step.

For integrating source functions in the internal energy, such as
radiative cooling, we also apply operator splitting, with the cooling
term applied twice per timestep, with a half-step at the beginning,
and a further half-step at the end. This also has the advantage that
completed timesteps (which may be saved as snapshots) always include
an updated collisional ionization balance. It is recommended that
feedback energy input is incorporated in a similar fashion.

\section{Parallelization strategy}  \label{secParallel}

The high performance of modern computer hardware arises from the
combined processing power of many, often independent compute cores.
Making full use of this parallel processing power requires the
identification and usage of opportunities for concurrent calculation,
which is one of the main challenges in contemporary high performance
computing, especially for tightly coupled problems such as those
commonly occurring in astrophysics. There are many different strategies
for expressing parallelism in a simulation code and for mapping it to
the available hardware. The approach of {\small GADGET-4} is
relatively traditional in that it relies on well established methods
for distributed (through MPI) and shared memory parallelization,
combined with a standard programming language (C++). While we include
explicit vectorization at a few select places, we presently do not use
special parallel programming languages, task-based scheduling systems,
or accelerators such as GPUs (although we have experimented with
GPU-accelerated versions of our gravity calculation routines, which is
work that remains in progress).

In this section we discuss our algorithmic approaches for different
aspects of the parallel code, and we motivate some of the choices we
made. We begin by discussing our domain decomposition and work-load
balancing approach, which is central to the parallelization scheme of
{\small GADGET-4}. We also discuss various communication patterns and
shared memory parallelization strategies employed by the code.

\subsection{Domain decomposition}

Aside from the goal of reaching high computational speed, another
important reason that motivates parallelization lies in the large
simulation sizes needed in cosmology. Calculations of cosmic structure
formation regularly involve much more data than can be stored in the
main memory of a single compute node, i.e.~they are not only CPU-time
limited but also memory-limited. Hence, parallelization on distributed
memory machines is essential in order to allow the use of the combined
memory of a large number of compute nodes.  It is clear then that
duplication of data needs to be avoided as much as possible in order
to be able to scale up the feasible simulation sizes. We thus aim for
a data decomposition strategy that makes optimum use of the combined
memory with as little redundancy as possible, and without significant
imbalance in the memory-use across nodes.\footnote{For definiteness,
  we assume compute nodes with an equal amount of memory and processor
  cores, which is the most relevant case in practice. As the maximum
  amount of memory that can be used on any given node is strictly
  limited, preventing large upward excursions in the memory use per
  node is essential. Also, only when the ratio of maximum to average
  memory use can be kept small and well under control, close to all of
  the available memory on the employed set of nodes can actually be filled
  by the target simulation.} Technically, we employ the standard
Message Passing Interface (MPI) for realizing distributed memory
parallelism.

We employ a spatial domain decomposition in which the simulated volume
is subdivided into disjoint regions that are then mapped to individual
MPI ranks.  This mapping is intended to be done such that the
resulting work-load {\em and} memory-load balance is as even as
possible. In addition, we demand that the distributed computational
algorithm does not change the semantics of the calculation, i.e.~all
results should be completely unaffected by the fact that such a
subdivision occurred in the first place. The latter is a relatively
strong demand. It precludes, for example, algorithms where the way the
domains are cut influences the geometry of the tree nodes and hence
the force errors obtained from walking the tree. In fact, it also
requires that the set of multipole expansions seen by any particle is
completely invariant under the domain decomposition.  We do, however,
not go as far as requiring that the results are binary invariant when
the number of MPI ranks (or equivalently the number of domains) is
changed, i.e.~differences caused purely by changes in floating point
round-off because mathematical operations are carried out in
different order are allowed. Trying to eliminate those would exclude a
large number of computationally efficient algorithmic possibilities.

But in case a simulation is repeated with identical settings
(including an equal number of MPI ranks), we ideally would like that
the delivered results are binary invariant. In other words, a fully
deterministic outcome between repeated simulation runs should be
achievable, which precludes certain strategies for adjusting the
domain decomposition or the work-load during a simulation. For
example, resorting to on-the-fly direct timing measurements on
different CPUs as a means to dynamically improve the work-load
balance, or using a central scheduler for asynchronously sending out
work packages depending on current CPU load, will in general not yield
results that are exactly reproducible from run to run, because such
strategies can typically influence the order in which certain
computations are done, for example by shuffling the order of floating
point additions when force contributions for a given particle are
summed up.  This in turn invariably introduces differences in
round-off error that will destroy the exact reproducibility of an
individual simulation run \citep{Genel:2019aa}.  Hence, we only
consider work-load measures and associated algorithmic solutions for
the domain decomposition that are compatible with strictly
reproducible behaviour.

The main virtue of deterministic outcomes lies in the ability to
reproduce code behaviour in a well-defined manner, something that is
of great value for code validation and debugging, especially in
parallel code with intricate communication patterns. While it is clear
that the particular invariant result selected in this way is not
fundamentally more accurate than those delivered by a semantically
correct algorithmic solutions where the round-off errors enter
differently each time the code is run, exact reproducibility is a
highly desirable trait of a scientific simulation code. Unfortunately,
this can not always be achieved, especially when asynchronous
parallelization techniques are used where the exact order of certain
operations may vary.

In our shared-memory parallelization approach that will be discussed
later in full, we allow imported tree nodes from foreign nodes to be
used by all other MPI ranks on the local nodes. Because these imports
do not necessarily always arrive in the same order, subtle differences
in the order in which interactions with such nodes are evaluated can
in principle occur, so that differences in numerical round-off break
the binary invariance. Fortunately, this can be easily suppressed, if
desired, by repeating tree walks that trigger such node imports,
rendering the design of {\small GADGET-4} in principle still binary
invariant in the above sense, despite the use of shared memory methods
that synchronize via atomic variables and spin locks. For maximum
speed, the extra work required to guarantee binary identical outcomes
on reruns of the code can however also be disabled.

\begin{figure*}
\resizebox{16cm}{!}{\includegraphics{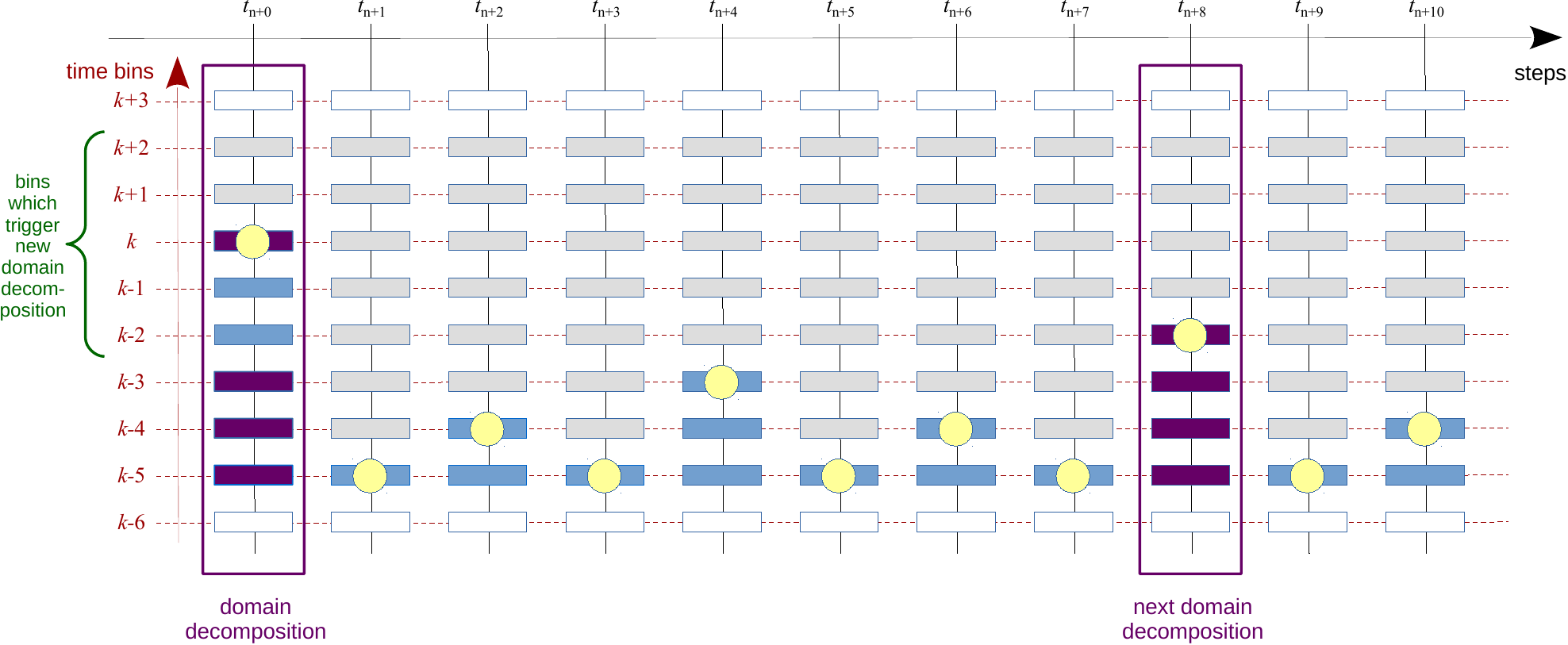}}
\caption{Sketch of the relation between work-load balancing
  and local time integration schemes. Filled
  rectangles mark the timebins that are occupied by
  particles. At each step of the simulation code, a certain number of
  timebins are synchronized, and the corresponding `active' particles
  require force calculations. Usually, the particle count of this
  active set (which is the union of all synchronized timebins) is 
  dominated by the highest synchronized timebin, which
  is marked with circles in the sketch. As the code hops from step to
  step (i.e. from $t_n$ to $t_{n+1}$, $t_{n+2}$, etc.), the set of
  active bins (marked in blue and red) and the corresponding
  particle configuration changes strongly, 
  such that  each step poses different
  work-load requirements for an optimum domain decomposition. 
  But since the low timebins are
  typically thinly populated, it is often not efficient to carry out a full
  domain decomposition for each of them. We therefore select a subset
  of the higher timebins, and only carry out a domain decomposition if
  at least one of them is active at the current synchronization time.
  In such a case, the corresponding domain
  decomposition (such as the one at $t_n$) should  however `think ahead',
  and also balance subsequent steps to the extent possible, until the
  next decomposition will be
  carried out (which is anticipated for $t_{n+8}$). The decomposition
  therefore needs to seek an optimum compromise for balancing the
  steps corresponding to cases where the highest active timebins are
  marked in red, taking into account also the number of
  times they will need to be executed until the next domain
  decomposition is carried out.
  \label{FigDomainBalancing}}
\end{figure*}

\begin{figure*}
\resizebox{15.5cm}{!}{\includegraphics{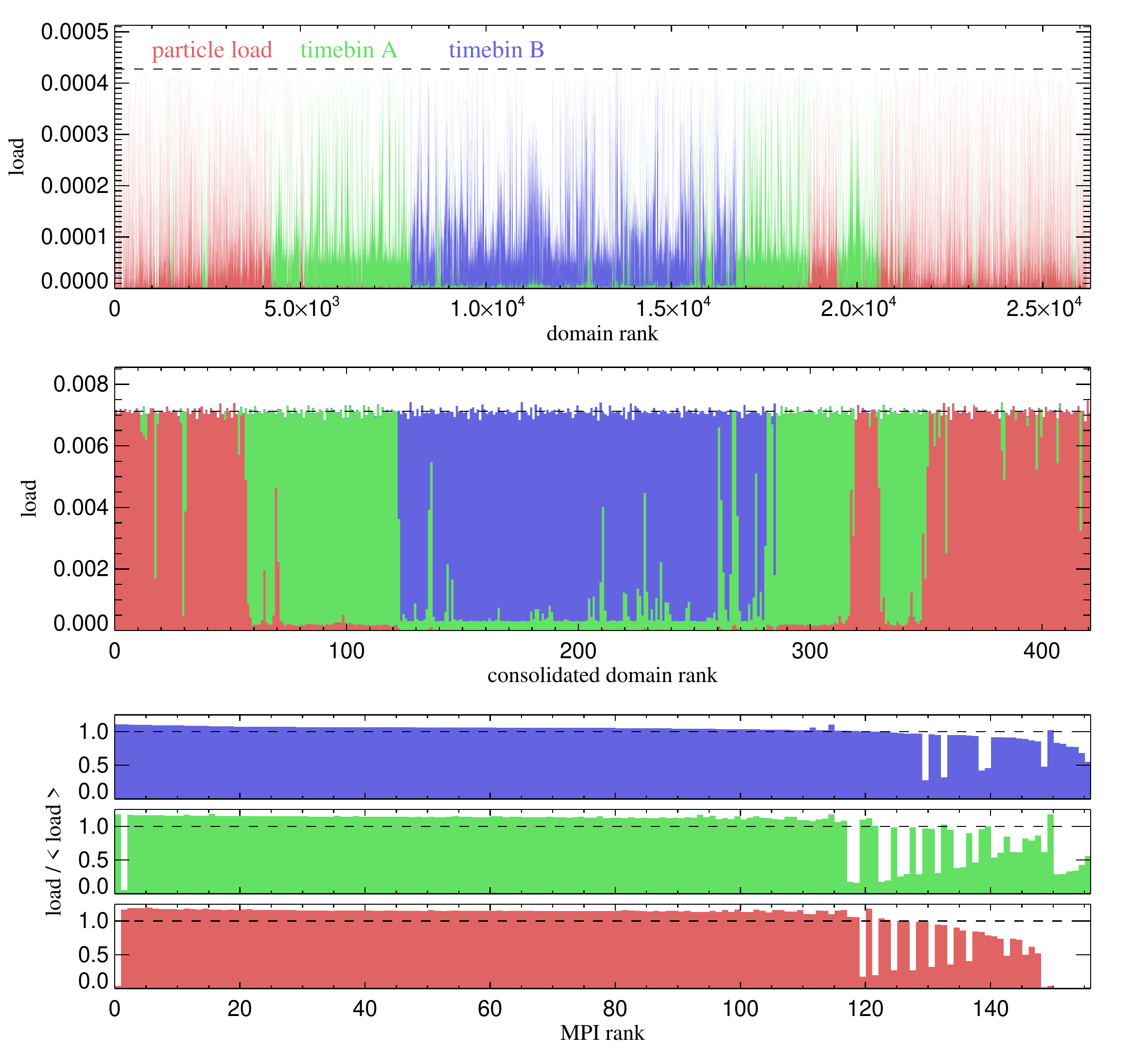}}
\caption{Real-world example of the domain balancing algorithm, here
  operating for two timesteps deep in the timestep hierarchy of an
  aggressive cosmological zoom simulation.  The red colour represents
  the cost factor `particle load', while the green and blue colours
  represent the computational cost of two different timebins that
  should be balanced simultaneously together with the particle
  load. In the top panel, we visualize the result of recursively
  subdividing the volume into about $\sim 2.5\times 10^4$ fine
  segments along a Peano-Hilbert curve that traverses the simulation
  volume, corresponding to the outcome of the first step in {\small
    GADGET-4}'s domain decomposition algorithm (the dashed vertical
  line marks the limit imposed for the maximum allowed combined load
  of the three cost items in each domain piece). These pieces are
  visualized in the graphics as small vertical columns of height
  proportional to the cost factors, and with a rank number (used as
  horizontal coordinate) assigned along the spatial Peano-Hilbert
  curve. We see that some pieces have many particles (red) but produce
  little or no computational load for the active timebins (green and
  blue), or vice versa. In the next step, the algorithm consolidates
  {\em adjacent} domain pieces into bigger chunks of approximately
  equal total load. We here have $f_{\rm mult} = 3$ (since there are
  three cost factors), $N_{\rm cpu} = 156$, and $N_{\rm extra} = 47$
  (the code tries different values for $N_{\rm extra}$ but
  we here only show the one eventually adopted because it yields the
  best final result). The outcome of this
  step is shown in the middle panel, with the horizontal line now
  showing the targeted average load equal to
  $f_{\rm mult}/(f_{\rm mult} N_{\rm cpu} - N_{\rm extra})$ for these
  consolidated domain pieces. These $N_{\rm domain} = 421$ domain
  pieces (each a single segment of the Peano-Hilbert curve) can still
  be either dominated by particle load, one of the timebin costs, or a
  combination thereof. Finally, in the third step, these domain pieces
  are mapped to the actual number of MPI ranks
  by combining several in such a way that each of the cost items is
  very close to be balanced individually (bottom three panels). The critical
  measure is here that the maximum cost or load assigned to any given
  MPI rank should not be much larger than the average cost or load for
  a uniform distribution among the ranks. If needed for an acceptable
  compromise, the algorithm may also opt to assign fewer domain pieces
  to certain MPI ranks than to others, as seen on the right of the
  the bottom panels. Importantly, however, none of the MPI ranks
  exceeds the average cost in any of the three categories (i.e. force
  calculation cost for two different timebins, and total particle
  load) by more than $\sim 18\%$ or so, representing a good balance
  given the extreme spatial inhomogeneity of the particle load, and
  even more so the computational cost, in this example.
  \label{FigDomainExampleBalancing}}
\end{figure*}
  
For constructing the domain decomposition, we adopt an algorithm with
three stages. In the first stage, we use an oct-tree covering the
simulated volume to identify a set of cubical nodes that tessellate
the simulated volume. The leaf nodes of this ``top-level tree'' do not
necessarily need to be all of the same size, instead they are selected
such that each amounts to at most a certain maximum computational load
(how this load is defined and measured will be discussed further
below). In stage two of the algorithm, we then consolidate the leaf
nodes into a set of domains, each consisting of several leaf nodes
that are adjacent to each other (because they are forming a
consecutive piece along the Peano-Hilbert curve that wraps through the
simulated volume). This is done to arrive at domains that all have a
reasonably similar total load, but which are still allowed to have
different characteristics in terms of whether they are dominated by
memory- or work-load, or from which timebins their work-load is mostly
created.  Multiple timebins need to be balanced when the domain
decomposition is not newly carried out for every single step, but
instead is retained for a sequence of sequential steps that traverse
the timestep hierarchy, as illustrated in
Figure~\ref{FigDomainBalancing}.  Also, for force computations in the
hierarchical time integration scheme, it is desirable to balance
multiple timebins all at once in a single domain decomposition.  Since
the depth and relative occupancy of the timestep hierarchy often
varies strongly throughout the simulation volume, simultaneously
balancing different timebins is generally not possible by just using
simply connected regions in the domain decomposition, but it can be
attained by combining pieces from different regions on the same
CPU. This is why we map these domains in the third stage of the
algorithm to the (smaller) set of MPI ranks. This mapping allows a
balancing of multiple criteria in a jigsaw style fashion all at
once. We now describe these three stages in turn.

\subsubsection{Creation of leaf nodes for the top-level tree}

Let $q_i$ be a gravitational cost factor associated with particle $i$,
meant to be a proxy for the CPU-time needed to compute the tree- or
FMM-based gravity calculation for this particle on the CPU it resides
on. We simply approximate this with the number of multipole
interactions the particle experiences in its gravity calculation, as
measured for its last force calculation.  We can then define a total
cost factor for an oct-tree node $j$ and timebin $b$ as
\begin{equation}
g_j^{b} = \sum_{\substack{i\, \in\, {\rm node}\, j \\ i \,\in\, {\rm timebin}\, b}} q_i.
\end{equation}
where only particles are counted that are active on bin $b$.  The
total gravitational cost that should be balanced by the domain
decomposition for timebin $b$ can hence be characterized by
\begin{equation}
 \tilde{g}^b = \sum_{j \,\in\,{\rm leaves}}  g_j^n
\end{equation}
where the sum extends over the leaf-nodes of the tree, which
tessellate the simulation volume.  We are further defining a total cost
factor $K_j$ of a leaf node as
\begin{equation}
K_j = \frac{N_j}{N_{\rm tot}} +  \sum_{b\, \in\, {\rm active\, bins}}
w_{b} \frac{g_j^{b}}{\tilde{g}^b} ,
\end{equation}
where $N_j$ is the particle number in node $j$, and $N_{\rm tot}$ is
the total particle number. The sum over the bins goes here over the
set of timebins the domain decomposition is supposed to simultaneously
balance, along with the particle load, which is always included. The
meaning of the weight parameter $w_b$ will become clear later on.

The construction of the top-level tree in stage one is now done
iteratively. At each iteration, all leaf nodes with load $K_j$ above some
threshold $K_{\rm max}$ are refined into 8 daughter nodes, replacing
these leaves. As limiting threshold we pick
\begin{equation}
 K_{\rm max} = \frac{1}{f_{\rm top} f_{\rm mult} N_{\rm CPU}},
\end{equation}
where $f_{\rm mult}$ is set equal to the number of different cost
categories we want to balance (so if two timebins are active, we would
set $f_{\rm mult}=3$, because we also have the particle load),
$N_{\rm CPU}$ is the number of MPI ranks for which we want to reach
the balance, and $f_{\rm top} \simeq 2-5$ is a parameter that can be
varied to make the decomposition finer if desired.
Note that the sum over the leaf nodes represents a disjoint
partitioning of the full particle set, i.e.~the sum of the
corresponding loads of the domains will be
\begin{equation}
 K_{\rm tot}  = \sum_{j\, \in \,{\rm leaf\, nodes}} K_j =  1 +   \sum_{b\, \in\, {\rm
     active\, bins}} w_{b}  .
\end{equation}
Our goal with this subdivision is to obtain a sufficiently fine set of
leaf nodes such that their load in each cost category is
less than $K_{\rm max}$. 

However, if a timebin is sparsely loaded, this may become impossible
to achieve if the number of particles is of order
$f_{\rm top} f_{\rm mult} N_{\rm CPU}$ or less. Then there are simply
too few particles to achieve a balanced decomposition for the number
of  cores. To cope with this situation, we use the weight
factors $w_{b}$.  We first determine
\begin{equation}
g_{{\rm max}}^{b} = \max_{  i
    \,\in\, {\rm timebin}\, b} q_i , 
\end{equation}
and then set
\begin{equation}
  w_{b} = \min\left(1, K_{\rm max} \frac{\tilde{g}^b}{g_{\rm max}^{b}} \right).
\end{equation}
This means that for the normal case of well-occupied timebins,
we will have $w_{b} =1$.  Otherwise, the gravity cost of the most
expensive particle in the bin will be reduced such that it enters 
with a cost $w_b \, g_{{\rm max}}^{b} / {\tilde{g}^b} = K_{\rm max}$
in its parent leaf node, so once the code has isolated this particle
in a leaf node, it will not dominate the total cost.

In passing, we note that the hydrodynamic work- and particle-load can
be added in a straightforward fashion to this approach.  Combined with
the further steps below, this then allows a simultaneous balancing of
gravitational and hydrodynamic costs and loads (which can in general
vary differently in space). We refrain from explicitly adding this to
the equations for clarity, but have implemented this in {\small
  GADGET-4}, adopting the approximation that each SPH particle
requires a similar interaction count for its density and
hydrodynamical force calculations.

\subsubsection{Consolidation of leaf nodes into domains} 

We next exploit a spatial Peano-Hilbert ordering of the leaf nodes that
effectively puts them into a one-dimensional list of nodes. Segmenting
this list of nodes into $N_{\rm domain}$ pieces by introducing
$N_{\rm domain}-1$ cuts then creates small groups of spatially
adjacent leaf nodes that are characterized by a small surface to
volume ratio. The cuts are done such that the total load of each group
of leaf nodes is as even as possible. Since we know the cost $K_j$ of
each leaf node from above, as well as their total cost $K_{\rm tot}$,
this consolidation into equal pieces of cost
$\simeq K_{\rm tot} /N_{\rm domain} $ can be achieved with good
accuracy by simply considering the cumulative cost function along the
Peano-Hilbert curve and carrying out corresponding cuts.  Note that
the factor $f_{\rm top}$ influences the average number of leaf nodes
available per domain, and thus the degree to which the grouping
procedure can smooth out the residual unevenness in the load
distribution of the leaf nodes.

Using directly $N_{\rm domain} = N_{\rm CPU}$ in this approach will
however not necessarily yield good results when the computational cost
associated with individual particles varies strongly within the
computational volume. In particular, it would then not be possible in
general to balance {\em both} memory-load and gravitational work-load
well at the same time, because the distribution of particles onto the
timestep hierarchy as well as the particle clustering (which influences
the gravitational cost factors) will typically show substantial
spatial variations, with some timebins being active only in small
subsets of the particle distribution. This means that a good balance
in all of these different cost categories can not be constructed if
only simply connected regions are used.

\subsubsection{Mapping domains to processors}

To improve on this, we therefore use a third stage in our domain decomposition
algorithm.  This starts with the idea to combine several domains of
different load characteristics on a single processor, with the
intention to average out the unevenness in their relative load
characteristics. Because all domains after stage two above have
constant total cost, an uneven distribution in the particle load
implies that there must be an uneven distribution in the gravitational
cost as well. It thus should be possible to combine a domain which has
many particles but low gravitational cost with one that has few
particles and high gravitational cost, achieving something that is
fairly balanced overall. This forms the basic idea.

We realize this approach with a heuristic algorithm that mimics how
one may perhaps pack a set of moving boxes. Here, a common strategy is
to first place the biggest item left into an empty box, and then
combine it with the largest remaining item that still fits in. Using
this as a guiding principle, we sort the domains in order of
decreasing load in each cost category. We then pick the next empty
processor, assign the domain with the most uneven cost that is left,
and then sift through the other domains that are most unevenly loaded
according to the \textit{other cost items} and see whether they still fit
in. Since we want to balance $f_{\rm mult}$ different cost items, we
expect that $N_{\rm domain} = f_{\rm mult} N_{\rm CPU}$ may be a good
starting point, with the idea to then combine $f_{\rm mult}$ domain
pieces on a single processor.

However, enforcing a fixed number of domain pieces per processor is
inflexible and makes the approach prone to worst case
outliers. Because at the end of the day, it unfortunately does not
help at all if almost all of the processors have a very uniform load,
apart from one exception that deviates and has a significant
excess. If, for example, this one processor has a load that lies 50\%
above the average, then it will hold up {\em all other processors},
independent of their number, creating a lever arm for losing a huge
amount of CPU time. So the figure of merit is to make the imbalance
\begin{equation}
B = \frac{\max L_i}{\left< L_i \right>}
\end{equation}
for each of the cost items as small as possible, where $L_i$ is the
load of a particular cost item on CPU $i$. 

To avoid becoming dominated by unwieldy domains that do not fit well
with others, we instead use the ansatz
\begin{equation}
N_{\rm domain} = f_{\rm mult} N_{\rm CPU} - N_{\rm extra},
\end{equation}
with $N_{\rm extra}$ being some number between 0 and $N_{\rm
  CPU}$. This would for example allow us to put $f_{\rm mult}$ domains
onto $N_{\rm CPU} - N_{\rm extra}$ cores, and $f_{\rm mult}-1$ domains
on the remaining $N_{\rm extra}$ CPUs. While this will induce an
imbalance of at least
\begin{equation}
B_{\rm min} \simeq \frac{ f_{\rm mult} N_{\rm CPU}}{ f_{\rm mult} N_{\rm CPU} - N_{\rm extra}},
\end{equation}
it may nevertheless allow for a better packing solution than for the
case $N_{\rm extra}=0$, simply because we have now the freedom to leave
some problematic domains from the tail of the distribution paired with
no or fewer other domains.

To pick a nearly optimum solution, we use the following algorithm.
For a given value of $N_{\rm extra}$, we first carry out stage two of
the above algorithm to obtain a set of domains with balanced total
cost.  We then select a fiducial load threshold
\begin{equation}
l_{\rm max} = \frac{B_{\rm thresh}}{N_{\rm CPU}} ,
\end{equation}
starting with $B_{\rm thresh} = B_{\rm min}$.  We stop packing a
processor with further domains when we cannot avoid filling it above
this maximum load for any of the cost categories that we want to
balance simultaneously. If this happens, we move to the next empty
processor. It can now happen that we run out of processors before we
have distributed all domains. In this case, we have aimed for a too
aggressive $l_{\rm max}$. We therefore try again by allowing for a
larger value of $B_{\rm thresh}$. Furthermore, we may retry the whole
packing procedure for different values of $N_{\rm extra}$.  

In practice, we carry out all these trials in parallel, with each
processor evaluating a different trial mapping that tests a binary
tuple $(N_{\rm extra}, B_{\rm thresh})$. This allow us to then adopt
the solution that yields the smallest overall imbalance value $B$.

In Figure~\ref{FigDomainExampleBalancing}, we show an example of the
outcome of this complex domain decomposition algorithm, here carried
out with two lower timebins in a high-resolution zoom simulation of a
dark matter halo. For definiteness, we pick the Aq-A-2 halo of the
Aquarius project, and consider it around $z\simeq 0$.  There are
$6.07\times 10^8$ particles in total in the simulation box, which is
relevant for the load balancing. Only a subset of $1.85\times 10^7$ is
active in timebin A, and far fewer still, $6.92 \times 10^5$, in
timebin B. These timebins are populated by particles in the centre of
the halo the simulation is zooming on, and occupy only an extremely
small region of the full volume of the simulation box. Making a
spatially disjoint subdivision of space such that each processor gets
roughly an equal number of particles in total, and has equal particles
in timebin A as well as timebin B is rather challenging. As the figure
illustrates, {\small GADGET-4}'s domain decomposition algorithm is
nevertheless reasonably successful in this. It pushes down the
maximum relative imbalance in any of these three cost categories to
about $18\%$ in this example.

\begin{figure*}
  \begin{center}
    \resizebox{16cm}{!}{\includegraphics{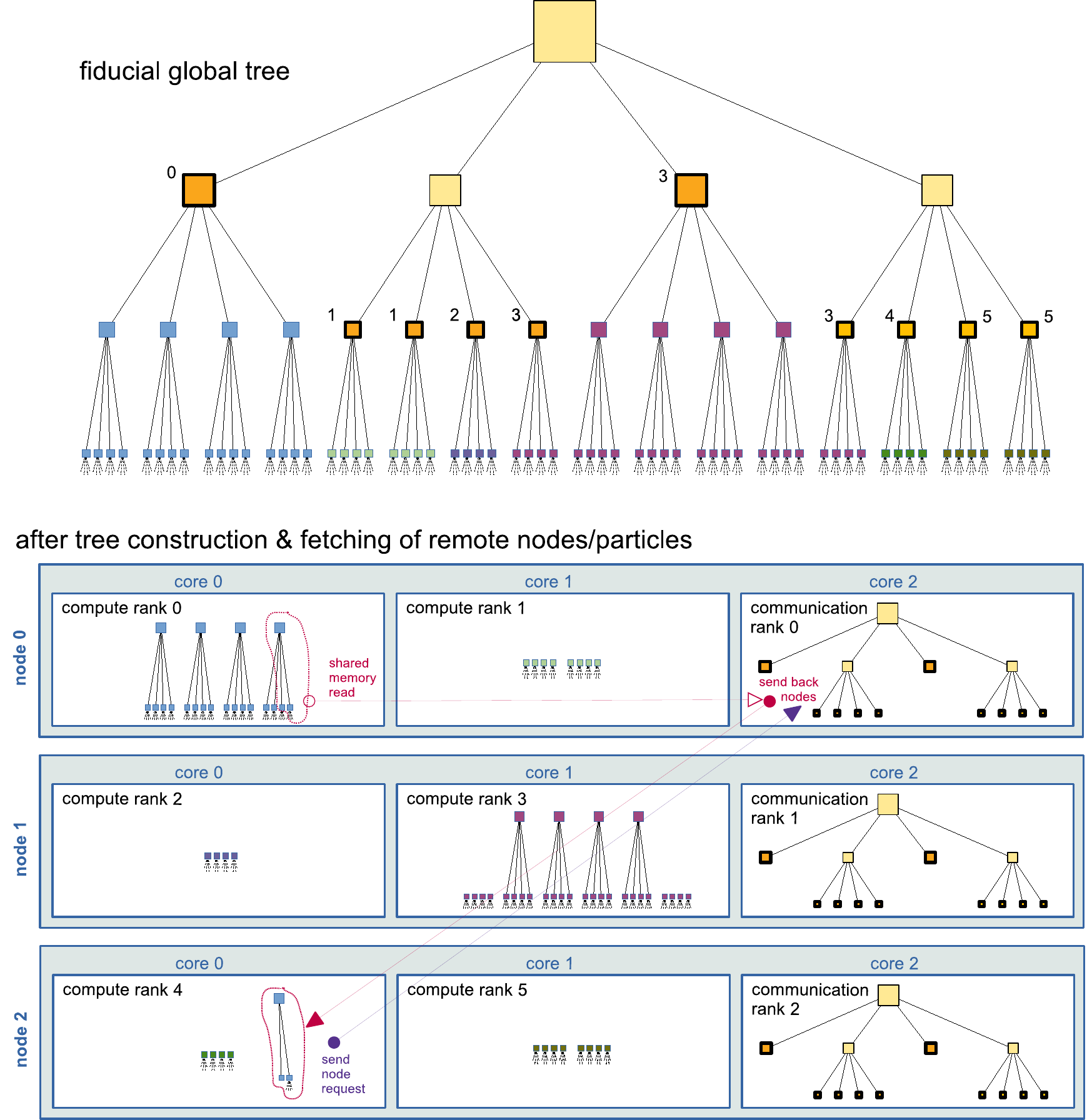}}%
  \end{center}
  \caption{Organization of parallel tree calculations in {\small
      GADGET-4}
    using a hybrid MPI/shared-memory parallelization
    approach. The fiducial oct-tree sketched in the top half covers the full simulation domain and
    contains all its particles. It is comprised of a `top-level' tree
    that is constructed in the domain decomposition and which ends in a set of leaf
    nodes, marked with bold borders. The leaf nodes form a
    non-overlapping tessellation of the simulated volume and are mapped to
    individual compute ranks, as indicated by the digits 0 to 5 in the
    sketch. In the example, the code runs
    on three compute nodes with three cores each. In total
    six compute ranks remain, because
    on each of the
    nodes, one core is set aside to act as a communication rank and
    for holding a local copy of the top-level tree.
    All branches below the top-level leaves are stored initially
    without redundancy on the compute ranks. If a compute
    rank needs further branch data from a remote node (because it
    wants to open the node),
    this data is fetched from the corresponding remote node and
    integrated into the tree on the local node. In the sketch, this is indicated by
    compute rank 4 needing a  node and its children from a branch
    stored by compute rank 0. To get it, rank 4 sends a request to the
    dedicated communication rank of the shared-memory node on which compute
    rank 0 resides, as
    sketched. The corresponding communication core constantly listens for incoming
    requests. It attends to an arriving message immediately by reading
    the requested data from compute rank 0 via a shared memory
    access (independent of whether or not rank 0 is busy with
    calculations) and sends it back to compute rank 4. There, it
    becomes available not only to the originally requesting rank, but
    also to all other compute cores on the same node (e.g.~rank 5 in the example).
    \label{FigParallelTree}}
\end{figure*}

\subsection{Hybrid parallelization}

The number of compute cores on powerful supercomputers is growing
rapidly, and much of their performance increase nowadays comes from a
larger number of cores, not an improved single core
performance. Unfortunately, subdividing a given problem size over an
ever larger number of MPI ranks becomes rapidly difficult. This
increases the cost for the domain decomposition algorithm,
amplifies work-load imbalance losses, and may expose problems of
scalability in the MPI software stack itself, for example if memory
consumption for MPI buffer spaces grows faster than linear in
the number of ranks, causing an eventual depletion of the memory left
for the science application.

In this situation it can be advantageous to combine shared-memory with
distributed-memory parallelization. One idea is to place only one (or
a few) MPI ranks per multi-core compute node.  The memory of the node
is then used in full by the MPI rank's process, running on a single
core, and the other cores are utilized with parallelization techniques
for shared memory. This can, for example, be done through code that
explicitly implements multi-threading with the pthreads
library. Alternatively, one can use compiler extensions introduced by
the OpenMP standard, which has become a widely available technique on
current platforms.

In practice, the combined use of MPI and OpenMP can reduce the number
of required MPI ranks to reach a given total memory size and core
number. It can also help to reduce work-load imbalance losses, limit
redundancies in data storage, and lower overhead due to the MPI
software stack. However, despite these advantages, it can often be
difficult to reach the same or a higher raw speed with MPI+OpenMP
compared with a pure MPI code for the same number of cores.  For
example, \citet{Lee:2021aa} recently reported results for a new hybrid
MPI-OpenMP version they developed for the {\small RAMSES} code,
reaching a speed-up of a factor of 8.5 when 64 threads are used on
nodes equipped with Intel Xeon Phi 7250 68C many-core processors. This
is still far from the theoretical speed up of 64 that one may expect
if each core was running its own MPI process and perfect
parallelization was achieved.  One reason lies in cache-utilization
and memory access patters. Because the OpenMP threads are not isolated
from each other, one needs to avoid that they interfere with each
other's cache lines in write operations, that they access memory banks
attached to remote sockets on the same node, and that they spend a lot
of time in congested locks, otherwise very significant performance
losses can result.  A more subtle problem can arise when one requires
strict reproducibility of simulation results when the same run is
repeated with the same setup, as this prevents the use of elementary
OpenMP reduction operations that compute sums of floating point
numbers in an unspecified order.

Another difficulty is that \emph{all} routines that consume relevant
amounts of CPU-time need to be very efficiently parallelized with
OpenMP, otherwise the additional cores occupied by the OpenMP threads
are only partially utilized, limiting scalability.  In simple cases,
the required OpenMP parallelization can be accomplished by annotating
the serial code with compiler pragmas that guide loop-level
parallelization done by the compiler. However, in less trivial cases
(which is unfortunately the common situation in {\small GADGET-4})
code rearrangements or algorithmic changes are also required to make
this really effective. One then faces a double challenge of
parallelization; one needs to devise an algorithm amenable to very
good distributed memory parallelization, which on top can be further
parallelized well at the loop-level or through task-based algorithms.

After pursuing this OpenMP approach for some time with mixed success
(our implementations worked extremely well for our tree-based gravity
and SPH calculation loops, but much less well for PM- and FMM-gravity,
as well as auxiliary code parts such as the domain decomposition and
the group finding algorithms), we eventually abandoned it for another
shared-memory parallelization strategy that we describe in the
following. An important motivation for this was primarily that the
OpenMP ansatz does not help at all with the need to parallelize across
distributed-memory compute nodes -- this remains to be addressed by
MPI.  And here our communication algorithms were still suffering from
losses due to insufficient synchronization, i.e.~when an MPI rank
wanted to fetch remote data from another process, it would often have
to wait for this other process to call an MPI function -- only then
the message exchange could proceed.

We have first tried strategies like asynchronous communication or
MPI's one-sided communication routines to mitigate these problems, but
only with limited success. A central problem proved to be that most
existing MPI libraries do not really guarantee asynchronous progress
of MPI message exchanges while the cores are busily executing
computational loops. In practice, this severely impairs the idea to
overlap communication and computation, and makes the performance
achievable with hardware-assisted remote memory access (RMA) highly
system dependent. Disappointingly, while one-sided communication
relying on the new functionality embedded in MPI-3 is fully portable,
the performance achieved with this is not
\citep[e.g.][]{Schuchart:2018aa}, simply because RMA accesses are not
possible on all systems with similar performance, or are implemented
by the MPI libraries in very different ways that require different,
library-dependent optimization strategies.

To address this in a different way, we are using in {\small GADGET-4} a
new feature of MPI-3, which allows the allocation of shared memory
that can be jointly accessed by the MPI ranks residing on the same
shared memory node. This allows the possibility to replace MPI send-
and receive-operations within a compute node by direct read and write
memory accesses using shared memory semantics, thereby effectively
allowing true one-sided communication within a shared memory node
that does not require any immediate cooperation of the other process.

We extend this concept to multi-node runs by setting aside one MPI
rank on each node to be solely responsible for dealing with one-sided
communication requests. When an MPI-rank wants to access remote data
on another MPI-rank on a different compute node, it does not send a
communication request directly to the target rank, but instead to the
target node's designated communication rank, which then fetches the
data via a shared memory access (the target rank's process doesn't
have to cooperate for this, i.e.~this is readily possible even if the
process is busy doing computations) and relays it back as
needed. Since the designated communication rank is doing nothing but
constantly waiting for such incoming requests, they can be answered
with minimal latency. This eliminates the synchronization losses in
our core computational algorithms we alluded to above, whereas OpenMP
parallelization only managed to somewhat reduce them by lowering the
number of MPI ranks that are in use.

The price one pays for this that at least one core per node needs to
be set aside for handling the MPI communications in this form. Typical
HPC systems now have often $\sim$24 -- 64 cores per node, with a clear
trend towards further growth of this number. This means that only a
few percent of the raw performance have to be set aside for the
communication rank, which appears insignificant if this yields an
overall faster and more scalable code. We note that this way of doing
MPI-based shared memory programming also allows other important
savings, like storing data that is equal on all MPI-ranks (e.g.~the
top-level tree, or large look-up tables) only once per compute node,
and not for all MPI ranks separately.

\subsection{Data layout and communication scheme in parallel tree walks}

In Figure~\ref{FigParallelTree}, we sketch how {\small GADGET-4} uses
the above concepts for organizing its parallel tree-based
calculations. The top-level tree constructed in the domain
decomposition is equal for all the MPI-ranks in a shared-memory
node. It is hence only stored once on each shared-memory node (we use
the memory associated with the dedicated communication rank for this,
but one of the compute ranks could also be used). The compute ranks
only store the tree branches assigned to them by the domain
decomposition, without redundancy. When the tree is walked, each
MPI-rank can freely access all nodes from the top-level tree, its own
branches, as well as those of the other MPI-ranks on the same compute
node, via shared-memory accesses\footnote{In doing this, the C++11
  memory model for multi-threaded execution needs to be
  respected.}. If an MPI-rank needs to open a tree node whose daughter
nodes or particles are stored on a remote compute node, this data is
fetched from the remote node by contacting its designed communication
rank, which then returns the requested data with minimal latency,
side-stepping the MPI-synchronization losses mentioned above. We
however still buffer such node-requests and accumulate them into
larger packages to avoid creating a large number of tiny
messages. Imported node data augments the locally stored tree, and
becomes immediately accessible also to all other MPI-ranks of the same
shared-memory node. After all tree walks are carried out by them, the
locally held tree data thus corresponds to a ``locally
essential tree'' (which is a subset of the fiducial global tree)
needed to fully carry out the tree walks of all particles mapped to the
local shared memory node.

We note that one side-effect of the buffering approach mentioned above
can be that it breaks the deterministic outcome at the binary floating
point level, which is undesirable as discussed earlier. This can
happen when a tree walk is continued despite a node's children need to
be fetched. Instead of waiting for arrival of the data, the node
interaction may be temporarily cached and revisited later once the
data has arrived. However, another MPI-rank may have been quick enough
and already imported the node's children, in which case the tree walk
would have been able to continue right away by opening this node,
altering the order in which the partial contributions are added up. A
tell-tale sign that this type of problem can occur is that the
insertion of imported nodes into the locally held tree needs to be
protected by a spin-lock\footnote{We use the efficient
  std::atomic\_flag available in C++11 for this, and also employ
  atomic variables at a few other places, consistent with the C++
  memory model for shared memory access \citep{Williams:2012aa}.}  to
avoid a data-race condition.

However, binary invariance can be easily re-established by carrying
out the actual tree calculations only after the locally essential tree
has been fully assembled. This requires a slightly higher
computational cost as it entails some tree-walking overhead in the
form of repeated evaluations of node-opening decisions, and is thus
made an optional feature in {\small GADGET-4}.

Note that the concept of constructing a locally essential tree, which
is followed by {\small GADGET-4} here and has been previously adopted
by other tree codes \citep{Dubinski:1996aa}, is notably different from
the communication strategy adopted in previous {\small GADGET}
versions, where instead of importing remote branch data to process a
local particle the particle was sent to the MPI-rank holding the
remote branch, which would then do the computational work and send
back the result. The latter approach works well for TreePM and SPH,
where the interactions are limited to a small neighbourhood around the
local domain, but it has a less favourable scaling behaviour with
respect to the data volume that needs to be communicated when a pure
Tree or FMM approach are used. Its main downside is that remote
processes need to actively contribute computational cycles to
processing local particles -- getting them to do that requires
synchronization of some kind, creating ample opportunity for
parallelization losses due to wait times. On the other hand, the
memory requirements of the locally essential tree approach can become
very challenging in the limit of tiny opening angles. If the opening
angle is approaching zero, the final locally essential tree becomes
ever closer to the fiducial global tree holding all particles,
i.e.~then each node would have to hold the full global tree. However,
as in this limit the force-calculation degenerates to a direct
summation algorithm anyhow, this is fortunately not a particularly
relevant limitation for practical applications.

\subsection{Parallel FFT calculation}  \label{secparfft}

For the PM calculations in {\small GADGET-4}, we use real-to-complex
and complex-to-real discrete fast Fourier transforms in 3D. The work
and memory need of such transforms scales rapidly as
$N_{\rm grid}^3\ln(N_{\rm grid}^3)$. In {\small GADGET-2}, the
transforms were implemented using the FFTW-2 library. This library has
since been deprecated and replaced with FFTW-3, which features a
different application interface and some slight improvements. Both
versions of the library use a slab-based decomposition for distributed
memory parallel FFTs. This does not allow the efficient use of more
MPI tasks than there are grid slabs, hence scalability is strictly
limited to the regime $N_{\rm cpu} \le N_{\rm grid}$. This is very
problematic for today's problem sizes in cosmology. Because FFTW-3
also uses additional internal memory allocations in its distributed
transforms (something we seek to avoid to have full control of the
application memory), we have developed our own version of 3D parallel
FFT routines based on FFTW-3's one-dimensional FFTs.

\begin{figure}
\resizebox{8cm}{!}{\includegraphics{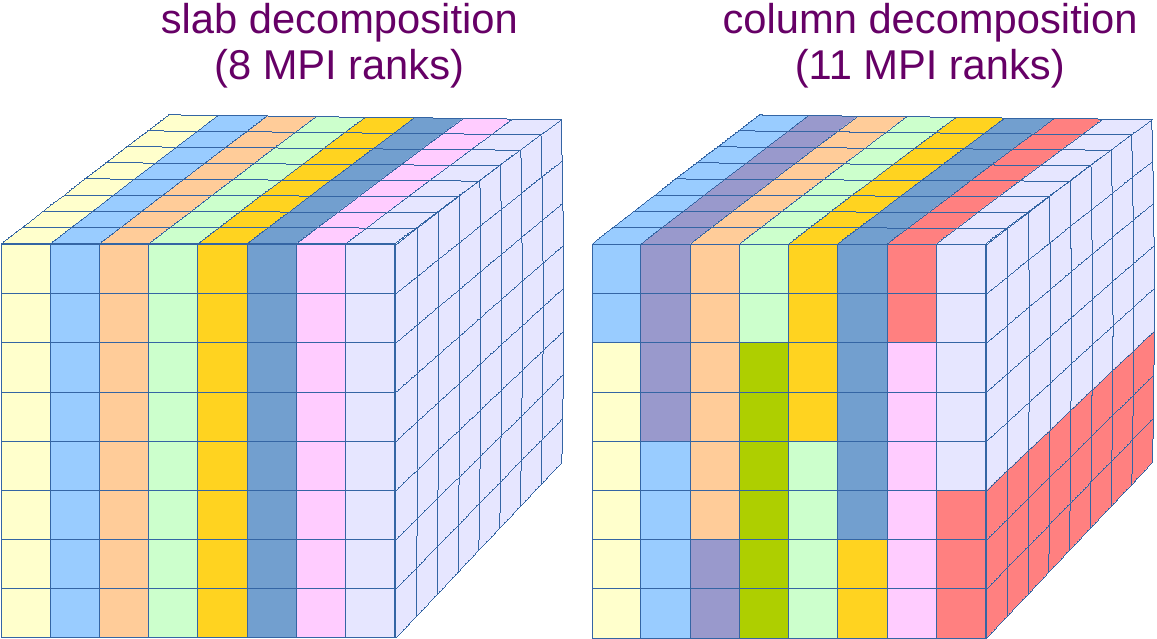}}
\caption{Slab-based versus column-based data decomposition for
  distributed 3D FFTs in {\small GADGET-4}. If the number $N_{\rm grid}$ of slabs is
a multiple of the number $N_{\rm cpu}$ of MPI ranks, a slab-based
decomposition is a good, balanced choice (as in the example sketched on the
left). However, this limits the scalability of the code, particularly
for large problem sizes. If the number of MPI ranks is much larger
than $N_{\rm grid}$, a more flexible decomposition is needed. {\small GADGET-4}
implements a column-based decomposition where the total number of
columns (typically $N_{\rm grid}\times N_{\rm grid}$ for cubical
transforms, but the three dimensions may also be different),
is distributed as evenly as possible among
the MPI ranks, thereby greatly enlarging the scalability regime. There
are no restrictions on permissible numbers $N_{\rm grid}$ and $N_{\rm cpu}$,
i.e.~also odd choices are readily possible, such as  $N_{\rm grid}=8$ and
$N_{\rm cpu}=11$ shown in the sketch on the right (different color shades
refer to different MPI ranks).
\label{FigColumnFFT}}
\end{figure}

To extend the range of scalability of the PM calculations, we allow
both for a slab-based or alternatively for a column-based distribution
of the data cube, as sketched in Figure~\ref{FigColumnFFT}. We allow
for an arbitrary number of MPI ranks in relation to the grid size, and
the number of columns assigned to each CPU is made as even as possible
to guarantee an optimum memory balance. In particular, this means that
the set of columns assigned to a MPI process does not need to have a
rectangular footprint in the $N_{\rm grid} \times N_{\rm grid}$
grid. Note, however, that the column-based approach requires two
global transpose operations per FFT, even allowing for a different
data layout in the result, whereas the slab-based approach (which we
also support) needs only one. Because the data shuffling of these
transpose operations is often the dominating part of the distributed
memory FFT algorithm, it is more efficient to use the slab-based
approach in the regime where this still scales reasonably well. We
note that the final result of our FFT algorithms is binary floating
point invariant with respect to whether the slab-based or column-based
version is used, and to what number of MPI ranks is employed. The code
is also general enough to allow different grid dimensions in each
spatial dimension, so we can employ it for stretched boxes as well.

We support two different communication algorithms to map the particle
data to a density grid, and to bring back the forces to the
particles. One of them is designed to be ideal for homogeneously
sampled boxes. Here the particle coordinates and masses are sent to
the processor holding the target column/slab, and this CPU then does
the binning work. For a zoom calculation, this would become fairly
imbalanced as the vast majority of particles is concentrated in a tiny
volume in this case, which may fall largely onto just a single
processor's slab or column. If this MPI rank had to do all the binning
work, it would form a parallel bottleneck. To rectify this, {\small
  GADGET-4} offers an alternative approach that is optimized for this
scenerio. In it, each processor bins the particles to local grid cells
of the density field (just those that are actually touched), and then
only the resulting density values are sent to the right target
processor where they are assembled to the complete density field.

\subsection{Generic parallel tree walks}

There are a few places in {\small GADGET-4} where parallel tree walks
are needed that are not best addressed by building a locally essential
tree, but rather by exporting a particle to the MPI-rank holding a
required remote tree branch, which then carries out the remaining walk
there and reports the result back, where it can be accumulated. This
is for example the case in our FOF group finding algorithm to connect
up groups that stretch across domain boundaries, or in the spherical
overdensity method to measure virial masses of halos by determining
the enclosed mass in large spherical apertures. It can also be useful
for certain types of feedback algorithms that spread energy (or
metals) from an SPH particle to its neighbours.

All these tree walks can be realized with the same generic
communication pattern. First, the code starts on each MPI rank with
walking its active particles. This accounts for all local
interactions, but some tree branches cannot be opened. This fact is
noted in an export buffer, together also with the exact information at
which tree node the local walk could not be continued. Once all local
particles have been processed (or the export buffer is full), the
particle coordinates and foreign node addresses stored in the export
buffer are sent to the corresponding target processor, while the local
processor itself receives incoming requests of the same kind. After
this data exchange, which is normally carried out with a robust
hypercube pattern with pairwise synchronous data exchange, branches of
the local tree are walked to produce partial results, which are then
sent back to the originating process and added in to the local
results. If needed, the cycle is repeated until all active particles
are processed.

Compared to older versions of {\small GADGET}, the newly implemented
approach of this generic pattern in {\small GADGET-4} is more
efficient because it includes the index (or several indices) of the
starting node of the partial tree walk required on the foreign
processor, instead of finding this by starting again from the root
node. This means that the actual tree-walking cost is invariant when
the number of CPUs is changed. We also automatically use all available
memory for the export buffer in a robust way, without a need to
specify the size of this buffer explicitly. Furthermore, we have
encapsulated the communication routines in a C++ class that puts them
completely out of view in the core routines of the code.  This makes
it much easier to verify and modify these core routines.

\subsection{Vectorization}
\label{subsecVectorization}
Modern CPUs offer increasingly powerful vector instructions (such as
the SSE2, AVX, AVX2, AVX512 instruction sets introduced by Intel)
where several arithmetic operations can be executed in parallel
through a single instruction. Reaching a substantial fraction of the
peak performance of these CPUs requires extensive use of vector
instructions. One approach to employ them is to rely on the
`auto-vectorization' features of modern compilers, which try to
identify opportunities to emit vector instructions in code that is
formulated with a serial syntax.  Unfortunately, this does not work
very effectively in {\small GADGET-4}, due to its complicated and
irregular memory access and calculation patterns. Better results
require a rewrite of the code with vectorization in mind. One
effective way to ensure use of the vector instructions is to inline
low-level vector intrinsics (which are essentially machine
instructions expressed as small C functions or macros) directly in the
code.  In order to maintain better readability and portability of the
code, we prefer to encapsulate vector intrinsics through the C++ {\em
  vectorclass}
library\footnote{http://www.agner.org/optimize/\#vectorclass}. It
defines new vector-sized data types and uses meta- and template
programming techniques to allow the writing of fully vectorized code
using the syntax or ordinary code.  As an additional benefit, this
allows the compilation of vectorized code on platforms that, e.g.,
feature different vector lengths or only support SSE2, because the
meta-programming of this library can automatically replace long vector
instructions with multiple short ones if only those are available.

As a demonstrator and test of the potential of this approach we have
created an alternative version of the innermost SPH-loop for the
density and the hydrodynamic forces based on the vectorclass library,
targeting 256-bit wide AVX instructions where 4 double precision
operations can be done in parallel.  Our approach in this compute
kernel has been to always work on four neighbour particles in
parallel, applying exactly the same operations to each of them. One
complication in this is the evaluation of the standard cubic SPH
kernel, which involves branching due to its formulation in terms of
piece-wise polynomials.  We here compute both versions and then mix
them using vectorized comparison and select operations. This is not
necessary for the Wendland kernels, which is helpful for a more
efficient vectorization (see our timings in
Section~\ref{subsecShocktube}).  We follow a similar approach in the
calculation of the artificial viscosity, which we compute for every
particle and then use a select operation for deciding whether we
really want to apply this term.

Compared to the ordinary serial version of the same routine (which is
still available in the code and used as default) we have only reached
a very modest speed-up of order 5-10\% on a compute cluster with Intel
Xeon Gold 6138 CPUs (which are in principle capable of AVX512
instructions). This is certainly less than what one may have (perhaps
naively) hoped for. It remains to be seen whether this is due to
limitations in our particular approach that can be overcome with more
clever coding, or whether it rather reflects the principal difficulty
to get high floating point throughput with SIMD instructions in
situations where scattered data needs to be fetched from memory, and
only few floating point operations are done per data item read from
memory, because this is unfortunately the regime of our SPH
loops. Furthermore, there can also be subtle side-effects due to the use
of AVX instructions, such as an associated temporary reduction of the
clock frequency of the executing core directly after such instructions.

We note that other authors have also reported somewhat mixed results
when explicitly using SIMD instructions in N-body codes.
\citet{Kodama:2019aa} have used the AVX2 instructions (in assembly) to
accelerate the evaluation of tree interactions with quadrupole moments
in an extension of the Phantom-GRAPE \citep{Tanikawa:2012aa,
  Tanikawa:2013aa} library for calculating gravity in N-body systems,
finding speedups relative to the pseudo-particle method of the library
by factor of 1.1. For the use of AVX-512 they estimate a speed-up of
1.08. This appears similar to our results here. On the other hand,
\citet{Ishiyama:2012aa} describe a rather successful SIMD based
parallelization of the inner force loop of their TreePM code for the
K-supercomputer, and \cite{Yoshikawa:2018aa} report good performance
gains with AVX-512 for the Phantom-GRAPE library. Similarly,
\citet{Potter:2017aa} reported substantial gains in {\small PKDGRAV-3}
due to explicit optimizations for SIMD instructions.

\section{On the fly analysis and other features} \label{secprocessing}

\subsection{Parallel group and  subhalo finder}

{\small GADGET-4} has a built-in Friends-of-Friends ({\small FOF})
group finder \citep{Davis:1985aa}, an implementation of the {\small
  SUBFIND} algorithm \citep{Springel:2001ac} for finding
gravitationally bound substructures in positional space in a single
timeslice, and a new variant, {\small SUBFIND-HBT}, that takes past
information about group membership into account. All three
implementations are capable of processing very large simulations as
well as high-resolution zoom calculations with extremely large
particle numbers in single halos. The outputs created by {\small
  SUBFIND} or {\small SUBFIND-HBT} can be created either on-the-fly
while a simulation is running, or in postprocessing, and they can be
combined into merger tree construction if desired.

The {\small FOF} algorithm uses link-lists to organize the halos,
starting out with each particle making up a group of its own.  The
code then finds all particle pairs with a distance smaller than a
prescribed linking length $l$. In cosmological simulations, the most
common choice for $l$ is to make it equal to $0.2$ times the mean
particle spacing, leading to halos that are bound by a density
contour that approximates the overdensity expected for virialized
structures according to the top-hat collapse model. Each pair that is
found with a sufficiently small distance causes the groups the two
particles belong to to be linked into a joint group.

\begin{figure}
\begin{center}
\resizebox{8cm}{!}{\includegraphics{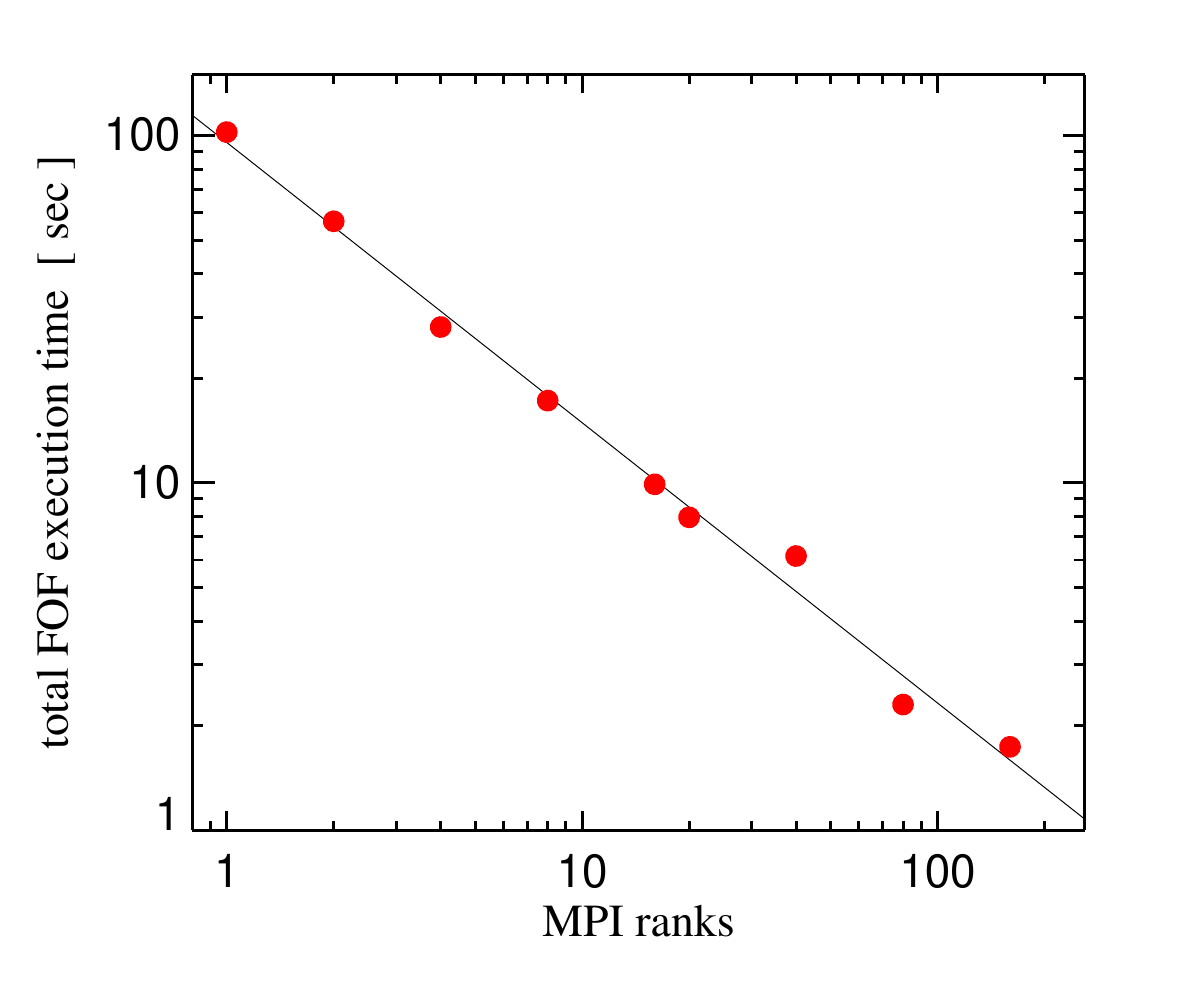}}\\
  \caption{Raw speed and parallel scalability of the inlined MPI-parallel FOF group
    finder in {\small GADGET-4}.  We show measurements for different
    numbers of MPI ranks when the algorithm is applied to the $z=0$
    output of one of the lower
    resolution DM-only
    variants of the TNG50 simulation model of the IllustrisTNG
    project.
    Specifically, we here used
    the $270^3$ run with close to 20 million particles. The largest
    found group has about 840000 particles, and there are about 18800
    groups with at least 32 particles. We have checked
    that the outcome of the group finder
    is invariant when the number of MPI ranks is changed, as required. The
    execution time closely follows strong scaling (solid line) over
    the measured range,   implying a raw processing speed of around $2 \times 10^5$ particles
    per second and core. 
    \label{FOFSpeed}}
\end{center}
\end{figure}

For finding neighbouring particles within a distance $l$, the code
uses a range searching technique based on an oct-tree, similar to the
neighbour finding in SPH. Significant speed-ups are realized by
flagging tree nodes which contain only particles already in the same
group, or which are small enough that all particles in the node are
guaranteed to be in the same node. If a target particle encounters
such a node in its neighbour search and sees at least one of the
node's particles as a neighbour, the other particles in the node need
not be checked any further. Compared to reported speeds of other
{\small FOF} algorithms in the literature, {\small GADGET-4}'s
implementation appears to be quite fast.  For example,
\citet{Creasey:2018aa} have argued that tree-based {\small FOF} group
finders would be slow, suggesting instead an algorithm based on
spatial hashing. Our results do not support this assertion, as our
algorithm appears to be at least as fast as the one described by
\citet{Creasey:2018aa}. Also, our method works well in parallel on
distributed-memory machines \citep[see also][]{Roy:2014aa}, and is
efficient for the demanding regime of extreme zoom calculations.  In
Figure~\ref{FOFSpeed} we show speed and scaling measurements
supporting these findings. We note that the speed of our parallel
algorithm for finding FOF groups is also substantially faster than the
{\small Parallel-HOP} method of \citet{Skory:2010aa}

\subsection{Finding subhalos with {\small SUBFIND}}

When one is interested in dark matter substructures or the
gravitationally bound part of halos, the {\small FOF} group finder can
be combined with the {\small SUBFIND} algorithm. The latter decomposes
each found halo into a set of disjoint, gravitationally bound
substructures. They are identified based on an excursion set algorithm
as described in \citet{Springel:2001ac}. First, the total mass density
field is found by adaptive kernel estimation, allowing the
identification of isolated density peaks. These density peaks are
grown with an excursion set algorithm (similar to a watershed
approach) until saddle points are encountered that connect two
isolated density peaks. Typically the smaller peak is then registered
as a substructure candidate, and is subsequently subjected to a
gravitational unbinding procedure. As one improvement compared to
previous versions of {\small SUBFIND}, the selection of which of the
two peaks is the substructure and which is background can now
optionally take into account information about previous subhalo
membership of the particles making up the peaks.  In the on-the-fly
processing, this is done by letting each particle carry a cumulative
sum of the subhalo lengths of all its previous subhalo memberships,
while in the postprocessing mode of {\small SUBFIND}, the same
information is gathered by processing the snapshots in their temporal
order. The peak having the larger sum of all these previous sizes is
then identified as the background peak, not necessarily the one with
the larger particle number (typically in less than ~1\% of the
saddle-point decisions this picks the smaller of the two peaks). A
similar approach has recently also been developed by
\citet{Angulo:2020aa}. This procedure substantially reduces temporal
variations of the bound mass assigned by {\small SUBFIND} to subhalos,
which often originate in mistakingly swapping the identity of
background and substructure peaks at the saddle points. As a further
side-effect, it also reduces the occurrence of swaps between central
and satellite systems in the tracking of subhalos in merger trees.

The calculation of the gravitational potential in the unbinding
procedure is carried out with a tree algorithm using the same
multipole order and accuracy as in the ordinary force calculation of
the simulation, without taking any short-cuts or additional
approximations. The unbinding procedure is iteratively repeated to
account for the changes in the potenial due to removal of unbound
particles, and only ends when all remaining particles are bound to
each other, or their number has fallen below a prescribed detection
threshold for a group (typically chosen as something in the range
$\sim$20-64 particles).  The gravitational unbinding is also applied
to the remaining ``background'' structure, defined as the particles
left in the {\small FOF} group after all bound substructures have been
extracted. Thus, every {\small FOF} halo is effectively segmented into
a set of gravitationally bound subhalos, and possibly some particles
not bound to any of the subhalos. The most massive of these subhalos
corresponds to the `background halo', also called the central subhalo,
while the others are referred to as satellites. Note that sometimes
the two most massive subhalos can emerge with comparable mass from the
decomposition, for example in an ongoing major merger, or if the
{\small FOF} algorithm has linked two nearby halos via a feeble
particle bridge, making the distinction between central and satellite
somewhat ambiguous.

In terms of parallelization, both {\small FOF} and {\small SUBFIND}
can cope with high-resolution zoom simulations that contain groups
much larger than are able to fit on individual MPI ranks. After the
{\small FOF} group finder has been run, the groups are ordered by
their size. If the largest group is too large to be processed by a
single MPI process, it gets assigned several MPI ranks, as many as
needed to fit it into their combined memory, likewise for the second
group, the third group, and so on. Each such set of processes gets its
own MPI communicator and works in parallel on the one halo assigned to
it. Halos that are small enough to fit on individual processors are
assigned in a round-robin fashion to the remaining MPI ranks, so that
these MPI processes each work on several smaller halos.

If the group finding algorithms {\small FOF} and {\small SUBFIND} are
used in an on-the-fly processing mode while a simulation runs, they
are always executed directly {\em before} a snapshot is written to
disk. This allows a convenient storage format of the timeslice data,
which we output in the order of the found groups and subhalos. Any
group can hence be loaded through random access to the data files, and
without the need to load more data than needed. In addition, this
eliminates the need to separately list membership particles in groups,
for example through their IDs, and therefore reduces the required
storage volume. In case {\small SUBFIND} is enabled, the particles are
additionally stored within each group in a nested fashion ordered by
decreasing subhalo size, and within each subhalo in the order of their
binding energy. We have introduced this type of storage format first
for the Illustris simulation, and refer to its public data release
\citep{Nelson:2015aa} for further documentation.  We note that if HDF5
storage of data is used (which is highly recommended), group
catalogues, like all other {\small GADGET-4} output, are stored in
HDF5 as well.

\begin{figure*}
\begin{center}
\resizebox{14cm}{!}{\includegraphics{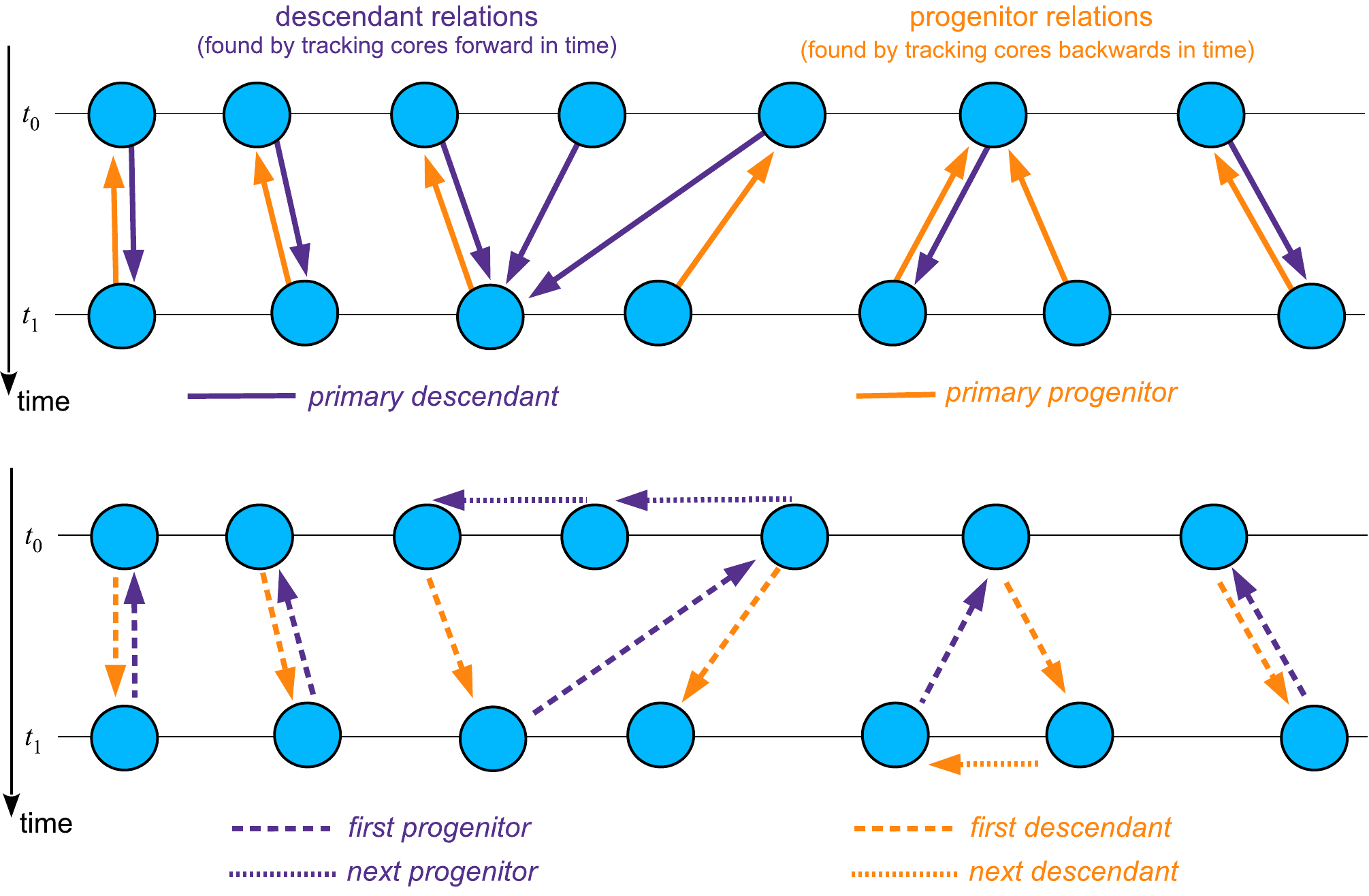}}
\end{center}
\caption{Descendant and progenitor relations at the stage
  of the linking of two subhalo catalogues. The diagram on top
  illustrates the fundamental tracking operation. For each subhalo, we
  identify a most probable {\em primary descendant} by tracking where its
  most bound particles end up. Additionally, we track the cores of
  subhalos back in time, leading to the identification of a {\em
    primary progenitor}. This is illustrated in the top half of the sketch.
In order to be able to identify all subhalos that have a given subhalo
as their primary descendant, we define an auxiliary chaining list,
with a {\em first progenitor} identifying the first of these, and {\em
  next progenitor} links any subsequent ones, if any. Likewise, we
introduce {\em first descendant} and {\em next descendant} links to be
able to efficiently enumerate all subhalos that have a common {\em
  primary progenitor}. These induced links are illustrated in the
bottom half of the sketch. 
  \label{FigMergerTreeLinking}}
\end{figure*}

\begin{figure*}
\begin{center}
\resizebox{16cm}{!}{\includegraphics{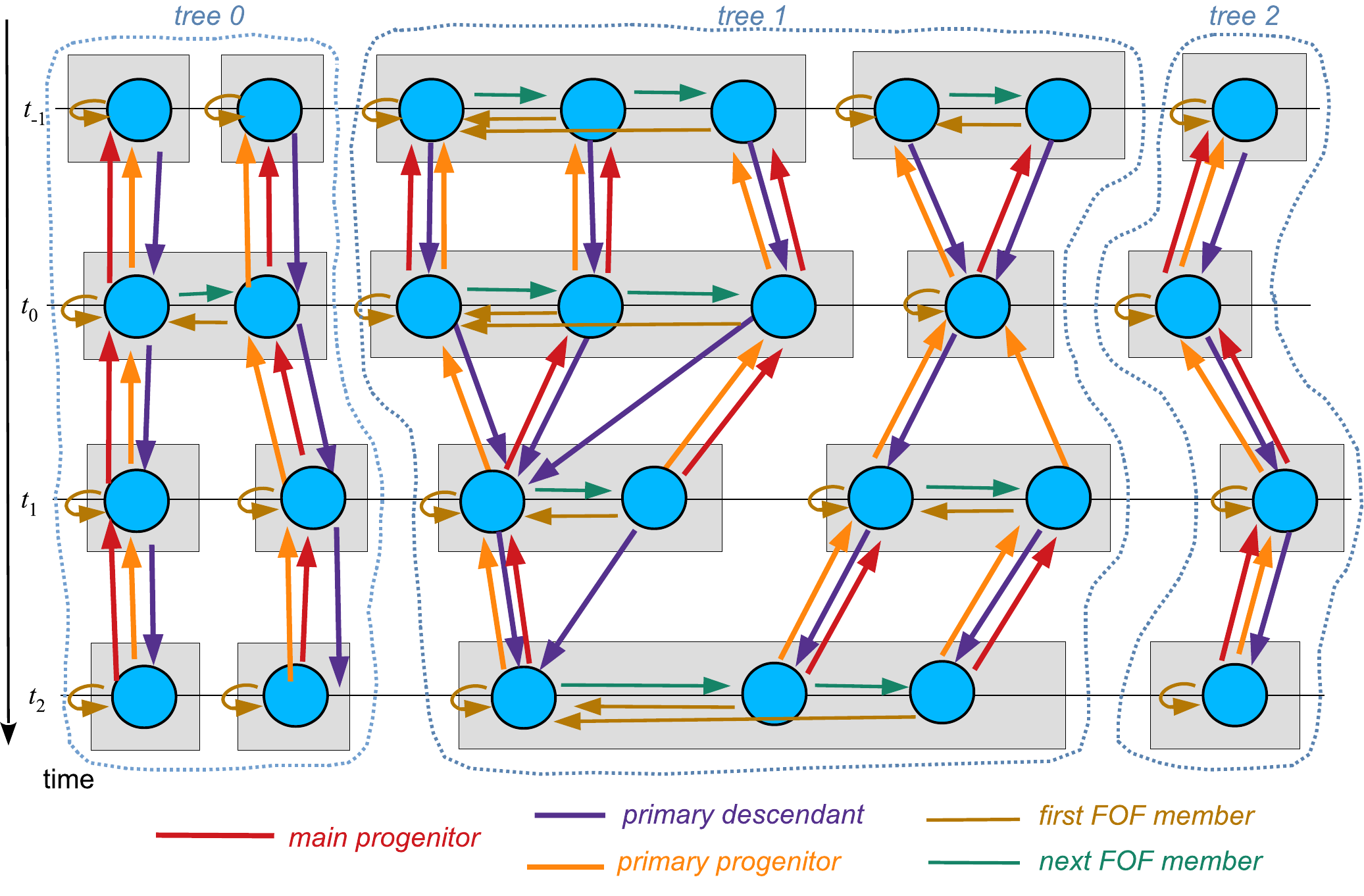}}
\end{center}
\caption{Sketch of the merger tree structure produced by
{\small GADGET-4}. The fundamental building blocks are the {\em
  primary descendant} and {\em primary progenitor} links that are formed by tracking
subhalo cores forward and backwards in time. These links also induce
enumerations of all subhalos that are descendants of a given subhalo,
which can be accessed by the {\em first progenitor} and {\em next
  progenitor} links shown in Fig.~\ref{FigMergerTreeLinking}. For
clarity, these links are omitted in the present sketch, but they are contained in the tree
structure produced by the code. Likewise, there are {\em first descendant} and {\em next
  descendant} pointers that provide access to all subhalos that have a
given subhalo as their {\em primary progenitor}. The tree structure
also contains a {\em main progenitor} pointer, which selects the subhalo
on the most massive history branch relative to the current
subhalo. The chain of links established by {\em first FOF member} and
{\em next FOF member} allows that each subhalo can easily enumerate all
subhalos within the same FOF group (schematically indicated by the
grey boxes in the sketch). For convenience, the code subdivides all
subhalos into disjoint sets of {\em trees}. All pairs of subhalos that are linked via a
progenitor or descendent relation, or are in the same FOF group, are
guaranteed to be found in the same tree. Note that this means that a
given tree can account for more than one FOF group at the final time,
as seen for example in the leftmost tree 0, where two FOF halos are
temporarily joined at time $t_0$.
\label{FigMergerTree}}
\end{figure*}

\subsection{Tracing  bound structures with {\small SUBFIND-HBT}}

The identification of gravitationally bound substructures in
configuration space based on a single time-slice, which {\small
  SUBFIND} introduced in a fully adaptive way, is computationally
demanding and can have problems in robustly identifying all matter
belonging to a substructure, especially close to pericenter of a
subhalo orbiting in a larger halo, as here only the densest part of
the substructure sticks out over the background density field
\citep{Knebe:2011aa, Muldrew:2011aa}. One approach to find this
material is to look at the clustering of particles in phase-space,
which prompted the development of group and subhalo finders working
with the phase-space distribution. Recent examples include {\small
  HST} \citep{Maciejewski:2009ab, Maciejewski:2009aa}, {\small
  ROCKSTAR} \citep{Behroozi:2013ab} and {\small VELOCIRAPTOR}
\citep{Elahi:2011aa,Elahi:2019ab}.  Finding a suitable metric for
distance in phase-space (to allow a 6D-FOF algorithm) is one of the
central requirements for these approaches. Sometimes these methods
neglect a gravitational unbinding step, making them particularly fast
(akin to the FOF algorithm), but arguably at the price of a less
securely defined physical nature of the identified structures. Even
phase space finders are however not free of difficulties from
ambiguities in the descendent identification, especially during
mergers and deep encounters, causing problems in merger trees through
missing or incorrect links \citep[e.g.][]{Srisawat:2013aa,
  Avila:2014aa}.

An alternative is to rely on past information for subhalo
identification. If one has somehow identified a set of particles that
are gravitationally bound to each other at a given time and fall as a
substructure into a bigger system, one can simply continuously check
for boundedness of these particles. This assumes that a genuine
substructure is only stripped of particles with time, but does not
grow in mass any further in any appreciable way while on orbit in a
bigger system. This is a fairly realistic approximation for the
dynamics of hierarchically growing cold dark matter structures. A
first realization of this idea goes back to the seminal work of
\citet{Tormen:1998aa}. Recently, it has been significantly refined in
the hierarchically bound tracing (HBT) algorithm of
\citet{Han:2012aa,Han:2018aa}.

We have implemented in {\small GADGET-4} a variant of the HBT+
algorithm of \citet{Han:2018aa}. Because it can in practice act as a
drop-in replacement for {\small SUBFIND} with identical output
structure and merger tree format, we call this approach {\small
  SUBFIND-HBT}. It only works for a sequence of simulation snapshots
that need to be processed in temporal order, i.e.~to identify the
substructures at any given time, our algorithm uses the known group
and subhalo catalogue of the previous output time (this is either
available when the algorithm is run on-the-fly, or loaded in
postprocessing). The method first finds new {\small FOF} groups at the
current time. Next, substructure candidates are identified within each
{\small FOF} group based on the previous substructure membership of
all its constituent particles. The largest substructure candidate
found in this way is then discarded for the moment, because our
approach identifies it with the background halo of the group, which
also may grow in mass. In contrast, the other substructure candidates
are subjected to a gravitational unbinding procedure, using their new
phase-space coordinates. By construction, this can at most recover the
mass the substructure had in the previous output, while in general,
the remaining bound mass may reduce. Finally, all particles of the
group not ending up still bound to any of these subhalo candidates
will finally be subjected to a further gravitational unbinding
procedure. Note that this particle set includes the particles that
were in the biggest substructure candidate which was initially
dropped. If the FOF group has grown in size, this gives the background
subhalo a corresponding opportunity to grow in mass.

Compared to the ordinary {\small SUBFIND} algorithm, the approach
followed by {\small SUBFIND-HBT} is evidently considerably simpler
and computationally cheaper. This is because it does not have to
compute an adaptively smoothed density field, and especially, these
densities do not have to be processed in an inherently serial approach
from high to low density to find all saddle points. Furthermore, there
are no spurious subhalos that are only eliminated by the comparatively
costly gravitational unbinding procedure.  The main advantage of
{\small SUBFIND-HBT} is however that it is capable to recover the full
mass of substructures also close to pericenter on their orbit. This
yields a more robust tracking of substructures, and an overall higher
quality of the corresponding merger trees. The superior properties of
the {\small HBT} algorithm have also been confirmed by
\citet{Behroozi:2015aa} and also show up in the results of
\citet{Elahi:2019ab}. Despite the conceptional simplicity of the
approach, the tracking delivered by HBT is highly robust, largely
making it superfluous to fix glitches in merger trees such as those
resulting from the occasional tracking problems of essentially all
other substructure finding methods. We therefore expect that merger
trees constructed with {\small SUBFIND-HBT} are a highly competitive
alternative to other algorithms.

\begin{figure*}
\begin{center}
  \resizebox{8.0cm}{!}{\includegraphics{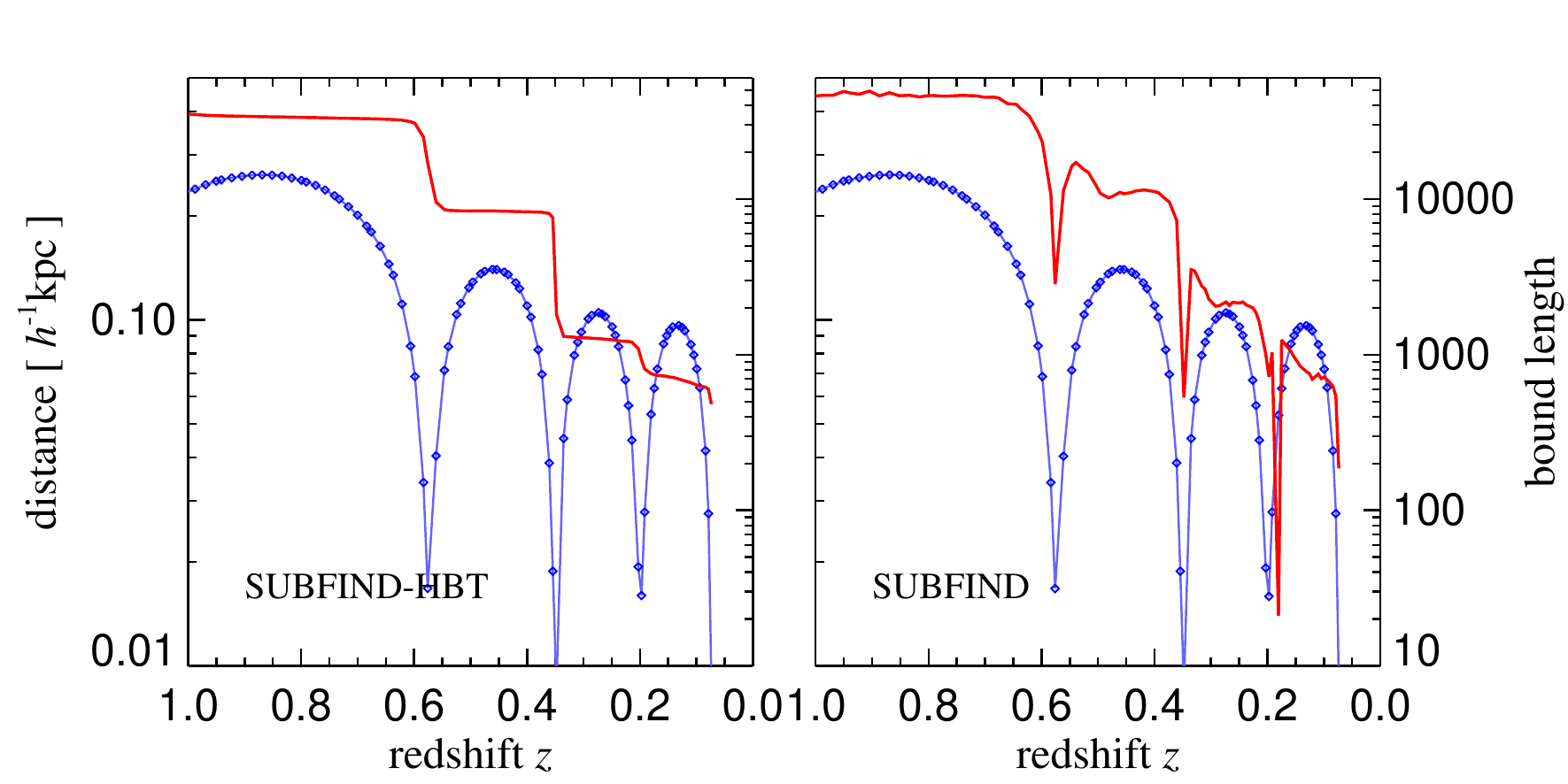}}\hspace*{0.8cm}%
  \resizebox{8.0cm}{!}{\includegraphics{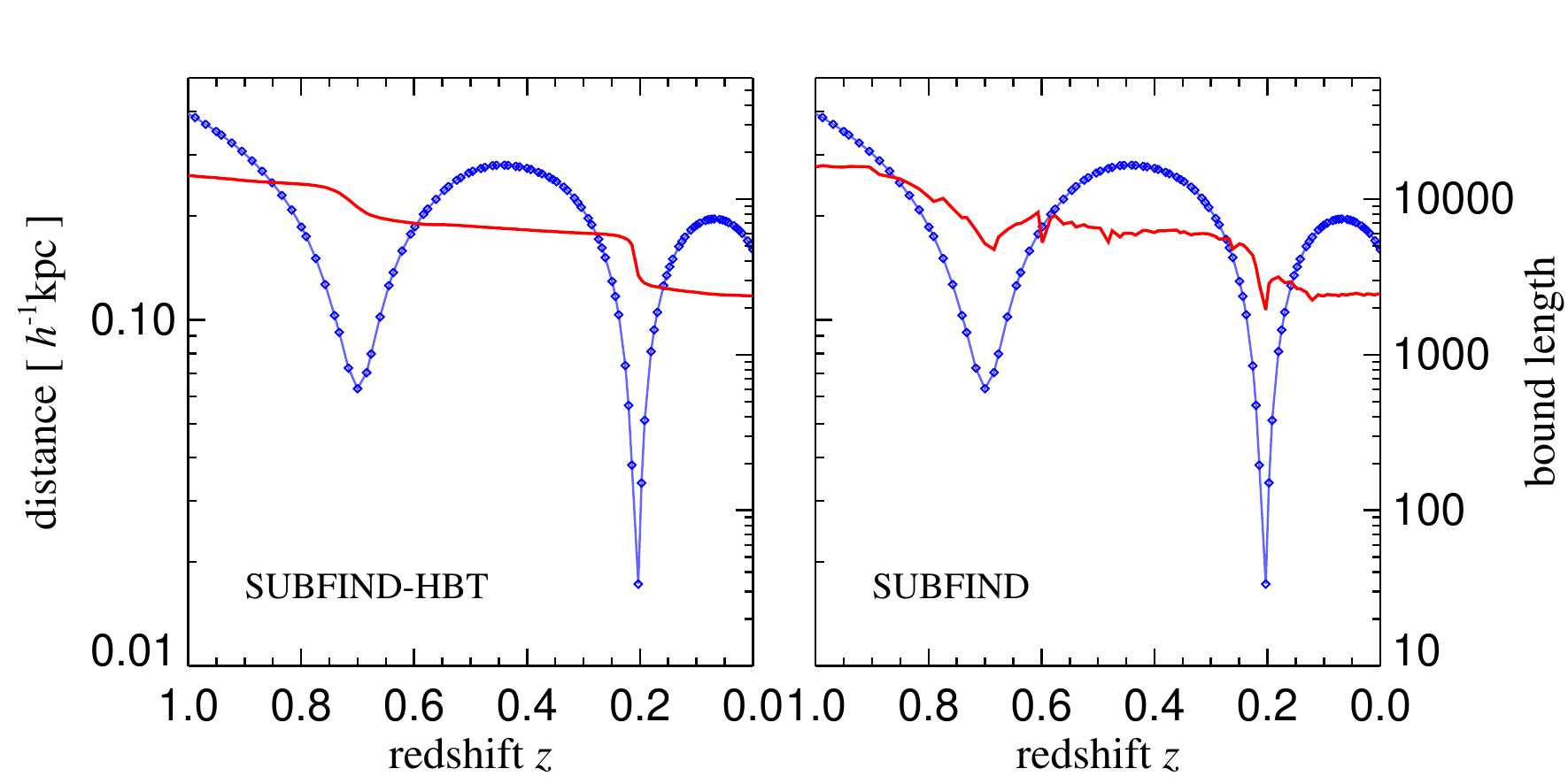}}\\%
  \resizebox{8.0cm}{!}{\includegraphics{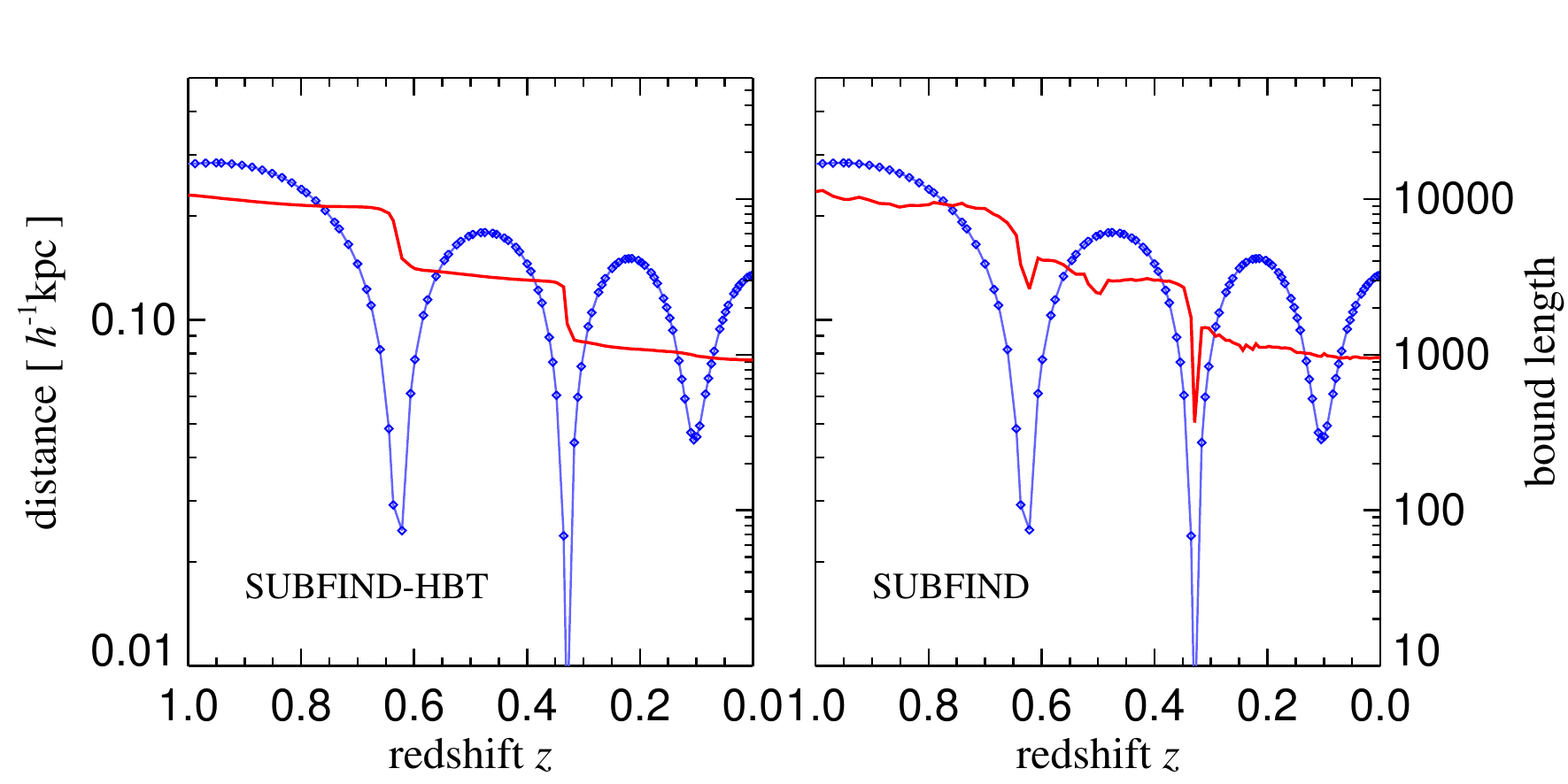}}\hspace*{0.8cm}%
  \resizebox{8.0cm}{!}{\includegraphics{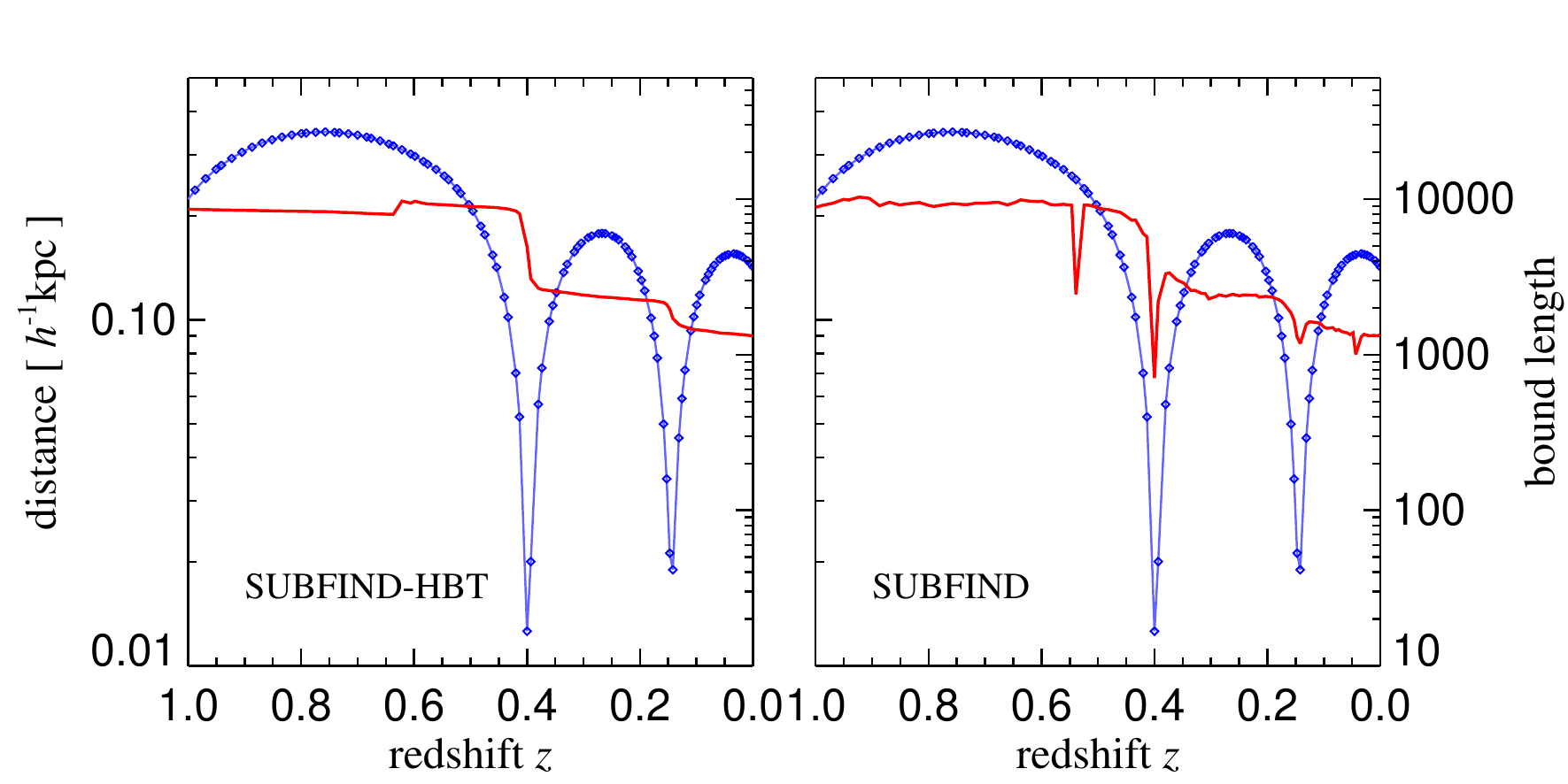}}\\%
  \resizebox{8.0cm}{!}{\includegraphics{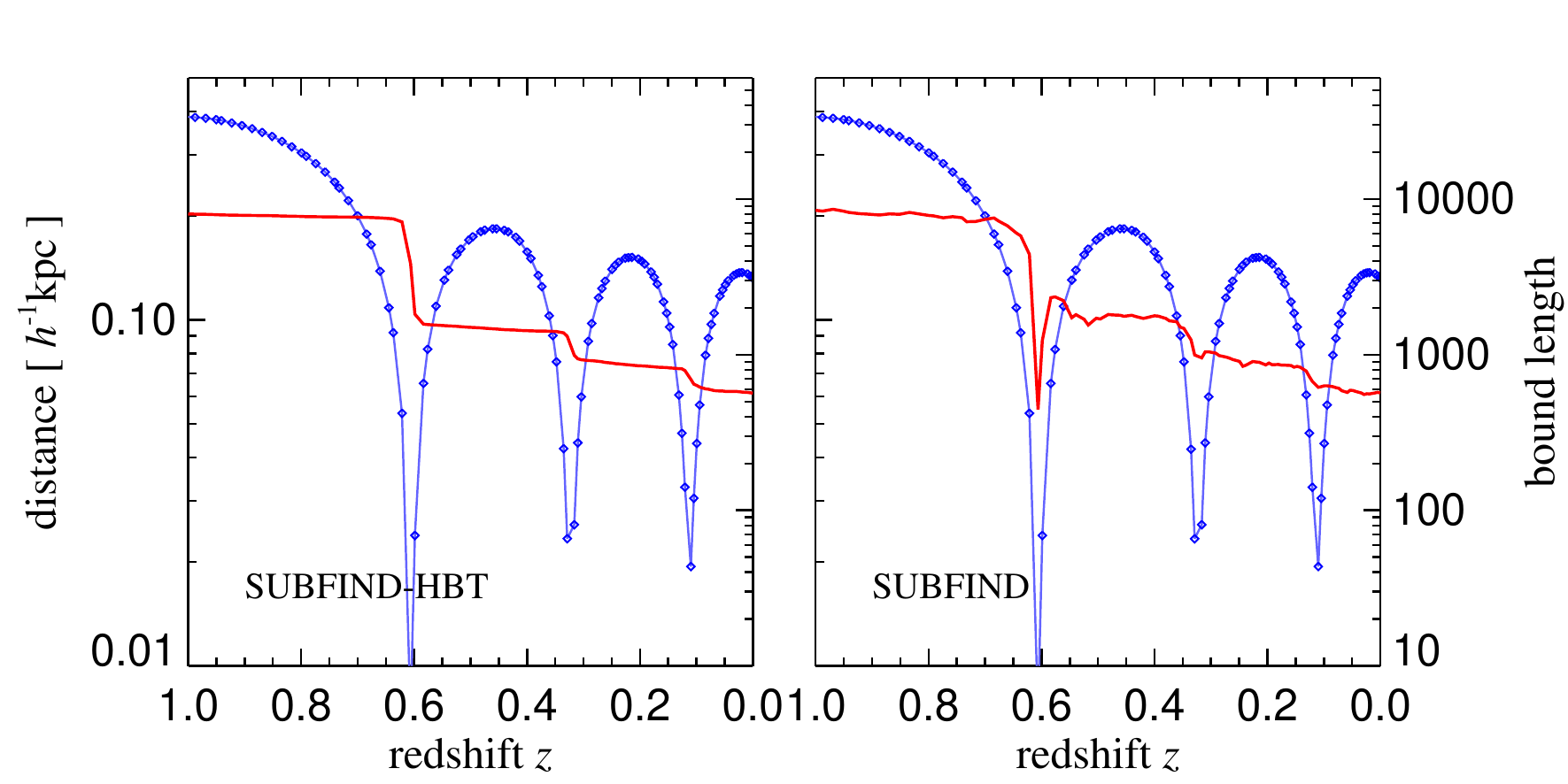}}\hspace*{0.8cm}%
  \resizebox{8.0cm}{!}{\includegraphics{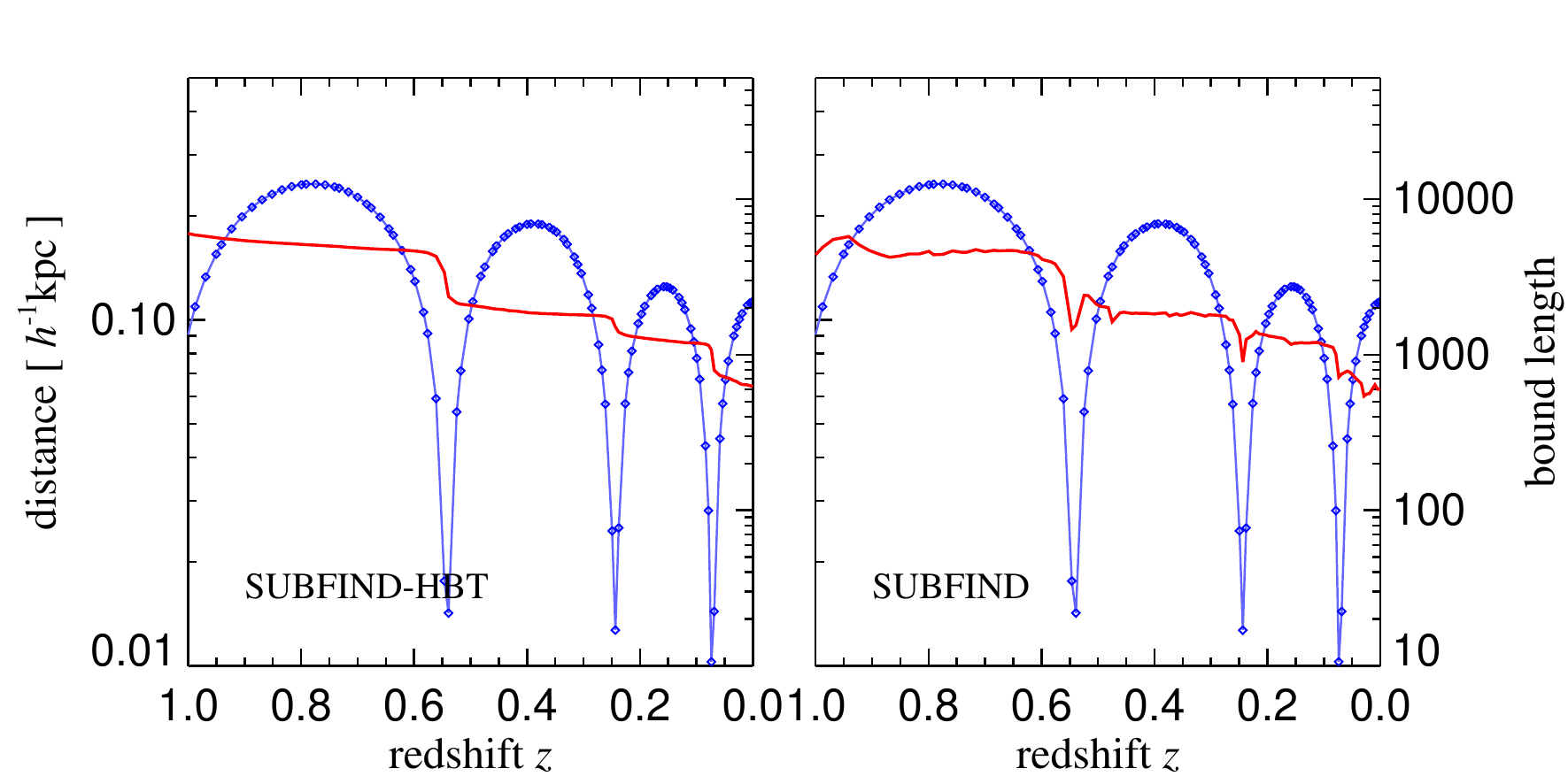}}\\%
  \resizebox{8.0cm}{!}{\includegraphics{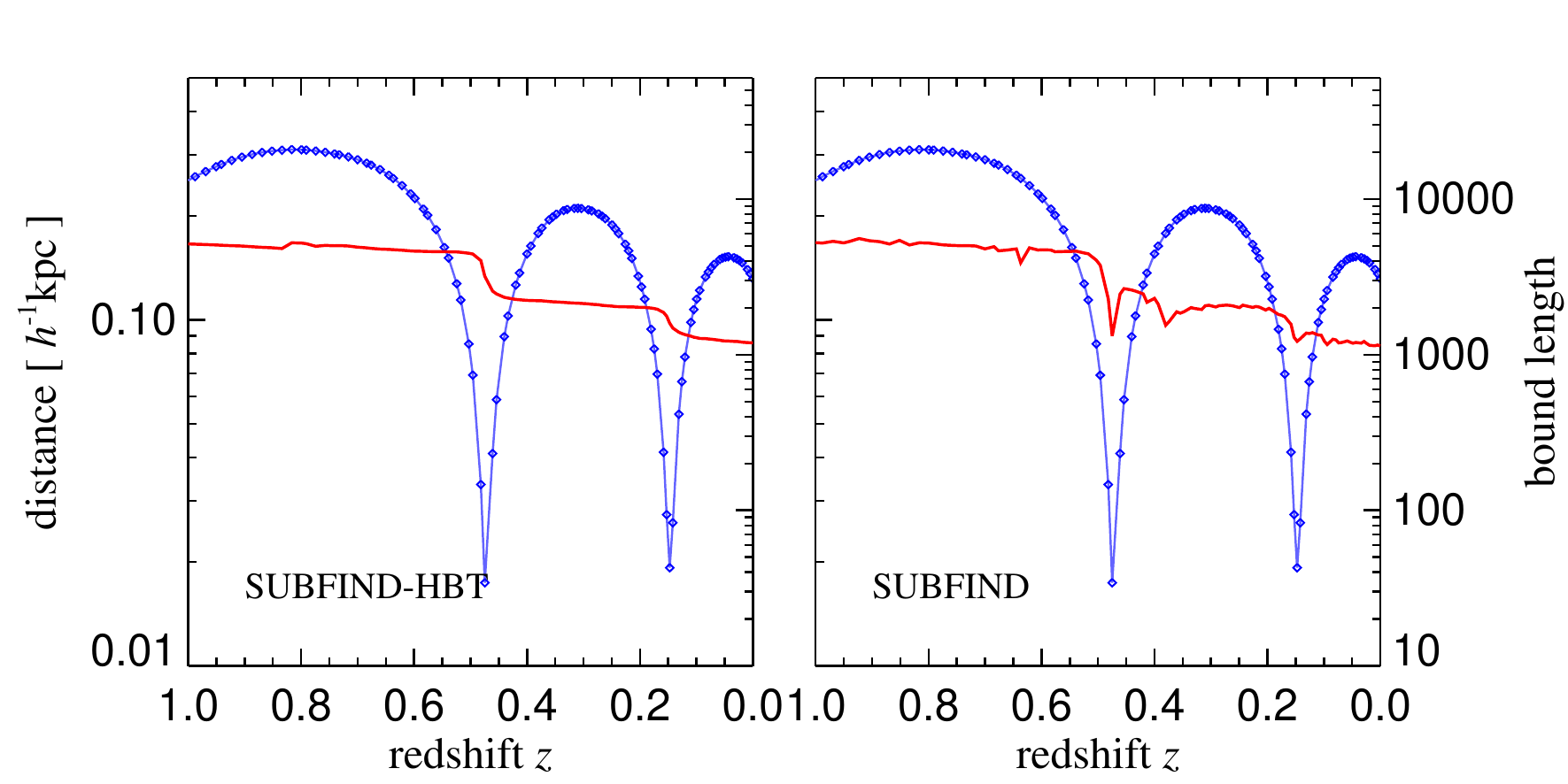}}\hspace*{0.8cm}%
  \resizebox{8.0cm}{!}{\includegraphics{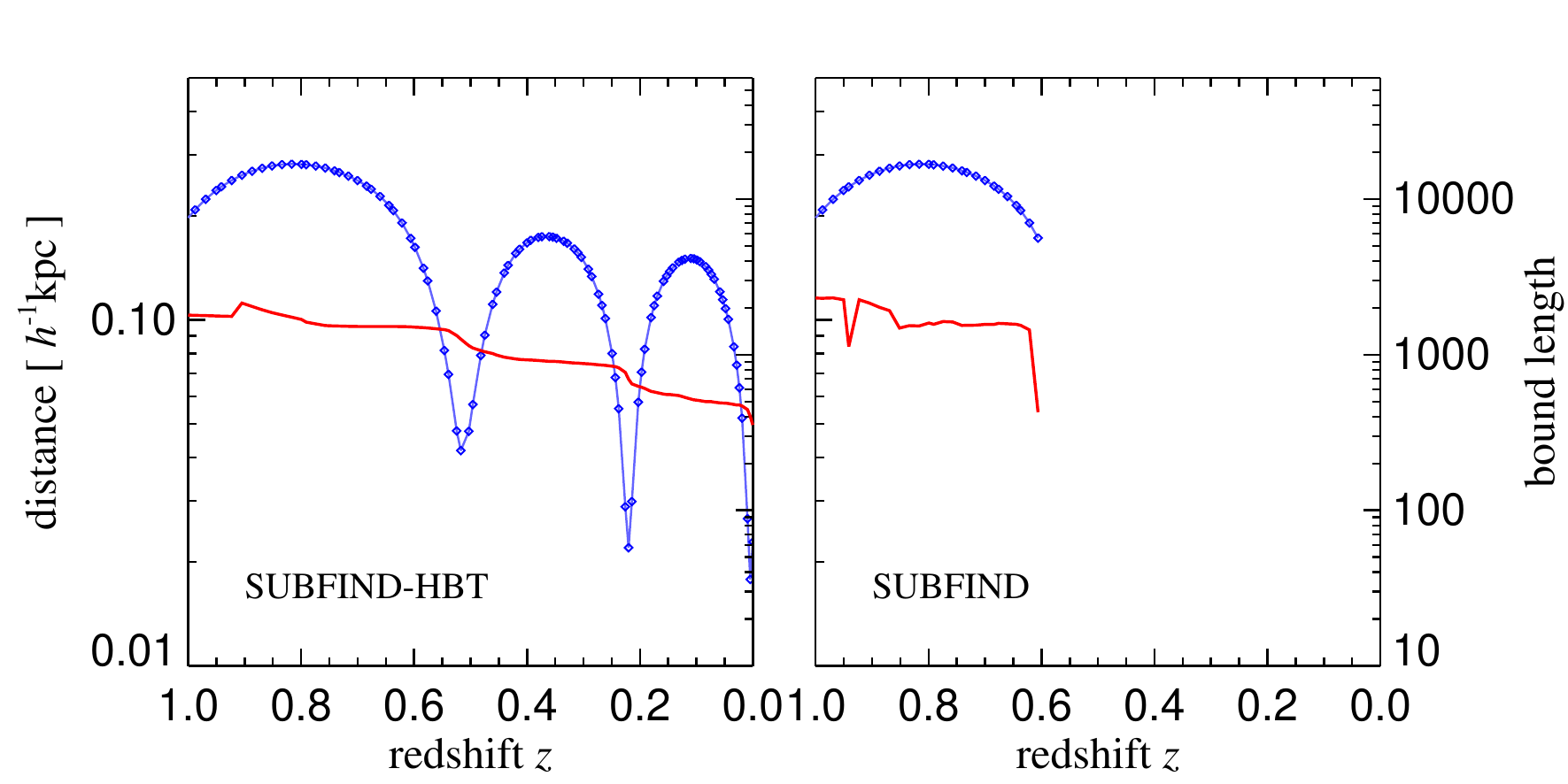}}\\%
  \resizebox{8.0cm}{!}{\includegraphics{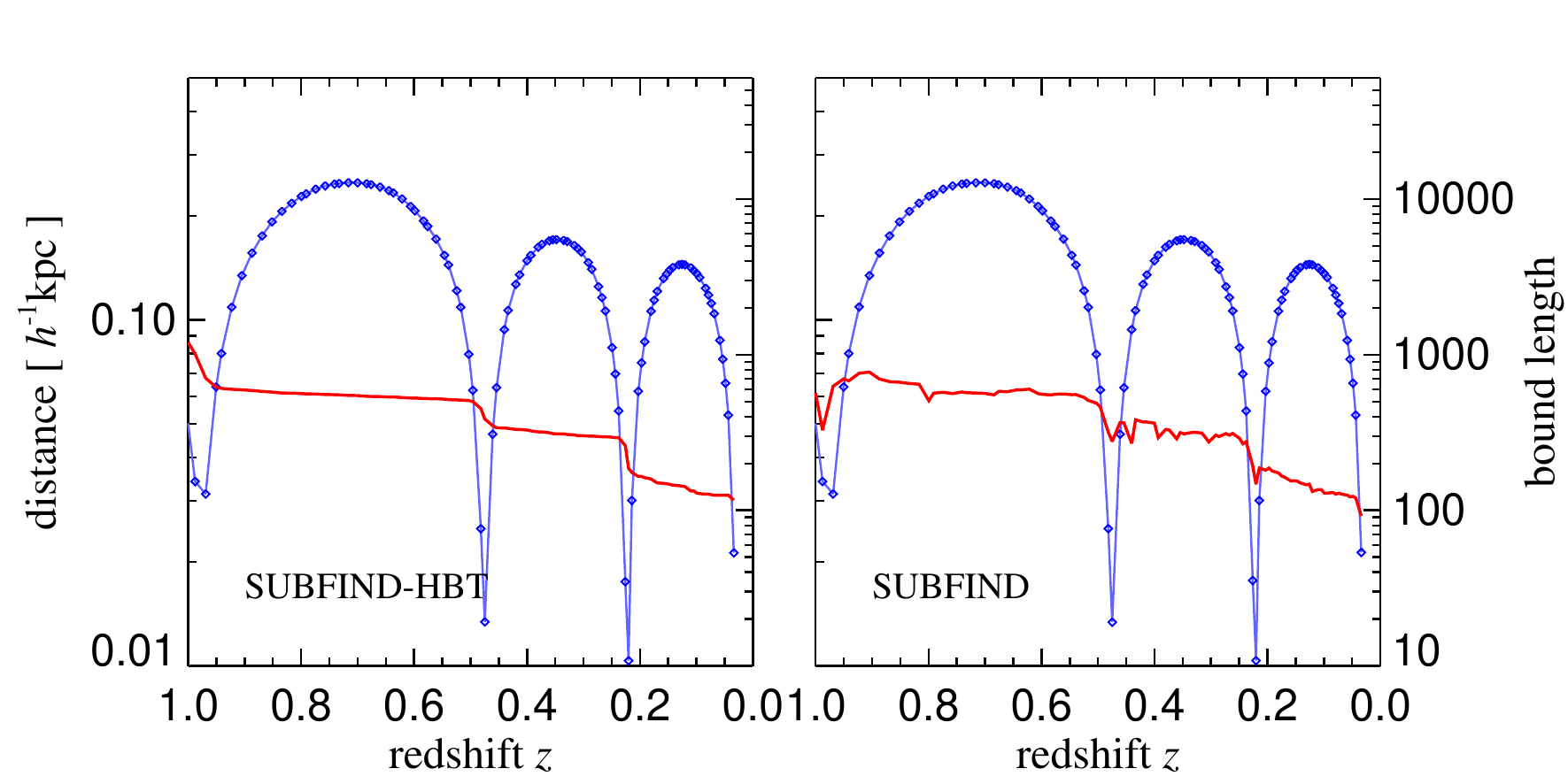}}\hspace*{0.8cm}%
  \resizebox{8.0cm}{!}{\includegraphics{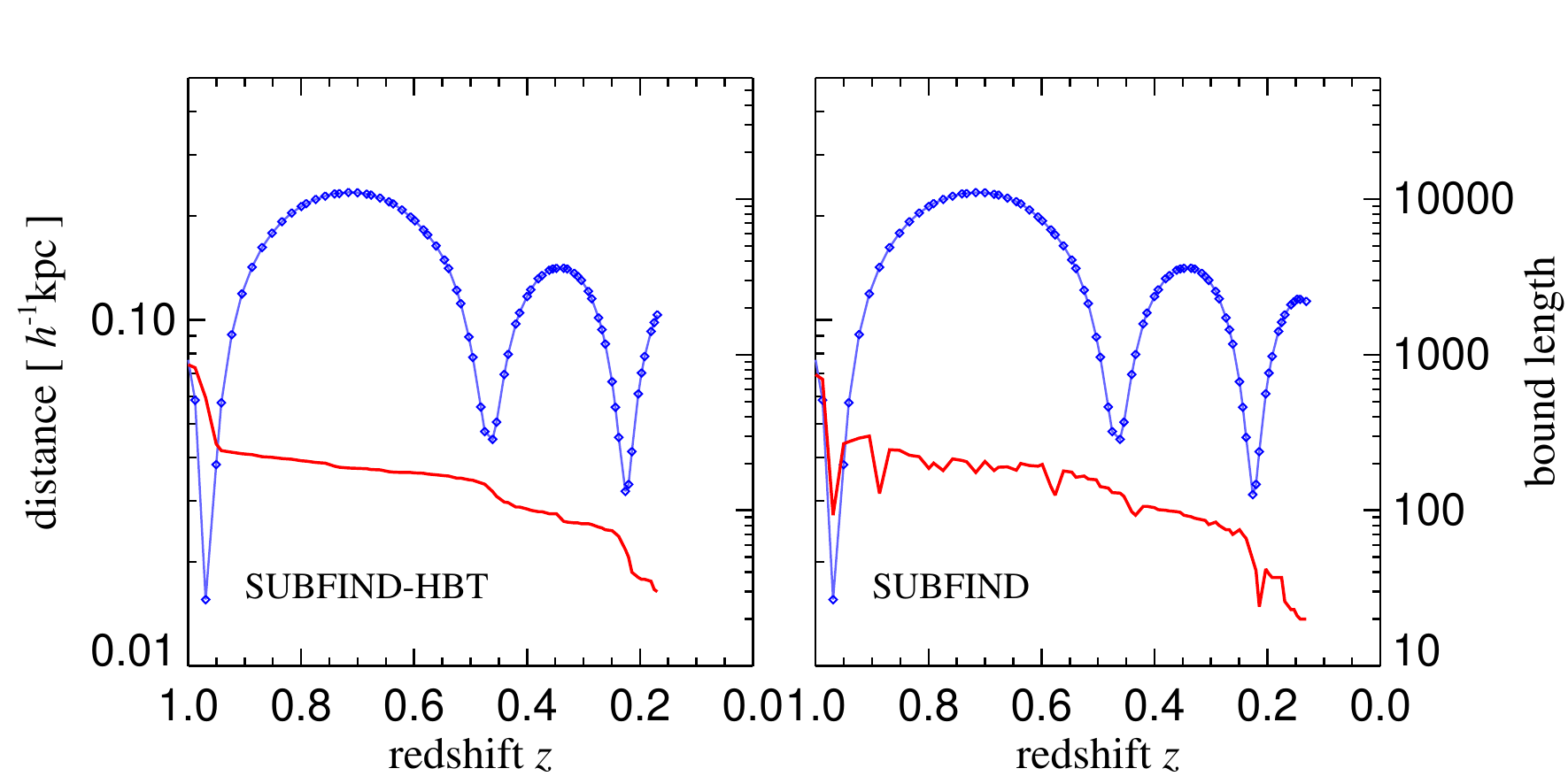}}\\%
\end{center}
\caption{Comparison of the {\small SUBFIND} and {\small SUBFIND-HBT}
  approaches when applied to subhalo tracking  
  forward in time, for a cosmological zoom-in simulation of
  a Milky Way-sized dark matter halo. For definiteness, we consider a
  rerun of the Aq-A-4 halo of the Aquarius project
  \citep{Springel:2008aa}, and identify halos and subhalos twice,
  once with the traditional {\small SUBFIND} algorithm
\citep{Springel:2001ac}
  and once with {\small SUBFIND-HBT}, which is
our implementation of the approach of
\citet{Han:2018aa}.
We select a set of 10 subhalos at $z=1$ and track their fate forward
in time based on the merger tree algorithm built into {\small
  GADGET-4}. In each pair of panels, the same subhalo is examined,
in the left it is found with {\small SUBFIND-HBT}, and on the right with
{\small SUBFIND}. Blue lines give the distance to the
halo centre (symbols mark the discrete output times),
while the red lines show the gravitationally bound
particle number retrieved by the substructure finders. One can clearly
see temporary depressions of the substructure size reported by {\small
  SUBFIND} around pericenter passages, and in some cases the substructure tracking is lost earlier in
{\small SUBFIND} than with  {\small SUBFIND-HBT}. Overall, both
approaches
make reassuringly similar predictions for the mass
loss experienced by the orbiting subhalos. 
\label{FigTrackingAquarius}}
\end{figure*}

\begin{figure}
\begin{center}
  \resizebox{8.5cm}{!}{\includegraphics{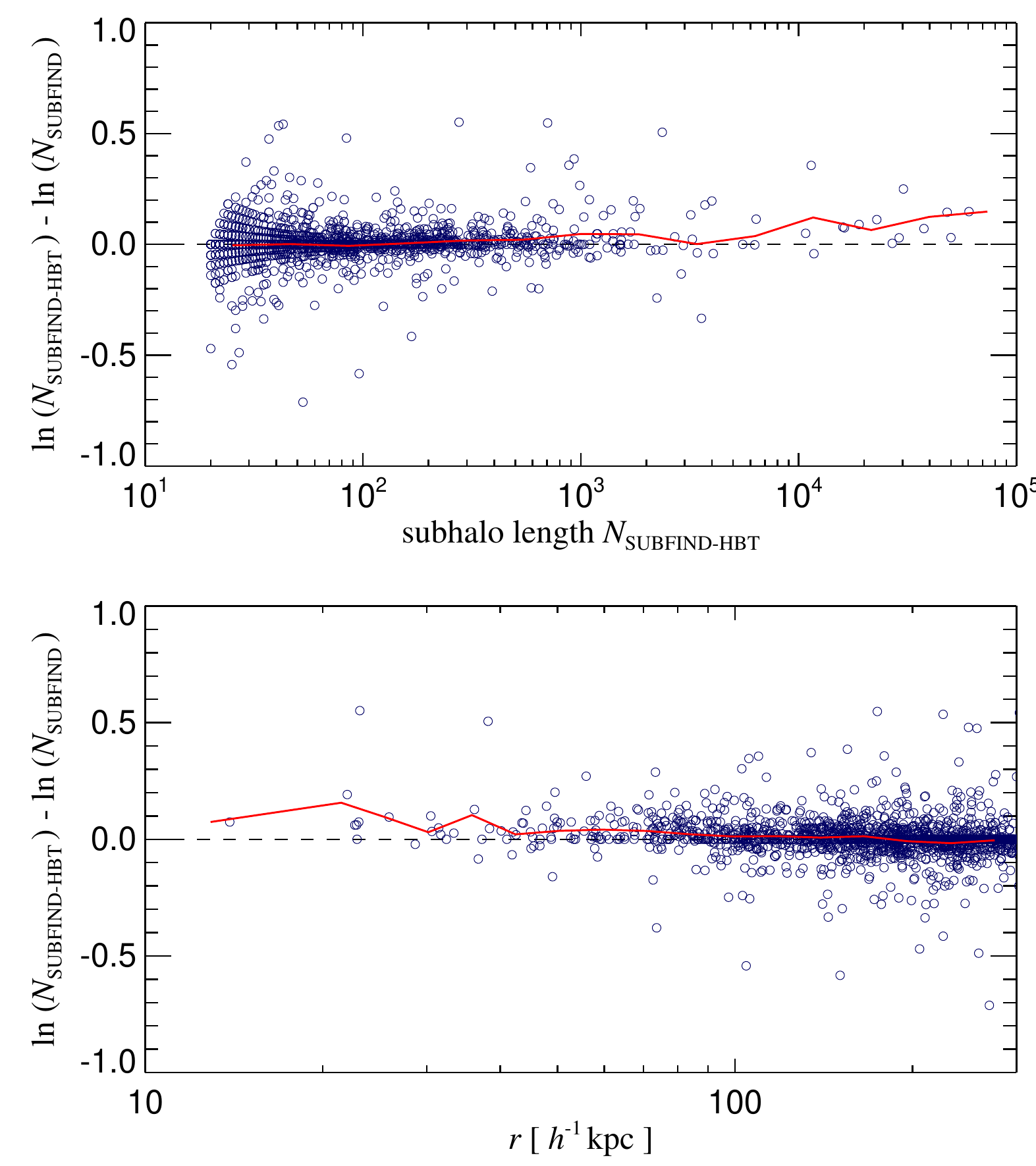}}
\end{center}
\caption{Comparison of subhalo sizes detected with the {\small SUBFIND-HBT} and
  {\small SUBFIND} algorithms, in a cosmological simulation of a Milky
  Way-sized halo at $z=0$ (a rerun of the Aq-A-4 simulation). The top
  panel shows the difference in the sizes of matching subhalos (symbols), as a
  function of their size, with an overplotted running median (solid
  red line). The bottom panel plots the same quantity, but now as a
  function of distance to the halo centre. On average, the sizes agree
  rather well, but there is a weak trend towards slightly larger sizes
  found with {\small SUBFIND-HBT} in the inner parts of the halo, and
  for the largest subhalos.
\label{FigSubhaloMassComparisonAquarius}}
\end{figure}

\begin{figure}
\begin{center}
  \resizebox{8.5cm}{!}{\includegraphics{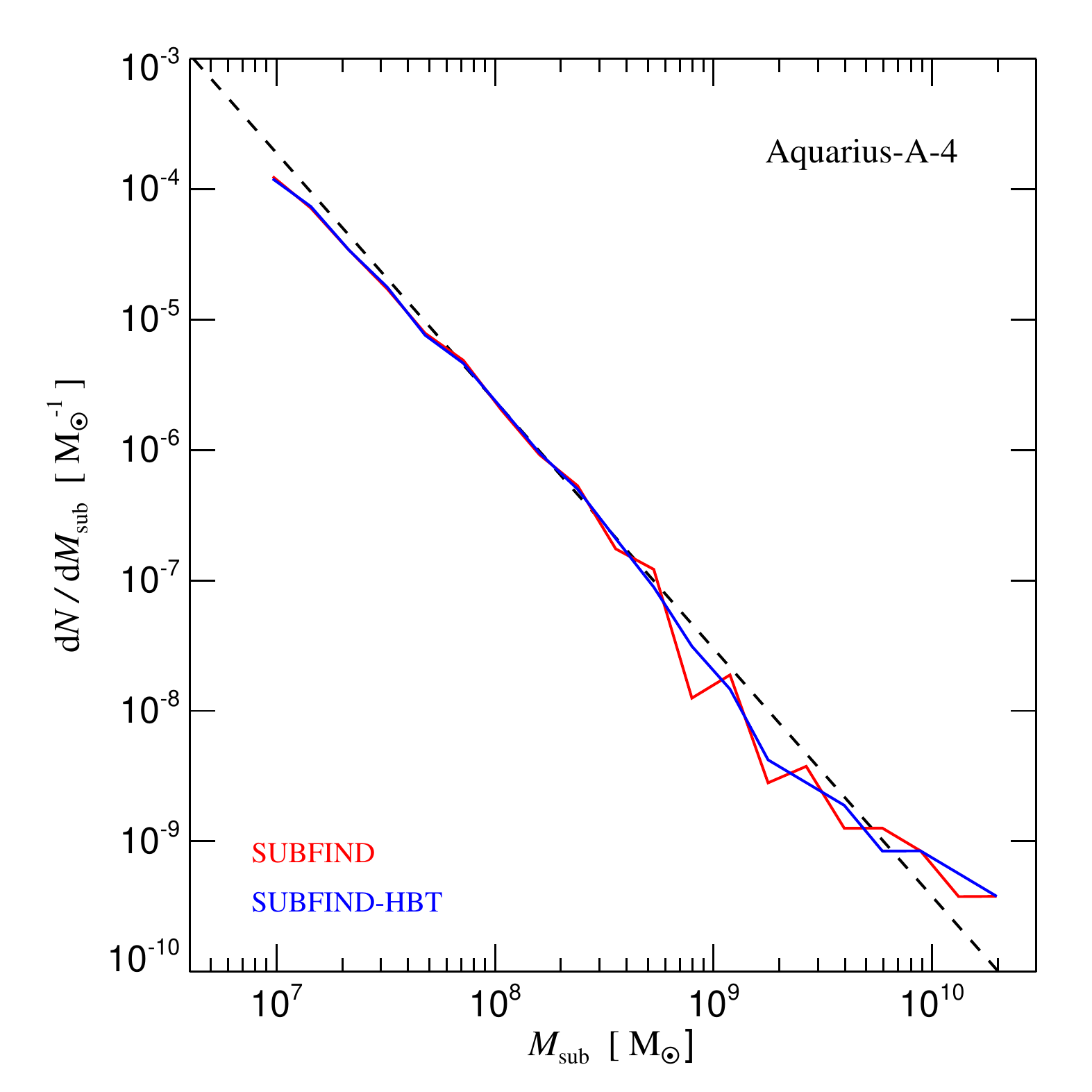}}
\end{center}
\caption{Comparison of the differential subhalo mass function for a
  re-run of the Aquarius Aq-A-4 halo with {\small GADGET-4}, using two
  different substructure detection algorithms, the `classic' {\small
    SUBFIND} algorithm, and the hierarchical bound tracing method
  {\small SUBFIND-HBT}. For further comparison, the dashed line gives
  the fit to the result reported in
  \citet{Springel:2008aa}. Reassuringly, the results are extremely
  close between the two subhalo finders, even though they detect
  substructure candidates in fundamentally different ways. The result
  is also in good agreement with the
  older findings in \citet{Springel:2008aa}, based at the time
  on an early version of
  {\small GADGET-3}.
  \label{FigAquariusMassFunction}}
\end{figure}

\subsection{Merger tree construction}

For many scientific applications, merger trees are an essential input
as they give the ability to track individual halos through cosmic
time. This is, for example, needed for semi-analytic models of galaxy
formation that follow physical models of galaxy evolution on the dark
matter backbone, and predict how galaxies populate the
structures. Traditionally, merger tree building is done in
postprocessing, because it involves a complex matching exercise
between many simulation outputs at different times. Because all
produced snapshots and group catalogues are involved, this
post-processing operation is also a rather data-intensive
operation. For simulation sizes that in select cases already reach
over a trillion particles, this poses a serious challenge.  For
example, for the recent Uchuu simulation of \citet{Ishiyama:2020aa}
merger tree construction was done by splitting the simulation volume
into 8000 boxes and sub-catalogues, as the calculation could otherwise
not be processed with the available {\small ROCKSTAR} halo finder
software. In addition, it basically has become infeasible for
simulations of that size to produce a large amount of finely spaced
snapshot outputs just for the sake of constructing merger trees, due
to the prohibitive data volume involved.

\begin{figure}
\begin{center}
  \resizebox{8.5cm}{!}{\includegraphics{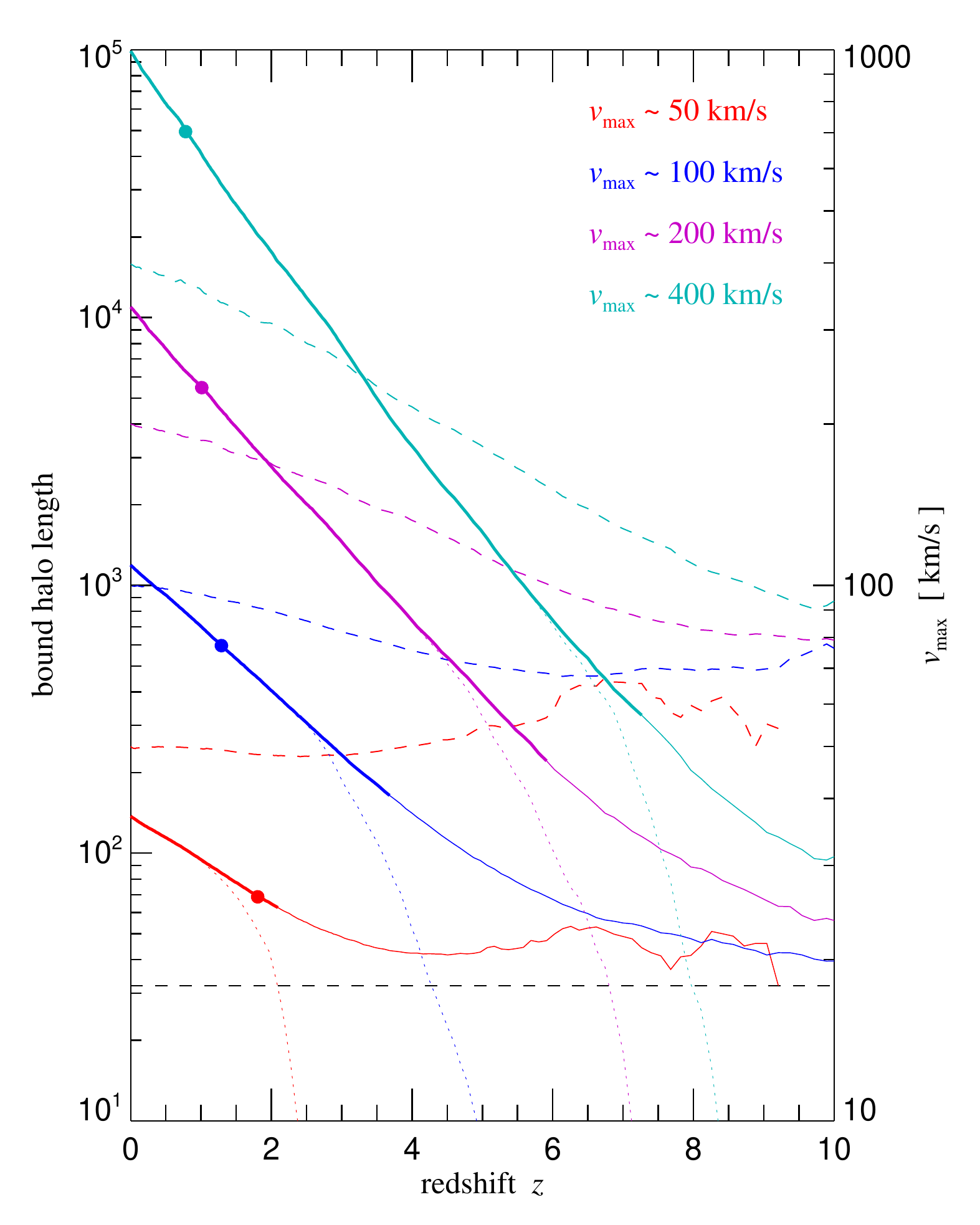}}
\end{center}
\caption{Backtracking of central halos selected by different maximum
  circular velocity at $z=0$, in a cosmological dark matter-only simulation with
  $648^3$ particles in a $75\,h^{-1}{\rm Mpc}$ box.
  The solid lines show the average bound particle
  number in the main progenitor stems of the selected halos as they
  are tracked back to high redshift. If a halo cannot be tracked any
  more, it is discarded in the averaging, that is why towards high
  redshift only a subset of the original halos can still be tracked,
  so that averages eventually become biased towards early formation
  times.
  For the thick part of
  the solid lines more than 50\% of the halos can still be tracked,
  while the dotted lines show the result
if the
  ones that cannot be tracked any more are considered in the averaging
  with zero length. The dashed lines
  give the average maximum circular velocity of the (still trackable)
  progenitor halos, with the corresponding quantitative scale indicated on
  the vertical axis on the right. The solid circles mark the times when the
  mass of the main progenitor has dropped to half the halo mass at $z=0$, which is
  a commonly employed definition of the formation time of a halo. The corresponding
  redshifts are $z=1.8$, $1.3$, $1.0$, and $0.8$, in increasing order of
  final mass for our four samples, reflecting the
  hierarchical growth of structure.
  \label{FigBackTrackingAverage}}
\end{figure}

\begin{figure}
\begin{center}
  \resizebox{9.0cm}{!}{\includegraphics{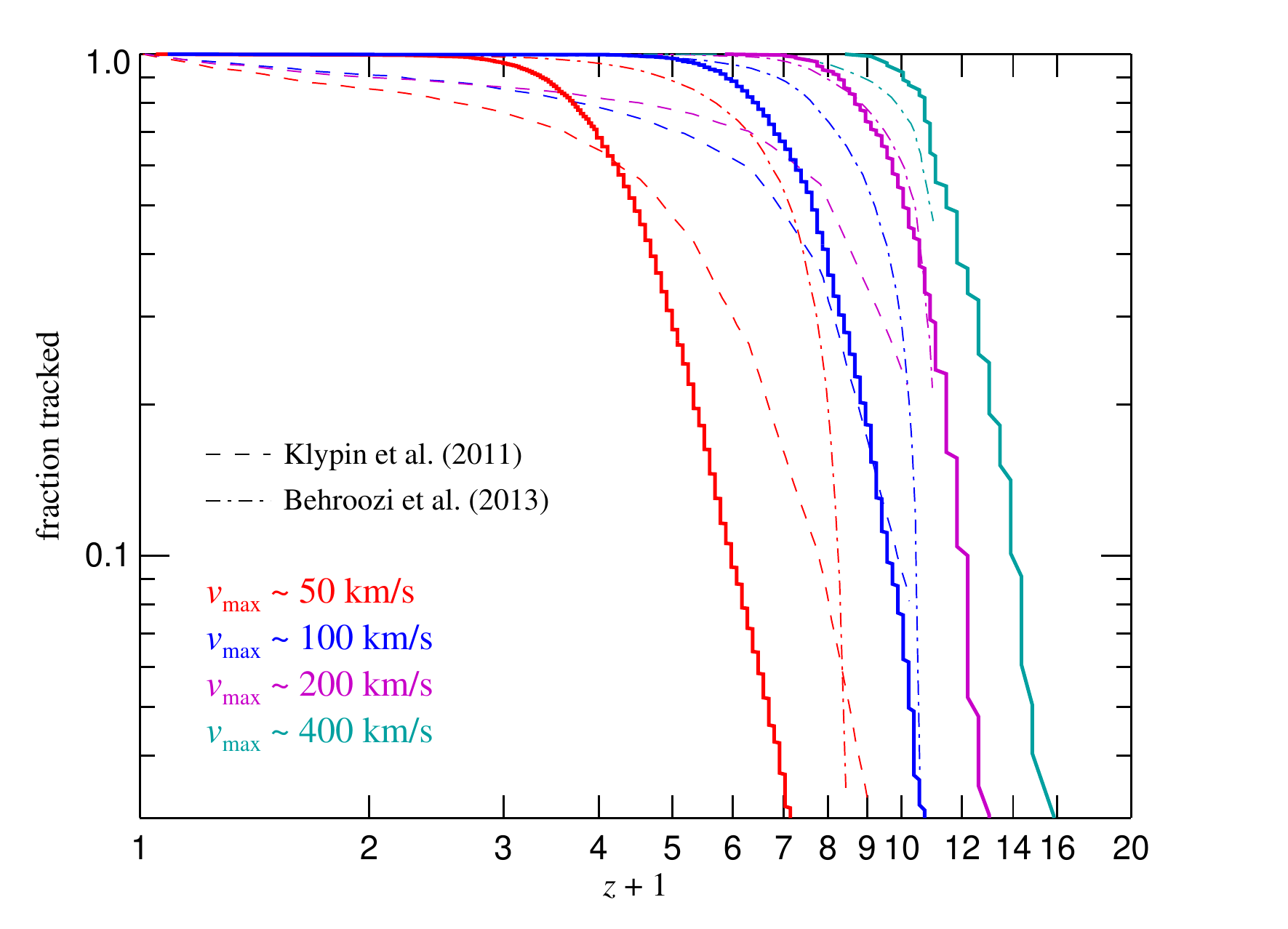}}
\end{center}
\caption{Fraction of holes selected at $z=0$ that can be tracked back
  to a given redshift (solid lines), for four different maximum
  circular velocity choices, as labelled.  For comparison, we also
  show results reported by \citet{Klypin:2011aa} and
  \citet{Behroozi:2013aa}, respectively, for the Bolshoi simulation,
  which has a very similar mass resolution (within $1\%$) as the test
  run considered here (with $648^3$ particles in a
  $75\,h^{-1}{\rm Mpc}$ box). Our tracking results, as well as those
  of \citet{Behroozi:2013aa}, clearly outperform the approach in
  \citet{Klypin:2011aa}. Our results also indicate an advantage over
  the combination of {\small ROCKSTAR} groupfinder and {\small 
  CONSISTENT-TREES} merger tree builder of \citet{Behroozi:2013aa} for
large halos, while the opposite appears to be the case for small
halos. The latter could however be simply related to our comparatively
conservative minimum halo size of 32 particles.
  \label{FigBackTrackingFraction}}
\end{figure}

It would thus be very helpful if the data-intensive part of the merger
tree finding could already be done on the fly, making it possible to
construct merger trees that are finely resolved in time without ever
having to output the full particle data. In {\small GADGET-4}, we
achieve this by executing FOF and {\small SUBFIND} (or {\small
  SUBFIND-HBT}) on the fly at a prescribed set of output times. But
instead of storing raw particle data to disk, only the group and
subhalo catalogues are saved, which contain information about
collective properties of each group and subhalo, like mass, position,
bulk velocity, spin, etc., but do not list each particle making up the
group. Correspondingly, the data volume of these group catalogues is
quite small. To establish the necessary links of halos in time (and
hence between different group catalogues) we let the code retain
membership information of each particle in a certain group and subhalo
while the simulation continues. Right after a new group catalogue is
determined, the membership information in the previous group catalogue
is then still present, allowing us to link the group catalogues with
each other.  This is done similarly as first described in
\citet{Springel:2001ac}, but with some notable improvements described
below. After the linking information has been determined, it is output
to disk (this again is very small in data volume compared with the
full particle set, and represents only a minor addition to the halo
catalogues) alongside with the new group catalogue. Finally, the
membership record of each particle is updated with that of the new
group catalogue; this will then be used when the next group catalogue
at a slightly later time is computed and the corresponding links to
this new one are determined.

We establish the links as follows. Let us consider two subhalo
catalogues at subsequent times $t_0 < t_1$, with a total number of
$N_0$ and $N_1$ subhalos, respectively. Note that we use the term
{\em subhalo} also for the main gravitational background halo of a
FOF-group in which secondary subhalos may be embedded, as is standard
nomenclature for {\small SUBFIND}. A particle can only be member of
one subhalo at a given time, hence the subhalos form gravitationally
bound disjoint sets of particles at a given time. Let's assume the
subhalos are numbered in some fashion\footnote{In practice, we number
  them consecutively, with the subhalos occupying the largest
  FOF-group coming first, followed by those of the second largest
  FOF-group, etc., and within each FOF group, the subhalo numbers are
  assigned in decreasing order of subhalo size.}, for example
$0, 1, \ldots, N_0-1$ at time $t_0$. For each particle $i$ that is
bound in a subhalo at
this time, we assign a rank $r^{(0)}_i$ based on
ordering the
particles according to increasing binding energy within the
corresponding subhalo. The particle with rank $r^{(0)}_i=0$ is
hence the gravitationally most bound particle in its subhalo. Based on
the rank, we assign a score
\begin{equation}
  S^{(0)}_i  =
  \left\{
    \begin{matrix}
        \;\left[1+[r^{(0)}_i]^{q}\right]^{-1} & \mbox{for}\;\;  r^{(0)}_i
        < N_{\rm core} \\
0 & \mbox{otherwise}
\end{matrix}
\right.
\end{equation}
to the particle $i$, where $N_{\rm core}$ is the maximum number of
most-bound particles that are used to link subhalos (we use
$N_{\rm core}=16$ as our default choice), and $q$ is a parameter that
influences how much more weight should be given to more strongly bound
particles in the tracking (we use $q=0.5$ as our default choice and
find that our results depend at most extremely weakly on this
choice). Particles not bound in a subhalo are assigned a zero
score. In an analogous fashion, we assign scores $S^{(1)}_i$ to the
particles based on the group catalogue at time $t_1$.

Next we determine for each pair of subhalos $(n^{(0)}, n^{(1)})$ from
the two group catalogues the sum of the scores of those particles that
are members of both of these subhalos. Specifically, we define the
descendant score $D(n^{(0)}, n^{(1)})$ of a halo $n^{(0)}$ at $t_0$
with respect to a halo $n^{(1)}$ at $t_1$ as
\begin{equation}
D(n^{(0)}, n^{(1)})  = \sum_i S^{(0)}_i \;\; \mbox{where}\; i \in n^{(0)}
\;\mbox{and}\;   i \in n^{(1)},
\end{equation}
i.e.~we count the particles from $n^{(0)}$ that end up in $n^{(1)}$,
weighted by how bound they were in $n^{(0)}$. Likewise, we define a
progenitor score
\begin{equation}
P(n^{(1)}, n^{(0)})  = \sum_i S^{(1)}_i \;\; \mbox{where}\; i \in n^{(0)}
\;\mbox{and}\;   i \in n^{(1)},
\end{equation}
which effectively counts how many particles in $n^{(1)}$ came from
$n^{(0)}$, weighted by how bound they are in $n^{(1)}$.  We now call a
subhalo $n^{(1)}$ a potential descendant of $n^{(0)}$ if
$D(n^{(0)}, n^{(1)}) > 0$. The subhalo with the largest value of $D$
for a given $n^{(0)}$ is called the {\em primary descendant}, and we
retain this link.  Likewise, we call a subhalo $n^{(0)}$ a potential progenitor
of subhalo $n^{(1)}$ if $P(n^{(1)}, n^{(0)}) > 0$, and denote the
subhalo with the largest value of $P$ for a given $n^{(1)}$ the {\em
  primary progenitor}. We retain this link as well.

In order to enumerate all descendants and progenitors defined in this
way, we use a chain of {\em next descendant} and {\em next progenitor}
pointers, and {\em first descendant} and {\em first progenitor} links
to the heads of the corresponding chains, as sketched in
Figure~\ref{FigMergerTreeLinking}. The sets of subhalos enumerated in
this are specified unambiguously by the primary links.  In particular,
the {\em primary descendant} links induces {\em first progenitor} and
{\em next progenitor} whereas {\em primary progenitor} gives rise to
by {\em first descendant} and {\em next descendant}.  Note however
that {\em first descendant} and {\em first progenitor} select an
arbitrary subhalo among the corresponding sets of descendants and
progenitors.

The above links constitute the basic information stored to link two
subsequent group catalogues. It can be output on-the-fly by the
simulation code, together with the catalogues themselves, such that
storing the full particle data for building a merger tree can be
completely avoided in this procedure.  {\small GADGET-4} can
alternatively compute the links also in post-processing, but only if
particle snapshot files have been stored along the group catalogues
for the involved output times, thus allowing the particle IDs making
up each subhalo to still be inferred.

Note that the above link structure generalises the formalism
introduced by \citet{Springel:2001ac,Springel:2005ab}, which has been
extensively used over the years for the {\small L-GALAXIES}
semi-analytic model \citep[see][for a recent
update]{Henriques:2020aa}. In this older tree-building approach,
only the {\em primary descendant} has been kept, without doing an
additional backwards linking. Physically, this
corresponds to the assumption that subhalos can never ``split up'',
and operationally, that one always manages to correctly track the
bound core of a surviving substructure as the descendant and not, for
example, accidentally material that has been tidally stripped out
of the core. The more general information retained by
{\small GADGET-4} allows a better treatment of rare special
cases, for example, if two subhalos physically collide without
merging, they could be identified temporarily as one object before
separating again. Such a situation can be handled with the extended
linking information, while it leads to a subhalo with a truncated
progenitor history in the merger-tree formalism of
\citet{Springel:2001ac,Springel:2005ab}.

The group catalogues and the enumerations of descendents and
progenitors for each pair of subsequent outputs are already sufficient
to track any subhalo forward and backward in time, and thus to examine
its fate and history. However, analysing the merger tree of a given
object using the above data structures is cumbersome as it requires
random access to a potentially very large set of subhalos distributed
across a large number of files. Also, for applying semi-analytic
models it may be necessary to have a more convenient access to all
subhalos contained in the same {\small FOF} halo, and to allow an easy
distinction between `central' and `satellite' halos. Furthermore, it
would be hard or impossible if one had to apply a semi-analytic model
to {\em all} subhalos of a very large simulation all at once. Instead,
a more natural processing unit would be a {\em tree} that only
contains subhalos that are actually connected in a physically
meaningful way, for example by a descendant or progenitor
relation. Then the semi-analytic model can be computed one tree at a
time.

To construct such trees, {\small GADGET-4} can be applied as a
postprocessing tool that turns all the group catalogues plus the
linking files between two subsequent group output times into a set of
trees. Two subhalos are placed into the same tree if they are in a
primary progenitor or primary descendant relation with each other. In
addition, if two subhalos are members of the same FOF group, they are
put into the same tree (this is done in order to allow semi-analytic
codes to easily access all subhalos contained in a common FOF groups),
and a chaining-link structure is added that allows each subhalo to
retrieve all other subhalos in the same FOF group. Finally, all
subhalos that contain a particle that was once a most-bound particle
of another subhalo are guaranteed to be found in the same tree as this
other subhalo (this is done to allow an `orphan-tracking' in the style
of {\small L-GALAXIES}).

Finally, when the trees are constructed, {\small GADGET-4} defines one
additional link to simplify navigating the tree.  This addresses the
common practical problem that one often wants to follow the ``main
stem'' of an object back in time, for example to define its formation
time. While consecutively following the {\em primary progenitor} for
this purpose works most of time, it is not always robust if there are
multiple progenitors of quite similar mass, a situation in which the
(classic) {\small SUBFIND} algorithm may not always select the same
substructure as the background halo, so that its identity can jump
between different cores. To address this, we define the {\em main
  progenitor} of a subhalo as that subhalo at the previous output time
that has the maximum {\em tree-branch mass} among all subhalos that
are in a primary progenitor of primary descendant relation with the
subhalo in question.  The {\em maximum tree-branch mass} of a subhalo
is recursively defined as the maximum of the tree branch masses among
all subhalos at the previous output time that are in a progenitor or
descendant relation with the subhalo in question, plus the mass of the
subhalo itself. Hence the selection of the {\em main progenitor}
selects the tree branch that is dominant in terms of total mass (here
defined as the mass of all subhalos summed to where the branch ends)
among all possible progenitor branches.  This selection gives a fairly
robust definition of the ``main stem'' of a merger tree \citep[see
also][]{De-Lucia:2007aa}, and is thus particularly useful for following a subhalo
back in time. In practice, {\em main progenitor} only occasionally
differs from {\em primary progenitor}.

In Figure~\ref{FigMergerTree}, we show a schematic sketch of the
different links between subhalos that define the merger tree structure
produced by {\small GADGET-4}. The trees are stored in one or several
tree files, and can be individually accessed, most conveniently of
course if the HDF5 output file format is selected. For convenience,
the subhalos making up a tree are renumbered, and the various links are
simply indices to an array of subhalos making up the tree. The
code also outputs additional bookkeeping information that augments
each group catalogue, informing about the tree in which each subhalo
is to be found. Conversely, the trees contain information about the
origin of each subhalo they contain, i.e.~the output time and subhalo
number they have in the original group catalogue in which they were
originally identified.

How well does all of this work in practice? Is there a significant
difference between identifying subhalos with {\small SUBFIND-HBT} and
then tracking them? In Figure~\ref{FigTrackingAquarius} we show
results for a rerun of the Aq-A-4 halo of the Aquarius project
\citep{Springel:2008aa}, a high-resolution dark matter only zoom
simulation of a Milky Way-sized dark matter halo. We focus on the fate
of a few randomly selected subhalos identified at $z=1$ in the halo,
and check how well they are tracked forward in time to $z=0$. We
repeat this twice based on the same simulation, except that we
identify and track substructures once with the classic {\small
  SUBFIND} algorithm, and once with {\small SUBFIND-HBT}. Each pair of
panels in the plots compares the results obtained with the two
substructure finders. The orbits (shown in blue) are identical over
the common time the substructures can be tracked. Close to pericenter,
{\small SUBFIND} shows characteristic dips in the bound mass (red
lines) found for the subhalos, but for most of the time, the
differences in mass to {\small SUBFIND-HBT} appear to be small, and
the overall mass loss with time is captured in a similar way. There
are some instances, however, where the tracking of the subhalo is lost
prematurely in {\small SUBFIND} compared to what is possible with
{\small SUBFIND-HBT}.

In Figure~\ref{FigSubhaloMassComparisonAquarius} we examine
quantitatively how similar the bound masses of subhalos identified
with the two substructure finders are. To this end we examine a fixed
time, redshift $z=0$ in this case, and directly compare the found
subhalo masses \textit{for the same substructures} in both cases. The
matching was here done by lining up the most bound particles. While
there is some scatter in the subhalo masses returned by the two
different techniques, a systematic difference in the mean subhalo
masses is only detectable for the inner parts of the background halo
(where it reaches $\sim 10\%$), and for relatively massive
subhalos. This can be readily understood and is largely expected based
on the detection principle of {\small SUBFIND}, which can only find
the part of a substructure that rises above the background density. If
the latter becomes high, outer parts of substructures will no longer
be found. The difference seen in
Fig.~\ref{FigSubhaloMassComparisonAquarius} is however reassuringly
small compared to what may have been expected based on findings
reported in substructure comparison studies using toy simulations
\citep{Knebe:2011aa, Muldrew:2011aa}.

Another way to look at this is to simply count substructures as a
function of mass, which is shown in
Figure~\ref{FigAquariusMassFunction}. In this subhalo mass function
comparison, also differences in the number of detected substructures
enter, besides potentially systematic differences in the detected
masses. Reassuringly, the results obtained for {\small SUBFIND} and
{\small SUBFIND-HBT} are extremely similar, apart from small scatter
at the massive end. Furthermore, the results also agree well with the
older results obtained in \citet{Springel:2008aa} with an early
version of the {\small GADGET-3} code, and a much older version of
{\small SUBFIND}. This speaks for the robustness of inferences
obtained from these simulations, for example for dark matter
annihilation \citep{Springel:2008ab}, which appear to be unaffected by
the substructure finding technique.

In Figure~\ref{FigBackTrackingAverage} we now turn to an example of
tracking halos back in time, using {\small SUBFIND-HBT} and the
built-in merger tree functionality in {\small GADGET-4}. We select
samples of subhalos with different maximum circular velocity today in
a cosmological simulation with $648^3$ particles within a
$75\,h^{-1}{\rm Mpc}$ box, and follow them back in time along their
main progenitor. The bound particle mass of the (sub)halos drops
quickly in time towards high redshift, but more rapidly for bigger
halos, so that the nominal formation time, where halos have assembled
$50\%$ of their final mass, lies at higher redshift for smaller
objects, as expected from hierarchical structure growth in
$\Lambda$CDM. The maximum circular velocity, however, shows a much
weaker evolution with time. Since we impose a minimum particle number
of 32 in these subhalo samples, the mean maximum circular velocity
eventually found among all still trackable subhalos at the highest
redshift approaches a common value, because they then all linger just
above the particle number detection threshold.

Finally, in Figure~\ref{FigBackTrackingFraction}, we extend the
analysis of how well substructures are tracked by asking what fraction
of halos identified with different $v_{\rm max}$ at $z=0$ can still
can be tracked back to a given redshift. Our results agree favourably
with those reported by \citet{Klypin:2011aa} for the Bolshoi
simulation. For massive halos they also show an advantage over the
elaborated {\small CONSISTENT-TREES} approach by
\citet{Behroozi:2013aa}, which invokes the insertion of fiducial halos
to improve the tracking, while the opposite is the case for small
halos. The latter could however be simply related to the more
conservative minimum halo particle number used here.

\subsection{Light-cone output}

For a direct comparison to deep extragalactic surveys, snapshots at
fixed time are in principle not ideal. Rather, a proper output of the
matter as it crosses the backwards lightcone would be highly
desirable. This has been implemented in the {\small GADGET-4} code as
a generally available feature. Such lightcone outputs have been
pioneered with the Hubble Volume simulation \citep{Evrard:2002aa}, and
in recent years been used in projects such as MICE
\citep{Fosalba:2015aa} and Horizon-AGN \citep{Gouin:2019aa}.

\begin{figure}
\begin{center}
\resizebox{7cm}{!}{\includegraphics{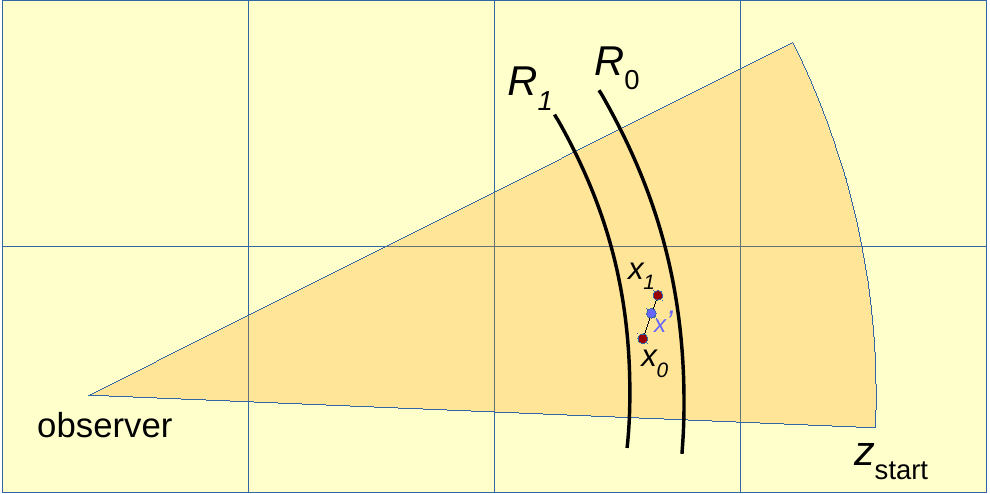}}
\end{center}
\caption{Geometry of the lightcone test. When a particle is drifted
  from comoving coordinate $\vec{x}_0$ to $\vec{x}_1$ over a time step from
  scale factor $a_0$ to $a_1$, {\small GADGET-4} checks whether it is overrun by the
  backwards lightcone of a fiducial observer. Here $R_0$ is the
  comoving distance to redshift $z_0 = 1/a_0 -1$, and $R_1$ to
  redshift $z_1 = 1/a_1 -1$. If an intersection with the lightcone
  occurs, an interpolated coordinate $\vec{x}'$ corresponding to the
  crossing time is registered in the lightcone output. The simulation box (yellow squares) may
  need to be periodically replicated to cover the full comoving volume
  of the lightcone, as sketched here. The code allows for multiple
  lightcones that may have a variety of 
  angular geometries (full sky, octants, pencil beams, etc.),
  and variable redshift depth.
  \label{FigLightCone}}
\end{figure}

Figure~\ref{FigLightCone} sketches the basic principle for how
particles are put onto a lightcone output in {\small GADGET-4}. An arbitrary
coordinate within the simulation box can be selected as an observer
position. One or several lightcone geometries can then be defined
through their starting redshift $z_{\rm start}$, their ending redshift
$z_{\rm end}$ (which can be zero as in the sketch of
Fig.~\ref{FigLightCone}, in which case the lightcone extends all the
way to the observer), and a geometric angle selection on the
sky. Currently, the code supports an all-sky geometry for this (full
solid angle), a selection through ranges in spherical polar
coordinates (this can for example be used to define an octant), or a
pencil beam (i.e.~a certain opening angle around an arbitrary
direction). Other geometric angular selections could be easily added
if desired.

\begin{figure}
  \begin{center}
    \resizebox{8.0cm}{!}{\includegraphics{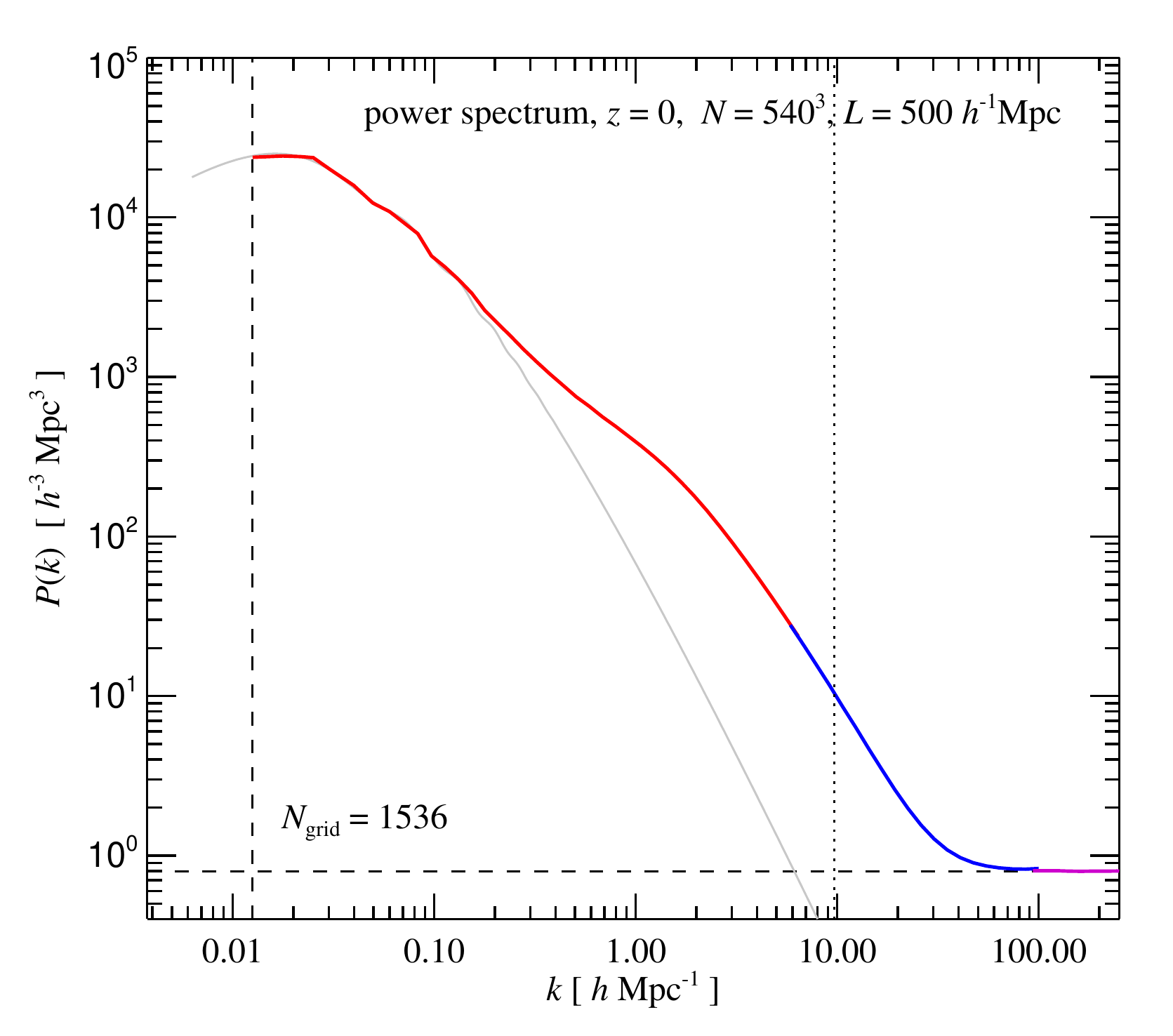}}\\%
    \resizebox{8.0cm}{!}{\includegraphics{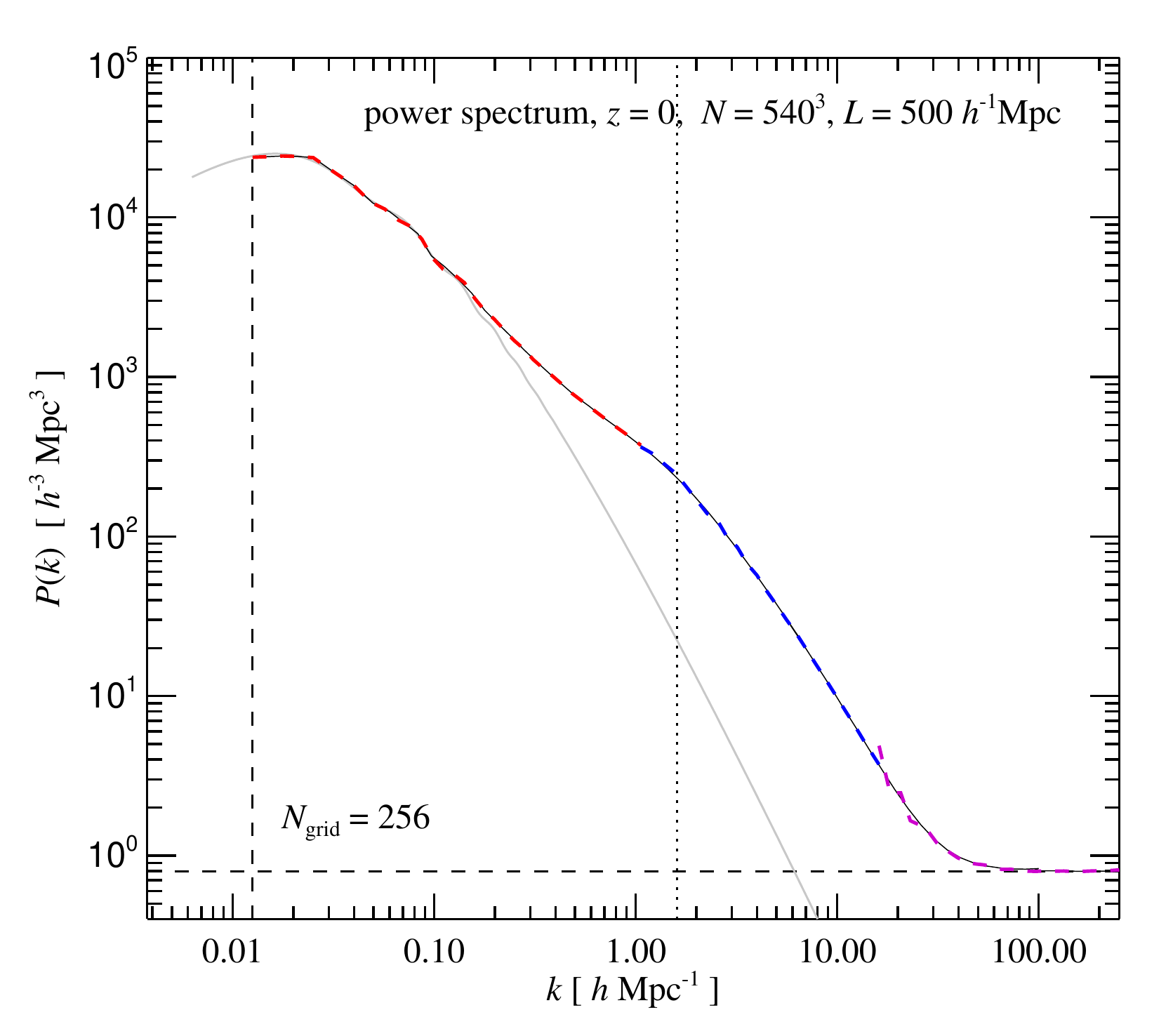}}\\%
\caption{Example for a dark matter power spectrum measurement with the
  inlined routines in {\small GADGET-4},  which employ the folding
  technique to go
  significantly beyond the Nyquist frequency of a single mesh. The top
  panel shows a measurement with an already quite large $1536^3$ base
  grid (red lines), augmented with two further measurements with different folding
  factors to obtain results at
 progressively higher $k$ (blue and magenta lines). Even though the 
  underlying simulation has only $512^3$ particles, a single
 $1536^3$ measurement would have at most reached the Nyquist frequency
 of that mesh (dotted line), and
 hence would be insufficient to measure the power on the smallest
 resolved scales and down to the shot noise limit (horizontal dashed
 line). The vertical dashed line on the left indicates the fundamental
 mode of the box, and the thin grey line is the linear theory power
 spectrum. In the bottom panel, the measurement is repeated, but this
 time using only a Fourier grid with $256^3$ cells (coloured dashed lines). Without the
 folding technique, the dynamic range of the measurement
 would be seriously limited. Using it, however, the same result as for the
 $1536^3$ mesh (which is repeated with a thin solid line in the
 background) is recovered in a computationally cheap way.
  \label{FigPowerSpecMeasurement}}
\end{center}
\end{figure}

Whenever a particle is drifted in position space (which happens
exactly in one place in the code), it is subjected to a lightcone
test. Suppose the particle is moved from comoving coordinates
$\vec{x}_0$ to $\vec{x}_1$ over the course of a timestep interval
going from scale factor $a_0$ to $a_1 > a_0$. Corresponding to these
times are comoving distances $R_0$ and $R_1 < R_0$ of the outer edge
of the backwards lightcone (which reaches the observer at the
present). If the observer is located at coordinate $\vec{q}$, then the comoving
distance vectors to the observer are given by
\begin{equation}
\vec{d}_{0/1} = \vec{x}_{0/1} - \vec{q} + \vec{n}\,L,
\end{equation}
where $\vec{n}$ is an (arbitrary) integer triplet accounting for
periodic replicas of the simulation box. The particle can be placed onto
the lightcone output during this step if and only if
\begin{equation}
 R_0 >
|\vec{d}_{0}| \;\;\;\mbox{and}\;\;\; R_1 < |\vec{d}_{1}|.
\end{equation}
This condition can in principle be fulfilled multiple times for
different $\vec{n}$, hence it is important to test for all
possibilities arising from translations of the particle box.  The
maximum extension $z_{\rm max}$ can help here to reduce the comoving
volume covered by the lightcone.  If one detects in this way that the
particle is overrun by the backwards lightcone, we determine the
comoving coordinate and time (i.e.~scale factor) where precisely this
happens through linear interpolation ($\vec{x}'$ in the sketch). The
resulting interpolated particle phase-space coordinates are then 
written into a temporary buffer.

Whenever a full timestep is completed, we check whether the temporary
buffer is filled by more than some nominal number of particles, in
which case the lightcone data is dumped and the buffer is cleared. The
same also happens when the simulation finishes, at which point the
remaining lightcone data still in the buffer is flushed in a lightcone
dump. If such a lightcone dump occurs, the particle coordinates are
tested against the angular mask of one or several lightcones that have
been defined, and against their redshift boundaries $z_{\rm start}$
and $z_{\rm end}$. Only if a particle matches this selection, it is
really written to a particle lightcone file, if appropriate also to
several of them. Like all other outputs of the code, the I/O itself can
be done in parallel (spread across multiple files), and uses HDF5 as
preferred file format.

Our lightcone code and the built-in group finders also support the
possibility to determine halos directly on the lightcone, using the
lightcone particles themselves. With this option, one can forego
outputting the particles on the lightcone, but rather do this for
halos only, in order to substantially save on storage volume. This
nevertheless allows later applications such as HOD mocks or
gravitational lensing that are free of any discontinuities or
ambiguities in the lightcone construction, which typically occur in
schemes that instead patch together a set of time-slices
\citep{Hilbert:2007aa}.

To avoid edge effects, the temporary lightcone buffer internally
always uses an all-sky geometry, even if only narrower
particle-lightcones or halo-lightcones are constructed. Whenever
lightcone data is flushed to disk and groups on the lightcone are
desired, {\small GADGET-4} first runs a FOF group finder directly on
the temporary particle lightcone data, followed optionally by a run of
{\small SUBFIND}. The actual code that is executed is the same that is
also used for the ordinary group finding (these are templated C++
routines that are simply applied to a different particle data
structure), such that the same set of group and subhalo properties are
automatically computed in a consistent way. This also carries over to
the output itself, i.e.~if particle data for the lightcone is produced
(and not only halos on the lightcone are output), the particle data on
the lightcone is stored in group order, allowing one to easily and
selectively retrieve the lightcone particles making up a given
substructure, just like for ordinary timeslices.  We note that the
code also takes care that groups are not split up by subsequent dumps
of lightcone particle data.  This is accomplished by computing for
each identified group a nearest distance to the fiducial observer. If
this is not further away than the inner boundary of the current dump
plus the linking length, the group may not yet be complete. It is then
ignored in the present lightcone dump, and all of its particles are
kept in the temporary buffer such that the group will be reassessed in
the next lightcone dump after having had a chance to grow further.

\begin{figure*}
\begin{center}
\resizebox{6.0cm}{!}{\includegraphics{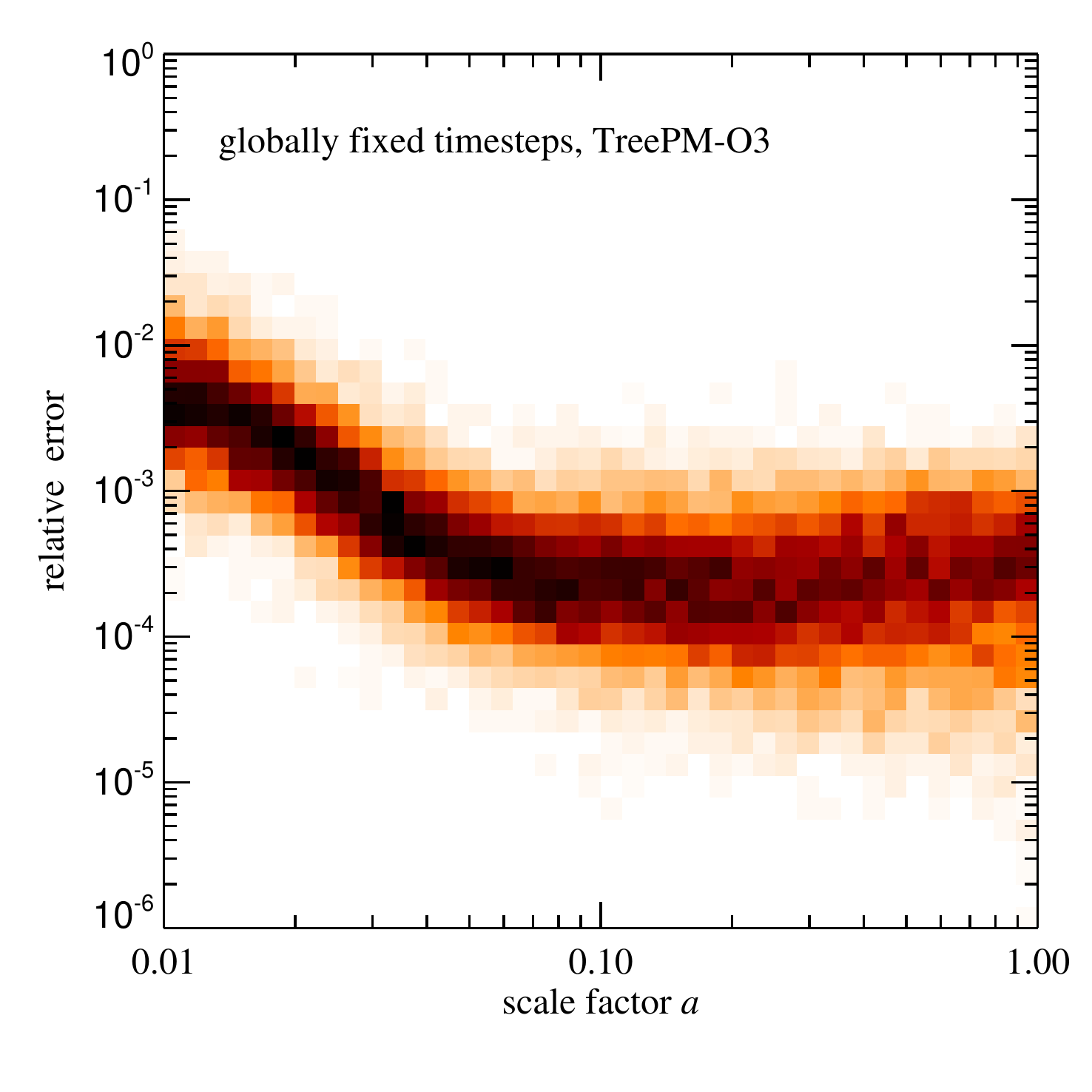}}%
\resizebox{6.0cm}{!}{\includegraphics{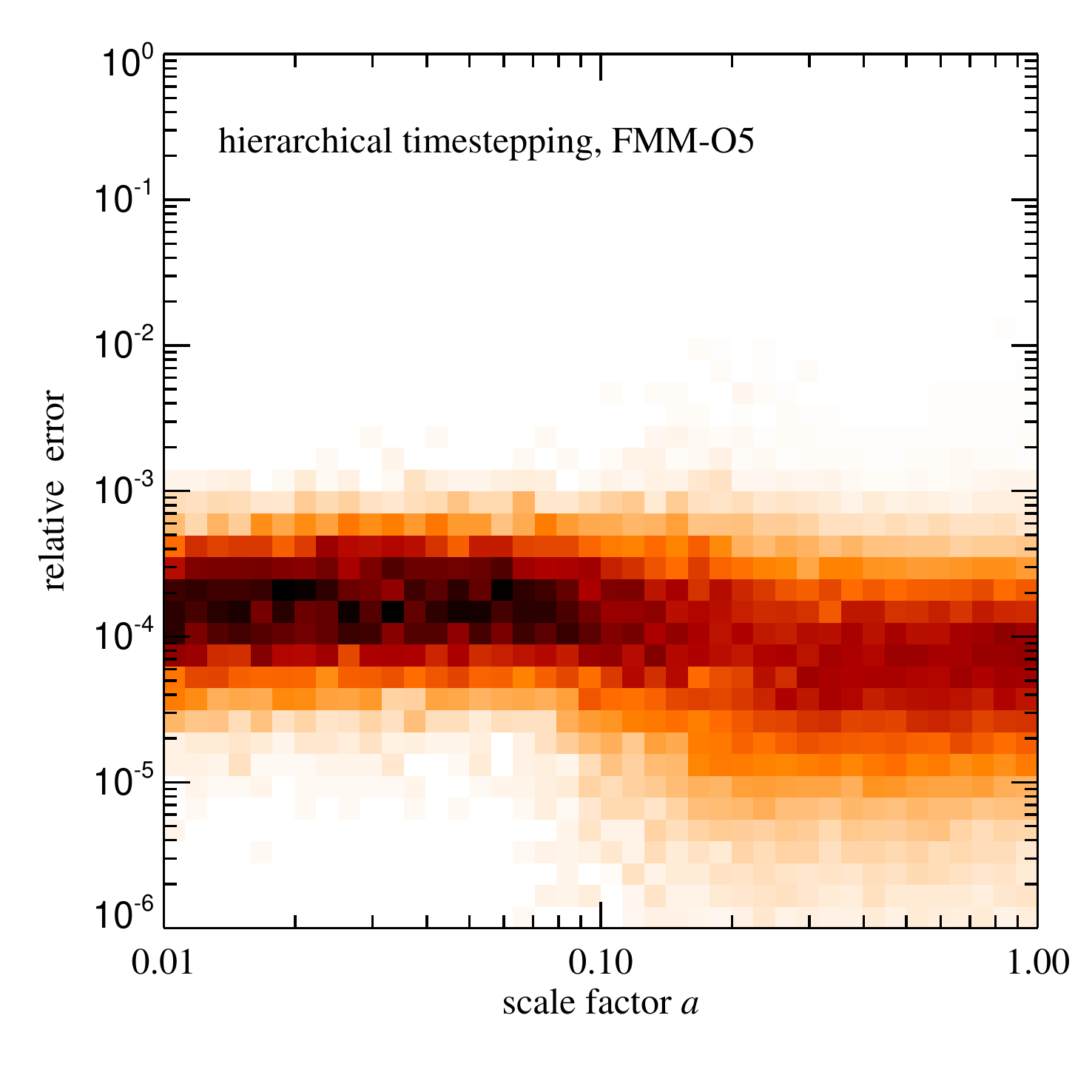}}%
\resizebox{6.0cm}{!}{\includegraphics{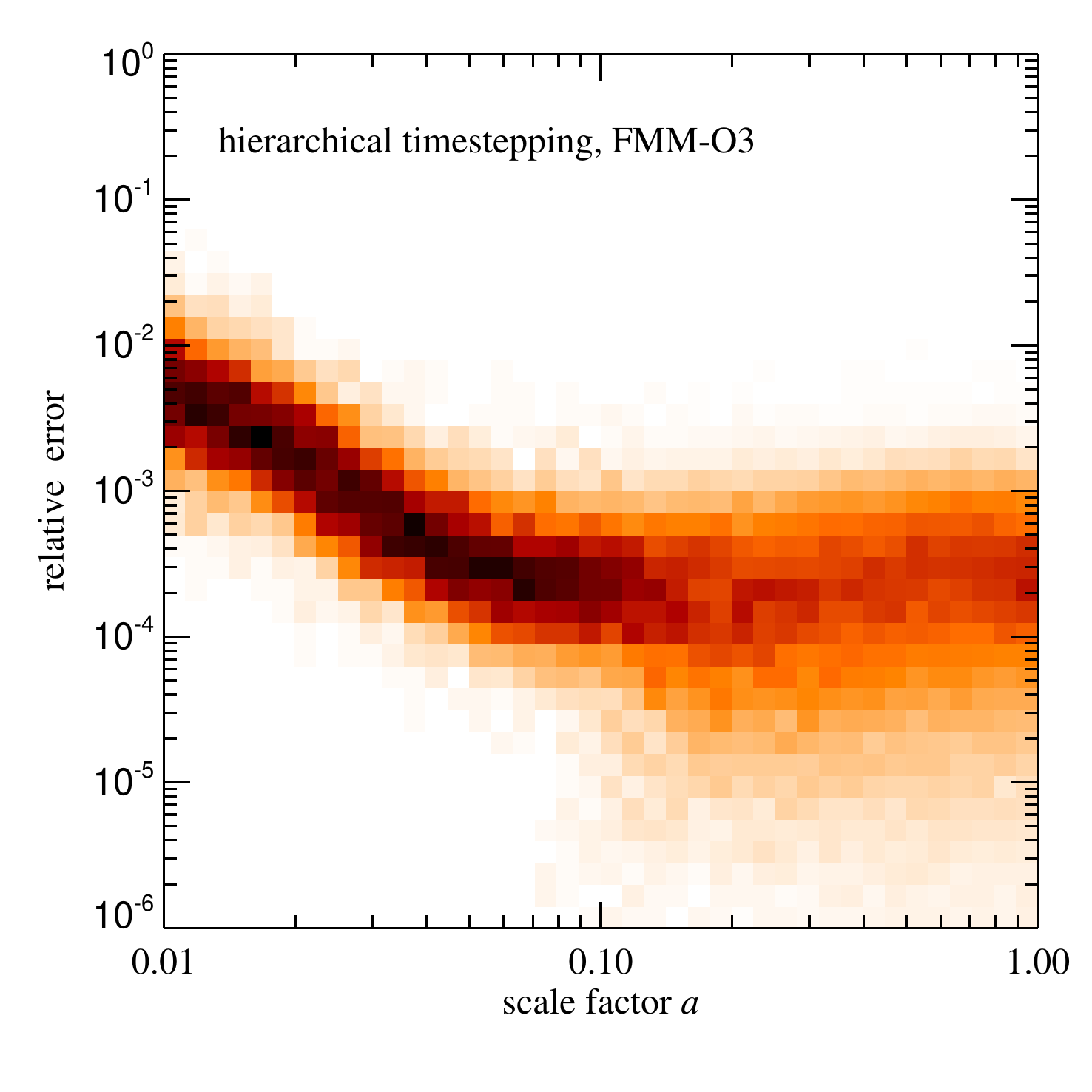}}\\
\resizebox{6.0cm}{!}{\includegraphics{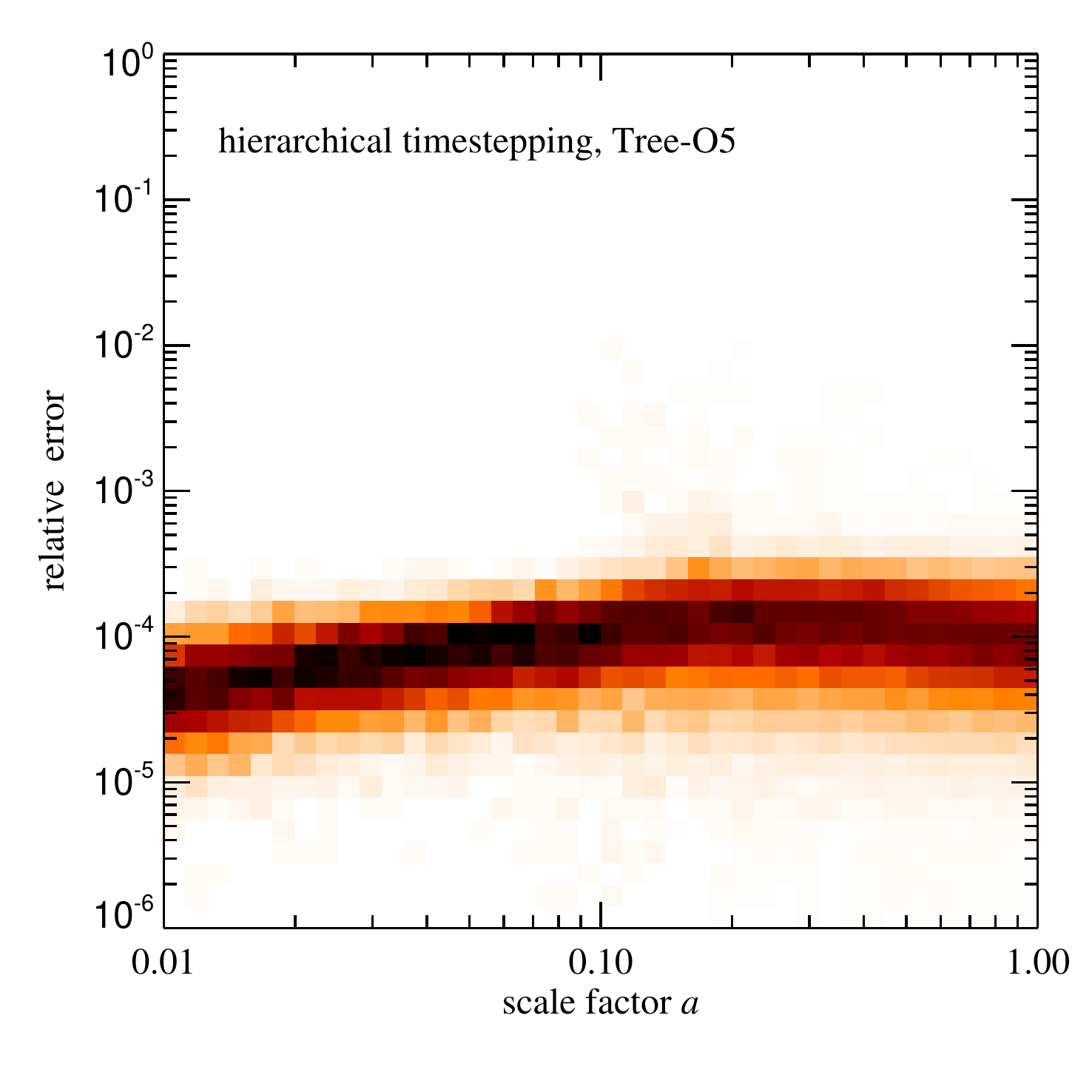}}%
\resizebox{6.0cm}{!}{\includegraphics{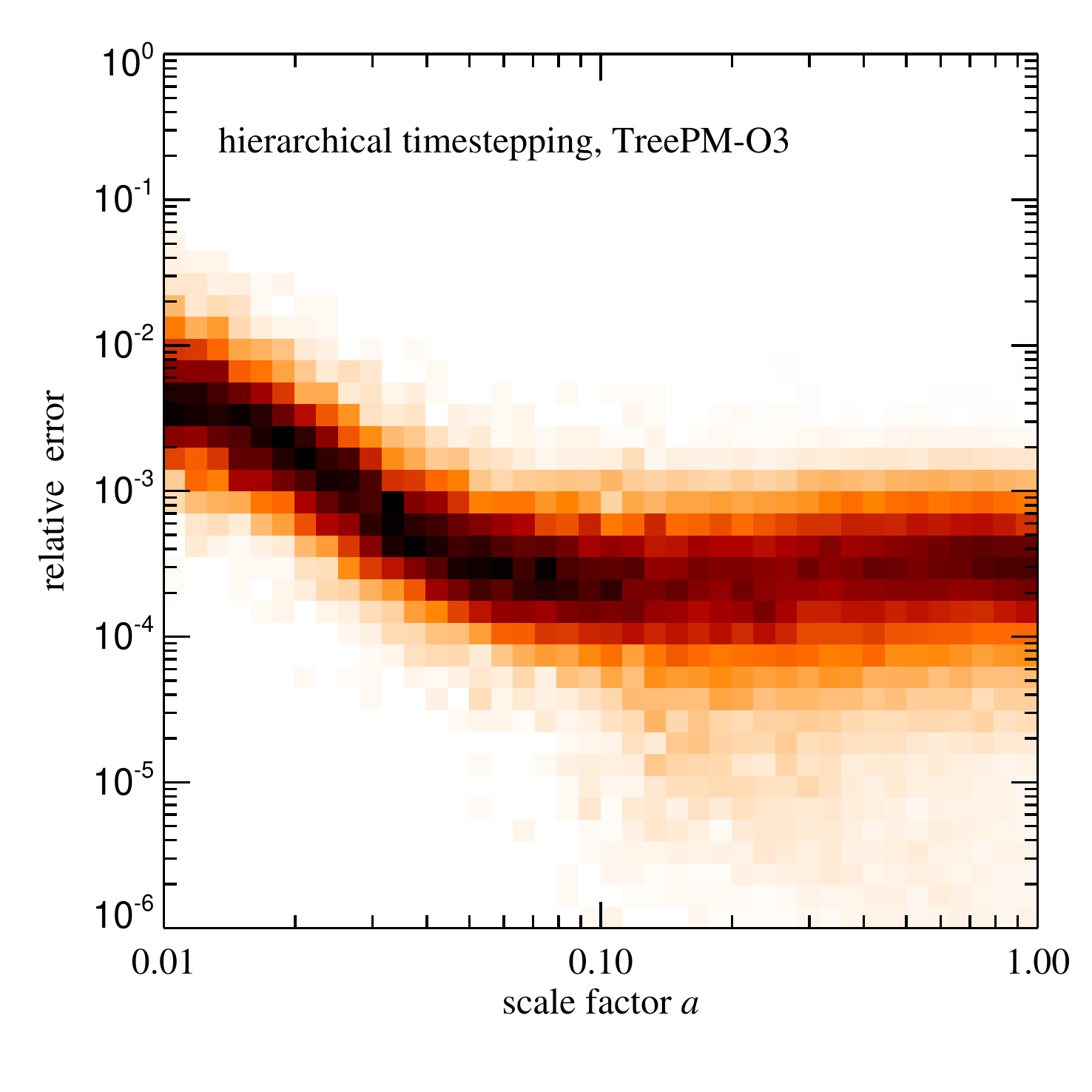}}%
\resizebox{6.0cm}{!}{\includegraphics{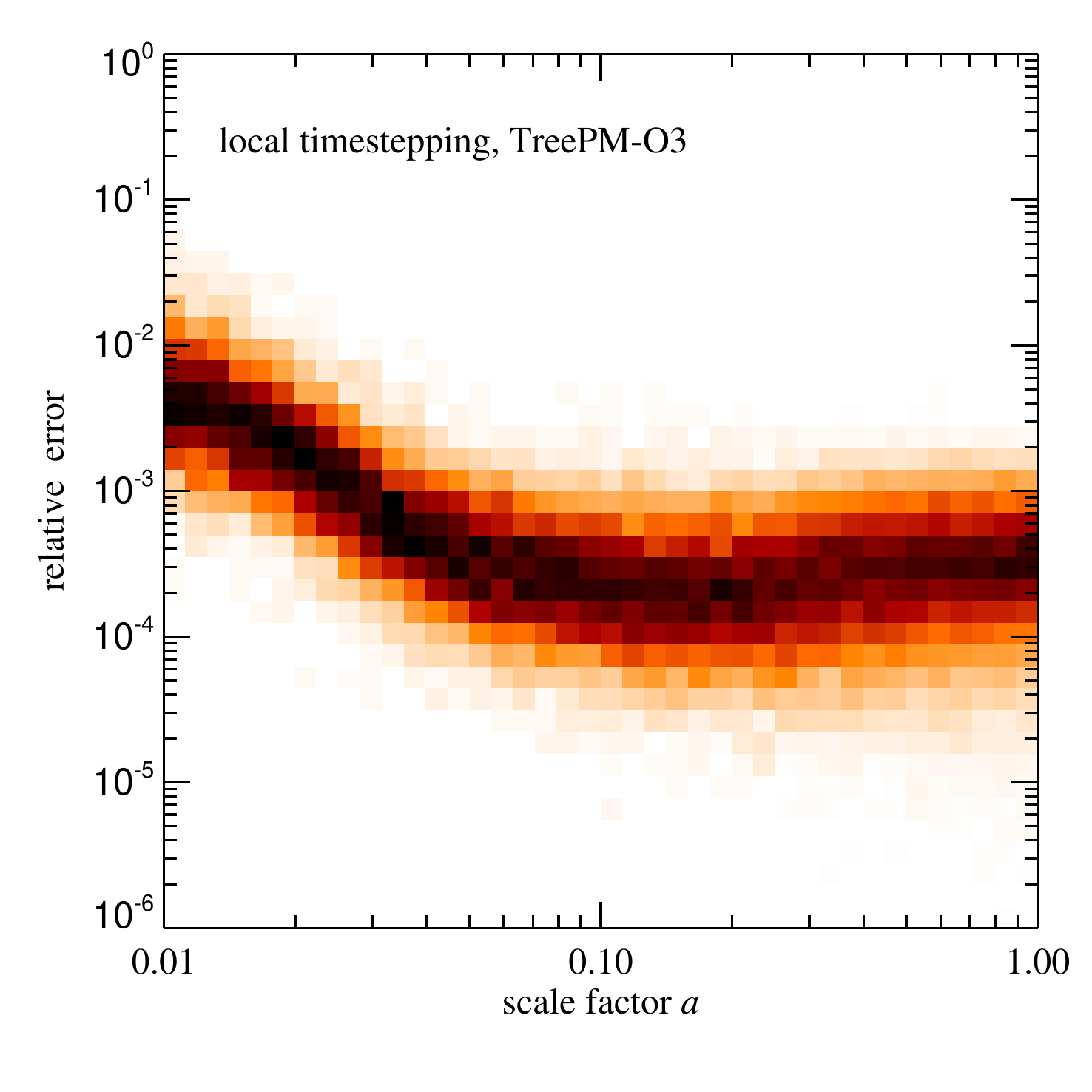}}
\end{center}
\caption{Force error distributions as a function of scale factor
  during cosmological simulations with $256^3$
  particles and a starting redshift of $z=99$. The different panels
  show on-the-fly measurements during {\small GADGET-4} runs with
  different force
  calculation algorithms and different timestepping schemes, as
  labelled. The force accuracy is assessed by selecting
  small random subsets of particles 
   after every force computation, followed by computing the force for
   them by
   direct summation. Doing this over the course of a full simulation
   can reveal
   temporal variations of the force accuracy due to the evolving
   clustering state, and by inlining the measurements in this way
   into the actual timestepping procedure, the semantics of the 
   code also for partially occupied timebins is explicitly tested.
\label{FigForceErrorAsAFunctionOfScaleFactor}}
\end{figure*}

Finally, a further feature of {\small GADGET-4} is the possibility to
produce particle lightcones composed only of those particles that have
at some point been a most-bound particle at a previous group/subhalo
catalogue. This is useful to construct mock galaxy light-cones based
on physical galaxy formation models computed with semi-analytical
models.  For example, in semi-analytic codes like {\small L-GALAXIES},
these particles mark the locations of semi-analytic galaxies
that do not have an associated dark matter subhalo any more, but are
instead `orphaned' and now tracked by a formerly most-bound particle
of a subhalo that has dropped below the mass threshold for tracking
due to tidal truncation.

In practice, the idea of this is to compute group catalogues for
(many) time-slices on-the-fly, which are used to compute detailed
merger trees. They are also used to mark those particle IDs that have
at some point been a most-bound particle in one of those groups, and
only those particles are output to particle lightcones. In a
postprocessing step, the semi-analytic code then evolves a galaxy
population along the merger trees. At any given time, this population
features a unique most-bound particle ID for every galaxy. It thus
becomes a task of matching these IDs to those appearing on the
lightcone at the particular lookback time stored for the lightcone
particle to obtain a high quality mock galaxy catalogue. We note that
this procedure in principle does not require the creation of any
particle dumps on disk at all, which we argue is a great
simplification and may even be a prerequisite to create extremely
large mock catalogues in this way.  Note also that the use of
semi-analytic models is preferable on physical grounds to HOD, SHAM,
or random sampling from SAM merger trees that only take the halo mass
into account, because it allows effects such as assembly bias to be
properly taken into account \citep[see for example][for an analysis of
HOD limitations]{Hadzhiyska:2020aa}.

\subsection{Power spectrum estimation}

The PM gravity solver of {\small GADGET-4} provides the basis to a
built-in measurement routine for the matter power spectrum. To vastly
extend the dynamic range to scales below the nominal grid scale of the
employed grid, we employ the `self-folding' trick described in
\citet{Jenkins:1998aa}, allowing to reach down to the gravitational
softening lengths or even below, if desired.  This amounts to
measuring several power spectra over different frequency ranges, one
is for the normal periodic box of size $L$ that is covered by the
PM-grid of dimension $N_{\rm grid}$. Further spectra are computed by
folding the box onto a power-of-two integer subdivision of itself,
i.e.~effectively the simulation box is stacked on itself by imposing a
smaller periodicity scale, for example $L/16$. This means that only
every 16-th mode in $k$-space is measured when the power spectrum is
determined from this folded density field. But this downsampling is
irrelevant for the high-$k$ regime we are interested in with this
measurement. In fact, by using the full box on large scales and the
folded box (or boxes) on small scales, a continuous and accurate power
spectrum measurement over a very wide range is reached.

As discussed in \citet{Springel:2018aa}, to cover the full dynamic
range of the largest cosmological simulations presently done,
computing three folded spectra with folding factors of 16 represents a
good practical compromise, so this is implemented in {\small
  GADGET-4}. In Figure~\ref{FigPowerSpecMeasurement}, this is
illustrated for a small cosmological box.  The code supports an
automatic measurement of the power spectrum whenever a snapshot is
written, but one can also apply {\small GADGET-4} to an existing
snapshot for measuring the power spectrum in postprocessing. Power
spectra separated by particle type (say for example dark matter and
baryons) are also possible, and are computed besides the total matter
power spectrum. The shot-noise is computed in each case, if needed by
taking variable particle masses into account.

\begin{figure*}
  \begin{center}
    \resizebox{8.5cm}{!}{\includegraphics{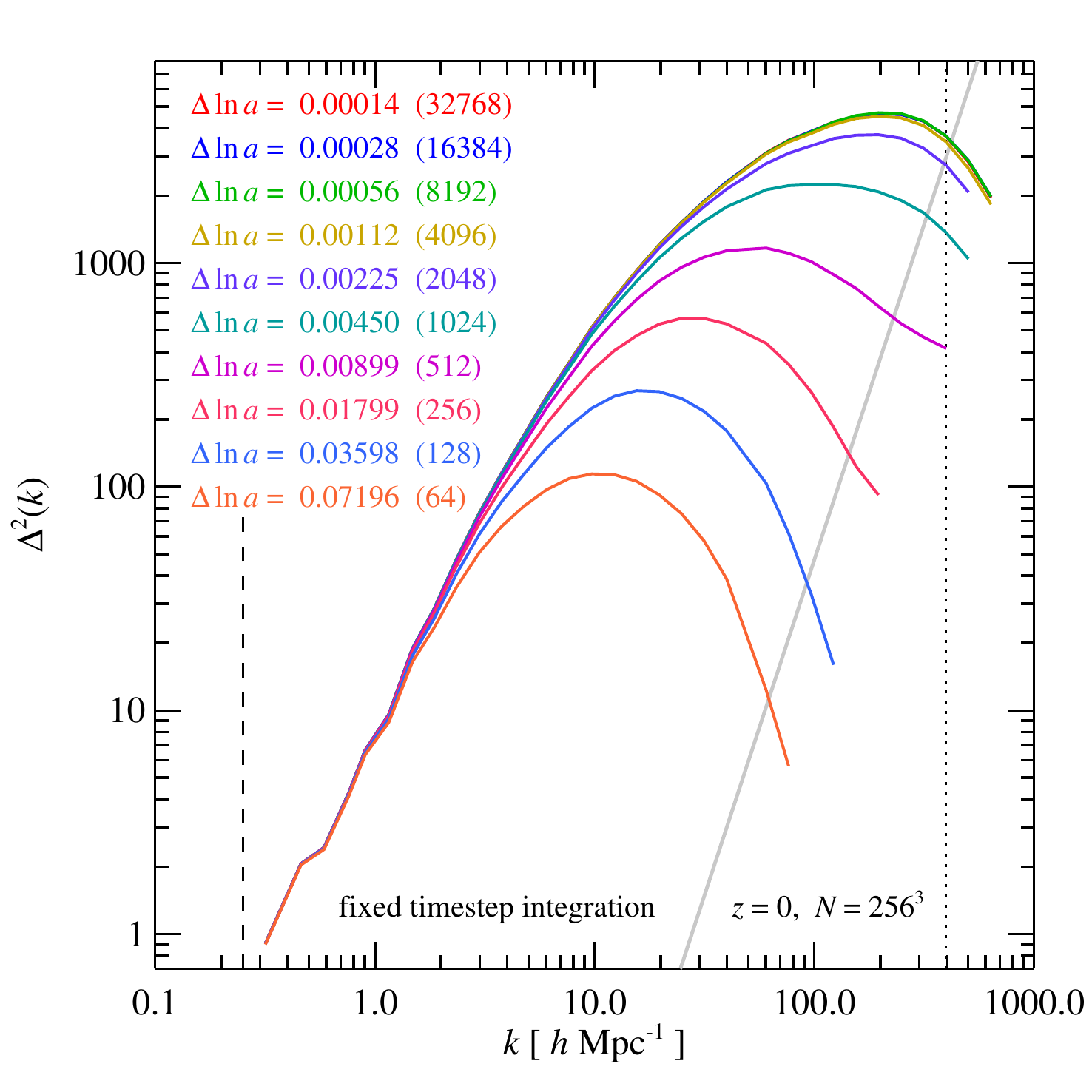}}%
    \resizebox{8.5cm}{!}{\includegraphics{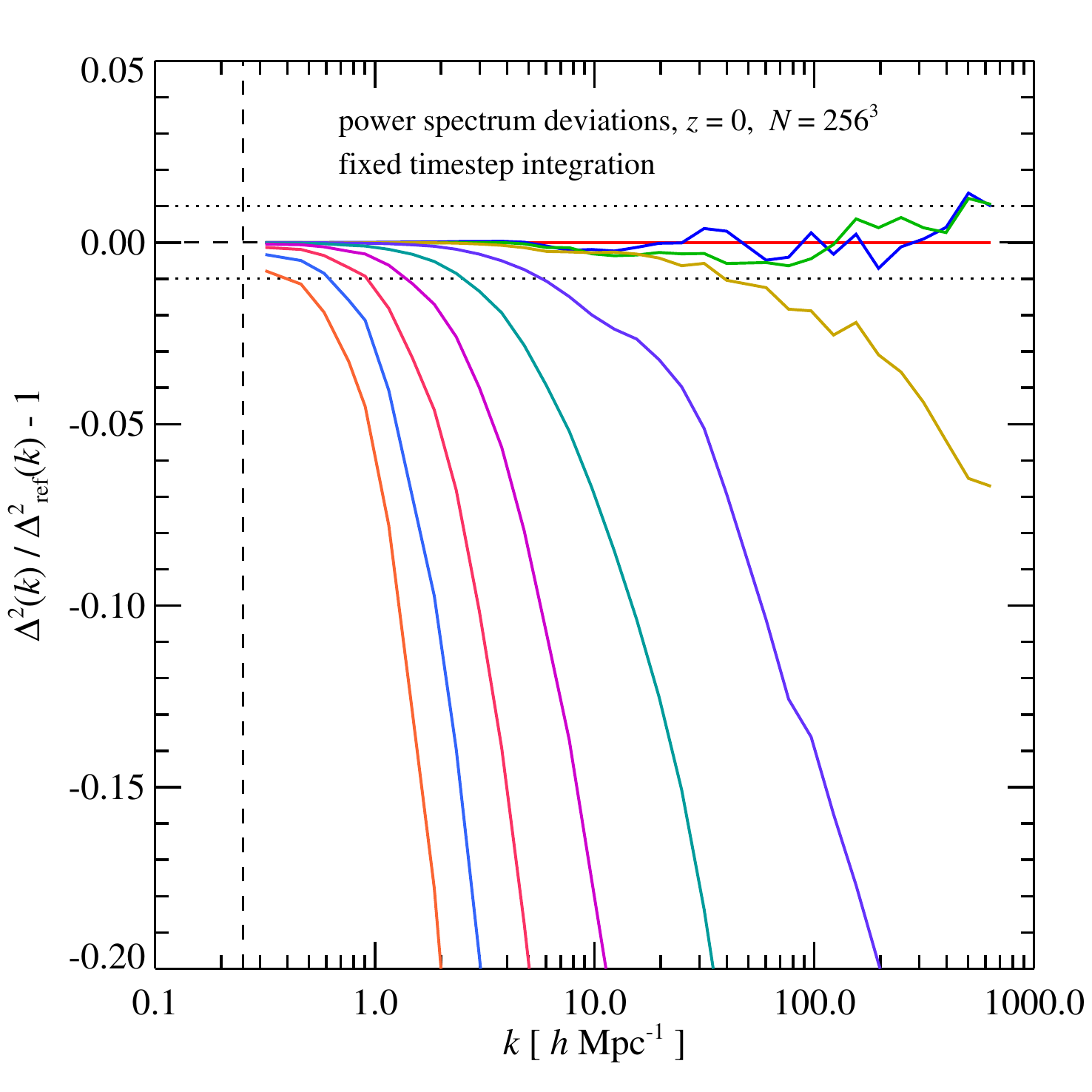}}%
  \end{center}
  \caption{Non-linear power spectra obtained at $z=0$ for
    a cosmological simulation when different fixed timestep sizes $\Delta
    \ln a$ (left panel, as labelled) are used. The simulations followed $256^3$ particles in
    a periodic box of $L=25\,h^{-1}{\rm Mpc}$, with  
    a starting redshift of $z=99$.
    In all the calculations, the force accuracy was chosen
    conservatively, and assured to
    be better than a maximum relative error of order $10^{-3}$.
    Spatial randomization after every step has been used, too, i.e.~the results 
    should be to good approximation unaffected by force errors, and
    thus only test the time
    integration accuracy. 
    The inclined grey line in the panel on the left shows the shot noise limit, which
    has been subtracted from the measurements. The dashed vertical
    line indicates the fundamental mode of the boxsize, and 
    the numbers in
    parentheses after the timestep sizes give the total
    number of time steps
    taken by the corresponding simulation.
    The panel on the right hand side shows the 
    deviations of the power spectra at $z=0$ relative to the
    simulation with the smallest timestep, which we identify as
    a fiducial simulation that is
    fully converged in terms of timestepping accuracy. We see that
    timesteps that are 2 or 4 times larger than used for this run give essentially
    indistinguishable results; the maximum difference of their power
    spectra stays within 1\% (dotted horizontal lines), and no
    systematic trend in these deviations is apparent. This is
    different for the simulation with an 8 times larger stepsize
    (i.e.~the run with only 4096 steps in total). Its power on the
    smallest scales starts to show a systematic deficit of up to $\sim
    5\%$
    for the
    highest resolved $k$. For still
    larger timesteps, this deficit rapidly becomes very large in the
    non-linear regime, rendering these simulations hopelessly
    inaccurate.
    Interestingly, modes
    that stay in the linear regime (for this box-size
    this is basically only the fundamental mode)  
    are reasonably well followed even for just 64 steps. 
    \label{FigDiffPSForFixedDt}}
\end{figure*}

\begin{figure*}
  \begin{center}
    \resizebox{8.5cm}{!}{\includegraphics{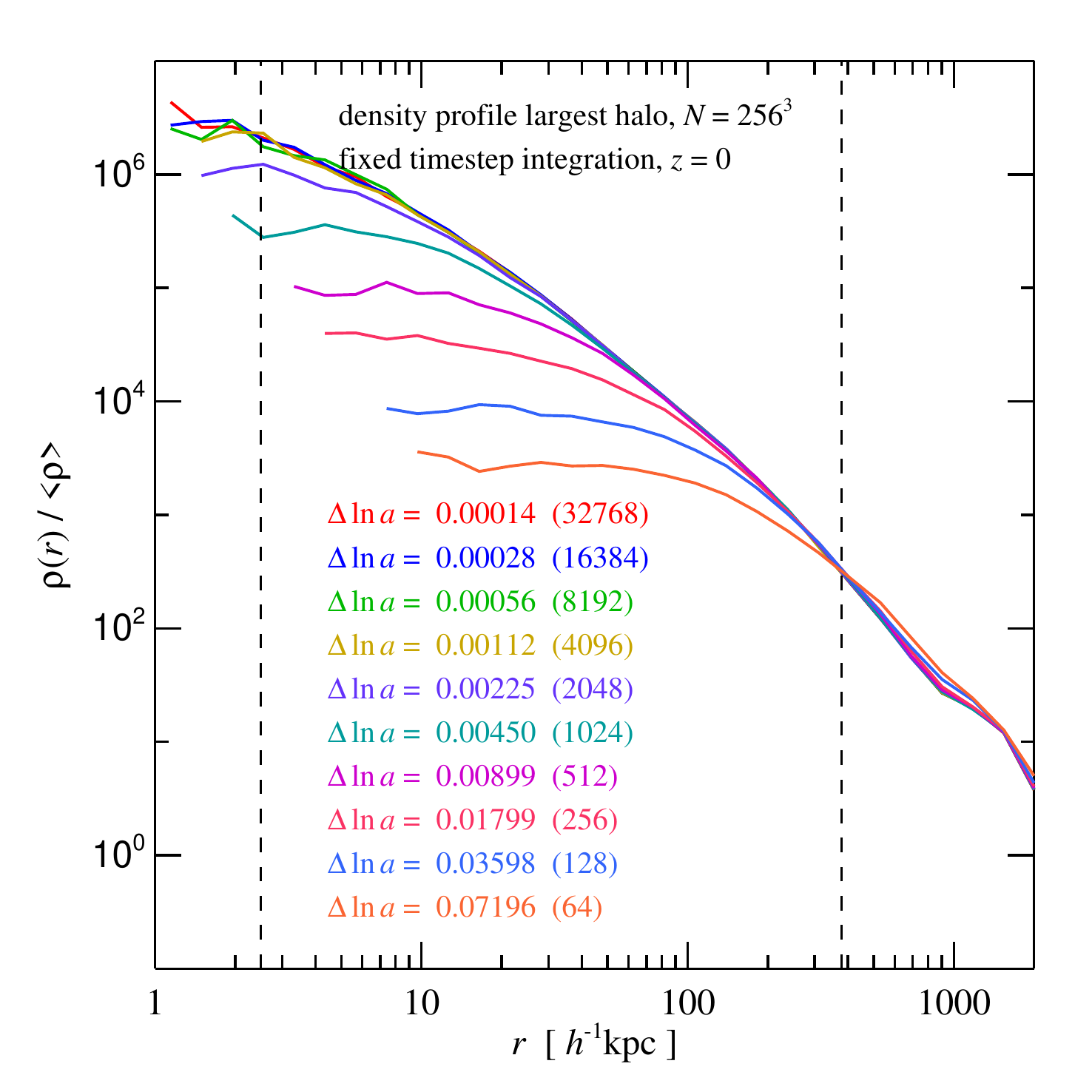}}%
    \resizebox{8.5cm}{!}{\includegraphics{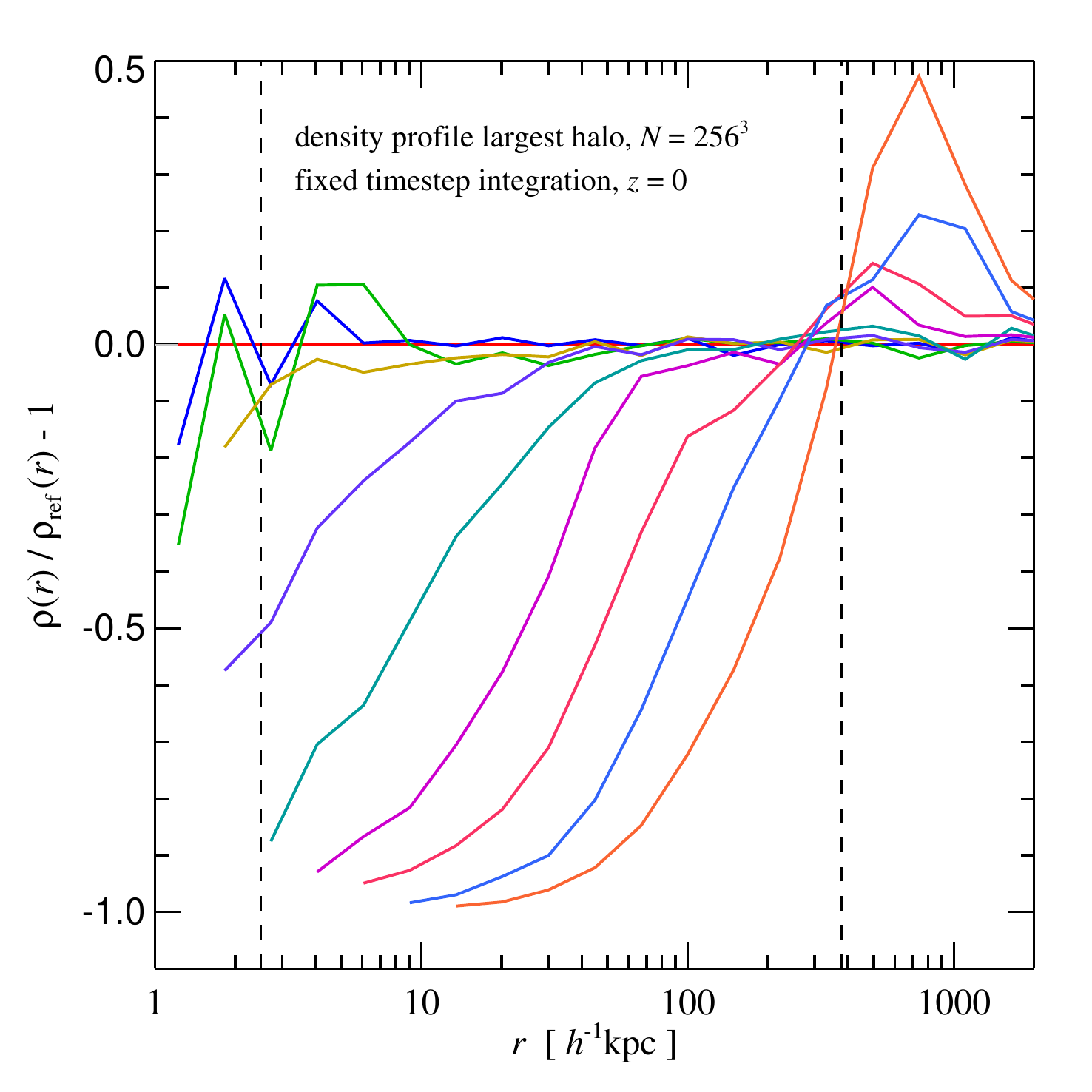}}
  \end{center}
  \caption{Spherically averaged density profile of the largest
    halo in simulations integrated with different fixed timesteps $\Delta \ln a$
    (the same runs are shown in Fig.~\ref{FigDiffPSForFixedDt}),  evolved
    from a starting redshift of $z=99$ to $z=0$. These simulations
    have been run with high force accuracy so that they are only
    affected by errors from the time integration. The vertical dashed
    lines mark the softening length and the virial radius (enclosing
    200 times the critical density) of the target halo, respectively.
    The left panel shows the plain density profiles, whereas the
    right panel gives deviations with respect to the result obtained with the
    highest time integration accuracy. We again see that that the runs
    with $\Delta \ln a = 0.0028$ and $\Delta \ln a = 0.0056$ are still
    converged, whereas $\Delta \ln a = 0.00112$ begins to show a
    central density suppression. Worse time integration manifests
    itself in reduced central halo densities, as expected.
    \label{FigDiffHaloForFixedDt}}
\end{figure*}

\begin{figure*}
  \begin{center}
    \resizebox{8.5cm}{!}{\includegraphics{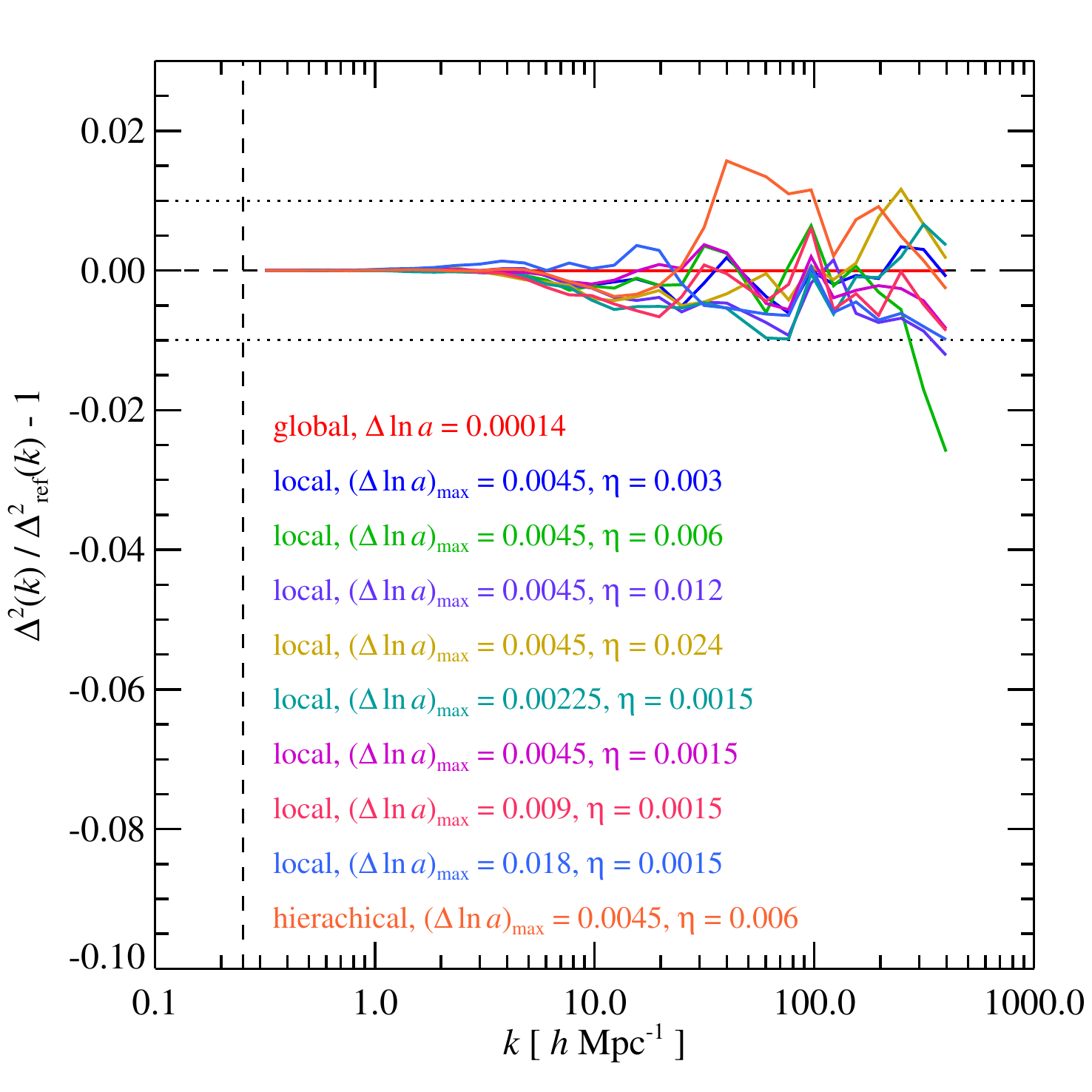}}%
    \resizebox{8.5cm}{!}{\includegraphics{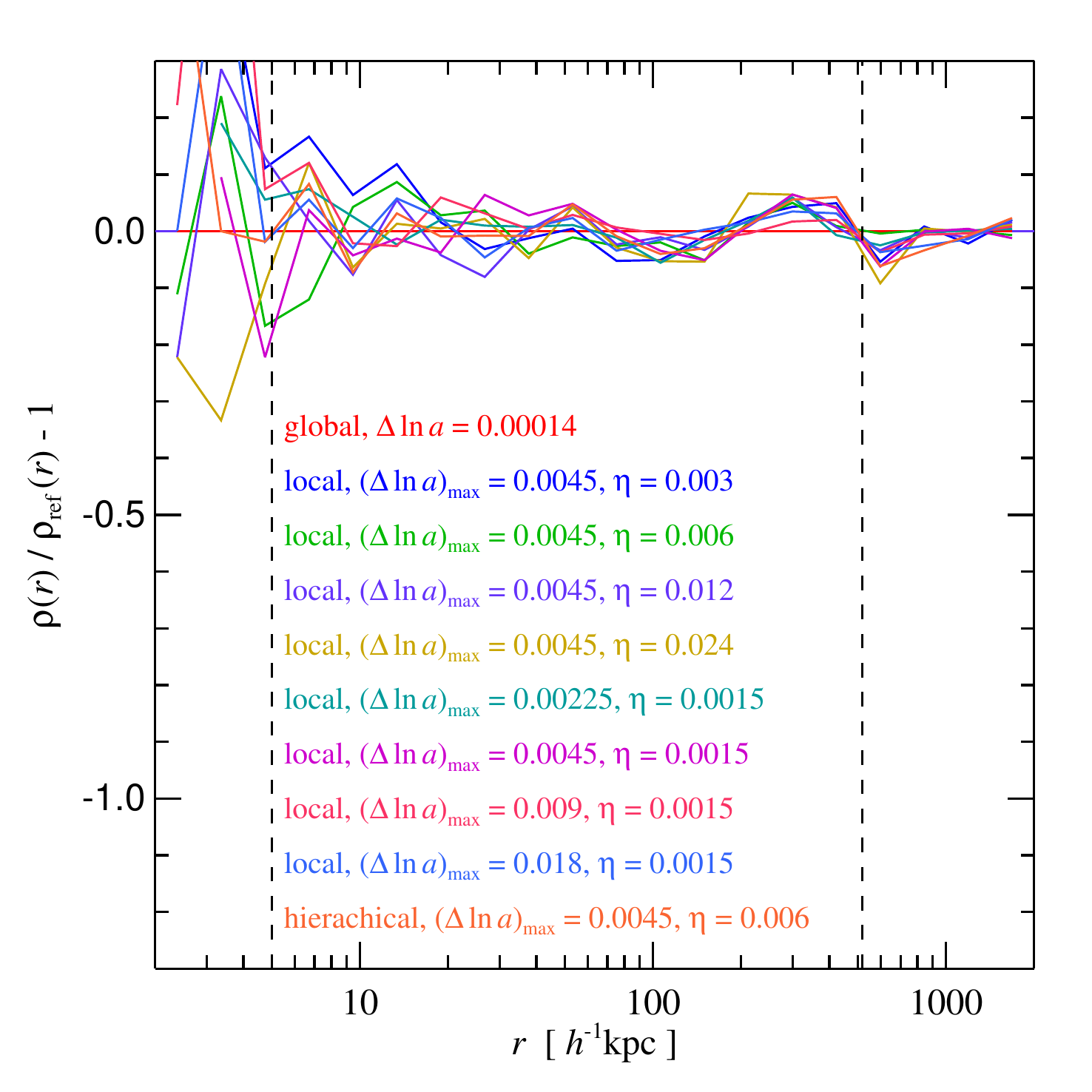}}\\
  \end{center}
  \caption{Relative difference of matter power spectrum (left panel) and the
    density profile of the largest
    halo (right panel) of simulations carried out with different local timestepping
    settings when compared 
    to a run done with a very small global timestep. The simulation
    specifications are the same as for the runs considered in
    Fig.~\ref{FigDiffHaloForFixedDt}, and we in fact use the
    run with a fixed $\Delta \ln a = 0.0014$ from that series as our fiducial
    `ground truth' result here. The local timestepping schemes are
    specified by a parameter $\eta$ controlling the timestep sizes of
    individual particles,
    plus an additional maximum allowed timestep size $(\Delta \ln
      a)_{\rm max}$. We consider 8 different choices for these
    settings, and also include one realization of the hierarchical
    time integration scheme. Within the noise, all runs give
    results of acceptable accuracy, without evidence for clear
    systematic deviations, suggesting a reassuring robustness
    of these time integration schemes. 
    \label{FigLocalSteppingN128}}
\end{figure*}

\subsection{Cosmological initial conditions creation}

Another feature we integrated into {\small GADGET-4} is the ability to
create cosmological initial conditions, a functionality we previously
provided in the stand-alone {\small N-GENIC} code. In particular, it
is now possible to create the initial conditions at start-up of a
simulation run, avoiding the need to create them first in a separate
step. This also avoids the need to store them on disk (although this
can be done, if desired). It is also possible to create initial
conditions in stretched cuboids, not only in cubic periodic boxes.

Technically, the code uses either the Zel'dovich approximation or
second-order Lagrangian perturbation theory \citep{Scoccimarro:1998aa,
  Scoccimarro:2012aa}, and sets up a realization of the power spectrum
in Fourier space. It thus implements similar functionality as the
{\small 2LPTIC} code\footnote{https://cosmo.nyu.edu/roman/2LPT}
originally based on {\small N-GENIC}. Both Cartesian grids and glass
distributions \citep{White:1996aa} for the initial particle load are
supported. For the latter, one can also opt to enable a deconvolution
filter that accounts for the smoothing effect of interpolating the
displacement field from a grid to the irregular particle
distributions. By default, a given initial conditions seed ties down
the density field for a given box everywhere, i.e.~if the simulation
resolution is increased later on, the large-scale phases are retained
and one gets the same large objets and cosmic web as before. Note that
the maximum size of Fourier meshes that can be realized in practice
limits the dynamic range of this approach for zoom simulations. It
would thus be worthwhile to integrate the much more sophisticated
{\small PANPHASIA} approach \citep{Jenkins:2013aa} for defining the
phases in real-space in the future.  The initial conditions creation
is fully parallel and able to create essentially arbitrarily large
initial conditions, either with dark matter only, or with dark matter
and gas. If desired, the variance suppression technique by
\citet{Angulo:2016aa} can also be enabled.

\begin{figure*}
  \begin{center}
    \resizebox{8.5cm}{!}{\includegraphics{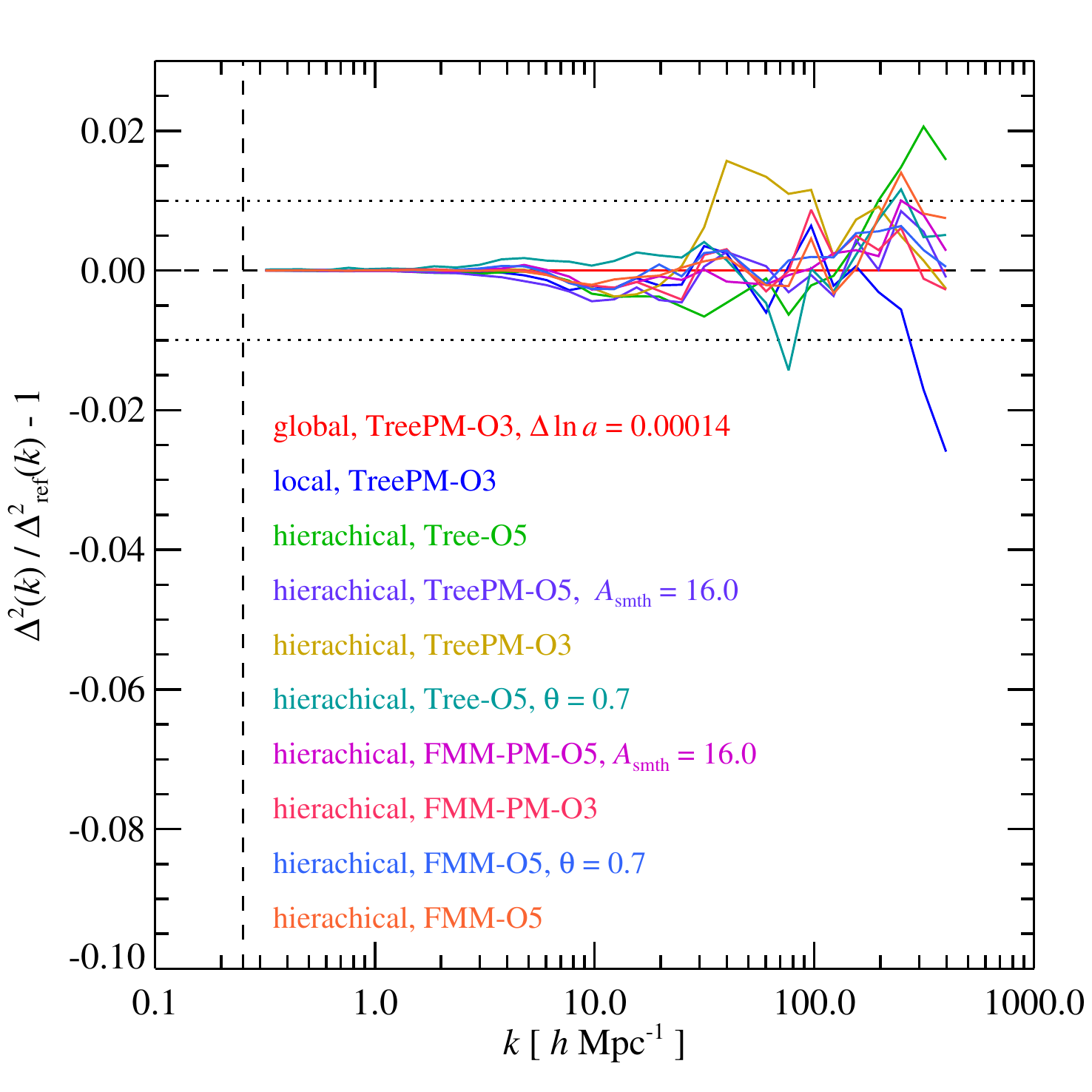}}%
    \resizebox{8.5cm}{!}{\includegraphics{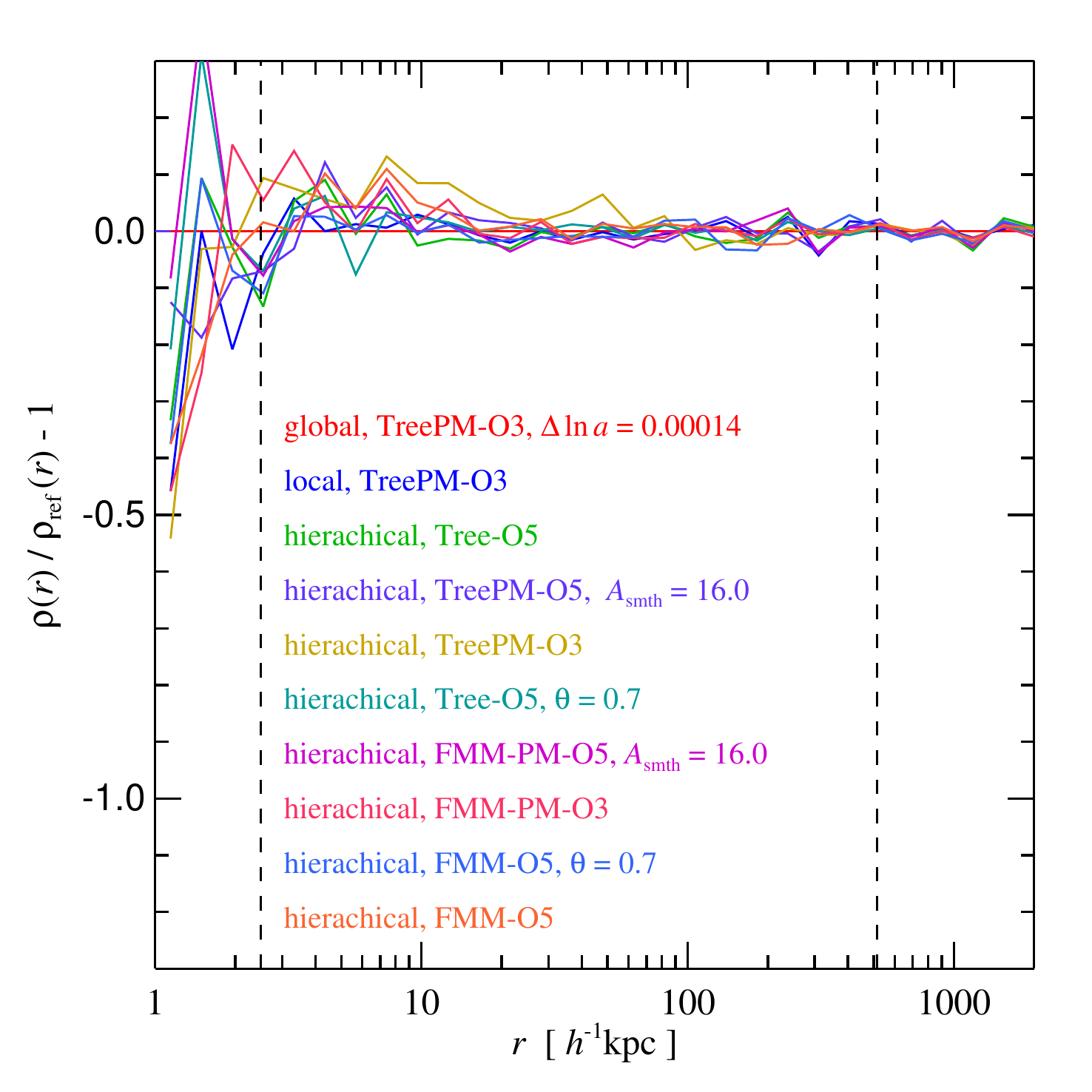}}\\
  \end{center}
  \caption{Similar to Fig.~\ref{FigLocalSteppingN128}, we here consider differences in the
    matter power spectrum (left panel) and the
    density profile of the biggest
    halo (right panel) when run with different code settings, but we
    now examine variations induced by the use of different force calculation schemes. The
    results are
    evaluated  relative to a run done with small global timesteps and
    high force accuracy, as
    before. All other runs are variations therefore carried out with different
    force calculation schemes,  as labelled, and have 
    either used local or hierarchical timestepping with $\eta = 0.006$ and a maximum timestep of  
    $(\Delta \ln a)_{\rm max} = 0.0045$. Again, we find
    the results to be robust under these changes.
    \label{FigForceSchemesN128}}
\end{figure*}

\begin{figure*}
\begin{center}  
\resizebox{4.5cm}{!}{\includegraphics{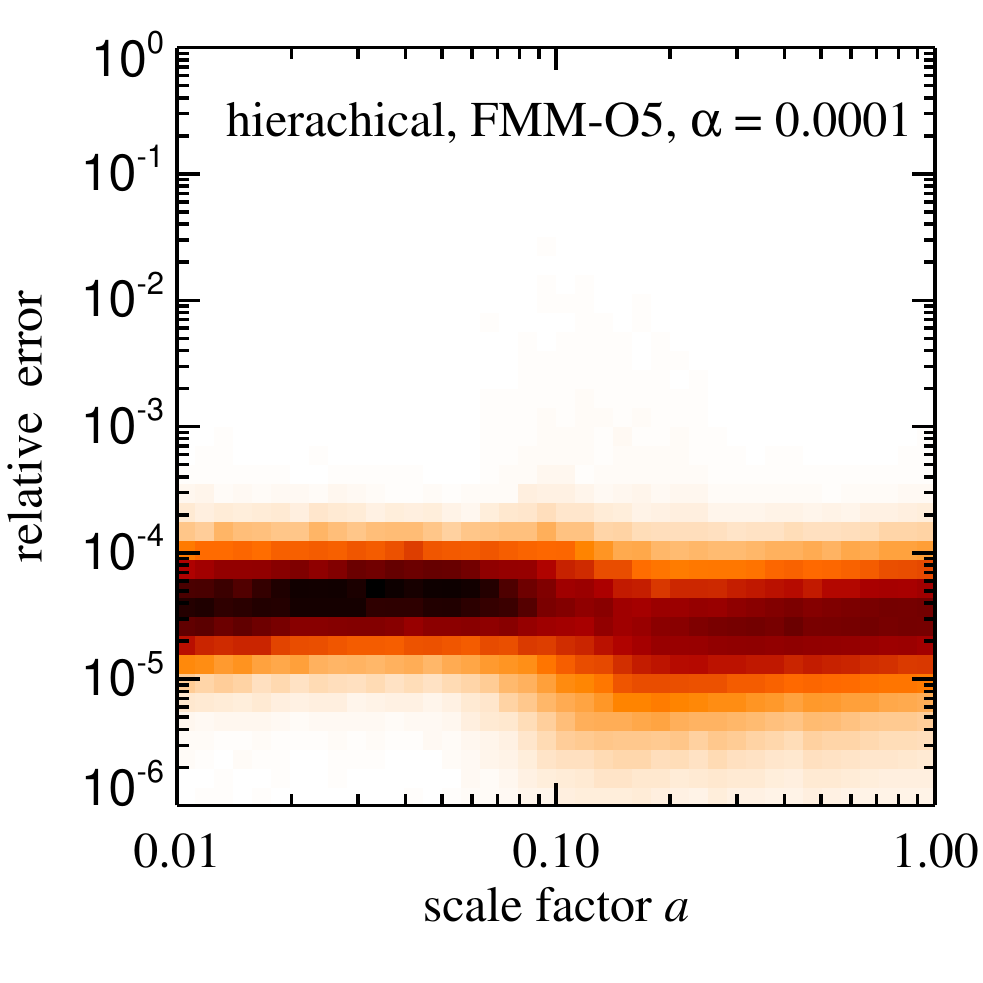}}%
\resizebox{4.5cm}{!}{\includegraphics{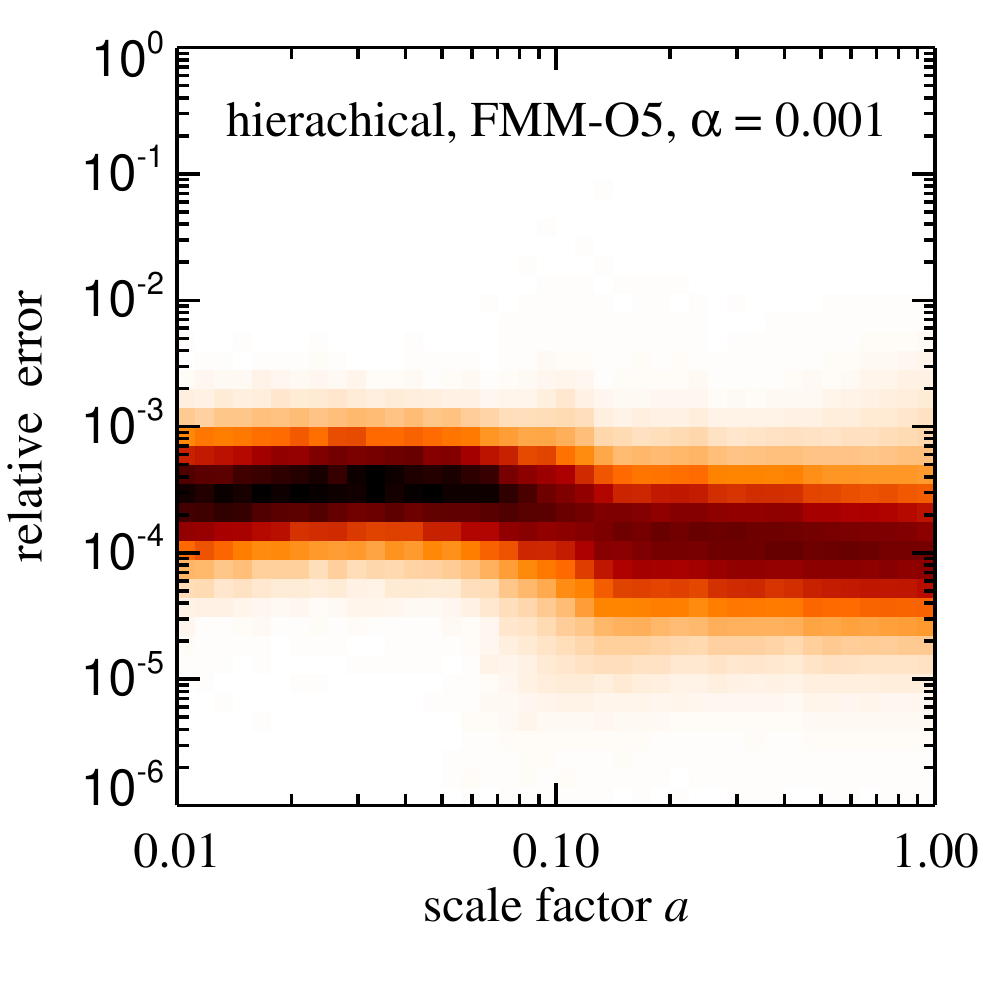}}%
\resizebox{4.5cm}{!}{\includegraphics{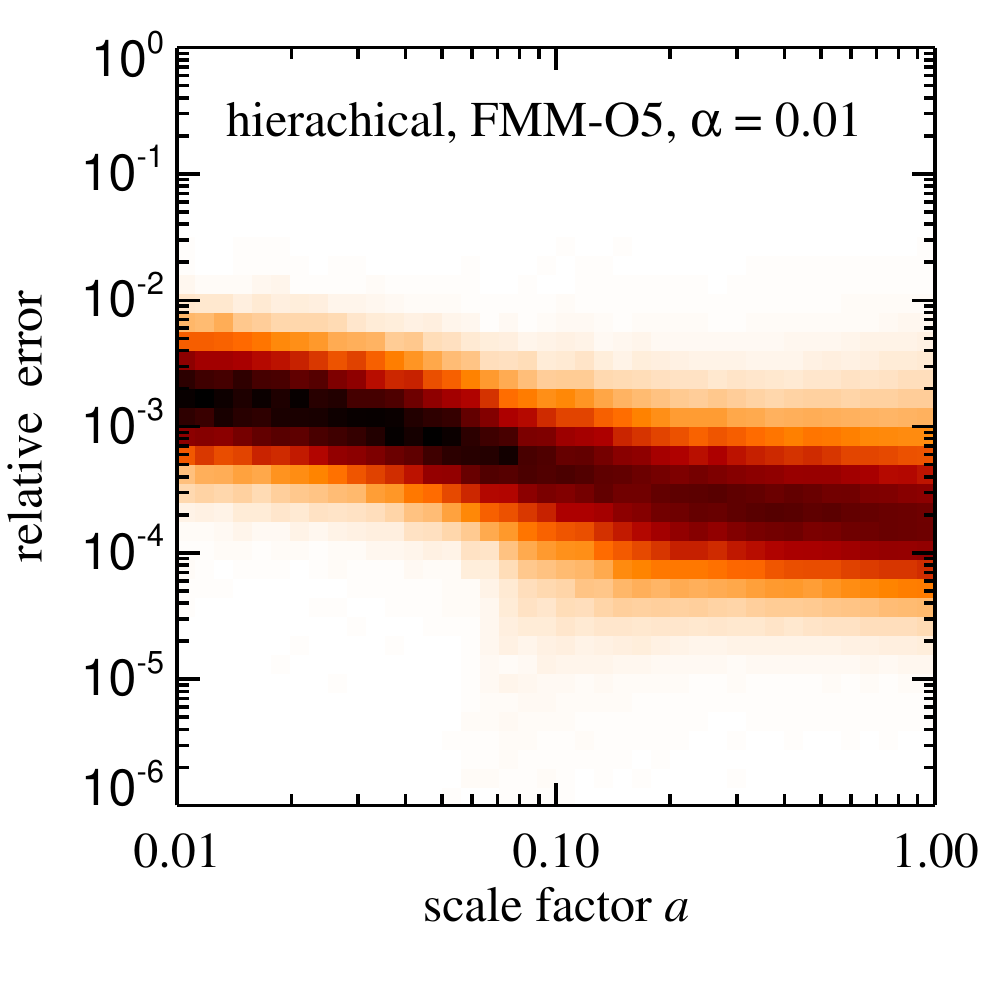}}%
\resizebox{4.5cm}{!}{\includegraphics{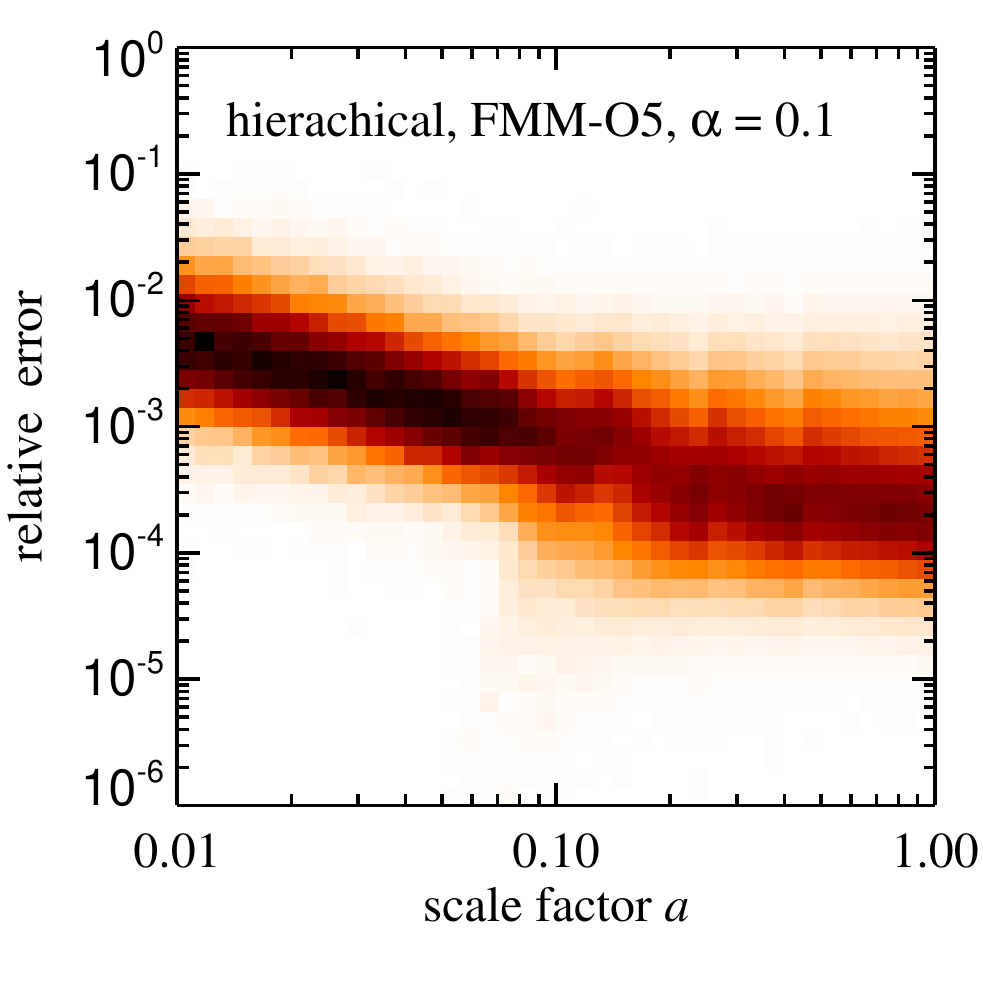}}\\
\resizebox{9.0cm}{!}{\includegraphics{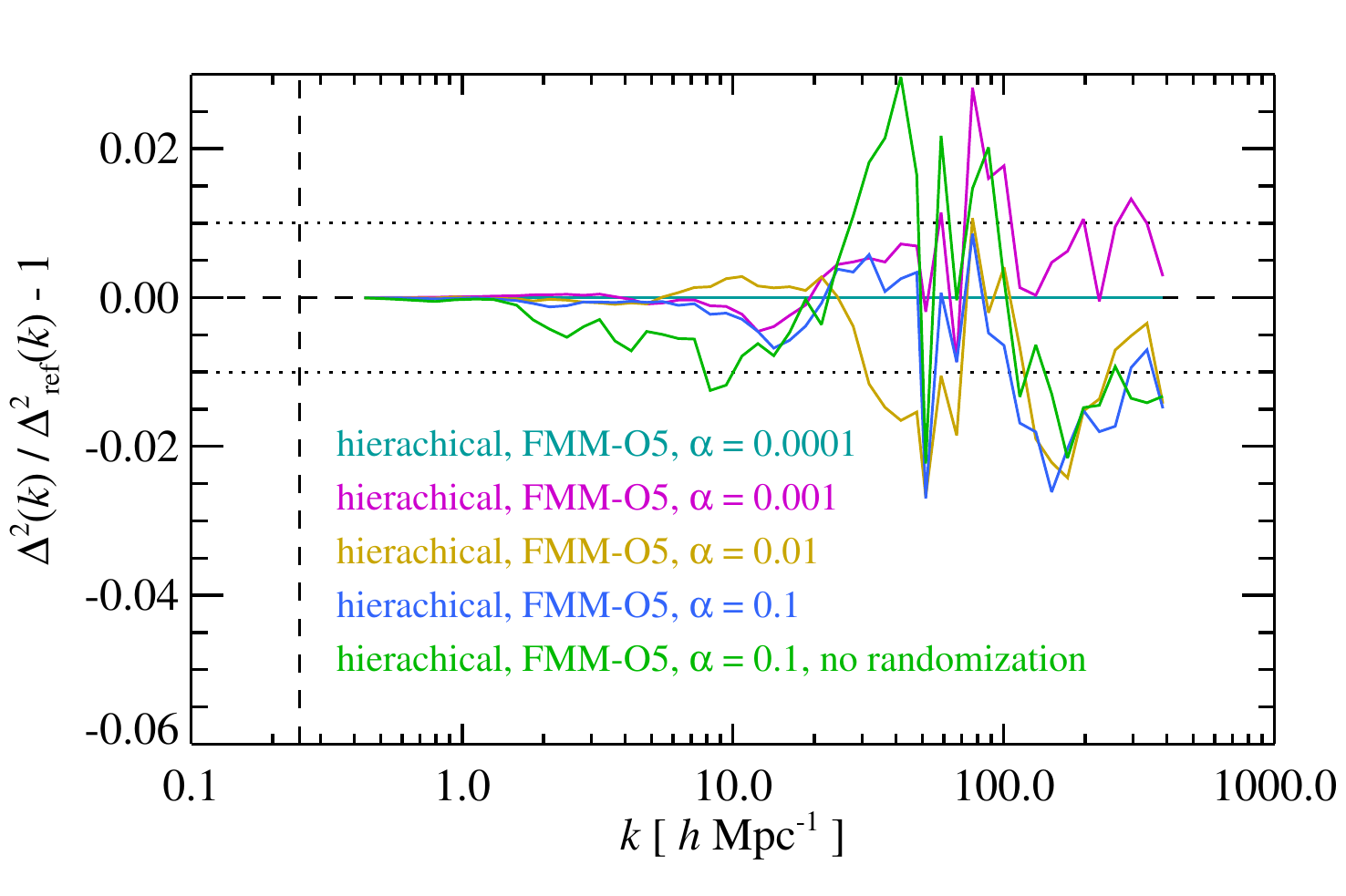}}%
\resizebox{9.0cm}{!}{\includegraphics{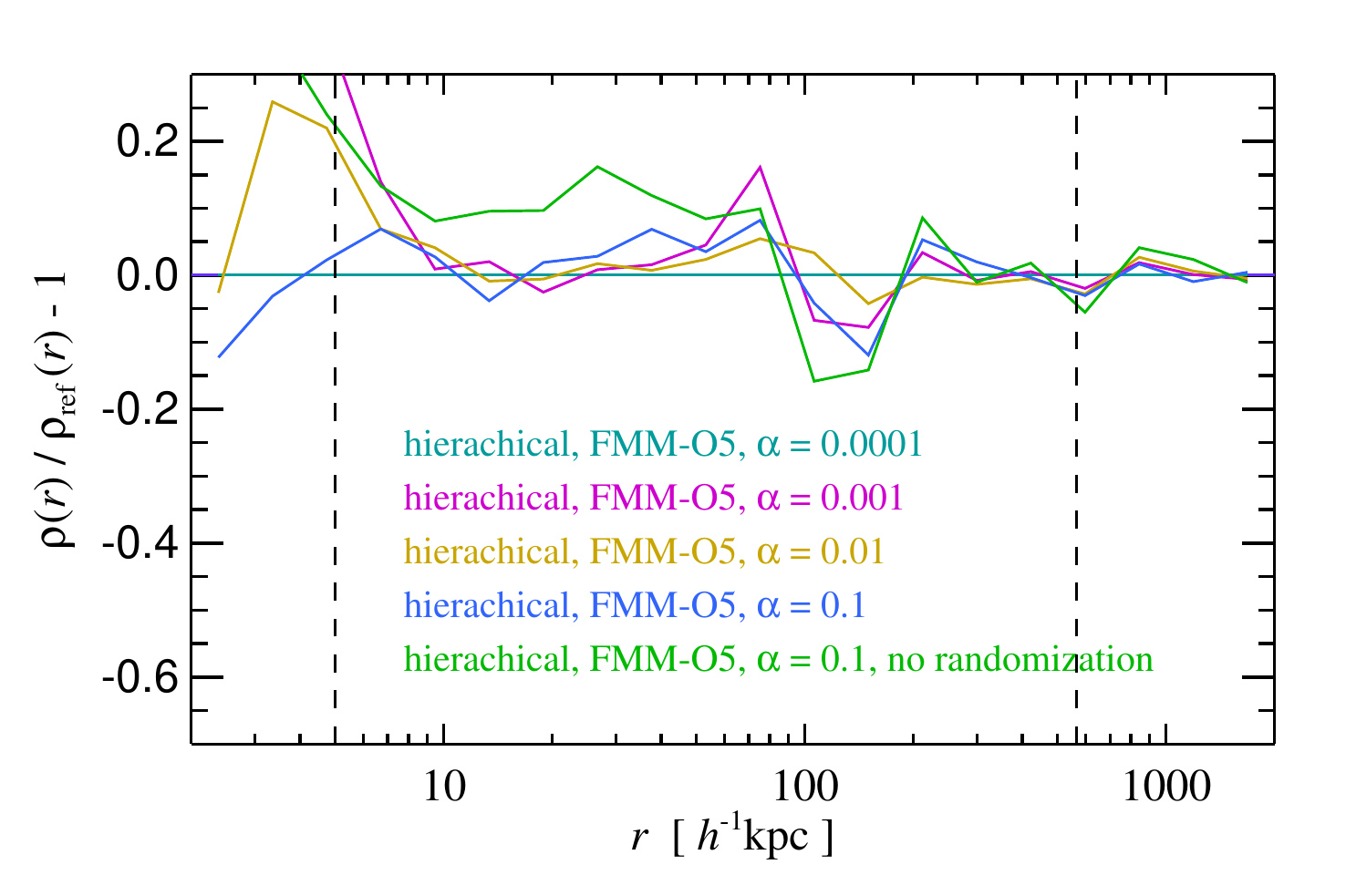}}
\end{center}
\caption{Sensitivity of the results of a cosmological simulation with
  $256^3$ particles to force accuracy. We here consider simulations
  carried out with the FMM-O5 scheme and hierarchical time
  integration. Our fiducial reference run uses very high force
  accuracy with $\alpha=0.0001$, and we then systematically degrade
  the force accuracy in several steps with values of $\alpha=0.001$,
  $\alpha=0.01$, and $\alpha=0.1$. Random translations are applied
  after every step with a new domain decomposition to reduce temporal
  correlations of force errors, except for a second run with
    $\alpha=0.1$ where no shifts of the particle distribution are
    applied, as labelled.  The top panels give the force accuracy as
  a function of scale factor for the the four settings of
    $\alpha$, measured on-the-fly by comparing to direct summation
  forces for random sub-sets of particles that are newly drawn every
  step. Here the two runs with $\alpha=0.1$ with and without
    randomization are indistinguishable.  The bottom left panel
  compares the power spectrum of the four lower accuracy
  runs with the highest accuracy run in the series, while the bottom
  right panel carries out this comparison for the density profile of
  the most massive halo in the box. The residual differences among the
  simulations are consistent with expectations from Poisson particle
  noise, except for the run with $\alpha=0.1$ without
    randomizations, which shows significant systematic
    differences with respect to the highest force accuracy run.  This
  confirms that unbiased and random
  force errors are ultimately less important for collisionless dynamics than time
  integration errors, and that random shifts of the particle
    set are helpful in establishing sufficiently random distributions
    of the residual force errors.
    \label{FigReducedForceAccuracy}}
\end{figure*}

\begin{figure}
  \resizebox{8.5cm}{!}{\includegraphics{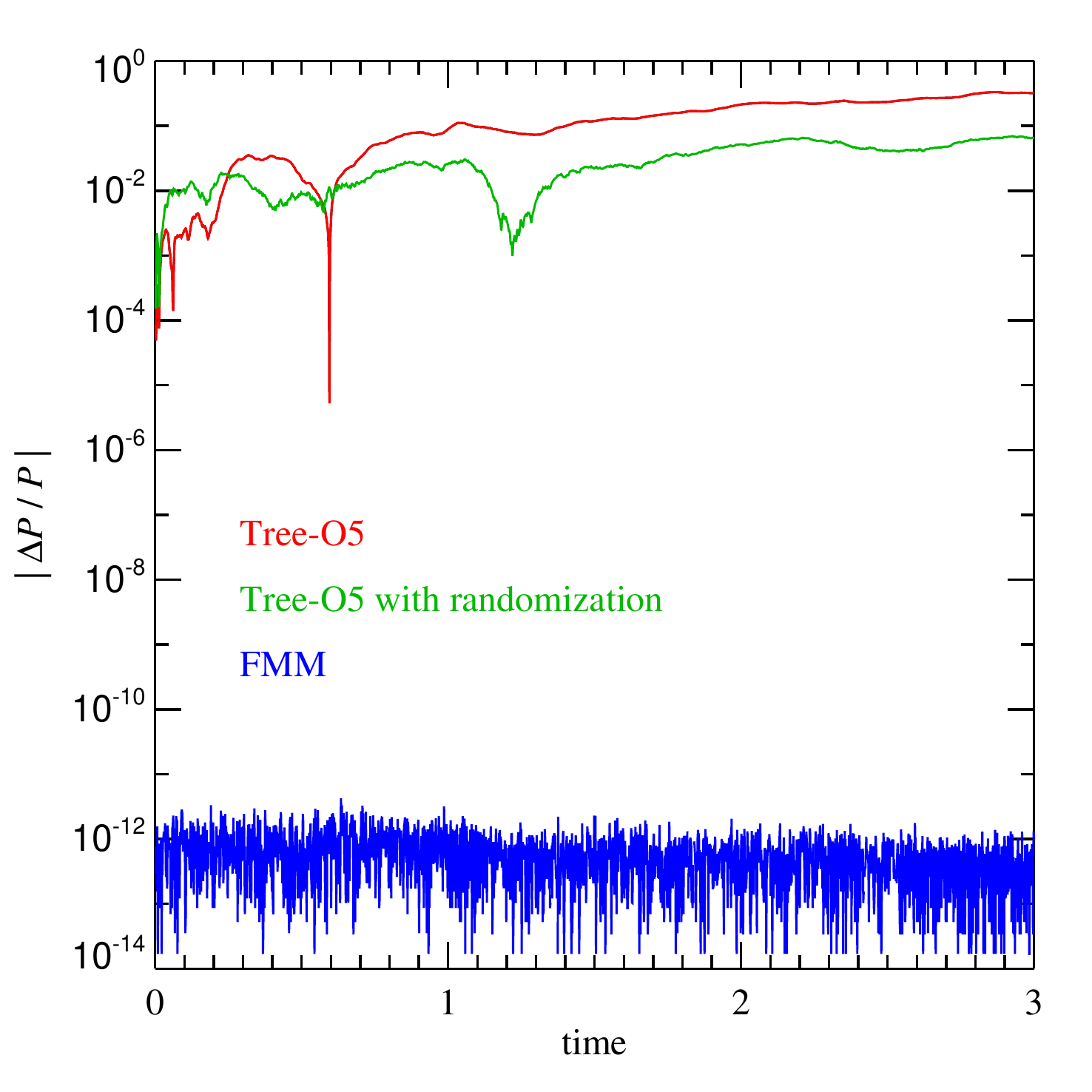}}%
\caption{Total momentum error in FFM and tree-based calculations of a
  galaxy merger simulation. All three simulations use the hierarchical
  time integration scheme with local timesteps, and an expansion order
  $p=5$. Despite one-sided momentum updates being avoided in this case,
  the tree-based calculation still shows a secular drift of the total
  momentum of the system, due to residual small force errors that do not
  all add up to zero. Note
  that the total momentum of the system is very small here (the system is meant to
  be at rest), so measuring its relative error provides a particularly
  sensitive test. 
  If random translations of
  the particle set relative to the root node are carried out
  after every step, the size of the error found for the Tree-based
  calculation is
substantially reduced, but it remains present at a reduced level. In
contrast,
with FMM, the error
vanishes entirely and
the total momentum is
preserved to machine precision even in the presence of local timestepping,
as expected.
\label{FigMomentumDrift}}
\end{figure}

\begin{figure}
  \begin{center}
    \resizebox{8.5cm}{!}{\includegraphics{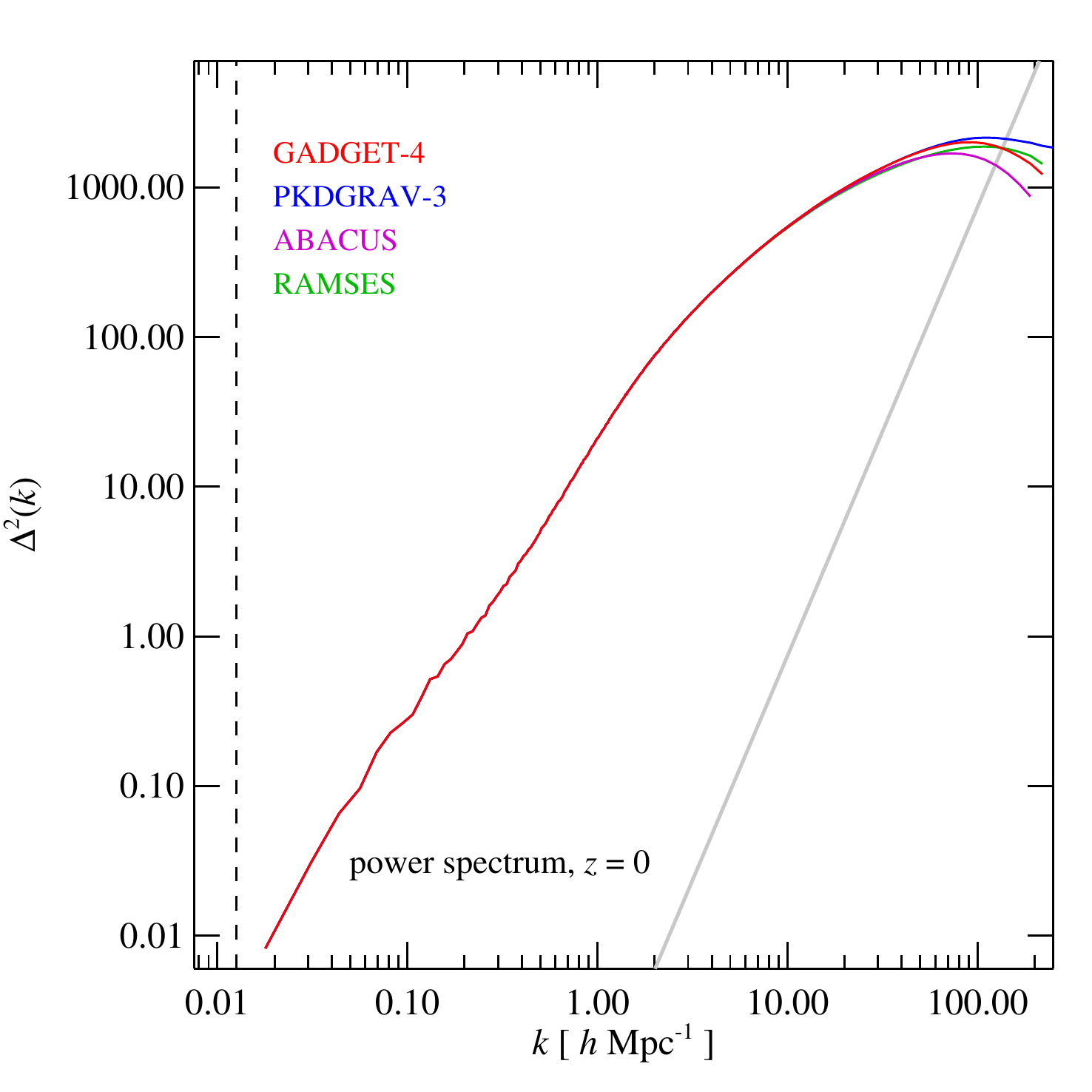}}\\%
    \resizebox{8.5cm}{!}{\includegraphics{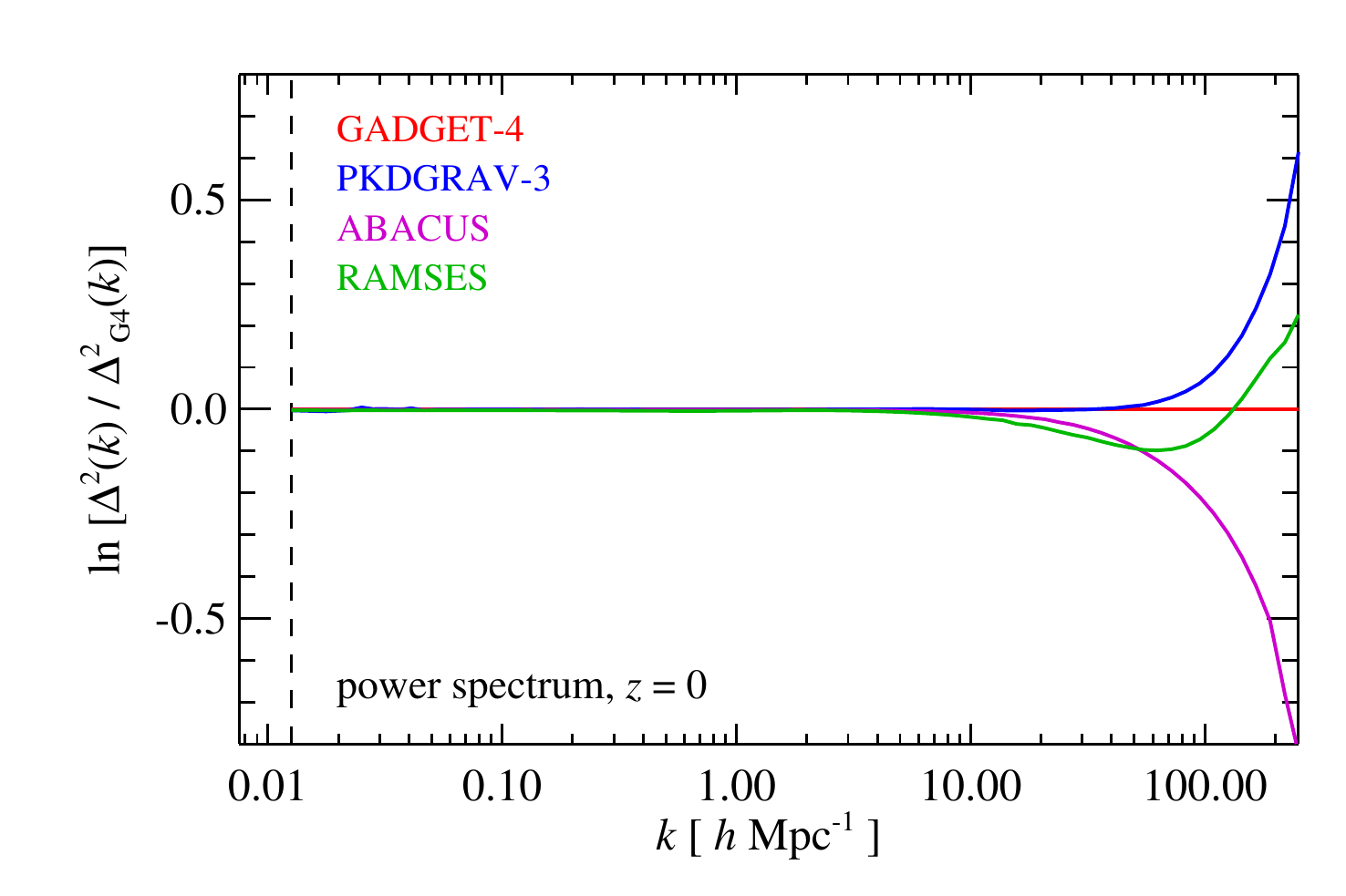}}\\%
    \resizebox{8.5cm}{!}{\includegraphics{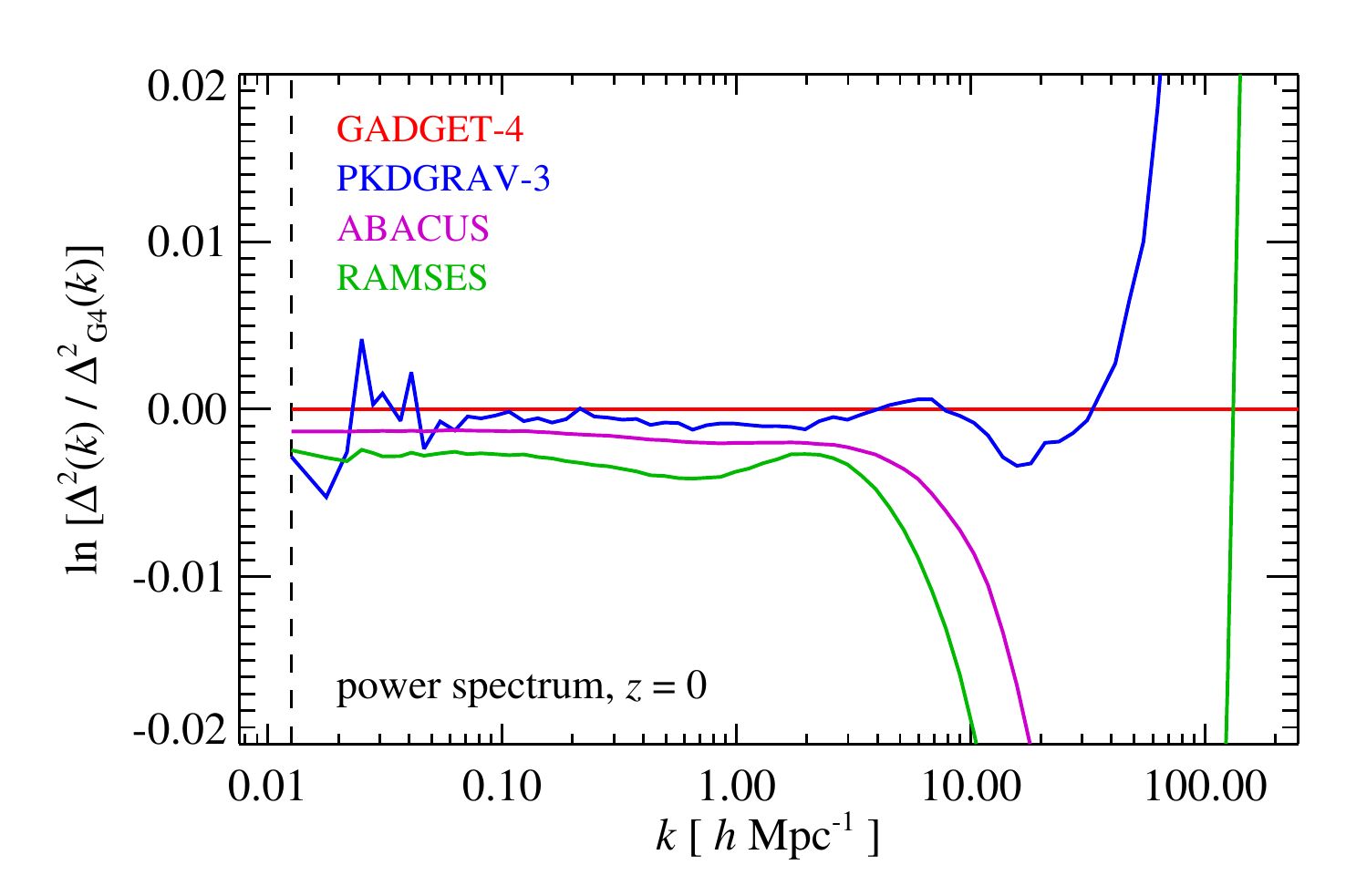}}
\end{center}
\caption{Power spectrum of the `Euclid test simulation' discussed by 
\citet{Schneider:2016aa}. We compare our measurement for {\small
  GADGET-4} to the results reported for {\small PKDGRAV-3} and {\small
  RAMSES} in that paper, and we also include a recent recalculation
by \citet{Garrison:2019aa} with their code {\small ABACUS}. All power spectra
have been remeasured by us in identical ways from the $z=0$ particle
data of the different runs. The top panel shows the power spectra of
the different runs with the shot noise subtracted, the bottom two panels
give the fractional differences with respect to the {\small
  GADGET-4} result, with the vertical axis range adjusted to focus on
small-scales (middle panel) and large-scales (bottom), respectively.
\label{FigEuclidReferenceSim}}
\end{figure}

\begin{figure*}
  \begin{center}
    \resizebox{6.0cm}{!}{\includegraphics{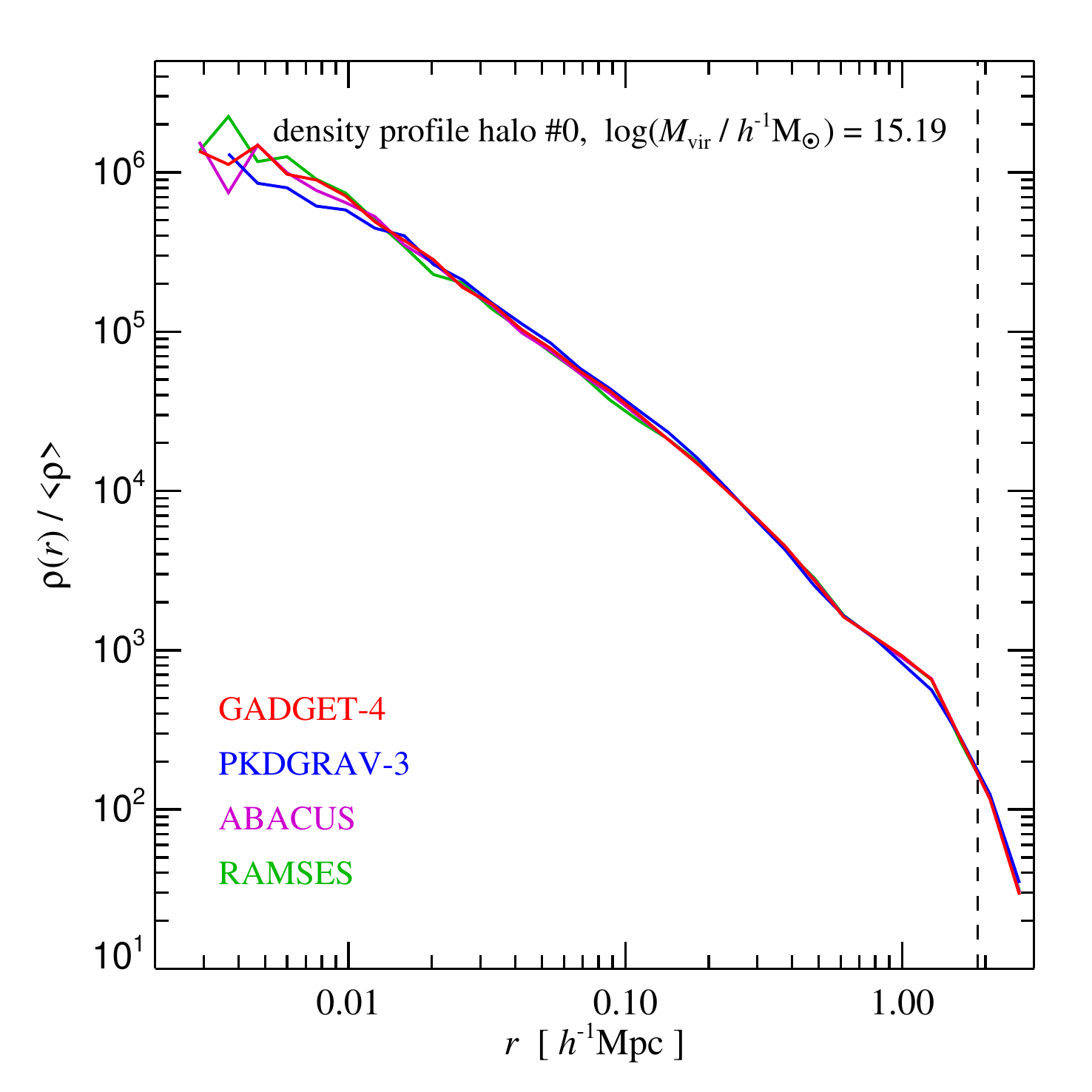}}%
    \resizebox{6.0cm}{!}{\includegraphics{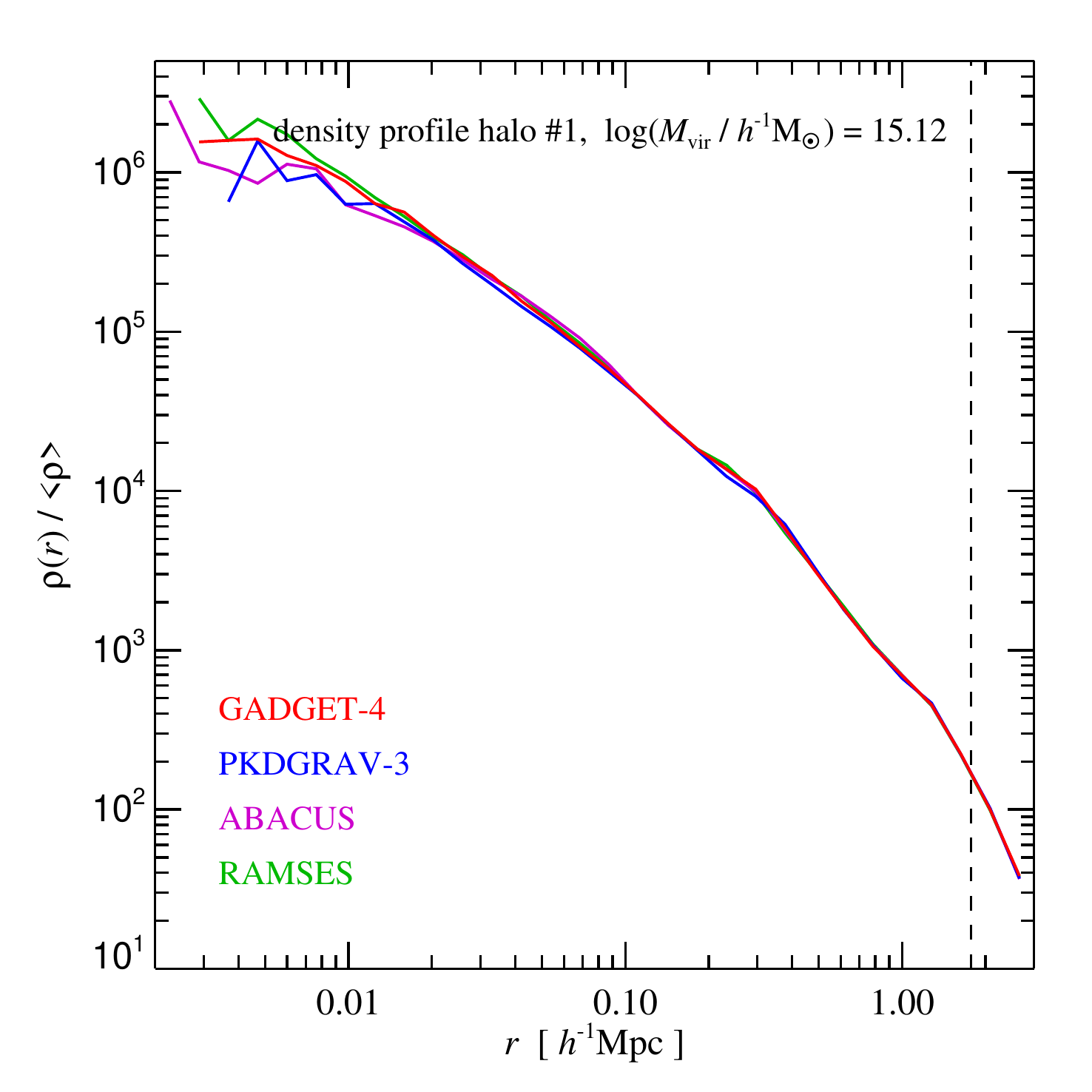}}%
     \resizebox{6.0cm}{!}{\includegraphics{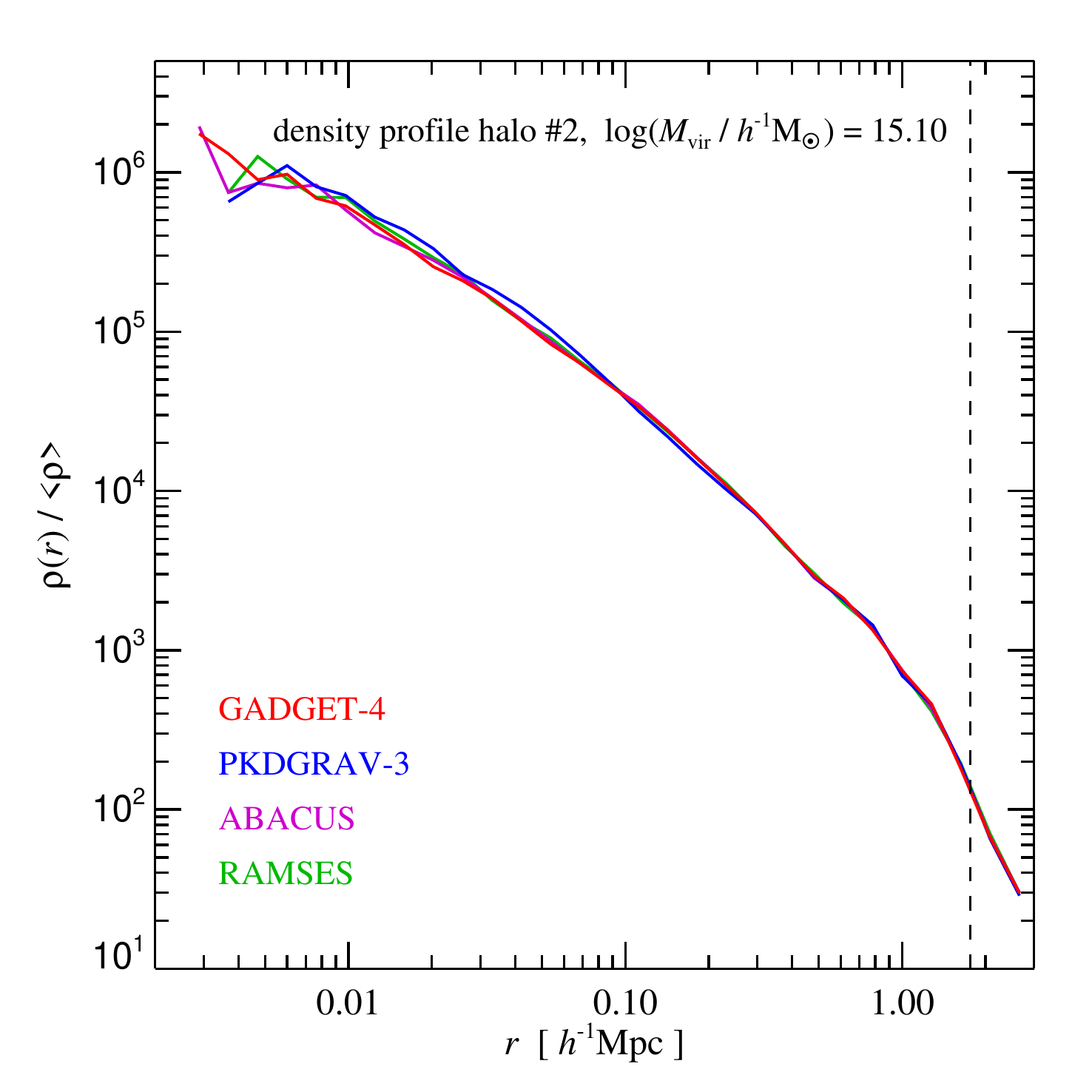}}\\%
    \resizebox{8.5cm}{!}{\includegraphics{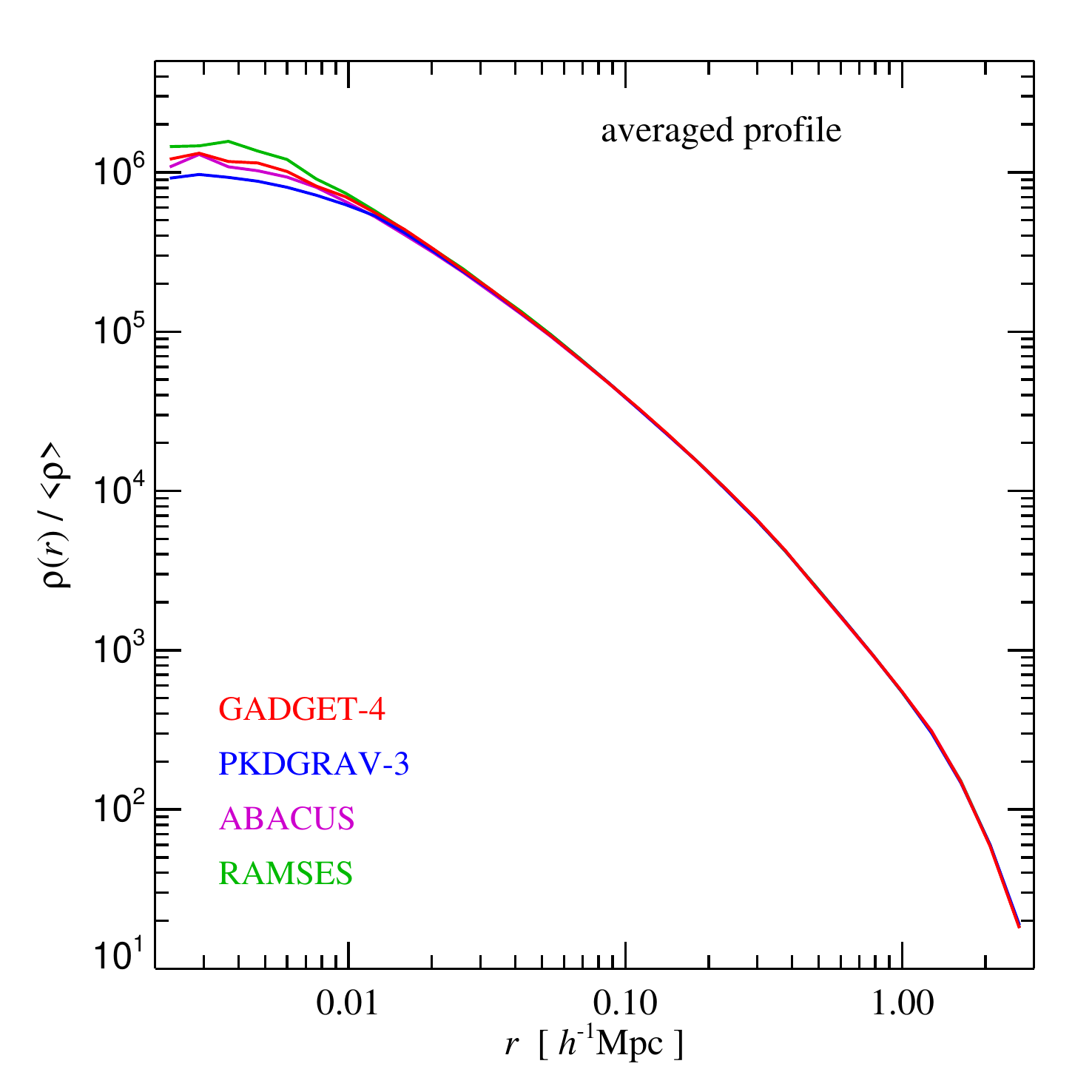}}%
    \resizebox{8.5cm}{!}{\includegraphics{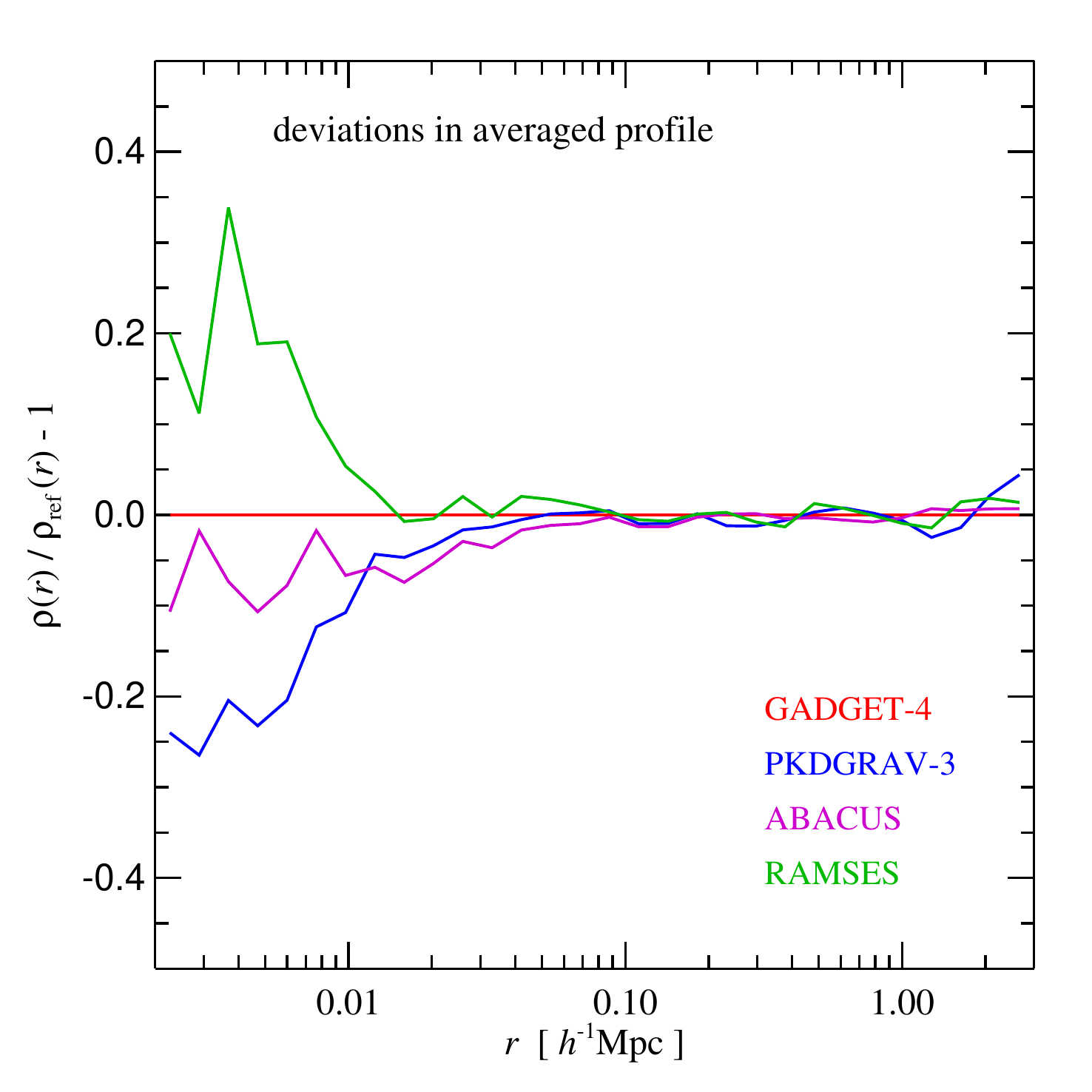}}\\%
\end{center}
\caption{Comparison of halo density profiles for the `Euclid test
  simulation' \citep{Schneider:2016aa, Garrison:2019aa}, using
  different cosmological codes.  We have computed FOF group catalogues
  for the $z=0$ particle data of simulations carried out with the
  {\small GADGET-4}, {\small PKDGRAV-3}, {\small RAMSES}, and {\small
    ABACUS} codes. Group centres were determined as the potential
  minima of the groups in all cases, and the profiles were binned in equal logarithmic
  shells. The top row panels show the density profiles
  for the three halos with the largest virial mass (defined here as
  the mass enclosing a mean overdensity 200 times the critical
  density). The dashed vertical lines indicate their corresponding virial
  radii. In the bottom top left panel, we show the {\em averaged profile} for the 25
  most massive halos to highlight the systematic differences between
  the codes. The deviations of the mean profiles relative to the
  {\small GADGET-4} result are shown in the bottom right panel.
   By matching halo centers, we made sure that the
  same halos are analysed in the four simulations in all cases. For
  simplicity, only particles contained in the FOF halos were included
  in the binning; if all particles would be included, the fall-off
  around and outside the virial radii would be slightly less steep.
  \label{FigEuclidDensityProfiles}}
\end{figure*}

\begin{figure}
\begin{center}
  \resizebox{8.3cm}{!}{\includegraphics{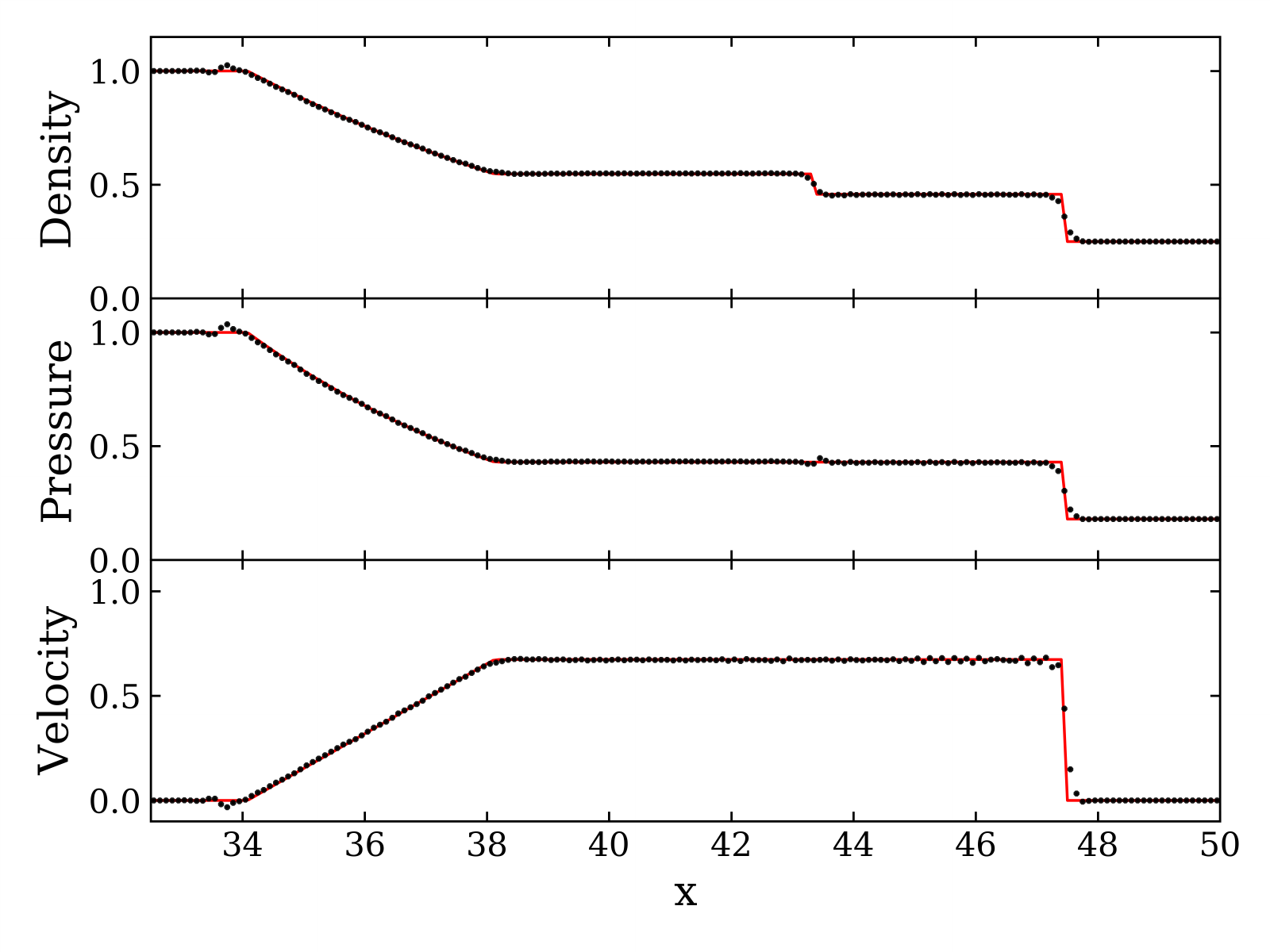}}\\%
  \end{center}
\caption{Shock tube test at time $t=5$ with the Wendland C6 kernel, a
  time-dependent artificial viscosity and density-based SPH. The
  resolution is given by initially $N=1440$ particles in
  the $x$-direction (with $L_x = 80$). For visual clarity, we only show the
  average values in $x$-bins of size $ \Delta x = 0.1$.
  \label{shocktube_result}}
\end{figure}

\begin{figure}
\begin{center}
  \resizebox{8.3cm}{!}{\includegraphics{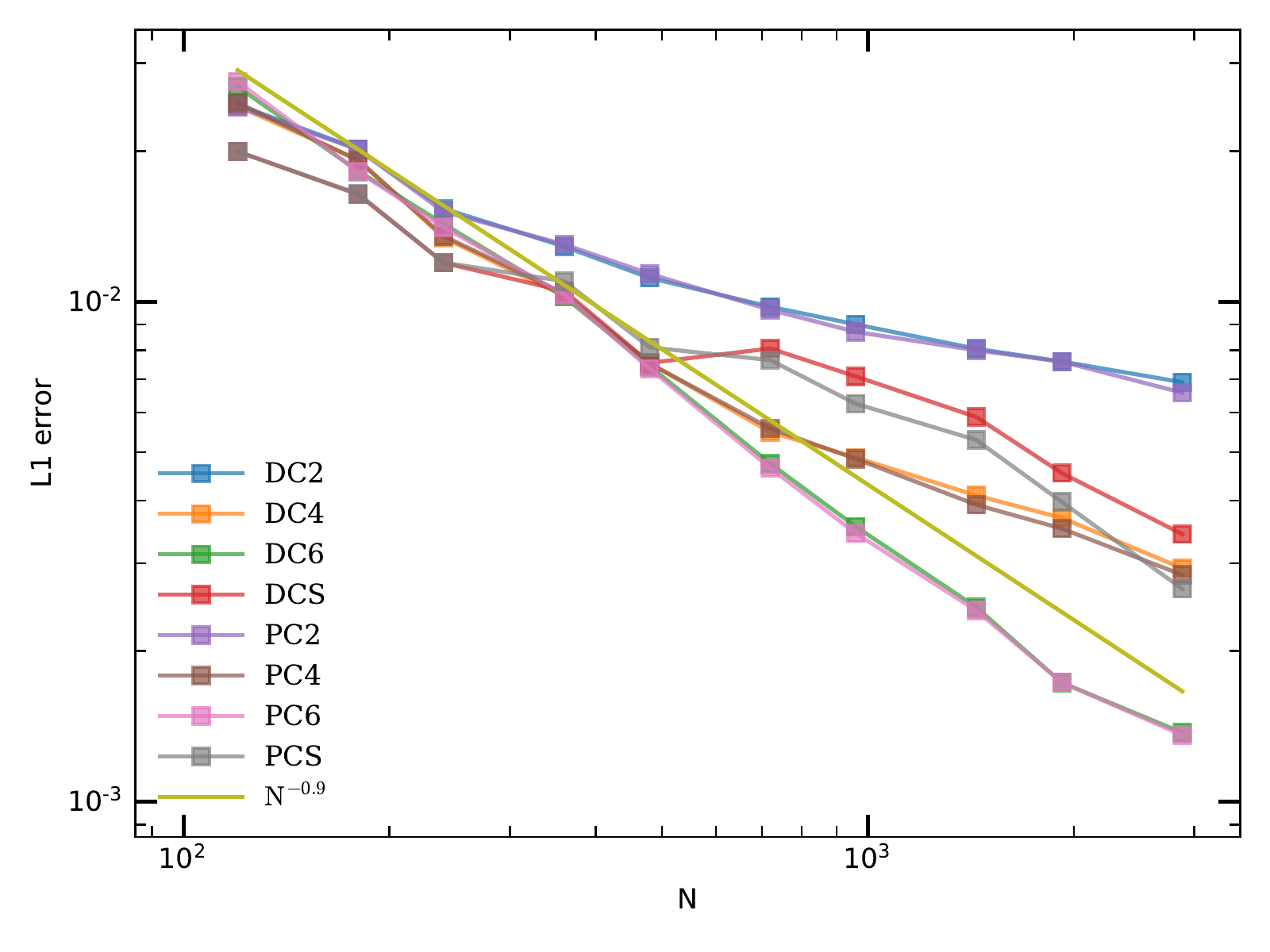}}\\%
  \end{center}
\caption{Convergence rate of the Sod shock tube test problem illustrated
  in Fig.~\ref{shocktube_result}. The L1 error gives the difference to the
  analytic solution at time $t=5$ as a function of resolution 
  for different combinations of kernels and SPH flavours.
  Here $N$ is the initial number of particles in the $x$-direction.
  All simulations were run with time-dependent viscosity, but the
  results for a fixed viscosity are very similar.
  We find a convergence rate $L1 \propto N^{-0.9}$ for the Wendland C6
  kernel, which is close to the optimal value of $L1 \propto N^{-1}$
  \citep{Springel:2010aa} excepted for this problem. The other
  kernels, especially the standard cubic spline kernel, show a poorer
  convergence  rate due to their higher residual particle noise.
  The naming convention of the different results is as follow.
  The first character gives the flavour of SPH (D for density- and P
   for pressure-based), and the two other characters define the
   used kernel (CS stands for cubic spline, while C2, C4 and C6 refer
   to the Wendland kernels of different order).
\label{L1ConvergeErrorDensity}}
\end{figure}

\begin{figure*}
  \resizebox{15.0cm}{!}{\includegraphics{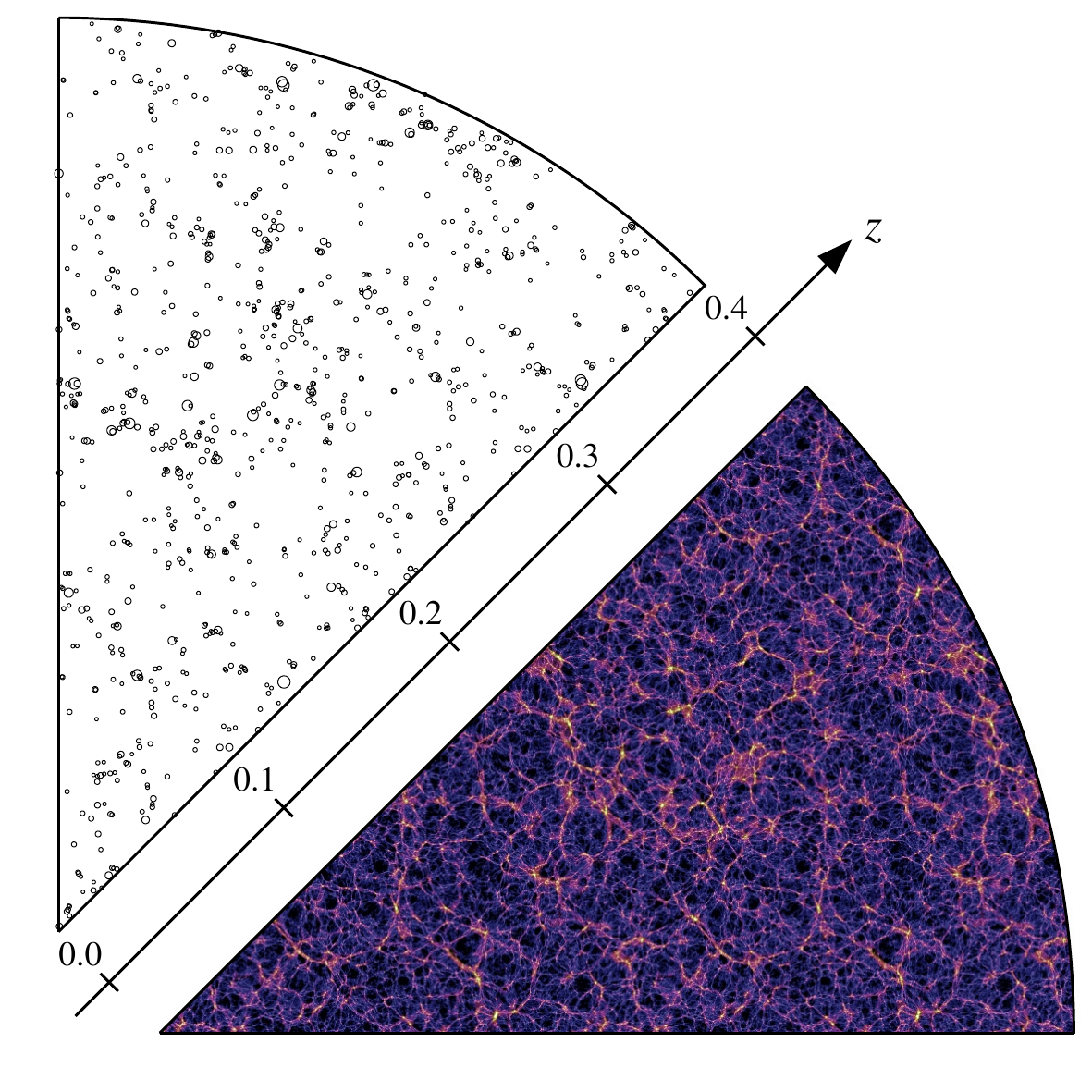}}%
  \caption{Example for lightcone output that can be produced by {\small
      GADGET-4}.  We show here a 45 degrees wedge-like region of
    constant comoving thickness of $5\,h^{-1}{\rm Mpc}$ excised from
      the full lightcone. The dark matter density is projected onto the
      plane and shown in color-scale, with comoving distance used as
      radial coordinate. The axis gives the corresponding
      look-back redshift, which at these low redshifts is still
      linearly related to comoving distance to good accuracy. The
      wedge in the upper left shows the same region, but this time
      small circles are drawn to indicate halos identified by the group
      finder of {\small GADGET-4}, which was applied on-the-fly directly to the
particles on the lightcone data. For visual clarity, only halos with virial mass above
      $10^{13}\,h^{-1}{\rm M}_\odot$ are shown, and drawn with circles
      that have a
      radius four times the actual virial radius.
      The base simulation of this example used a box
      size of $500\,h^{-1}{\rm Mpc}$ on a side with $1080^3$
      particles, so the code replicated the box automatically a few times in each
      dimension to cover the lightcone (the comoving distance out to
      $z=0.4$ is about
      $1083\,h^{-1}{\rm Mpc}$), a fact that is recognizable by the
      repetition of structures (albeit at different evolutionary
      state) in the image in the horizontal direction.
      \label{FigPartLightCone}}
\end{figure*}

\begin{figure}
  \resizebox{8.5cm}{!}{\includegraphics{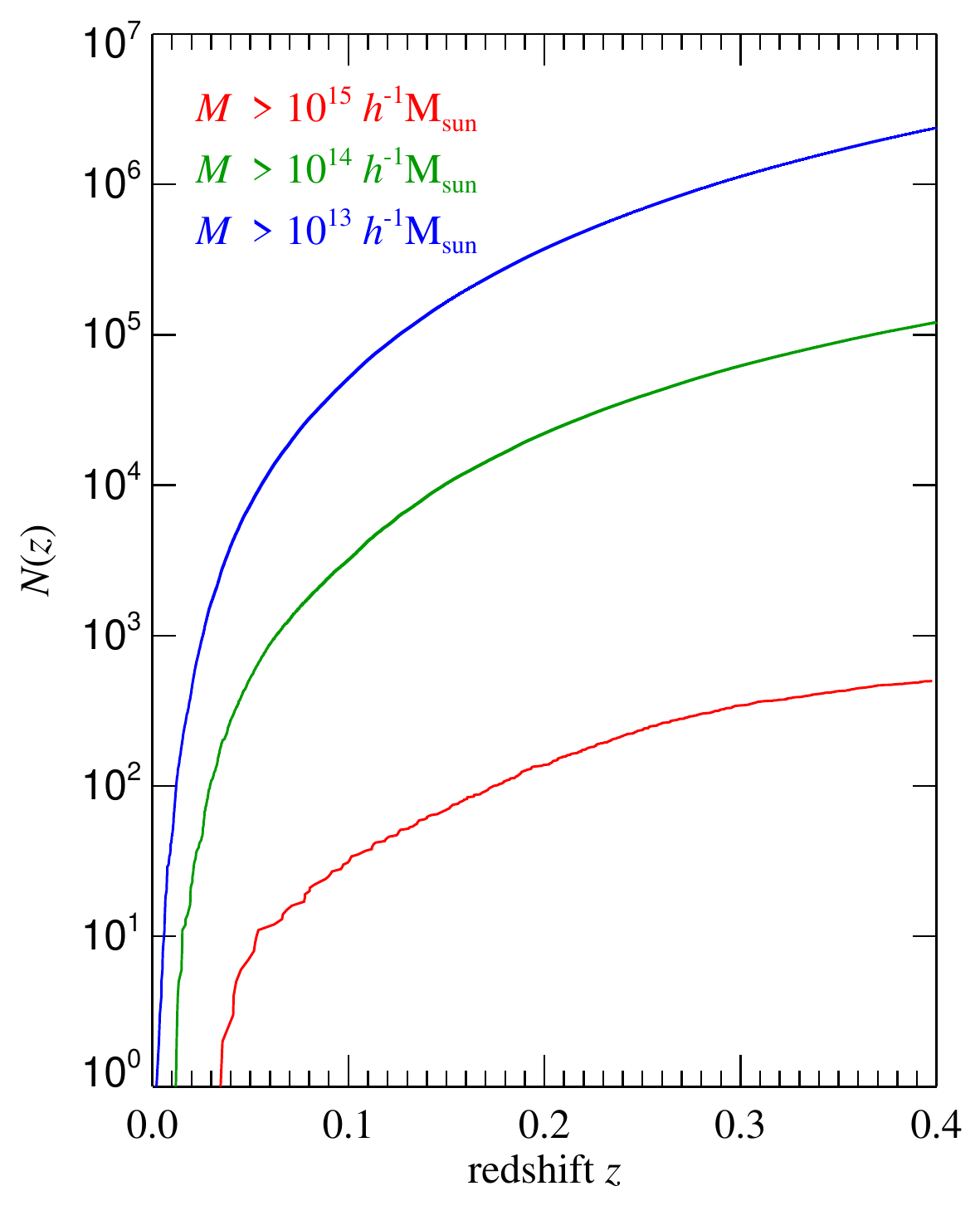}}%
  \caption{Cumulative mass function of halos on the all-sky backwards
    lightcone of a $L=500 \,h^{-1}{\rm Mpc}$ test simulation (the one shown in
    Fig.~\ref{FigPartLightCone}), for different mass thresholds (as labelled), as
    a function of the lookback redshift. This corresponds to the number
    of objects larger than the given mass that are in principle observable out to a
    certain redshift. Note that the absolute counts for rich clusters are expected to be
    biased low in this example, due to missing large-scale power in the
    intermediate-size box that was simulated here. 
    \label{FigLightConeMassFunc}}
\end{figure}

\subsection{I/O organization}

The recommended default format of all output in {\small GADGET-4} is
HDF5, but the code continues to support the legacy binary formats of
{\small GADGET} (called `format 1' and `format 2') as well. For
parallel I/O, and to limit the sizes of individual files (useful for
easier transfer and archival), the output can be distributed over
multiple files that are written or read simultaneously. If multiple
files are written in parallel, the number of simultaneous writes can
be limited to avoid overburdening a filesystem. The I/O operations are
scheduled automatically such that as many distinct compute nodes as
possible are active at any given time within the imposed bounds, which
can help to maximize the achievable I/O bandwidth on a given system.

All HDF5 output files contain a record of the configuration options
and the parameterfile that was used to run the code, as well as a git
hash key encoding the code version that was used to produce the data.
All output fields are annotated with their units and auxiliary scaling
factors needed to convert to physical values. Output data may be
stored in single precision (or even in half-precision for velocities)
to save disk storage space.  Upon input, the code automatically
detects the floating point type of the data and converts it as needed
to the internal representation (this also works for the two legacy
binary formats of {\small GADGET}, not only for HDF5). Another useful
feature is that {\small GADGET-4}'s I/O routines support on-the-fly
lossless compression of output fields, which is fully transparent to
analysis scripts thanks to the HDF5 library. The achievable
compression ratios can be quite high for some of the fields, in
particular for integer data related to merger tree bookkeeping, or for
the ID field used to tag simulation particles.

\section{Code validation and convergence tests} \label{seconvergence}

We now want to discuss the question which code settings are in
practice required to reach converged answers for the non-linear mass
distribution, and whether this can be achieved robustly for all the
many combinations of gravity solvers and time-stepping approaches
supported by {\small GADGET-4}.  For a semantically correct code,
stable results that are invariant under small changes of the
integration parameters should be obtained if there are negligible
force errors and negligible errors due to the time integration. In the
absence of hideous bugs in the code, this limit should be reached for
sufficiently small settings of the force accuracy parameters and of
the timestep sizes. But note that requesting recovery of identical
individual phase space positions of particles makes no sense in
practice even in this limit, because individual particle orbits tend
to be chaotic in the non-linear regime
\citep[e.g.][]{Thiebaut:2008aa, Keller:2019ab, Genel:2019aa}
and are merely Monte Carlo tracers of the underlying collisionless
fluid. Rather, statistical quantities that depend on a collection of
particles, such as halo density profiles, the halo mass function, or
the matter power spectrum are of interest here, and can be expected to
converge to a unique answer for high force accuracy and sufficiently
accurate time-stepping, modulo sampling noise present due to the
finite number of particles or modes used.

It depends on the particular quantity under study at what point a
simulation model is giving a converged answer (for example, getting
the total mass of a halo right is easier than getting its central
density profile right). For definiteness, we in the following pick the
matter power spectrum, and the inner regions of halos for our
convergence experiments. For our test simulations, we typically
consider the relative deviations to a fiducial ``fully converged''
simulation, which we in practice identify with the simulation with our
most conservative integration settings. In the absence of analytic
solutions, this provides a heuristic approach to test whether the
simulation outcomes are robust to variations of the integration
settings.

\subsection{Validating the time integration schemes}

We begin by aiming to establish that under conditions of negligible
force errors, integrations carried out either with a global timestep,
with standard local timesteps, or with the hierarchical
timestepping introduced here, all give the same results provided
sufficiently conservative timestep sizes are chosen.

To this end, we carry out simulations with conservatively high force
accuracy settings and explicitly verify during the simulation run that
forces in all parts of the algorithms have a high relative
accuracy. This is done by randomly picking a small subset of particles
in every step, and computing a direct summation force for them for
comparison. A result of this test is illustrated in
Figure~\ref{FigForceErrorAsAFunctionOfScaleFactor}, where we look at
the force error distributions as a function of time in a cosmological
simulation with $256^3$ particle in a $25\,h^{-1}{\rm Mpc}$ box, for a
variety of typical setups. We show six examples of runs carried out
with different force calculation algorithms (using Tree and FMM
schemes of different order, and optionally accelerated with the TreePM
or FMM-PM schemes), and different timestepping approaches. As we saw
also earlier, it is more challenging to obtain low relative force
errors at high redshift for the tree-based algorithms, especially for
moderate expansion order. This is primarily due to the small absolute
sizes of the peculiar accelerations there. However, all algorithms can
be pushed to relative force errors well below $10^{-2}$ even
there. Importantly, there are no outliers towards much large force
errors, serving as an important ``in-situ'' validation of the
implementation of the parallel force computation algorithms. Combined
with our scheme for decorrelating force errors in time through random
translations, we expect these force errors to be small enough to
affect the following results only at negligible levels, something that
we also verified with additional tests to be discussed later on with
still lower residual force errors.

Next, we turn to the time integration schemes, with a view to
establish how small the steps at least need to be to obtain converged
results. We first consider simulations with fixed global timesteps, in
order to see the general impact of time-stepping accuracy on
collective results. We begin with deliberately coarse stepping, and
then subsequently improve it by factors of two. In
Figure~\ref{FigDiffPSForFixedDt}, we compare the power spectra of the
resulting sequence of simulations and their deviations from the run
with the shortest timestep, which we identify with the most accurate
result in this series. These test simulations again use $256^3$
particles in a relatively small box of $25\,h^{-1} {\rm Mpc}$ in order
to still have a relatively high mass resolution and a high degree of
non-linearity comparable to current high-resolution simulations done for
large-scale structure studies. We see that for a {\em fixed} timestep
size of $\Delta \ln ( a) = 5.6 \times 10^{-4}$ or lower, the power
spectrum is converged to better than $1\%$ accuracy all the way to
scales corresponding to the softening length. This corresponds to at
least 8192 steps from redshift $z=99$ to $z=0$. For the fundamental
mode, which is starting to leave the linear regime for this small box
size, a step size as low as $\Delta \ln ( a) = 7.2 \times 10^{-2}$ can
already be sufficient (corresponding to just 64 steps) to limit
deviations from a converged answer to less than $1\%$, but this is
inadequate already for the mildly non-linear regime, and hopelessly
inaccurate for the highly non-linear regime.

In Figure~\ref{FigDiffHaloForFixedDt}, we extend this comparison the
spherically averaged density profile of the largest halo found in the
simulation box. Consistent with the power spectrum findings, here the
$\Delta \ln (a) \le 5.6 \times 10^{-4}$ setting provides a density
profile that is indistinguishable within the noise from runs that use
considerably finer timestepping, whereas twice as large steps show
already small but systematic reductions of the density around
$r\sim 10\,h^{-1}{\rm kpc}$. Insufficient time integration accuracy
generically shows up as a reduction in the central halo density, as
expected.

\begin{figure*}
  \resizebox{7.5cm}{!}{\includegraphics{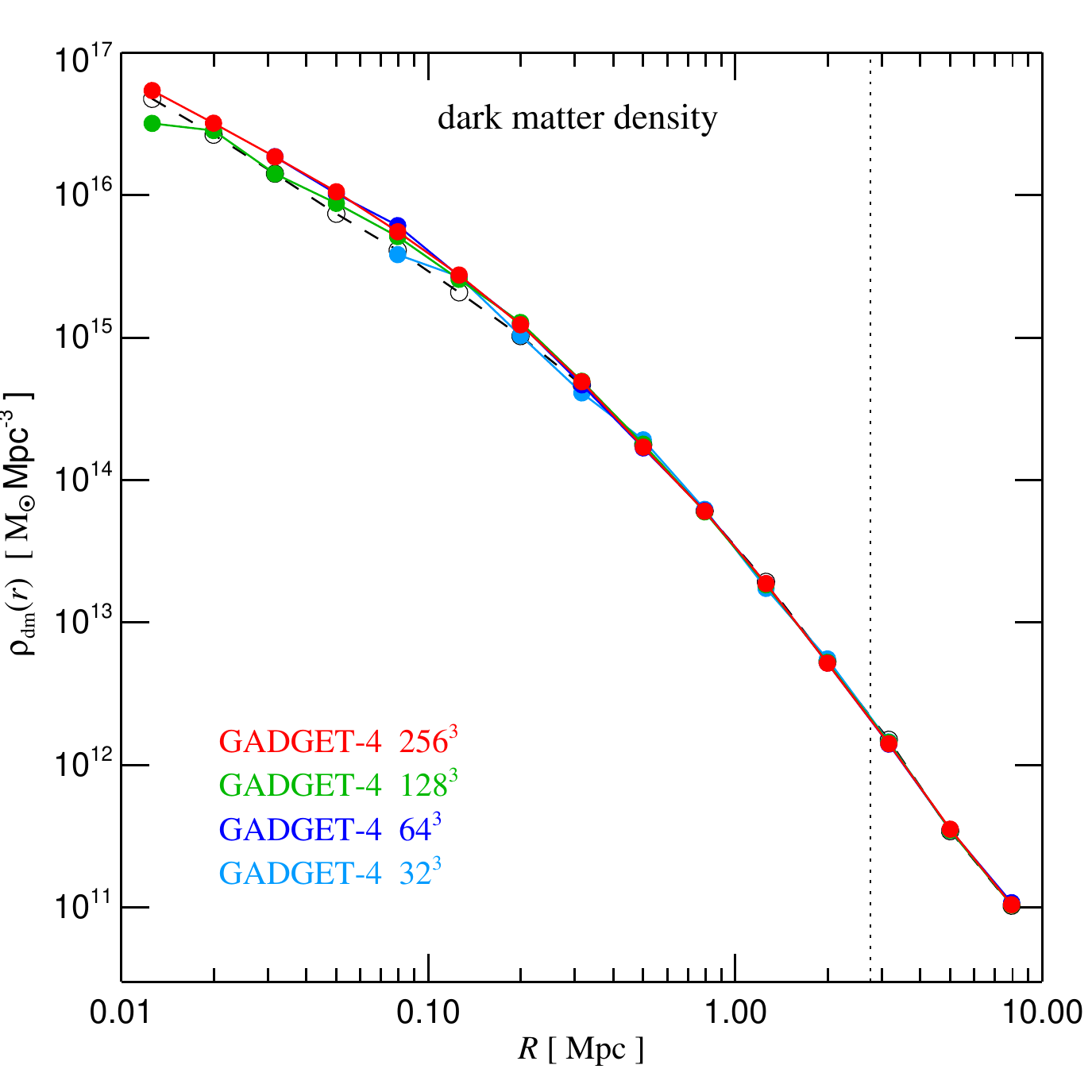}}%
  \resizebox{7.5cm}{!}{\includegraphics{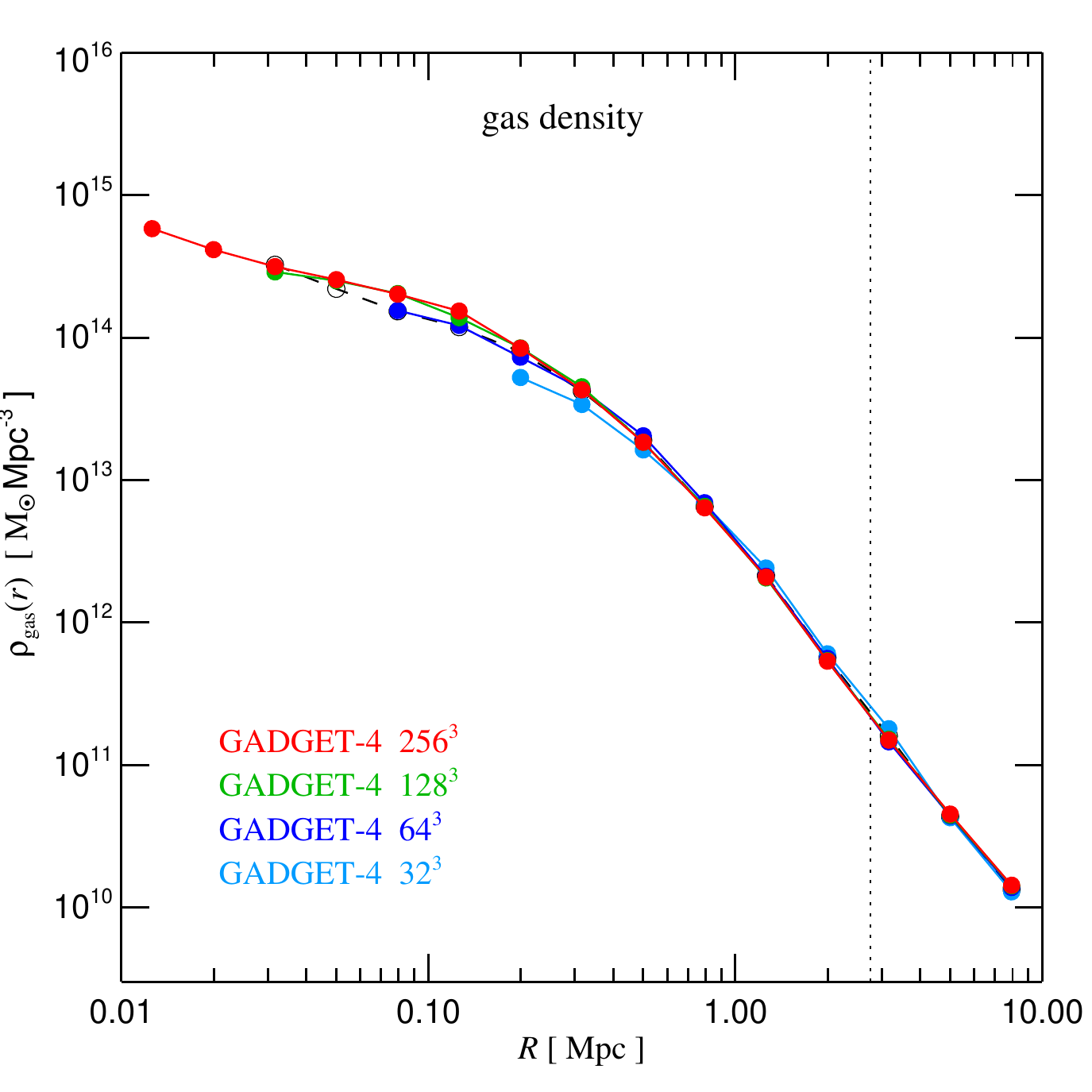}}\\%
  \resizebox{7.5cm}{!}{\includegraphics{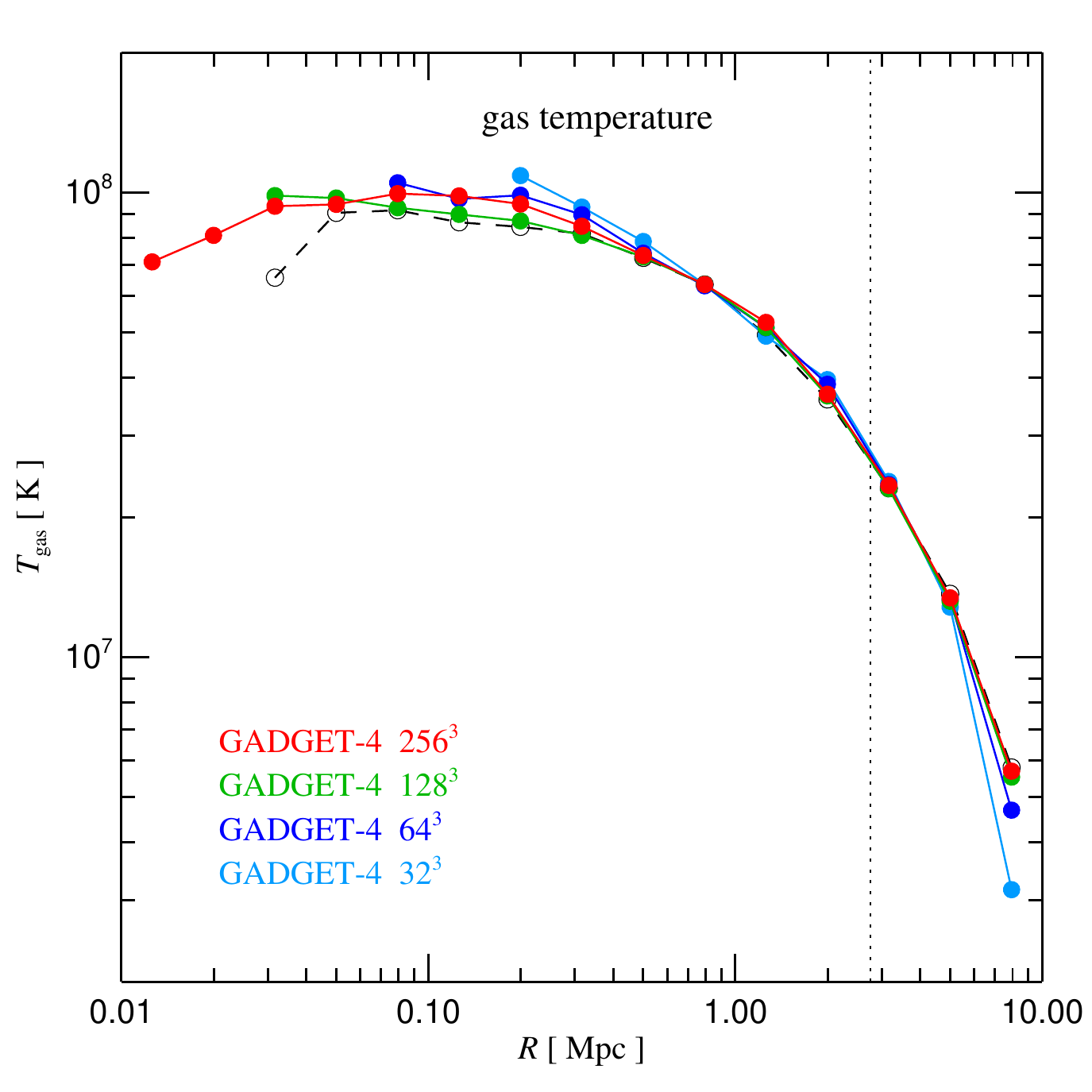}}%
  \resizebox{7.5cm}{!}{\includegraphics{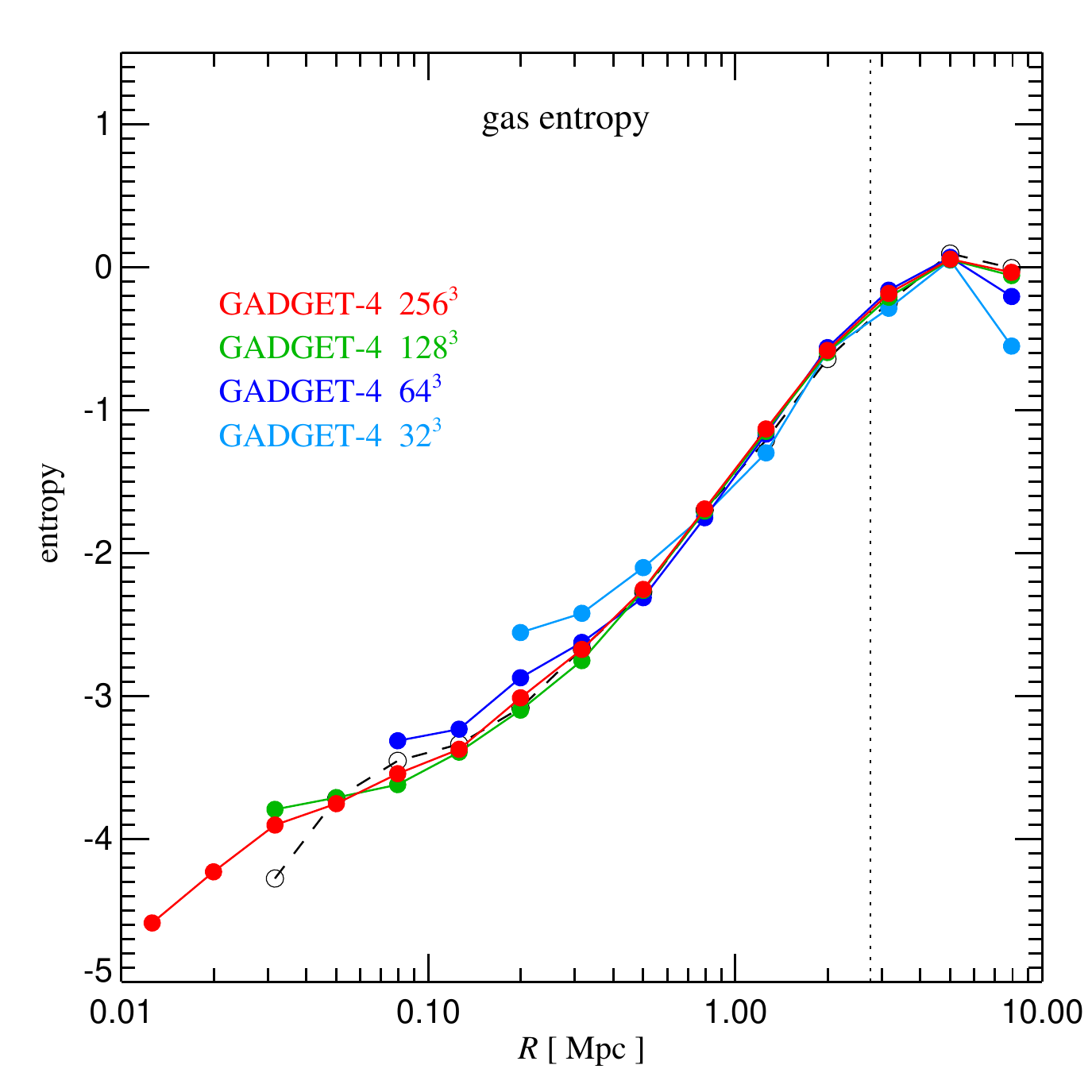}}\\%
\caption{Dark matter density, gas density, temperature, and entropy
  profiles of the cluster studied in the Santa Barbara cluster
  comparison project \citep{Frenk:1999aa}, here computed at the maximum $2\times 256^3$
  resolution that was created (but not actually used) at that time. We compare
  the highest resolution result to runs carried out at $2\times 128^3$, $2\times 64^3$, and $2\times 32^3$
  resolutions, and to the average of all the different simulation
  results
  reported in \citet{Frenk:1999aa}, which are shown as dashed lines.
  Our simulations were here computed with the pressure-based SPH version. Compared
  to mesh-based calculations and some alternative SPH formulations
  that try to explicitly account for mixing, the results show quite
  low entropy in the centre. The density-entropy formulation
  \citep{Springel:2002ab}
  tends to give still slightly lower central entropies,
  presumably reflecting a spurious suppression of
  turbulent mixing at the resolution scale.  
  \label{FigSantaBarbara}}
\end{figure*}

It is now interesting to compare time integration schemes that employ
local timestepping with the `converged' result obtained for a globally
fixed timestepping. This is considered in
Figure~\ref{FigLocalSteppingN128}, where the left panel shows relative
differences in the power spectrum with respect to the fiducial
``ground truth'' result (global timestep with 32768 steps), and the
right panel shows this comparison for the density profile of the
largest halo. We consider a variety of combinations of a maximum
timestep size (to give good accuracy for small modes $k$ that stay in
the linear regime) with a local timestepping parameter $\eta$ that
controls the accuracy within dense halos.  Clearly, for this softening
setting, a value of $\eta = 0.006$ appears required to obtain fully
consistent results, while $\eta = 0.012$ is not quite as good as some
systematic reduction in power of $\sim 0.5\%$ becomes visible in the
range $k\simeq 10-100 h\,{\rm Mpc}^{-1}$. We have also included a test
with hierarchical timestepping which produces consistent
results. Overall, the results provide an important confirmation that
local timestepping for collisionless dynamics in structure formation
is viable and a highly accurate approach to accelerate the simulations
\citep[see also][]{Power:2003aa}. Note that even here, where the
dynamic range in timescales is still fairly limited, this already
entails a significant computational saving compared to the use of a
global timestep.

\subsection{Validating the force accuracy of different schemes}

For a simulation that is converged with respect to timestepping, we
can now go back and ask whether the different force calculation
algorithms implemented in {\small GADGET-4} still give the same
results if larger force errors are tolerated, as is often done in
practice to keep the computational cost at bay.

This is addressed in Figure~\ref{FigForceSchemesN128}, where we show
simulations done with different force computation schemes, again
compared to the fiducial ``ground truth'' fixed timestep run. Here, we
consider a variety of different force calculation schemes possible
with {\small GADGET-4}, using a set of force accuracy parameter
settings, and we also include a number of modifications of the time
stepping scheme at the same time, thus covering a range of code
settings that may be encountered in practice.

Reassuringly, the results overall suggest that the code indeed
produces consistent results for high enough force accuracy and high
enough time integration settings, pretty much independent of which of
the code's force computation algorithms or time-stepping schemes is
used. This degree of robustness is important, as it offers a certain
degree of protection (but unfortunately no guarantee) against a poor
quality of results due to the adoption of non-ideal settings.

The above also begs the question how much the force accuracy could be
degraded -- assuming good time integration accuracy is preserved --
before significant differences in the results show up, and if they do,
where they become noticeable first. This is obviously important
because computing forces with very small error is computationally
expensive. Aiming for something that is excessively conservative here
has the potential to needlessly increase the computational cost, just
like excessively small timesteps would do. The crux is of course that
erring on the other side is even more problematic, because there is
often little advantage in obtaining results quickly if they are of
poor quality, and worse, not being aware of this could lead in extreme
cases to misleading scientific conclusions. On the other hand,
depending on the scientific question that is asked, there are of
course sometimes situations where approximate answers are perfectly
sufficient. It is hence important to be fully aware of the accuracy of
the results one obtains.

To partially address this question, we show in
Figure~\ref{FigReducedForceAccuracy} results where the force accuracy
is systematically reduced while the time integration accuracy is kept
high. For the lowest force accuracy setting, we carry out the
simulation a second time by disabling the random spatial translation
of the particle set in every domain decomposition. We see that the
results are rather stable, even for relatively coarse settings,
provided that randomizations are applied. This confirms that small
force errors that are {\em random} and {\em uncorrelated in time}, do
not seriously impair collisionless dynamics. Time integration accuracy
is in that sense the more important category, as errors introduced
here invariably show up in the results. However, the run with a low
force accuracy setting and without random translations shows clear
systematic differences relative to the high force accuracy simulation,
both in the matter power spectrum and the mean density profiles of the
largest halos. This confirms that random shifts of the particle set
are helpful in establishing sufficiently random distributions of the
residual force errors.

Finally, as an aside, we check whether the total momentum is indeed
conserved to machine precision when the FMM scheme and hierarchical
time integration are used. In Figure~\ref{FigMomentumDrift}, we
consider the total linear momentum in the simulation box as a function
of scale factor, comparing FMM and tree-based simulations. Furthermore,
we repeat this test without the randomization approach for reducing
correlated force errors. The results confirm our expectations. While
FMM with hierarchical (or global) timestepping manifestly confirms
momentum to machine precision, this is not the case for any variant of
the normal Tree algorithm. However, translational randomization of the
particle set in every step does already reduce the size the corresponding
secular error can build up to.

 \begin{figure*}
  \begin{center}
  \resizebox{16.0cm}{!}{\includegraphics{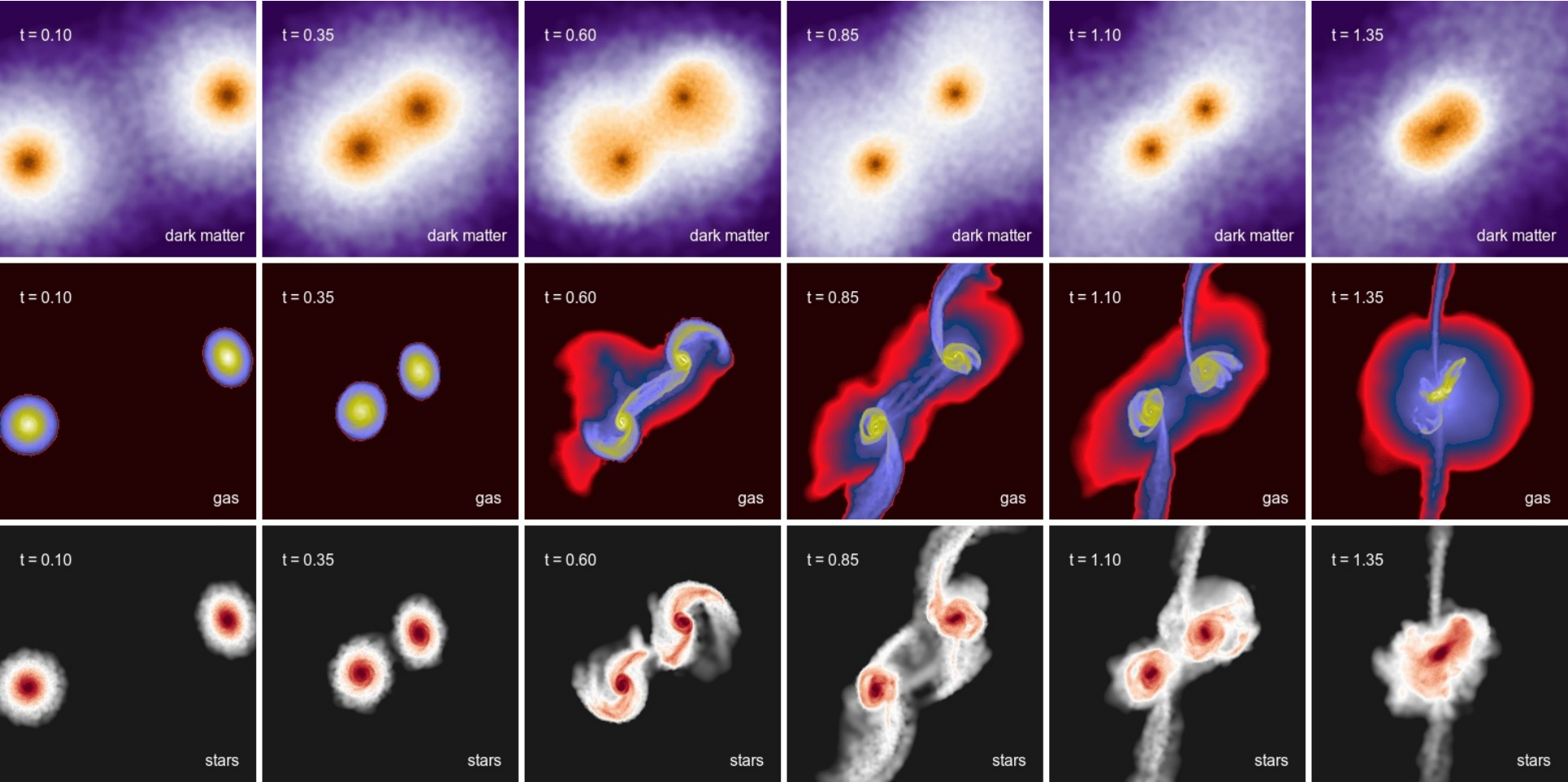}}\\%
  \caption{Visualization of a galaxy merger simulation carried out
    with {\small GADGET-4}. The top row illustrates the
    evolution of the 
    projected  
    dark matter density in a major merger of two equal mass disk
    galaxies that collide on a prograde, zero-energy orbit. The middle row shows the gas phase,
    which initially comprises 33\% of the disk mass of the galaxy
    models, and is actively forming new stars.
    The other 66\% of the initial disk mass are made up of preformed
    stellar material, whose evolution is
    illustrated in the bottom row. 
    All panels
    are $160\,{\rm kpc}$ on a side, and the time of the individual images is
    given in Gyr in the labels.
    \label{FigMergerTimeSequence}}
\end{center}
\end{figure*}

\begin{figure}
\resizebox{8.3cm}{!}{\includegraphics{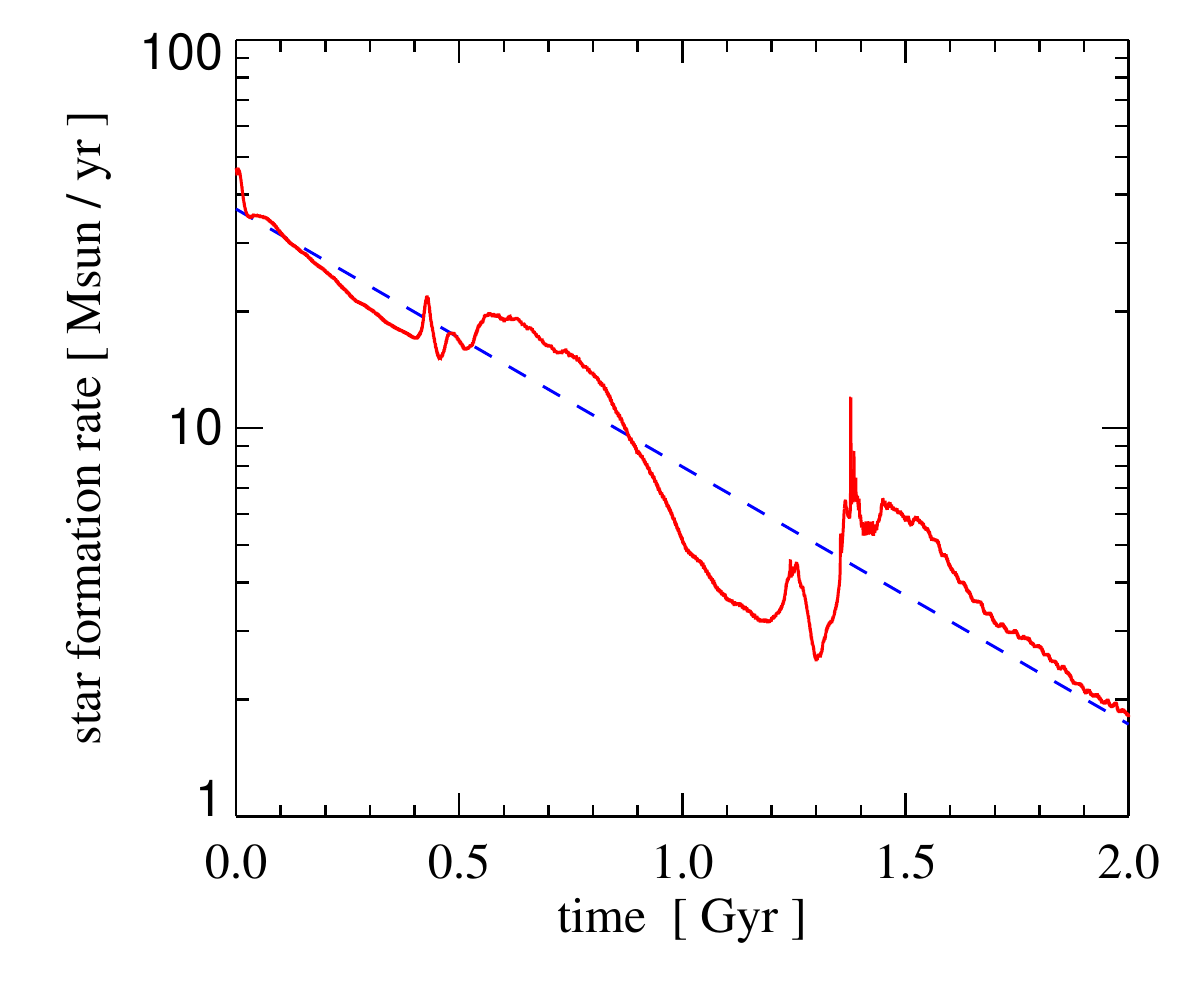}}\\%
\caption{Evolution of the star formation rate in a galaxy merger
  simulation (the one illustrated in
  Fig.~\ref{FigMergerTimeSequence}). The overall trend is that the gas
  mass is used up to make stars with a consumption timescale of around
  $1\,{\rm Gyr}$, as indicated by the dashed line. This is modulated
  by the merger, which first induces a mild enhancement around the
  first encounter at $t\sim 0.6,{\rm Gyr}$, followed by a reduction
  due to tidal removal of gas, and a mild starburst at
  $t\sim 1.4\,{\rm Gyr}$, when the galaxies collide a second time and
  then quickly coalesce.
  \label{FigSFRGalaxyMerger}}
\end{figure}

\subsection{Comparing results to other N-body codes}

It is also important to compare N-body results with other codes in
order to assess remaining systematic uncertainties in their
predictions, especially in the non-linear regime of cosmic structure
formation, where definitive analytic results are not available. The
exquisite accuracy demanded by modern cosmological surveys that target
dark energy, such as EUCLID, provides additional motivation for this,
because here simulation results need to reach a nominal accuracy of 1
percent or better to meet the statistical needs of these surveys. It
is not readily clear whether current codes agree to this precision,
and whether this is reached for standard settings of numerical
nuisance parameters that control timestep size, force accuracy, or
softening length \citep{Smith:2014aa}.

\cite{Schneider:2016aa} presented in this context a comparison of the
cosmological codes {\small PKGRAV-3}, {\small RAMSES}, and {\small
  GADGET-3}, focusing on the non-linear power spectrum in a large
$2048^3$ simulation in a $500\,h^{-1}{\rm Mpc}$ box, the ``Euclid test
simulation''.  While the runs showed good agreement on large scales,
there were a few noticeable difference at high $k$. In particular,
{\small GADGET-3} came out a bit lower than the other two codes,
whereas {\small RAMSES} was slightly higher. While overall this
agreement can be deemed satisfactory, it was arguably less good on
small scales than one may hope for, given that the comparison stopped
already at $k\simeq 10 \,h\, {\rm Mpc}^{-1}$, i.e.~was only done from
the fundamental node to a scale equal to the mean particle
spacing. This is basically where the highly non-linear regime only
begins; the softening scale is still a factor $\sim 40$ smaller than
the mean particle spacing, and in principle the simulations should
agree over most of this range, too.

The EUCLID test simulation of \cite{Schneider:2016aa} has recently
been revisited by \citet{Garrison:2019aa} using the {\small ABACUS}
code, which is a novel out-of-core N-body code that uses
GPU-accelerated direct summation for short-range forces, and an
FMM-based convolution method with a fixed grid to compute the
long-range forces. The authors emphasize that the high-force accuracy
resulting from this method should yield a high fidelity result for the
test run. Their simulation was however carried out with a global
timestepping scheme using a comparatively coarse step size.  While this
proved sufficient to give good agreement with the power spectra
obtained by \cite{Schneider:2016aa} up to the mean particle spacing,
according to our earlier tests on timestepping convergence, it is not
clear whether the fidelity of the result can be expected to extend
into the highly non-linear regime at the centres of halos.

In any case, it is interesting to also apply {\small GADGET-4} to the
corresponding initial conditions of \citet{Schneider:2016aa} and see
what we get.  We show our results in
Figure~\ref{FigEuclidReferenceSim}, where the top panel gives the
shot-noise subtracted power spectrum, this time measured accurately to
much smaller scales than considered in \cite{Schneider:2016aa} and
\cite{Garrison:2019aa}. The relative differences in the power spectrum
are given in the two bottom panels. We include here the original
simulation particle data from \citet{Schneider:2016aa} and
\citet{Garrison:2019aa}, but use our own power spectrum routines for
consistency in the measurements.

Interestingly, while the simulations agree again reasonably well over
the range $k<10 \,h\,{\rm Mpc}^{-1}$, the differences in the highly
non-linear regime quickly become very large. {\small GADGET-4} agrees
best with {\small PKDGRAV} out to about $k\sim 40 \,h\,{\rm Mpc}^{-1}$,
but then {\small PKDGRAV} gives considerably higher small-scale power.
In contrast, {\small RAMSES} and especially {\small ABACUS} show a
deficit of power on nonlinear scales, which in the case of {\small
  RAMSES} is however rather a dip relative to {\small GADGET-4} that
eventually turns around and becomes an excess at the smallest scales.
The systematically low power of {\small ABACUS} on very small scales
may well reflect the more limited time-integration accuracy of the
corresponding run due to use of global timestepping, as suggested by
the findings discussed earlier in this section. The differences of
{\small GADGET-4} with respect to {\small RAMSES} and {\small PKDGRAV}
on very small scales could instead be explained by different force
softening laws. Another interesting effect is that some $k$-modes at
the largest scales show small fluctuating differences in growth in
{\small PKDGRAV}, and to a lesser degree in {\small ABACUS}, relative
to {\small RAMSES} and {\small GADGET-4}. This probably reflects the
lower accuracy of a pure real-space evaluation of the large-scale
gravitational field compared to a spectral methods.  We remark that we
have also repeated a run with {\small GADGET-3}. Curiously, this shows
a result very close to {\small GADGET-4}, and hence gives slightly
higher small-scale power than the run reported in
\citet{Schneider:2016aa}. This was also independently found by Angulo
et al. using the {\small L-GADGET-3} variant of {\small GADGET}. The
original result reported in \citet{Schneider:2016aa} for {\small
  GADGET-3} thus seems to be slightly anomalous, for a still
unidentified reason, but a slightly high softening setting or a poor
timestepping accuracy setting could be a possible explanation.

In Figure~\ref{FigEuclidDensityProfiles}, we extend this comparison to
the density profiles of the largest halos in the box. In the top three
panels we consider in an exemplary fashion the density profiles of the
largest three halos, while in the bottom panels we examine the
averaged profile of the 25 most massive halos (bottom left).  The
systematic differences in the mean profile predicted by the codes are
shown in the lower right panel. We made sure that really the same
halos are identified in this comparison in all cases. While the
profiles agree quite well down to a scale of
$\sim 30 \,h^{-1}{\rm kpc}$, there are noticeable systematic
differences in the innermost regions of the halos. With respect to the
{\small GADGET-4} result, {\small RAMSES} finds higher dark matter
density by up to $30\%$ at around the softening scale, while the
opposite is the case for {\small ABACUS} and {\small PKDGRAV}. Part of
this is likely driven by different effective softening laws and
time-stepping accuracy, but it will be interesting to identify and
eliminate the root cause of these relatively small but systematic code
differences in the future.

\subsection{Convergence in a basic SPH shock tube problem}
\label{subsecShocktube}

Finally, we test our different SPH implementations presented in
Section~\ref{sechydro} by comparing their convergence rate in an often
used standard Sod shock-tube problem \citep{Springel:2005aa,
  Springel:2010ab, Hu:2014aa, Hopkins:2015aa}.  We consider a
two-dimensional stretched periodic box with sizes $L_x = 80$ and
$L_y = 10$, filled with an ideal gas initially at rest with adiabatic
index $\gamma = 1.4$.  The half-space for $x < 40$ is filled with gas
of unit density and unit pressure ($\rho_1=1$, $P_1=1$) while the half
space for $x > 40$ is filled with low-density ($\rho_2 = 0.25$) and
low-pressure ($P_2 = 0.1795$) gas.  We let the system evolve until
time $t=5$, which is enough time to stretch the self-similar wave
structure over a significant spatial range while avoiding overlap with
the waves generated from the equivalent problem at the periodic box
boundaries.  To easily create initial conditions with different
resolutions, we use Cartesian grids for the initial particle
distribution in both half-spaces.

In Figure~\ref{shocktube_result} we illustrate the result obtained for
the Wendland C6 kernel with time-dependent viscosity and density-based
SPH for a resolution of $N=1440$ particles in the $x$-direction.
Overall, the simulation agrees well with the analytical solution but
shows small bumps at the rarefaction wave and smoothes out the shock
discontinuity, as expected.  It also shows a small ``pressure blip''
at the contact discontinuity which can be avoided by using the
pressure-based SPH formulation.  In the velocity profile, one can
observe small oscillations around the theoretical value in the
post-shock region, which are not entirely suppressed by the artificial
viscosity.

To check for convergence of our implementation to the correction
solution, we define an L1 error as
\begin{equation}
L1 = \frac{1}{N_b} \sum_i^{N_b} \left|\bar{\rho}_i -\bar{\rho}(x_i)\right|,
\end{equation}
where $N_b$ is the number of spatial bins, $\bar{\rho}_i$ is the
arithmetic mean of the particle densities in the bin, and
$\bar{\rho}(x_i)$ gives the mean analytic solution in bin $i$.  In
Figure~\ref{L1ConvergeErrorDensity}, we show the results of our
convergence study, using a bin size of 0.05 units in the
$x$-direction.  We include only results for the time-dependent
viscosity since they match those with constant viscosity closely.  The
results for both SPH flavours are also quite similar, i.e.~the
convergence behaviour mainly depends on the used kernel, and therefore
indirectly on the employed number of neighbours.

The best results are obtained with the Wendland C6 kernel using 66
neighbours, yielding a convergence rate of $L1 \propto N^{-0.9}$ that
agrees well with the results from previous studies
\citep{Springel:2010aa, Read:2012aa, Hu:2014aa}, and is close to the
theoretical optimum of $L1 \propto N^{-1}$ for this problem, which is
limited to first order convergence due to the presence of physical
discontinuities that dominate the total error \citep[see
also][]{Springel:2010aa}.

Note however that although larger neighbour numbers can help to
decrease the $E_0$-error \citep{Read:2012aa} of SPH, they also come
with an increased computational cost. This is reflected in the total
run time of these test simulations, which we give in
Table~\ref{TabRuntimeSPH} for different kernel and parallelization
choices, in all cases using a constant resolution of initially
$N=1440$ particles in the $x$-direction.  Whether or not a
time-dependent viscosity is used, or the density- or pressure-based
SPH is adopted, has only a small effect on the total runtime, which in
turn depends however sensitively on the chosen SPH kernel function and
neighbour number.  As mentioned in Section~\ref{subsecVectorization},
our explicit vectorization approach for SPH is able to realize a small
efficiency gain for the Wendland kernels, as they are not defined in a
piecewise fashion, but this is largely lost again when a
time-dependent viscosity is used, except for the large neighbour
numbers involved with the C6 kernel. Overall, this underlines that the
comparatively small number of floating point operations per memory
access in the inner loops of SPH makes it challenging to reach a high
fraction of the theoretically possible peak floating point performance
of the CPUs that are currently in widespread use.

\begin{table}
  \begin{center}
\begin{tabular}{c|c|c|c}
\hline
Kernel & Artificial Viscosity & Vectorized& Runtime(s)\\
\hline
Cubic Spline & constant & no& 1661\\
Wendland C2 & constant & no& 1831\\
Wendland C4 & constant & no& 2251\\
Wendland C6 & constant & no& 3042\\
Cubic Spline & time-dependent & no& 1761\\
Wendland C2 & time-dependent & no& 1903\\
Wendland C4 & time-dependent & no&  2333\\
Wendland C6 & time-dependent & no&  3124\\
Cubic Spline & constant & yes& 1708\\
Wendland C2 & constant & yes& 1742\\
Wendland C4 & constant & yes&  2229\\
Wendland C6 & constant & yes& 2881 \\
Cubic Spline & time-dependent & yes& 1837\\
Wendland C2 & time-dependent & yes& 1914\\
Wendland C4 & time-dependent & yes& 2339 \\
Wendland C6 & time-dependent & yes&  3032\\
\hline
\end{tabular}
\end{center}
\caption{Total runtimes for the two-dimensional shocktube tests
  until $t=5$, for a resolution of initially $N=1440$ in the
  $x$-direction
  and for different kernels and artificial viscosity options. We give
  results both for SPH kernels coded without or with explicit
  vectorization used SIMD intrinsics.
  The choice between density and pressure-based SPH has only a
  small effect on the total runtime, and hence we
  only report results for the density-based SPH runs. 
  All simulations were run using one node on
  the ``Freya'' compute cluster at MPA
  (the nodes are equipped with dual Intel Xeon Gold 6138 CPUs,
  each with 20 physical cores at 2.0 GHz).
  \label{TabRuntimeSPH} }
\end{table}

\section{Exemplary applications} \label{sectests}

In this section, we discuss a small set of different types of
simulations possible with the {\small GADGET-4} code. These examples
are mostly meant to illustrate and test certain features of the code.
We first consider a cosmological dark matter simulation with a
continuous lightcone output, then turn to a cosmological simulation
with baryons (where we revisit the Santa Barbara cluster for
definiteness), followed by a simulation with cooling and star
formation (illustrated with a galaxy merger in isolation), and
finally, we consider self-gravity of isothermal gas in a stretched box
with periodicity in the gravitational field imposed only in two out of
three dimensions.

\subsection{Cosmological simulation with lightcone output}

In this example, we consider a $L=1\,h^{-1}{\rm Gpc}$ periodic box,
sampled with $768^3$ particles in a standard $\Lambda$CDM
cosmology. The initial conditions are created upon start-up of the
code at redshift $z=99$. In terms of outputs, we ask the code to
produce group catalogues for a full-sky particle lightcone between
redshifts $z=0.4$ and $z=0$, but without actually outputting the
particle data itself, except for a thin disk of comoving thickness
$5\,h^{-1}{\rm Mpc}$ excised from the lightcone. 

In Figure~\ref{FigPartLightCone} we show part of this disk region of
the lightcone in terms of a 45 degree wide wedge, in two different
views. The first view gives a projection of the dark matter particles
in the disk region, adaptively smoothed and shown with a grey-scale
that encodes the logarithm of the density. The other view displays the
groups identified by the code in the full lightcone data. We only
include groups whose centres fall within the geometric bounds of the
wedge, and illustrate them with little circles with a radius four
times as large (for visual clarity) as their virial radii. It would be
easily possible to apply HOD modelling to make predictions for the
galaxy population of these halos in order to generate galaxy mock
catalogues for the lightcone. A more sophisticated approach would be
to model galaxy formation physically by applying semi-analytic galaxy
formation techniques to the (sub)halo merger trees that {\small
  GADGET-4} can also produce on the fly, such as implemented in the
{\small L-GALAXIES} code, and then to interpolate the predicted galaxy
properties to the halo positions on the lightcone.

Note that the halo catalogues on the lightcone exactly correspond to
what is in principle observationally accessible. Plotting a mass
function for the lightcone halo catalogue therefore directly gives
absolute sky counts of those objects that are detectable with ideal
observational capabilities. In Figure~\ref{FigLightConeMassFunc}, we
show such an ``observable mass function'' of halos, plotted here as
the number of objects that can be seen above a given mass threshold,
out to a certain lookback redshift $z$. For example,
$10^{15}\,h^{-1}{\rm M}_\odot$ halos grow in number out to
$z\sim 0.5$, at which point a plateau at a terminal abundance of
around 800 objects is reached, because at still higher redshifts these
massive clusters become quickly rarer due to their late formation
time, and the drop in number density even overwhelms the rapid growth
of the comoving volume seen on the lightcone towards higher
redshift. Note that the finite box size we use here causes a deficit
of large-scale power that biases the number of large clusters in this
simple example slightly low. This is less of a problem for lower mass
halos, where one can generally see many more objects, not only because
they are more abundant at any given given redshift, but also because
their earlier formation time greatly enhances the effective volume on
which they can be found on the backwards lightcone.

\begin{figure}
 \resizebox{8.5cm}{!}{\includegraphics{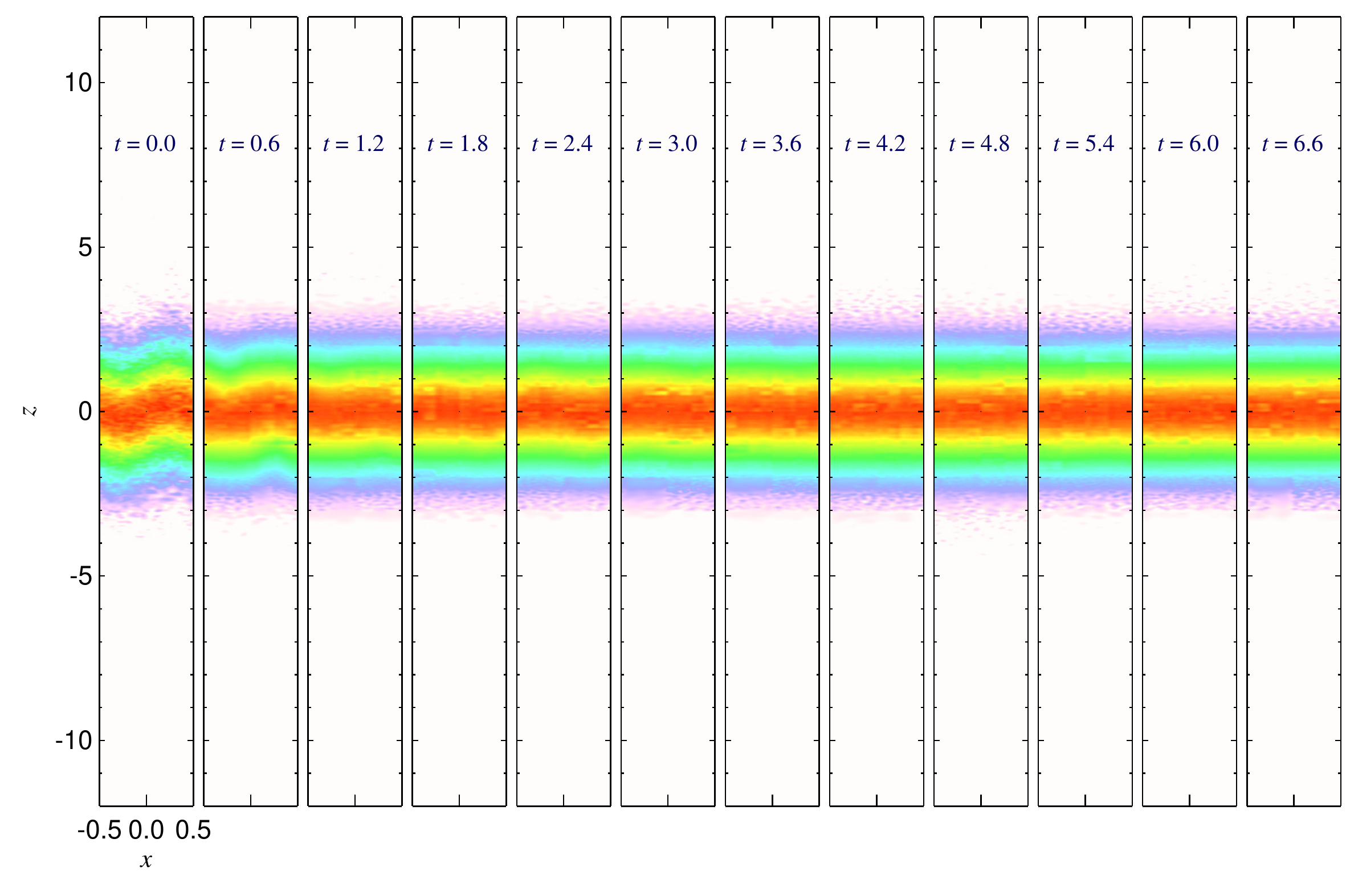}}\\%
 \resizebox{8.5cm}{!}{\includegraphics{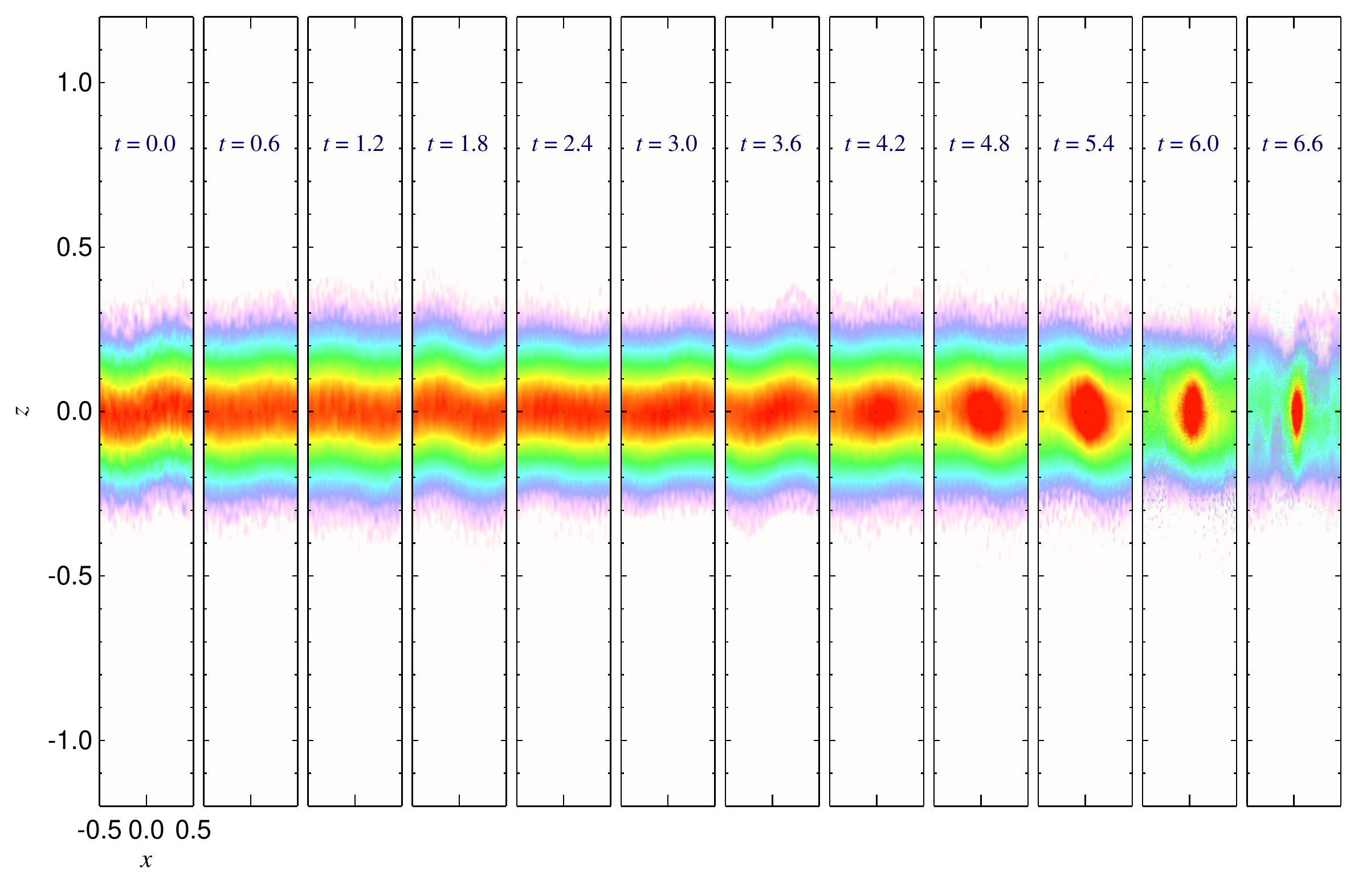}}%
 \caption{Time evolution of an isothermal gas sheet under self-gravity when
   periodicity is imposed only in the $x$- and $y$-directions. The
   upper simulation follows a disk that is thick enough relative to the
   box width to guarantee that
   transverse perturbations cannot have a wavelength larger than
   $2\pi$ times the scale-height, thus it should be stable.
   In contrast, the sheet simulated in the
   bottom panel
   is a factor of 10 thinner in units of the transverse box size, and
   thus should be prone to instabilities. The stability is tested
   explicitly by imposing in the initial conditions a weak sinusoidal perturbation with
   wavelength equal to the box width. This yields results consistent
   with the theoretical expectations, as the upper simulation settles
   into a stable sheet, while the lower one breaks up.
   \label{FigStretchedBox}}
\end{figure}

\begin{figure}
 \resizebox{8.0cm}{!}{\includegraphics{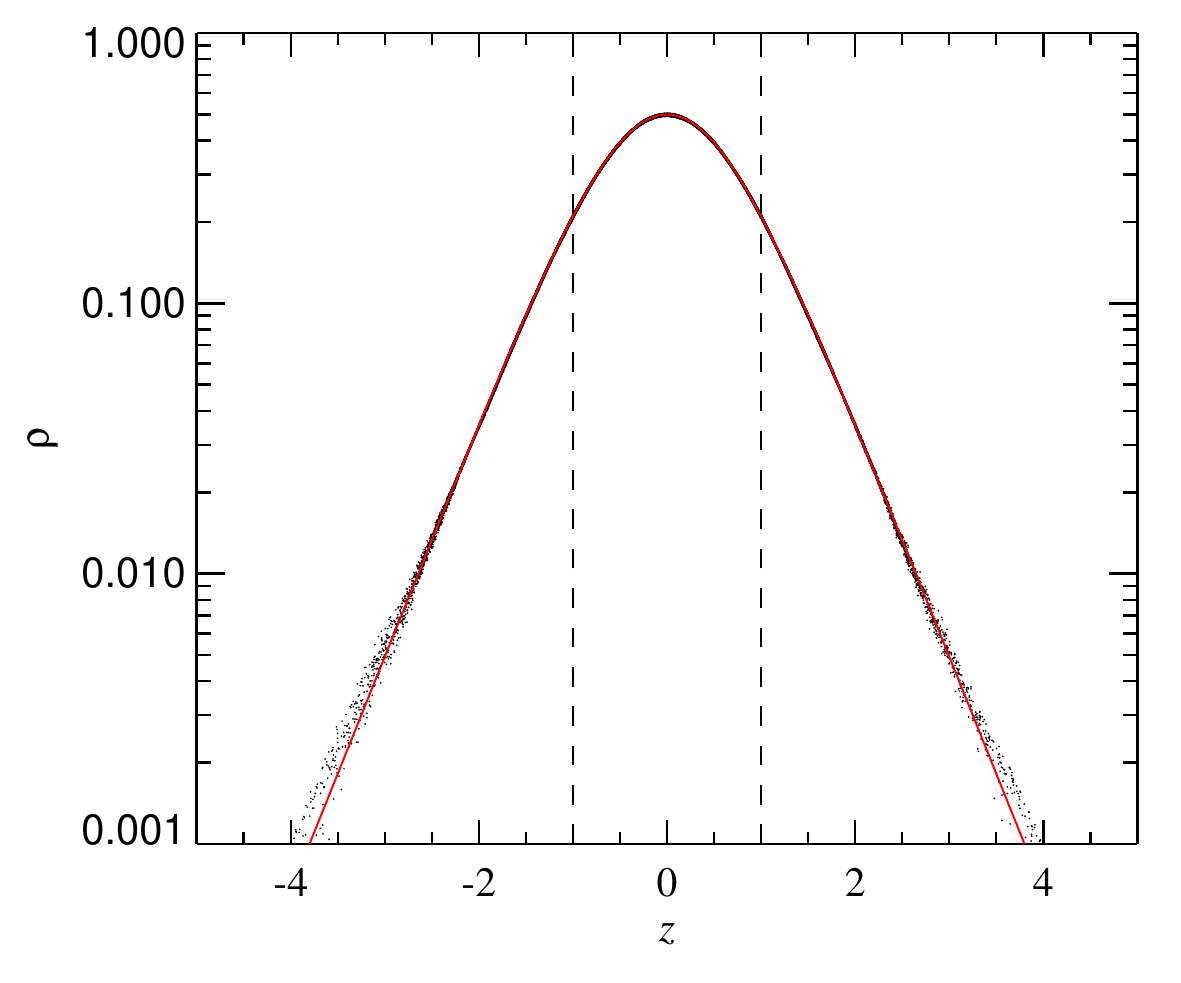}}
 \caption{Equilibrium state reached by the `thick' isothermal sheet
   shown in Figure~\ref{FigStretchedBox} when compared to the analytic
   solution (solid red line). The good agreement is an indirect
   validation of {\small GADGET-4}'s gravity solver with mixed boundary conditions. The dashed vertical lines indicate one
   scale-height below and above the mid-plane. 
   \label{FigIsothermDensity}}
\end{figure}

\begin{figure*}
  \begin{center}
    \resizebox{8.5cm}{!}{\includegraphics{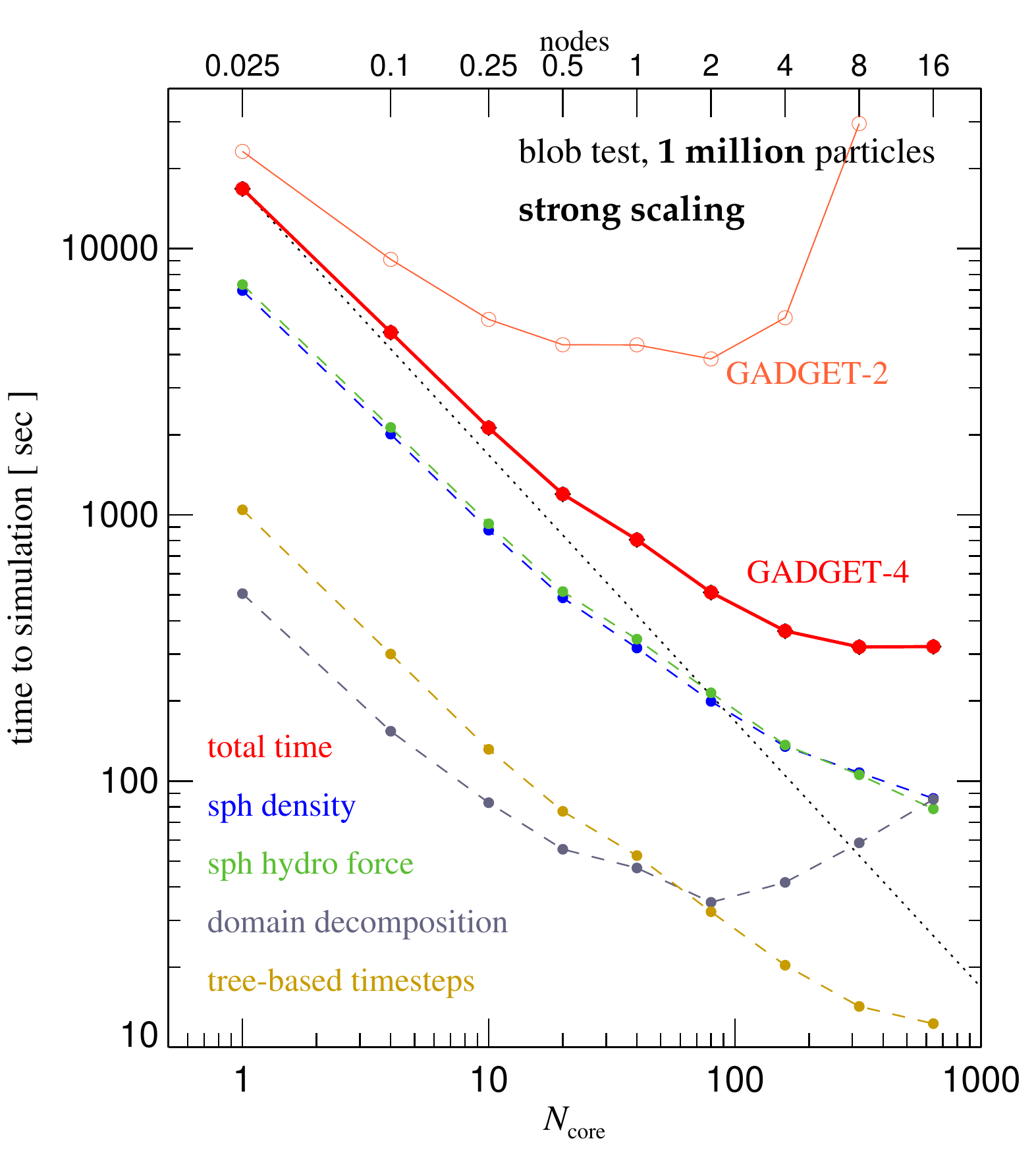}}%
    \resizebox{8.5cm}{!}{\includegraphics{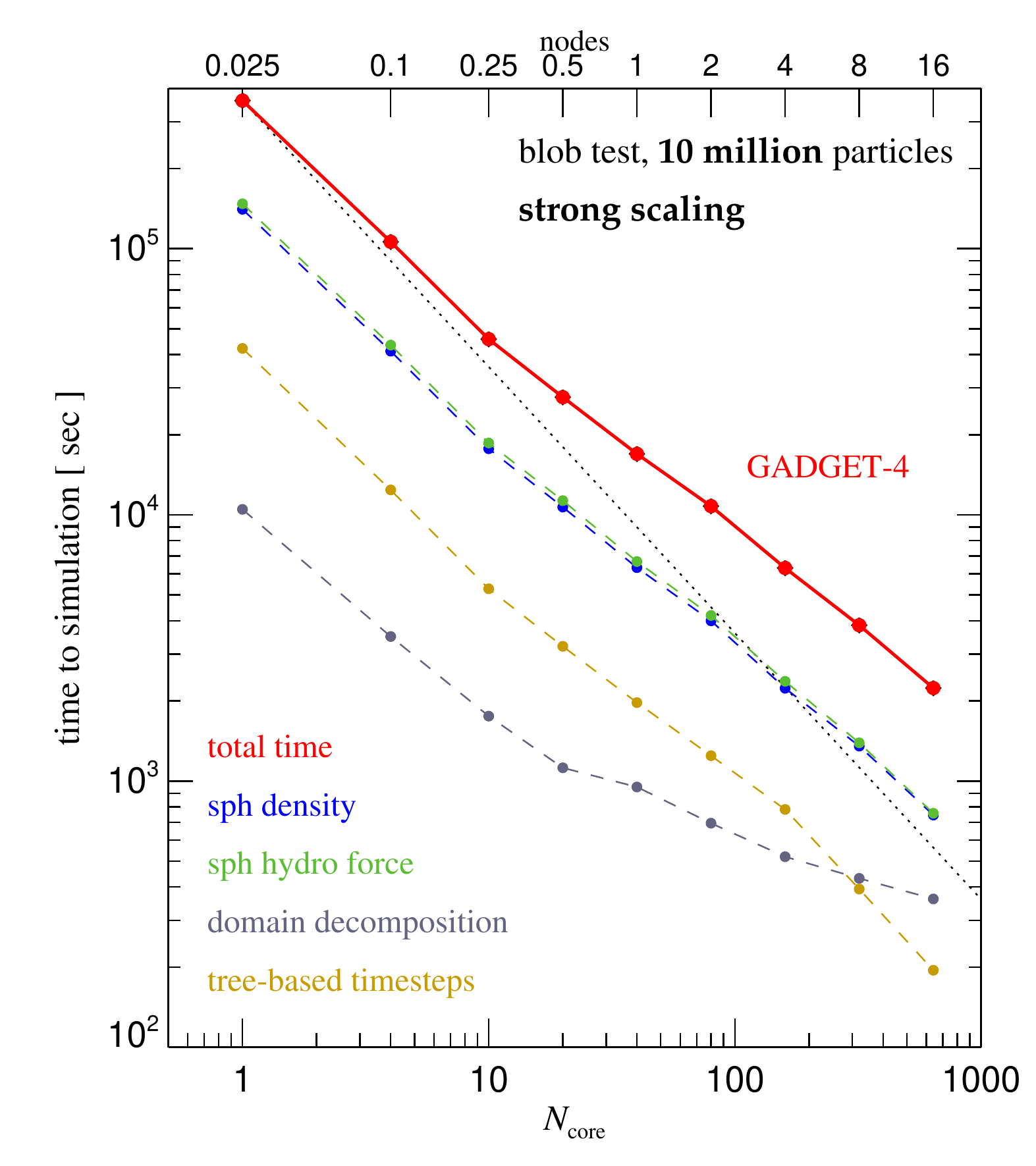}}
\end{center}
\caption{Evaluation of strong scalability and raw computational speed
  of a pure SPH calculation without self-gravity, the `blob test'
  analyzed in \citet{Agertz:2007aa}.  We have used the ``Freya''
  cluster at MPA for these tests (the nodes are equipped with dual
  Intel Xeon Gold 6138 CPUs, each with 20 physical cores at 2.0 GHz),
  and the original initial conditions at resolutions of 1 million
  (left panel) or 10 million (right panel) prepared by
  \citet{Agertz:2007aa} for the blob test. In both cases, we have run
  simulations in serial (1 core) and parallel with up to 640 cores (16
  full nodes). The filled circles show the wall-clock time consumed by
  different code parts of {\small GADGET-4} over the full course of
  the simulation up to time $t=5.0\, \tau_{\rm KH}$ (where
  $\tau_{\rm KH}$ is the Kelvin-Helmholtz time). The solid red line
  marks the total elapsed time. Perfect strong scalability corresponds
  to the dotted line. For the small problem size, strong scaling is
  lost at around 8 nodes; at this point it does not make sense any
  more to add more compute resources. The main culprit responsible for
  this is the domain decomposition algorithm, while the actual SPH
  calculations still perform well at this point, although imbalance
  and communication losses also become noticeable. Still, {\small
    GADGET-4} performs much better than {\small GADGET-2} (open
  symbols and thin red line), which has been run for comparison with
  the same set-up and loses scalability much earlier.  As is well
  known, larger problem sizes allow for better scalability in
  practice. This is explicitly demonstrated in the right hand panel,
  where the corresponding test results for the simulation with 10
  million SPH particles are reported. Here {\small GADGET-4} shows
  good strong scaling all the way to 640 cores.
\label{FigBlobTest}}
\end{figure*}

\subsection{The Santa Barbara cluster}

As a further example and useful comparison to other hydrodynamical
codes, we use the Santa Barbara cluster simulation, which goes back to
a community code comparison effort carried out by
\citet{Frenk:1999aa}. Since then, the Santa Barbara cluster has
repeatedly been used in the literature as an important validation
check of cosmological hydro codes, and also as a means to study
fundamental issues related to entropy creation and the treatment of
mixing in different hydrodynamical discretization techniques. Here we
focus on a basic validation test of the cosmological SPH
implementation in the {\small GADGET-4} code, by carrying out the
Santa Barbara simulation at a set of different resolutions, including
$2\times 256^3$, which is higher than considered in the original
comparison project.

In Figure~\ref{FigSantaBarbara}, we show radial profiles of the dark
matter density, the gas density, the gas temperature and the entropy
of the main cluster halo at $z=0$.  We here show the pressure-based
formulation of SPH, but note that the density-based formulation gives
very similar results.  The agreement between different resolutions in
the dark matter and gas properties shows the expected
systematics. Diverse answers have been reported for the central
entropy profile between different hydrodynamical codes \citep[see,
e.g.,][]{Ascasibar:2003aa, Springel:2005aa, Almgren:2013aa,
  Hopkins:2015aa, Saitoh:2016aa}. The low entropies we find here for
SPH in the central cluster region can be interpreted as arising from a
low degree of mixing in the central cluster region, likely resulting
from an unphysical suppression of turbulent mixing, an effect that
tends to be slightly more pronounced in the density-entropy
formulation of SPH. Most mesh codes predict considerably higher
central entropies, likely too high ones in some cases, as these codes
can easily produce excess entropy associated with truncation errors
and numerical diffusivity \citep{Mitchell:2009aa, Springel:2010ab,
  Vazza:2011aa}. A number of modifications to SPH have been suggested
that also produce entropy cores \citep{Sembolini:2016aa}, but
ultimately the correct answer for this problem is not fully
established yet.

\subsection{Merger-induced starburst}

We next consider a simple simulation of a galaxy merger including
radiative cooling and star formation, here modelled with the explicit
subgrid model introduced by \citet{Springel:2003aa}, which is
presently the only approach available in {\small GADGET-4} to account
for feedback from star formation. For definiteness, we set-up two
identical compound galaxy models which have initially $3\times 10^5$
particles in the dark matter halo, $1.5 \times 10^5$ particles in an
initial exponential stellar disk, $3\times 10^5$ SPH particles for the
gas component, and $5\times 10^4$ in a central stellar bulge.  The
virial velocity of the halo is chosen as $160\, {\rm km\,s^{-1}}$,
giving the galaxy a total mass close to
$10^{12}\,h^{-1}{\rm M}_\odot$. We choose a total disk mass
fraction of 0.05, with 33\% of this in the gas component, the other
66\% in the preformed stellar disk. The initial disk scale length is
$2.53\,h^{-1}{\rm kpc}$, and we additionally put a mass fraction of
0.02 into a central, spherically symmetric stellar bulge.

The galaxies are set-up in quasi-equilibrium with the {\small
  MAKENEWDISK} code, which implements a method described in
\cite{Springel:2005ac} that can also account for gas pressurized by
the ISM model, unlike our more sophisticated {\small GALIC} code
described in \citet{Yurin:2014aa}.  We then put them onto a collision
orbit that is parabolic, and has a nominal minimum separation at first
encounter equal to $3\,h^{-1}{\rm kpc}$ if the galaxies were point
masses. The merger simulation is started when the galaxy centers are
$160\,h^{-1}{\rm kpc}$ apart. At this time, the dark matter halos
already have started to overlap, but the initial orbit has been
corrected such that the galaxies fall together on the same path as if
they had started from a still larger separation. We also incline the
disks slightly relative to the orbital plane, by 10 and 40 degrees,
respectively, making this a prograde merger.

In Figure~\ref{FigMergerTimeSequence}, we show a time sequence of this
high-resolution, slightly ``wet'' galaxy merger. We have used
hierarchical time integration here, but the results are essentially
invariant with respect to the details of the integration settings.
The time evolution shows the typical dynamics of a major merger in
$\Lambda$CDM. The galaxies experience a strong tidal shock in their
first encounter, which induces tidal arms and drives some gas into the
central regions, but also some gas away out of the disks. Due to
braking by dynamical friction in the dark matter halos, the galaxies
fall together for a second encounter and eventual coalescence. This
largely destroys the disks, triggers a moderate starburst, and creates
a hot gaseous halo by shock heating some of the gas in the collision.

This evolution is also reflected in the star formation rate, which we
show in Figure~\ref{FigSFRGalaxyMerger}. Globally, the gas is consumed
on a consumption timescale of about $1\,{\rm Gyr}$, as illustrated by
the dashed line in the figure. The merger modulates this evolution by
first triggering an enhancement in the star formation in the first
encounter. Eventually, the star formation is diluted while the
galaxies linger around at their turn-around radii, before they engage in
a second encounter that quickly leads to a full merger. The latter
triggers a moderate starburst, before the residual gas in the remnant
is consumed again on the consumption timescale of $\sim 1\,{\rm Gyr}$
imposed by the relatively stiff equation of state model used here.

\subsection{Isothermal sheet}

To demonstrate and validate {\small GADGET-4}'s feature to allow
self-gravity with periodicity only in two-dimensions, we consider the
problem of an isothermal gaseous sheet with periodicity in the $x-$
and $y$-directions, and open boundaries in the $z$-direction.  For
this case of translational symmetry parallel to the sheet, there is an
analytic solution, Spitzer's isothermal sheet. The density profile in
the $z$-direction follows
\begin{equation}
  \rho(z) = \rho_c \,{\rm sech}^2\left(\frac{z}{z_0}\right)
\end{equation}
with central density $\rho_c =  {\sigma}/({2\,z_0})$.
The vertical scale height $z_0$ is given by
\begin{equation}
  z_0 = \frac{c_s}{ ( 2\pi G \rho_c)^{1/2}},
\end{equation}
where $c_s$ is the sound-speed, and the gravitational potential relative to
the mid-plane of the sheet is given by
\begin{equation}
  \phi(z) = 2 c_s^2 \ln \left[ {\rm cosh}
    \left(\frac{z}{z_0}\right)\right].
\end{equation}

The isothermal sheet is unstable to perturbations transverse to the
$z$-directions if their wavelength exceeds
\citep{Ledoux:1951aa,Hunter:1972aa}
\begin{equation}
\lambda_{\rm crit} = c_s \left(\frac{2 \pi}{G \rho_c}\right)^{1/2}.
  \end{equation}
 This implies $\lambda_{\rm crit} = 2 \pi z_0$. We thus expect that
the isothermal sheet can only stay stable if the periodic boxsize
in the directions transverse to $z$ is at most $2 \pi $ times the
scale height.

To verify our gravity solver with periodicity in two out of three
dimensions, and to test for this bound, we carry out two simulations.
For definiteness we pick a box with dimensions $L_x= L_y = 1$, and
$L_z = 32$. In one of the runs, we adopt a scale-height of $z_0=1$
with a surface density of $\sigma = 1$, in the other we choose a
thinner disk with $z_0=0.1$ and $\sigma = 0.1$, such that the
dynamical timescales of both disks are the same. With these
parameters, the thicker disk should be in the stable regime, whereas
the thinner is expected to be unstable to perturbation transverse to
the $z$-direction. To trigger this potential instability in a
controlled fashion, we create the particle positions in the initial
conditions by randomly sampling the isothermal sheet and then
displacing them around the midplane by
$\Delta z(x) = 0.2 \, z_0 \sin(2 \pi x /L_x)$.

In Figure~\ref{FigStretchedBox}, we show the time evolution obtained
for these two sheets, which can be considered `thick' and `thin' in
units of the transverse box size.  The expectations on their stability
are borne out by the time evolution of the sheets; whereas the initially
imposed perturbation in the $z$-direction is dissipated away for the
thick sheet, it grows for the thin sheet, leading to a fragmentation
of the sheet into a set of spherical blobs.

The equilibrium state assumed by the first sheet accurately follows
Spitzer's isothermal sheet profile. This is illustrated in
Figure~\ref{FigIsothermDensity}, which compares the SPH particle
densities in the final state with the analytic solution. Indirectly
this also confirms that the gravity solver produces accurate results
for this case of mixed boundary conditions.

\begin{figure*}
\begin{center}
  \resizebox{18.2cm}{!}{\includegraphics{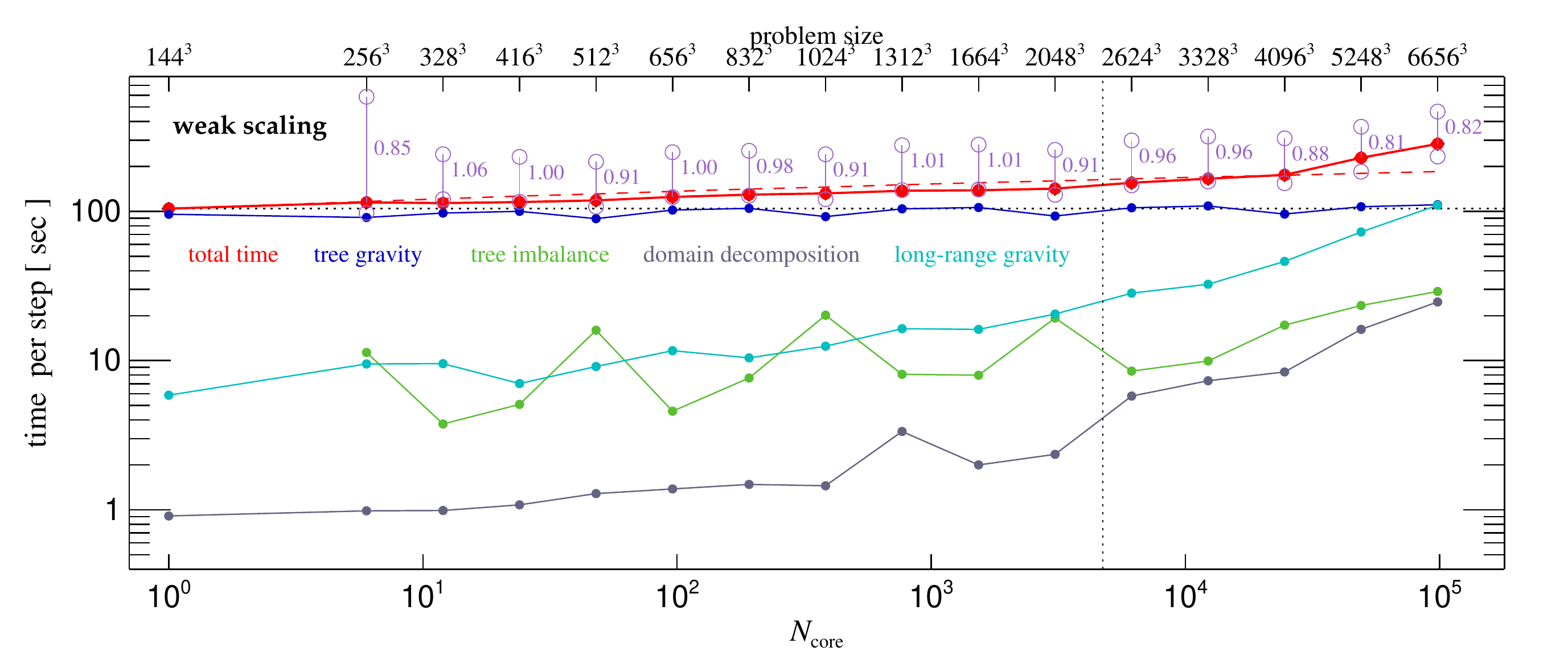}}
\end{center}
\caption{Weak scaling test of {\small GADGET-4} on SuperMUC-NG (which
  consists of nodes with two 24-core Intel Xeon Platinum 8174 CPUs at
  3.1 GHz) at the Leibniz Supercomputing Centre. We hold the load per
  MPI rank fixed, and run typical cosmological simulations in the
  high-redshift regime where the box size, particle resolution, and
  PM-grid size are all increased in lock-step, while the mass
  resolution is kept fixed. We start with $144^3$ particles on a
  single core, and go all the way to $6656^3$ particles on 98304
  cores, corresponding to 2048 compute nodes of SuperMUC-NG.
  The PM grid is always twice the particle grid per
    dimension, i.e.~$288^3$ for the smallest test, and going up to
    $13312^3$ for the largest. The different coloured lines show the
  wall-clock time consumed per step in the most expensive code parts,
  averaged over three full steps. We see that the most expensive code
  part, the tree-based short-range force calculation, shows
  essentially perfect weak scaling over the full range. In particular,
  the work-load imbalance losses in the tree calculation are
  negligible compared to the total cost, thanks to the high-quality
  domain decomposition and our special shared-memory one-sided
  communication strategy. The domain decomposition and long-range
  force calculation show a less ideal scaling; towards the largest
  processor number considered, the PM-force becomes increasingly
  expensive and will eventually overtake the short-range gravity
  calculation due to a saturation of the bandwidth of the
  communication backplane (which for the large partitions suffers from
  the need to use several islands of the machine with an associated
  reduction of the available total bisectional
  bandwidth). Nevertheless, the total cost (red solid line) shows
  excellent weak scaling, especially when compared to the
  theoretically expected optimum (red dashed line) that takes into
  account that the cost of the force calculation algorithm scales as
  $N\,\log(N)$, where $N$ is the particle number. The two purple
  hollow circles shown for each processor number $N_{\rm core}$ give,
  for the lower circle, the total cost assuming perfect weak scaling
  relative to the previous run with $N_{\rm core}/2$, or for the upper
  circle, the expected run time if no speed improvement results
  despite using twice as many cores. The inlined numbers give the
  fraction of the speed-up that is actually realized by the code
  thanks to the increase of the number of cores relative to the
  previous run.  Multiplying all these numbers together for the whole
  scaling series, one obtains 0.37, which can be interpreted as the
  relative speed of the $10^5$ core run with respect to a fiducial
  speed based on the time needed by the serial code, assuming perfect
  scalability and {\em ignoring} the $\log(N)$ factor in the
  theoretically expected scaling. If the latter factor is included as
  well, one obtains 0.66. This means that we see a parallelization
  loss of only about $33\%$ compared to what is theoretically
  possible, even though we parallelize here a tightly coupled physical
  problem over $\simeq 10^5$ cores. We consider this to be
  reassuringly good. Note that by choosing a smaller PM-grid size, the
  scalability could be improved even further, albeit at the expense of
  making the tree calculation slightly more costly. Finally, we note
  that the dotted vertical line marks the simulation size where a
  slab-based FFT algorithm looses scalability, because for larger
  problem sizes the number of mesh slabs becomes smaller than the
  number of processors. Our column-based FFT algorithm used here
  eliminates this scaling limitation.
\label{FigWeakScaling}
}
\end{figure*}

\section{Code performance and scalability} \label{secscalability}

The quest of achieving maximum computational speed often leads to
highly specialized algorithms and computer codes that are only
applicable to a particular class of problems. {\small GADGET-4}
deliberately aims to be a multi-purpose code that can be applied
comparatively flexibly to a range of problem types, without being tied
to special hardware. It therefore may not necessarily be the fastest
code, but the hope is that it instead excels in reliability and
flexibility. But also with these goals in mind, the code still needs
to show competitive performance and good scalability to be useful in
practice.

In this section we thus examine the scalability of the code for
realistic problems, including very demanding ones such as extreme
zooms. The latter tend to be harder than tests for uniform particle
load, and are arguably more interesting for the methodological goal to
allow future simulations to better address multi-scale, multi-physics
problems. In the following, we shall first consider the strong
scalability of the code for a pure SPH simulation. We then turn to the
important problem of weak and strong scalability of homogeneously
sampled cosmological dark matter simulations, which is directly
relevant for timely studies of cosmic large-scale structure, which
increasingly call for very large N-body simulations. We then turn to
strong scaling tests of a cosmological hydrodynamical simulation, and
finally to a zoom simulation of a Milky Way-sized high resolution dark
matter halo.

\begin{figure*}
  \resizebox{9.1cm}{!}{\includegraphics{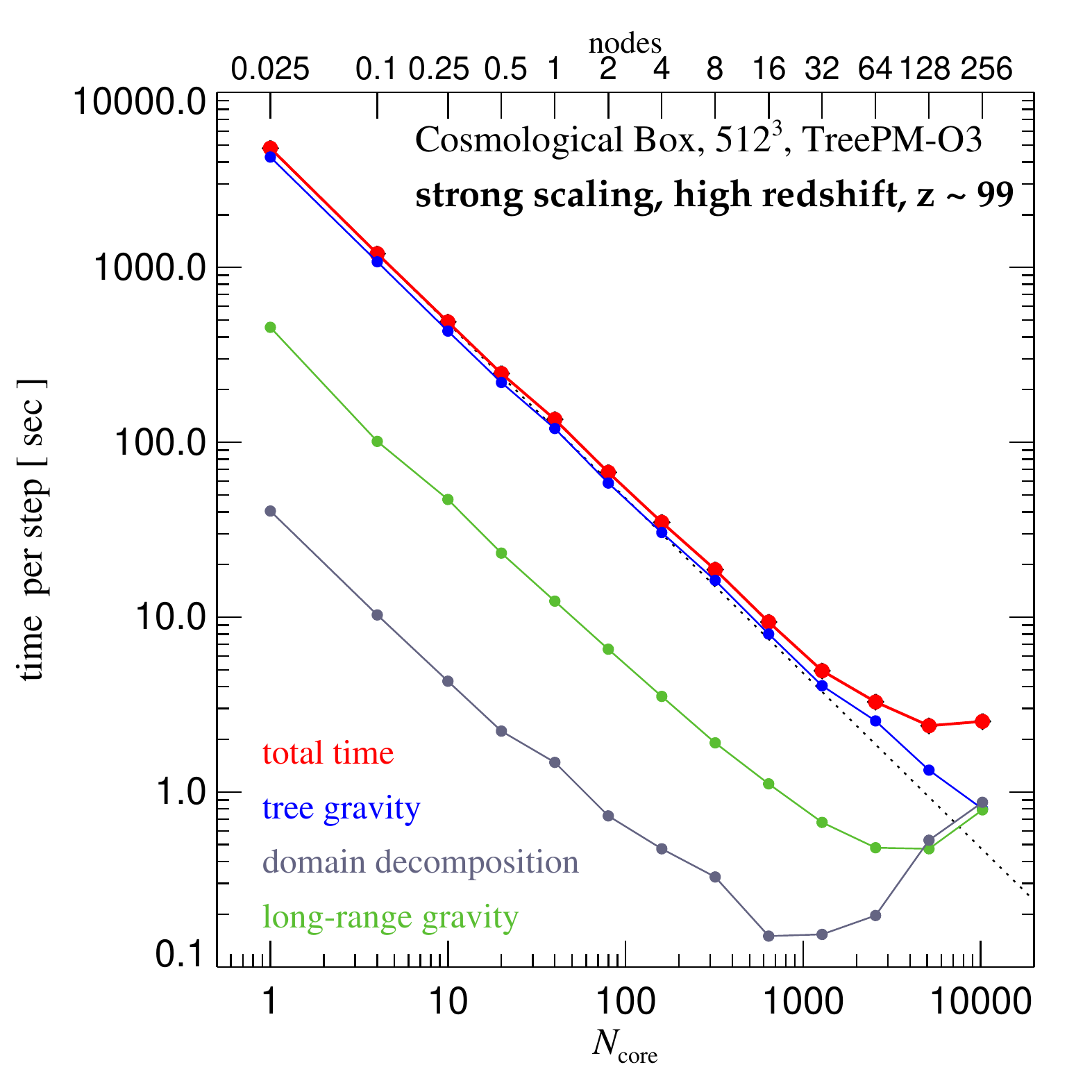}}%
  \resizebox{9.1cm}{!}{\includegraphics{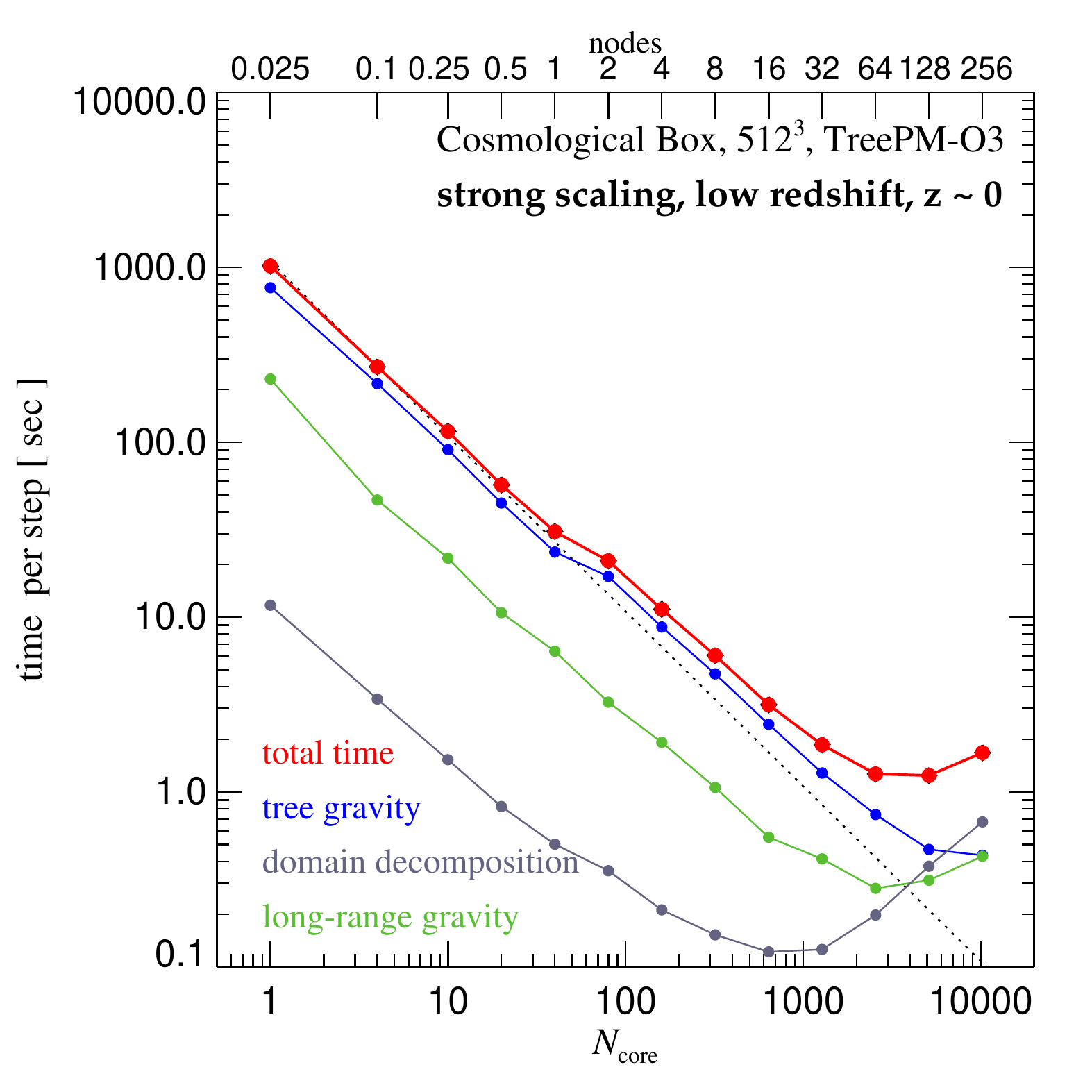}}%
  \caption{Strong scaling of {\small GADGET-4} for a homogeneously sampled
    cosmological simulation. As a test simulation,
    we use a $512^3$ run in a $L= 100\,h^{-1}{\rm Mpc}$ box.
    The tests were run on ``Cobra'', a large parallel
    compute cluster operated by the Max Planck Computing and Data Facility. Each node is equipped with
    two  Intel Xeon Gold 6148 CPUs with 20 physical cores at 2.4
    GHz. The lowest number of cores we run on is 1 (i.e.~we carry it
    out as a serial  calculation), and we extend
    this up to 10240 cores (256 nodes), using identical code parameters in each
    case (except for the amount of memory that may be allocated per
    MPI-rank). As the clustering state changes strongly between the
    high redshift and low redshift regimes, we examine the scaling in
    both regimes separately, with high redshift shown in the left
    panel, and low redshift in the right panel. The reported wall-clock times 
    for different code parts are averaged times per step; at high-$z$
    the averaged timestep size is $\Delta \ln a = 4.5\times 10^{-3}$, at low-$z$ it is
   $\Delta \ln a = 1.78\times 10^{-4}$.
   For this problem size, we lose strong scaling slightly below $\sim
    10000$ cores, at which point the load per rank
    has already dropped to $\sim 13000$ particles. The tree calculation
    still scales at this point, but both the domain decomposition and
    the PM-mesh calculation (here done with a $1024^3$ grid) do not
    scale anymore for this low load per core.
\label{FigStrongScaling}
}
\end{figure*}

\begin{figure}
\begin{center}
  \resizebox{8.5cm}{!}{\includegraphics{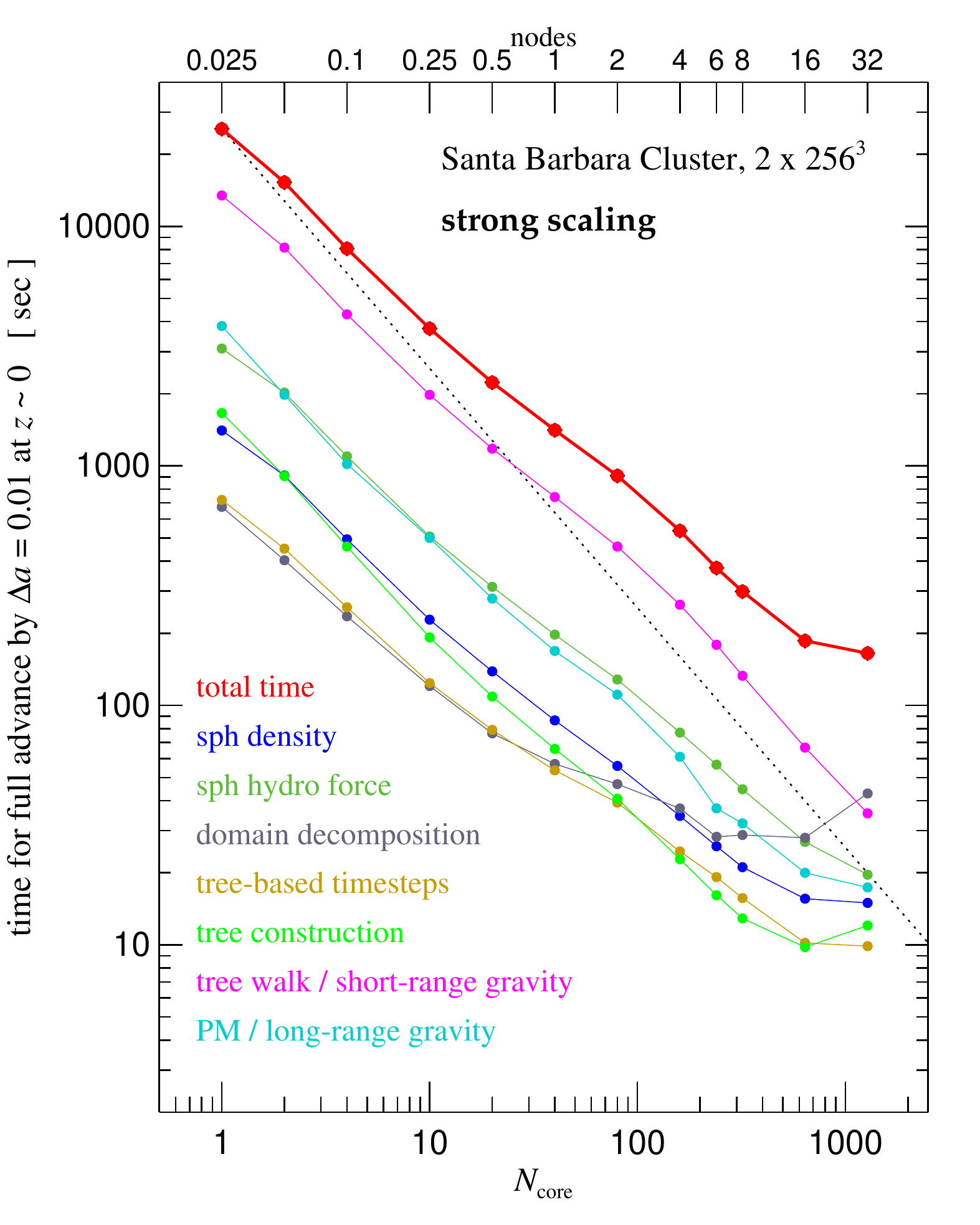}}
\end{center}
\caption{Strong scaling test of the Santa Barbara
  Cluster \citep{Frenk:1999aa}, using the $2\times 256^3$ realization
  discussed earlier in Fig.~\ref{FigSantaBarbara}.
We here consider the elapsed time needed in different code parts of
{\small GADGET-4} to evolve the simulation at $z=0$ for a time
interval corresponding to $\Delta a = 0.01$ (i.e.~for 1\% in
expansion factor), which averages over the full timestep hierarchy.
We use different numbers of cores on MPA's
``Freya'' cluster, starting at a serial computation, and then going parallel
by using up to 32 nodes with 40 cores each. We see that for this problem, strong scaling to around
1000 cores is obtained, at which point the domain decomposition starts
to spoil further speed-ups when more nodes are added.
\label{FigStrongScalingSB}
}
\end{figure}

\begin{figure}
\begin{center}
  \resizebox{8.5cm}{!}{\includegraphics{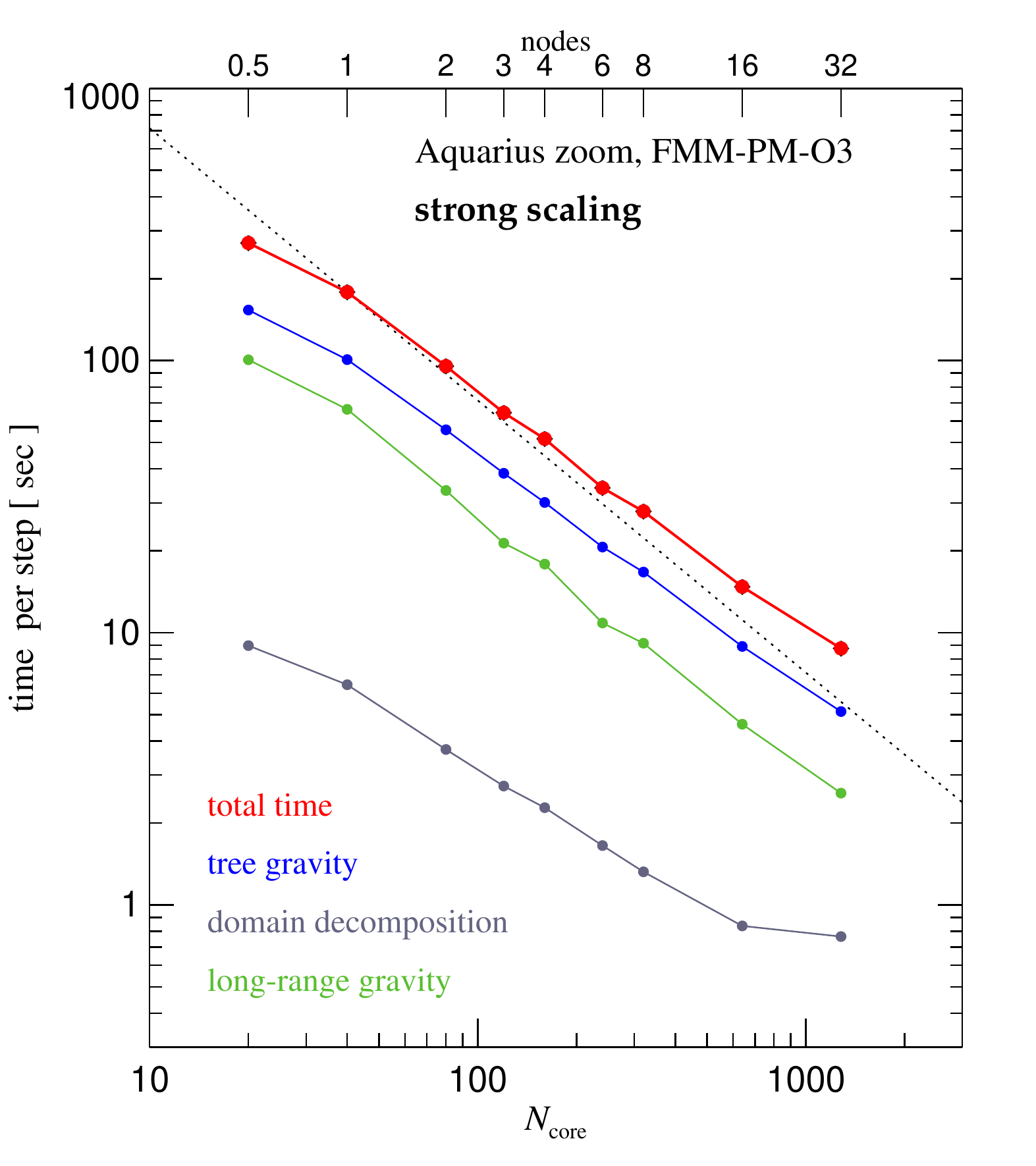}}
\end{center}
\caption{Strong scaling of a high-resolution zoom simulation of a
  Milky Way-sized dark matter halo on ``Cobra''. For definiteness, we consider the
  Aq-A-2 model of the Aquarius project, at redshift $z\simeq 0.03$, i.e.~when the highest degree of
  clustering is reached.
This simulation contains about 607 million particles in total, of
which around 184 million are within the virial radius, defined as enclosing 200 times the mean
density of the box. The corresponding sphere occupies a fraction of
just $3.1 \times 10^{-8}$ of the simulated volume, highlighting the
extremely inhomogeneous clustering of the particle distribution.
  We employ the FMM-PM approach  with $p=3$  and $N_{\rm grid}=1536$ in this
  test case, combined with the
  hierarchical timestepping approach
  (not because this is necessarily the fastest, but rather just for
  variety compared to the other tests done previously). The reported times for different code parts
  are the average wall-clock time per step (which are of size $\Delta \ln a =
  1.7\times 10^{-5}$) averaged over 3 full cycles through the timestep
  hierarchy.  Clearly, the strong scaling
  of  {\small GADGET-4} for this demanding problem is remarkably good.
  \label{FigG4AquariusScaling}}
\end{figure}

\subsection{Pure hydrodynamical problem}

We here examine a pure SPH problem, the `blob test' of
\citet{Agertz:2007aa}, in order measure the speed and scalability of
the SPH implementation in {\small GADGET-4}, with results given in
Figure~\ref{FigBlobTest}. The left panel shows speed measurements in
different code parts when the problem is studied with 1 million
particles, while the right panel is for the same problem with 10
million particles instead. In this set-up, a cold dense sphere of gas
is exposed to a hot, supersonic background wind in order to study the
shredding of the clump by fluid instabilities created in the strong
shear flow around the clump.  Periodic boundary conditions are imposed
on all sides of a stretched box, gravity or cooling processes are not
included, and ordinary 'vanilla' SPH is used. We use local
timestepping for the integration. The shortest occurring timesteps are
about a factor of $\sim 10$ smaller than the longest, so that up to
four timebins of the timestep hierarchy are used.

We find that the code shows generally good strong scalability for this
problem. Scalability is lost once the load per core has dropped to a
few thousand SPH particles, at which point our domain decomposition
algorithm becomes too costly in relation to the little work it tries
to distribute over more and more shoulders. For a larger problem size,
the limit for scalability is accordingly pushed to a larger number of
cores (see the right panel). This highlights an important general
finding that we expect to be true for all cosmological simulation
codes -- good scalability to a certain processor number requires a
sufficiently large problem size. Importantly, as the direct comparison
to {\small GADGET-2} shows (see the left panel in
Fig.~\ref{FigBlobTest}), the new code {\small GADGET-4} is less
demanding in this respect than older versions of {\small GADGET}, and
thus shows a significantly improved scalability.

\subsection{Homogeneously sampled cosmological simulations}

We now consider the weak scaling of the code for cosmological DM-only
simulations. This is shown in Figure~\ref{FigWeakScaling} for a mass
resolution resolution of
$m\simeq 5.9 \times 10^8 \, h^{-1}{\rm M}_\odot$, carried on the
SuperMUC-NG computer at the Leibniz Supercomputing Centre in Garching.
In this test, we increase the used partition size, the PM-mesh size,
and the size of the particle grid in lock step, such that the particle
load per core stays constant.  We begin with $144^3$ particles on a
single core, and proceed up to $6656^3$ particles on 98304 cores
(which are 2048 compute nodes of SuperMUC-NG).  The calculations use
the column-based FFT and the TreePM algorithm for $p=3$.  The
slab-based FFT would be slightly faster (by up to a factor of 2 for
the PM part) for low core numbers, but ceases to scale once the number
of mesh-planes becomes less than the number of cores, and then quickly
loses out against the column-based FFT algorithm.

Overall, the code achieves very good weak scalability. With respect to
the theoretically expected mathematical cost of the calculation, which
scales as $N\log N$ with particle number $N$, the code realizes a
parallel efficiency of about $66\%$ on $10^5$ cores relative to
carrying out the calculation on a \textit{single core}, or in other
words, a speed-up by a factor of 66000 is reached -- despite the
coupling of every particle with every other particle through
long-range gravity, and despite the fact that the simulations with the
largest core sizes have to be spread over several ``islands'' of the
SuperMUC-NG computer, which brings about a significant drop in the
available point-to-point communication bandwidth.

Next, we turn to the problem of the strong scalability of a
homogeneously sampled cosmological simulation. We examine this for a
$512^3$ problem size, both for the situation at high redshift close to
the initial conditions, and for the highly clustered state at redshift
$z=0$, where the code traverses a timestep hierarchy.  Results for
these tests are shown in Figure~\ref{FigStrongScaling}, again starting
at a single core, and now going up to $10^4$ cores. Accordingly, the
particle load per core drops from more than 134 million to little more
than 10000. We find quite good strong scalability of the code for all
relevant code parts up to about a load of $10^5$ particles per
core. Then the domain decomposition starts to increase in cost, and
soon seriously impacts further scalability. This would thus be the
first target for future work on further improving the scalability of
{\small GADGET-4}. The angle of attack for this is in fact relatively
obvious -- unlike the gravitational algorithm itself, the domain
decomposition does not yet exploit the fact that typically a sizeable
number of MPI ranks runs on a shared-memory node. Instead it assumes
the ranks are all on distributed memory-memory machines, making the
domain decomposition a more difficult and laborious task than really
necessary. Implementing a version of the domain decomposition that
takes this hierarchy into account has the potential to speed it up by
about an order of magnitude on typical nodes of today's compute
clusters, which would come in handy for pushing the scalability
further.

We also note that the scalability of {\small GADGET-4} is rather
similar for the TreePM and FMM-PM variants of the force computation
algorithms. But of course, the absolute and relative speeds vary
depending on the expansion order that is chosen, force accuracy
parameter settings, etc. Also note that using an extremely small
opening angle for the multipole algorithms can make the relative cost
of other code parts irrelevant. This would formally improve the
scalability while increasing the overall runtime. In the tests shown
here, we have however rather aimed to adopt settings that are typical
for production runs with {\small GADGET-4}.

Finally, for completeness, we examine a strong scaling experiment for
a hydrodynamical cosmological simulation. For definiteness, we
consider the Santa Barbara Cluster simulation at a resolution of
$2\times 256^3$ dark matter and SPH particles. In
Figure~\ref{FigStrongScalingSB}, we show the speed of different code
parts when the simulation is evolved in the highly clustered state at
$z\sim 0$, through several full cycles of the timestep hierarchy. We
consider runs carried out in serial on 1 core and compare with parallel execution
on up to 1280 cores. Over this range, good strong scalability can be observed,
but it starts to break down at around 1000 cores, again due to the
domain decomposition raising its expensive head.

It is interesting to compare these scaling results with other parallel
performance measurements reported in the literature for cosmological
codes.  Recently, \citet{Borrow:2018aa} discussed the strong and weak
scaling of the new {\small SWIFT} code for cosmological simulations,
which employs task-based parallelization controlled with a custom
scheduler. For the strong scaling of a problem with $6\times 10^7$
particles (hence about twice as large as the Santa Barbara cluster)
using 512 cores, they achieve a speed up of $\sim 33.1$ relative to a
single-threaded execution. Using a very large particle load of
$7\times 10^6$ per core, \citet{Borrow:2018aa} report weak scaling to
4096 threads with an overall loss of $25\%$ efficiency. These scaling
results appear somewhat worse than what we find for {\small GADGET-4},
but their test problem was also characterized by a deep time-step
hierarchy.  We note, however, that {\small SWIFT} has been claimed
\citep{Schaller:2016aa, Gonnet:2013aa} to offer a substantially higher
base speed than {\small GADGET}, so it could be a faster code
nevertheless, at least for a moderate number of cores.

Our weak scaling results also appear to be better than the ones
reported by \citet{Wunsch:2018aa} for the {\small FLASH} code. There,
even the hydrodynamics shows already a factor of 2 slow-down in a weak
scaling to 1536 cores, while this approaches a factor of 10 for the
tree-based gravity in {\small FLASH}, and is presumably still worse
for {\small FLASH}'s multigrid gravity solver, judging at least from
its relatively limited range of strong scalability.

At face value, our scaling results are also better than the ones
reported for {\small CHANGA} \citep{Menon:2015aa}, a cosmological
N-body code employing the Charm++ framework for task-based
parallelization.  They also appear better than the weak scaling
reported by \citet{Harnois-Deraps:2013aa} for {\small CUBEP3M}, who
found a weak scaling efficiency of 58\% when going from 864 to 21952
cores, a dynamic range in  core number of a factor of
$\sim 25$, considerably less than the wide dynamic range of $10^5$ we
have examined here.

Our scaling results are however not as impressive as the ones reported
by \citet{Habib:2016aa} for the {\small HACC} code framework, who
found essentially perfect weak scaling on a BlueGene/Q system up to
$1$~million cores.  Similarly, \citet{Winkel:2012aa} documented very
good weak scalability of their parallel implementation of a tree-code
on a BlueGene/P system up to 256000 cores.

\subsection{Collisionless zoom simulations}

Finally, let us also consider the computational problem of an
aggressive zoom simulation. The extreme clustering and much higher
dynamic range of such simulations represent a quite demanding regime
in which good parallel scalability is far harder to achieve than in
homogeneously sampled boxes. As a specific test problem we consider a
simulation from the Aquarius project \citep{Springel:2008aa},
specifically halo Aq-A-2, which we evolve for a couple of full
timestep cycles at $z\simeq 0$, when the clustering is strongest. This
simulation features about 532 million high resolution particles, of
which around one third are contained in the virial radius of a single
halo, which occupies far less than a millionth of the simulated
volume. In addition, there are about 75 million lower resolution
particles of mass progressively growing with distance from the zoomed
halo, filling the rest of the volume.

In Figure~\ref{FigG4AquariusScaling}, we show strong scaling
measurements with {\small GADGET-4} for this demanding situation, this
time using the FMM-PM-O3 algorithm with hierarchical time integration
for the sake of showing a variety of tests, not because it scales
particularly well (in fact it does not, the Tree-PM variants tend to
scale a bit better).  We explore the scalability from 20 cores (half a
node) to 1280 cores (32 nodes), finding a quite satisfactory strong
scaling efficiency over this range. Again, this is a welcome
improvement over older versions of {\small GADGET}, which do less well
for this type of problem.

\section{Public code release} \label{secpublic}

More than a decade ago, we described the {\small GADGET-2} code
\citep{Springel:2005aa}, and made it publicly available for download
on the internet as a packaged source-code archive file, as it was
commonly done at the time. The code had been released as is, and no
simple process was foreseen for the community to feedback suggestions
for code changes or improvements. In fact, although such developments
have happened at many places, such feedback hardly happened, and the
public version of {\small GADGET-2} remained essentially static,
modulo a relatively small number of fixes.

The development of {\small GADGET} continued in a decoupled way from
this public version, leading to a major transformation of the code
around 2007, and the creation of {\small GADGET-3} (this became the
basis of the Aquarius project). Subsequently, a large number of people
became involved in developing this code further.  Unfortunately, it
was never made publicly available and properly documented, but it has
been widely shared and distributed by the author upon request. It has
also been used as a basis for developing further codes, as described
in the introduction section. In recent years, besides our own efforts,
two other groups have independently pushed this code to much better
scalability, in Munich by Klaus Dolag and collaborators, and in
Pittsburgh by Tiziana di Matteo and collaborators. The largely
uncoordinated development of {\small GADGET} has inspired many
creative solutions and enabled a large amount of science, but it also
created a fair amount of duplication of effort and a jungle of
different versions of {\small GADGET-3} that is quite hard to see
through at this point.  This of course complicates code validation and
the reproducibility of results, and the repetition of effort involved
ultimately slows down progress.

With the public release of {\small GADGET-4} through a more modern,
collaborative, git-based coding
platform\footnote{https://gitlab.mpcdf.mpg.de/vrs/gadget4} that
integrates version control, bug tracking, user registration, etc., and
the hosting of the code on a standard platform, we hope to be able to
avoid this situation in the future. Instead of downloading a static
source archive file, users can fetch the source code repository from
the version control system that contains the commit history of the
code. This has become the standard for most open source software
projects, and has also been picked up successfully by other community
codes in the field, such as {\small ENZO} \citep{Bryan:2014aa},
{\small RAMSES} \citep{Teyssier:2002aa} , {\small PHANTOM}
\citep{Price:2018aa}, {\small AREPO} \citep{Weinberger:2020aa}, among
others. We expect this to more easily allow people to contribute
suggestions for improvements of {\small GADGET-4} (as a patch or a
pull request) that can be integrated into the main code base after
approval of a team of moderators. At the same time, users can branch
the code for their own private developments, and share these
developments with others according to their own policies.

In addition, we have implemented measures in the code that are
designed to protect against basic forms of ``code rot'', in particular
against adding features that are undocumented. If a new feature
(controlled through a compile time option) or a new run-time parameter
(set through a parameterfile) is introduced, it must be documented as
well, otherwise the code will simply refuse to compile with a
corresponding error message. This is realized through a special control
script as part of the build-process. Likewise, if a documented feature
is removed in the actual code, the corresponding documentation needs
to be purged as well, so that consistency is always ensured. We hope
this simple measure encourages better discipline when modifying the
code, and promotes a culture where all code features are always well
documented.

The code uses a sophisticated internal memory management system that
excludes the possibility of accidental memory leaks while at the same
time preventing memory fragmentation. Also, this allows one to always
determine precisely how much memory is used, and which code lines are
responsible for the different allocations. The code also protects
automatically against overuse of the physical memory available on a
machine, and hence against the possibility to drive a compute node
into swapping ('threshing'). Further stability features include an
optional health test upon code start-up that checks for homogeneous
CPU-performance across the assigned partition, and evaluates the
communication speed between different CPUs. This can for example
detect whether a compute node is slowed down by a rogue process
(something that can have a huge impact on a parallel job), or whether
a single compute node shows reduced network bandwidth that would
similarly lead to a skewed performance of the whole run.

Finally, we want to note that we have moved to C++ as default language
in {\small GADGET-4}, while older versions were written in C. Because
this has in part been achieved by modifying older C code, {\small
  GADGET-4} is hardly in a style that would emerge from redesigning it
in C++ from the outset. Still, the code has been segmented in a set of
different classes with well defined purposes, and which also hold the
code's data, thereby improving modularity. Inheritance and templating
have been used where appropriate to avoid duplication of similar type
of code at different places. For example, there is only one basic tree
construction routine in the code, and it is used to construct
different trees for the gravity calculation, neighbour search, or for
computing the potential in the gravitational unbinding in {\small
  SUBFIND}. Also, the tensor arithmetic needed in the higher-order
multipole expansions as part of the Tree and FMM calculations can be
neatly encapsulated into a corresponding class that supplies tensor
objects and overloaded binary operators for doing arithmetic with
them. Complicated expressions with tensors can then be expressed very
compactly in the code, without any index clutter, thereby greatly
improving readability and thus reducing opportunities for coding
errors.

\section{Discussion and conclusions} \label{secdiscussion}

In this work, we have described a new version of the {\small GADGET-4}
code, which we publicly release to the community. Our goal has been to
modernize {\small GADGET-2} from the ground-up, and to provide a
reference code that primarily excels in accuracy, versatility and
robustness when applied to gravitational dynamics in cosmic structure
formation. We hope that this will be a useful code base for numerical
cosmology, especially for the high accuracy simulations needed for
future large-scale structure work, but also for calculations with very
high dynamic range, because the code has been improved significantly
in scalability, and in its ability to carry out very large problem
sizes.

Compared to previous versions of {\small GADGET}, the following
features are arguably most noteworthy. The code now offers a Fast
Multipole Method as an alternative to a plain tree code, and both
algorithms can optionally be run with higher-order multipole moments,
or as TreePM or FFM-PM hybrid methods. The FFM approach is momentum
conserving and can in some situations be substantially faster at a
given force accuracy than the pure tree code. When combined with a new
hierarchical time integration scheme for gravity, the manifest
momentum conservation of FMM can even be retained for local and
adaptive timestepping.

We have improved the domain decomposition algorithm, letting it
balance different constraints at the same time and establish an
optimum compromise between them. In particular, this prevents memory
imbalance as a side-effect of attempts to balance the computational
work, which otherwise has often shown up as a problem in past work,
especially for zoom-in calculations. For small problem sizes, the cost
of the domain decomposition itself is presently usually the limiting
factor for strong scalability. However, our current algorithm for the
domain decomposition does not yet take note of the fact that easy
communication within nodes is possible through the use of shared
memory. Rather it subdivides the domains as if each MPI rank lived in
a separate distributed memory node. Refining this into a hierarchical
approach that takes the mapping of MPI ranks onto shared-memory
compute nodes into account has great potential to significantly reduce
the cost of the domain decomposition, and thus to extend the
scalability of {\small GADGET-4} further.

With respect to smoothed particle hydrodynamics, the new code offers
an improved version of vanilla SPH, and as an alternative one flavour
of pressure-based SPH. The new code structure allows much easier
extension of the SPH routines despite their more sophisticated
communication patterns. In particular, {\small GADGET-4} now uses
hybrid shared-memory parallelization with an MPI-3 enabled one-sided
communication approach, which we think is a promising coding style for
the powerful many-core shared-memory nodes that have become the
primary building of today's supercomputers.  Efficiency for a large
dynamic range is also further supported by the fact that the timesteps
of hydrodynamics and gravity can be split, with hydrodynamics
subsampling the gravitational dynamics.  We note that the public
version of the code also contains simple prescriptions for radiative
cooling and star formation, which are primarily intended to facilitate
the development of improved physics models by interested users.

The improved scalability of {\small GADGET-4} is further assisted by a
removal of certain barriers that existed in older versions of {\small
  GADGET}, such as the limits imposed by a slab-based approach to
parallel distributed FFTs. The code now offers an optional column
based approach that is completely flexible. It also allows arbitrarily
stretched boxes, as well as the possibility to restrict the
periodicity of gravity to just two dimensions. Other limits that have
been addressed and lifted are issues like the former restriction of
the halo numbers to 2 billion objects, or the fact that single MPI
commands cannot transmit more than 2 billion data elements. The code
detects such situations and uses several communication calls if
needed.

The code offers for the first time support for built-in merger tree
construction and lightcone outputs, optionally also featuring halo
identification directly on the lightcone. This promises to make
{\small GADGET-4} particularly useful for populating the backwards
lightcone with halos and/or galaxies predicted by a suitable
model. The built-in power spectrum measurement routines are another
feature that should be useful to quickly diagnose and analyse
simulation results. Usability of the code is also improved by
extensive code performance metrics that are produced and reported
during a calculation, and the support of HDF5 output throughout. 

In first test applications of the code we have shown that it provides
converged solutions for publicly available test problems that are
robust to changes of timestepping and force accuracy. Because we can
independently demonstrate that the forces are correct and unbiased, we
have good reason to believe that {\small GADGET-4} indeed gives
correct results when applied carefully, thereby providing reference
solutions that may be useful in many scientific applications. With the
public release of {\small GADGET-4} we hope to make a useful
contribution to the development of more powerful and accurate
simulation codes in astrophysics. We would be pleased if this work
helps to stimulate further work in this direction.

\section*{Acknowledgements}  

We thank the referee, Mathieu Schaller, for an insightful and
constructive report that helped to improve the paper.  We are thankful
for discussions and input from Raul Angulo, Federico Marinacci,
Andreas Bauer and Kevin Schaal. We would like to thankfully
acknowledge support by the Max-Planck Computing and Data Facility
(MPCDF) and the Leibniz Supercomputing Centre (LRZ) for providing and
supporting excellent computing infrastructures that were instrumental
for the test simulations of this paper. In particular, we acknowledge
CPU-time awarded through the Gauss Centre for Supercomputing (GCS)
through project-id `pn34mo' on SuperMUC-NG at LRZ.  We also
acknowledge financial support in early phases of the development of
this code through subproject EXAMAG of the Priority Programme 1648
`SPPEXA' of the German Science Foundation, and through the European
Research Council through ERC-StG grant EXAGAL-308037.

\section*{Data availability}  

The code discussed in this study has been made publicly available at
\url{https://wwwmpa.mpa-garching.mpg.de/gadget4}.
Data of specific test simulations can be obtained upon
reasonable request from the corresponding author. 

\bibliographystyle{mn2e}
\bibliography{paper}

\appendix
\onecolumn

\section{Multipole extensions at different order}  \label{secappendixA}

The following expressions spell out the potential approximation in our
FMM scheme (see subsection~\ref{subsecfmm}) at different order $p$:
\begin{eqnarray}
\Phi_{p=2}(\vec{x}) & \simeq & -G\left[ {Q}_0{D}_0 +  \frac{1}{2}  \vec{Q}_2
  \cdot \vec{D}_2  \right.  
-   {Q}_0 \vec{D}_1 \cdot\vec{a} 
 \left. 
+\frac{1}{2}  {Q}_0 \vec{D}_2 \cdot \vec{a}^{(2)} \right],
\end{eqnarray}

\begin{eqnarray}
\Phi_{p=3}(\vec{x})  & \simeq & -G \left[{Q}_0 {D}_0 + \frac{1}{2} \vec{Q}_2
                          \cdot \vec{D}_2 +\frac{1}{6} \vec{Q}_3 \cdot \vec{D}_3
  \right. 
 - ({Q}_0 \vec{D}_1  +\frac{1}{2}\vec{Q}_2 \cdot \vec{D}_3) \cdot \vec{a}
 + \frac{1}{2} {Q}_0\, \vec{D}_2 \cdot \vec{a}^{(2)} 
\left.
- \frac{1}{6}
    {Q}_0 \vec{D}_3\cdot \vec{a}^{(3)}
\right],
\end{eqnarray}

\begin{eqnarray}
\Phi_{p=4}(\vec{x})  & \simeq & -G \left[{Q}_0 {D}_0 + \frac{1}{2} \vec{Q}_2
                          \cdot \vec{D}_2 +\frac{1}{6} \vec{Q}_3 \cdot
                                \vec{D}_3  +\frac{1}{24} \vec{Q}_4 \cdot \vec{D}_4
  \right. 
- ({Q}_0 \vec{D}_1  +\frac{1}{2}\vec{Q}_2 \cdot \vec{D}_3  +\frac{1}{6}\vec{Q}_3 \cdot \vec{D}_4) \cdot \vec{a}
                        + \frac{1}{2} \left ( {Q}_0\, \vec{D}_2   + \frac{1}{2} \vec{Q}_2\, \vec{D}_4     \right)\cdot \vec{a}^{(2)}
\nonumber\\
& & 
 - \frac{1}{6}
    {Q}_0 \vec{D}_3 \cdot \vec{a}^{(3)}  \left.
   + \frac{1}{24}
    {Q}_0 \vec{D}_4 \cdot \vec{a}^{(4)}
\right].
\end{eqnarray}

\begin{eqnarray}
\Phi_{p=5}(\vec{x})  & \simeq & -G \left[{Q}_0 {D}_0 + \frac{1}{2} \vec{Q}_2
                          \cdot \vec{D}_2 +\frac{1}{6} \vec{Q}_3 \cdot
                                \vec{D}_3  +\frac{1}{24} \vec{Q}_4
                                \cdot \vec{D}_4
+\frac{1}{120} \vec{Q}_5 \cdot \vec{D}_5
  \right. 
- ({Q}_0 \vec{D}_1  +\frac{1}{2}\vec{Q}_2 \cdot \vec{D}_3  +\frac{1}{6}\vec{Q}_3 \cdot \vec{D}_4  +\frac{1}{24}\vec{Q}_4 \cdot \vec{D}_5) \cdot \vec{a}
                        \nonumber \\
& & + \frac{1}{2} \left ( {Q}_0\, \vec{D}_2   + \frac{1}{2} \vec{Q}_2\, \vec{D}_4  + \frac{1}{6} \vec{Q}_3\, \vec{D}_5     \right)\cdot \vec{a}^{(2)}
 - \frac{1}{6} \left( 
    {Q}_0 \vec{D}_3  + \frac{1}{2}     \vec{Q}_2 \vec{D}_5 \right)\cdot \vec{a}^{(3)} 
\left.
   + \frac{1}{24}
    {Q}_0 \vec{D}_4 \cdot \vec{a}^{(4)} \right. 
\left.  
- \frac{1}{120}
    {Q}_0 \vec{D}_5 \cdot \vec{a}^{(5)} 
\right].
\end{eqnarray}

\ \\
\noindent Explicit forms for the derivatives of $g(r)$ that enter
equations~(\ref{eqngderiv0})-(\ref{eqngderiv}):
\begin{eqnarray}
g_0(r) &=& g \\
g_1(r) &=& \frac{1}{r} \, g'\\
g_2(r) &=& \frac{1}{r^2} \left(g'' - \frac{g'}{r}\right)\\
g_3(r) &=& \frac{1}{r^3} \left(g''' - 3 \frac{g''}{r} +  3\frac{g'}{r^2}\right)\\
g_4(r) &=& \frac{1}{r^4} \left(g'''' - 6 \frac{g'''}{r} +  15\frac{g''}{r^2}  - 15 \frac{g'}{r^3}\right)\\
g_5(r) &=& \frac{1}{r^5} \left(g''''' - 10 \frac{g''''}{r} +
           45\frac{g'''}{r^2}  - 105 \frac{g''}{r^3} + 105
           \frac{g'}{r^4} \right)\\
g_6(r) &=& \frac{1}{r^6} \left(g'''''' - 15 \frac{g'''''}{r} +
           105\frac{g''''}{r^2}  - 420 \frac{g'''}{r^3} + 945
           \frac{g''}{r^4}  - 945
           \frac{g'}{r^5}  \right)\\
g_7(r) &=& \frac{1}{r^7} \left(g^{(7)} - 21 \frac{g^{(6)}}{r} +
           210\frac{g'''''}{r^2}  - 1260 \frac{g''''}{r^3} + 4725
           \frac{g'''}{r^4}  - 10395
           \frac{g''}{r^5} + 10395 \frac{g'}{r^6} \right)
\end{eqnarray}
In the Newtonian case, we simply have $g_0=1/r$, 
$g_1=-1/r^3$, $g_2=3/r^5$, $g_3=-15/r^7$, $g_4=105/r^9$,
$g_5=-945/r^{11}$, $g_6 = 10395/r^{13}$, and  $g_7 = -135135/r^{15}$.

\ \\
The following expressions for $\vec{D}_6$ and $\vec{D}_7$ give the
next higher order derivatives of the interaction kernel, extending
equations~(\ref{eqngderiv0})-(\ref{eqngderiv}):
\begin{eqnarray}
(\vec{D}_6)_{ijklmn} & = &  g_3 \left[ 
\delta_{mn} \delta_{ij}\delta_{kl} + \delta_{mn} \delta_{jk}\delta_{il} +
                          \delta_{mn} \delta_{ik}\delta_{jl} +  
\delta_{ln} \delta_{ij}\delta_{km} + \ldots \mbox{(15 terms)} \right] 
+ \nonumber\\
         & &   g_4 \, \left [ r_lr_m \left( \delta_{ij}\delta_{kn} + \delta_{jk}\delta_{ik} +
                        \delta_{ik}\delta_{jk} \right)+\ldots
             \mbox{(15 x 3 terms)} \right]  + \nonumber \\
& &      g_5\, \left[  \delta_{ij} \,r_k r_l r_m r_n + \ldots
             \mbox{(15 terms)} \right]  + \nonumber \\
& &      g_6\, r_i r_j r_k r_l r_m r_n
\end{eqnarray}

\begin{eqnarray}
(\vec{D}_7)_{ijklmno} & = &  g_4 \, \left [ r_o
                            \delta_{ij}\delta_{kl}\delta_{mn} +\ldots
             \mbox{(15 x 7 terms)} \right]  + \nonumber \\
& &      g_5\, \left[ r_o r_l r_m (\delta_{ij}\delta_{kn} +
    \delta_{ik}\delta_{jn} + \delta_{in}\delta_{jk}) + \ldots
             \mbox{(35 triples x 3 kronecker terms)} \right]  + \nonumber \\
& &      g_6\, \left[ r_i r_j r_k r_l r_m  \delta_{no}  + \ldots
    (\mbox{21 terms}) \right] +\nonumber \\
& &      g_7\, r_i r_j r_k r_l r_m r_n r_o
\end{eqnarray}

\label{lastpage}

\end{document}